# CEPC

# *Conceptual Design Report*

## Volume II - Physics & Detector

The CEPC Study Group

# CEPC

# *Conceptual Design Report*

## Volume II - Physics & Detector

The CEPC Study Group

# ACKNOWLEDGMENTS

The CEPC Physics and Detector Conceptual Design Report (CDR) was prepared and written by the CEPC Study Group. The study was organized and led by scientists from the Institute of High Energy Physics (IHEP) of the Chinese Academy of Sciences (CAS), and from many universities and other institutes in China and abroad.

This work was supported by the National Key Program for Science and Technology Research and Development of the Ministry of Science and Technology (MOST); Yifang Wang's Science Studio of the Ten Thousand Talents Project; the National Natural Science Foundation of China; the Center for Excellence in Particle Physics, Chinese Academy of Science; the Key Research Program of Frontier Sciences, Chinese Academy of Science; the 1000 Talent Foreign Expert Project of State Administration of Foreign Experts Affairs; the Scientific Innovation Plan, Chinese Academy of Science; the International Partnership Program, Chinese Academy of Sciences; the CAS/SAFEA International Partnership Program for Creative Research Teams; the Innovation Grant of Institute of High Energy Physics of Chinese Academy of Science; the National 1000 Talents Program of China; the Hundred Talent programs of Chinese Academy of Science; the Xie Jialin Fund of Institute of High Energy Physics of Chinese Academy of Science; the Ta-You Wu Memorial Award; the H2020 Project AIDA2020; and the Corporation fund of Wuxi Toly Electric Works Co. Ltd..

We are thankful to the members of the international review committee:

Claudia Cecchi, INFN Perugia, Italy
Mogens Dam, Niels Bohr Institute, Copenhagen, Denmark
Sasha Glazov, DESY, Hamburg, Germany
Christophe Grojean, DESY Hamburg and Humboldt U. Berlin, Germany

Liang Han, University of Science and Technology, China
Tao Han, University of Pittsburgh, USA
Bill Murray, Warwick University and RAL, UK
Maxim Perelstein, Cornell University, USA
Marcel Stanitzki, DESY, Hamburg, Germany
Marcel Vos (chair), IFIC UV/CSIC, Valencia, Spain
Hitoshi Yamamoto, Tohoku University, Sendai, Japan

for their review of this volume of the CEPC CDR during a three-day review meeting on 13-15 September 2018 in Beijing. We are also grateful to Dmitri S Denisov (FNAL), Aleandro Nisati (Roma I), Abe Seiden (UCSC), Walter Snoeys (CERN), Pippa Wells (CERN), and Akira Yamamoto (KEK) for their comments and help reviewing the document.

# EDITOR LIST

**General Editors**
Joao Guimaraes da Costa,[1] guimaraes@ihep.ac.cn,
Yuanning Gao,[2] gaoyn@mail.tsinghua.edu.cn,
Shan Jin,[3] jins@ihep.ac.cn,
Jianming Qian,[13] qianj@umich.edu,
Christopher Tully,[10] cgtully@princeton.edu,
Charles Young,[9] young@slac.stanford.edu

**Chapter 1: Introduction**
Joao Guimaraes da Costa,[1] guimaraes@ihep.ac.cn,
Jianming Qian,[13] qianj@umich.edu

**Chapter 2: Overview of the Physics Case for CEPC**
Lian-Tao Wang,[11] liantaow@uchicago.edu

**Chapter 3: Experimental Conditions, Physics Requirements and Detector Concepts**
Joao Guimaraes da Costa,[1] guimaraes@ihep.ac.cn,
Manqi Ruan,[1] ruanmq@ihep.ac.cn,
Hongbo Zhu,[1] zhuhb@ihep.ac.cn

**Chapter 4: Tracking System**
**4.1: Vertex Detector**
Mingyi Dong,[1] dongmy@ihep.ac.cn,
Yunpeng Lu,[1] yplu@ihep.ac.cn,
Qun Ouyang,[1] ouyq@ihep.ac.cn,
Zhigang Wu,[1] wuzg@ihep.ac.cn

**4.2: Time Projection Chamber and Silicon Tracker**
**4.2.1: Time Projection Chamber**
Zhi Deng,[2] dengz@mail.tsinghua.edu.cn,
Yulan Li,[2] yulanli@mail.tsinghua.edu.cn,
Huirong Qi,[1] qihr@ihep.ac.cn

**4.2.2: Silicon Tracker**
Meng Wang,[4] mwang@sdu.edu.cn

**4.3: Full Silicon Tracker**
Chengdong Fu,[1] fucd@ihep.ac.cn,
Wei-Ming Yao,[12] wmyao@lbl.gov

**4.4: Drift Chamber Tracker**
Franco Grancagnolo,[14] franco.grancagnolo@le.infn.it

**Chapter 5: Calorimetry**
**5.3: Particle flow oriented electromagnetic calorimeter**
Jianbei Liu,[5] liujianb@ustc.edu.cn,
Tao Hu,[1] hut@ihep.ac.cn

**5.4: Particle flow oriented hadronic calorimeter**
Haijun Yang,[6,7] haijun.yang@sjtu.edu.cn

**5.5: Dual-readout calorimeter**
Franco Bedeschi,[16] bed@fnal.gov,
Roberto Ferrari,[15] roberto.ferrari@cern.ch

**Chapter 6: Detector Magnet System**
Wei Zhao,[1] zhaow@ihep.ac.cn,
Zian Zhu,[1] zhuza@ihep.ac.cn

**Chapter 7: Muon Detector System**
Paolo Giacomelli,[17] paolo.giacomelli@cern.ch,
Liang Li,[6,7] liangliphy@sjtu.edu.cn

**Chapter 8: Readout Electronics, Trigger and Data Acquisition**
Fei Li,[1] lifei@ihep.ac.cn,
Zhenan Liu,[1] liuza@ihep.ac.cn,
Christopher Tully,[10] cgtully@princeton.edu,
Kejun Zhu,[1] zhukj@ihep.ac.cn

**Chapter 9: Machine Detector Interface and Luminosity Detectors**
Suen Hou,[19] suen@sinica.edu.tw,
Ivanka Bozovic Jelisavcic,[18] ibozovic@vinca.rs,
Hongbo Zhu,[1] zhuhb@ihep.ac.cn

**Chapter 10: Simulation, Reconstruction and Physics Object Performance**
Gang Li,[1] li.gang@mail.ihep.ac.cn,
Jianming Qian,[13] qianj@umich.edu,
Manqi Ruan,[1] ruanmq@ihep.ac.cn

**Chapter 11: Physics Performance with Benchmark Processes**
**11.1: Higgs boson physics**
Yaquan Fang,[1] fangyq@ihep.ac.cn,
Qiang Li,[8] qliphy0@pku.edu.cn,
Jianming Qian,[13] qianj@umich.edu,
Manqi Ruan,[1] ruanmq@ihep.ac.cn

**11.2: W and Z boson physics**
Maarten Boonekamp,[20] maarten.boonekamp@cea.fr,
Zhijun Liang,[1] zhijun.liang@cern.ch,
Fulvio Piccinini,[15] fulvio.piccinini@pv.infn.it

**Chapter 12: Future Plans and R&D Prospects**
Joao Guimaraes da Costa,[1] guimaraes@ihep.ac.cn,
Xin Shi,[1] shixin@ihep.ac.cn

1. Institute of High Energy Physics, Chinese Academy of Sciences, Beijing
2. Department of Engineering Physics Department, Tsinghua University, Beijing
3. Department of Physics, Nanjing University, Nanjing
4. Institute of Frontier and Interdisciplinary Science and Key Laboratory of Particle Physics and Particle Irradiation, Shandong University, Qingdao
5. University of Science and Technology of China, Hefei
6. Department of Physics and Astronomy, Shanghai Jiao Tong University, Shanghai
7. Tsung-Dao Lee Institute, Shanghai Jiao Tong University, Shanghai
8. School of Physics, Peking University, Beijing
9. SLAC National Accelerator Laboratory, Menlo Park, CA
10. Princeton University, Princeton, NJ
11. Department of Physics, University of Chicago, Chicago, IL
12. Lawrence Berkeley National Labotatory, Berkeley, CA
13. Department of Physics, University of Michigan, Ann Arbor, MI
14. INFN - Sezione di Lecce and University of Lecce
15. INFN - Sezione di Pavia and University of Pavia
16. INFN - Sezione di Pisa, Universita' di Pisa and Scuola Normale Superiore
17. INFN - Sezione di Bologna and University of Bologna
18. Vinca Institute of Nuclear Sciences, University of Belgrade, Belgrade
19. Institute of Physics, Academia Sinica, Taipei
20. IRFU, CEA, Universite Paris-Saclay, Paris

# EXECUTIVE SUMMARY

The discovery of the Higgs boson in 2012 by the ATLAS and CMS Collaborations at the Large Hadron Collider (LHC) at CERN has ushered a new era in particle physics. The Higgs boson has a special role in our quest to answer some of the most profound questions in physics. These questions include the nature of the electroweak phase transition that governed the evolution of the early Universe and why the gravitational force is so weak compared with other forces in nature.

The Higgs boson, with its low mass, can be produced at a circular electron-positron collider of a modest energy. Unlike in proton-proton collisions at the LHC, the study of the Higgs boson in $e^+e^-$ collisions is practically free of systematic uncertainties that limit the measurements at the LHC and its upgrade, the High-Luminosity LHC (HL-LHC). Precise measurements of the Higgs boson properties, along with those of the mediators of the weak interaction, the $W$ and $Z$ bosons, will provide critical tests of the underlying fundamental physics principles of the Standard Model (SM) and are vital in the exploration of new physics beyond the SM (BSM). Such a precision physics program will be a critical component of any worldwide road map of particle physics in the coming decades.

The Circular Electron Positron Collider (CEPC) is a large international scientific facility proposed by the Chinese particle physics community in 2012 to explore the aforementioned physics program. The CEPC, to be hosted in China in a circular underground tunnel of approximately 100 km in circumference, is designed to operate at around 91.2 GeV as a $Z$ factory, at around 160 GeV of the $WW$ production threshold, and at 240 GeV as a Higgs factory. The CEPC will produce close to one trillion $Z$ bosons, 100 million $W$ bosons and over one million Higgs bosons. The vast amount of bottom quarks, charm quarks and $\tau$-leptons produced in the decays of the $Z$ bosons also makes the CEPC an effective $B$-factory and $\tau$-charm factory. The CEPC offers an unmatched opportunity for precision measurements and searches for BSM physics.

The CEPC will measure the Higgs boson properties in greater detail and in a model-independent way, in comparison with the (HL)-LHC. The CEPC will also reach a new level of precision for the measurements of the $W$ and $Z$ bosons properties. An order of magnitude or more improvement in precision is expected for most of the Higgs measurements and many electroweak observables. The clean collision environment of the CEPC will allow the search for potential unknown decay modes that are impractical at the (HL-)LHC. Through these measurements, the CEPC could uncover deviations from the SM predictions and reveal the existence of new particles that are beyond the reaches of direct searches at the current experiments. The precision Higgs boson measurements could potentially reveal crucial physics mechanism that determines the nature of the electroweak phase transition. It will be another milestone in our understanding of the early history of our Universe, and could hold the key to unlock the origin of the matter and anti-matter asymmetry in the Universe. These results could test the ideas that explain the vast difference between the energy scales associated with the electroweak and gravitational interactions.

The CEPC will also search for a variety of new particles. Running as both a Higgs factory and a $Z$ factory, the exotic decays of Higgs and $Z$ bosons are sensitive vehicles for the search of new physics, such as those with light new particles. The dark matter can be searched for through its direct production and its indirect effects on the precision measurements. The CEPC, as $B$ and $\tau$-charm factories, can perform studies that help to understand the origin of different species of matter and their properties. The CEPC is also an excellent facility to perform precise tests of the theory of the strong interaction.

To deliver the physics program outlined above, the CEPC detector concepts must meet the stringent performance requirements. The detector designs are guided by the principles of large and precisely defined solid angle coverage, excellent particle identification, precise particle energy/momentum measurement, efficient vertex reconstruction, and superb jet reconstruction and measurement as well as the flavor tagging. Two primary detector concepts are described, a baseline with two approaches to the tracking systems, and an alternative with a different strategy for meeting the jet resolution requirements. The baseline detector concept incorporates the particle flow principle with a precision vertex detector, a Time Projection Chamber (TPC) and a silicon tracker, a high granularity calorimetry system, a 3 Tesla superconducting solenoid followed by a muon detector embedded in a flux return yoke. A variant of the baseline incorporates a full silicon tracker without the TPC. The alternative concept is based on a precision vertex detector, a drift chamber tracker, a dual readout calorimetry, a 2 Tesla solenoid, and a muon detector. The baseline detector concept has been studied in detail through realistic simulation and the results demonstrate that it can deliver the performance necessary to achieve the physics goals of the CEPC.

To develop the detector concepts into full-scale technical designs for the planned two detectors, a set of critical R&D tasks has been identified. Prototypes of key detector components will be built and tested. Mechanical integration, thermal control and data acquisition schemes must be developed. Industrialization of the detector component fabrication will be pursued. International collaborations will need to be formed before the detector designs can be finalized and the technical design reports can be developed.

The CEPC will be a world-class multifaceted scientific facility for research, education, and international collaboration. It will be a center for discoveries and innovation and a magnet for attracting top scientists from all over the world to work together to understand the fundamental nature of our Universe. The CEPC will also provide leading educational opportunities for universities and research institutions in China and around the world. The CEPC together with its possible upgrade, the Super proton-proton Collider, will firmly place China at the forefront of the cutting-edge research and exploration in fundamental physics for the next half century. Such a facility will have profound impacts on science, economy and society that will reverberate across the world.

This document is the second volume of the CEPC Conceptual Design Report (CDR). It presents the physics case for the CEPC, describes the conceptual detectors and their technological options, highlights the expected detector and physics performance, and discusses future plans for detector R&D and physics investigations. The first volume, recently released, describes the design of the CEPC accelerator complex, its associated civil engineering, and strategic alternative scenarios. A Preliminary Conceptual Design Report (Pre-CDR) was successfully published in March 2015.

# CONTENTS

# CHAPTER 1

# INTRODUCTION

The discovery of the Higgs boson in 2012 by the ATLAS and CMS Collaborations [1, 2] at the Large Hadron Collider (LHC) at CERN has ushered a new era in particle physics. Due to its low mass, the Higgs bosons can be produced in the relatively clean environment of a circular electron-positron collider with a reasonable luminosity at an affordable cost. The Higgs boson is a crucial cornerstone of the Standard Model (SM). It is at the center of the biggest mysteries of modern particle physics, such as the large hierarchy between the weak and Planck scales and the nature of the electroweak phase transition. Precise measurements of the properties of the Higgs boson along with those of the $W$ and $Z$ bosons, will provide critical tests of the underlying fundamental physics principles of the SM. These measurements are also vital in the exploration of physics beyond the SM (BSM). Such a physics program will be a critical component of any road map for particle physics in the coming decades.

The Circular Electron Positron Collider (CEPC) is a large international scientific project initiated by and to be hosted in China. The collider with a circumference of 100 km is designed to operate at center-of-mass energies ($\sqrt{s}$) of 240 GeV (Higgs factory), around 91.2 GeV ($Z$ factory or $Z$ pole), and around 160 GeV ($WW$ threshold scan). It will produce large samples of Higgs, $W$ and $Z$ bosons to allow precision measurements of their properties as well as searches for BSM physics. The CEPC was first presented to the international community at the ICFA Workshop "Accelerators for a Higgs Factory: Linear vs. Circular" (HF2012) in November 2012 at Fermilab [3]. A Preliminary Conceptual Design Report (Pre-CDR) [4, 5] was published in March 2015.

This document is the second volume of the Conceptual Design Report (CDR). The first volume [6], released in July 2018, describes the design of the CEPC accelerator complex, its associated civil engineering, and strategic alternative scenarios. This volume explores the physics potential of the CEPC, presents possible detector concepts and discusses the corresponding R&D program. It describes the main features of the detectors that are required to realize the full physics potential of the CEPC. It aims to demonstrate that a wide range of high-precision physics measurements can be made at the CEPC with detectors that are feasible to construct in the next 12–15 years. The Higgs factory operation is used as a benchmark to illustrate the detector requirements and illuminate the CEPC physics potential. Consideration is also given to the $WW$ threshold and high-rate $Z$ pole operation.

This report consists of 12 chapters. The next chapter presents an overview of the physics case for the CEPC, where the physics potential for both precision measurements and searches for BSM physics is highlighted. Chapter 3 introduces the CEPC accelerator and the experimental environment and outlines the detector requirements that must be met to achieve the CEPC physics goals. This chapter ends with the introduction of the CEPC detector concepts proposed to satisfy these physics requirements. The detector subsystems are then described in detail in the subsequent chapters. Chapter 4 describes the tracking systems including the vertex detectors. Chapter 5 presents the calorimeter options. Chapter 6 outlines the design of the detector solenoid, and Chapter 7 describes the muon system concepts. A summary plan for the readout electronics and data acquisition system is presented in Chapter 8. Results from detailed full simulation and test beam studies are presented when available. The challenging design of the interaction region is described in Chapter 9, together with the beam backgrounds and plans for the luminosity measurement. The overall performance of the CEPC baseline detector concept is presented in Chapters 10 and 11. Chapter 10 introduces the detector software used in the studies and details the physics object performance, taking into account full detector simulation and reconstruction. Chapter 11 demonstrates CEPC's physics potential through selected benchmark physics results. Finally, Chapter 12 ends this report with an overview of plans on future detector R&D and physics studies towards the Technical Design Report.

In what follows we present a short summary of the CDR content, highlighting the CEPC physics case, the collider and experimental environment, the detector concepts, and the detector performance and physics benchmarks.

**Physics Case** The precision measurements of the Higgs boson properties will be a critical component of high energy physics research in the coming decades. The Higgs boson provides a unique sensitive probe of BSM physics which may manifest itself as observable deviations in the Higgs boson couplings relative to the SM expectations. The couplings, and other electroweak physics parameters, can be measured at the CEPC with unprecedented precision. Such measurements can be used to address important open questions of the electroweak symmetry breaking such as the large difference between the electroweak scale and the fundamental Planck scale. The ideas of naturalness has been crucial in seeking solutions of this so-called hierarchy problem. At the CEPC, it is possible to test the ideas of naturalness to an unprecedented level. The precision measurements can be used to probe fine-tuning down to the percent level in the conventional scenarios such as Supersymmetry and Composite Higgs. They are also sensitive to the signals of a range of newly developed ideas. In addition, precision Higgs boson coupling measurements can probe the global structure of the Higgs potential, and reveal the nature of the electroweak phase transition. Understanding the electroweak phase transition will mark another concrete step forward in our knowledge of the early universe, and it could hold the key to solve the problem of the asymmetry between matter and anti-matter in our universe. In addition to the Higgs boson self-coupling, models with the first-order electroweak phase transition generically predict significant deviations in other Higgs boson couplings. An important example is the modification of the coupling of the Higgs boson to the $Z$ boson, which can be measured with a sub-percent level accuracy at the CEPC.

The CEPC can also be used to search for a variety of new particles. Running as both a Higgs factory and a $Z$ factory, the exotic decays of the Higgs and $Z$ bosons can be used to search for new physics, such as those associated with a light dark sector. The CEPC can also search for dark matter, through both its direct production and its indirect effects on the precision electroweak measurements. There is also the possibility of the direct production of a right handed neutrino which will probe a class of see-saw models. Finally, both the direct searches and the indirect measurements can look for signals of a possible extended Higgs sector.

An electron-positron collider is an excellent facility to perform precise QCD measurements to further our understanding of the strong interaction. Possible topics, include the measurement of the strong coupling constant $\alpha_s$, jet and event shapes and their utility in probing Yukawa couplings of light quarks. The CEPC can also produce close to $10^{12}$ $Z$ bosons from which billions of bottom quarks, charm quarks and $\tau$-leptons will be produced in the decays. Hence, the CEPC can be a powerful $B$-factory and $\tau$-charm factory with excellent physics potential. For example, new physics may show up as rare flavor-changing $Z$ boson decays.

**Collider and the Experimental Environment** The CEPC is a double-ring $e^+e^-$ collider with a 100 km circumference and two interaction points (IP). It will operate in three different modes, corresponding to three different center-of-mass energies: Higgs factory at $\sqrt{s} = 240$ GeV for the $e^+e^- \to ZH$ production, $Z$ factory at $\sqrt{s} = 91.2$ GeV for the $e^+e^- \to Z$ production and $WW$ threshold scan at $\sqrt{s} \sim 160$ GeV for the $e^+e^- \to W^+W^-$ production. The instantaneous luminosities are expected to reach $3 \times 10^{34}$, $32 \times 10^{34}$ and $10 \times 10^{34}$ cm$^{-2}$s$^{-1}$, respectively, as shown in Table 1.1. The current tentative operation plan will allow the CEPC to collect one million or more Higgs bosons, close to one trillion $Z$ bosons, and over one hundred million $W^+W^-$ events.

| **Operation mode** | $\sqrt{s}$ **(GeV)** | $L$ **per IP** ($10^{34}$ **cm**$^{-2}$**s**$^{-1}$) | **Years** | **Total** $\int L$ (**ab**$^{-1}$**, 2 IPs**) | **Event yields** |
|---|---|---|---|---|---|
| $H$ | 240 | 3 | 7 | 5.6 | $1 \times 10^6$ |
| $Z$ | 91.2 | 32 (*) | 2 | 16 | $7 \times 10^{11}$ |
| $W^+W^-$ | 158–172 | 10 | 1 | 2.6 | $2 \times 10^7$ (†) |

**Table 1.1:** CEPC operation plan at different center-of-mass energies ($\sqrt{s}$), and corresponding anticipated instantaneous luminosity ($L$), total integrated luminosity ($\int L$) and event yields. (*) The maximum instantaneous luminosity achievable at the $Z$ factory operation is dependent on the detector solenoid magnet field. The value reported here assumes a 2 Tesla solenoid. (†) Additional $9.4 \times 10^7$ $W^+W^-$ events will be produced during the Higgs factory operation.

The detectors will record collisions in beam conditions presented in Table 3.1. Several of these parameters impose important constraints on the detectors. The bunch spacing of the colliding beams differ greatly in the three operational modes (680 ns, 25 ns, and 210 ns, respectively) as do the background levels and event rates. The three most important sources of radiation backgrounds are (1) synchrotron radiation photons from the last bending dipole magnet, (2) $e^+e^-$ pair production following the beamstrahlung process, and (3) off-energy beam particles lost in the interaction region. These backgrounds generate a hit density in the first vertex detector layer ($r = 1.6$ cm) of about 2.2 hits/cm$^2$ per bunch crossing when running at $\sqrt{s} = 240$ GeV and tolerable levels of the total ionizing energy and non-ionizing energy loss. The event rate reaches ~32 kHz for the $Z$ factory operation from the $Z$ boson and Bhabha scattering events.

**Detector Concepts** The CEPC detector concepts are based on the stringent performance requirements needed to deliver a precision physics program that tests the Standard Model and searches for new physics over a wide range of center-of-mass energies and at high beam luminosities. These specifications include large and precisely defined solid angle coverage, excellent particle identification, precise particle energy/momentum measurements, efficient vertex reconstruction, excellent jet reconstruction and flavor tagging.

The physics program demands that all possible final states from the decays of the intermediate vector bosons, $W$ and $Z$, and the Higgs bosons need to be separately identified and reconstructed with high resolution. In particular, to clearly discriminate the $H \to ZZ^* \to 4j$ and $H \to WW^* \to 4j$ final states, the energy resolution of the CEPC calorimetry system for hadronic jets needs to be pushed quite beyond today's limits. The $H \to \gamma\gamma$ decay and the search for $H \to \mathrm{invisible}$ decays impose additional requirements on energy and missing energy measurement resolutions. To measure the coupling of the Higgs boson to the charm quark, the CEPC detectors are required to efficiently distinguish $b$-jets, $c$-jets, and light jets from each other. To achieve excellent sensitivity for the $H \to \mu^+\mu^-$ decay the the momentum resolution is required to achieve a per mille level relative accuracy. The latter two requirements drive the performance of the vertex detector and tracking systems.

Two primary detector concepts were studied, a baseline detector concept with two approaches to the tracking systems, and an alternative detector concept with a different strategy for meeting the jet resolution requirements. The baseline detector concept incorporates the particle flow principle with a precision vertex detector, a Time Projection Chamber, a silicon tracker, a 3 Tesla solenoid, and a high granularity calorimeter followed by a muon detector. A variant of the baseline detector concept incorporates a full silicon tracker. An alternative detector concept is based on dual readout calorimetry with a precision vertex detector, a drift chamber tracker, a 2 Tesla solenoid, and a muon detector. The different technologies for each detector subsystem are being pursued actively with R&D programs and provide many opportunities to leverage leading advances in detector development in the coming years.

**Performance and Physics Benchmarks** Precise measurements of the Higgs boson properties and the electroweak observables at the CEPC place stringent requirements on the performance of the CEPC detectors to identify and measure physics objects such as leptons, photons, jets and their flavors with high efficiencies, purities, and precision. The performance of the CEPC baseline detector concept have been investigated with full simulation. Electrons and muons with momenta above 2 GeV and unconverted photons with energies above 5 GeV can be identified with efficiencies of nearly 100% and with negli-gible backgrounds. Jets from Higgs, $W$, and $Z$ boson decays can be measured with an energy resolution of 3–5%, allowing an average of $2\sigma$ or better separation of hadronic decays of these bosons. Heavy-quark jets can be tagged with unprecedented efficiencies and purities. $K^{\pm}$ with momenta up to 20 GeV can be distinguished from $\pi^{\pm}$ with a significance better than $2\sigma$. These performance results can be further improved with more optimizations and better calibrations.

Many new physics models predict deviations of Higgs boson couplings to particles at the sub-percent level, beyond those achievable at the (HL-)LHC. The CEPC complements the LHC and will be able to study the properties of the Higgs boson in great detail with unprecedented precision. With over $10^6$ Higgs bosons produced, most of the relevant Higgs boson couplings can be measured with precision at a percent level or better, in particular the coupling to the $Z$ boson can be determined with a relative precision of 0.25%. More importantly, the CEPC will be able to measure many of the key Higgs boson properties such as the decay branching ratios and hence the total width in a model-independent way. The clean collision environment of the CEPC will allow the search for potential unknown decay modes that are impractical at the LHC.

The CEPC will reach a new level of precision for the measurements of the properties of the $W$ and $Z$ bosons. With samples of $10^8$ $W$ bosons and $10^{12}$ $Z$ bosons at the CEPC, an order of magnitude improvements in precision are expected for many electroweak observables. Precise measurements of the $W$ and $Z$ boson masses, widths, and couplings are critical to test the consistency of the SM. These measurements could discover deviations from the SM predictions and reveal the existence of new particles that are beyond the reaches of the direct searches at the current experiments. These new particles are predicted by many extensions of the SM.

This report provides a snapshot of the current studies, many of them are ongoing and more analyses are needed to fully explore the physics potential of the CEPC. Nevertheless, the performance results presented have either already met or are close to meet the requirements of the CEPC experiments. Studies of physics benchmark processes suggest that the CEPC has the potential to characterize the Higgs boson, similarly to what LEP did to the $Z$ boson, and significantly improve the precision of electroweak measurements, ultimately shedding light on potential new physics.

**Future Plans** The CEPC construction is expected to start in 2022 and be completed in 2030, followed by the commissioning of the accelerator and detectors. A tentative operational plan covers ten years of physics data: seven years for the Higgs boson physics, two years at the $Z$ pole and one year for the $WW$ threshold scan. Prior to the construction, there will be a five-year R&D period (2018–2022). During this period, two international collaborations will be formed to produce Technical Design Report, build, and operate two large experiments. Prototypes of key-technical detector components will be built, and worldwide infrastructure established for industrialization and manufacturing of the required components.

The CEPC is an important part of the world plan for high-energy particle physics research. It will support a comprehensive research program by scientists from all over the world and provide leading educational opportunities for universities and research institutes in China and around the world. Physicists from many countries will work together to explore the science and technology frontiers, and to bring a new level of understanding of the fundamental nature of matter, energy and the universe.

**CHAPTER 2**

# OVERVIEW OF THE PHYSICS CASE FOR CEPC

After a very brief summary of the projections of the precision of Higgs coupling and electroweak measurements, the next part of this chapter describes the potential of using these measurements to address important open questions of the electroweak symmetry breaking. First, the most important question about the electroweak symmetry is to explain the size of the weak scale, which is much smaller than some of the fundamental scales, such as the Planck scale. The idea of naturalness (Section 2.2.1) has been crucial in constructing solutions of this so called hierarchy problem. At the CEPC, it is possible to test the idea of naturalness to an unprecedented level. They can be used to probe fine-tuning down to the percent level in the conventional scenarios such as Supersymmetry and Composite Higgs. They are also sensitive to the signals of a range of newly developed ideas, from the neutral naturalness to the relaxion. In addition, precision Higgs coupling measurements can help probing the global feature of the Higgs potential, and reveal the nature of the electroweak phase transition. Understanding the electroweak phase transition (Section 2.2.2) marks another concrete step forward in our knowledge of the early universe, and it could hold the key to the problem of asymmetry between matter and anti-matter in our universe. It is argued that in addition to the triple Higgs coupling, models with first order electroweak phase transition generically predict significant deviations in other Higgs couplings. An important example is the modification of the coupling of the Higgs boson to the $Z$ boson which can be measure with sub-percent level accuracy at the CEPC. This is demonstrated in a representative scenario, a singlet extension of the Higgs sector.

The CEPC can also search for a variety of new physics. Section 2.3 contains a set of such examples. Running as both a Higgs factory and a $Z$-factory, the exotic decays of the Higgs boson and the $Z$ boson can be used to search for new physics, such as those associated with a light dark sector. CEPC can also search for dark matter (Section 2.3.3), both through direct production and through its indirect effects on electroweak precision measurements. There is also an exciting possibility of direct producing the right handed neutrino and probing a class of see-saw models (Section 2.3.4). Finally, both direct search and indirect measurement can look for signals of a possible extended Higgs sector.

A lepton collider is an excellent place to perform precise Quantum ChromoDynamics (QCD) measurement, and further our understanding of the strong interaction. Possible topics, covered in Section 2.4, include measurement of $\alpha_s$, jet, event shapes and their utility in probing light Yukawa couplings, are summarized in the next part of the section.

The CEPC can produce close to $10^{12}$ $Z$ bosons. Hence, it can be a powerful $B$-factory and $\tau$-factory with excellent potential. At the same time, new flavor physics may show up as rare flavor violating $Z$ decays. Section 2.5 is an overview of such physics, and provides an estimate of the potential of the CEPC.

## 2.1 CEPC: THE PRECISION FRONTIER

The discovery of a Higgs boson in 2012 by the ATLAS and CMS collaborations [1, 2] at the Large Hadron Collider (LHC) has opened a new era in particle physics. Subsequent measurements of the properties of this new particle have indicated compatibility with the predictions of the SM. While the SM has been remarkably successful in describing experimental phenomena, it is important to recognize that the SM is not a complete theory. In particular, the SM does not *predict* the parameters of the Higgs potential, such as the Higgs boson mass. The vast difference between the Planck scale and the weak scale remains a major mystery. In addition, there is not a complete understanding of the nature of the electroweak phase transition. The discovery of a spin zero Higgs boson, the first elementary particle of its kind, has only sharpened these questions, and their resolution will necessarily involve new physics beyond the SM. In this respect, the discovery of the Higgs boson marks the beginning of a new era of theoretical and experimental exploration.

The precision measurement of Higgs boson properties will be a critical component of any road map for high energy physics in the coming decades. In addition to motivating new physics beyond the SM, the Higgs boson provides a uniquely sensitive probe of new physics. In particular, new physics beyond the SM can lead to observable deviations in Higgs couplings relative to SM expectations. These deviations $\delta$ are generically of order

$$\delta = c \frac{v^2}{M_{\rm NP}^2} , \qquad (2.1)$$

where $v$ and $M_{\rm NP}$ are the vacuum expectation value of the Higgs field and the typical mass scale of new physics, respectively. The size of the proportionality constant $c$ is model-dependent, but it should not be much larger than $\mathcal{O}(1)$. The current and upcoming LHC runs will measure Higgs couplings to about the 5% level [3], while direct searches at the LHC will test many new physics scenarios from a few hundreds of GeV to at least a TeV. Equation (2.1) implies that probing new physics significantly *beyond* the LHC's reach requires measuring Higgs couplings with sub-percent-level accuracy. Achieving such a level of precision will require new facilities, for which a lepton collider operating as a Higgs factory is a natural candidate.

In this section we explore the physics potential of the CEPC, translating the potential precision of Higgs coupling measurements into implications for a variety of motivated scenarios for physics beyond the SM. Projections for the precision in Higgs coupling mea-surements and electroweak observables attainable by the CEPC are summarized below. The details of the analysis underpinning these projections are presented in Section 11.1

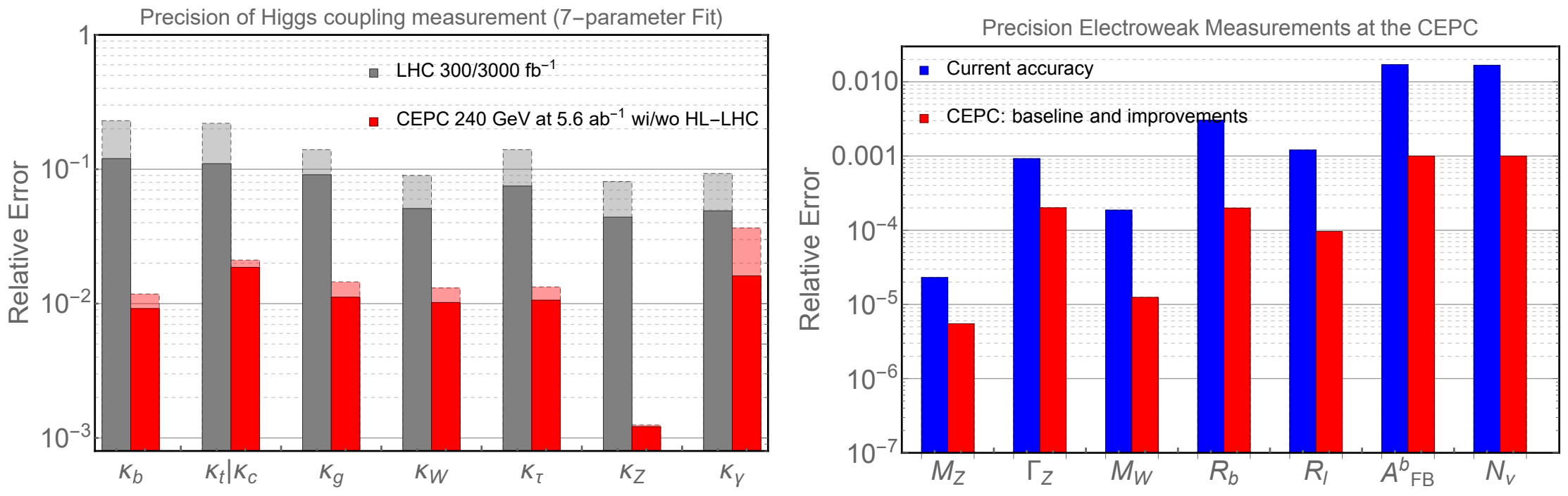

(a) (b)

**Figure 2.1:** (a) Higgs coupling extraction in the $\kappa$-framework. (b) Projection for the precision of the $Z$-pole measurements.

The CEPC will operate primarily at a center-of-mass energy of $\sqrt{s} \sim 240$ GeV. The main mode of Higgs boson production is through $e^+e^- \to ZH$ process, and with an integrated luminosity of $5.6\,\text{ab}^{-1}$, over one million Higgs bosons will be produced. At CEPC, in contrast to the LHC, Higgs boson candidate events can be identified through a technique known as the recoil mass method without tagging its decay products. This allows Higgs boson production to be disentangled from Higgs boson decay in a model-independent way. Moreover, the cleaner environment at a lepton collider allows much better exclusive measurement of Higgs boson decay channels. All of these give CEPC impressive reach in probing Higgs boson properties. The resulting precision attainable by CEPC in measurements of Higgs couplings is shown in the left panel of Figure 2.1(a) in terms of the $\kappa$ framework [4]. The results can be further improved by including additional measurements. For example, $\kappa_Z$ and $\kappa_W$ would be tightly constrained to be very close to each other by the electroweak precision measurements.

Several aspects of the precision attainable at CEPC stand out. The CEPC will be able to measure the Higgs coupling to the $Z$ boson with an accuracy of $0.25\%$[1], about a factor of 10 better than the reach of the High Luminosity upgrade of the LHC (HL-LHC). Such a precise measurement gives CEPC unprecedented reach into interesting new physics scenarios which are very difficult to probe at the LHC. The CEPC also has strong capability in detecting invisible decays of the Higgs boson. For example, with $5.6\,\text{ab}^{-1}$, it can improve the accuracy of the measurement of the Higgs boson invisible branching ratio to $0.3\%$, also more than 10 times better than the projected precision achievable by the HL-LHC. In addition, it is expected to have excellent sensitivity to exotic decay channels which are swamped by backgrounds at the LHC. It is also important to stress that an $e^+e^-$ Higgs factory can perform *model independent* measurement of the Higgs boson width. This unique feature in turn allows for the determination of the Higgs couplings without assumptions about Higgs boson decay channels.

[1]This is the result from a 10-parameter fit. In particular, it includes the Higgs boson width as a free parameter. The result shown in Figure 2.1 is from a more constrained 7-parameter fit. See Section 11.1 for a full set of results and more detailed explanations.

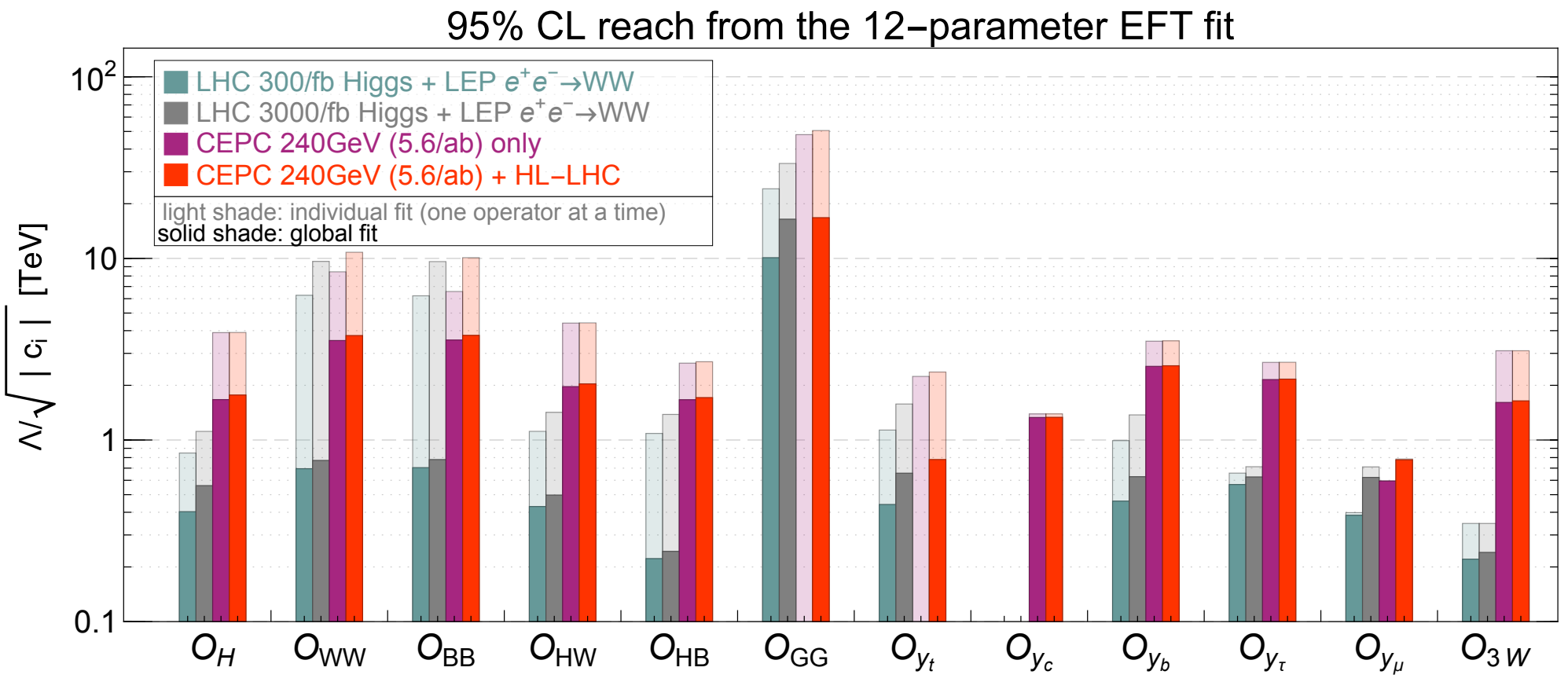

(a)

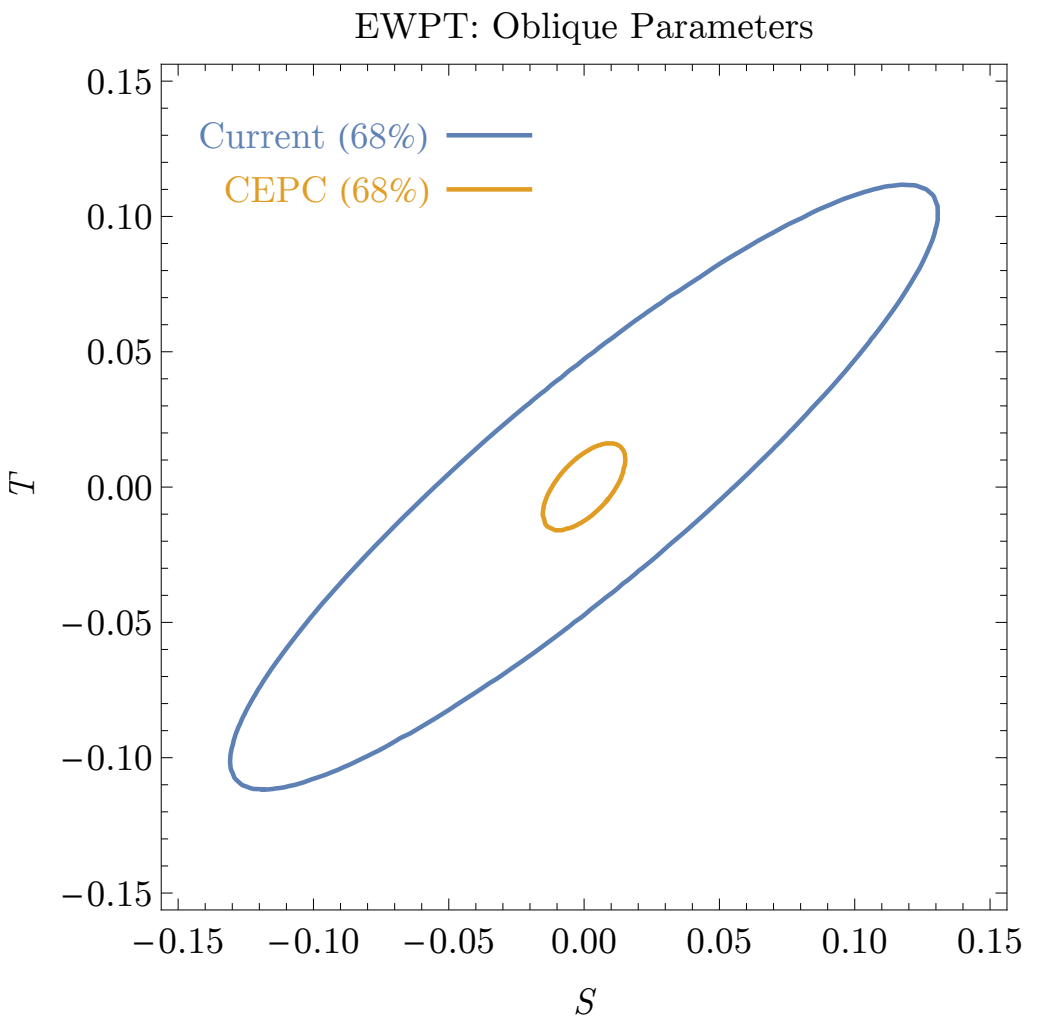

(b)

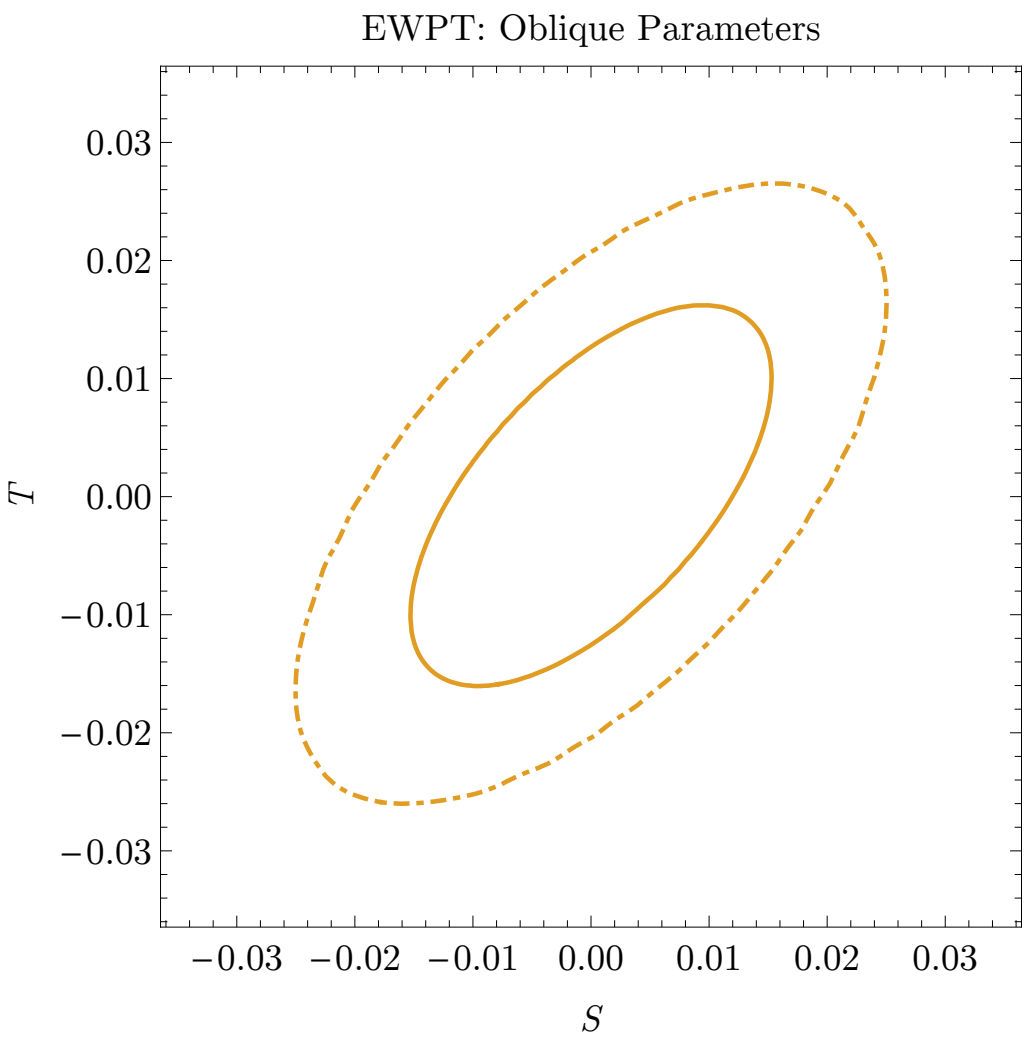

(c)

**Figure 2.2:** (a) The reach of the Higgs measurement on the size of effective field theory operators, normalized as $c_i(\mathcal{O}_i/v^2)$. (b) the CEPC limit on the oblique parameters in comparison with the current precision. (c) 68% (dash-dot) and 95% (solid) contours from CEPC measurement.

The CEPC is also designed to run at the $Z$ pole (with about $10^{12}$ $Z$ bosons) and near the $W^+W^-$ threshold (with about $10^7$ $W$ pairs). This enables a robust program of electroweak precision measurements to complement the Higgs precision program. The projected precision for a set of such observables is shown in on the Figure 2.1(b). CEPC can improve the current precision by about one order of magnitude.

The combination of precision Higgs and electroweak measurements at CEPC is particularly powerful. This is most readily apparent in the potential for CEPC to constrain departures from the Standard Model parametrized in the language of Effective Field Theory (EFT). The reach of CEPC Higgs measurements in constraining Wilson coefficients of select dimension-6 operators in the SM EFT is shown in Figure 2.2, while the reach of CEPC electroweak precision measurements in terms of the so-called oblique parameters (likewise expressible in terms of Wilson coefficients of dimension-6 operators in the SM EFT) is shown in the lower panel of Figure 2.2. The significant improvement of CEPC relative to both current and projected LHC measurements is apparent. Later in this sec-tion, we will explore in detail the implications of the precision measurements at CEPC for important open questions of the Standard Model.

CEPC, running as both Higgs factory and $Z$-factory, will also probe interesting new physics, offer an excellent opportunity of studying flavor physics, and allow precise QCD measurements. We will also elaborate on these later in this section. To set the stage, we briefly comment on the CEPC operation scenarios assumed in the results presented in this section. While the plan for the Higgs factory has been fixed, the plan for the $Z$-factory run is still preliminary. The total number of $Z$s with different scenarios is in the range of $10^{11-12}$. To give an characterization of the full potential of the CEPC, we will use $10^{12}$ $Z$s (Tera $Z$) in our estimates.

## 2.2 HIGGS BOSON AND ELECTROWEAK SYMMETRY BREAKING

### 2.2.1 NATURALNESS

The appearance of large numerical hierarchies in fundamental theories has long been a source of discomfort, articulated in the modern era by Dirac [5] and subsequently refined in the context of quantum field theory by Wilson [6], Susskind [7], 't Hooft [8], and others. In the context of quantum field theory, dimensionless parameters of a quantum field theory are naturally expected to be $\mathcal{O}(1)$, while the dimensional parameters are naturally the size of the fundamental scale at which the theory is defined. An exception arises when a symmetry is manifested in the limit that a parameter of the theory is taken to zero. In this case, it is "technically natural" for some parameters to remain smaller than others, in the sense that they are protected from large quantum corrections, though even in this case one is left to find an explanation for the dynamical origin of the small parameter. This notion of naturalness has been reinforced by the widespread successes of effective field theory and diverse realizations in both particle physics and condensed matter physics.

Famously, all of the observed parameters of the Standard Model satisfy the naturalness criterion in some form, with the exception of the Higgs boson mass parameter and the strong $CP$ angle. The naturalness of these parameters remains an open question, and in each case a natural explanation entails a significant extension of the Standard Model. Of these, the naturalness of the weak scale is perhaps the most pressing, as it is drawn into sharp relief by the discovery of an apparently elementary Higgs boson at the LHC. Evidence for a natural explanation for the value of the weak scale has yet to appear, with null results across a suite of experimental searches imperiling many preferred candidates. But the LHC is not capable of decisively deciding the naturalness of the weak scale, insofar as there are completely natural theories for the weak scale that may be consistent with the full LHC data set. This provides strong motivation for colliders that complement LHC sensitivity to natural new physics.

The oft-cited quadratically divergent radiative corrections to the Higgs boson mass parameter,

$$\delta m_H^2 \sim \frac{3y_t^2}{8\pi^2}\Lambda^2 \,, \tag{2.2}$$

where $\Lambda$ is a cut-off, are not the naturalness problem in and of themselves, but rather an indication of the problem. Such divergences indicate that the Higgs boson mass parameter is precisely that – a parameter – and incalculable in the Standard Model. But the robust expectation is that the Higgs boson mass and other parameters of the Standard Model are fully calculable in a fundamental theory. In this case, the quadratically divergent contributions to the Higgs boson mass parameter in the Standard Model are replaced by finite contributions dictated by the fundamental theory. The Higgs boson mass in terms of underlying parameters will take the form

$$m_H^2 = a\Lambda_h^2 + b\frac{3y_t^2}{8\pi^2}\Lambda_h^2 + \dots \tag{2.3}$$

where $a, b, \dots$ are dimensionless constants and $\Lambda_h$ is an underlying mass scale of the fundamental theory. If the Higgs boson mass is *natural*, the parameters $a$ and $b$ will be $\mathcal{O}(1)$, up to possible manifestations of technical naturalness associated with symmetries in the underlying theory. In this case, one expects $m_H \sim \Lambda_h$, corresponding to the appearance of new physics near the weak scale. Alternatively, $m_H \ll \Lambda_h$ points either to *fine-tuning* among fundamental parameters, or to a correlation between ultraviolet and infrared aspects of the theory with no known counterpart in effective field theory.

The most promising strategy for rendering the weak scale natural in a more fundamental theory is to extend the Standard Model to include additional symmetries that render the Higgs boson mass parameter technically natural. In four dimensions, the available symmetries are supersymmetry and global symmetry. In the former case, the fields of the Standard Model are extended into complete supersymmetric multiplets, and supersymmetry is softly broken to accommodate the non-degeneracy of Standard Model fields and their partners [9–11]. The Higgs boson is related to a fermionic partner, thereby rendering the Higgs boson mass technically natural by the same chiral symmetries that protect the fermion masses. In the latter case, the Higgs boson is a pseudo-Nambu-Goldstone boson (pNGB) of a spontaneously broken global symmetry, with a mass parameter protected by the corresponding shift symmetries. The scale of global symmetry breaking in such theories must itself be rendered natural, leading to e.g. composite Higgs models [12] and little Higgs models [13] (for an excellent recent review, see [14]).

In both cases, these symmetries predict an abundance of new physics near the weak scale. Although this new physics may be searched for efficiently at the LHC, such searches typically leverage ancillary properties of the new physics unrelated to the naturalness of the weak scale. For example, searches for the scalar top partners predicted by supersymmetry typically leverage QCD quantum numbers of the stop and decay modes unrelated to the stop-Higgs coupling. That is to say, these searches generally exploit both production modes and decay modes for top partners that have nothing to do with the top partner's role in stabilizing the Higgs potential. The sensitivity of LHC searches to inessential features of the new physics makes them imperfect probes of electroweak naturalness.

In this respect, a Higgs factory provides the ideal context for probing natural new physics via precision Higgs couplings. The same couplings and diagrams that control the size of the Higgs boson mass in a natural theory generate radiative corrections to its couplings. As such, precision tests of Higgs boson properties directly probe natural physics in a way that is complementary to, and less subject to caveats than, direct searches at the LHC.

Signatures of natural new physics in precision Higgs boson measurements take a variety of forms. In most symmetry solutions, there are Higgs coupling deviations due to tree-level mixing with additional Higgs-like states. However, these tree-level deviations need not be the leading effect. Radiative corrections are also significant, due to both the size of Higgs couplings and the proximity of new particles to the weak scale. In theories where new physics associated with naturalness carries Standard Model quantum numbers, such as conventional supersymmetric and composite models, the most distinctive radiative corrections modify loop-induced Higgs couplings to gluons and photons. In addition, all symmetry solutions – whether or not they involve new states charged under the Standard Model – radiatively modify Higgs couplings through effective wavefunction renormalization of the physical Higgs scalar, an effect that may be observed in loop-level corrections to tree-level Higgs couplings.

Although our discussion of naturalness has focused on symmetries, they are not the only mechanism for explaining the value of the weak scale. The most notable alternative is to lower the cutoff in Equation (2.3), the avenue realized by technicolor [7, 15] and large [16, 17] or warped [18, 19] extra dimensions. However, these solutions typically do not predict a significant mass gap between the Higgs boson and additional degrees of freedom, making them more susceptible to LHC null results. More recent proposals, such as relaxation of the weak scale [20], can potentially lead to $m_H \ll \Lambda_h$ without fine-tuning, and remain interesting targets for exploration. Nonetheless, these alternatives still involve new particles coupling to the Higgs boson, and may leave their imprint on Higgs couplings or exotic decays.

## SUPERSYMMETRY

Supersymmetric extensions of the Standard Model have the virtue of rendering the weak scale natural with an elementary Higgs scalar, consistent with properties observed thus far at the LHC. While searches for supersymmetric partner particles at the LHC have excluded large regions of the natural supersymmetric parameter space, significant blind spots remain that are best covered by precision Higgs coupling measurements.

**Tree-level modifications to Higgs boson properties** Supersymmetric extensions of the Standard Model necessitate more than one Higgs doublet. Mass mixing between the $CP$-even neutral Higgs scalars leads to tree-level deviations in Higgs properties. In the limit that the additional Higgs scalars are heavy and may be integrated out, this leads to dimension-six operators that shift Higgs couplings to fermions and dimension-eight operators that shift Higgs couplings to massive vectors. As a result, deviations are largest in Higgs couplings to fermions, particularly those in the down quark and lepton sectors. Percent-level CEPC sensitivity to modifications of the Higgs coupling to bottom quark enables indirect tests of the Minimal Supersymmetric Standard Model (MSSM) Higgs sector to the TeV scale, as illustrated in Figure 2.3. More broadly, CEPC sensitivity to tree-level effects in extended Higgs sectors such as the MSSM is studied comprehensively in [21]. However, due to the decoupling properties of the MSSM Higgs sector, heavy Higgs states may remain above the TeV scale without increasing the fine-tuning of the weak scale. In this respect, tree-level modifications to Higgs properties arising in supersymmetric theories represent a discovery opportunity but not an irreducible constraint.

**Loop-level modifications to Higgs properties** The plethora of new partner particles predicted by supersymmetric extensions of the Standard Model leads to a wealth of loop-level contributions to Higgs couplings. These contributions are typically largest in the stop sector, due to the large coupling to the Higgs boson required by supersymmetry, but may be significant for any of the partners of third-generation fermions. The most distinctive consequences are modifications to the loop-level Standard Model couplings of the Higgs boson to gluons and photons, though radiative corrections to tree-level couplings arise as well and may be used to cover blind spots arising in the loop-level couplings. The potential for CEPC to probe a suite of loop-level corrections to Higgs and electroweak observables in supersymmetric models is comprehensively studied in [24].

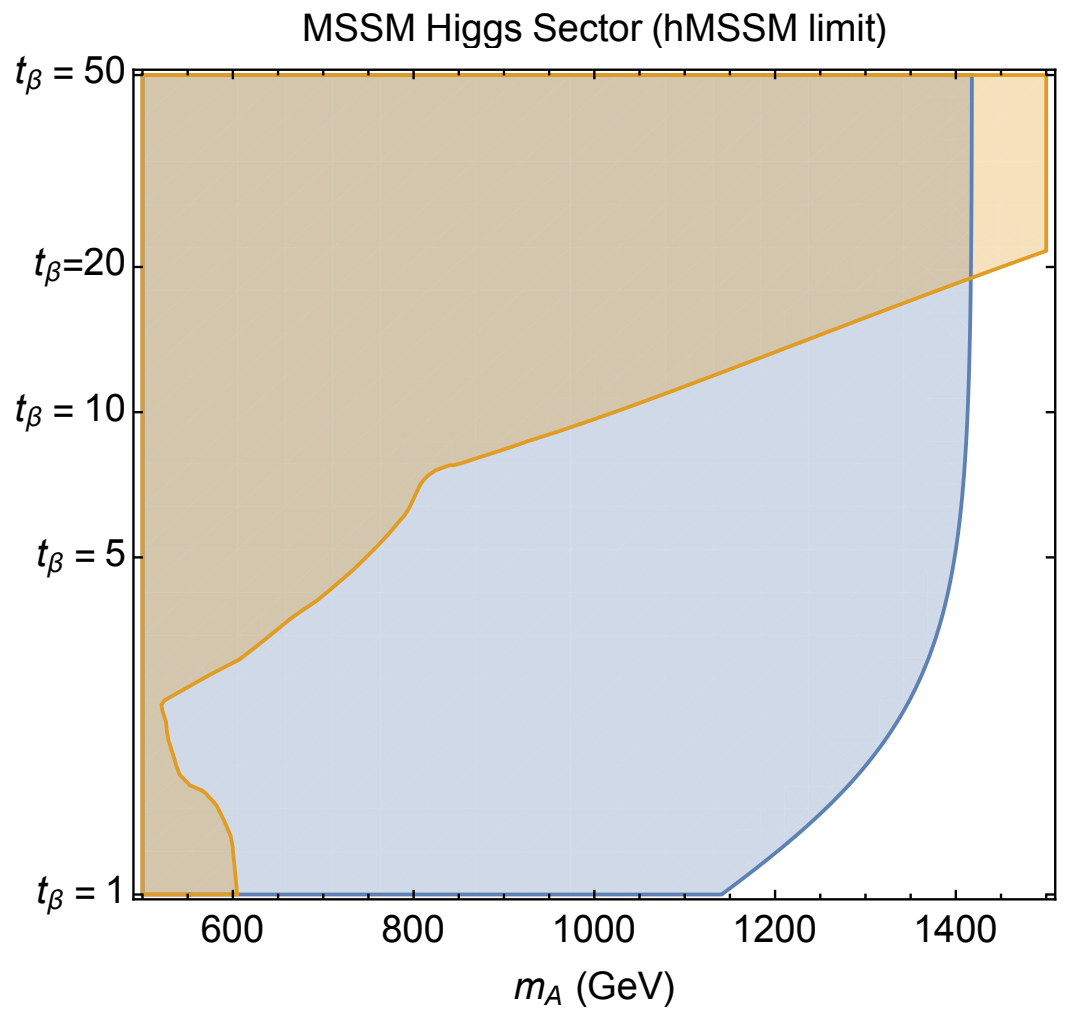

**Figure 2.3:** Potential coverage of the MSSM Higgs sector in the hMSSM limit [22] at CEPC is shown in blue. Sensitivity is driven largely by modifications of the Higgs coupling to bottom quarks, which are of the form $\kappa_b \sim 1 + 2m_Z^2/m_A^2$ at large $\tan\beta$. Projected HL-LHC coverage of the MSSM Higgs sector in the same limit due to direct searches for heavy Higgs states is shown in orange [23], for which sensitivity improves with $\tan\beta$ due to the growth of the heavy Higgs couplings to the Standard Model.

For simplicity, here we will focus on the loop-level consequences in the stop sector, corresponding to the scalar partners of both the right-handed and left-handed top quarks. In the limit that the stops are significantly heavier than the Higgs boson, the correction to gluons and photons is proportional to

$$\frac{1}{4}\left(\frac{m_t^2}{m_{\tilde{t}_1}^2} + \frac{m_t^2}{m_{\tilde{t}_2}} - \frac{m_t^2 X_t^2}{m_{\tilde{t}_1}^2 m_{\tilde{t}_2}^2}\right) \tag{2.4}$$

where $m_{\tilde{t}_1}, m_{\tilde{t}_2}$ are the stop mass eigenstates and $X_t = A_t - \mu \cot\beta$ is the off-diagonal mixing parameter in the stop mass matrix. The mixing parameter is bounded from above by the avoidance of tachyonic stops, and from below by precision measurements of the Higgs coupling to gluons and photons. A robust bound may be placed on the stop sector whenever the minimum value exceeds the maximum value [25]. The strongest constraints arise in the degenerate limit when $m_{\tilde{t}_1} = m_{\tilde{t}_2}$, in which case CEPC is capable of probing stop masses close to the TeV scale, with significant reach even away from equality; this is illustrated in the Figure 2.4(a). However, the modification of Higgs couplings is highly sensitive to the mixing in the stop sector, and blind spots arise when the mixing leads to vanishing deviations in the Higgs coupling to gluons and photons [24, 26]. However, as illustrated in the Figure 2.4(b), these blind spots may be covered by precision measurements of the $ZH$ cross section, which is sensitive to loop-level corrections to the tree-level $HZZ$ coupling that are generically nonzero in the gluon/photon blind spot [26].

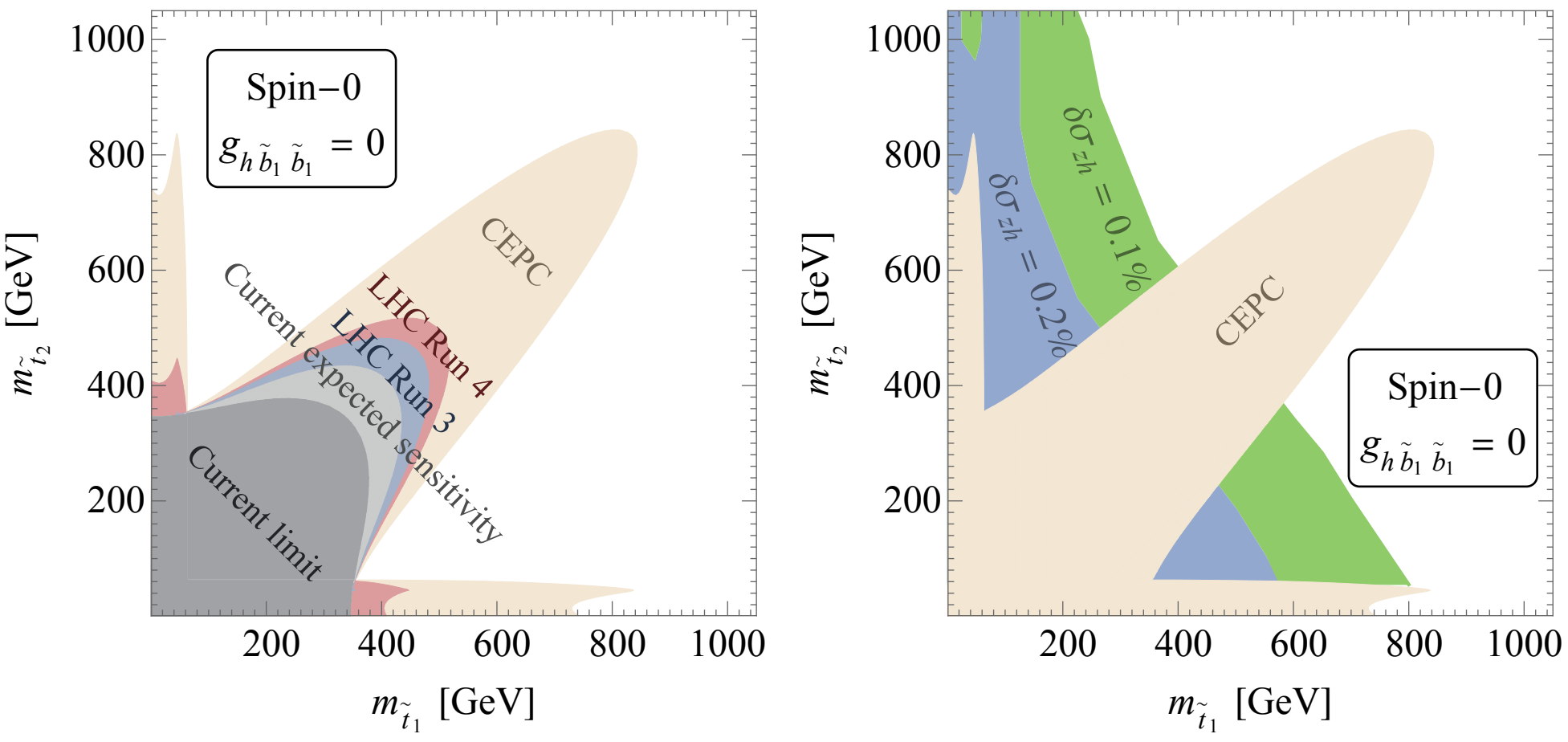

(a) (b)

**Figure 2.4:** (a) LHC and CEPC precision Higgs constraints in the $m_{\tilde{t}_1} - m_{\tilde{t}_2}$ plane from Higgs couplings to gluons and photons. Here $\tan\beta = 1$ and the mixing parameter $X_t$ is allowed to vary over all values consistent with the physical stop masses; the excluded area is that for which no allowed value of $X_t$ is consistent with Higgs coupling measurements. Larger values of $\tan\beta$ lead to qualitatively similar coverage. (b) Coverage of blind spots including precision measurement of the $ZH$ cross section. Figures adapted from [27].

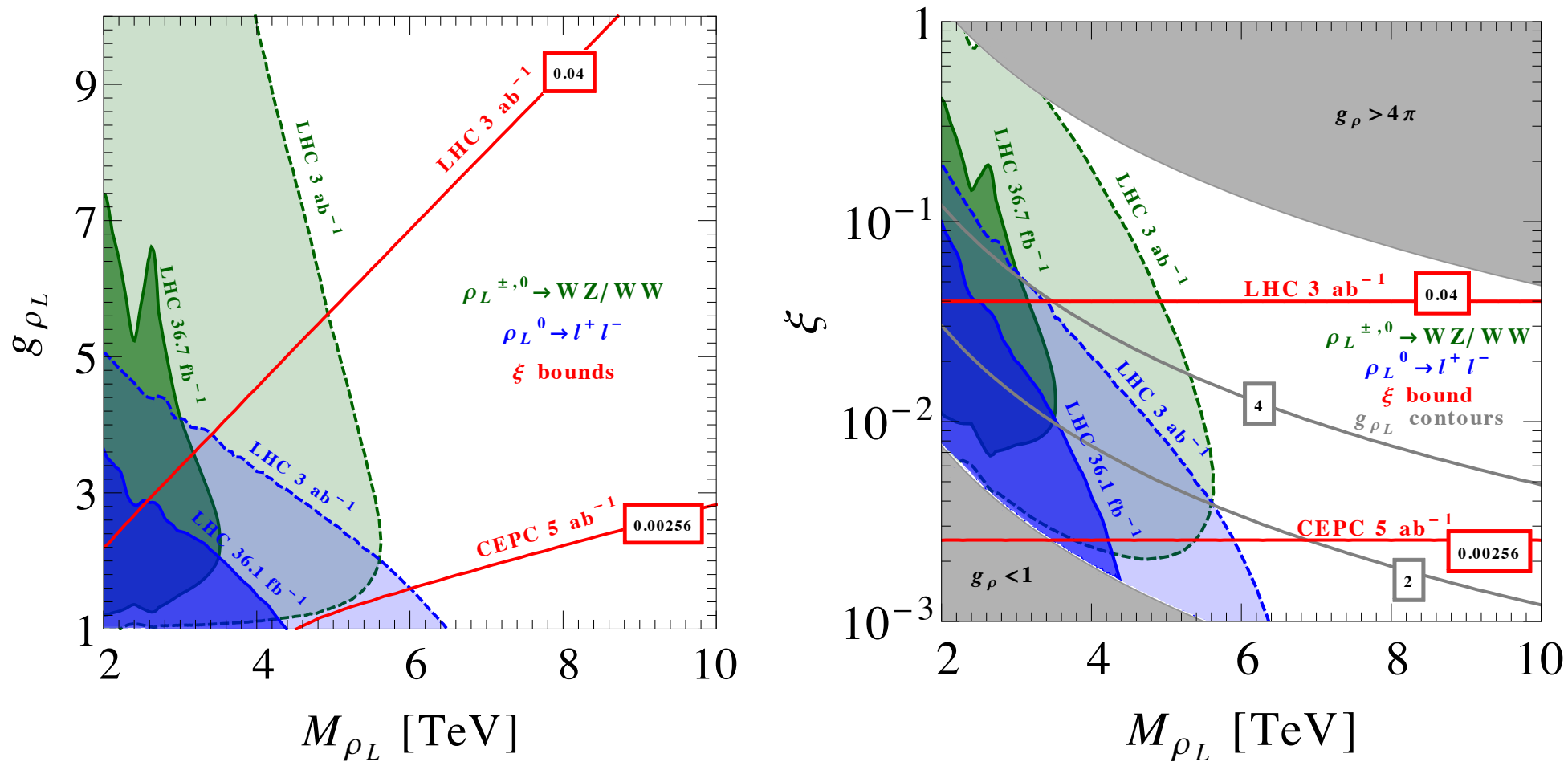

(a) (b)

**Figure 2.5:** Potential coverage of composite-type global symmetry models in terms of resonance mass $m_\rho$ and coupling parameter $g_{\rho_L}$ (a) or mixing parameter $\xi \equiv v^2/f^2$ (b) via direct searches at the LHC (blue and green shaded regions) and precision Higgs measurement constraints (red lines).

## GLOBAL SYMMETRY

Global symmetry approaches to the weak scale cover a vast array of specific models and UV completions, but share the common features of an approximately elementary Standard Model-like Higgs boson mixing with heavier resonances and further influenced by the presence of light fermionic excitations.

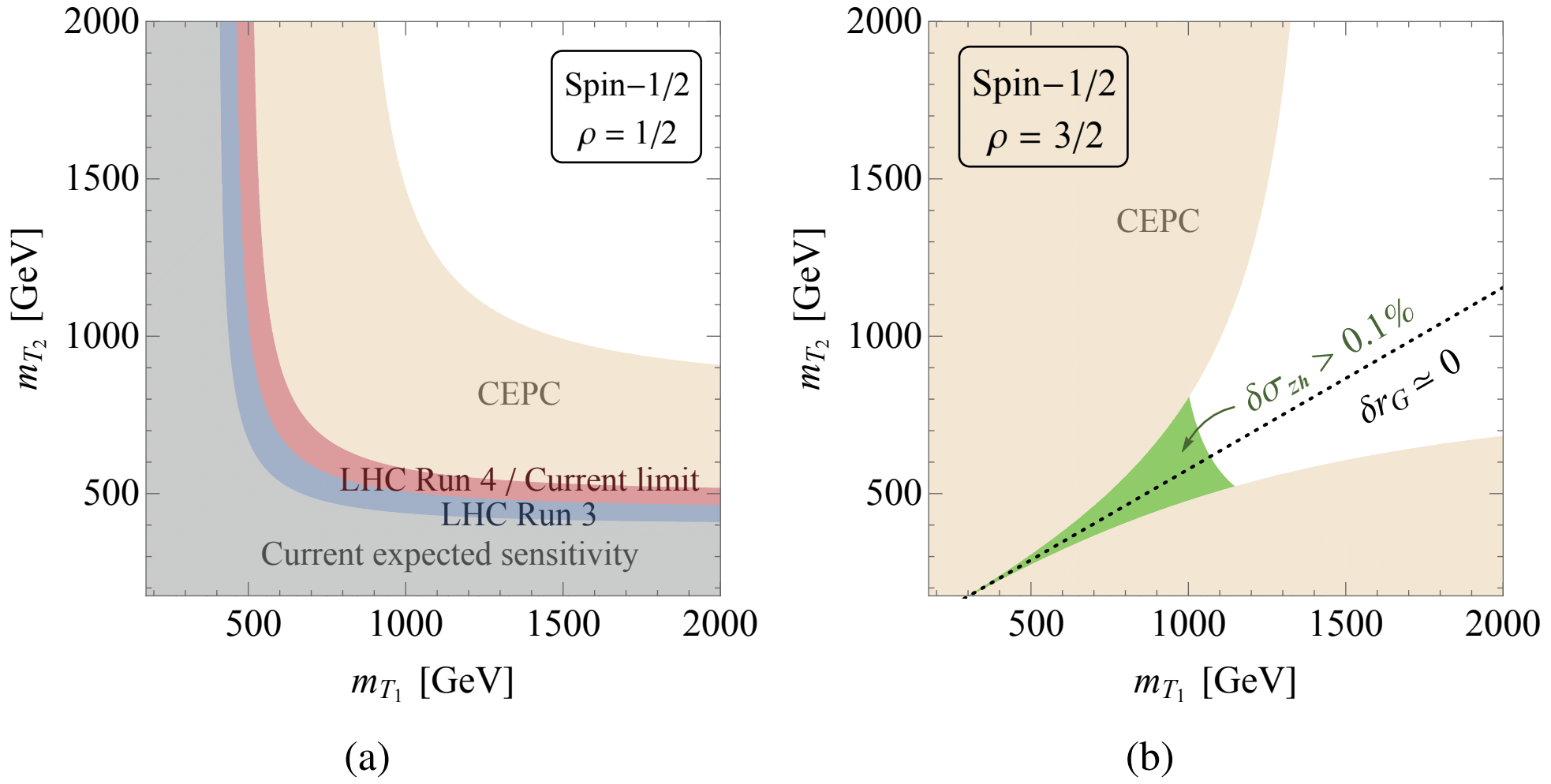

**Figure 2.6:** (a) LHC and CEPC precision Higgs constraints in the $m_{T_1} - m_{T_2}$ plane from Higgs couplings to gluons and photons assuming equal couplings. (b) Coverage of blind spots including precision measurement of the $ZH$ cross section. Figures adapted from [27].

**Tree level** In global symmetry solutions, the Higgs boson is a pNGB of a spontaneously broken global symmetry. This invariably implies tree-level corrections, which can be interpreted as arising from mixing between the Standard Model-like Higgs boson and heavy states associated with the spontaneously broken global symmetry. This mixing is typically proportional to $v^2/f^2$, where $f$ is the decay constant associated with the broken global symmetry (see e.g. [28] for a comprehensive parameterization), although precise corrections may vary between Higgs couplings to fermions and gauge bosons depending on the model. As shown in Figure 2.5, the precision attainable at CEPC probes this mixing to better than one part in one hundred, translating to an energy reach of several TeV. In the simplest composite realizations of global symmetries, bounds on $v^2/f^2$ translate directly into lower bounds on the tuning of the electroweak scale, but this tuning may be avoided in Little Higgs models and related constructions. The complementarity between precision measurements of Higgs couplings and direct searches at future colliders in probing global symmetry approaches to the hierarchy problem is explored in detail in e.g. [29].

**Loop level** Global symmetry approaches to naturalness likewise feature a plethora of new states near the weak scale, albeit with the same statistics as their Standard Model counterparts. While corrections to Higgs couplings from loops of these new particles are typically sub-dominant compared to tree-level corrections, they provide a more immutable test of naturalness. As with supersymmetry, the largest corrections are typically due to the fermionic top partner sector, due to the large coupling of these partners to the Higgs boson and their proximity to the weak scale. As such partners typically carry Standard Model quantum numbers, the most striking corrections are to the loop-level couplings of the Higgs boson to gluons and photons.

For the sake of definiteness, consider a theory involving two top partners $T_1, T_2$ whose couplings are dictated by the global symmetry protecting the Higgs boson mass. In this case corrections to the Higgs coupling to gluons and photons are proportional to [27]

$$-\left(\rho\frac{m_t^2}{m_{T_1}^2}+(1-\rho)\frac{m_t^2}{m_{T_2}^2}\right) \tag{2.5}$$

where $\rho$ parametrizes the "fraction" of the quadratic divergence cancellation coming from the $T_1$ field, which is directly reflected in the modification of Higgs couplings. In the case of equal couplings, CEPC is capable of probing fermionic top partners above the $\mathrm{TeV}$ scale, as shown in the Figure 2.6(a). Note that the existence of more than one fermionic top partner allows for the possibility of a blind spot to arise when $\rho > 1$ (which can arise in situations wherein top and $T_2$ loops contribute to the Higgs with the same sign, see [27]). This may be constrained by radiative corrections to the $ZH$ cross section (shown in the Figure 2.6(b)) in analogy with the stop blind spot in supersymmetry. A comprehensive exploration of CEPC's potential to constrain radiative corrections to Higgs couplings arising in global symmetry solutions to the hierarchy problem may be found in [27].

## NEUTRAL NATURALNESS

While it is entirely possible that the naturalness of the weak scale is explained by conventional symmetries that have thus far evaded LHC detection, LHC null results may indicate that the weak scale is stabilized by less conventional symmetries that do not lead to partner particles carrying Standard Model quantum numbers. This form of "neutral naturalness" [30] can occur, for example, when only discrete symmetries are operative at the weak scale. To date both opposite-statistics and same-statistics examples of neutral naturalness have been constructed. The former case is exemplified by Folded Supersymmetry [31], which features new partner particles carrying electroweak quantum numbers but no irreducible tree-level corrections. The latter case is exemplified by the Twin Higgs [32], which features new partner particles entirely neutral under the Standard Model, as well as significant tree-level Higgs coupling deviations. Examples also exist of theories with entirely neutral scalar top partners [33, 34] and electroweak-charged fermionic top partners [35], both of which share the tree-level modifications to Higgs couplings of the Twin Higgs.

The primary phenomenological consequences of neutral naturalness are (1) a significant weakening of direct search limits due to the paucity of states charged under the Standard Model, and (2) the reduction of loop-level corrections to loop-level Higgs couplings. However, these models still lead to distinctive patterns of Higgs coupling deviations that may be first revealed at a Higgs factory.

**Tree level** Many theories of neutral naturalness, most notably the Twin Higgs [32], feature significant tree-level mixing between the Standard Model-like Higgs boson and an additional $CP$ even scalar state. Much as with conventional global symmetries, this leads to $\mathcal{O}(v^2/f^2)$ deviations in Higgs couplings. In contrast to conventional global symmetries, however, these corrections are typically universal in the sense that they are the same for Higgs couplings to both vectors and fermions. Bounds on $v^2/f^2$ may be translated directly into bounds on the mass of the twin top partner, as shown in Figure 2.7. In such cases, CEPC can probe multi-TeV scales and test the efficacy of neutral naturalness down to the percent level.

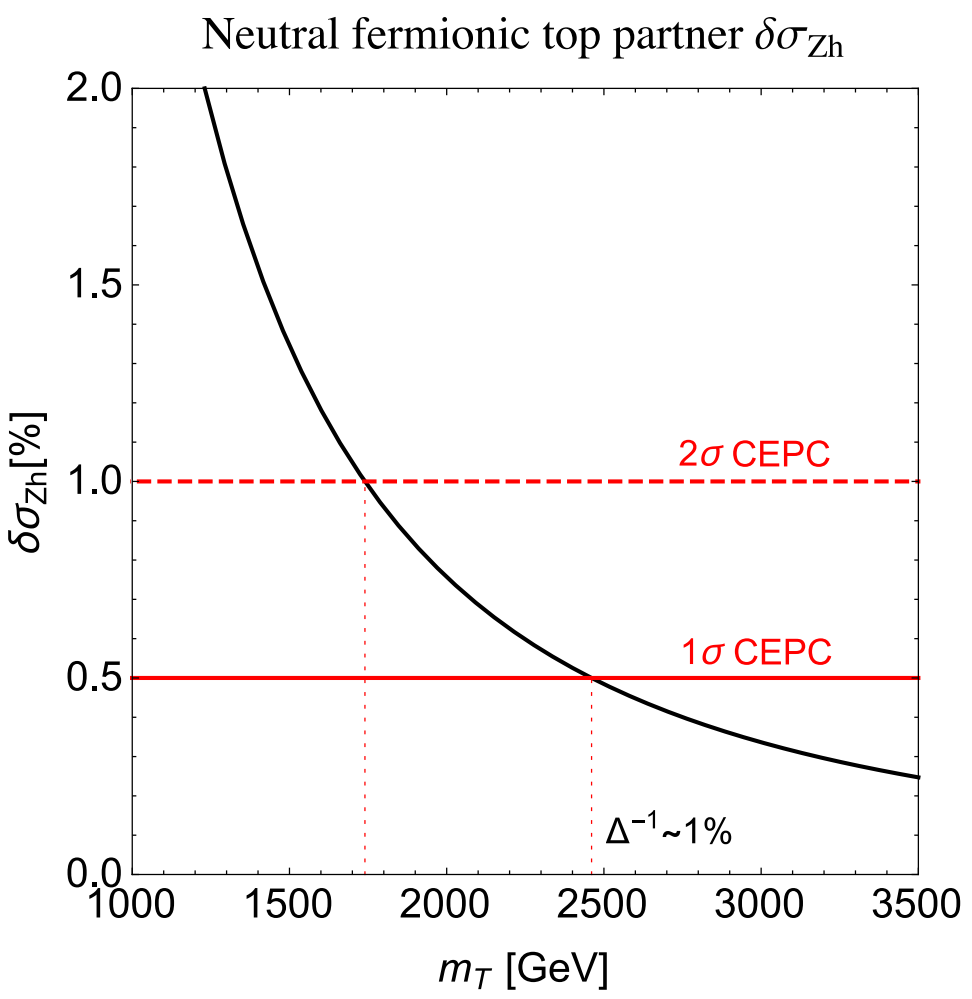

**Figure 2.7:** CEPC reach in the mass scale of neutral fermionic top partners due to tree-level mixing corrections to $\sigma_{ZH}$. Here $m_T = (f/v)m_t$.

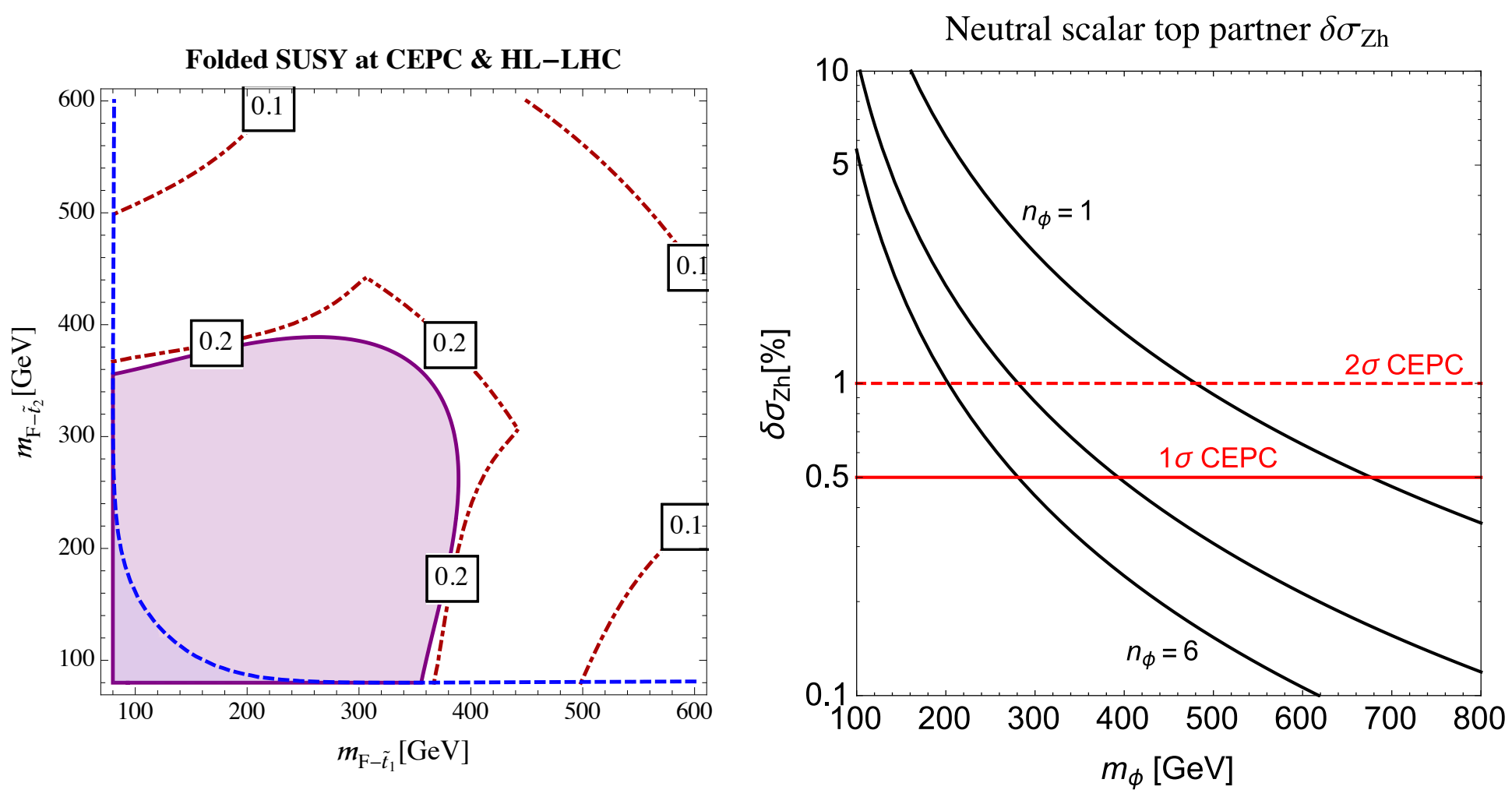

(a) (b)

**Figure 2.8:** (a) CEPC reach for color-neutral folded stops in Folded SUSY from Higgs couplings to photons, from [24]. (b) CEPC reach in the mass scale of neutral scalar top partners due to loop-level corrections to $\sigma_{ZH}$, adapted from [36].

**Loop level** While all models of neutral naturalness feature loop-level corrections to Higgs properties, they are the leading effect in many opposite-statistics models such as folded supersymmetry. New partner particles in these models still carry electroweak quantum numbers, leading to loop-level deviations in the Higgs coupling to photons, as shown in Figure 2.8. This allows CEPC to place constraints on the mass scale of folded partner particles in the hundreds of GeV, probing tuning of the weak scale to the 20% level in these theories.

It is also possible that the weak scale is stabilized by scalar top partners entirely neutral under the Standard Model without accompanying tree-level Higgs coupling deviations. In this case, all of the distinctive direct search channels and corrections to loop-level Higgs couplings are absent. However, a precision measurement of the $ZH$ cross section is still sensitive to the wavefunction renormalization of the physical Higgs scalar induced by loops of the scalar top partners [36]. In general, $n_\phi$ scalars $\phi_i$ coupling via the Higgs portal interaction $\sum_i \lambda_\phi |H|^2|\phi_i|^2$ leads to a correction to the $ZH$ cross section of the form

$$\delta\sigma_{ZH} = \frac{n_\phi|\lambda_\phi|^2}{8\pi^2}\frac{v^2}{m_H^2}\left[1 + \frac{1}{4\sqrt{\tau(\tau-1)}}\log\left(\frac{1-2\tau-2\sqrt{\tau(\tau-1)}}{1-2\tau+2\sqrt{\tau(\tau-1)}}\right)\right] \tag{2.6}$$

where $\tau = m_H^2/4m_\phi^2$. This leads to the sensitivity shown in Figure 2.8, for which CEPC is able to place constraints in the hundreds of GeV on a scenario that is otherwise largely untestable at colliders.

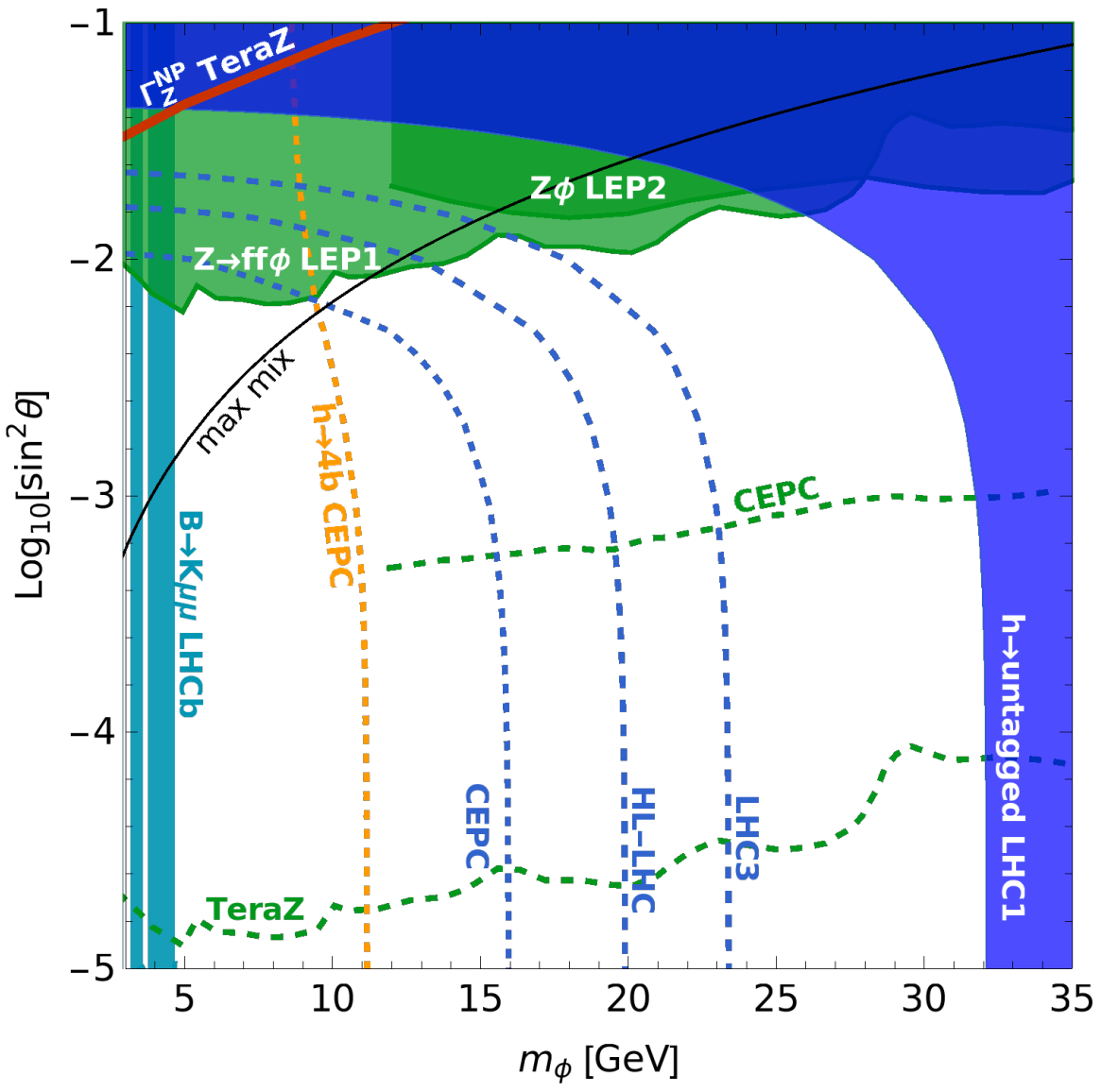

**Figure 2.9:** Constraints on the relaxion mass $m_\phi$ and relaxion-Higgs mixing angle $\sin\theta$ from the non-Standard Model decay of the Higgs boson into relaxion pairs, adapted from [37]. Shaded regions indicate current exclusions from LEP and the LHC. Dashed blue lines indicate the reach of CEPC and future operation of the LHC in searches for untagged non-Standard Model decays of the Higgs boson, while the orange dashed line indicates the reach of CEPC in searches for $H \to \phi\phi \to 4b$. The green dashed line indicates the reach of CEPC's $Z$-pole run in searches for $e^+e^- \to Z\phi$.

## OTHER SOLUTIONS

Symmetries are not the only mechanism for explaining the origin of the weak scale, though other solutions may not be manifestly natural in the same way. However, even non-symmetry explanations for the value of the weak scale (excepting anthropic ones) generically entail some degree of coupling between new degrees of freedom and the Higgs boson itself. This typically leads to deviations in Higgs couplings, new exotic decay modes of the Higgs boson, or a combination thereof.

A compelling example of non-symmetry solutions is the relaxion [20], in which the value of the weak scale is set by the evolution of an axion-like particle across its potential in the early universe. The relaxion necessarily couples to the Higgs boson in order for its evolution to influence the Higgs boson mass. This leads to a variety of signatures that may be tested via precision Higgs measurements [37, 38].

The most promising signature is that of new exotic Higgs boson decays, most notably into the relaxion itself. This signature arises in most relaxion models as a generic conse-quence of the backreaction of electroweak symmetry breaking onto the relaxion potential. The mixing angle between the Higgs boson and relaxion in these scenarios is parametrically of order

$$\sin\theta \approx \frac{\Lambda_{\rm br}^4}{v f m_H^2} \tag{2.7}$$

where $\Lambda_{\rm br}$ is the confinement scale inducing a potential for the relaxion (identifiable with $\Lambda_{\rm QCD}$ in the most minimal models) and $f$ is the relaxion decay constant. This leads to the decay of the Higgs boson into pairs of relaxions $\phi$, which in turn decay back into Standard Model states via Higgs-relaxion mixing.

The CEPC can constrain these scenarios through both direct searches for processes such as $H \to \phi\phi \to 4b$ and indirect limits on exotic Higgs boson decays coming from precision Higgs measurements, as shown in Figure 2.9. While substantial regions of relaxion parameter space remain outside of CEPC's reach (most notably for relaxions lighter than the GeV scale, which are optimally probed by lower-energy experiments), this nonetheless represents a significant improvement in experimental sensitivity. This exemplifies the considerable power of CEPC in identifying natural explanations for the weak scale, even in the absence of additional symmetries, by virtue of its broad sensitivity to new particles interacting with the Higgs boson.

### 2.2.2 ELECTROWEAK PHASE TRANSITION

The discovery of the Higgs boson marks the culmination of a decades-long research program to understand the source of ElectroWeak Symmetry Breaking (EWSB). We have known since the mid-$20^{\rm th}$ century that this symmetry is not realized in nature and that the weak gauge bosons are massive. Now measurements at the Large Hadron Collider (LHC) have provided overwhelming evidence that EWSB results from the recently-discovered Higgs boson. With the Higgs boson discovery we have learned *why* the electroweak symmetry is broken in nature, but we still do not understand *how* it is broken dynamically — this is the question of the *electroweak phase transition*.

The nature of the ElectroWeak Phase Transition (EWPT) is controlled by the properties and interactions of the Higgs boson. For instance the Higgs mass sets the temperature scale of the phase transition to be roughly $T \sim m_H \simeq 125$ GeV. The more detailed and interesting features of the phase transition depend also upon the interactions of the Higgs boson with itself, with other Standard Model particles, and with possible new physics. The nature of these interactions will not be determined very precisely at the LHC, where we have only just begun to study the Higgs boson. Rather, if we want to understand the nature of the electroweak phase transition, we require precision measurements of Higgs physics. As we will discuss in detail in the rest of this section, a dedicated Higgs factory experiment like CEPC can reveal the important clues which will help us answer this question.

#### FIRST ORDER PHASE TRANSITION OR CONTINUOUS CROSSOVER?

Despite years of careful study at the LHC, we still have such a poor understanding of the Higgs boson that it is impossible to determine even the *order of the electroweak phase transition*. In general, these two scenarios are used to classify symmetry-breaking phase transitions:

- A *first order phase transition* proceeds through the nucleation of *bubbles* that grow, coalesce, and eventually fill the system.

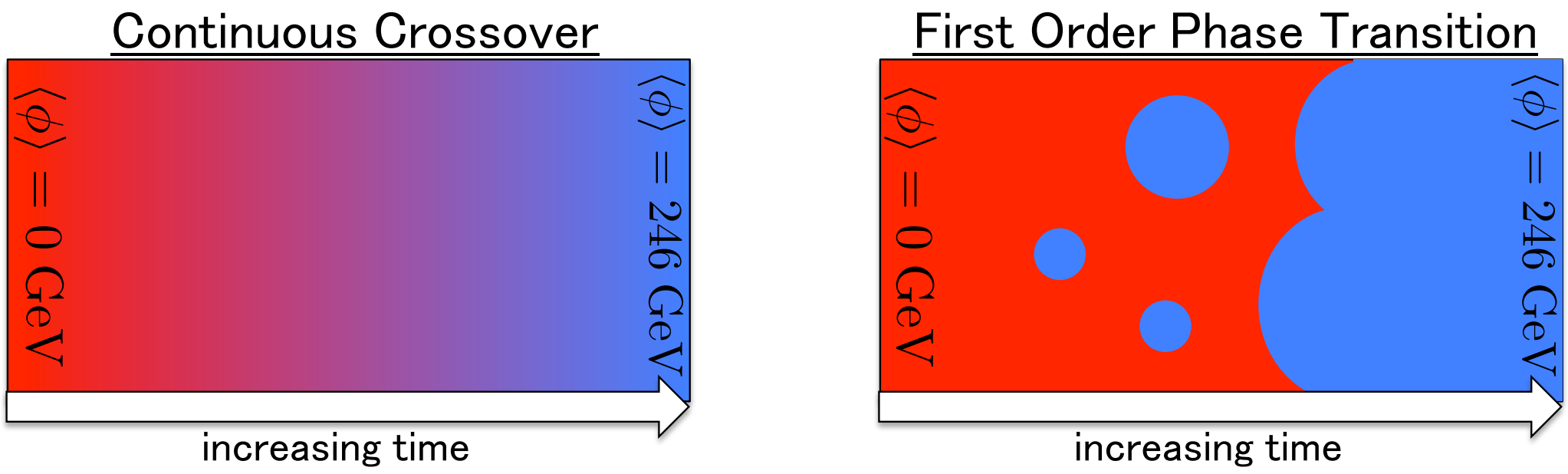

**Figure 2.10:** An illustration of a continuous crossover (left) and a first order phase transition (right).

- By contrast, a *continuous crossover* occurs smoothly throughout the system.

See also Figure 2.10. If the phase transition is determined to be first order, there would be profound implications for early-universe cosmology and the origin of the matter-antimatter asymmetry. Moreover, determining the order of the EWPT is simply the first step in a much richer research program that deals with other aspects of the phase transition including its latent heat, bubble wall velocity, and plasma viscosity.

## THE HIGGS POTENTIAL

The order of the EWPT is intimately connected to the shape of the *Higgs potential energy function*. For each value of the Higgs field, $\phi$, there is an associated potential energy density, $V(\phi)$. During the electroweak phase transition, the Higgs field passes from $\phi = 0$ where the electroweak symmetry is unbroken to $\phi = v \simeq 246$ GeV where the electroweak symmetry is broken and the weak gauge bosons are massive. Thus the order of the phase transition is largely determined by the shape of $V(\phi)$ in the region $0 < \phi < v$.

For instance, if the Higgs potential has a barrier separating $\phi = 0$ from $\phi = v$, then electroweak symmetry breaking is accomplished through a first order phase transition with the associated bubble nucleation that we discussed above. If there is no barrier in $V(\phi)$, the transition may be either first order or a crossover depending on the structure of the *thermal effective potential*, $V_{\text{eff}}(\phi, T)$.

Currently we know almost nothing about the shape of the Higgs potential. This situation is illustrated in Figure 2.11 and the following discussion. When we make measurements of the Higgs boson in the laboratory, we only probe small fluctuations of the potential around $\phi = v$. By measuring the strength of the weak interactions, $G_F = (\sqrt{2}v^2)^{-1} \simeq 1 \times 10^{-5}\ \text{GeV}^{-2}$, we learn that the Higgs potential has a local minimum at $v \simeq 246$ GeV. By measuring the Higgs boson's mass, we learn that the local curvature of the potential at its minimum is $(d^2V/d\phi^2)\big|_{\phi=v} = m_H^2 \simeq (125\ \text{GeV})^2$. This is the extent of what we know today about the Higgs potential. Even the third derivative, which is related to the Higgs boson's cubic self-coupling, is completely undetermined!

Measurements of the Higgs boson thus far are consistent with the predictions of the Standard Model of particle physics. The Standard Model asserts that the Higgs potential has the form

$$V(\phi) = \frac{1}{2}\mu^2\phi^2 + \frac{1}{4}\lambda\phi^4\ , \tag{2.8}$$

which only depends on the two parameters $\mu^2$ and $\lambda$. Taking $\lambda > 0$ and $\mu^2 < 0$ induces a Vacuum Expectation Value (VEV) for the Higgs field and triggers electroweak symmetry breaking. At the minimum of the potential $v = \sqrt{-\mu^2/\lambda}$ gives the Higgs field VEV and $m_H^2 = -2\mu^2$ gives the Higgs boson's mass. Thus, having measured both $v \simeq 246$ GeV and $m_H \simeq 125$ GeV in the laboratory, the Standard Model completely predicts the shape of the Higgs potential. For these values of the Higgs boson mass and VEV, the electroweak phase transition is expected to proceed via a continuous crossover in the absence of additional physics beyond the Standard Model.

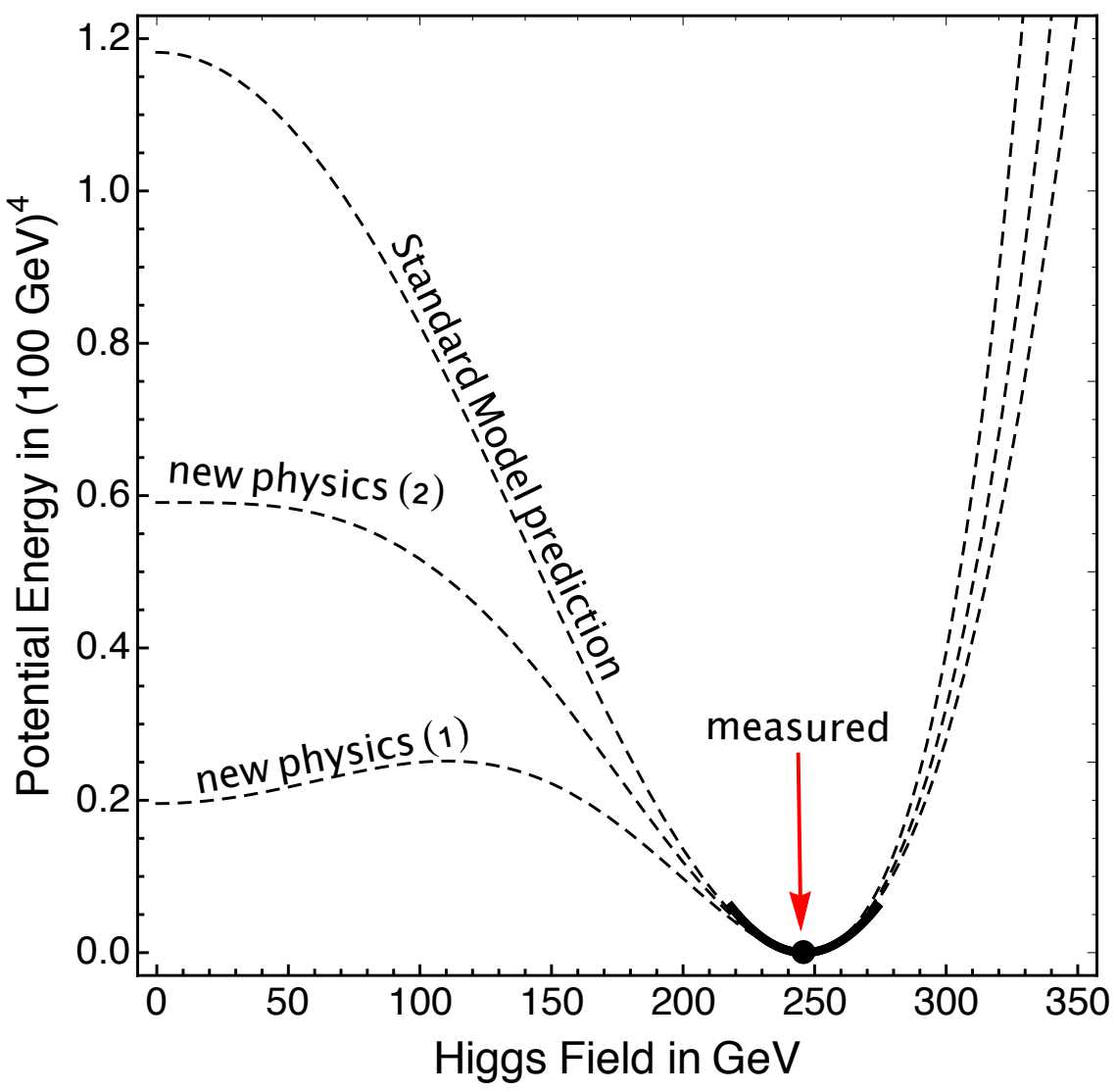

**Figure 2.11:** The Higgs potential energy function. All we know about the shape of the Higgs potential is the local curvature at its minimum. These observations are consistent with the Standard Model, but they are also consistent with models containing new physics that can dramatically change the nature of the electroweak phase transition.

However the presence of new physics can dramatically change the shape of the Higgs potential without disrupting the measurements of $v$ and $m_H$. For example, a simple generalization of Equation (2.8) is to include a sextic term and write the Higgs potential as [39–41]

$$V(\phi) = \frac{1}{2}\mu^2\phi^2 + \frac{1}{4}\lambda\phi^4 + \frac{1}{8\Lambda^2}\phi^6 . \tag{2.9}$$

A potential of this form arises if new, heavy particles are coupled to the Higgs boson, and then $\Lambda$ is related to the mass scale of the new particles. This potential has enough structure to support two local minima with a barrier between, which we see in Figure 2.11 for the curve labeled "new physics (1)." The nature of the electroweak phase transition in this model is expected to be very different from the Standard Model due to the barrier [42–44]. Alternatively the new physics can manifest through a non-analytic term in the Higgs potential, such as the one proposed by Coleman and Weinberg [45],

$$V(\phi) = \frac{1}{4}\lambda\phi^4 \log\frac{\phi^2}{\Lambda^2} . \tag{2.10}$$

Such a potential arises when new physics is coupled to the Higgs boson and leads to a strong running in the Higgs quartic self-coupling [46]. As shown by the curve labeled "new physics (2)" in Figure 2.11, this potential is very flat near the origin allowing thermal corrections to induce a barrier and thus a first order phase transition.

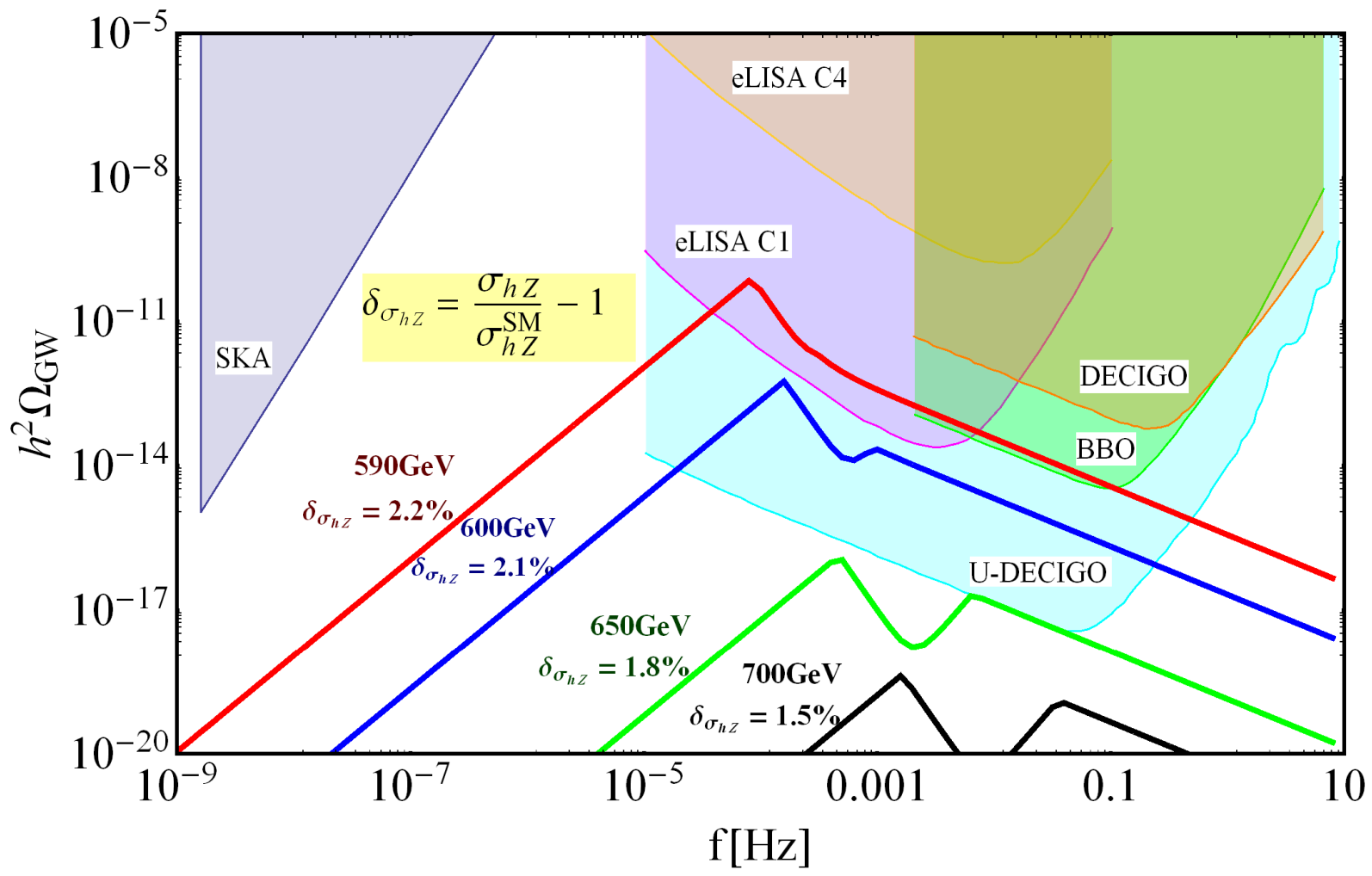

**Figure 2.12:** The spectrum of gravitational waves generated during a first order electroweak phase transition for the model described in Equation (2.9). Colored curves show the predicted spectrum for different models as the scale of new physics, $\Lambda$, is varied. The figure is reproduced from Ref. [44].

Precision measurements of the Higgs boson's interactions with itself and other particles give us crucial information about the shape of the potential energy function and thereby provide much-needed experimental input to test the order of the electroweak phase transition.

## COSMOLOGICAL IMPLICATIONS

Since we cannot reproduce the high-temperature conditions of the electroweak phase transition in the laboratory, the question of the EWPT has the most relevance for studies of the early universe. Most cosmologists expect that a thermal EWPT occurred soon after the Big Bang when the universe was filled with a very hot plasma. If the early universe EWPT was first order, it may have left behind interesting cosmological relics that could be accessible to observations today.

**Gravitational Waves.** During a first order electroweak phase transition, gravitational waves are produced from the collisions of bubbles, the decay of magnetohydrodynamic turbulence, and the damping of sound waves [47]. Today these gravitational waves would look like a stochastic and isotropic "noise" from all directions on the sky. As we see in Figure 2.12 the predicted gravitational wave spectrum falls within reach of future space-based interferometer experiments, including LISA, DECIGO, BBO, Taiji, and TianQin. The detection of these gravitational waves would provide direct evidence that the cosmological EWPT was a first-order one. A future collider like CEPC, through precision measurements of the Higgs couplings, can uncover clues of the nature of the new physics that explains *why* the EWPT is first order.

**Matter-Antimatter Asymmetry.** A first order cosmological EWPT provides the right environment to explain the Universe's excess of matter over antimatter through the mechanism of *electroweak baryogenesis* [48]. This mechanism uses the fact that baryon number is violated in the Standard Model through reactions mediated by the electroweak sphaleron. Before the cosmological EWPT, the sphaleron efficiently converts matter into antimatter, but during the electroweak phase transition the sphaleron-mediated reactions are shut off. If this shutoff is sufficiently abrupt, then an excess of matter over antimatter can be generated. This requires that the electroweak phase transition is *strongly first order* in the sense that

$$\frac{v(T_{\rm pt})}{T_{\rm pt}} \gtrsim 1.0 \qquad \text{("strongly first order" electroweak phase transition)} \tag{2.11}$$

where $v(T_{\rm pt})$ is the value of the Higgs field inside of the bubbles during the phase transition at temperature $T_{\rm pt}$.

Electroweak baryogenesis is not viable in the Standard Model, because the electroweak phase transition is a continuous crossover, $v(T_{\rm pt}) = 0$, and thus the observed excess of matter over antimatter is an irrefutable motivation for physics beyond the Standard Model. In general the new physics can take many forms, but in the context of electroweak baryogenesis, it is clear that the new physics must couple to the Higgs boson so that the sphaleron-suppression condition in Equation (2.11) is satisfied. Therefore this condition directly quantifies the required departure from Standard Model physics.

## NEW PHYSICS AND THE ELECTROWEAK PHASE TRANSITION

The Standard Model predicts that the EWPT is a continuous crossover, but we have seen in the discussion of Figure 2.11 that even minimal extensions of the Standard Model can drastically change the predictions for electroweak symmetry breaking. Thus for any model with new physics coupled to the Higgs boson, it is necessary to ask: What is the nature of the electroweak phase transition?

In the years before the LHC started running, much of the work was focused on the *light stop scenario* of the Minimal Supersymmetric Standard Model (MSSM) [49, 50]. Early LHC data determined that this scenario is ruled out [51, 52], because the light stops, which are colored and charged particles with spin-0, should have been easy to produce and detect at the LHC. However, if the new scalar particles were not charged or colored, the electroweak phase transition could still be first order while evading collider constraints; to leading order, the electroweak phase transition only cares about couplings with the Higgs boson, not quantum numbers [53]. Therefore in order to assess the unique power of CEPC to test new physics that leads to a first order electroweak phase transition, it is useful to consider models with uncharged and uncolored particles, which are very difficult to probe at the LHC [54].

A viable model with a first order EWPT is found in even the most minimal extension of the Standard Model with a real, scalar singlet field $S$ [55–57]. The relevant Lagrangian is written as

$$\mathscr{L} = \left(D_\mu H\right)^\dagger \left(D^\mu H\right) + \frac{1}{2}\left(\partial_\mu S\right)\left(\partial^\mu S\right) - \mu_H^2 H^\dagger H - \lambda_H \left(H^\dagger H\right)^2 \\ - \frac{\mu_S^2}{2} S^2 - \frac{a_S}{3} S^3 - \frac{\lambda_S}{4} S^4 - \lambda_{HS} H^\dagger H S^2 - 2 a_{HS} H^\dagger H S \tag{2.12}$$

where $H(x)$ denotes the Higgs doublet field. The last two operators in Equation (2.12) correspond to the so-called Higgs portal interactions. The Higgs field acquires a vacuum expectation value, $\langle H \rangle = (0\,,\, v/\sqrt{2})$ that breaks the electroweak symmetry. In general the singlet field may acquire a vacuum expectation value, $\langle S \rangle = v_S$, and it can mix with the Higgs boson, which is parametrized by an angle $\theta$. The spectrum of this theory contains two scalars with masses $m_H \simeq 125$ GeV and $m_S$.

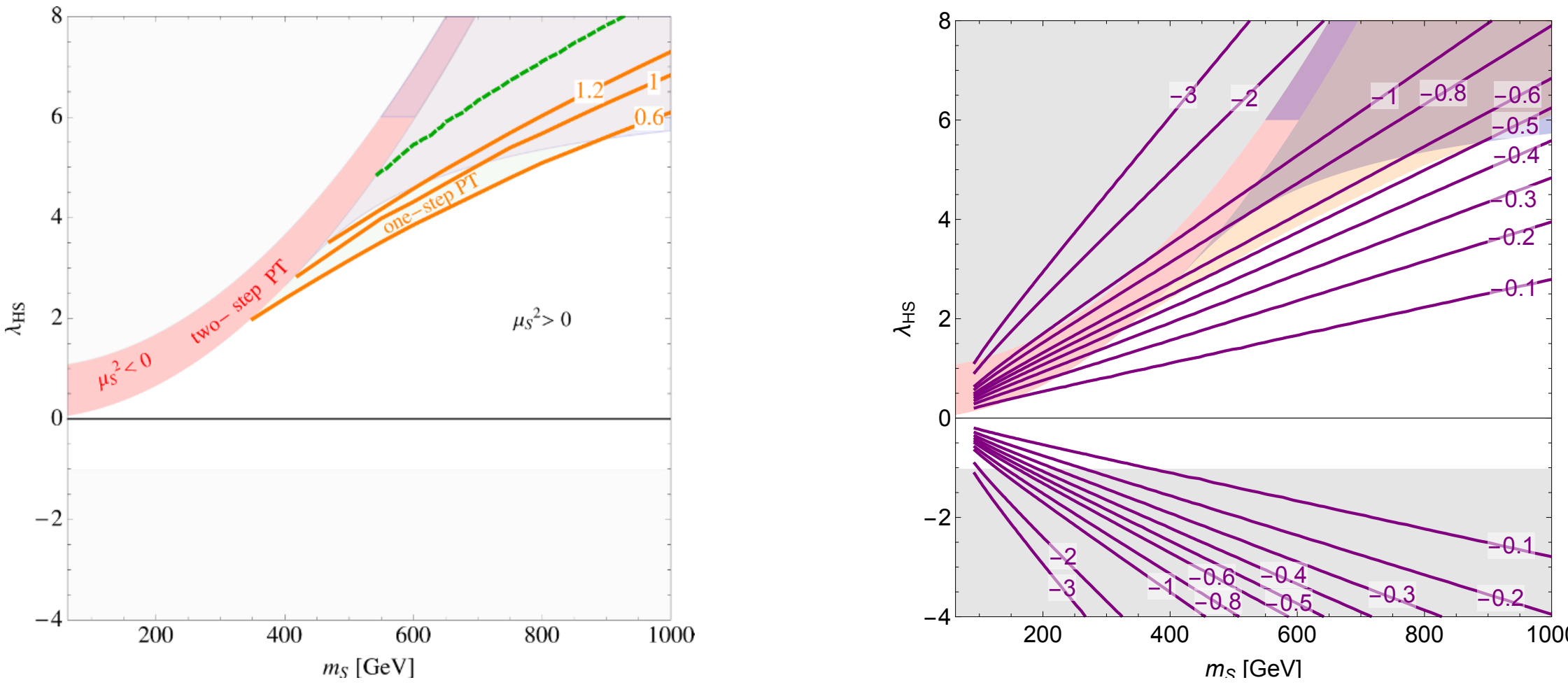

(a) (b)

**Figure 2.13:** Parameter space of the real scalar singlet model with $\mathbb{Z}_2$ symmetry. (a) Regions of parameter space that lead to a first order electroweak phase transition that proceeds in one or two steps. The orange curves show the strength of the electroweak phase transition, $v(T_c)/T_c$, in the one-step region. (b) Purple curves show the fractional change to the $ZH$ production cross section relative to the SM prediction in percent; these values are $2\,\delta g_{HZZ}$ using the notation in the text (2.14). The figures are taken from Ref. [59]. (Also see Ref. [60].)

It is also interesting to consider the model that is obtained by imposing a $\mathbb{Z}_2$ symmetry on Equation (2.12). This symmetry transformation, $S(x) \to -S(x)$, enforces $a_{HS} = a_S = 0$, and it is conventional to also assume that $v_S = 0$.

The singlet extension of the Standard Model allows for a first order electroweak phase transition in a variety of ways [53]. If the singlet particle is heavy, $m_S \gg m_H$, then it can be integrated out of the theory generating an effective potential for the Higgs field. In the regime where the $a_S$ and $\lambda_S$ terms are negligible and $\mu_S^2 \gg \lambda_{HS} v^2$, the Higgs potential takes the form

$$V(\phi) = \frac{1}{2}\mu_H^2\phi^2 + \frac{1}{4}\left(\lambda_H - \frac{2a_{HS}^2}{\mu_S^2}\right)\phi^4 + \frac{\lambda_{HS}\, a_{HS}^2}{2m_S^4}\phi^6\,, \tag{2.13}$$

which has the same structure as the one that we encountered in Equation (2.9). The two potentials are matched by taking $\Lambda^2 = m_S^4/(4\lambda_{HS}a_{HS}^2)$. For smaller $\Lambda$ the shape of the Higgs potential begins to deviate more from the Standard Model prediction, and the phase transition becomes first order. This example illustrates the intuition that models with a first order electroweak phase transition require new, light particles with a large coupling to the Higgs boson. If the singlet particle is so light that we are not justified to integrate it out ($m_S \sim m_H$) the analysis above is inapplicable, but the phase transition can still be made first order due to the presence of large loop corrections to the Higgs potential [46], large thermal corrections, and/or a multi-step phase transition [58]. Some of these scenarios are illustrated in the Figure 2.13(a) for the $\mathbb{Z}_2$-symmetric singlet extension.

In general the presence of new particles coupled to the Higgs boson modifies how strongly the Higgs boson couples to itself and to the other Standard Model particles. It is precisely the goal of Higgs factory experiments, like CEPC, to measure these couplings with high precision. Therefore, if the electroweak phase transition is first order, we ex-pect that the measurements of these couplings must deviate from their Standard Model predictions.

The coupling that will be measured most precisely at CEPC and future lepton colliders is the Higgs-$Z$-$Z$ coupling. We can parametrize deviations in this parameter away from the Standard Model prediction with the variable

$$\delta g_{HZZ} \equiv \frac{1}{2}\left(\frac{\sigma(e^+e^- \to HZ)}{\sigma_{\rm SM}(e^+e^- \to HZ)} - 1\right)\Bigg|_{s=(240\text{ GeV})^2} = \frac{g_{HZZ}}{g_{HZZ,\rm SM}} - 1\Bigg|_{s=(240\text{ GeV})^2}. \tag{2.14}$$

In the singlet extension model, the strength of the $HZZ$ coupling is suppressed compared to the SM prediction. The leading-order suppression arises from the Higgs-singlet mixing, and the sub-leading effect arises from Higgs wavefunction renormalization [36] and the Higgs triple self-coupling [61]. Combining these effects, the fractional suppression is written as [59, 62]

$$\begin{aligned}\delta g_{HZZ} =& \big(\cos\theta - 1\big) - 2\frac{|a_{HS} + \lambda_{HS} v_S|^2}{16\pi^2} I_B(m_H^2; m_H^2, m_S^2) \\ &- \frac{|\lambda_{HS}|^2 v^2}{16\pi^2} I_B(m_H^2; m_S^2, m_S^2) + 0.006\left(\frac{\lambda_3}{\lambda_{3,\rm SM}} - 1\right)\end{aligned} \tag{2.15}$$

where $\theta$ is the Higgs-singlet mixing angle, and $I_B$ is a loop function. The Higgs triple self-coupling $\lambda_3$ also deviates from the Standard Model prediction due to the Higgs-singlet mixing. Then the self-coupling is predicted to be [63]

$$\begin{aligned}\lambda_3 = &\big(6\lambda_H v\big)\cos^3\theta + \big(6a_{HS} + 6\lambda_{HS} v_S\big)\sin\theta\cos^2\theta \\ &+ \big(6\lambda_{HS} v\big)\sin^2\theta\cos\theta + \big(2a_S + 6\lambda_S v_S\big)\sin^3\theta\,.\end{aligned} \tag{2.16}$$

In the Standard Model we have $\lambda_3 = \lambda_{3,\rm SM} \equiv 3m_H^2/v \simeq 191$ GeV. If the singlet is light, $m_S < m_H/2$, then the Higgs boson acquires an exotic decay channel, $H \to SS$, which may be invisible depending on the stability of $S$. The rate for this decay is

$$\Gamma(h \to SS) = \frac{\lambda_{211}^2}{32\pi m_H}\sqrt{1 - \frac{4m_S^2}{m_H^2}} \tag{2.17}$$

where

$$\begin{aligned}\lambda_{211} = &\big(2a_{HS} + 2\lambda_{HS} v_S\big)\cos^3\theta + \big(4\lambda_{HS} v - 6\lambda_H v\big)\sin\theta\cos^2\theta \\ &+ \big(6\lambda_S v_S + 2a_S - 4\lambda_{HS} v_S - 4a_{HS}\big)\sin^2\theta\cos\theta + \big(-2\lambda_{HS} v\big)\sin^3\theta\end{aligned} \tag{2.18}$$

is the effective tri-linear coupling of the mass eigenstates. Measurements of the Higgs boson at the LHC already strongly constrain the invisible decay channel, which requires $\lambda_{211} \ll 1$ or $m_S > m_H/2$.

The complementarity between a first order electroweak phase transition and precision Higgs observables is shown in Figure 2.14 for the singlet extension of the Standard Model. Orange points correspond to models with a first order phase transition, $v(T_{\rm pt})/T_{\rm pt} \neq 0$. Blue points correspond to models with a strongly first order phase transition, $v(T_{\rm pt})/T_{\rm pt} \gtrsim 1$, which is a necessary requirement for electroweak baryogenesis (2.11). Red points correspond to models with a very strongly first order phase transition that can potentially be probed by the space-based gravitational wave interferometer telescope LISA.

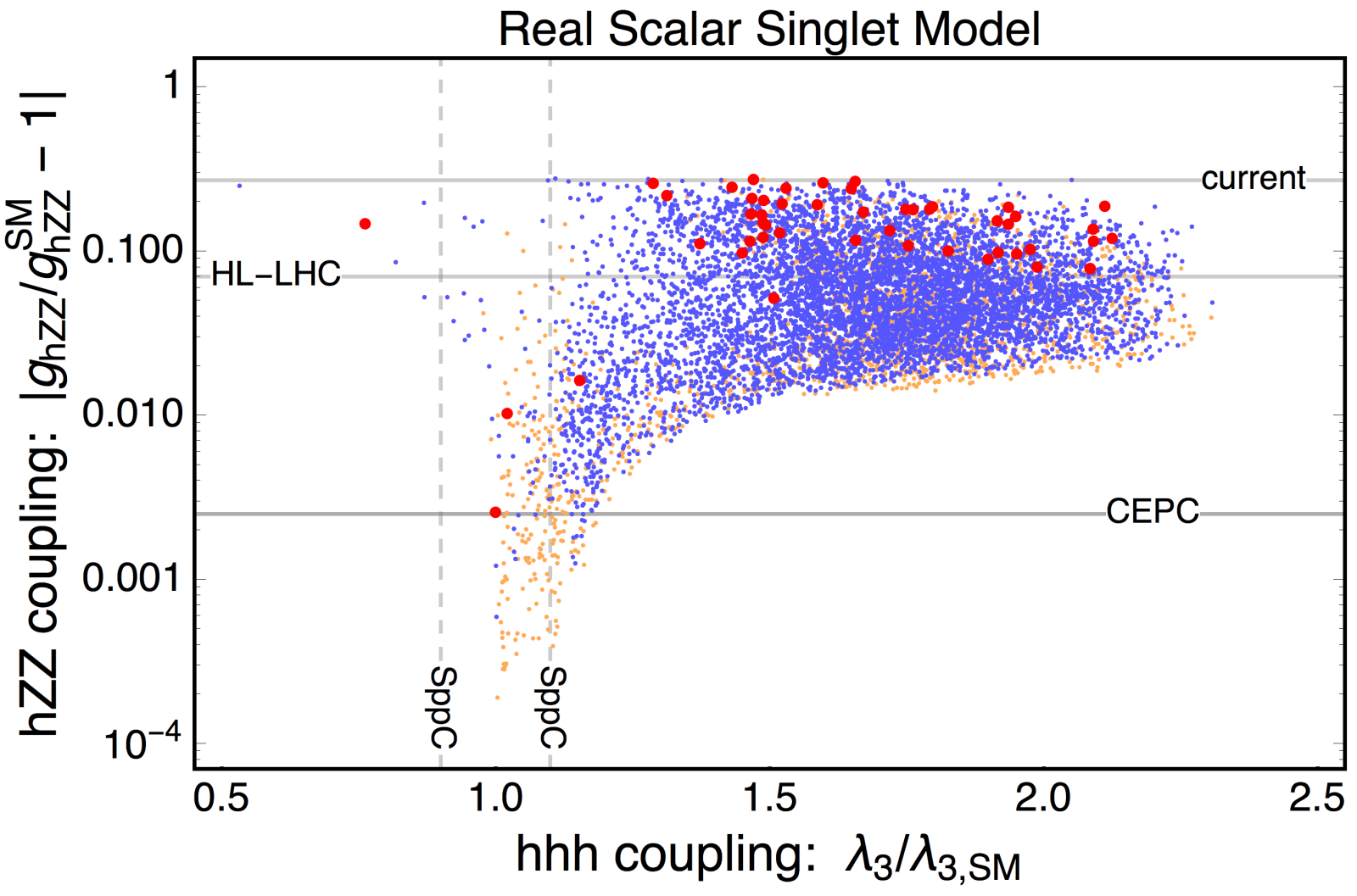

**Figure 2.14:** Collider observables in the real scalar singlet model. Points in theory space with a first order phase transition are shown in orange, points with a strongly first order phase transition are shown in blue, and points with a strongly first order phase transition that also produces detectable gravitational waves are shown in red. The funnel region at $\lambda_3/\lambda_{3,\rm SM} \approx 1$ corresponds to a "blind spot" where a first order phase transition is obtained despite having SM-like couplings. The figure is reproduced from Ref. [62].

Figure 2.14 shows that the models with a first order phase transition (all colored points) also generally predict large deviations in the $HZZ$ coupling. For the models with a strongly first order phase transition (blue and red points) the effect on $g_{HZZ}$ is large enough to be tested by CEPC. Additionally, most of the parameter points also predict a large enhancement to the Higgs trilinear self-coupling that can be probed by a future $100$ TeV hadron collider experiment, like the proposed SppC. The funnel region of orange points at $\lambda_3/\lambda_{3,\rm SM} \approx 1$ corresponds to a "blind spot" where the Higgs-singlet mixing vanishes. Thus, apart from the blind spot, the reach of CEPC is sufficient to probe a first order electroweak phase transition across the entire parameter space of these models.

The blind spot mentioned above corresponds to two scenarios. The Higgs-singlet mixing could vanish, because of an accidental cancellation between $a_{HS}$ and $\lambda_{HS} v_S$. This corresponds to an artificially fine-tuned parameter space, that is not theoretically appealing. Alternatively, the mixing vanishes identically in the $\mathbb{Z}_2$ symmetric limit of the singlet extension. In this case, the relevant parameter space is shown in Figure 2.13. The right panel shows the predicted deviation in the $HZZ$ coupling away from the Standard Model expectation, which is comfortably within reach of CEPC's projected sensitivity.

Another representation of the parameter space appears in Figure 2.15, which shows a correlation between the phase transition temperature and the Higgs cubic self-coupling. For a similar analysis see also Ref. [64], but note that this article was published before the Higgs boson mass was determined.

Among all possible new physics that renders the electroweak phase transition to be first order, we focus on the singlet extension here, because it is the most challenging to test with collider experiments. To illustrate this point, one can allow the new scalar particles to carry an electric charge (similar to a two-Higgs doublet model). An analysis of this model has been performed in Ref. [62], and the results are shown in Figure 2.16. The CEPC has enough sensitivity to test the entire interesting parameter space, and much of the space will also be tested by measurements at the LHC.

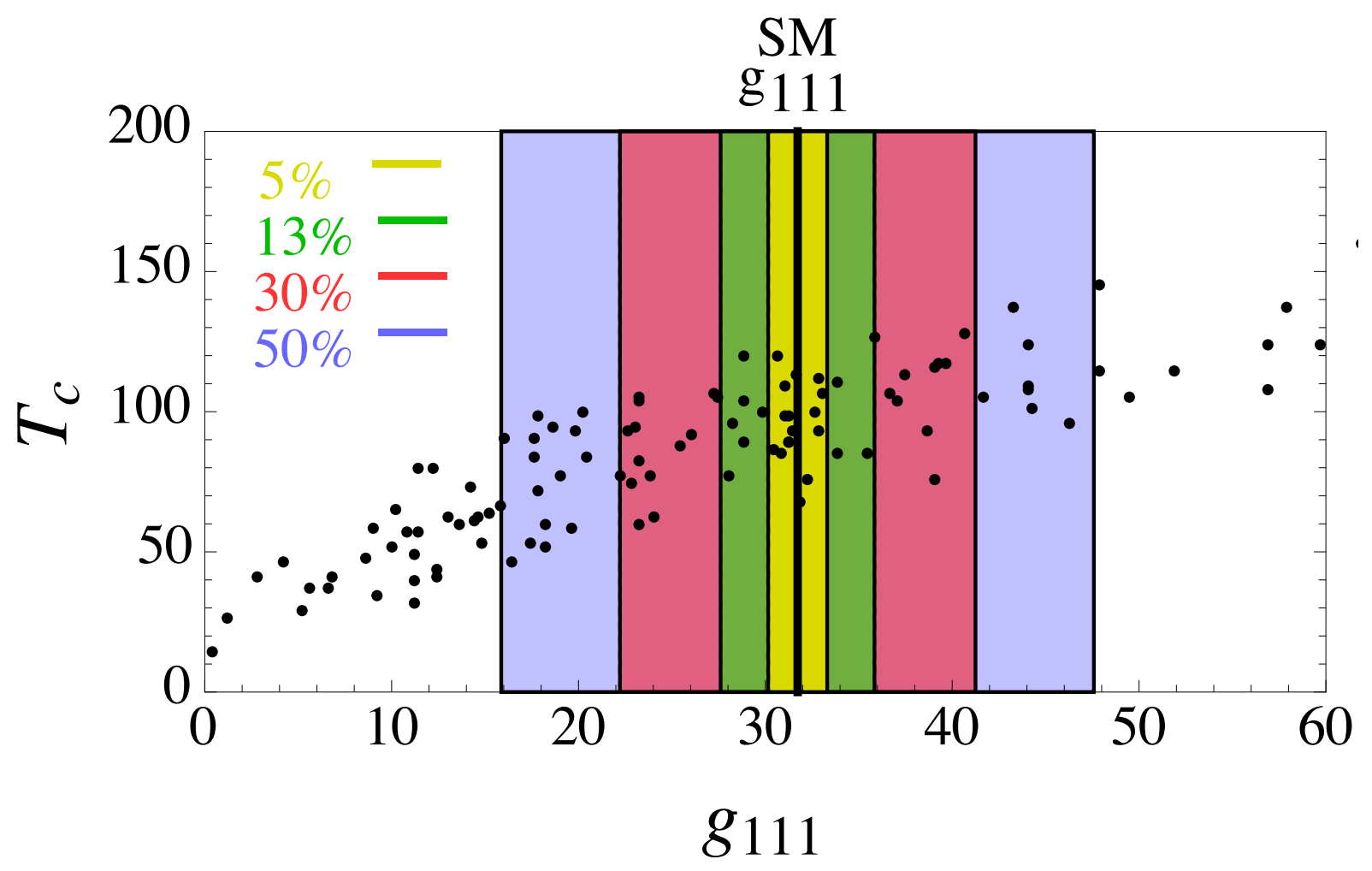

**Figure 2.15:** A correlation between the cubic self-coupling of the SM-like scalar boson and the critical temperature of the first order electroweak phase transition. To connect with the notation in the text, $g_{111} \to \lambda_3/(6\ \mathrm{GeV})$ and $T_c \to T_{\rm pt}/\mathrm{GeV}$. The figure is reproduced from Ref. [63].

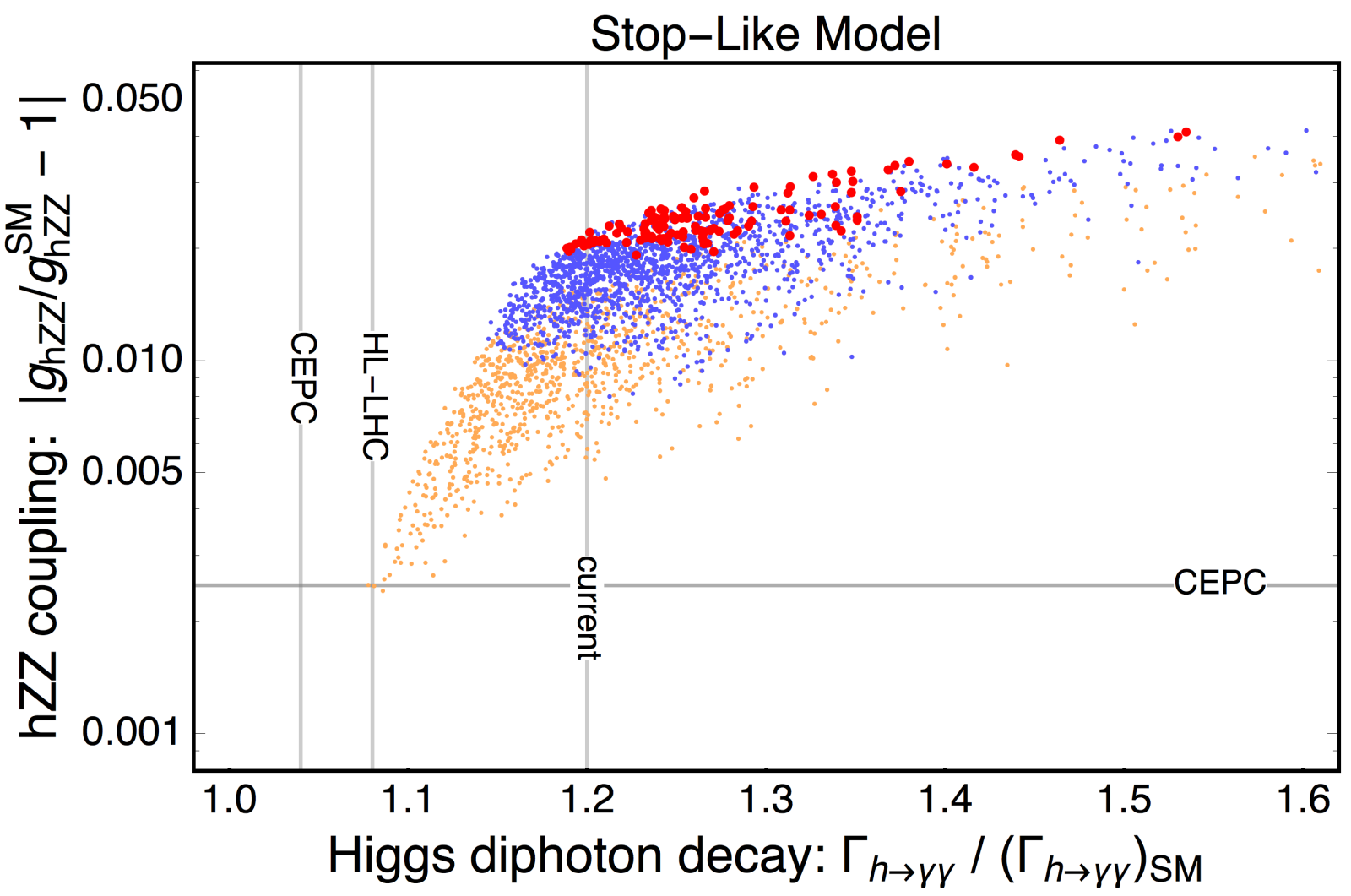

**Figure 2.16:** A model in which the new scalar particles are charged and uncolored. Such a model can be tested by CEPC, but it is already strongly constrained by the LHC's measurement of the Higgs diphoton decay width. The figure is reproduced from Ref. [62].

## WHAT WILL WE LEARN FROM CEPC?

The CEPC will probe the Higgs boson with unprecedented precision. While the LHC has taught us that the Higgs is responsible for electroweak symmetry breaking, measurements at CEPC provide a unique opportunity to learn *how* electroweak symmetry breaking occurs. The nature of the electroweak phase transition is a question that we cannot settle using only measurements at the LHC and its upgrades. Simple and compelling extensions of the Standard Model can have a dramatic effect on the shape of the Higgs potential and the nature of the electroweak phase transition, while remaining extremely challenging to probe by the LHC. However, the presence of new particles coupled to the Higgs boson must affect the way that the Higgs boson couples to itself and to other Standard Model particles, such as the Z-boson. Therefore precision measurements of the Higgs boson's couplings are precisely what's required to expose the new physics. Whereas a Higgs factory experiment like CEPC is not well suited to measuring the Higgs boson's self-coupling, which would provide a direct probe of the shape of the Higgs potential, it is instead ideally situated to make precision measurements of the strength of the $hZZ$ coupling. Measurements at the $0.1\%$ level by CEPC, will serve as an excellent litmus test for a first order electroweak phase transition.

## 2.3 EXPLORING NEW PHYSICS

Exotic new physics could interact with the Standard Model in multiple ways that could be tested at CEPC. Here we summarize and classify different possible scenarios, which are discussed in more detail in the following sections:

1. Exotic particles carry Standard Model charges. The classic example in the dark matter context is dark matter in electroweak multiplets: although dark matter must be neutral, it could be part of an $\mathrm{SU}(2)$ multiplet that also contains charged particles. Because CEPC is primarily a machine for Higgs and electroweak physics, it would be sensitive to such electroweak multiplets.

2. Renormalizable Standard Model portals: if there are no new particles with Standard Model gauge interactions and no new gauge groups that the Standard Model particles are charged under, exotic particles in the hidden (dark) sectors can still interact with the Standard Model via the gauge-singlet operators $H^\dagger H$ ("Higgs portal") [65–72], $B_{\mu\nu}$ ("hypercharge portal" or kinetic mixing) [73–79], and $HL$ ("neutrino portal") [80–86].

3. Portals with additional Standard Model sector physics or new gauge groups that the Standard Model is charged under: if some exotic particle itself carries no Standard Model gauge charges, it may nonetheless interact with the Standard Model via unknown new particles with Standard Model charges. For instance, the existence of a second Higgs doublet that couples dominantly to leptons can make models of "leptophilic" dark matter possible. The second possibility is that there exists some new gauge group, e.g. $\mathrm{U}(1)'$, that (some) Standard Model particles are charged under. Then there is a renormalizable coupling between the new gauge boson and the current made of the Standard Model particles. If the new gauge group is anomalous with the Standard Model particle content, there could also be a Wess-Zumino type interaction between the $Z$ and the new gauge boson [87–96].

4. Effective theory and high dimensional operators: this approach is agnostic to which of the above three scenarios we consider. The theory only contains certain light exotic particles and the Standard Model. The other new physics that generates the coupling between them is not identified and is only encoded in Wilson coefficients. Examples include an Axion-Like Particle (ALP) interacting with the $Z$ boson or photon through dimension-five operators [97–109] and magnetic inelastic dark matter and Rayleigh dark matter models [110–114], in which the dark sector interacts with $Z$ via even higher dimensional operators.

These different scenarios may result in modifications to precision Higgs and $Z$ observables or to exotic Higgs and $Z$ boson decays. The first type of signal has been discussed in Section 2.2. In Sections 2.3.1 and 2.3.2, we will discuss the potential of CEPC for measuring exotic Higgs and $Z$ boson decays. Then in Section 2.3.3, we will focus on the implications for dark matter and dark sectors. In Sections 2.3.4 and 2.3.5, we will discuss the potential of measuring exotic physics connected to neutrino and flavor physics.

### 2.3.1 EXOTIC HIGGS BOSON DECAYS

Higgs boson can be an important portal to new physics beyond the Standard Model. Such new physics could manifest itself through Higgs boson exotic decays if some of the degrees of freedom are light. The Higgs boson BSM decays have a rich variety of possibilities. Two-body Higgs boson decays into BSM particles $H \to X_1 X_2$, where the BSM particles $X_i$ are allowed to subsequently decay further, are considered here. These decay modes are classified into four cases, schematically shown in Figure 2.17. These processes are well-motivated by BSM models such as singlet extensions of the SM, two-Higgs-doublet-models, SUSY models, Higgs portals, gauge extensions of the SM, and so on [115–117]. In this study, only prompt decays of the BSM particles are considered. For Higgs decays into long-lived particles, novel search strategies can be developed in future studies utilizing the advancement in detector development [118].

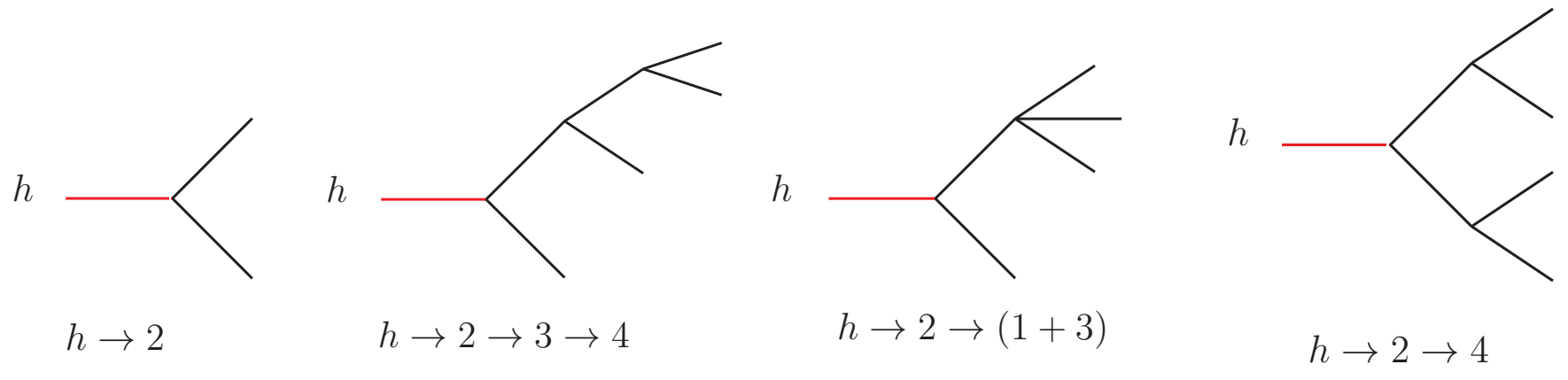

**Figure 2.17:** The topologies of the SM-like Higgs boson exotic decays.

For CEPC running at the center of mass energy 240 GeV, the most important Higgs boson production mechanism is $Z$-Higgs associated production $e^+e^- \to Z^* \to ZH$. The $Z$ boson with visible decays enables Higgs boson tagging using the "recoil mass" technique. A cut around the peak of the recoil mass spectrum would remove the majority of the SM background. Further selection and tagging on the Higgs boson decay product can hence achieve high signal efficiency, and the major background would be from the Higgs boson SM decays. The details of these analysis can be found in Ref. [117].

The set of Higgs boson exotic decays with their projected LHC constraints and limits from the CEPC with $5.6\,\text{ab}^{-1}$ integrated luminosity are summarized in Table 2.1. For the LHC constraints, both the current limits and projected limits on these exotic decay channels from various references are tabulated. The comparison are performed for particular benchmark points to demonstrate the qualitative difference between the (HL-)LHC and CEPC.

The exotic Higgs boson decay channels summarized in Table 2.1 and the corresponding Figure 2.19 are among the most difficult modes to constrain at the LHC and exemplify the considerable sensitivity of the CEPC. The red bars in Figure 2.19 correspond to a recoil mass analysis that only uses leptonic decays of the $Z$-boson that is produced in association with the Higgs boson. The inclusion of hadronic decays of the $Z$-boson provides around ten times more statistics and would lead to substantially improved reach. Based upon the study of Higgs boson decays $H \to WW^*$, $ZZ^*$ and invisible particles, hadronically decaying $Z$-bosons are conservatively assumed to provide a limit comparable to the limit from leptonic $Z$-bosons, and hence improve the limits by around 40% when combined. These extrapolated results are shown in yellow bars in Figure 2.18.

| Decay Mode | 95% CL limit on BR | | |
|---|---|---|---|
| | LHC (current) | LHC (projections) | CEPC |
| $E_{\mathrm{T}}^{\mathrm{miss}}$ | 0.23 | 0.056 | 0.0030 |
| $(b\bar{b}) + E_{\mathrm{T}}^{\mathrm{miss}}$ | – | [0.2] | $1\times10^{-4}$ |
| $(jj) + E_{\mathrm{T}}^{\mathrm{miss}}$ | – | – | $4\times10^{-4}$ |
| $(\tau^+\tau^-) + E_{\mathrm{T}}^{\mathrm{miss}}$ | – | [1] | $8\times10^{-5}$ |
| $b\bar{b} + E_{\mathrm{T}}^{\mathrm{miss}}$ | – | [0.2] | $2\times10^{-4}$ |
| $jj + E_{\mathrm{T}}^{\mathrm{miss}}$ | – | – | $5\times10^{-4}$ |
| $\tau^+\tau^- + E_{\mathrm{T}}^{\mathrm{miss}}$ | – | – | $8\times10^{-5}$ |
| $(b\bar{b})(b\bar{b})$ | 1.7 | (0.2) | $6\times10^{-4}$ |
| $(c\bar{c})(c\bar{c})$ | – | (0.2) | $8\times10^{-4}$ |
| $(jj)(jj)$ | – | [0.1] | $2\times10^{-3}$ |
| $(b\bar{b})(\tau^+\tau^-)$ | [0.1] | [0.15] | $4\times10^{-4}$ |
| $(\tau^+\tau^-)(\tau^+\tau^-)$ | [1.2] | $[0.2 \sim 0.4]$ | $2\times10^{-4}$ |
| $(jj)(\gamma\gamma)$ | – | [0.01] | $1\times10^{-4}$ |
| $(\gamma\gamma)(\gamma\gamma)$ | $[7\times10^{-3}]$ | $4\times10^{-4}$ | $8\times10^{-5}$ |

**Table 2.1:** The current and projected limits on Higgs boson exotic decay modes for the (HL-)LHC and CEPC with 5.6 ab$^{-1}$ integrated luminosity, based upon results from Ref. [117]. In the first column, the particles in the same parenthesis are decay products of an intermediate resonance. The projections for the future runs of the LHC are collected in the third column, where the limits for 100 fb$^{-1}$ and 300 fb$^{-1}$ alone are shown in parentheses and square brackets, respectively.

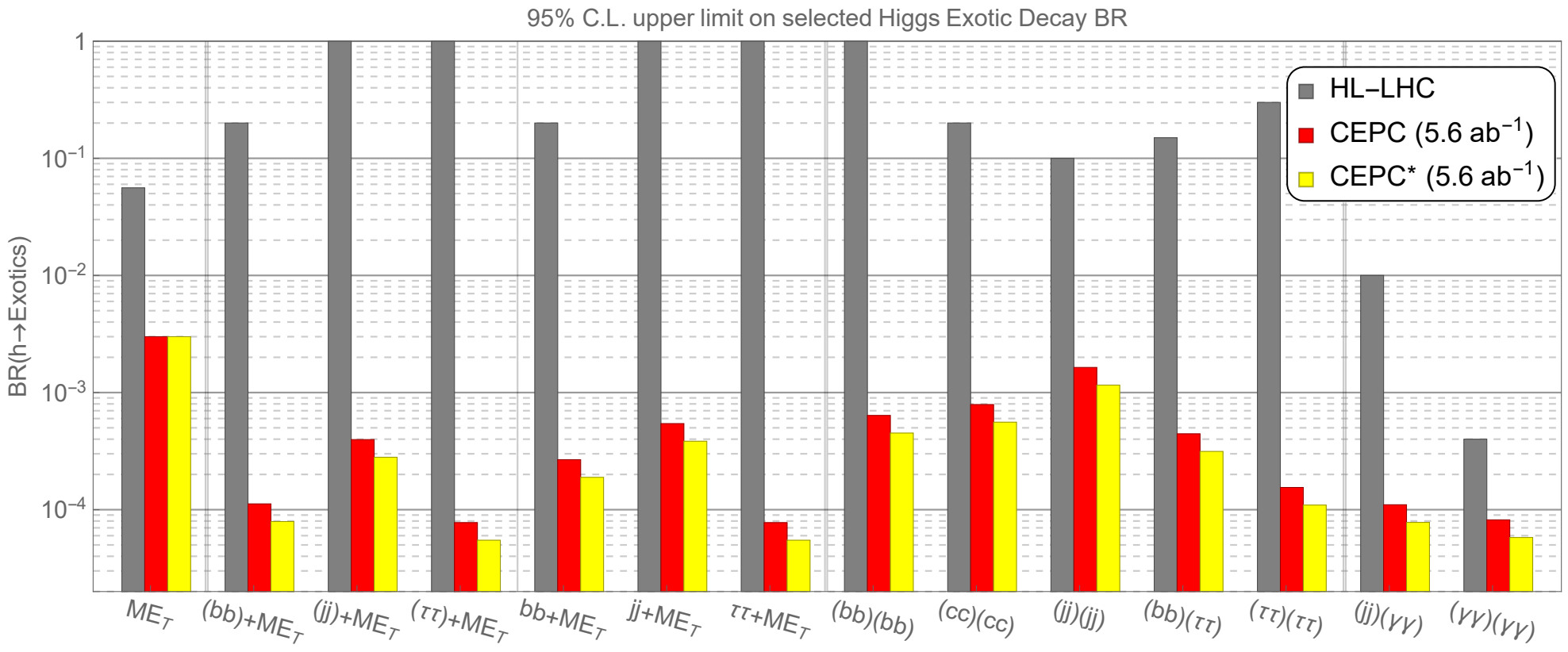

**Figure 2.18:** The 95% CL upper limit on selected Higgs exotic decay branching fractions at HL-LHC and CEPC, based on Ref. [117]. The benchmark parameter choices are the same as in Table 2.1. The red bars correspond to the results using only leptonic decays of the spectator $Z$-boson. The yellow bars further include extrapolation with the inclusion of the hadronic decays of the spectator $Z$-boson. Several vertical lines are drawn in this figure to divide different types of Higgs boson exotic decays.

In comparison with the HL-LHC, the improved coverage of Higgs boson exotic branching fractions is significant, varying from one to four orders of magnitude for the channels under consideration. For the Higgs boson exotic decays into hadronic final states plus missing energy, $b\bar{b} + E_{\mathrm{T}}^{\mathrm{miss}}$, $jj + E_{\mathrm{T}}^{\mathrm{miss}}$ and $\tau^+\tau^- + E_{\mathrm{T}}^{\mathrm{miss}}$, CEPC improves on the HL-LHC sensitivity for these channels by three to four orders of magnitude. These significant improvements benefit from low QCD backgrounds and the Higgs boson tagging from recoil mass reconstruction at CEPC. As for the Higgs boson exotic decays without missing energy, the comparative improvements vary between two to three orders of magnitude, as LHC performance in these channels is improved by reconstruction of the Higgs boson mass from visible final state particles and reduced QCD backgrounds in events with leptons and photons.

### 2.3.2 EXOTIC $Z$ BOSON DECAYS

The CEPC's $Z$-pole run will offer unique possibilities to test new physics that allows the $Z$ boson to decay through new, exotic channels. Figure 2.19 summarizes the sensitivity of CEPC to exotic $Z$ decays, and it compares CEPC's sensitivity to that of the high-luminosity LHC (HL-LHC) and a proposed Tera $Z$ upgrade. Exotic $Z$ decay channels are classified by final states, the number of intermediate resonances, and different topologies. The final states considered here include $Z \to \not{E}+\gamma$, $\not{E}+\gamma\gamma$, $\not{E}+\ell^+\ell^-$, $\not{E}+jj$, $(jj)(jj)$ and $\gamma\gamma\gamma$. Each pair of photons, charged leptons, or jets can form a resonance, denoted with $()$. All six categories of final states are represented in Figure 2.19; several representative decay topologies are chosen for each category and correspondingly labeled on the bar-chart. For CEPC and Tera $Z$, the sensitivity reach for exotic $Z$ decay branching ratios $(\mathrm{BR})$ are plotted as blue and red bars. These projections include kinematic cuts, namely general $p_T$ and angular cuts on reconstructed objects, as well as an appropriate invariant mass cut if there is a resonance in the pair of particles (including dark matter particles). The cuts are optimized for each topology by checking the kinematic variable distributions. The sensitivity reach for the HL-LHC at 13 TeV with $\mathcal{L} = 3\ \mathrm{ab}^{-1}$ has been computed in a similar way. Details of the simulation can be found in Ref. [119].

The sensitivity to final states with missing energy reaches branching ratios of $10^{-6}$ to $10^{-9.5}$ for CEPC and $10^{-7}$ to $10^{-11.5}$ for Tera $Z$. For each topology, the light blue and red shaded regions indicate the range from varying the model parameters, like mediator or dark matter mass. The light color regions with dashed boundary show the optimal sensitivity, while the dark color regions with solid boundary show the pessimistic benchmark of the model. In all the channels, future $Z$ factories improve the sensitivity by several orders of magnitude above those of the HL-LHC.

In general, CEPC has several advantages compared to a hadron collider like the HL-LHC. First, an $e^+e^-$ collider has a much cleaner environment compared to a hadron collider with a huge QCD background. Second, in the Drell-Yan production of a $Z$ boson at a hadron collider, the decay products tends to be soft because the $Z$ boson mass is small compared to the beam energy, which makes them hard to detect at the HL-LHC. Therefore, it is natural that CEPC has better sensitivity compared to the HL-LHC and provides a better opportunity to investigate dark sector physics through exotic $Z$ decays.

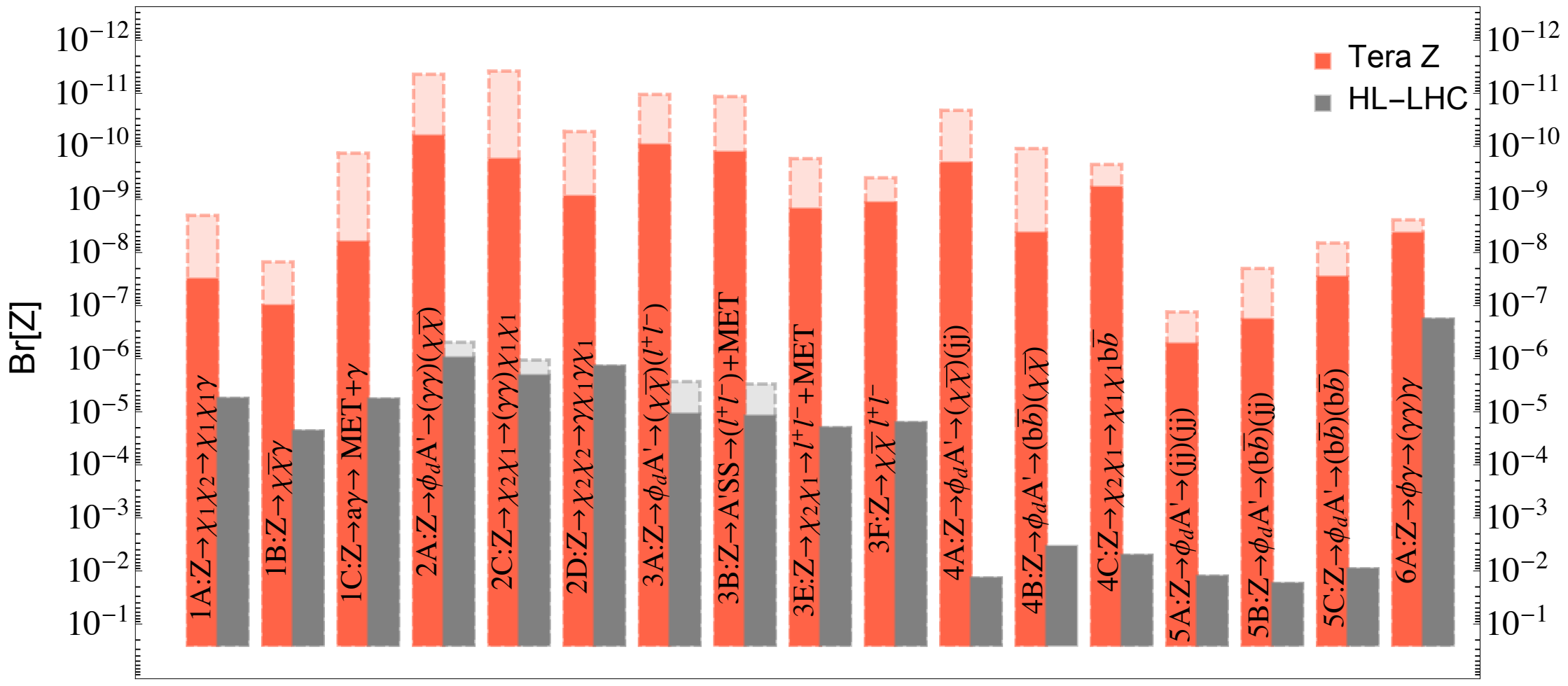

**Figure 2.19:** The sensitivity reach in the $Z$ branching ratio for various exotic $Z$ decay topologies at CEPC ($10^{12}$ $Z$), and the high luminosity LHC at 13 TeV with $\mathcal{L} = 3\ \mathrm{ab}^{-1}$. Adapted from Ref. [119].

Two specific benchmark scenarios demonstrate the significant power of exotic $Z$ decays to probe different dark (hidden) sectors [119]. (Further discussion of a variety of exotic $Z$ decays appears in [120].) The first model contains fermionic dark matter interacting with a singlet real scalar $S$, which mixes with the Standard Model Higgs boson. The possible exotic $Z$ decay channel in this case is $Z \to \tilde{s}Z^* \to (\bar{\chi}\chi) + \ell^+\ell^-$, where $\tilde{s}$ is the light scalar mass eigenstate (mostly the dark Higgs $S$) and $\chi$ is the fermionic dark matter. The second model is an axion-like particle $a$ coupling to the Standard Model $\mathrm{U}(1)_Y$ gauge field $B_\mu$. Then the exotic $Z$ decay is $Z \to a\gamma \to (\gamma\gamma)\gamma$. The final state is $3\gamma$ and in the case that $m_a$ is too small to separate the two photons, the final state is $2\gamma$. The sensitivity of exotic $Z$ decays (as well as other possible probes) to key parameters in these two models is summarized in Figure 2.20.

Projections for CEPC and Tera $Z$ reach in the first model are shown in the Figure 2.20(a). There are two free parameters, namely the Higgs mixing angle $\sin\alpha$ and dark Higgs boson mass $m_{\tilde{s}}$. The other two parameters related to dark matter are fixed. One is the dark matter mass, fixed close to half of $m_{\tilde{s}}$, which only affects the dark matter relic abundance but not other limits. The other one is the Yukawa coupling between dark matter $\chi$ and the dark Higgs $\tilde{s}$, which is taken to be $y_\chi = 0.1$ for illustrative purposes. Limits are projected for the exotic $Z$ decay process $Z \to \ell^+\ell^-\tilde{s} \to \ell^+\ell^-(\bar{\chi}\chi)$, which has been labeled as an orange solid line for CEPC ($10^{10}$ $Z$) option and a red dot-dashed line for the Tera $Z$ ($10^{12}$ $Z$) option, and compared with the LEP result with an integrated luminosity $114\ \mathrm{pb}^{-1}$ [121] labeled as "LEP-Zs-inv".

The dark Higgs in this benchmark scenario can also be constrained by the modification of SM Higgs couplings proportional to the mixing angle $\sin\alpha$, independent of the scalar mass $\tilde{s}$. The global fit to Higgs data at the LHC 7 TeV and 8 TeV runs can constrain the single scaling factor to Higgs interactions, giving $\sin\alpha < 0.33$ [122]; this is labeled as "$\tilde{h}$ current global fit (LHC)". The HL-LHC can extend this reach to $\sin\alpha < 0.28$ $(0.20)$ using 300 fb$^{-1}$ (3 ab$^{-1}$) luminosity [4]. At CEPC, the precision measurement of the Higgs Bremsstrahlung cross-section $\sigma(ZH)$ can reach the accuracy of $\mathcal{O}(0.3\% - 0.7\%)$ expected from $5 - 10$ ab$^{-1}$ [123–125], which can probe the scalar mixing down to $0.055 - 0.084$ [126]; this is labeled as "$\delta\sigma(ZH)$". In addition, there are constraints coming from the invisible decay of the SM Higgs boson. The current LHC limits from the Run I combination of ATLAS and CMS data constrains $\mathrm{BR}(h \to \mathrm{inv}) \leq 0.23$ at 95% CL [127, 128]. Following the $\tilde{h}$ invisible decay branching ratio in the Higgs portal dark matter model, the limit on the mixing angle $\sin\alpha$ is labeled as "$\mathrm{BR}^{\tilde{h}}_{\rm inv} < 0.23$". We also add the HL-LHC (3 ab$^{-1}$) and future $e^+e^-$ collider projections on invisible Higgs boson search, which lead to 95% CL limits $\mathrm{BR}^{\tilde{h}}_{\rm inv} \lesssim 0.08 \sim 0.16$ [129, 130] and $\mathrm{BR}^{\tilde{h}}_{\rm inv} \lesssim 0.003$ [124, 131] at ILC and CEPC. There are also constraints based on dark matter assumptions. The dark matter relic abundance [132] is satisfied on the dashed gray line, while the direct detection limits on spin-independent cross-sections (XENON1T [133], LUX [134], PANDAX-II [135], and CRESST-II [136]) exclude the region within the dashed green line.

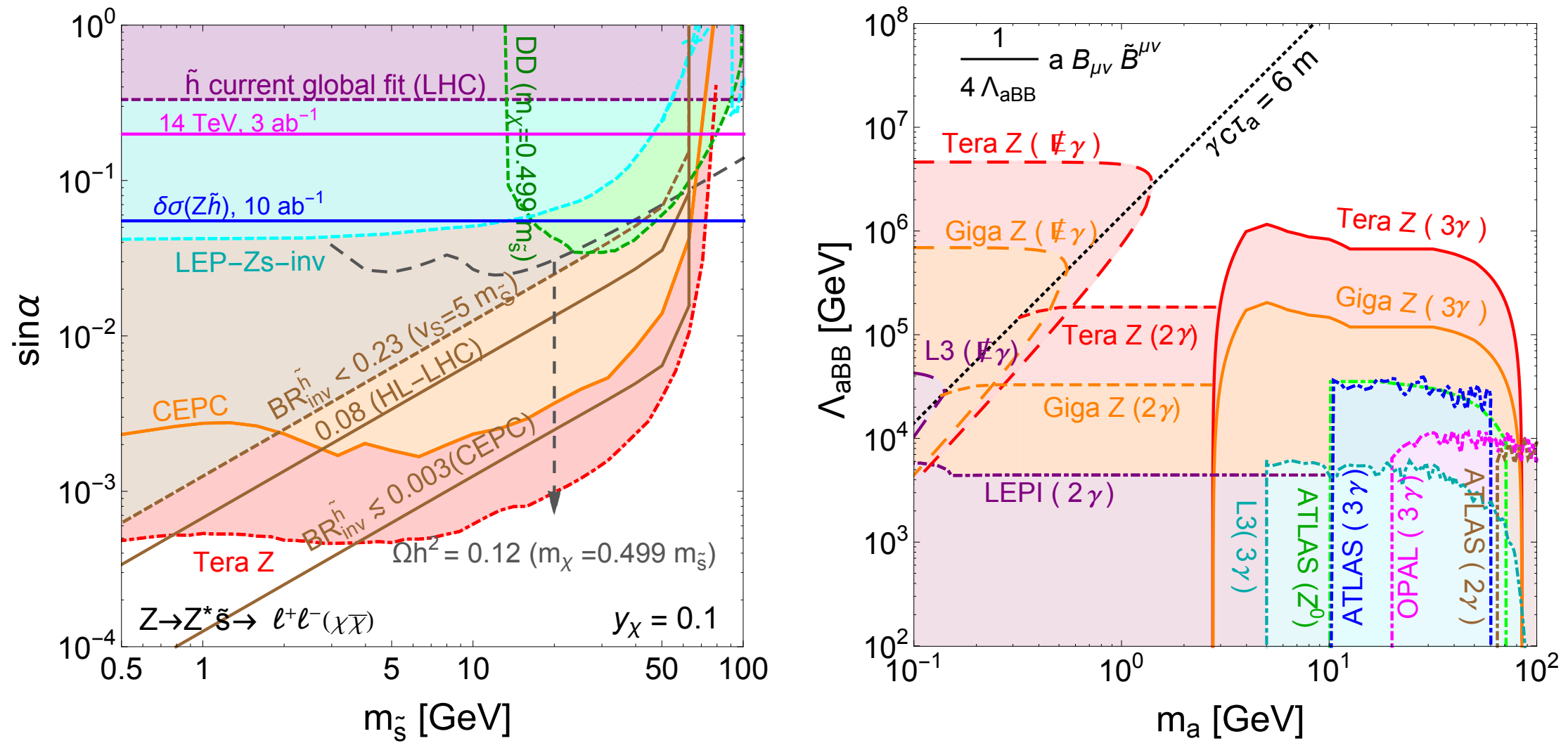

(a) (b)

**Figure 2.20:** The reach for rare $Z$ decays at CEPC in two benchmark scenarios, adapted from Ref. [119]. (a) the sensitivity to the dark Higgs mixing angle $\sin\alpha$ at CEPC ($10^{10}$ $Z$) and at a Tera $Z$ option ($10^{12}$ $Z$) in a Higgs portal dark matter model, using the process $Z \to \ell^+\ell^-\tilde{s} \to \ell^+\ell^-(\bar{\chi}\chi)$. The model points on the gray dashed contour have correct thermal relic abundance under a specific assumption about the masses of the dark matter and the dark Higgs, as indicated by the arrow in the figure. (b) the sensitivity to the coupling $\Lambda_{\rm aBB}$ for an axion-like particle (ALP) model as a function of the ALP mass $m_a$, where B is the hypercharge gauge field. The signal process is $Z \to \gamma a$, where $a$ can decay to a pair of photons ($3\gamma$), be detected as one photon due to high boost ($2\gamma$), or be detected as missing energy due to its long lifetime ($\gamma\not{E}$).

Projections for CEPC and Tera $Z$ reach in the second model are illustrated in the Figure 2.20(b), focusing on the exotic $Z$ decay $Z \to \gamma a$ followed by $a \to \gamma\gamma$. In the $3\gamma$ signal, the ALP mass is heavy enough that the two photons are well separated and detectable. When the mass of the ALP is below $\mathcal{O}(1)$ GeV, the boost of the axion makes the two photons from the axion decay close enough together that they cannot be resolved, leading to signals in the $2\gamma$ search channel. The current constraints on the two cases are given by LEP and LHC photon searches. In Figure 2.20, the LEP I [137] constraint uses an inclusive diphoton search $e^+e^- \to 2\gamma + X$ covering the small mass region. In the higher mass region, the boost of the axion decreases and the $3\gamma$ channel is considered. The LEP II (OPAL) constraints have $2\gamma$ and $3\gamma$ data [138], which are employed to put bounds on the process $e^+e^- \to \gamma/Z^* \to a\gamma \to 2\gamma + \gamma$. ATLAS $3\gamma$ and $Z \to 3\gamma$ [139, 140] searches can be translated to an ALP bound, as derived in [141]. There is also the possibility that the ALP decays outside of the detector, which is relevant for a $\not{E} + \gamma$ search. In this case the strongest bound comes from the LEP L3 collaboration with $137\ \mathrm{pb}^{-1}$ data at the $Z$ pole [142], which constrains the branching ratio of the exotic decay $Z \to \gamma\not{E}$ down to $1.1 \times 10^{-6}$ if the photon energy is greater than $\sim 30$ GeV. It directly excludes $\Lambda_{\mathrm{aBB}} < 4.3 \times 10^4$ GeV for $Z \to \not{E} + \gamma$ decay, and is labeled as "L3 ($\not{E}\gamma$)" in the Figure 2.20(b). The sensitivity curves are plotted as an orange solid line for CEPC ($10^{10}$ $Z$) and a red dot-dashed line for a Tera $Z$ ($10^{12}$ $Z$) option, demonstrating the significant reach of CEPC and Tera $Z$ in this scenario.

These comparisons show that searches for exotic $Z$ decays at CEPC (and a possible Tera $Z$ extension) can provide the leading sensitivity to a range of motivated extensions of the Standard Model, substantially exceeding the reach of dark matter direct detection experiments, current limits from collider searches, and estimated sensitivities of the high luminosity run of the LHC (HL-LHC).

### 2.3.3 DARK MATTER AND HIDDEN SECTORS

Observations tell us that the majority of matter in the universe is Dark Matter (DM). Because the abundance of dark matter in the universe is within an order of magnitude of the abundance of ordinary matter, it is natural to suspect that dark matter and ordinary matter should be related in some way. A variety of models, including the classic thermal relic weakly interacting massive particle (WIMP), attempt to explain the abundance of dark matter in terms of its interactions with ordinary matter. In some models, there is a richer "dark sector" consisting not only of dark matter itself but of new force-carrying particles that can mediate self-interactions between dark matter particles or interactions of dark matter with ordinary matter.

Different classes of possibilities for how dark matter interacts with the Standard Model have been summarized in Section 2.3. Below we discuss each of these possibilities in turn. This categorization of studies may be useful in the future for identifying DM scenarios at CEPC that have not yet been fully studied.

There are major efforts underway to search for dark matter via direct detection, indirect detection, and searches at the LHC and lower-energy-but-high-luminosity collider and fixed-target experiments. It is possible that one of these experiments will discover a dark matter signal before CEPC operates. Even in that case, CEPC can play a crucial role in discovering the *nature* of the dark matter particle. Direct detection, for example, may tell us a spin-independent scattering rate, but without knowledge of the local dark matter density or whether the particle we are seeing constitutes all of the dark matter or is just a component, limited knowledge of particle physics would be gleaned from the discovery. The role of CEPC in such a case could be to tell us that dark matter interacts directly with the Higgs boson or weak gauge bosons, for instance. Below we will emphasize both cases in which CEPC can *measure* dark matter properties and supplement other experiments and cases in which CEPC could play the crucial role in *discovering* a DM signal for the first time.

## DARK MATTER IN ELECTROWEAK MULTIPLETS

The CEPC's strength is electroweak physics, both through precision measurements of properties of the $W$ and $Z$ bosons and through its primary role as a Higgs factory. Studies of CEPC's capabilities for detecting new electroweak physics include Refs. [21, 24, 42, 143–149]. Hence, the most natural place to begin is with CEPC searches for dark matter particles that are in electroweak multiplets (e.g. doublets or triplets of $\mathrm{SU}(2)_L$) or mixtures of electroweak multiplets (including admixtures of a singlet). Studies on this topic include Refs. [150–155].

One question is whether other, dedicated dark matter experiments will cover the full parameter space of dark matter in electroweak multiplets. Dark matter direct detection experiments, like the currently-operating Xenon1T [133] and PandaX [156], are currently probing much of the parameter space for spin-independent dark matter scattering on nucleons mediated by Higgs exchange. The current bound on the DM-nucleon cross section of a few times $10^{-46}$ $\mathrm{cm}^2$ corresponds to an $h\chi\chi$ coupling in the Lagrangian with coefficient of order $10^{-2}$. Future experiments like DARWIN [157] will potentially push the search down to the neutrino floor, corresponding to $h\chi\chi$ couplings of order $10^{-3}$, which is smaller than its typical size. This will probe a large swath of the parameter space for electroweak dark matter.

As noted above, CEPC could help to measure DM properties even if a direct detection experiment makes the discovery first. Still more interesting are possibilities in which electroweak DM could be *missed* by direct detection experiments but seen by CEPC. There are two main scenarios to consider where this could happen. The first is if DM is a nearly pure electroweak multiplet, such as a pseudo-Dirac higgsino. Such particles have very small interactions with the Higgs boson, so their direct detection rate is loop-suppressed and at about the level of the neutrino floor [158]. These particles would also be very difficult to detect at the LHC [159]. Indirect detection may constrain them, but at low mass their thermal abundance is low, and even a significant non-thermal abundance may fall below current constraints [160, 161]. A second possibility is that DM lies in a mixed electroweak multiplet with couplings to the Higgs boson, but the coupling of the lightest mass eigenstate has a small coupling to the Higgs boson, either accidentally or due to an approximate symmetry. This is referred to as a *blind spot* for direct detection [162, 163]. For instance, a mostly-wino dark matter particle in a supersymmetric theory has vanishing tree-level coupling to the Higgs boson if $M_2 = -\mu \sin(2\beta)$. In some cases, a spin-independent blind spot may be covered by spin-dependent scattering. Blind spots might also be uncovered by collider searches [164].

Robust blind spots for both spin-dependent and spin-independent scattering arise in some theories due to approximate *parity* or *custodial* symmetries. In the MSSM, this occurs for higgsino dark matter at $\tan\beta = 1$ and $\mathrm{sign}(\mu M_{1,2}) < 0$. In closely related theories, these blind spots have been understood to result from custodial symmetries [151]. These robust direct detection blind spots are excellent opportunities for CEPC to play a role in dark matter physics, so let us explain the physics in somewhat more detail. They arise for pseudo-Dirac DM, i.e. theories with a Dirac mass term of the form $\mu\chi_1\chi_2$ which can be written as a sum of two Majorana mass terms, $\mu(\chi_+\chi_+ - \chi_-\chi_-)$ where $\chi_\pm =$ $(\chi_1 \pm \chi_2)/\sqrt{2}$. In such a theory the $Z$ boson couples off-diagonally, $Z_\mu(\chi_+^\dagger \overline{\sigma}^\mu \chi_- + \text{h.c.})$. Mixing or higher-dimension operators can split the mass eigenstates, but in the custodially symmetric limit, the eigenstates remain $\chi_+$ and $\chi_-$ rather than mixtures thereof. There is a parity symmetry under which $\chi_+$ and the $Z$ are odd but $\chi_-$ and $h$ are even, which forbids an $h\chi_+\chi_+$ coupling. Hence when $\chi_+$ is the lighter mass eigenstate, both spin-dependent and spin-independent scattering are turned off.

A number of studies have been carried out on two particular models of electroweak dark matter, the doublet–singlet and doublet–triplet models (e.g. [165–167]). The doublet–singlet model introduces a singlet fermion $S$ (with zero hypercharge) with Majorana mass $-(m_S/2)SS$ and two electroweak doublet Weyl fermions $D_{1,2}$ with opposite hypercharges $\mp 1/2$ and Dirac mass $-m_D \epsilon_{ij} D_1^i D_2^j$, together with mixing through the SM Higgs boson:

$$y_1 H S D_1 - y_2 H^\dagger S D_2 + \text{h.c.}\,. \tag{2.19}$$

The doublet–triplet model introduces the same doublet fields as well as an $\mathrm{SU}(2)$ triplet with zero hypercharge, $T$, with a Majorana mass $-(m_T/2)T^iT^i$ and mixing with the doublet through the Higgs boson:

$$y_1 (H\sigma^i D_1)T^i - y_2(H^\dagger \sigma^i D_2)T^i + \text{h.c.}\,. \tag{2.20}$$

Both of these models have blind spots for both spin-independent *and* spin-dependent direct detection in the pseudo-Dirac case when $m_D < m_{S,T}$ (all mass parameters taken to be positive) and $y_1 = y_2$. An explicit rewriting of the Lagrangian that makes a custodial symmetry manifest in this limit has been given in [151]. This blind spot can also be understood in terms of a parity symmetry at the point $y_1 = y_2$ along the lines explained in the previous paragraph.

In the SUSY context we can identify the fields $S$, $D$, and $T$ with the bino, higgsino, and wino. In this case the couplings $y_1$ and $y_2$ are equivalent to $g^{(\prime)} \cos\beta$ and $g^{(\prime)} \sin\beta$ in the doublet–triplet (doublet–singlet) case. These relatively small couplings tend to lead to small signals at CEPC. However, it is also interesting to consider extensions of the MSSM with an *additional* doublet and singlet that mix to serve as dark matter. Such theories can help to explain why the observed Higgs boson mass is heavier than expected in the simplest SUSY theories [168], which offers a motivation for considering the larger values of $y_{1,2}$ that could be probed at CEPC.

Precision electroweak physics at the $Z$ pole is most sensitive to the $S$ and $T$ parameters. Although these operators appear in studying the propagators of gauge fields, they originate from new physics that couples to the Higgs boson. For instance, in the basis of Ref. [169], the $S$ parameter is related to the operators $H^\dagger \sigma^i H W^i_{\mu\nu} B^{\mu\nu}$, $(H^\dagger \sigma^i \overleftrightarrow{D}_\mu H) D^\nu W^i_{\mu\nu}$, and $(H^\dagger \overleftrightarrow{D}_\mu H)\partial^\nu B_{\mu\nu}$; the $T$ parameter, to $(H^\dagger \overleftrightarrow{D}_\mu H)^2$. These operators are generated in the doublet–singlet or doublet–triplet model because the fermions mix by coupling to the Higgs boson. On the other hand, for a pure electroweak multiplet like the pseudo-Dirac higgsino, Higgs couplings are very small and $S$ and $T$ are suppressed. The $T$ parameter is also suppressed in models with a good approximate custodial symmetry. In such theories, other electroweak precision observables like the $W$ and $Y$ operators $(D^\mu W^i_{\mu\nu})^2$ or $(\partial^\mu B_{\mu\nu})^2$ may be relatively important, though they are generated with small coefficients and are harder to probe. In this case, observables at $240$ GeV from processes like $e^+e^- \to \mu^+\mu^-$ [170] or $e^+e^- \to W^+W^-$ [152, 171] may be more effective probes of electroweak dark matter than $Z$-pole observables.

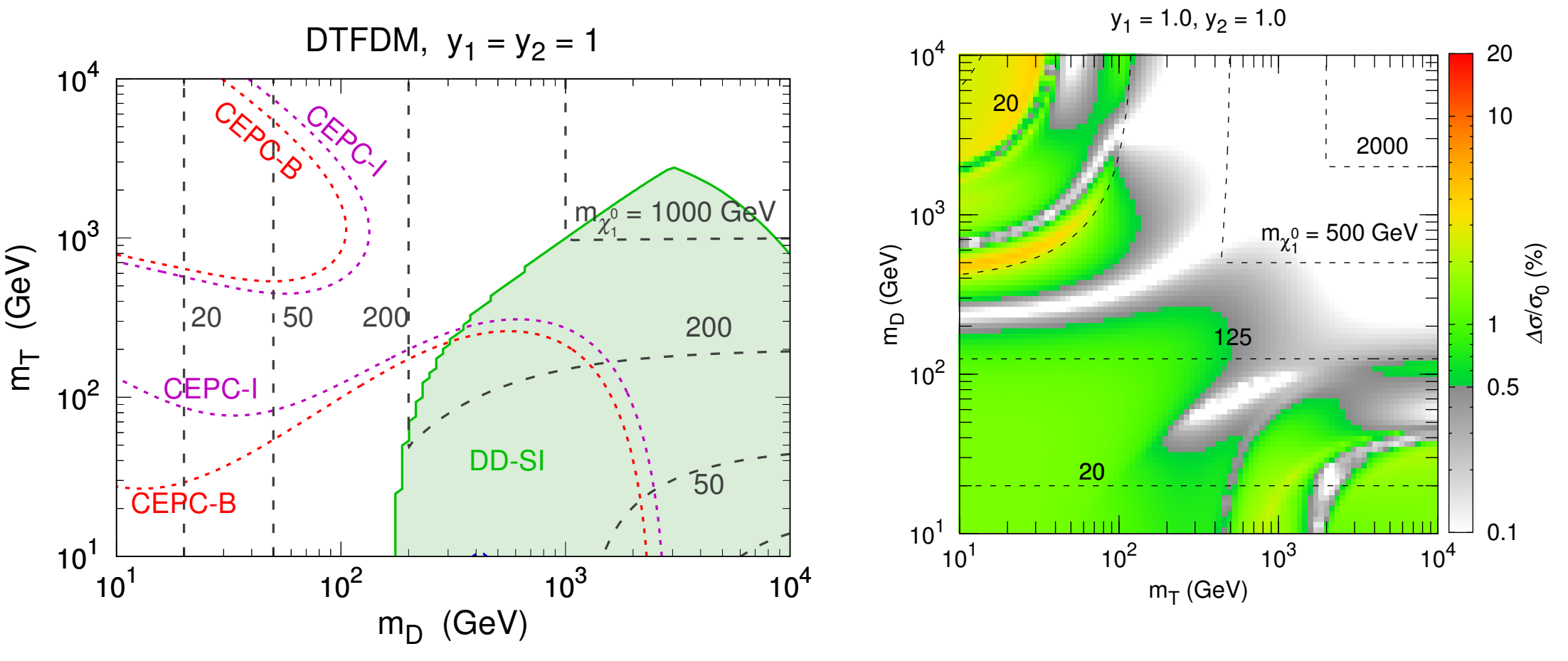

(a) (b)

**Figure 2.21:** (a) The CEPC electroweak precision $(S, T)$ fit probe of the doublet–triplet model at the custodially symmetric point $y_1 = y_2 = 1$, taken directly from Figure 5a of Ref. [151]. When the dark matter particle is mostly triplet ($m_D \gg m_T$), spin-independent direct detection is a powerful probe (shaded green region). When the dark matter particle is mostly doublet, the tree-level direct detection rate vanishes but CEPC's measurement of the $S$ parameter becomes a powerful probe (dashed contours). (b) CEPC's sensitivity to the same model via the Higgsstrahlung cross section $\sigma(ZH)$, taken directly from Figure 11b of Ref. [154]. We see that in a large part of parameter space with $m_T \gg m_D$, where the direct detection rate is low due to custodial symmetry, there are observable (percent-level or higher) deviations in the $ZH$ cross section.

The doublet–singlet and doublet–triplet models at CEPC have been discussed in Ref. [151], which focuses on the $S$ and $T$ parameters (and also discusses a quadruplet–triplet model with similar properties).[2] They have shown that CEPC can probe a large region of parameter space where the dark matter mass is below $200$ GeV, and certain regions of parameter space with even larger masses. In particular, the $S$ parameter allows a probe of the custodially symmetric region that is hidden from direct detection. We show some results from this paper in the Figure 2.21(a). A related study in Ref. [154] considers effects of doublet–singlet and doublet–triplet dark matter on Higgs observables, including the $ZH$ cross section, the $h \to \gamma\gamma$ decay rate, and the Higgs boson invisible width. Away from the custodially symmetric point in the doublet–singlet model, when $y_1 = 0.5$ and $y_2 = 1.5$, CEPC's measurement of the total $ZH$ cross section probes the lightest neutralino mass up to $200$ GeV. For $y_1 = y_2 = 1$, with custodial symmetry, deviations are smaller and $m_D$ is probed only up to about $125$ GeV. In the doublet–triplet case, the region of parameter space bounded by the $ZH$ measurement is illustrated in the Figure 2.21(b). Aspects of a slightly different doublet–singlet model, with the singlet taken to be a Dirac fermion, have also been discussed in Ref. [150]. They focus on the region with mostly singlet DM, in which case the doublet may be thought of as allowing a completion of a "Higgs portal" model. In this case, the most important constraints come from the $T$ parameter. They also present results for a wider range of doublet and singlet masses including cases where dark matter is mostly doublet.

[2]Earlier papers discussing electroweak and Higgs constraints on similar models include [172–176].

In the case in which DM resides in a nearly pure electroweak multiplet, the $S$ and $T$ parameters and the $H \to \gamma\gamma$ rate are no longer useful probes. For the case of nearly pure higgsinos, Ref. [152] has studied the prospects of an $e^+e^- \to W^+W^-$ measurement at CEPC as a constraint. This measurement is sensitive not only to corrections to the photon and $Z$ propagators but to loop corrections to the triple gauge coupling vertex. Ref. [152] claims that a $0.1\%$ precision measurement of $e^+e^- \to W^+W^-$ at CEPC could probe higgsino dark matter up to about $210$ GeV. However, the scatter plot in Figure 1 of that reference suggests that many models with even heavier higgsinos will be accessible. A more detailed future exploration of the parameter space probed by the $W^+W^-$ measurement would be useful. The rate of $e^+e^- \to \mu^+\mu^-$ at $240$ GeV can also be a sensitive probe of deviations in the propagators of photons and $Z$ bosons; in particular, for new physics contributing to the $W$ and $Y$ parameters but not to $S$ and $T$, it may be superior to electroweak precision studies on the $Z$ pole thanks to the larger center-of-mass energy. A detailed study of this probe of electroweak physics has been carried out in Ref. [170]. Their conclusion is that if systematic uncertainties can be controlled to achieve a $0.1\%$ precision on the rate, pseudo-Dirac higgsinos may be excluded up to a mass of about $200$ GeV. This is encouraging, since pseudo-Dirac doublets are among the most difficult electroweak particles to probe in any experiment. In particular, the LHC is not expected to reach far above $200$ GeV (though this will depend in part on how well systematic uncertainties can be understood). The results of Ref. [170] may not apply directly to CEPC due to their assumptions about beam polarization, so a further dedicated CEPC study of this process is warranted.

Another interesting possibility is that of light singlet dark matter mixing with heavier electroweak-charged particles. A particular example arises for mostly-bino dark matter in the MSSM [177], $\tilde{\chi}^0_1$, which could have a non-thermal relic abundance. Because the bino is a pure singlet, it couples to the Standard Model only through small mixing parameters and is difficult to detect directly. However, in some cases it can be either detected or constrained through the invisible width of the Higgs boson. The parameter space probed by dark matter direct detection and CEPC is shown in the Figure 2.22(b). This figure illustrates that CEPC could probe the region allowed by the current direct detection with a sensitivity to BR$(H \to \tilde{\chi}^0_1\tilde{\chi}^0_1) \gtrsim 0.24\%$.

## STANDARD MODEL PORTALS

If the dark matter does not reside in an electroweak multiplet, it may still interact with the SM particles through gauge-invariant "portal" operators. The portal operators include

$$H^\dagger H\,, \quad B^{\mu\nu}\,, \quad \text{and} \quad HL\,, \tag{2.21}$$

where $H$ is the SM Higgs doublet, $B^{\mu\nu}$ is the hypercharge field strength tensor, and $L$ is a SM lepton doublet. These three portals are usually referred to as the Higgs portal, the kinetic mixing (or hypercharge) portal, and the lepton (neutrino) portal. These simple portal dark matter scenarios predict rich phenomenology and a plethora of experimental signatures. They have been established as well-defined dark matter benchmarks and experimental targets, in addition to the traditional electroweak WIMP scenario.

The many powerful direct and indirect probes available at the CEPC mean that it could play an important role in detecting and testing these SM portals to dark matter. Below we will present estimates of the CEPC potential for the Higgs and kinetic mixing portals based on the studies in the existing literature. The neutrino portal is discussed further in Section 2.3.4.

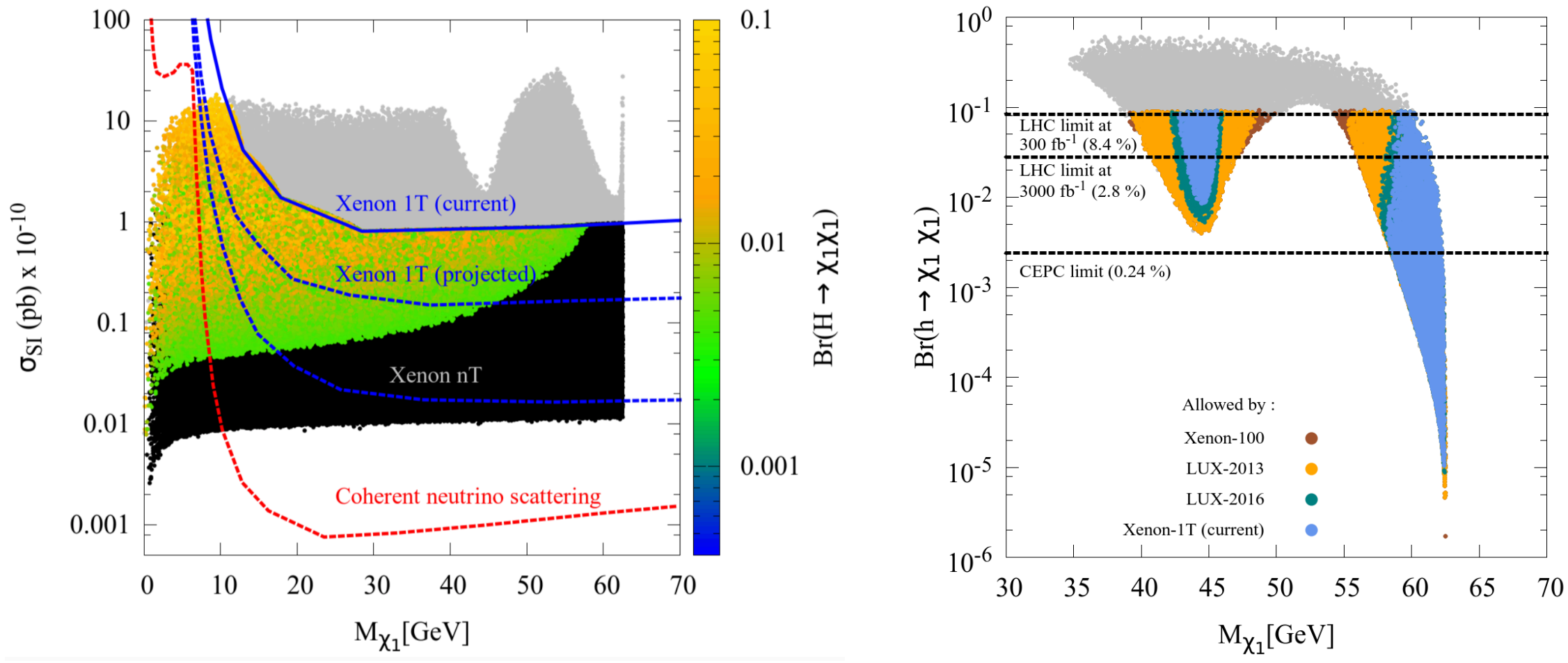

(a) (b)

**Figure 2.22:** These figures, which are adapted from Ref. [177], show constraints on nonthermal neutralino dark matter and invisible Higgs boson decays. (a) The spin-independent WIMP-nucleon cross-section vs $M_{\tilde{\chi}_1^0}$ for all points allowed by collider and relic density constraints. The color code characterizes the value of $\mathrm{BR}(H \to \tilde{\chi}_1^0 \tilde{\chi}_1^0)$, while black points have $\mathrm{BR}(H \to \tilde{\chi}_1^0 \tilde{\chi}_1^0) < 0.4\%$. The solid blue line shows the current limit from LUX-2016 [178], and the dashed blue line shows the reach for Xenon1T [179] and Xenon-nT [179]. (b) The Higgs boson to invisible branching ratio $\mathrm{BR}(H \to \tilde{\chi}_1^0 \tilde{\chi}_1^0)$ vs. the LSP mass $M_{\tilde{\chi}_1^0}$. The gray (colored) points distinguish the points allowed before (after) the Higgs signal strength constraints. Blue, green, yellow, red points are allowed by the current limits on SI WIMP-nucleon cross-section from Xenon1T, LUX-2016, LUX-2013, and Xenon-100. From top to bottom, the black-dashed line represents the reach of the LHC with 300 fb$^{-1}$, the LHC with 3000 fb$^{-1}$, and CEPC.

In a simple example of the Higgs portal model, the dark matter (DM) is assumed to be either a real scalar ($S$) or a Majorana fermion ($\chi$), with the following interaction terms with the Higgs field [180, 181]

$$\mathcal{L} = -H^\dagger H \left( \frac{\lambda_{\mathrm{DM}}}{4} S^2 + \bar{\chi} \frac{y_{\mathrm{DM}} + i y_{\mathrm{DM}}^P \gamma_5}{\sqrt{2} v} \chi \right) . \tag{2.22}$$

The couplings between a single Higgs boson particle and the dark matter fields are thus given by

$$\mathcal{L} = -\frac{\lambda_{\mathrm{DM}} v}{4} H S^2 - \frac{y_{\mathrm{DM}}}{\sqrt{2}} H \bar{\chi} \chi - \frac{i y_{\mathrm{DM}}^P}{\sqrt{2}} H \bar{\chi} \gamma_5 \chi \, . \tag{2.23}$$

For dark matter masses smaller than $m_H/2$, the decay channel $H \to SS/\bar{\chi}\chi$ is open, which produces the signal of Higgs boson invisible decays. As shown in Section 11.1, the CEPC could reach a sensitivity of 0.31% (at 95% CL) on the branching ratio of Higgs boson invisible decays.[3]

[3] Here we only include the Higgs boson invisible decay to BSM particles. If the SM decay $H \to ZZ \to \nu\bar{\nu}\nu\bar{\nu}$ is also included, the bound on the Higgs boson invisible branching ratio becomes 0.42% instead. See Section 11.1 for more details.

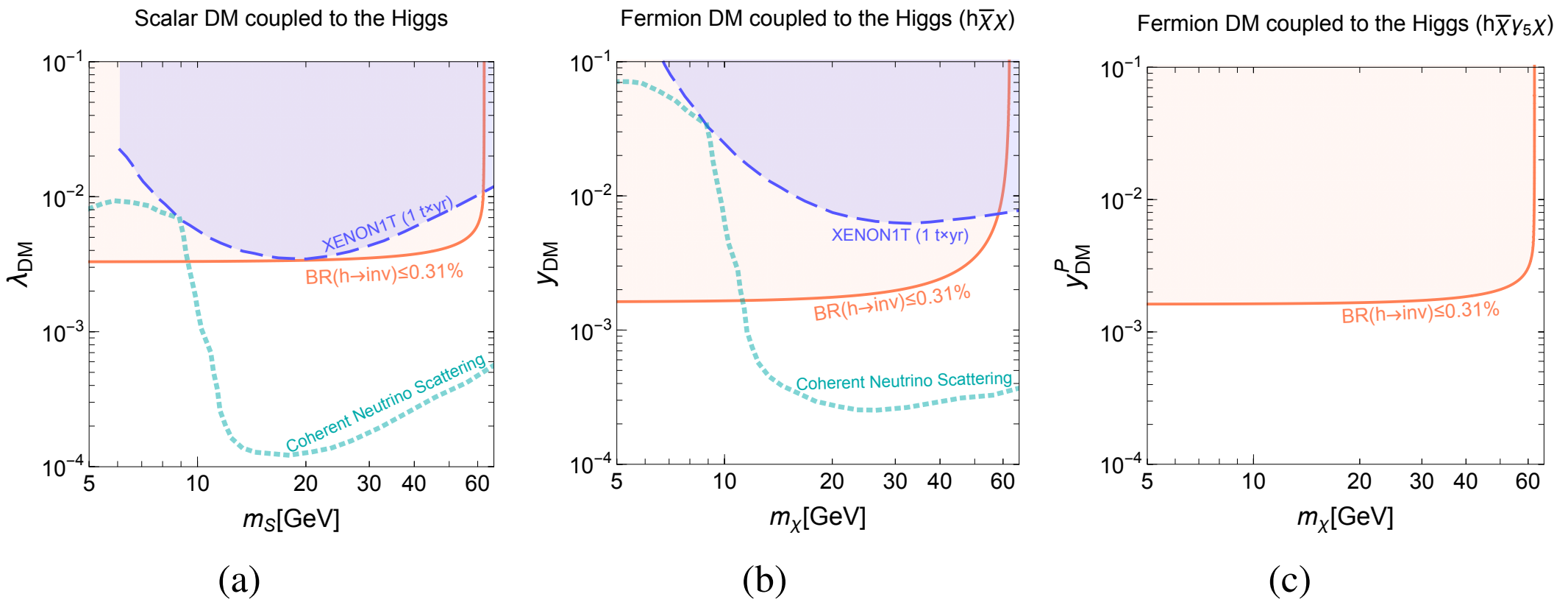

**Figure 2.23:** The *mass-coupling* plane for the Higgs portal models of Equation (2.23) with a scalar DM (a), a Majorana fermion DM with a scalar coupling $y_{\rm DM}$ (b) and a Majorana fermion DM with a pseudo-scalar coupling $y^P_{\rm DM}$ (c). The orange region is excluded by the invisible Higgs boson decay measurements at the CEPC, which constrains the branching ratio to be below 0.31% at 95% CL. The blue region is excluded by the most recent result from XENON1T [182]. The cyan dotted curve corresponds to the discovery limit set by the coherent-neutrino-scattering background, adapted from Ref. [183].

This provides considerable sensitivity to Higgs portal models with a dark matter mass below $m_H/2$, which can be competitive with the reaches of current and future direct detection experiments. To illustrate this, we make a comparison between the reach of the CEPC and the one from the most recent result of XENON1T [182] in the *mass-coupling* plane for both the scalar and Majorana fermion DM. For the fermion DM, we consider two separate scenarios, one with a purely scalar coupling ($y_{\rm DM}$) and the other with a purely pseudo-scalar coupling ($y^P_{\rm DM}$), as shown in Equation (2.23). We also assume that the correct relic abundance is achieved regardless of the model parameters. The results are shown in Figure 2.23. For the three scenarios in consideration, the CEPC bound on the Higgs boson invisible branching ratio, 0.31%, corresponds to a sensitivity to the Higgs-DM coupling of around $10^{-3}$ for DM mass smaller than $m_H/2$. For the scalar DM and Majorana fermion DM with coupling $y_{\rm DM}$, this clearly surpasses the reach of XENON1T in this mass region. Even for future direct detection experiments, the reach could not go beyond the so-called "neutrino floor" (shown by the cyan dotted curve) due to the coherent-neutrino-scattering background [183], while the CEPC could still probe a significant part of the region below the neutrino discovery limit in the region $m_{\rm DM} \lesssim 10\,{\rm GeV}$. The pseudo-scalar coupling $y^P_{\rm DM}$ only produces a spin-dependent WIMP-nucleon interaction which is suppressed by the transferred momentum. The constraints on the fermion DM with $y^P_{\rm DM}$ from direct detection experiments are thus much weaker, while the reach of the CEPC still remains strong. In addition to these bounds, the CEPC's sensitivity to fermionic Higgs portal dark matter through exotic $Z$ boson decays has been discussed in Section 2.3.2.

In Figure 2.24, the CEPC coverage of Higgs portal dark matter models for both scalar and fermionic DM ($y_{\rm DM}$) is converted to the corresponding spin-independent WIMP-nucleon cross-section, and compared to the coverage of direct detection experiments. In addition to the Xenon1T, the sensitivities of other experiments are also presented, including LUX (2017) [134] and PandaX-II (2017) [156], as well as future projections of PandaX4T with 5.6 $t \times yr$ data [184], XENONnT with 20 $t \times yr$ data [179], LUX-ZEPLIN (LZ) with $15.6\ t \times yr$ data [185] and a xenon experiment with $200\ t \times yr$ data [186] that corresponds to either DARWIN [157] or PandaX-30T. The current and future reaches of the LHC Higgs boson invisible decay measurements are also shown. The current bound, $\mathrm{BR}(H \to \mathrm{inv}) < 24\%$ at 95% CL, comes from the CMS analysis in Ref. [187]. The projection by the ATLAS collaboration on the reach of $\mathrm{BR}(H \to \mathrm{inv})$ at the HL-LHC is around 10% [188]. A study in Ref. [189] suggests that the reach could be improved to 3.5% with multivariate techniques. Both the current bound (24%) and the optimistic projection (3.5%) are plotted in Figure 2.24, which cover the possible range that the (HL-)LHC could reach in the future. Finally, the cyan dashed curve corresponds to the projected discovery limit from Ref. [183]. The region below this curve is inaccessible by direct detection experiments due to the coherent-neutrino-scattering background.

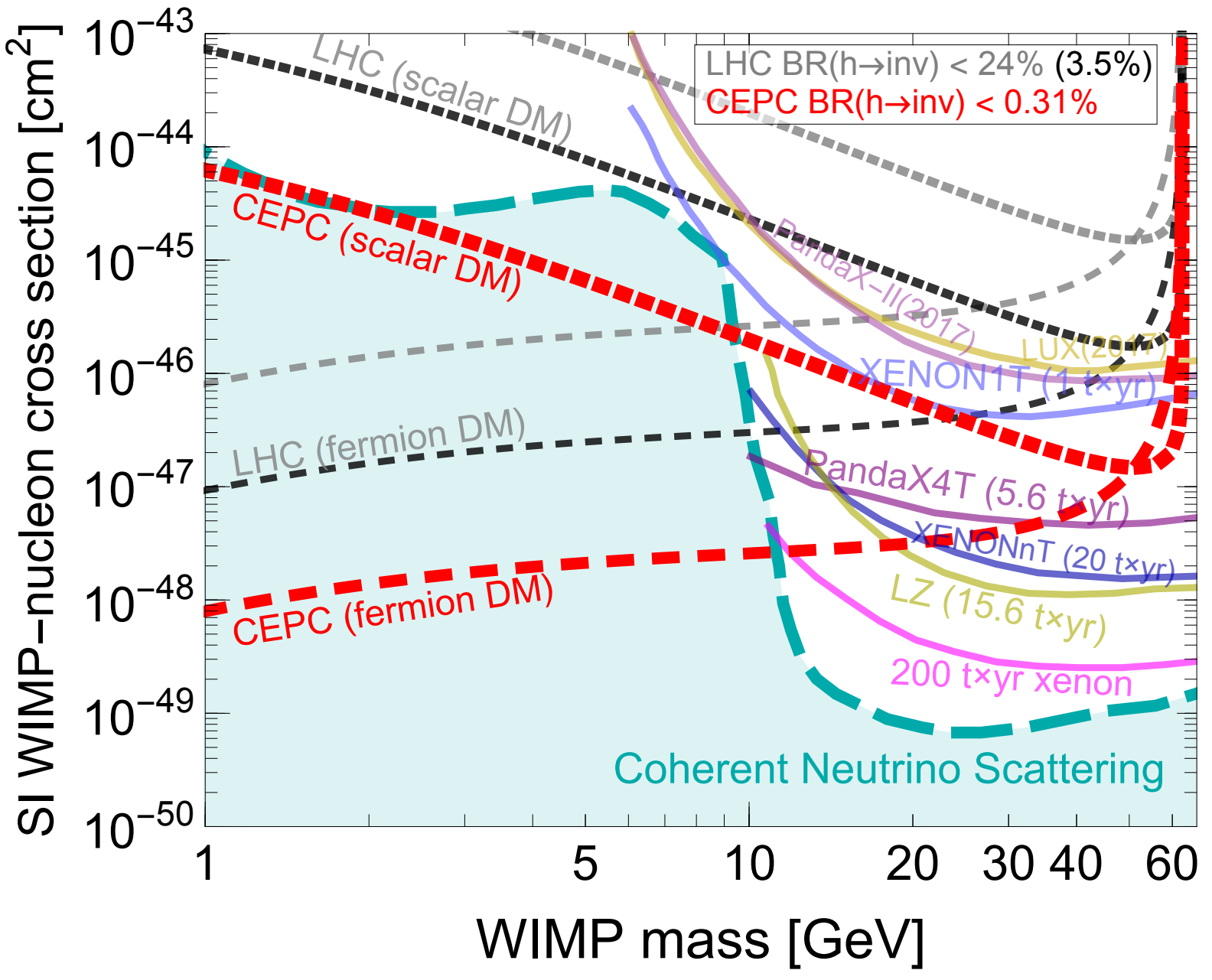

**Figure 2.24:** The sensitivity to the spin-independent WIMP-nucleon cross-section of current and future direct detection experiments, compared with the reach of Higgs boson invisible decay measurements at the LHC and CEPC in Higgs portal dark matter models. The direct detection limits are shown in solid lines, which include the most recent limits from LUX (2017) [134], PandaX-II (2017) [156], XENON1T [182] and future projections for PandaX4T [184], XENONnT [179], LZ [185] and a 200 $t \times yr$ xenon experiment [186]. For the Higgs portal models, the dark matter is assumed to be either a scalar or a Majorana fermion with a scalar coupling. The red dotted curves show the limits from CEPC which corresponds to a invisible Higgs boson branching ratio of $\mathrm{BR}(H \to \mathrm{inv}) < 0.31\%$ at the 95% CL. The gray dotted curves correspond to $\mathrm{BR}(H \to \mathrm{inv}) < 24\%$, the current limit at the LHC [187], and the black dotted curves correspond to $\mathrm{BR}(H \to \mathrm{inv}) < 3.5\%$, the projected reach at HL-LHC from Ref. [189]. The cyan dashed curve corresponds to the discovery limit set by the coherent-neutrino-scattering background, adapted from Ref. [183].

We see in Figure 2.24 that the sensitivity of the Higgs boson invisible decay measurements to the scalar DM and the Majorana fermion DM have different dependence on the mass. This is due to the following two reasons: first, the Higgs portal interaction with the scalar DM is a dimension-four operator, while the fermion one is of dimension five, which results in different mass dependence of the WIMP-nucleon cross-section; second, the Higgs boson decay rates are also different for the two cases, with $\Gamma(H \to SS) \propto (1 - 4m_S^2/m_H^2)^{1/2}$ and $\Gamma(H \to \bar{\chi}\chi) \propto (1 - 4m_\chi^2/m_H^2)^{3/2}$ , a result of the $s$ ($p$)-wave nature of the scalar (fermion). Nevertheless, for both scenarios, it is clear that the Higgs boson invisible decay measurements provide the strongest limit in the dark matter mass region below $\sim 10\,\mathrm{GeV}$. Not only do direct detection experiments become less efficient in this region due to the mass threshold, the "neutrino floor" is also higher in this region, which sets the limit for the reach of direct detection experiments regardless of the size and length of the experiment. For dark matter masses in the region $10\,\mathrm{GeV} \lesssim m_{\mathrm{DM}} < m_H/2$, the sensitivity of the Higgs boson invisible decay measurements is comparable with that of direct detection experiments. In particular, for fermion DM the CEPC still has sensitivity in regions not covered by PandaX4T, XENONnT or LZ. On the other hand, a $200\ t \times yr$ xenon experiment would fully surpass the reach of the CEPC in this region.

It bears emphasizing that, as mentioned earlier, the interaction term between the Higgs boson and the fermion DM in Equation 2.22 is of dimension five. Such a nonrenormalizable operator indicates that the theory is only an effective one, and needs to be UV completed at a higher scale. More specifically, this operator can be generated by integrating out a heavy mediator that connects the Higgs boson and the fermion DM. The validity of the effective theory thus requires the mediator to be heavier than the scale of the interaction. For direct detection experiments, the momentum exchange is in the nonrelativistic regime, and is at the MeV level. For the Higgs boson decay, the interaction scale is at the order of the Higgs boson mass. Our results for the fermion DM are thus only valid if the mediator is at least as heavy as the Higgs boson.

Next, let us consider the kinetic mixing portal scenario, in which the hidden sector containing the dark matter is charged under a broken dark Abelian gauge symmetry, $\mathrm{U}(1)_D$. The $\mathrm{U}(1)_D$ could mix with the SM hypercharge $\mathrm{U}(1)_Y$ through the operator

$$\frac{1}{2}\frac{\epsilon}{\cos\theta} Z_{D\mu\nu} B^{\mu\nu}, \tag{2.24}$$

where $\epsilon$ is the (dimensionless) kinetic mixing parameter and $\theta$ is the weak mixing angle. The heavy gauge boson associated with $\mathrm{U}(1)_D$, often called the dark photon, could be searched for at a lepton collider in a variety of ways. First, the dark photon introduces two effects in the fit of precision electroweak observables: a shift in the $Z$ mass observable and a shift of the $Z$ couplings to SM fermions. The $Z$-pole program at CEPC could improve the sensitivity to electroweak observables by a factor of 10 compared to LEP and push the reach of $\epsilon$ down to $\sim 10^{-3}$ for $m_{Z_D} < 90$ GeV [190]. A more powerful way is to search for dark photons directly through the radiative return processes such as $e^+e^- \to \gamma Z_D \to \gamma\mu^+\mu^-$. The search can be implemented by simply counting the number of events in the dimuon invariant mass spectrum in both the $Z$-pole and Higgs programs at CEPC. The direct searches probe $\epsilon \subset (3 \times 10^{-4} - 10^{-3})$ depending on $m_{Z_D}$ in the entire mass range up to 250 GeV that could be covered by CEPC [191], as illustrated in Figure 2.25. Another possible direct probe is the rare $Z$ decay: $Z \to h_D Z_D \to Z_D Z_D Z_D$, where $h_D$ is the dark Higgs. The reach of this search has been discussed in Section 2.3.2.

In the remainder of this subsection we will discuss a case study of a model with two renormalizable Standard Model–dark sector couplings, the Double Dark Portal model of Ref. [126]. This model rests on the observation that one possible origin for the mass of a $\mathrm{U}(1)_D$ dark gauge boson is through the VEV of a dark Higgs scalar $\Phi$ carrying $\mathrm{U}(1)_D$ charge. The $\mathrm{U}(1)_D$ gauge boson kinetically mixes with the photon (with mixing parameter

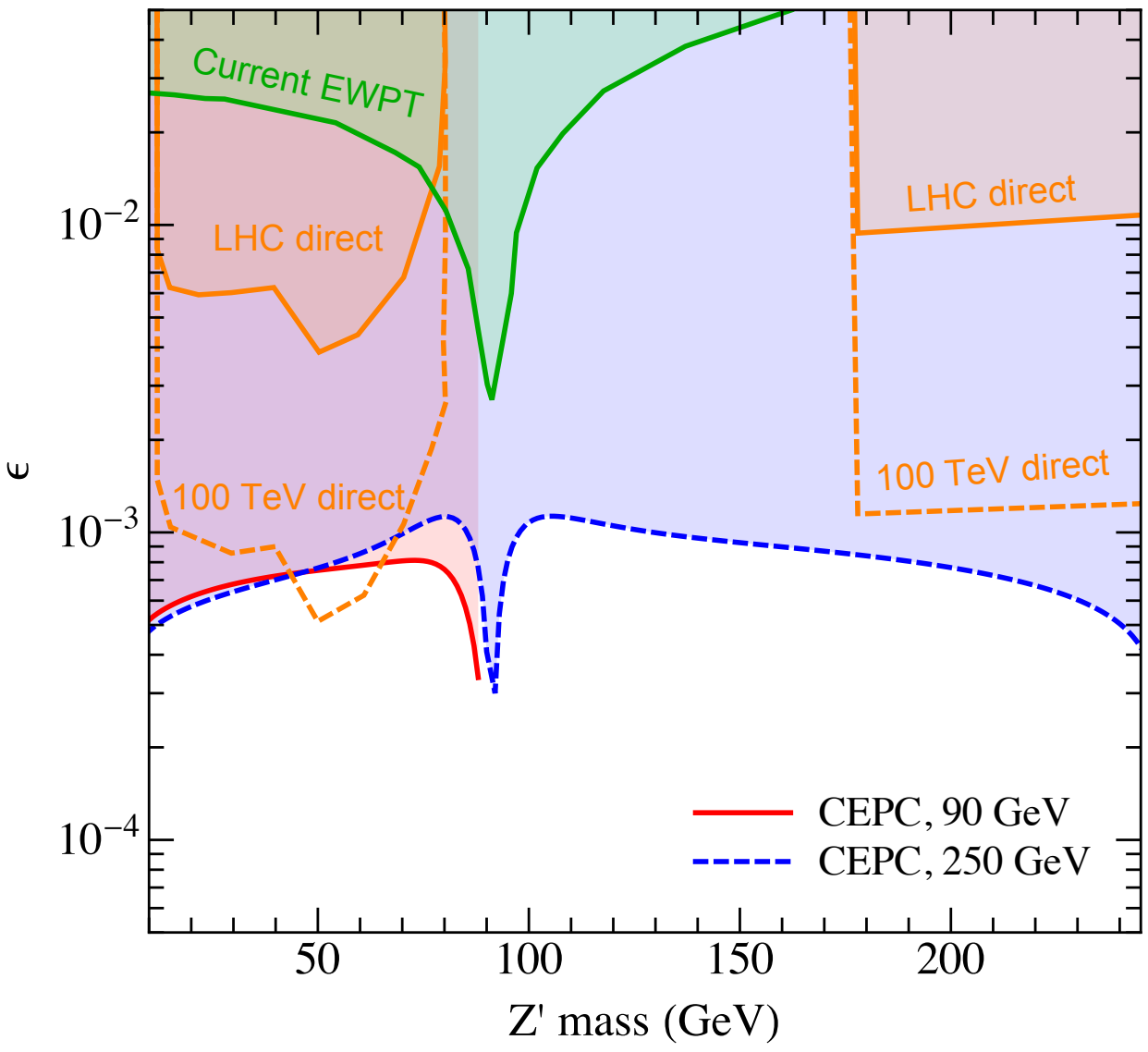

**Figure 2.25:** This figure illustrates CEPC's capacity to probe dark photons via radiative return. The red-solid and blue-dashed lines show the 95% CL projected sensitivity to the (hypercharge) mixing parameter, $\epsilon$, as a function of the dark photon's mass, $m_{Z'}$. The red curve corresponds to $\sqrt{s} = 90$ GeV and $\mathcal{L} = 0.5\ \text{ab}^{-1}$ while the blue curve shows $250$ GeV and $5\ \text{ab}^{-1}$. The figure is adapted from Refs. [190, 191].

| Parameter | Signal process | | Background (pb) | | Signal region |
|---|---|---|---|---|---|
| $\epsilon$ | $\tilde{Z}\tilde{K}$ | $\tilde{Z} \to \bar{\ell}\ell$, $\tilde{K} \to \bar{\chi}\chi$ | $\bar{\ell}\ell\bar{\nu}\nu$ | 0.929 | $N_\ell \geq 2$, $\vert m_{\ell\ell} - m_Z\vert < 10$ GeV, and $\vert m_{\text{recoil}} - m_{\tilde{K}}\vert < 2.5$ GeV |
| | | $\tilde{Z} \to \bar{\ell}\ell$, $\tilde{K} \to \bar{\ell}\ell$ | $\bar{\ell}\ell\bar{\ell}\ell$ | 0.055 | $N_\ell \geq 4$, $\vert m_{\ell\ell} - m_Z\vert < 10$ GeV, and $\vert m_{\ell\ell} - m_{\tilde{K}}\vert < 2.5$ GeV |
| | $\tilde{A}\tilde{K}$ | $\tilde{K}$ inclusive decay | $\gamma\bar{f}f$ | 23.14 | $N_\gamma \geq 1$, and $\vert E_\gamma - (\frac{\sqrt{s}}{2} - \frac{m_{\tilde{K}}^2}{2\sqrt{s}})\vert < 2.5$ GeV |
| | | $\tilde{K} \to \bar{\ell}\ell$ | $\gamma\bar{\ell}\ell$ | 12.67 | $N_\gamma \geq 1$, $N_\ell \geq 2$, $\vert E_\gamma - (\frac{\sqrt{s}}{2} - \frac{m_{\tilde{K}}^2}{2\sqrt{s}})\vert < 2.5$ GeV, and $\vert m_{\ell\ell} - m_{\tilde{K}}\vert < 5$ GeV |
| | | $\tilde{K} \to \bar{\chi}\chi$ | $\gamma\bar{\nu}\nu$ | 3.45 | $N_\gamma \geq 1$, $\vert E_\gamma - (\frac{\sqrt{s}}{2} - \frac{m_{\tilde{K}}^2}{2\sqrt{s}})\vert < 2.5$ GeV, and $\not{E} > 50$ GeV |
| | $\tilde{Z}H_0$ | $H_0 \to \tilde{K}\tilde{Z}$ with $\tilde{K} \to \bar{\chi}\chi$, $\tilde{Z} \to \bar{\ell}\ell$ | $\bar{\ell}\ell\bar{\ell}\ell\bar{\nu}\nu$ | $1.8 \times 10^{-5}$ | $N_\ell \geq 4$, $\vert m_{\ell\ell} - m_Z\vert < 10$ GeV, and $\vert m_{\text{recoil}} - m_{\tilde{K}}\vert < 2.5$ GeV |
| $\sin\alpha$ | $\tilde{Z}S$ | $\tilde{Z} \to \bar{\ell}\ell$ $S \to \tilde{K}\tilde{K} \to 4\chi$ | $\bar{\ell}\ell\bar{\nu}\nu$ | 0.87 | $N_\ell \geq 2$, $\vert m_{\ell\ell} - m_Z\vert < 10$ GeV, and $\vert m_{\text{recoil}} - m_S\vert < 2.5$ GeV |

**Table 2.2:** Double Dark Portal model: summary of the different vector + scalar and vector + vector production modes studied, along with the most salient cuts to identify the individual signals. All background processes include up to one additional photon to account for initial and final state radiation. Background rates are given for $\sqrt{s} = 250$ GeV, and visible particles are required to satisfy preselection cuts given in the main text of [126].

$\epsilon$) while the dark Higgs $\Phi$ mixes with the Higgs through a $\lambda_{HP}|\Phi|^2|H|^2$ quartic potential. A dark fermion $\chi$ with Dirac mass $m_\chi$ carrying $\text{U}(1)_D$ dark charge can play the role of dark matter. We denote the two scalar mass eigenstates of this model by $H_0$ (mostly Higgs) and $S$ (mostly $\Phi$) with mixing angle $\alpha$. We denote the vector mass eigenstates by $\tilde{Z}_\mu$ (mostly the SM $Z$ boson) and $\tilde{K}_\mu$ (mostly the dark photon). Both of the renormalizable portal couplings lead to attractive discovery prospects at CEPC from a variety of channels summarized in Table 2.2.

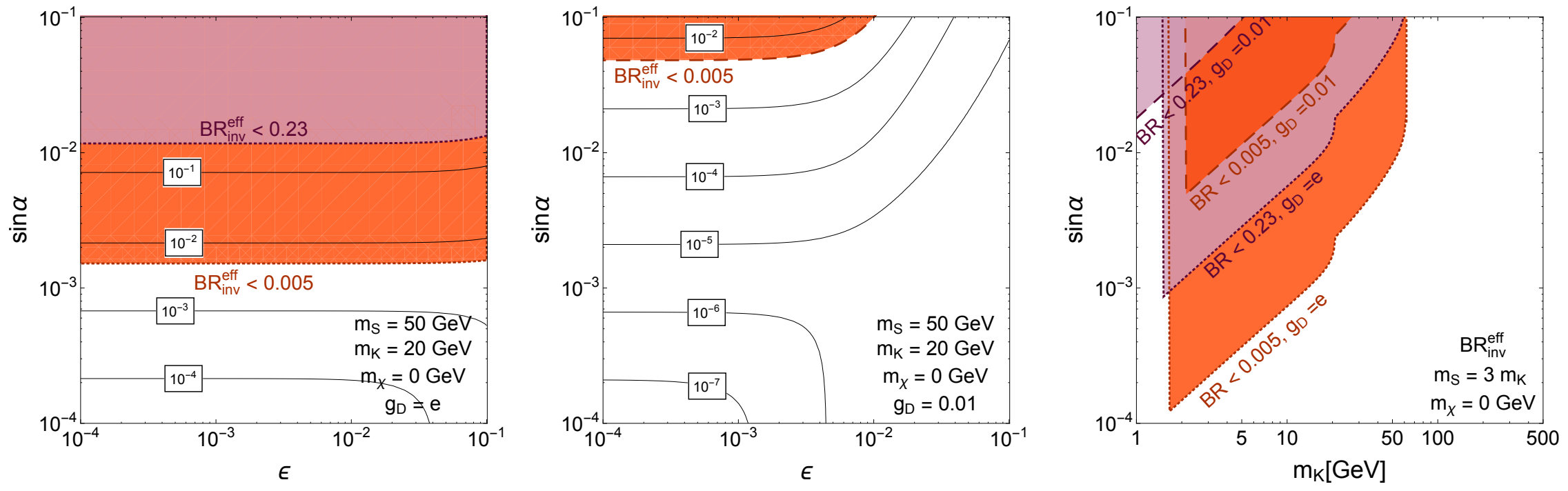

(a) (b) (c)

**Figure 2.26:** This figure shows the reach of CEPC to test the Double Dark Portal model [126] through invisible decay of the SM-like Higgs boson. (a, b) Rates for the invisible branching fraction of the 125 GeV Higgs boson in the $\sin\alpha$ vs. $\epsilon$ plane, setting $m_S = 50$ GeV, $m_K = 20$ GeV, and $g_D = e$ (left) and 0.01 (center). (c) Exclusion regions in the $\sin\alpha$ vs. $m_K$ plane from the search for an invisible decay of the 125 GeV Higgs boson by ATLAS and CMS giving $\mathrm{BR_{inv}} < 0.23$ [127, 128], and the projected reach from a future $e^+e^-$ machine giving $\mathrm{BR_{inv}} < 0.005$ [4, 123, 124, 192].

This model contains several couplings allowing transitions from the Standard Model to the dark sector, proportional to an insertion of a mixing parameter. Vertices proportional to $\alpha$ include $H_0SS$; $H_0H_0S$; $\tilde{K}_\mu\tilde{K}^\mu H_0$; and $\tilde{Z}_\mu\tilde{Z}^\mu S$. Vertices proportional to $\epsilon$ include $\tilde{Z}_\mu\tilde{K}^\mu S$ and $\tilde{Z}_\mu\tilde{K}^\mu H_0$. If $4m_\chi < 2m_{\tilde{K}} < m_S$, then both the dark photon $\tilde{K}$ and dark Higgs $S$ will dominantly decay invisibly, with visible branching ratios suppressed by $e^2\epsilon^2/g_D^2$ and $\tan^2\alpha/g_D^2$ respectively. Hence, the Double Dark Portal model contains invisible Higgs boson decay modes $H_0 \to SS \to 4\tilde{K} \to 8\chi$ and $H_0 \to 2\tilde{K} \to 4\chi$, in addition to the possible exotic decay $H_0 \to \tilde{Z}\tilde{K}$ which is either partially visible or invisible depending on the $\tilde{Z}$ decay channel. A precision measurement of the invisible branching fractions of the Higgs boson can significantly constrain the model, as summarized in Figure 2.26. Precision observation of the Higgsstrahlung rate with $\mathcal{O}(0.3\% - 0.7\%)$ accuracy [123–125] will constrain the scalar mixing angle at the level $\sin\alpha \lesssim 0.055 - 0.084$.

Direct searches for dark sector particles are possible in the channels $\tilde{Z}H_0$, $\tilde{Z}S$, $\gamma\tilde{K}$ and $\tilde{Z}\tilde{K}$. The sensitivity of CEPC searches for these signals and comparisons to existing constraints from BaBar, LEP, and LHC are summarized in Figure 2.27. The $\tilde{Z}\tilde{K}$ final state can be searched for using the recoil mass in events containing $Z \to \ell^+\ell^-$. The radiative return process $e^+e^- \to \gamma\tilde{K}$ allows a search for events with a monochromatic photon together with $\tilde{K} \to \bar{\chi}\chi, \ell^+\ell^-$. The Figure 2.27(a) shows that searches with invisible $\tilde{K}$ are more effective than those with $\tilde{K} \to \ell^+\ell^-$, due to the larger branching fraction. The figure also shows that a search for $H_0 \to \tilde{Z}\tilde{K}(\to \bar{\chi}\chi)$ is less effective. Finally, the Figure 2.27(b) shows the reach of a search for the $S$-strahlung process $e^+e^- \to \tilde{Z}S$ in the mixing angle $\sin\alpha$. This search is exactly analogous to the previous search at LEP-II for a purely invisible decaying Higgs boson [121]. Improved sensitivity could be obtained by varying the $\sqrt{s}$ of the collider to maximize the $\sigma(e^+e^- \to \tilde{Z}S)$ rate for the test $S$ mass (see also Ref. [193]).

## PORTALS WITH ADDITIONAL SM-SECTOR PHYSICS

While the renormalizable SM portals are simple, they are not the only possibilities. Portals between the dark and visible sectors could be formed by additional particles with Standard

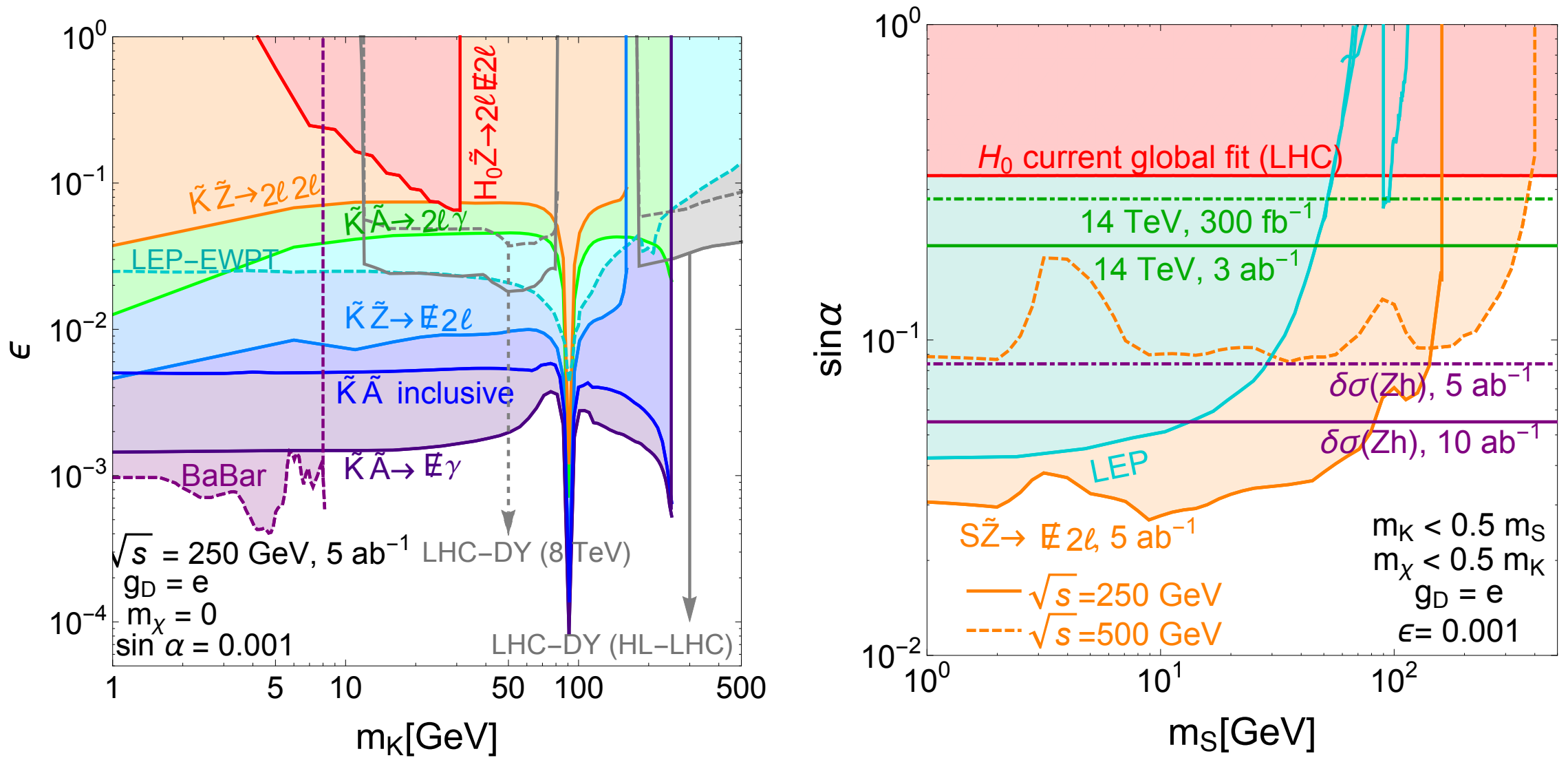

(a) (b)

**Figure 2.27:** This figure shows the reach of CEPC to test the Double Dark Portal model [126] through searches for dark-sector particles. (a) Projected exclusion regions in the $\epsilon$ vs. $m_K$ plane from multiple complementary searches of $\tilde{K}$ production. Solid lines enclose expected exclusion regions with $\mathcal{L} = 5$ ab$^{-1}$ of $\sqrt{s} = 250$ GeV $e^+e^-$ machine data. Dashed lines indicate existing limits from the LEP $e^-e^+ \to \ell^-\ell^+$ contact operator search, the LEP electroweak precision tests, the BaBar $\tilde{K}$ invisible decay search (BaBar), and the LHC Drell-Yan constraints (LHC-DY). The 3 ab$^{-1}$ HL-LHC projection for Drell-Yan constraints is also shown as a solid line. Note that $m_K$ is approximately the $m_{\tilde{K}}$ mass eigenvalue. (b) Exclusion reach from the $\tilde{Z}S$, $\tilde{Z} \to \ell^+\ell^-$ search in the recoil mass distribution for invisible $S$ decays in the $\sin\alpha$ vs. $m_S$ plane using 5 ab$^{-1}$ of $e^+e^-$ data at $\sqrt{s} = 250$ GeV or 500 GeV. We also show comparisons to the current fit, $\sin\alpha < 0.33$ [122], future LHC projections of 0.28 (0.20) using 300 fb$^{-1}$ (3 ab$^{-1}$) luminosity [4], and precision $\delta\sigma(ZH)$ measurements constraining 0.084 (0.055) using 5 ab$^{-1}$ (10 ab$^{-1}$) [123–125]. We plot the excluded region from LEP searches for invisible low mass Higgs boson in the $ZS$ channel in cyan [121, 194–196].

Model gauge charges. These can offer interesting variations on the renormalizable portal. One example of such a portal is the leptonic Higgs portal [197]. This model includes an elementary scalar, $S$, which only couples to the SM leptons, $g_\ell S\bar{l}l$.[4] Note that this operator is not SM gauge invariant and has to be UV completed. One possible simple UV completion is to couple a SM singlet to two Higgs doublets with one of the doublets only coupling to leptons and the other one only coupling to quarks. At a lepton collider, assuming that the couplings $g_\ell$ are proportional to the corresponding lepton mass, $S$ could be produced in association with $\tau$ leptons, $e^+e^- \to \tau^+\tau^- + (S \to e^+e^-, \mu^+\mu^-, \tau^+\tau^-)$. Current beam dump and lepton colliders only probe $m_S$ to a few GeV. CEPC could be capable of extending the sensitivity to much heavier $S$ up to $m_S \sim 250$ GeV. In the particular lepton-specific two Higgs doublet UV completion, the mixing between the singlet $S$ and the Higgs boson $H$ leads to exotic Higgs boson decays such as $H \to SS \to 4\tau, 2\mu 2\tau$. For the $4\tau$ final state, CEPC could test a branching fraction as small as $10^{-4}$ at 95% CL, improving the sensitivity by three orders of magnitude compared to even the HL-LHC [198]! This is translated to a factor of 30 improvement in testing the coupling $g_\ell$, fixing all the other parameters. Another similar possibility is a leptonic portal arising from some gauge bosons coupling to SM lepton-flavor currents [199].

[4]A variant of the model with $S$ dominantly coupling to the muon and proton with tiny couplings to the electron and neutron might explain the proton radius puzzle and the muon anomalous magnetic moment discrepancy.

In general, the dark matter portal models could give rise to exotic Higgs boson decays. A thorough review of the models leading to exotic Higgs boson decays and the status of LHC searches can be found in Ref. [115]. Supersymmetric exotic decays of the Higgs boson have been studied in Refs. [198, 200]. The potential of detecting exotic Higgs boson decays in 14 different final states at CEPC has been presented in Ref. [198]. In every final state, we expect at least one order of magnitude improvement in sensitivity compared to the HL-LHC and in quite a few channels, we expect 3-4 orders of magnitude improvement at CEPC. More details are discussed in Section 2.3.1.

A characteristic feature of many models that go beyond renormalizable portals is the possibility of new sources of flavor violation. For example, nonrenormalizable (dipole moment) operators could allow one SM fermion to decay to a dark photon and another SM fermion of different flavor, e.g. $\mu^\pm \to e^\pm \gamma_d$ or $t \to c\gamma_d$ [201]. Renormalizable completions of such models introduce new "messenger" particles that interact with the SM gauge groups and the dark photon. The induced flavor-violating decays could be searched for at CEPC.

Another possibility that could be tested at CEPC is flavor-violating dark matter in which dark matter couples dominantly to muons [202]. The dark multiplet contains a scalar and a vector-like fermion and couples to the muon through a Yukawa interaction. The neutral component of the scalar serves as the dark matter candidate. The interaction generates a loop correction to the $\gamma\mu^+\mu^-$ and $Z\mu^+\mu^-$ couplings that could be measured as deviations in the cross section of $e^+e^- \to \mu^+\mu^-$. Choosing the Yukawa coupling to be $\mathcal{O}(1)$ means that a 2% precision measurement of the cross section can probe dark matter mass within $20$ GeV around $120$ GeV. Related models include flavored dark matter [203, 204], in which the dark matter particle carries flavor quantum numbers and has renormalizable contact interactions with the SM fields. In particular, electron-flavored dark matter could be produced copiously at a lepton collider associated with a photon if its mass is below $\sim 120$ GeV.

## EFFECTIVE THEORY

So far, our discussion of dark matter has been organized based on details of the model. However, one could also take a portal-agnostic or "model-independent" approach, simply searching for a generic signal like a single photon plus missing energy [205]. This could arise if DM is part of an electroweak multiplet, due to loops of the charged $\mathrm{SU}(2)_L$ partners of dark matter and $W$ bosons. It could also arise if completely new charged particles, independent of DM, exist and couple to DM. Results can be expressed simply in terms of effective operators, without committing to a particular UV completion. A variety of studies of such signals at $e^+e^-$ colliders have been carried out, e.g. [206–209].

In an effective theory approach, such signals arise from dimension-7 effective operators coupling fermionic dark matter to pairs of SM gauge bosons. The operators that can be

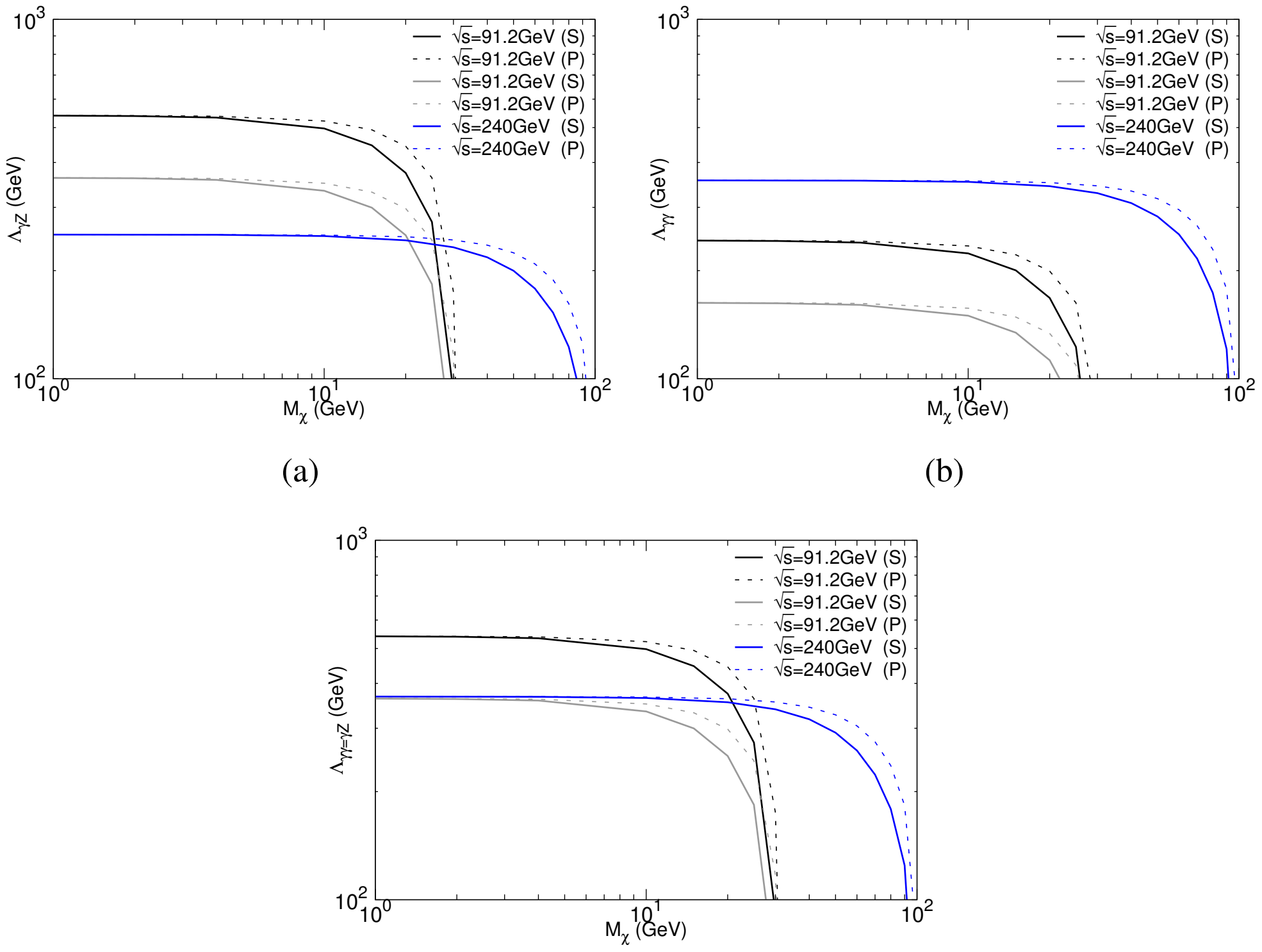

(a) (b)

(c)

**Figure 2.28:** CEPC's capacity to test whether dark matter couples to the SM photon and/or $Z$ boson through the operators in Equation (2.25). From (a) to (c), the three panels correspond to pure $\Lambda_{\gamma Z}$ interaction, pure $\Lambda_{\gamma\gamma}$ interaction, and $\Lambda_{\gamma Z} = \Lambda_{\gamma\gamma}$ interaction. The various curves show CEPC's $3\sigma$ projected sensitivity to the dark matter mass, $m_\chi$, and the energy scale of new physics, $\Lambda$. The black, gray, and blue lines refer to $\sqrt{s} = 91.2$ GeV with 2.5 ab$^{-1}$, 91.2 GeV with 25 fb$^{-1}$, and 240 GeV with 5 ab$^{-1}$, respectively. The photon is required to have $|\eta| < 3$ and a $p_T > 25\,(35)$ GeV for $91.2\,(240)$ GeV collision energy to optimize the sensitivity for a low $m_\chi$. The solid lines are for a scalar operator and the dashed lines for the pseudoscalar case. The figure is adapted from Ref. [210].

efficiently constrained by searches at CEPC are

$$\mathcal{L}_\mathrm{S} \supset \frac{1}{\Lambda^3_{\gamma\gamma}}\bar{\chi}\chi A^{\mu\nu}A_{\mu\nu} + \frac{1}{\Lambda^3_{\gamma Z}}\bar{\chi}\chi A^{\mu\nu}Z_{\mu\nu},$$
$$\mathcal{L}_\mathrm{P} \supset \frac{1}{\Lambda^3_{\gamma\gamma}}\bar{\chi}i\gamma_5\chi A^{\mu\nu}\widetilde{A}_{\mu\nu} + \frac{1}{\Lambda^3_{\gamma Z}}\bar{\chi}i\gamma_5\chi A^{\mu\nu}\widetilde{Z}_{\mu\nu}\ , \tag{2.25}$$

where the field strengths $A_{\mu\nu}$ and $Z_{\mu\nu}$ and their duals $\widetilde{A}_{\mu\nu}$ and $\widetilde{Z}_{\mu\nu}$ couple to the scalar (S) and the pseudoscalar (P) fermionic dark matter bilinears. The $\Lambda$ factors in the coefficients represent the approximate mass scale of new physics (up to loop factors). Similar operators can also be written for the $\mathrm{SU}(2)_L$ gauge fields, but the $WW$ couplings may not be as efficiently probed by $e^+e^-$ collisions at the $Z$ pole.

The diphoton operator dominates processes with low momentum transfer because the photon is massless. It is much more stringently constrained by direct detection than its DM-$\gamma Z$ and DM-$ZZ$ counterparts. For DM lighter than half of $m_Z$, indirect detection using diffuse gamma rays is also more sensitive to the diphoton operator. Collider searches, on the other hand, can more effectively probe $Z$ couplings. The high-luminosity $Z$-pole run at CEPC offers a unique opportunity to test the DM couplings to the $Z$ boson. For a light DM mass, the resonantly produced $\bar{\chi}\chi\gamma$ system is best searched for in the monophoton + missing energy channel.

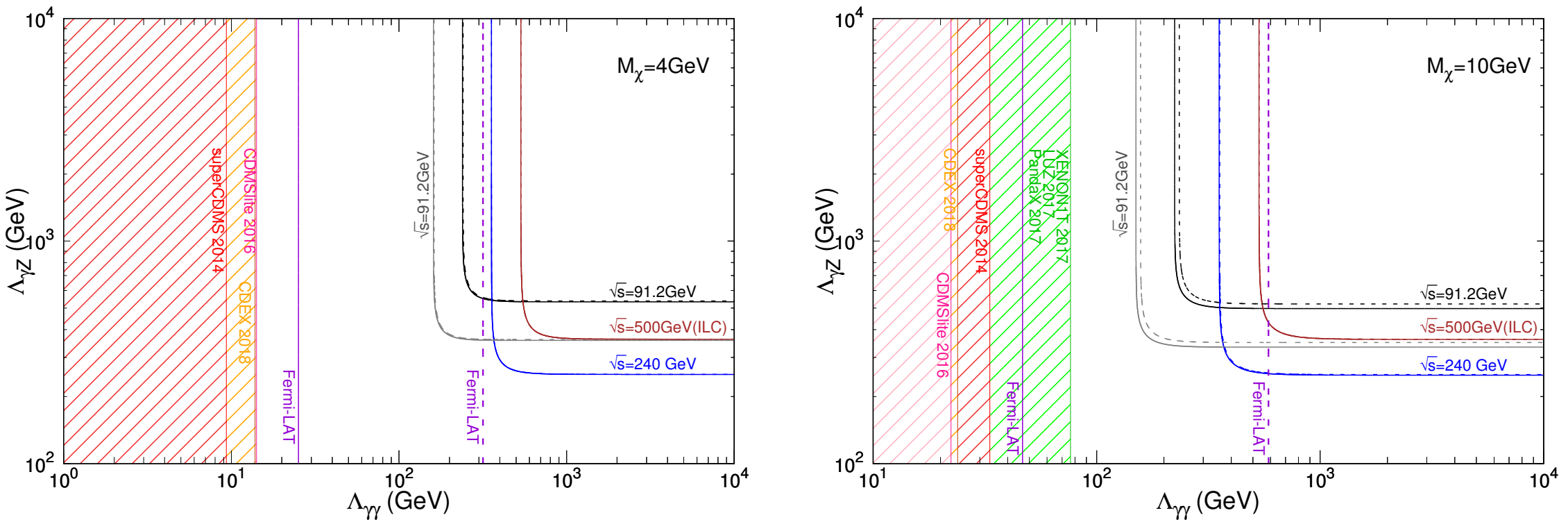

(a) (b)

**Figure 2.29:** CEPC's capacity to test whether dark matter couples to the SM photon and/or $Z$ boson through the operators in Equation (2.25). The (a) (b) panel shows a DM mass of $m_\chi = 4\,(10)$ GeV. The CEPC sensitivity is shown by the black, gray, and blue curves, which are defined in the caption of Figure 2.28. The brown line denotes the ILC $3\sigma$ sensitivity with an integrated luminosity of 500 fb$^{-1}$ at $\sqrt{s} = 500$ GeV with cuts $10° < \theta_\gamma < 170°$ and $p_T(\gamma) > 90$ GeV. Constraints from dark matter direct detection experiments are shown in red for SuperCDMS [211], orange for CDEX [212], pink for CDMSlite [213], and green for XENON1T [133], LUX [134], and PandaX [156] (which are in close proximity to each other). The purple-dashed line denotes the Fermi-LAT bound from the R3 region [214]. Note that the XENON1T/LUX/PandaX limit only appears in the $m_\chi = 10$ GeV case. The figure is adapted from Ref. [210].

The proposed $Z$-pole runs' prospective limits on effective DM-$\gamma Z$ and $\gamma\gamma$ couplings in the monophoton channel are studied in Ref. [210]. The major SM background, $e^+e^- \to \bar{\nu}\nu\gamma$, can be effectively controlled by optimizing the cut on the single photon's $p_T$. The corresponding constraints on $\Lambda$ are illustrated in Figures. 2.28 and 2.29. The best sensitivity is obtained for light dark matter mass. In case only one operator is considered, the projected sensitivity for $\Lambda_{\gamma Z}$ is 360 GeV and 540 GeV for 25 fb$^{-1}$ (giga Z) and 2.5 ab$^{-1}$ (tera Z) luminosities at the $Z$ pole, respectively. In comparison, $\Lambda_{\gamma\gamma}$ is best probed at higher energy runs, and a limit of 360 GeV is obtained for a 5 ab$^{-1}$ run at 240 GeV center-of-mass energy. In general, both $\Lambda_{\gamma Z}$ and $\Lambda_{\gamma\gamma}$ would be present and their relative size is model dependent.

Figure 2.29 further shows the direct and indirect detection limits together with CEPC's constraint in the $\Lambda_{\gamma\gamma} - \Lambda_{\gamma Z}$ plane. For direct detection, we adopt the calculation of the spin-independent scattering rate via the scalar operator from Ref. [113, 215], which takes into account the diphoton exchange that dominates over $\gamma Z$ contributions. We choose benchmark DM masses at 4 and 10 GeV that are accessible to major nuclear recoil experiments. For indirect detection, we show the 95% CL constraint from the gamma ray line search at Fermi-LAT [214]. The nonrelativistic DM annihilation cross section into two photons ($\bar{\chi}\chi \to \gamma\gamma$) is dominated by $\Lambda_{\gamma\gamma}$ for $m_\chi$ below $m_Z/2$. The $\Lambda_{\gamma Z}$ dependence only emerges in a tiny correction as part of the $\bar{\chi}\chi \to \gamma(\gamma^*/Z^* \to \bar{f}f)$ process, and can be ignored at the DM masses shown.

The channel of monophoton + missing energy would also be sensitive to effective interactions between dark matter and electrons. In this case, the photon arises from initial

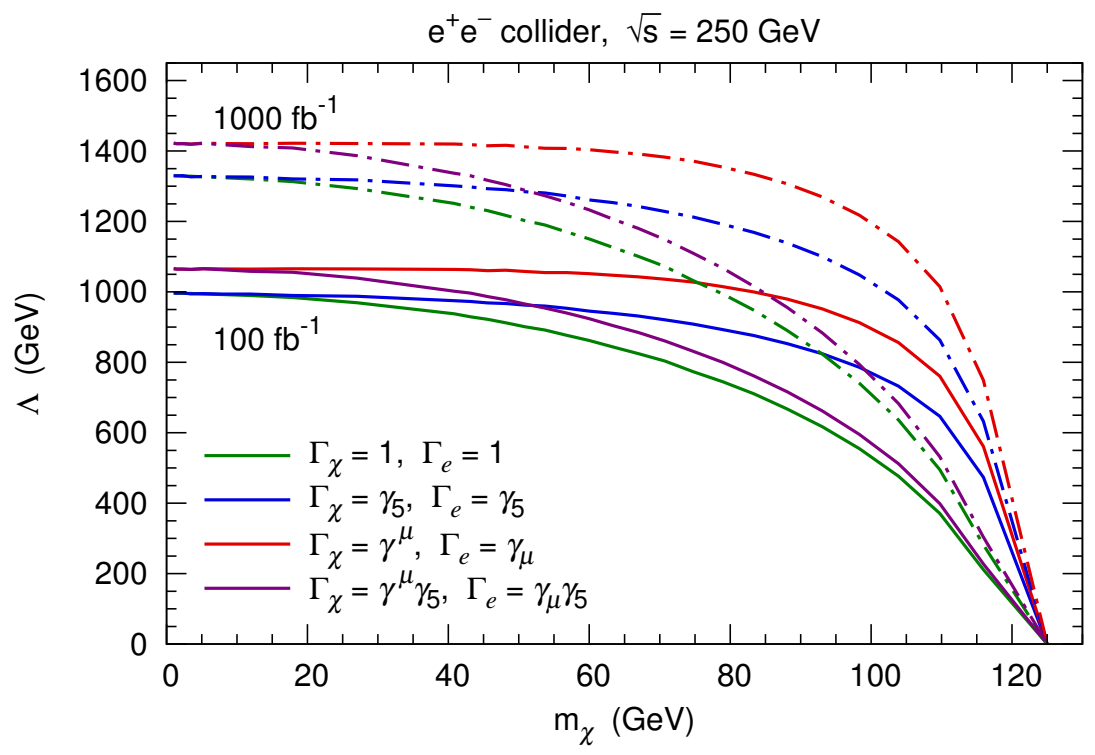

**Figure 2.30:** CEPC $3\sigma$ reach for several effective interactions between dark matter and electrons in the channel of monophoton + missing energy with integrated luminosities of $100\ \text{fb}^{-1}$ and $1\ \text{ab}^{-1}$.

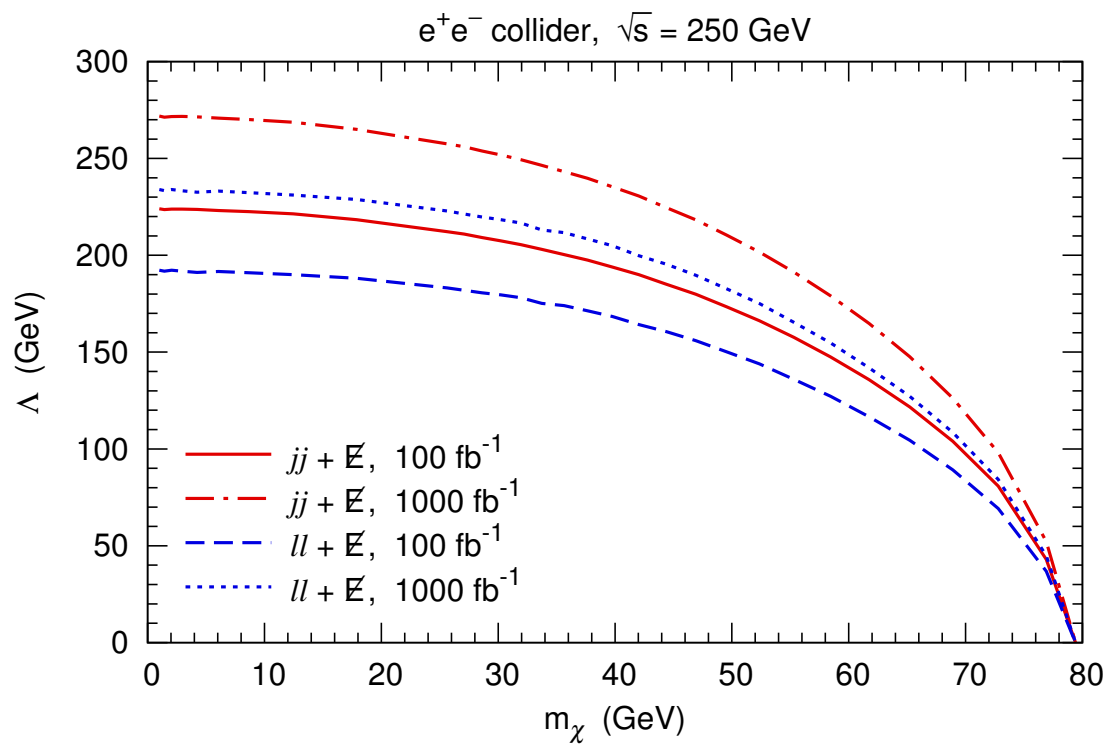

**Figure 2.31:** CEPC $3\sigma$ reach for the effective operator $\mathcal{L}_{\chi Z}$ in the channel of mono-$Z$ + missing energy, adapted from [209]. Both the hadronic ($jj + \not{E}$) and the leptonic ($\ell\ell + \not{E}$) modes are presented with integrated luminosities of $100\ \text{fb}^{-1}$ and $1\ \text{ab}^{-1}$.

state radiation. The related dimension-6 operators are

$$\mathcal{L}_{\chi e} = \frac{1}{\Lambda^2}\bar{\chi}\Gamma_\chi \chi \bar{e}\Gamma_e e, \tag{2.26}$$

where $\Gamma_\chi, \Gamma_e \in \{1, \gamma_5, \gamma^\mu, \gamma^\mu\gamma_5, \sigma^{\mu\nu}\}$. The CEPC reach at 250 GeV center of mass energy is demonstrated in the $m_\chi - \Lambda$ plane in Figure 2.30. For low masses, limits of $\sim 1.4$ TeV on $\Lambda$ could be achieved with an integrated luminosity of $1\ \text{ab}^{-1}$.

An analogous approach is to search for a signal in the channel of mono-$Z$ + missing energy. This channel is sensitive to effective operators like

$$\mathcal{L}_{\chi Z} = \frac{1}{\Lambda^3}\bar{\chi}\chi Z^{\mu\nu} Z_{\mu\nu}. \tag{2.27}$$

$Z$ bosons can be reconstructed by either two jets or two opposite-sign same-flavor leptons. Figure 2.31 shows the CEPC reach in the $m_\chi - \Lambda$ plane at 250 GeV center of mass energy. It is expected that the hadronic modes would provide a better sensitivity than the leptonic modes.

### 2.3.4 NEUTRINO CONNECTION

#### NEUTRINO MASS MODELS

The CEPC is an excellent tool to study the physics of neutrino mass generation as a portal to unknown new physics during both the $240$ GeV and the $Z$-pole runs. In this respect it can serve as a discovery machine for new physics that evades detection at hadronic colliders, including feebly coupled "hidden sector" extensions of the SM that can address fundamental questions in particle physics and cosmology.

The experimental observation of neutrino flavor oscillations [216, 217] indicates that neutrinos have a nonzero mass. Global fits to neutrino oscillation experiments (see e.g. [218, 219]) are sufficient to fix two neutrino mass-square differences and all mixing angles in the Pontecorvo-Maki-Nakagawa-Sakata matrix $V_\nu$ (assuming it to be unitary), while the absolute neutrino mass scale is constrained from cosmology to be in the sub-eV range (see e.g. [220]). These results raise a pair of pressing questions, namely why the neutrinos are so much lighter than all other fermions, and why the elements of the neutrino mass mixing matrix are so different from the quark mixing matrix.

Since the Standard Model of particle physics cannot account for nonzero neutrino masses in a renormalizable way, neutrino oscillations provide compelling experimental evidence for physics beyond the SM. While the origin of mass for the charged SM fermions (at least of the third generation) is well established by Higgs coupling measurements, the origin of mass for neutrinos is unknown and calls for a more fundamental theory of nature underlying the Standard Model. Moreover, neutrinos may be Majorana fermions [221], fundamentally different from their charged fermion counterparts, with consequences related to violation of lepton number that are potentially discoverable at colliders [222]. Lepton number violation may also be connected to an open question in cosmology, the Baryon Asymmetry of the Universe (BAU), i.e., the tiny excess $\sim 10^{-10}$ [132] of matter over antimatter.

Under the assumption that the scale of new physics $\Lambda$ associated with the mass of the lightest new particle involved in the generation of neutrino masses is much larger than the typical energy $E_\nu \sim \mathrm{MeV}$ in neutrino oscillation experiments,[5] the neutrino oscillations can be described in the framework of Effective Field Theory (EFT). The relevant operators $\mathcal{O}_i^{[n]}$ have mass dimension $n > 4$, are suppressed by powers of $\Lambda^{n-4}$, and have Wilson coefficients $c_i^{[n]}$ that are matrices in flavor space. In this framework the smallness of the neutrino masses can be a consequence of any combination of the following reasons:

I) *High-Scale Seesaw Mechanism:* Large values of $\Lambda$ automatically lead to small $m_i$. The three tree level implementations of the idea [225] are known as: Type-I Seesaw [80, 81, 226–229], involving the SM plus right-handed neutrinos $N$; Type-II Seesaw [229–233], involving the SM plus a scalar $\mathrm{SU}(2)_L$ triplet $\Delta_L$; and Type-III Seesaw [234], involving the SM plus a fermionic $\mathrm{SU}(2)_L$ triplet field $\Sigma_L$.

II) *Small numbers:* The $\mathcal{O}_i^{[n]}$ can remain small (for all values of $\Lambda$, including those accessible to CEPC) if the Wilson coefficients $c_i^{[n]}$ are small. In particular, if the neutrinos are Dirac particles their masses can be generated by the Higgs mechanism in exactly the same way as all other fermion masses with tiny Yukawa couplings. Tiny constants can be avoided e.g. when the neutrino interactions are created dynamically due to the spontaneous breaking of a flavor symmetry by flavons [235], or when the $\mathcal{O}_i^{[n]}$ are created radiatively (see e.g. [236–240]).

[5] Scenarios with $\Lambda < E_\nu$ are in principle feasible (see e.g. Refs. [223, 224] and references therein), but strongly constrained by the success of the high level of consistency in global fits to neutrino oscillation data that assume only three light neutrinos [218, 219].

III) *Low-Scale Seesaw Mechanism:* A low scale $\Lambda$ and $\mathcal{O}(1)$ couplings between the SM and the new particles can be realized when symmetries give rise to cancellations in the neutrino mass matrix. For instance the $B-L$ symmetry of the SM can keep the $\mathcal{O}_i^{[n]}$ small for $\Lambda$ below the $\mathrm{TeV}$ scale [241–243]. Specific models that implement this idea include the inverse [244–246] and linear [247, 248] seesaw, the Neutrino Minimal Standard Model [249, 250] and scale invariant models [251].

Here the terms "high scale" and "low scale" should be understood with respect to the CEPC collision energy; for values of $\Lambda$ far above $240\ \mathrm{GeV}$ the EFT treatment introduced here to describe neutrino oscillation experiments can also be applied to CEPC phenomenology, while lower values imply that the new particles can be found at CEPC and should be described dynamically.

The original setting for the seesaw mechanism was grand unified theories, based on $\mathrm{SO}(10)$ [227], and $\mathrm{SU}(5)$ [226], as well as the minimal Left-Right (LR) symmetric model [80, 81] and flavor/family symmetries [228]. The large scale of grand unification typically sets the mass scale $\Lambda$ related to neutrino physics beyond the direct reach of colliders, although parts of multiplets may lie well below the GUT scale. For example, the minimal $\mathrm{SU}(5)$ model with the addition of a $24_F$ multiplet requires a light fermionic triplet in order for gauge couplings to unify [252, 253], motivating Type III Seesaw searches at the $\mathrm{TeV}$ scale. Other well known examples are for instance $B-L$ symmetry, additional "neutrinophilic" Higgs doublets, and flavor symmetries. Such neutrino mass physics generally predicts the existence of new particles, which could at least in principle be discovered and studied at CEPC.

## LEPTON NUMBER VIOLATION

If neutrinos are Majorana particles, the mechanism that generates their mass can mediate Lepton-Number Violating (LNV) processes at colliders if the scale $\Lambda$ is below or near the collision energy [222]. A variety of signatures arise in specific models for neutrino mass generation.

**Type I Seesaw:** Observing the violation of lepton number from heavy neutrino mass eigenstates ($N_i$) in the process $e^+e^- \to N\nu$ at lepton colliders is possible in principle due to the different kinematics of LNV and LNC processes, as was demonstrated for the ILC [254]. In particular, for heavy neutrino $N_i$ masses $M_i > m_Z$, the process $e^+e^- \to \nu\ell jj$ is a promising signature at lepton colliders [255–257] and has been studied specifically for CEPC [258]. The subleading production process for heavy neutrinos at lepton colliders $e^+e^- \to N\ell^\pm W^\mp$ allows for same sign dileptons for $N \to \ell^\pm W^{(*)}$ and $W \to$ hadrons [257].

It is worth pointing out, however, that LNV in the Type I Seesaw mechanism is suppressed by the smallness of the light neutrino masses [242, 243]. It has been proposed that the suppression of LNV may be alleviated by the process of heavy neutrino-antineutrino oscillations, which occurs for heavy neutrinos with masses below the $W$ boson's mass, $m_W$, and with $U^2 < \mathcal{O}(10^{-5})$ [259–261].

**Type II Seesaw:** The triplet scalar multiplet $\Delta_L$ in the Type II Seesaw contains three complex fields, which are respectively neutral, singly charged, and doubly charged under electromagnetism. The appealing feature of the model is the direct connection between neutrino masses and mixing parameters [262, 263] and the Majorana Yukawa matrix $M_\nu = Y_\Delta \langle \Delta_L \rangle$, which may lead to charged lepton flavor violating signals [264].

Collider phenomenology is governed by the final state, which primarily depends on the triplet's vacuum expectation value (VEV) [265] and the mass splittings of its components [266]. If the masses are degenerate, the dominant decay mode is to leptons if the triplet VEV is smaller than $\sim 10^{-4}$ GeV. This decay mode tests the flavor structure of the neutrino mass matrix and leads to significant flavor-dependent bounds on the triplet scalar mass up to $870$ GeV at the LHC [267]. For the triplet VEV above $\sim 10^{-4}$ GeV, the states decay to pairs of gauge bosons. A relatively small mass splitting, consistent with precision electroweak constraints, triggers cascade decay modes [266] which produce soft hadronic and multi-lepton final states [268]. Signal in the $WW$ final state lead to weak lower bounds on doubly charged scalars at the LHC, $m_{\Delta_L^{++}} \gtrsim 90$ GeV [269] or less, depending on the lepton's flavor. Similarly, the cascade decays [266, 270] are not easy to look for in hadronic colliders [271]; however, they may be observable in cleaner lepton collisions [272].

At lepton colliders, the triplet components can be produced pair-wise through $e^+e^- \to S\overline{S}$ (where $S = \Delta_L^0, \Delta_L^\pm, \Delta_L^{\pm\pm}$ are the various charged states in the triplet) or in single production in association with two same-sign leptons $e^+e^- \to \Delta_L^{\pm\pm}\ell^\mp\ell^\mp$ [273, 274]. Another possible production mode is via vector-boson fusion $e^+e^- \to \ell\,\ell' S\,S'$, where $\ell, \ell' = e^\pm, \nu$, as discussed in [275].

The doubly charged scalar bosons $\Delta_L^{\pm\pm}$ can couple to the electrons and positrons directly and contribute to Bhabha scattering in the $t$-channel [273, 276]. Running the lepton colliders with same-sign beams may strongly enhance the production of the doubly charged components in the $s$-channel [273, 277], see [278] for more recent work.

**Left-Right Symmetric Model:** The mixing of the SM Higgs doublet with the $\mathrm{SU}(2)_R$ triplet Higgs that gives Majorana mass to right-handed neutrinos in the Left-Right Symmetric Model (LRSM) [279–282] may lead to LNV decays of $h \to NN$ [283]. The subsequent (and possibly displaced) decay of $N \to \ell^\pm jj$ can lead to a $\Delta L = 2$ LNV and potentially charged lepton flavor violating final state with two same sign-leptons and up to four jets. Due to the soft final states and displacement, such searches may be challenging at the LHC; however lepton colliders are much more suitable to detect such signals due to the absence of triggers and lower QCD backgrounds.

The presence of the mixing also allows for an enhanced production of the $\mathrm{SU}(2)_R$ triplet $pp \to \Delta_R^0 \to NN$ at the LHC [284] with varying kinematics, depending on its mass. Moreover, one may be able observe a truly exotic Higgs boson decay with $h \to \Delta_R^0\Delta_R^0 \to 4N$, where lepton number can be broken to up to four units [284]. The production at lepton colliders may proceed through the Higgs mixing $e^+e^- \to Z\Delta_R^0 \to NNZ$ for $\sqrt{s} \lesssim 100$ GeV and in the Vector Boson Fusion (VBF) channel that produces the $NN\nu\bar{\nu}$ final state with lepton number violation and missing energy [284]. At $\sqrt{s} = 240$ GeV and $\mathcal{L} = 5$ ab, one may expect from a few hundred to more than 5000 $NNZ$ events, depending on the masses of triplets and heavy neutrinos, as well as the Higgs-triplet mixing. Such events are essentially background free at lepton colliders because of the LNV final state, $Z$ tagging, and characteristic displacement. Similarly, the quadruple production of $N$'s can proceed through the Higgs-triplet triple vertex with the potential of observing $\mathcal{O}(10^4)$ events with the branching ratio of Higgs boson to $\Delta_R^0\Delta_R^0$ at the $1\%$ level.

## CHARGED LEPTON FLAVOR VIOLATION

Neutrino oscillations violate lepton flavor, which is transferred to the charged leptons via perturbation theory, such that the violation of the charged lepton flavor (cLFV) is a prediction [285]. This gives rise to a variety of distinctive processes that may be probed at CEPC. The most stringent constraints and the CEPC prospects in both the on-shell and off-shell modes are collected in Figure 2.32.

**Mixed flavor leptonic Higgs or $Z$ boson decays:** Observables at high energy that can measure cLFV are exotic decays of the $Z$ boson into two charged leptons of different flavor, $Z \to e^\pm\mu^\mp$, $e^\pm\tau^\mp$, $\mu^\pm\tau^\mp$ [286, 287]. Also the decays of the Higgs boson into two charged leptons of different flavor are possible [288, 289]. The processes $h \to e^\pm\mu^\mp$, $e^\pm\tau^\mp$, $\mu^\pm\tau^\mp$ are lepton flavor violating Higgs boson decays that can be measured at CEPC for branching ratios as small as $1.2 \times 10^{-5}$ to $1.6 \times 10^{-4}$ [290].

**Lepton universality violation in $W$ boson decays:** The branching ratios of the $W$ bosons should be identical for the three different leptons[6] due to the lepton flavor universality in the SM. Another probe of lepton universality is given by the decays of the $\tau$ lepton. Mixing of the active neutrinos with neutral fermions from the Type I or III Seesaw can lead to violations of lepton universality, see e.g. [292]. Charged scalar particles can affect the measurement of lepton-universality observables from $W$ boson branching ratios [293].

**Mixed flavor final states with and without resonance:** In addition to exotic decays of Higgs boson, $W$, and $Z$ bosons, an observable cLFV process at lepton colliders $e^+e^- \to \ell_\alpha^\pm\ell_\beta^\mp\ (+H)$. These processes receive contributions from electrically neutral scalars, for instance from neutrinophillic Two Higgs Doublet models, Type II-based Seesaw models, $B-L$, or left-right symmetry. A dedicated study of such cLFV processes involving neutral scalars can be found in Ref. [294].

## HIGGS BOSON PROPERTIES

The Higgs boson is a particularly sensitive probe of the mechanism of neutrino mass generation. Higgs boson-based signatures motivated by neutrino mass models include anomalous Higgs boson production mechanisms; invisible or exotic Higgs boson decays; lepton-flavor-violating Higgs couplings; and modified Higgs couplings, all of which may be probed at CEPC.

**Anomalous Higgs boson production:** In models with heavy neutrinos, for heavy neutrino masses $M_i > m_H$ additional Higgs bosons can be produced from heavy neutrino decays in processes $e^+e^- \to Z^* \to N\nu \to H\nu\nu$. This can yield an enhancement of the SM mono-Higgs channel of up to $\sim 2\%$ when applying "standard" filters [295, 296]. The CEPC sensitivity via additional Higgs bosons from dedicated analyses is shown by the yellow line in Figure 2.33.

[6] Current LEP data features a branching $\mathrm{Br}(W \to \tau\nu)$ that is larger than $\mathrm{Br}(W \to \ell_{e,\mu}\nu)$ by $\sim 2\sigma$ [291].

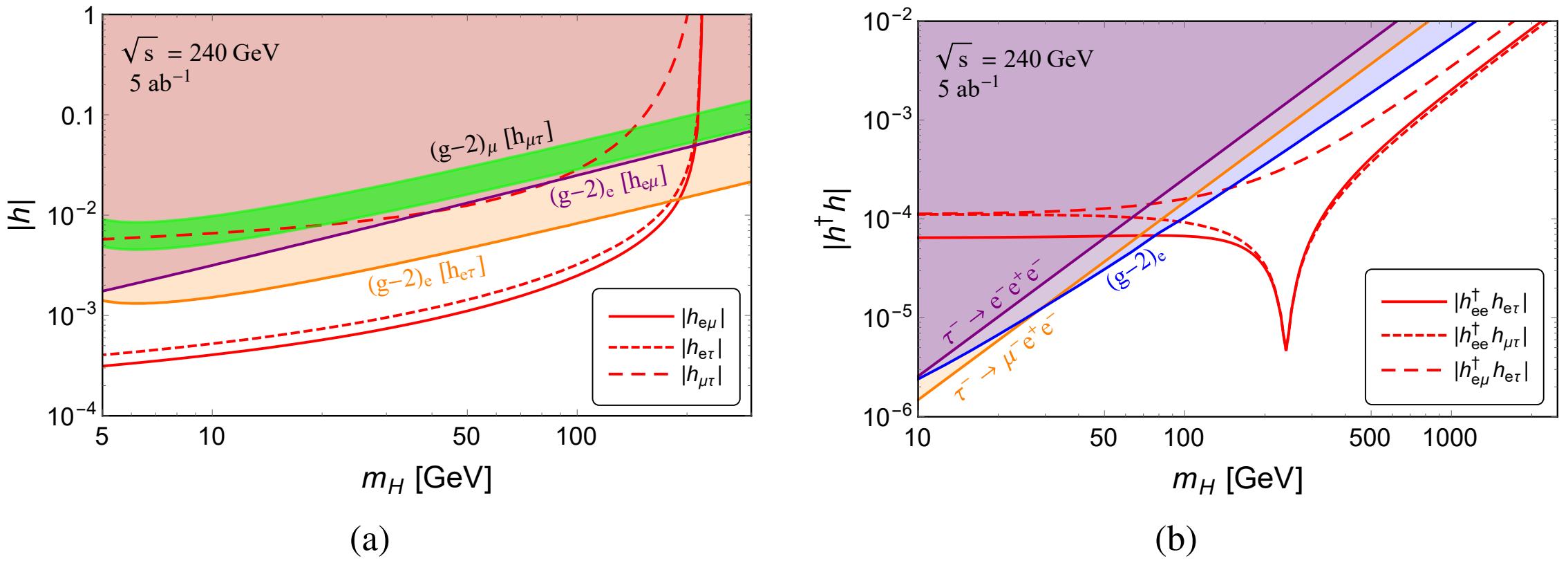

**Figure 2.32:** The CEPC's ability to probe charged lepton-flavor violation (cLFV) is illustrated here as a sensitivity to the cLFV couplings, $h_{\alpha\beta}$ ($\alpha \neq \beta$), and the mass of a new, electrically-neutral scalar particle, $m_H$. Searches for $e^+e^- \to \ell_\alpha^\pm \ell_\beta^\mp H$ (left) and $e^+e^- \to \ell_\alpha^\pm \ell_\beta^\mp$ (right) at CEPC with $\sqrt{s} = 240$ GeV and $\mathcal{L} = 5$ ab$^{-1}$ lead to projected sensitivities shown by the red curves (assuming 10 cLFV signal events). In (a) panel, the shaded regions are excluded by electron and muon $g-2$, but the green band could explain the $(g-2)_\mu$ discrepancy at the $2\sigma$ level. In (b) panel the shaded regions are excluded by rare $\tau$ lepton decays, $\tau \to eee$ and $\tau \to ee\mu$. See the text and Ref. [294] for more details.

**Invisible Higgs boson decays:** The $N_i$ can leave measurable imprints in precision measurements of the Higgs boson branching ratios. In the Type I Seesaw the Higgs boson can decay into a light and a heavy neutrino mass eigenstate when kinematically accessible, which can account for up to 30% of the Higgs boson decays [297] without violating present constraints [298]. CEPC sensitivity to this scenario from searches for the Higgs boson invisible branching ratio, considering the precision from Ref. [125], is shown by the red line in Figure 2.33.

**Leptonic Higgs boson decays with cLFV or LNV:** As mentioned previously, cLFV decays also add loop-induced additional channels to the total Higgs boson decay width, and processes where the Higgs boson couples to two $N_i$ can give rise to exotic LNV decay channels, all of which may be extensively probed by the precision Higgs program at CEPC.

**Higgs boson decays into two $N_i$:** In $B-L$ and $L-R$ symmetric models, additional neutral scalars can mix with the Higgs boson. This can give rise to additional decay channels into two $N_i$, which can be observable depending on their masses and lifetimes. Such signatures were studied in the context of LRSM [283, 284] and $B-L$ models [299, 300].

**Anomalous diphoton decays:** In the Type II Seesaw additional scalar particles couple directly to the Higgs boson, such that the singly and doubly charged components contribute to the loop-induced coupling of the Higgs boson to two photons [266, 301–303]. Similarly, the Type III Seesaw contains additional charged particles that can contribute to the Higgs-to-diphoton branching ratio, see e.g. [304]. In the LRSM, the doubly charged component of the $\mathrm{SU}(2)_R$ triplet couples rather strongly to the SM Higgs boson, leading to an $\mathcal{O}(100\ \mathrm{GeV})$ lower bound on its mass [305] from similar radiative corrections. The Higgs-to-diphoton in the SM could have non-trivial on-shell and off-shell interference ef-fect [306, 307], which can be used to constrain Higgs boson properties and help resolve higher dimensional operators.

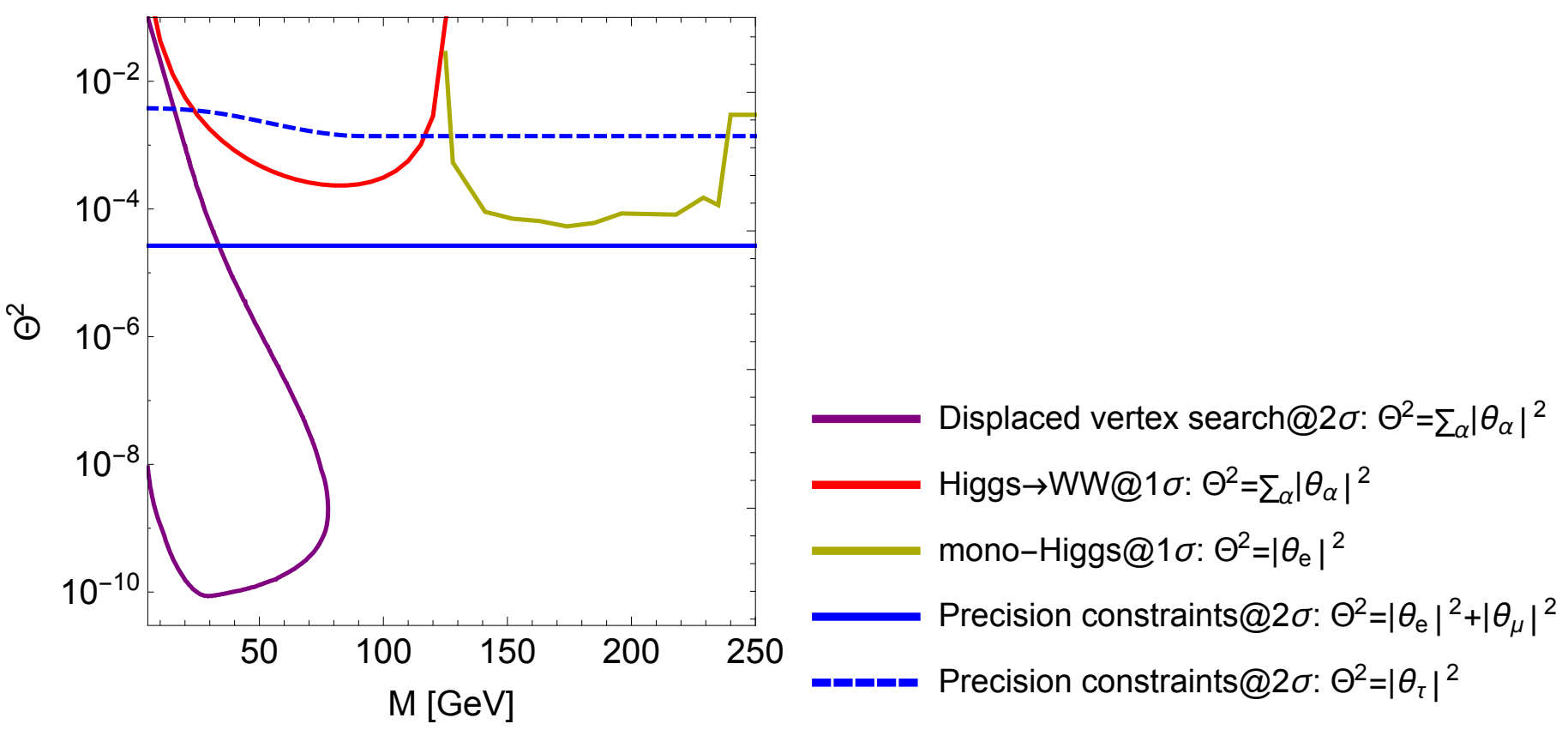

**Figure 2.33:** The CEPC's ability to probe heavy sterile neutrinos is expressed as a projected sensitivity on the active-sterile mixing angle, $\Theta$, and the sterile neutrino mass scale, $M$. The blue (solid and dashed) line denotes electroweak precision measurements [292, 297, 311, 312]. The purple line denotes displaced vertex searches [313] at the $Z$-pole run with an integrated luminosity of $10\ \mathrm{ab}^{-1}$. The yellow and red lines stem from the measurements of Higgs boson production [295, 296] and decay [297] for an integrated luminosity of $5\ \mathrm{ab}^{-1}$ at $\sqrt{s} = 240\ \mathrm{GeV}$.

**Modified Higgs self couplings:** In the Type I Seesaw the $N_i$ with masses $M_i$ of a few $\mathrm{TeV}$ can modify the trilinear Higgs self-coupling up to 30 percent [308]. This modification is also expected for the low-scale Type III Seesaw [309, 310]. CEPC sensitivity to the Higgs self-coupling via radiative corrections to Higgsstrahlung provides a promising probe of this effect.

## MODIFICATIONS OF ELECTROWEAK PRECISION OBSERVABLES

Neutrino mass physics can modify the theory predictions for electroweak precision observables, which may be measurable even if the new mass scale is above the CEPC center-of-mass energy. These can either occur due to virtual exchange of the new particles (which may be represented by higher dimensional operators in an EFT approach [314, 315]) or due to the production of new particles that mix with SM particles (e.g. with the active neutrinos or the SM Higgs boson).

In the context of the Type I Seesaw mechanism the mixings $\theta_{ai} = vY_{ai}/M_i$ of $n_s$ heavy right handed neutrinos with the SM neutrinos leads to an effective violation of unitarity in the $3 \times 3$ mixing matrix $V_\nu$, which is a submatrix of the $(3 + n_s) \times (3 + n_s)$ leptonic mixing matrix $\mathcal{U}$ [292, 316–318]. This affects all the electroweak precision observables. Such tests are mostly independent of the heavy neutrino masses $M_i$, and they test different combinations of the active-sterile mixing parameters [292, 297, 311, 312]. We show the corresponding possible sensitivity of CEPC by solid and dashed blue lines in Figure 2.33, considering a total integrated luminosity of $0.1\ \mathrm{ab}^{-1}$. In addition to the modified precision observables, one also expects violations of lepton universality and (apparent) violations of the unitarity of the Cabibbo-Kobayashi-Maskawa matrix [292, 298, 319–322].

In the context of Type II Seesaw, the electroweak precision observables are affected both by the triplet VEV, as well as by the mass splittings [266] that enter the oblique $T$ parameter. In the minimal LRSM, this splitting is predicted to be large and leads to a lower bound on the entire $\mathrm{SU}(2)_L$ triplet multiplet [305].

## DISPLACED SECONDARY VERTICES

The mechanism of neutrino mass generation can also give rise to truly exotic signatures in the form of long-lived particles whose decays produce displaced secondary vertices. Such displaced vertices are often poorly constrained at the LHC due to trigger and background limitations, whereas CEPC can provide significant sensitivity.

**Single displaced vertex in Type I Seesaw:** For masses below the $W$ bosons's mass, $m_W$, the lifetime of $N_i$ scales as $\tau_{N_i} \propto |\sum_a |\theta_{ai}|^2|^{-2} G_F^{-2} M_i^{-5}$ and their decays give rise to a visibly displaced secondary vertex in a large part of the allowed parameter space. Displaced vertex signatures have been studied in detail for the case of the Type I Seesaw, and the CEPC specific results from Refs. [261, 313] are shown in Figure 2.33 by the purple line. It is worth noting that with a longer $Z$-pole run the sensitivity for $M_i < m_Z$ can be significantly increased, see Figure 2.34. The sensitivity of a standard detector could be increased with additional detectors of the MATHUSLA [323, 324] or FASER [325] type.

**Long lived neutral scalars:** Due to mixing with the Higgs boson, the electrically neutral scalars in gauged $\mathrm{U}(1)_{B-L}$ [326] or the neutral scalar from $\mathrm{SU}(2)_R$ [327] can decay via the SM Yukawa couplings into the SM fermions. For masses in the $\mathrm{GeV}$ range, the resulting proper lifetimes can easily be $\mathcal{O}(1 \text{ cm})$, such that their decays give rise to displaced secondary vertices.

**Multiple displaced vertices:** Pair production of $N$ in exotic Higgs boson decays may lead to two displaced vertices, each containing a lepton and two jets at parton level, as pointed out in the context of LRSM [283, 284] and models with $B-L$ symmetry [299, 300]. Rare exotic decays of the SM-like Higgs boson to a pair of triplets with subsequent decay to 4 $N$s leads to up to four displaced vertices with rather soft final states, for which the CEPC is likely to be much better suited than the LHC.

Similarly, the associated production of the scalar triplet at $e^+e^- \to Z^* \to Z\Delta_R^0$ leads to two displaced vertices when $\Delta_R^0 \to NN$, while $Z$ decay gives additional prompt leptons/jets or missing energy.

## EXTRA GAUGE BOSONS

Extended theoretical frameworks generally predict more and stronger signals from heavy neutrinos. In particular the gauged $B-L$ symmetry, which contains an extra $Z'$ gauge boson, may give rise to a modified rate for the processes $e^+e^- \to \ell^+\ell^-$ at lepton colliders [331–333].

## LEPTOGENESIS

*Leptogenesis* refers to the idea that a matter-antimatter asymmetry is initially generated in the lepton sector [334] and then transferred into a baryon asymmetry via sphaleron processes [335]. Leptogenesis provides an explanation for the observed *baryon asymmetry of the universe* (BAU), i.e., the tiny excess $\eta_B \sim 10^{-10}$ [132] of matter over antimatter in the early universe over that formed the origin of the baryonic matter in the universe after mutual annihilation of all other particles and antiparticles, see e.g. [336]. Thus leptogenesis connects one of the deepest mysteries in cosmology to the properties of neutrinos.

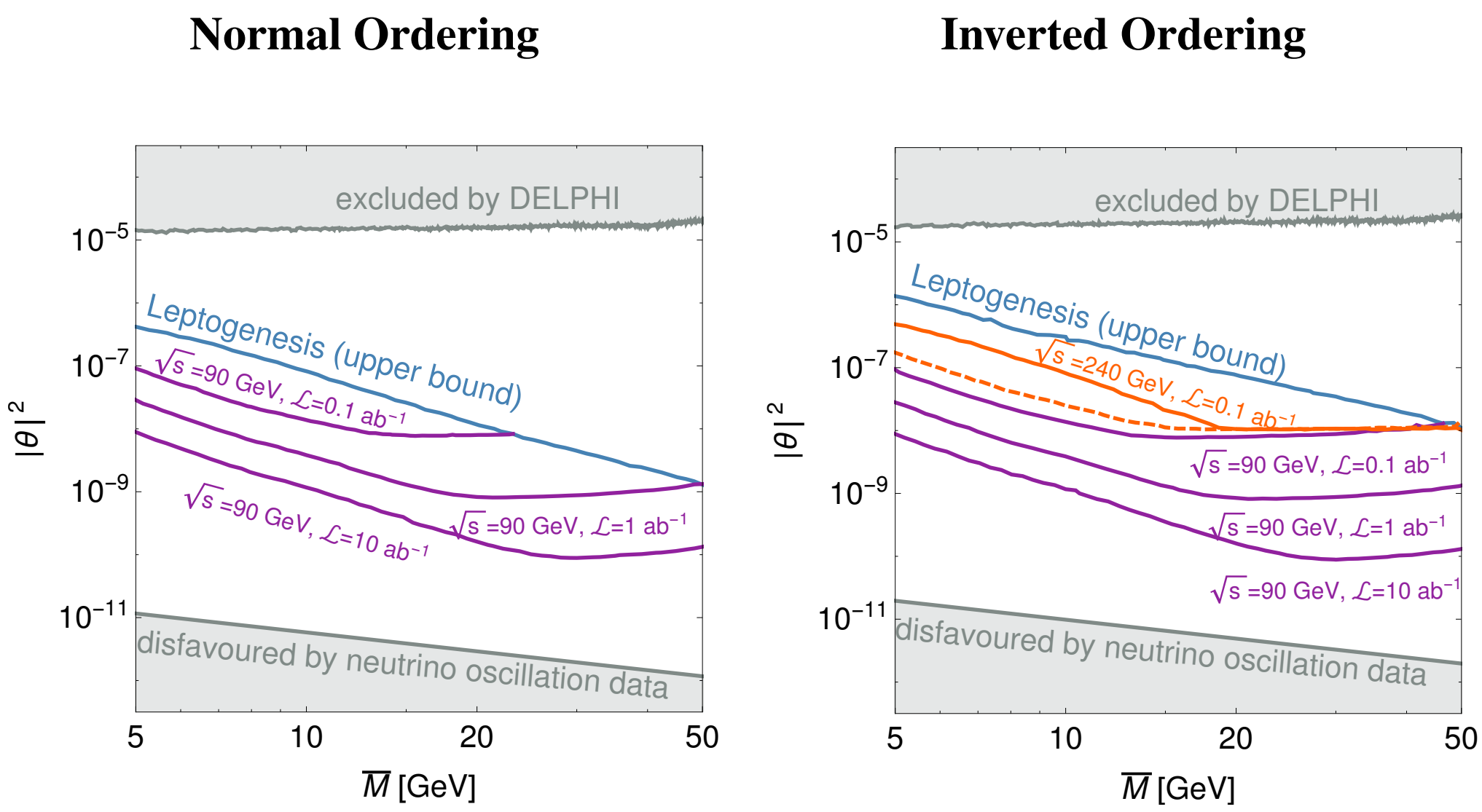

**Figure 2.34:** CEPC's capacity to test models of leptogenesis. The parameter space for a minimal Type I Seesaw model with $n_s = 2$ is shown; the two sterile neutrino masses, $M_1$ and $M_2$, are combined to form $\bar{M} = (M_1 + M_2)/2$ (with $|M_2 - M_1|/(M_2 + M_1) < 0.1$), and $\theta$ represents the active-sterile mixing angle. Models in the parameter space below the blue line are consistent with the observed baryon asymmetry of the universe through leptogenesis. Models above the orange lines are tested by CEPC at $\sqrt{s} = 240$ GeV, which is expected to observe at least four displaced vertex events. Models above the purple lines are probed by CEPC at the $Z$ pole. The gray areas are ruled out by the DELPHI experiment [328, 329] (top) and current neutrino oscillation data (bottom). The figure is based on Ref. [261]. Note that for $n_s = 3$ heavy neutrinos, the "leptogenesis" upper bound is expected to be much higher [330] and practically identical to the DELPHI constraint, so that CEPC at 240 GeV can enter the cosmologically interesting parameter region for both hierarchies.

**Motivation.** Global fits to present neutrino oscillation data prefer Charge-Parity ($CP$) violation in the leptonic sector at the 2 to 2.5 $\sigma$ level, see [218, 219]. This $CP$ violation in the leptonic sector may be related [337] to the observed BAU.

When the scale of new physics $\Lambda$ is above the collision energies at CEPC, it is impossible to discover the new particles responsible for the generation of the BAU via direct production. However, observing a combination of LNV and cLFV signatures at scales accessible to CEPC could still rule out such "high scale leptogenesis" scenarios because particles with LNV interactions near the electroweak scale could wash out baryon asymmetries that were produced at high scales [338, 339].

If, in contrast, $\Lambda$ is within reach of CEPC, one can directly probe the mechanism of leptogenesis by studying the properties of the new particles [340]. One of the best studied scenarios that accommodates leptogenesis is based on the low-scale Type I Seesaw model. The Yukawa couplings $Y_{ai}$ that couple the right-handed neutrinos $N_i$ to the Higgs boson and the left-handed neutrinos $\nu_{La}$ in general are complex and are a potential source of $CP$ violation. Hence, the $N_i$ may be the common origin on neutrino masses and baryonic matter in the universe.

If the mass range $M_i$ is around or below the collider-accessible $\mathrm{TeV}$ scale, leptogenesis can proceed in two different ways. For $M_i$ above the electroweak scale, the BAU can be generated during the freeze-out and decay of the $N_i$ [341] ("freeze-out scenario"). For masses below the electroweak scale the BAU can be generated in $CP$-violating oscilla-tions [250, 342] and Higgs decays [343] during the $N_i$ production ("freeze-in scenario"). The latter effectively also describes leptogenesis in the *Neutrino Minimal Standard Model* ($\nu$MSM) [249, 250], a complete model where a third heavy neutrino composes the Dark Matter [82, 83] and does not contribute significantly to neutrino mass generation and leptogenesis due to strong observational constraints [344]. Due to its minimality, part of the relevant parameter space of this model is in principle fully testable at colliders [345, 346], and significant fractions of the parameter space can be probed with CEPC [261]. For $M_i$ below the electroweak scale, this analysis could be done with an accuracy on the percent level at the $Z$ pole with $10\ \mathrm{ab}^{-1}$ [261].

**Lepton-number violation.** Lepton number violation is a crucial ingredient of any leptogenesis scenario. Typical signatures at CEPC may involve same sign dilepton final states, either in prompt or displaced decays. An observation of such processes in all three SM flavors or a combination of LNV in some channel and different cLFV signatures could potentially falsify high scale leptogenesis scenarios [338, 339].

Many low scale models rely on an approximate lepton-number conservation to explain the smallness of the neutrino masses in the regime of coupling constants that is accessible to CEPC [241–243], which parametrically suppresses the rate of LNV processes in prompt decays. For particles with quasi-degenerate masses and comparable lifetimes, as they e.g. appear in resonant leptogenesis scenarios of the $\nu$MSM, it has been proposed that this suppression may be overcome by the long time that they have to undergo coherent oscillations within the detector [259, 260, 347]. Since the amount of lepton number violation is proportional to the mass splitting, indirect measurements may be possible from a comparison of the rates for lepton number violating and conserving processes [259, 347] or by observing heavy neutrino-antineutrino oscillations in the detector [260] in displaced vertex searches at CEPC [261]. The reach of such searches at CEPC in the minimal seesaw model is shown in Figure 2.34, see also Figure 2.33.

**Lepton-flavor violation.** Measurements of cLFV are crucial to test high scale leptogenesis models at CEPC, because an efficient washout of the asymmetries in all flavors at temperatures above the electroweak scale is crucial to rule out such scenarios as the origin of the BAU [339].

Low scale leptogenesis scenarios typically rely on flavor effects and therefore tend to make predictions for the rates of cLFV. In the minimal Type I Seesaw with $n_s = 2$ (or the $\nu$MSM), leptogenesis significantly restricts the flavor mixing pattern of heavy neutrinos $N_i$ with experimentally accessible mixing angles [346]. The accuracy on the percent level at which the flavor mixing pattern can be probed in displaced vertex searches with $10\ \mathrm{ab}^{-1}$ at the $Z$ pole are sufficient to probe large fractions of the parameter region for which heavy neutrinos can be discovered.

**Displaced decays from long lived heavy neutrinos.** For heavy neutrino masses below the electroweak scale, where leptogenesis proceeds in the "freeze in" manner, the $N_i$ couplings should be comparably small to avoid a complete washout of the BAU in the early universe ($|\theta_{ai}|^2 < 10^{-8} \times (10\ \mathrm{GeV}/M_i)$ [348], where larger values can be allowed due to strong hierarchies in their couplings to individual SM flavors [330]). Hence, most of the parameter space of active-sterile neutrino mixing and masses that is compatible with low scale leptogenesis in this scenario gives rise to long lifetimes of the heavy neutrino mass eigenstates, which can be found with high sensitivity via displaced vertex searches at CEPC. The reach of such searches at CEPC is compared to the parameter region where leptogenesis is feasible in the minimal seesaw model in Figure 2.34.

### 2.3.5 EXTENDED HIGGS SECTOR

In many extensions of the Standard Model, the Higgs boson is embedded in a larger Higgs sector. Searching for new Higgs bosons is an important experimental target with a high priority. One of the most straightforward and well-motivated extensions is the two-Higgs-doublet model (2HDM) [349], in which there are five massive spin-zero states in the spectrum $(h, H^0, A^0, H^\pm)$ after electroweak symmetry breaking. Extensive searches for BSM Higgs bosons have been carried out, especially at the LHC [23, 350–360]. Null results in searches to date imply that either the non-SM Higgs bosons are much heavier and essentially decoupled from the SM, or the lightest $CP$-even Higgs boson mimics the SM Higgs boson by accident or symmetry while non-SM Higgs bosons are light as well [361–363]. In either case, it would be challenging to observe those states directly in experiments.

Complementary to the direct searches, precision measurements of the SM parameters and Higgs properties could also provide useful probes of new physics. High-precision measurements at future Higgs factories with about $10^6$ Higgs bosons, and $Z$-pole measurements with $10^{10} - 10^{12}$ $Z$ bosons [143, 364–367] would invariably shed light on new physics associated with the electroweak sector such as an extended Higgs sector. There is an extensive literature on the effects of the heavy Higgs states on the SM Higgs couplings, e.g. [21, 349, 368–376]. Identifying the light $CP$-even Higgs boson $h$ to be the experimentally observed 125 GeV Higgs boson, the couplings of $h$ to the SM fermions and gauge bosons receive two contributions: tree-level values, which are controlled by the mixing angles $\alpha$ of the two $CP$-even Higgs bosons and $\tan\beta$, ratios of the vacuum expectation values of two Higgs doublets, $\tan\beta = v_1/v_2$, and loop corrections from heavy Higgs bosons running in the loop. Of particular interest is the "alignment limit" with $\cos(\beta - \alpha) = 0$, in which the light $CP$-even Higgs couplings are identical to the SM ones at the tree-level, regardless of the other scalars' masses. Loop corrections, however, could lead to deviations of the couplings of $h$ to the other SM particles, even in the alignment limit.

We first consider tree-level corrections. The allowed region at 95% CL in the $\cos(\beta-\alpha)$ vs. $\tan\beta$ plane for various types of 2HDM (depending on how the two Higgs doublets are coupled to the quarks and leptons) are shown in Figure 2.35 including only tree-level effects. This is obtained via a global fit to the Higgs rate measurements at the LHC as well as CEPC, assuming that no deviation to the SM values is observed at future measurements. From the figure, one can see that $\cos(\beta - \alpha)$ in all four types is tightly constrained at both small and large values of $\tan\beta$, except for Type-I, in which constraints are relaxed at large $\tan\beta$ due to suppressed Yukawa couplings.

To fully explore the Higgs factory potential, both the tree-level deviation and loop corrections to the SM Higgs couplings need to be taken into account. Figure 2.36 shows the global fit results to all CEPC Higgs rate measurements in the Type-II 2HDM parameter space, including both tree level and loop corrections. Degenerate heavy Higgs boson masses $m_A = m_H = m_{H^\pm} = m_\Phi$ are assumed so that the $Z$-pole precision constraints are automatically satisfied. The left panel is in the $\cos(\beta-\alpha)$ vs. $\tan\beta$ plane with regions en-closed by curves allowed if no deviation from the SM prediction is observed. Black, red, blue, and green curves are for $\sqrt{\lambda v^2} = \sqrt{m_\Phi^2 - m_{12}^2/s_\beta c_\beta} = 0$, 100, 200, and 300 GeV, respectively. The global fit result with tree-level only corrections is shown by dashed black lines for comparison. In all scenarios, $|\cos(\beta-\alpha)|$ is typically constrained to be less than about 0.008 for $\tan\beta \sim 1$. For smaller or larger values of $\tan\beta$, the allowed range of $\cos(\beta-\alpha)$ is significantly reduced. Loop effects from the heavy Higgs bosons tilt the allowed $\cos(\beta-\alpha)$ towards negative values, especially when $\tan\beta$ is large.

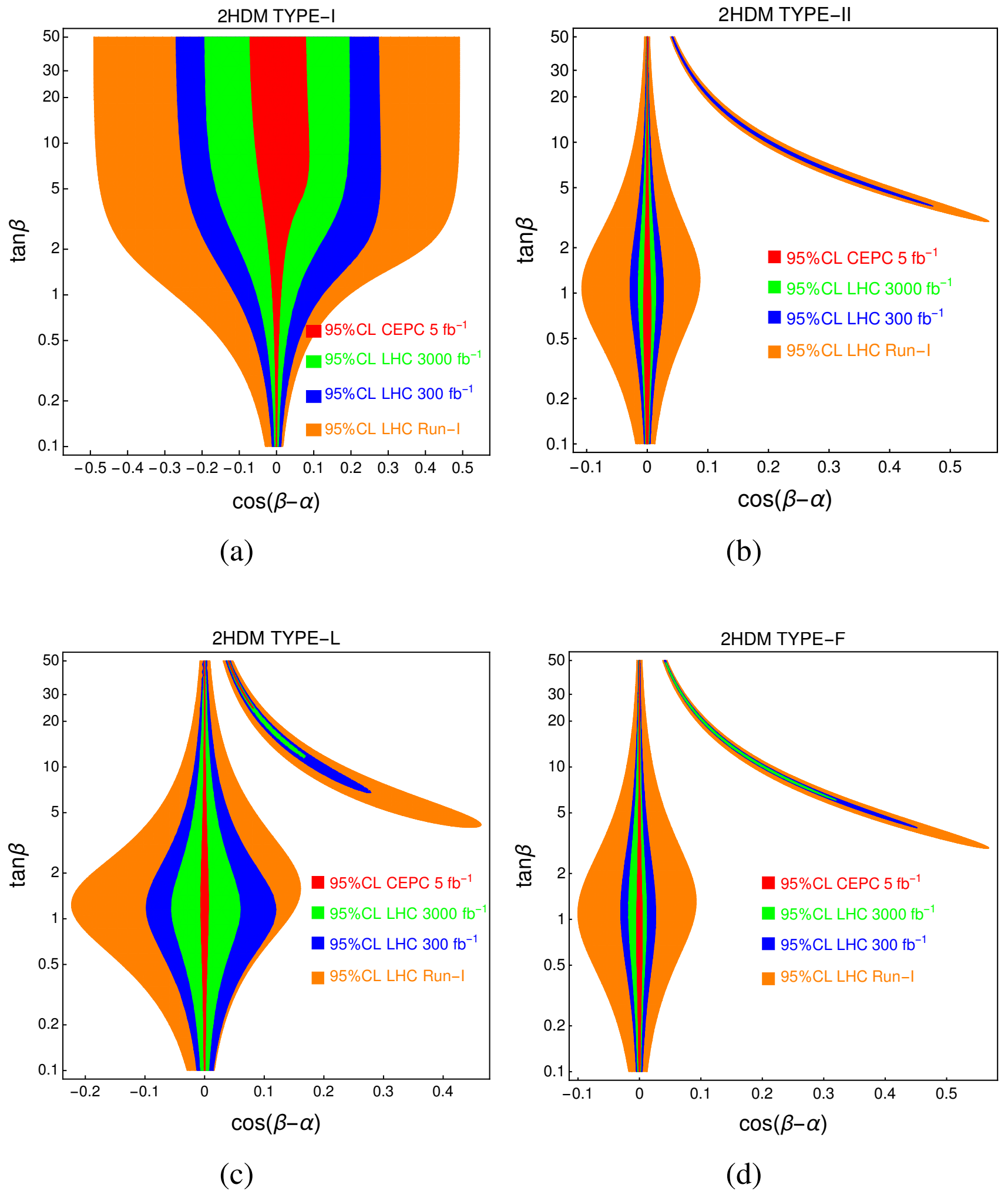

**Figure 2.35:** This figure shows the CEPC's capacity to test for new physics in the Higgs sector and its dramatic improvement over existing and projected limits from the LHC. Shaded regions show the viable parameter space assuming that the future measurements agree with SM predictions. The panels show the four types of two-Higgs-doublet models (2HDM). The special "arm" regions for the Type-II, L, and F 2HDMs are the wrong-sign Yukawa regions. Plots are taken from Ref. [21].

The Figure 2.36(b) shows the allowed region at 95% CL in the $m_\Phi$ vs. $\tan\beta$ plane, with $\cos(\beta-\alpha) = -0.005$ (green), 0 (blue), and 0.005 (red). In the alignment limit with $\cos(\beta-\alpha) = 0$, the mass of the heavy Higgs bosons $m_\Phi > 500$ GeV is still allowed when $\tan\beta \lesssim 10$. Once deviating away from the alignment limit, the constraints on the heavy Higgs boson mass get tighter. The reach in the heavy Higgs boson mass and couplings at future Higgs factories can be complementary to the direct search limits at the LHC [23, 350–360], especially at intermediate values of $\tan\beta$.

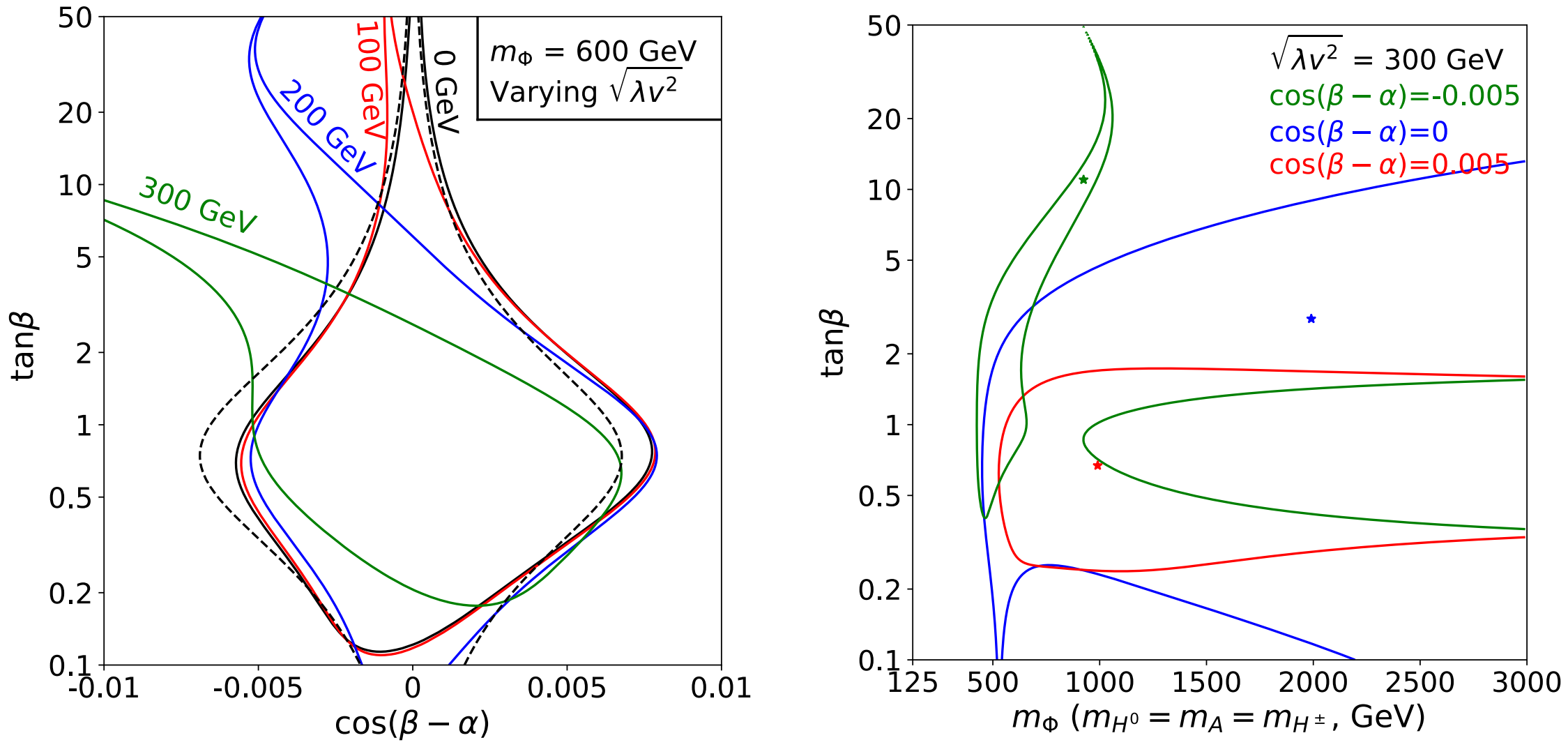

(a) (b)

**Figure 2.36:** The constraining power of the CEPC precision measurements are illustrated here using the Type-II 2HDM parameter space. Assuming that no deviation from SM predictions are observed, the allowed regions of parameter space (at 95% CL) are enclosed by the curves with the same style. These curves are calculated by a fit including both the tree-level and the loop correction to the SM Higgs couplings. The (a) is in the $\cos(\beta-\alpha)$ vs. $\tan\beta$ plane, with $m_A = m_H = m_{H^\pm} = m_\Phi = 600$ GeV. The parameter $\sqrt{\lambda v^2}$ is set to be 0 (black solid), 100 (red solid), 200 (blue solid), and 300 GeV (green solid). The global fit result with tree-level only correction is represented by the dashed black lines for comparison. The (b) is in the $m_\Phi$ vs. $\tan\beta$ plane with $\sqrt{\lambda v^2} = 300$ GeV. The values of $\cos(\beta-\alpha)$ are chosen to be $-0.005$ (green), 0 (blue), and 0.005 (red). The stars represent the corresponding best fit points. These plots are taken from Ref. [376].

Going beyond the degenerate mass case, both the Higgs and $Z$-pole precision measurements are sensitive to the mass splitting between the charged and neutral Higgeses, as well as the splitting between the neutral ones. Figure 2.37 shows the allowed region of $\Delta m_A = m_A - m_H$ and $\Delta m_C = m_{H^\pm} - m_H$ at 95% CL, for different choices of $\cos(\beta-\alpha)$. The Higgs and $Z$-pole precision constraints are presented separately in the left panel while the combined constraints are shown in the right panel, with $m_H = 600$ GeV and $\sqrt{\lambda v^2} = 300$ GeV. For the Higgs precision fit, in the alignment limit, $\Delta m_A$ and $\Delta m_C$ are bounded to be around 0 within a few hundred GeV. $\Delta m_A$ is constrained to be positive when $\cos(\beta-\alpha)$ takes a (small) positive value, and negative when $\cos(\beta-\alpha)$ is negative. The $Z$-pole precision measurements constrain either $\Delta m_C \sim 0$ or $\Delta m_C \sim \Delta m_A$, equivalent to $m_{H^\pm} \sim m_{H,A}$. In the small range of $\cos(\beta-\alpha)$ allowed by the current LHC Higgs precision measurements, the change of the $Z$-pole constraints due to different choices of $\cos(\beta-\alpha)$ is negligible. Combining both the Higgs and $Z$-pole precisions (right panel), the allowed $\Delta m_{A,C}$ is further constrained to be in a smaller region. From the plots, one can see that $Z$-pole measurements and Higgs measurements are complementary in constraining the heavy Higgs boson mass splittings.

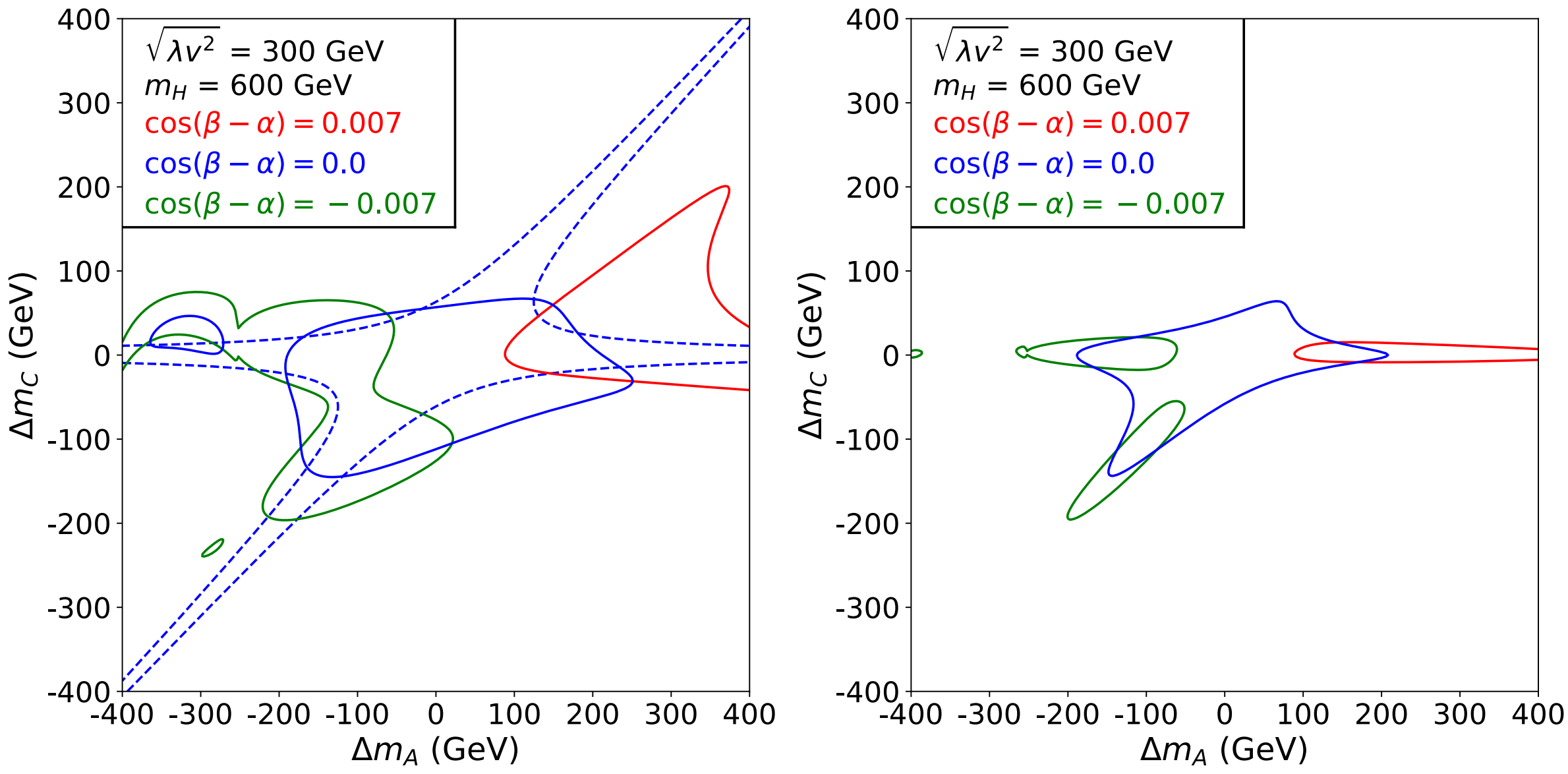

**Figure 2.37:** Allowed regions of $\Delta m_A = m_A - m_H$ and $\Delta m_C = m_{H^\pm} - m_H$ at 95% CL, for different choices of $\cos(\beta - \alpha)$. Left: Higgs precisions constraints for $\cos(\beta - \alpha) = 0.007$ (solid red), 0 (solid blue), and $-0.007$ (solid green) and $Z$-pole constraints (dashed blue). Note that the $Z$-pole constraints are the same for $\cos(\beta - \alpha) = 0.007$, 0, and $-0.007$. Right: constraints from combining both the Higgs and $Z$ pole measurements for $\cos(\beta - \alpha) = 0.007$ (solid red), 0 (solid blue), and $-0.007$ (solid green). Plots are taken from Ref. [377].

## 2.4 QCD PRECISION MEASUREMENT

As a fundamental force in nature, the strong force is primarily responsible for the generation of the proton's mass. The discovery in the 1970's of Quantum Chromodynamics (QCD) as a correct theory for describing the strong force marks a great achievement in the history of physics. Despite forty years of intense study and much progress, QCD remains the least understood quantum field theory of nature, particularly in its non-perturbative domain. Even at high energy where the strong force becomes weak due to the property of asymptotic freedom, it is still challenging to obtain a precise quantitative description of QCD phenomena. For example, the "fine structure constant" of QCD, $\alpha_s$, is eight orders of magnitude less constrained than the fine structure constant of Quantum ElectroDynamics (QED). In this respect, QCD is currently the least constrained fundamental force of nature, including gravity. Improving the precision in our understanding of QCD directly impacts our understanding of nature, ranging from the production and decay of the Higgs boson, the partonic structure of proton, and the stability of the Standard Model vacuum.

QCD can be studied at lepton, lepton-hadron, and hadron colliders. In recent years much efforts have been devoted to the study of QCD dynamics at hadron colliders, such as the Large Hadron Collider (LHC), because of its rich QCD phenomenology at diverse energy scales and the intimate relation between QCD prediction for the SM background and new physics searches. However, the strongly-interacting nature of the initial state adds additional complications to the description of hard scattering, including the need for the detailed knowledge of Parton Distribution Functionss (PDFs), as well as the re-moval/subtraction of the effects from multiple scattering or underlying events. These complications make precision study at hadron collider challenging, although remarkable progresses have been made. On the other hand, such complications are absent at a lepton collider, making it an ideal environment for studying QCD with the highest precision. In the past lepton colliders have played an important role in the study of QCD, from the direct observation of gluon jets to the precise extraction of $\alpha_s$. Compared with LEP, the largest $e^+e^-$ collider ever built, CEPC will have substantial improvement in statistics and theoretical systematics. Potential improvement in detector design, in particular particle identification capabilities, can also lead to better experimental systematics. Therefore CEPC will provide opportunities for QCD studies at unprecedented precision, as well as providing new constraints to our current understanding hadronization in high-energy processes. The increase in collision energy will also allow for the exploration of QCD phenomena in territory previously unaccessible at a lepton collider. Besides those well-known problems from the LEP era, many new directions in QCD and jet physics have been opened since the LHC era due increasing attention to the study of jet structure, either as a way to disentangle new physics from QCD backgrounds, or as a probe of QCD dynamics. CEPC will be an ideal machine to address many of these questions at high precision, due to the absence of complications from multiple scattering and underlying events.

Combined with the remarkable progress in QCD theory, ranging from new methods for efficient calculation of cross sections, to the development of effective field theory for collider processes, to new ideas for simulating scattering processes on the lattice, it is expected that CEPC will mark a new chapter in QCD research.

### 2.4.1 PRECISION $\alpha_S$ DETERMINATION

The strong coupling constant $\alpha_s$ is perhaps the most important parameter in QCD. It enters the perturbative predictions of QCD in every observable, in particular cross sections for scattering processes involving hadronic final states at CEPC. A precision determination of $\alpha_s$ at CEPC with unprecedented experimental uncertainties will be an important contribution to the world's effort to determine $\alpha_s$. At a lepton collider, $\alpha_s$ can be measured in a number ways. The represented ones include hadronic $Z$ decay, hadronic $\tau$ decay, QCD jet rates, and QCD event shape measurements. A summary of $\alpha_s$ determination from these observables using LEP data can be found in Ref. [291].

A distinct feature of CEPC compared with previous lepton colliders is the increase in center-of-mass energy, $Q$. The measurements which can benefit from increased energy are event shape observables, for which non-perturbative corrections typically scale as $c\Lambda_{\rm QCD}/Q$, where $c$ is an $\mathcal{O}(1)$ parameter that can not be calculated from first principle with our current understanding of QCD. There exist two different approaches in the modeling of non-perturbative hadronization effects for event shapes. One approach is based on corrections for non-perturbative hadronization effects using QCD inspired Monte Carlo tools [378–382], and the other is based on analytic modeling of the non-perturbative shape function [383–387]. Neither of the two treatments can be regarded as fully satisfactory. In the Monte Carlo approach, there is mismatch in the parton level definition of a Monte Carlo simulation and the fixed order calculation. In the analytic power correction approach, the associated systematics have not been fully verified. Therefore, by going to higher center-of-mass energy, the impact of hadronization effects and their associated uncertainties can be reduced.

As an example of $\alpha_s$ determination from event shape observables using analytic power correction, we quote the recent determination based on the $C$ parameter from Ref. [387],

$$\alpha_s(m_Z) = 0.1123 \pm 0.0002_{\text{exp}} \pm 0.0007_{\text{hadr}} \pm 0.0014_{\text{pert}} \,, \tag{2.28}$$

where hadronization effects and perturbative uncertainties are the main source of uncertainties contributing to $\alpha_s$ determination. While the perturbative uncertainties can be expected to be reduced further in the coming years, given the remarkable progress in the calculation of higher order corrections and in the resummation of large logarithms, the reduction of hadronization uncertainty will likely come from an increase of center-of-mass energy. It is interesting to observe that the value of $\alpha_s$ determined from Ref. [387], as well from thrust using similar methods [388], seem to be systematically lower than the world average. With large data sets for event shape and higher center-of-mass energy to suppress hadronization uncertainties, CEPC provides an excellent opportunity to address this discrepancy and deepen our quantitative understanding of QCD.

Currently, for thrust [384, 389], $C$ parameter [386, 387], and heavy-jet-mass distribution [390], the best theoretical predictions are at the level of N$^3$LL resummation matched to NNLO in fixed order perturbation theory. A notable recent development is the calculation of Energy-Energy Correlation (EEC) at NNLO. EEC is an event shape observable which exhibits the so-called rapidity divergence, and leads to additional logarithms to be resummed, compared with thrust and other observables. Very recently, a determination of $\alpha_s$ using NNLL resummation matched to NNLO, and Monte Carlo for the modeling of power corrections, has been done, with the result [391] being

$$\alpha_s(m_Z) = 0.11750 \pm 0.00018_{\text{exp}} \pm 0.00102_{\text{hadr}} \pm 0.00257_{\text{ren}} \pm 0.00078_{\text{res}} \,, \tag{2.29}$$

where hadronization effects are important source of uncertainties. Since the analysis in Ref. [391] only uses data at or below the $Z$ pole, it is expected that future data from CEPC at $250$ GeV can significantly reduce the hadronization uncertainty. Additional scale and resummation uncertainties can also be reduced in the future by incorporating N$^3$LL resummation [392].

### 2.4.2 JET RATES AT CEPC

Another distinct feature of CEPC compared with LEP is its unprecedented luminosity, in particular above the $Z$ pole. The higher luminosity opens the door for the precision study of multi-jet production at an $e^+e^-$ collider.

As an example, we show in Figure 2.38 the four-jet production cross sections at CEPC ($\sqrt{s} = 250$ GeV) with the Durham jet algorithm as a function of the resolution parameter $y_{cut}$, calculated using NLOjet++ [393]. The cross sections are at the level of a few $\text{pb}$ to tens of $\text{pb}$ for the range of $y_{cut}$ considered. The colored bands represent the scale variations calculated by varying the renormalization scale from $\sqrt{s}/2$ to $2\sqrt{s}$. The NLO predictions show a smaller scale variation as compared to the LO ones. The cross sections diverge for small resolution parameter where further QCD resummations are needed to stabilize the theoretical predictions. The right panel shows the projected statistical uncertainties assuming an integrated luminosity of $1$ and $5$ $\text{ab}^{-1}$. The statistical uncertainties are at the level of one per mille or better for $y_{cut}$ below $10^{-2}$ due to the large luminosity. The scale uncertainties of the NLO predictions are large in comparison and about $10\%$, which can be reduced with QCD resummation [393]. The $n$-jet rate has been employed to measure the strong coupling constant $\alpha_s$ at LEP [394]. The four-jet cross sections are proportional to $\alpha_s^2$ at leading order, thus the statistical uncertainties in the measurement of $\alpha_s$ are estimated to be well below one per mille. On the other hand, the theoretical uncertainties will play a dominant role and need further investigation. Currently, NNLO predictions for $e^+e^-$ to three jets are available [395–399]. Along this line there has been remarkable progress in the calculation of two-loop amplitudes with five external particles [400, 401] and its associated integrals [402, 403]. Although there is still substantial work to be done, an NNLO calculation for four jet production can be expected in the future. There has also been progress in resumming the large logarithms in jet rates. A Monte Carlo approach for resummation has been proposed and used to resum the large logarithms in two-jet rates in Ref. [404], which can achieve resummation at NNLL level. Ideally this approach can be extended to three and four jet rates.

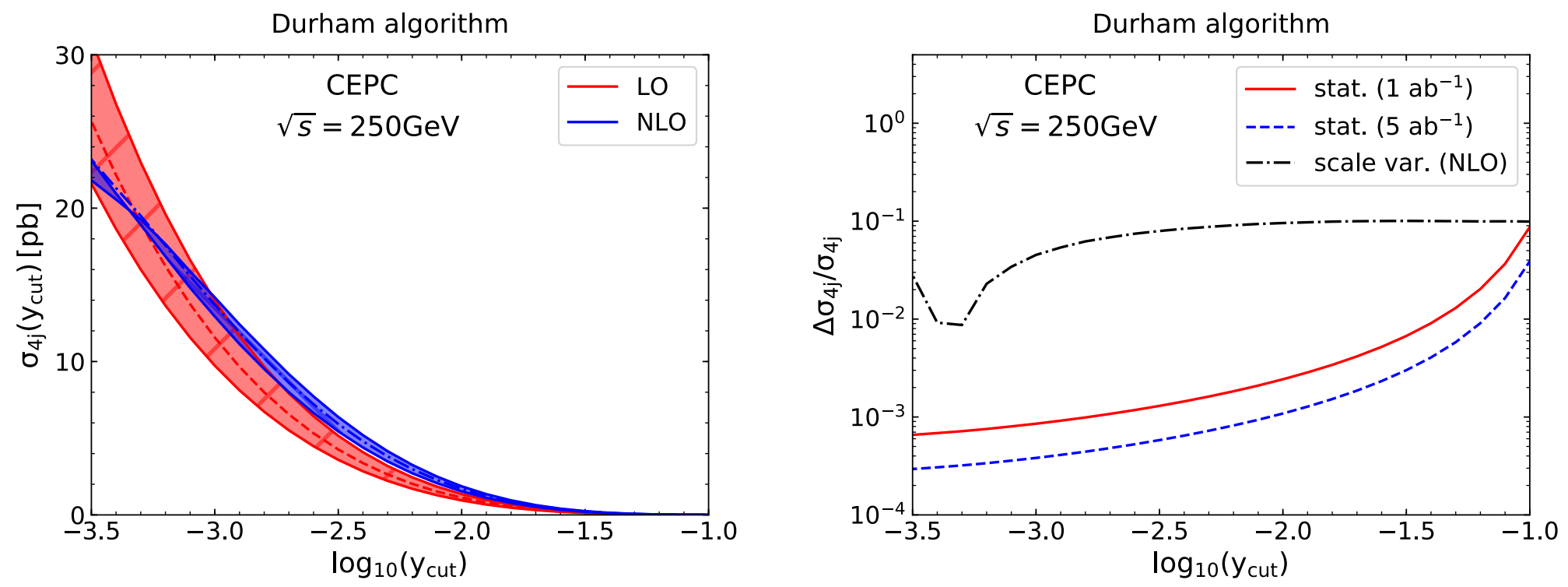

(a) (b)

**Figure 2.38:** (a) The four-jet production cross section at CEPC ($\sqrt{s} = 250$ GeV) with the Durham jet algorithm as a function of the resolution parameter $y_{cut}$. (b) The scale variation and expected statistical uncertainties for the same cross sections normalized to their central values.

### 2.4.3 NON-GLOBAL LOGARITHMS

Besides the precision extraction of $\alpha_s$ from jetty final states, there has also be significant interest in understanding some novel aspects of QCD dynamics from jet processes at a lepton collider. An important example is the study of Non-Global Logarithms (NGL) [405, 406].

Non-global logarithms are significant obstacles in the study of soft physics at high energy colliders (jet physics, energy flow measurements, hadronization, and so on). Therefore it is important to develop a theoretical framework to understand their structure. NGLs were first pointed out by Dasgupta and Salam in Ref. [405], where they developed a Monte-Carlo algorithm to resum leading-logarithmic(LL) NGLs in the large $N_c$ limit. After that work, based on the strong energy ordering limit, Banfi, Marchesini and Smye derived an integral-differential evolution equation that can also resum LL NGLs [406]. Since then, there has been a great effort to improve the theoretical predictions [407–412], including the sub-leading $N_c$ effects [413–415] and some fixed-order calculations [416, 417].

Recently, there have been several developments in this field [418–426]. One example is the effective field theory developed in Ref. [420]; this reference was the first to write down the factorization formula for non-global observables and to give an any-order renormalization group evolution equation for NGLs.

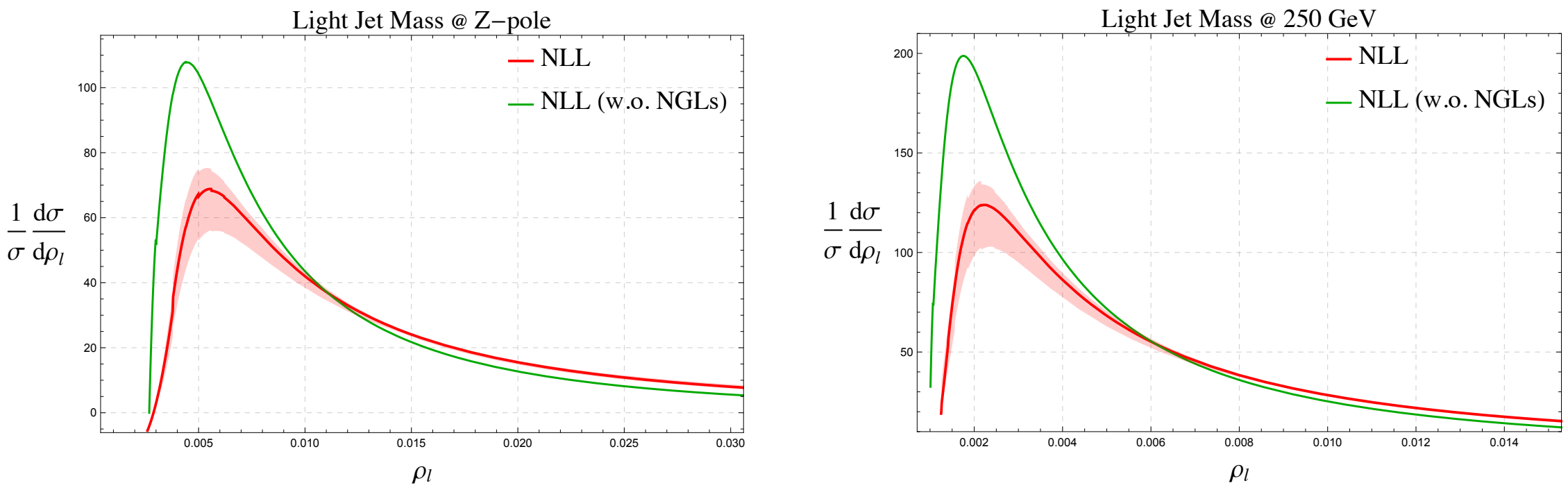

**Figure 2.39:** The normalized light-jet-mass distribution both at $Z$-pole (left) and at 250 GeV (right). Green curves are NLL results without NGLs, and red bands are full NLL results with scale uncertainties.

While NGLs can be studied with exclusive jet shape observables, their precision studies at hadron collider are difficult. This is because the environment at hadron collider is so complicated that it is difficult to isolate the NGL dynamics from underlying events and hadronization. As an electron-positron collider at high energy, CEPC will provide new opportunities of precisely measuring NGLs in many observables, where those unrelated effects are absent or suppressed. Figure 2.39 shows the normalized light-jet-mass distribution both at $Z$-pole (left) and at 250 GeV (right). Green curves are NLL results without NGLs, and red bands are full NLL results with scale uncertainties. Obviously, after including NGLs theoretical predictions are reduced significantly, and this reduction is especially magnificent at 250 GeV. Therefore CEPC will give us the first opportunity to measure NGLs.

### 2.4.4 QCD EVENT SHAPES AND LIGHT QUARK YUKAWA COUPLING

The SM Higgs boson decays dominantly to various hadronic final states with a total branching fraction of more than 80%. These hadronic decays provide a new source for QCD studies at CEPC (in its Higgs factory mode). In particular, Higgs decays produce a unique color-neutral digluon state. Table 2.3 summarizes the estimated number of events for different hadronic decay modes of the Higgs boson, assuming that the tagged $Z$ boson decays into electrons or muons.

At CEPC the traditional hadronic event shapes, e.g., thrust distribution, can be well measured due to the high statistics. At a lepton collider one can reconstruct the kinematics fully and then boost all final states back to the rest frame of the decaying Higgs boson. On the theory side those distributions can be calculated with high precision by QCD resummation matched with fixed-order results. There exist uncertainties from non-perturbative QCD effects, e.g. hadronization modeling, which are usually estimated by Monte Carlo event generators. The Figure 2.40(a) shows the normalized distribution of the variable thrust for several different hadronic decay channels of the Higgs boson, including $gg$, $q\bar{q}$, $b\bar{b}$, and $W(q\bar{q})W^*(q\bar{q})$ [427]. The distribution peaks at $\tau \sim 0.02$ for the light-quark decay channel. The peak shifts to $\tau \sim 0.05$ for the gluon channel, corresponding to a scaling of roughly $C_A/C_F$. The distribution is much broader for the gluon case due to the stronger QCD radiation. The distribution for the $b\bar{b}$ channel is very close to the $q\bar{q}$ case, except at very small $\tau$, where the mass and hadronization effects become important. For the $WW^*$ channel there already exist four quarks at leading order and the distribution is concentrated in the large-$\tau$ region.

| $Z(l^+l^-)H(X)$ | $gg$ | $b\bar{b}$ | $c\bar{c}$ | $WW^*(4h)$ | $ZZ^*(4h)$ | $q\bar{q}$ |
|---|---|---|---|---|---|---|
| BR [%] | 8.6 | 57.7 | 2.9 | 9.5 | 1.3 | $\sim 0.02$ |
| $N_{\text{event}}$ | 6140 | 41170 | 2070 | 6780 | 930 | 14 |

**Table 2.3:** This table shows branching ratios (BR) for decays of the SM Higgs boson in different hadronic channels [428] and the number of expected events ($N_{\text{event}}$) for $ZH$ production at CEPC ($\sqrt{s}$ = 240 GeV and $\mathcal{L}$ = 5 ab$^{-1}$) with the corresponding hadronic Higgs decay. In this table, $H$ represents the Higgs boson, $h$ represents any of the quarks except the top quark, and $q$ are light quarks.

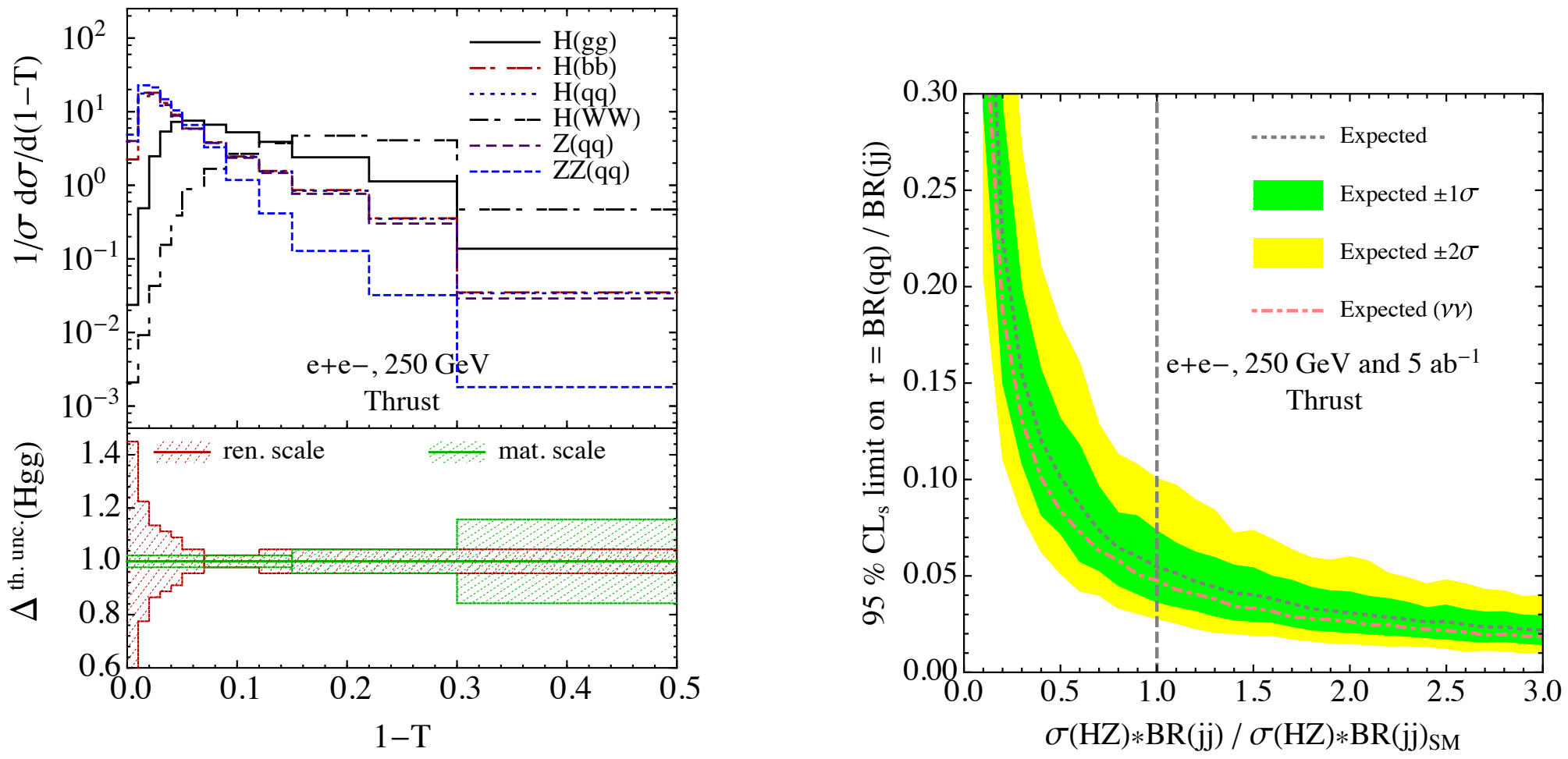

(a) (b)

**Figure 2.40:** (a) The normalized distributions of thrust in hadronic Higgs decay, in $e^+e^- \to q\bar{q}$ with $\sqrt{s} = 125$ GeV and in $e^+e^- \to Zq\bar{q}$ with $\sqrt{s} = 250$ GeV. (b) CEPC's capacity to probe the Higgs boson's decay into light quarks. The green and yellow bands show the expected 95% CL exclusion limit on $r = \text{BR}(qq)/\text{BR}(jj)$ as a function of the total cross section of the Higgs boson decay to $jj$ normalized to the SM value.

Different shapes of the thrust distribution from diquark and digluon final states motivates the idea of using global event shapes to probe the Yukawa couplings of light quarks [427], namely strange, up and down quarks. The provided discrimination can largely reduce background due to Higgs boson decays into two gluons while backgrounds from Higgs boson decays into heavy quarks can be suppressed with the usual heavy-flavor tagging algorithms. It is a great challenge to probe the light-quark Yukawa couplings since they are very small and the expected number of events with CEPC's full luminosity is only 14, as shown in Table 2.3. The expected exclusion limits on decay branching ratios of Higgs boson to light quarks are shown in the right plot of Figure 2.40, indicated by intersections with the vertical line and normalized to the branching ratio to digluon. The results can be translated into an upper limit of 0.48% on the decay branching ratios or 5 times of the Standard Model value for the Yukawa coupling of strange quark.

## 2.5 FLAVOR PHYSICS WITH THE $Z$ FACTORY OF CEPC

A high luminosity $Z$ factory that produces $10^{12}$ $Z$ bosons provides unique opportunities for various flavor measurements. In particular, the decay of $10^{12}$ $Z$ bosons will result in approximately $10^{11}$ $b$ hadrons, which is almost two orders of magnitude larger than the number of $B$ mesons produced at the $B$ factories BaBar and Belle and comparable to the number of $B$ mesons expected at Belle II.

As the $B$ factories are running mainly on the $\Upsilon(4S)$ resonance, they mostly produce $B^0$ and $B^\pm$ mesons; they also produce $B_s$ mesons but in much smaller numbers from shorter runs on $\Upsilon(5S)$. A machine running on the $Z$-pole on the other hand will not only produce a large number of $B^0$, $B^\pm$, and $B_s$ mesons, but also a large sample of $b$ baryons. In Table 2.4 we compare the expected numbers of produced $b$-hadrons corresponding to $10^{12}$ $Z$-boson decays to those produced with the $50\,\mathrm{ab}^{-1}$ run on $\Upsilon(4S)$ and the $5\,\mathrm{ab}^{-1}$ run on $\Upsilon(5S)$ of Belle II and at LHCb with $50\,\mathrm{fb}^{-1}$. The expected tagging efficiency at the CEPC may be conservatively be gauged by the $b$-tagging efficiency employed by the DELPHI collaboration at LEP, roughly $30\%$ [429], which is comparable to the one at Belle II [430]. For the tera-$Z$ we also list number of produced charmed hadrons and tau leptons (we use the known $Z$ branching fractions $\mathrm{BR}(Z \to b\bar{b}) = (15.12 \pm 0.05)\%$, $\mathrm{BR}(Z \to c\bar{c}) = (12.03 \pm 0.21)\%$, and $\mathrm{BR}(Z \to \tau^+\tau^-) = (3.3696 \pm 0.0083)\%$ [291] and the $b$ and $c$ hadronization fractions at the $Z$ pole from Refs. [431–433]). Using the large sample of produced $b/c$ hadrons and taus, the tera-$Z$ factory of CEPC will be able to access many rare decays of these particles, many with a precision beyond any of the ongoing or planned experiments. In addition, the $10^{12}$ $Z$ bosons would also allow measurements of flavor violating $Z$ decays with an unprecedented precision.

There are important key differences between a future circular electron–positron collider running at the $Z$ pole, the LHCb experiment, and the Belle II $B$ factory. Compared to LHCb, an electron–positron collider offers a much cleaner environment and, therefore, generally much smaller background levels. The low background allows reconstruction of final states with multiple neutrals like photons, neutral pions and kaons. Compared to the Belle II flavor factory, running at the $Z$-pole leads to a much larger boost of the $b$ hadrons and their decay products. On the one hand, the larger boost leads to a much larger displacement of secondary vertices, and can facilitate the reconstruction of decays with taus in the final state [434]. On the other hand, the larger boost leads to much more collimated decay products. This might be useful in constraining the kinematics of decays with missing energy, but might also make it more challenging to isolate the decay products. In this respect, a dedicated study would be very useful to obtain a quantitative handle of the balance of these effects and how they affect the statistics after the event selection.

In the following we will discuss the expected sensitivities of a tera-$Z$ factory to a number of flavor violating processes that are known to be sensitive probes of physics beyond the standard model. We stress that the sensitivities presented here are largely based on the scaling of existing measurements to the expected statistics at a tera-$Z$ factory. We rescale $Z$ pole measurements from LEP whenever they are available, which should lead to reasonable sensitivity estimates. When no LEP results are available, we base our estimates on Belle II sensitivity studies, keeping in mind the different kinematical regimes. All our sensitivities should be considered as rough estimates and need to be followed up by ded-icated studies that take into account reconstruction efficiencies, background systematics, etc.

| Particle | Tera-$Z$ | Belle II | LHCb |
|---|---|---|---|
| **$b$ hadrons** | | | |
| $B^+$ | $6\times10^{10}$ | $3\times10^{10}$ ($50\,\text{ab}^{-1}$ on $\Upsilon(4S)$) | $3\times10^{13}$ |
| $B^0$ | $6\times10^{10}$ | $3\times10^{10}$ ($50\,\text{ab}^{-1}$ on $\Upsilon(4S)$) | $3\times10^{13}$ |
| $B_s$ | $2\times10^{10}$ | $3\times10^{8}$ ($5\,\text{ab}^{-1}$ on $\Upsilon(5S)$) | $8\times10^{12}$ |
| $b$ baryons | $1\times10^{10}$ | | $1\times10^{13}$ |
| $\Lambda_b$ | $1\times10^{10}$ | | $1\times10^{13}$ |
| **$c$ hadrons** | | | |
| $D^0$ | $2\times10^{11}$ | | |
| $D^+$ | $6\times10^{10}$ | | |
| $D_s^+$ | $3\times10^{10}$ | | |
| $\Lambda_c^+$ | $2\times10^{10}$ | | |
| $\tau^+$ | $3\times10^{10}$ | $5\times10^{10}$ ($50\,\text{ab}^{-1}$ on $\Upsilon(4S)$) | |

**Table 2.4:** Collection of expected number of particles produced at a tera-$Z$ factory from $10^{12}$ $Z$-boson decays. We have used the hadronization fractions (neglecting $p_T$ dependencies) from Refs. [431, 432] (see also Ref. [433]). For the decays relevant to this study we also show the corresponding number of particles produced by the full $50\,\text{ab}^{-1}$ on $\Upsilon(4S)$ and $5\,\text{ab}^{-1}$ on $\Upsilon(5S)$ runs at Belle II [430], as well as the numbers of $b$ hadrons at LHCb with $50\,\text{fb}^{-1}$ (using the number of $b\bar{b}$ pairs within the LHCb detector acceptance from [435] and the hadronization fractions from [431]).

In Section 2.5.1 we discuss the prospects of measuring a number of rare $b$-hadron decays at the tera-$Z$ factory of CEPC: we cover leptonic decays, semi-leptonic decays, and decays with missing energy. Particular emphasis is laid on rare decays to final states with tau leptons, in which the sensitivity of the tera-$Z$ program of CEPC might be unparalleled. We also comment on possible implications of the current hints for lepton-flavor-universality violation in rare $B$ decays, that have been observed by LHCb. A discussion of tau decays follows in Section 2.5.2, where we discuss the prospects of CEPC to significantly improve lepton universality tests in leptonic tau decays as well as its prospects for measuring rare, lepton-flavor violating tau decays. Flavor violating $Z$ decays are discussed in Section 2.5.3.

### 2.5.1 RARE $B$ DECAYS

#### 2.5.1.1 LEPTONIC DECAYS $B^0 \to \ell^+\ell^-$ AND $B_S \to \ell^+\ell^-$

The purely leptonic $B_s \to \ell^+\ell^-$ and $B^0 \to \ell^+\ell^-$ decays are strongly suppressed in the Standard Model and therefore highly sensitive to new-physics contributions. Their branching ratios are known with high precision in the Standard Model [436]

$$\text{BR}(B_s \to e^+e^-)_{\text{SM}} = (8.54 \pm 0.55) \times 10^{-14} \,, \tag{2.30}$$
$$\text{BR}(B^0 \to e^+e^-)_{\text{SM}} = (2.48 \pm 0.21) \times 10^{-15} \,, \tag{2.31}$$
$$\text{BR}(B_s \to \mu^+\mu^-)_{\text{SM}} = (3.65 \pm 0.23) \times 10^{-9} \,, \tag{2.32}$$
$$\text{BR}(B^0 \to \mu^+\mu^-)_{\text{SM}} = (1.06 \pm 0.09) \times 10^{-10} \,, \tag{2.33}$$
$$\text{BR}(B_s \to \tau^+\tau^-)_{\text{SM}} = (7.73 \pm 0.49) \times 10^{-7} \,, \tag{2.34}$$
$$\text{BR}(B^0 \to \tau^+\tau^-)_{\text{SM}} = (2.22 \pm 0.19) \times 10^{-8} \,. \tag{2.35}$$

Presently, LHCb has provided the most sensitive measurement of the $\mu^+\mu^-$ decays with a precision at the level of $10^{-9}$ [437]. The current most stringent bound on the $e^+e^-$ modes is still coming from CDF [438]. With $50\,\text{fb}^{-1}$ of data, LHCb is expected to reach sensitivities of approximately $10^{-10}$ in the muonic modes and few$\times 10^{-10}$ in the electronic modes [435].

To estimate the sensitivity of a tera-$Z$ factory for the decays to electrons and muons we rescale the existing bound from the L3 collaboration [439] from the full LEP-I data sample, which corresponds to approximately $3 \times 10^5$, and $9 \times 10^4$ $B^0$'s, and $B_s$'s, respectively. A naive rescaling of these bounds accounting for the number of $B^0$'s and $B_s$'s produced at a tera-$Z$ factory gives the following sensitivities

$$\text{BR}(B_s \to e^+e^-)_{\text{tera-}Z} \sim 4 \times 10^{-10} \,, \tag{2.36}$$
$$\text{BR}(B^0 \to e^+e^-)_{\text{tera-}Z} \sim 8 \times 10^{-11} \,, \tag{2.37}$$
$$\text{BR}(B_s \to \mu^+\mu^-)_{\text{tera-}Z} \sim 3 \times 10^{-10} \,, \tag{2.38}$$
$$\text{BR}(B^0 \to \mu^+\mu^-)_{\text{tera-}Z} \sim 7 \times 10^{-11} \,. \tag{2.39}$$

Note, that this linear scaling with the number of $B$ mesons assumes that backgrounds can be kept under control also at the CEPC.

The comparison with the projections from LHCb [435] shows that LHCb with $50\,\text{fb}^{-1}$ will likely outperform the tera-$Z$ factory by a factor of few for the muonic modes. For the electronic modes the tera-$Z$ factory might be able to compete with LHCb. A detailed study would be required to establish the precise sensitivities at CEPC.

The rare $B$ decays to the $\tau^+\tau^-$ final state are experimentally still a largely uncharted territory. The existing bound from BaBar [440], $\text{BR}(B^0 \to \tau^+\tau^-) < 4.1 \times 10^{-3}$, is orders of magnitude away from the corresponding SM prediction, and contrary to the electronic and muonic case there are no measurements performed at LEP. Measurements of the $\tau^+\tau^-$ final states are highly challenging at LHCb. The current sensitivities are at the level of few$\times 10^{-3}$ [441] and could improve down to few$\times 10^{-4}$ [435]. Also Belle II will likely only reach sensitivities at the level of $10^{-4}$ for $B^0 \to \tau^+\tau^-$ [442]. The decays $B^0 \to \tau^+\tau^-$ and $B_s \to \tau^+\tau^-$ are thus prime examples of processes to which a circular electron–positron collider running at the $Z$ pole might have unique opportunities.

At the moment no dedicated study exists for the sensitivity of a tera-$Z$ factory to these decays and no searches for these decays have been performed at LEP. We can therefore only give a naive estimate. Due to the similar number of $B$ mesons available at CEPC and Belle II, we expect at the very least that CEPC should be able to reach sensitivities similar as Belle II, $\sim 10^{-4}$. However, the larger boost of the $B$ mesons from $Z$ decays leads to a much larger displacements of the decay vertices as compared to $B$ factories.

This should enable more efficient reconstruction methods [434] and might considerably improve the CEPC sensitivities. We estimate the CEPC sensitivities to the $B \to \tau^+\tau^-$ and $B \to \tau^+\tau^-$ decays by starting from the $B \to \mu^+\mu-$ sensitivities at CEPC discussed above and multiplying by the relative Belle II sensitivity between the $B \to \tau^+\tau^-$ and $B \to \mu^+\mu^-$ modes. The corresponding rough estimates read

$$\text{BR}(B^0 \to \tau^+\tau^-)_{\text{tera-}Z} < 4 \times 10^{-6} \,, \tag{2.40}$$

$$\text{BR}(B_s \to \tau^+\tau^-)_{\text{tera-}Z} < 2 \times 10^{-5} \,. \tag{2.41}$$

Based on these estimates it is conceivable that CEPC's tera-$Z$ factory might provide the most stringent measurements of the $B_s \to \tau^+\tau^-$ and $B^0 \to \tau^+\tau^-$ decays, improving the expected sensitivities at LHCb and Belle II by potentially more than an order of magnitude.

#### 2.5.1.2 SEMILEPTONIC DECAYS $B \to S(D)\ell^+\ell^-$

Semileptonic FCNC decays of $b$-mesons are not as theoretically clean as the $B_{s,d} \to \ell^+\ell^-$ decays. They are, however, *i)* less rare within the SM, which makes them experimentally more accessible, and *ii)* three-body or four-body decays resulting in multiple observables for a given mode, e.g., invariant-mass and angular distribution observables, $CP$ asymmetries, etc.

In recent years, the exclusive decays $B \to K^{(*)}e^+e^-$ and $B \to K^{(*)}\mu^+\mu^-$ have attracted a lot of attention due to the large number of LHCb measurements and in particular due to some persistent $\approx 2-3\sigma$ tensions between data and SM expectations in related observables, i.e., $R_{K^{(*)}}$ [443, 444] theoretically clean observables that tests for lepton-flavor-universality violation, and the angular observable $P_5'$ [445]. The tensions are present in LHCb's Run-1 data set of $3\,\text{fb}^{-1}$, with Run-2 results yet to be announced. We expect significant progress as LHCb collects a data set of $50\,\text{fb}^{-1}$. Belle II will also probe these exclusive modes. Its $50\,\text{ab}^{-1}$ run on $\Upsilon(4S)$ will provide measurements of these modes with a precision not much lower that LHCb with its full data set [442]. As the number of $B^0$ and $B^+$ mesons produced at the tera-$Z$ factory and at Belle II are approximately the same, we ultimately expect a similar precision at the two machines. In this respect, the CEPC's measurements of these modes will be invaluable, especially if the tensions in the exclusive $B \to K^{(*)}e^+e^-$ and $B \to K^{(*)}\mu^-\mu^+$ persist in the full data set of LHCb. In such a case, the CEPC's program would provide a new data set and should be able to independently confirm the existence of new-physics effects in the electronic and muonic decays.

Both Belle II and CEPC will also be able to access the inclusive decays $B \to X_s e^+e^-$ and $B \to X_s\mu^+\mu^-$, likely with comparable precision. Hadronic uncertainties are under better control in the inclusive modes and their measurements will complement the studies of the exclusive decays mentioned above.

Contrary to the $ee$ and $\mu\mu$ modes, little experimental information exists on the semi-tauonic modes $b \to s(d)\tau^+\tau^-$ so far. The only existing bound from BaBar [446], $\text{BR}(B \to K\tau^+\tau^-) < 3.3 \times 10^{-3}$, is approximately four orders of magnitude above the SM prediction and it is not clear whether LHCb will be able to improve the sensitivity substantially. The first major improvements are thus expected at Belle II. For instance, its $50\,\text{ab}^{-1}$ run will probe the branching ratio of $B^+ \to K^+\tau^+\tau^-$ at the level of $2 \times 10^{-5}$ [442]. A dedicated study is required to quantitatively assess the full potential of the tera-$Z$ factory, which is expected to outperform Belle II in modes as the ones in question, in which missing energy from the tau decays is present in the event. A study for the FCC-$ee$ program that investigates the $B^0 \to K^*\tau^+\tau^-$ decay [434], finds that approximately a thousand cleanly reconstructed events are expected from $10^{13}$ $Z$'s. We can thus expect up to hundred reconstructed $B^0 \to K^*\tau^+\tau^-$ events at the CEPC tera-$Z$ factory probing the SM branching ratio with a statistical uncertainty of $O(10\%)$ (which corresponds to a sensitivity to $B^0 \to K^*\tau^+\tau^-$ of $\sim 10^{-8}$).

We see that, similarly to the $B^0 \to \tau^-\tau^+$ and $B_s \to \tau^+\tau^-$ modes, also here the tera-$Z$ factory might provide the by far most accurate measurements. With hundred events even a partial angular analysis might be possible. Additionally, the large number of $B_s$ mesons and $\Lambda_b$ baryons produced at the tera-$Z$ factory should facilitate the first measurements of the corresponding decays, $B_s \to \phi\tau^+\tau^-$, and $\Lambda_b \to \Lambda\tau^+\tau^-$ at a similar level of precision. The measurements of the semi-tauonic decay will also open the path towards measurements of lepton-flavor-universality violation involving not only electrons and muons, but also taus, which will be of particular interest if the present tensions in the muon–electron data persist.

#### 2.5.1.3 DECAYS WITH MISSING ENERGY $B \to S(D)\nu\bar{\nu}$

The rare FCNC decays $B \to K^{(*)}\nu\bar{\nu}$ are widely recognized as important flavor probes, as they are not affected by non-factorizable corrections and thus theoretically cleaner compared to $b \to s\ell\ell$ transitions. The SM predictions for the branching ratios of these decays read [447]

$$\mathrm{BR}(B^+ \to K^+\nu\bar{\nu})_{\mathrm{SM}} = (4.68 \pm 0.64) \times 10^{-6}\,, \tag{2.42}$$

$$\mathrm{BR}(B^0 \to K^0\nu\bar{\nu})_{\mathrm{SM}} = (2.17 \pm 0.30) \times 10^{-6}\,, \tag{2.43}$$

$$\mathrm{BR}(B^+ \to K^{*+}\nu\bar{\nu})_{\mathrm{SM}} = (10.22 \pm 1.19) \times 10^{-6}\,, \tag{2.44}$$

$$\mathrm{BR}(B^0 \to K^{*0}\nu\bar{\nu})_{\mathrm{SM}} = (9.48 \pm 1.10) \times 10^{-6}\,, \tag{2.45}$$

with uncertainties in the theoretical predictions of roughly $10\%$, dominated by parametric and form-factor uncertainties. The accuracy of these predictions in combination with the fact that these modes have not yet been observed (current bounds are typically a factor of few away from the SM predictions, e.g., see Ref. [448, 449]), is the reason why these modes are prime candidates for disentangling small new-physics contributions. Since the neutrinos are never tagged in the experiments, such modes are not only relevant for searches for heavy new physics, but can also provide the leading constraints in models with light, long-lived particles with small flavor-violating couplings, e.g., the "axiflavon" [450].

With its full, $50\,\mathrm{ab}^{-1}$ data set run on $\Upsilon(4S)$, Belle II is expected to probe for the first time deviations from the SM predictions at a level of approximately $17\%$ [451]. The dominant uncertainties in such measurements are expected to be due to statistics. The related decays based on the $b \to d\nu\bar{\nu}$ transition, i.e., $B \to \pi\nu\bar{\nu}$ and $B \to \rho\nu\bar{\nu}$, are further suppressed in the SM by a factor of approximately 30. Limits at a level of $10^{-6}$ are expected at Belle II.

Given that the number of $B^0$ and $B^+$ particles produced with $50\,\mathrm{ab}^{-1}$ at Belle II are roughly the same at a tera-$Z$ factory, we naively expect similar statistical uncertainties there. Note, however, that the missing energy from the neutrinos and the hadronic decay products will be highly collimated at a collider running on the $Z$ pole. As long as calori-metric isolation does not become an issue, the tera-$Z$ factory could likely probe the SM predictions of $B \to K^{(*)}\nu\bar{\nu}$, $B \to \pi\nu\bar{\nu}$, and $B \to \rho\nu\bar{\nu}$ at the same level as Belle II, i.e. around $10^{-6}$. Scaling up an existing LEP bound BR$(B \to K^*\nu\nu) \lesssim 10^{-3}$ [452] to the statistics expected at a tera-$Z$ factory leads to a similar estimate.

As illustrated in Table 2.4 the tera-$Z$ factory will produce two orders of magnitude more $B_s^{(*)}$ mesons than a $5\,\text{ab}^{-1}$ run of Belle II on $\Upsilon(5S)$. Also, approximately $10^{10}$ $b$ baryons will be produced at the tera-$Z$ factory, whereas none can be produced at Belle II without (not planned) dedicated runs. The tera-$Z$ factory will thus for the first time have access to decay modes of $B_s$ mesons and $\Lambda_b$ baryons, like $B_s \to \phi\nu\bar{\nu}$ and $\Lambda_b \to \Lambda\nu\bar{\nu}$. Given the large statistical sample of $\Lambda_b$ baryons (only factor of six smaller than mesons), we expect the tera-$Z$ factory to probe these branching fractions at a level similar to the related $B_0$ and $B^+$ modes, i.e., branching fractions as low as approximately $10^{-6}$.

More than one higher-dimensional operator of the five-flavor effective theory can induce these decays. By probing multiple members of this whole family of decays, the measurements of the tera-$Z$-factory will probe more than a single linear combination of operators. For instance, the combination of the information from the pseudoscalar to pseudoscalar transitions ($B \to K\nu\bar{\nu}$), the pseudoscalar to vector transitions ($B \to K^*\nu\bar{\nu}$ and $B_s \to \phi\nu\bar{\nu}$), as well as the fermion to fermion transition ($\Lambda_b \to \Lambda\nu\bar{\nu}$) could be a way to disentangle possible new-physics contributions from right-handed currents.

### 2.5.1.4 PROBING NEW PHYSICS WITH $B \to S(D)\tau^+\tau^-$ DECAYS

As we have seen, CEPC might have unique sensitivities to FCNC $b$ decay modes with taus in the final state. There are many new-physics scenarios, e.g., models with extended Higgs sectors, or extended gauge sectors, or scenarios with leptoquarks, that could give rise to sizable effects in leptonic or semi-leptonic $\tau^+\tau^-$ modes, without violating constraints from the $e^+e^-$ and/or $\mu^+\mu^-$ channels. Model independently, tau-specific new physics in rare $B$ decays can be encoded in an effective Lagrangian

$$\mathcal{L}_{\text{NP}} = -\frac{G_F}{\sqrt{2}} V_{tb} V_{tq}^* \frac{e^2}{16\pi^2} \sum_i \Big( C_i O_i + C_i' O_i' \Big) \,, \quad q = s, d \,, \tag{2.46}$$

with the operators

$$\begin{aligned}
O_7 &= (\bar{q}\sigma_{\mu\nu} P_R b) F^{\mu\nu} \,, & O_7' &= (\bar{q}\sigma_{\mu\nu} P_L b) F^{\mu\nu} \,, \\
O_9 &= (\bar{q}\gamma_\mu P_L b)(\bar{\tau}\gamma^\mu \tau) \,, & O_9' &= (\bar{q}\gamma_\mu P_R b)(\bar{\tau}\gamma^\mu \tau) \,, \\
O_{10} &= (\bar{q}\gamma_\mu P_L b)(\bar{\tau}\gamma^\mu \gamma_5 \tau) \,, & O_{10}' &= (\bar{q}\gamma_\mu P_R b)(\bar{\tau}\gamma^\mu \gamma_5 \tau) \,, \\
O_S &= (\bar{q} P_R b)(\bar{\tau} P_L \tau) \,, & O_S' &= (\bar{q} P_L b)(\bar{\tau} P_R \tau) \,.
\end{aligned}$$

Constraining all possible $\tau^+\tau^-$ operators requires measurements of both the leptonic and semi-leptonic modes, as they have different blind directions in the parameter space of Wilson coefficients [453, 454]. Note, that also the decays with neutrinos, $b \to q\nu\bar{\nu}$, can constrain the operator-combinations that contain a left-handed tau current $O_9 - O_{10}$ and $O_9' - O_{10}'$, due to SU$(2)_L$ invariance. On the other hand, the neutrino modes are blind to the orthogonal directions $O_9 + O_{10}$ and $O_9' + O_{10}'$, which contain right-handed tau currents.

There are various new-physics models that can lead to non-standard effects in $b \to (d, s)\tau^+\tau^-$ decays. Interestingly, several models that address the LHCb anomalies in the $B \to K^*\mu^+\mu^-$ angular distribution or the hints for lepton-flavor-universality violation in $R_{K^{(*)}}$ [443, 444] or $R_{D^{(*)}}$ [455] predict characteristic non-standard effects in $b \to s\tau^+\tau^-$ transitions.

The model proposed in Ref. [456] is based on gauging the difference of muon- and tau-number, $L_\mu - L_\tau$. Given the current anomalies in $b \to s\mu^+\mu^-$, the model predicts that all semi-leptonic $b \to s\mu^+\mu^-$ decays are suppressed by approximately $25\%$ [457]. The $L_\mu - L_\tau$ symmetry implies that all semi-leptonic $b \to s\tau^+\tau^-$ decays are instead enhanced by a similar amount. However, the $B_s \to \tau^+\tau^-$ decay remains SM-like in the $L_\mu - L_\tau$ framework.

In the new-physics scenarios originally introduced in Refs. [458–460], the current $B$-physics anomalies are addressed by non-standard left-handed currents involving mainly the 3rd generation of quarks and leptons. In such scenarios, enhancements of $B_s \to \tau^+\tau^-$ and $b \to s\tau^+\tau^-$ rates by an order of magnitude compared to the SM predictions are possible. Left-handed currents also imply a strong correlation between $b \to s\tau^+\tau^-$ and $b \to s\nu\bar{\nu}$ decays, as well as enhanced $b \to s\nu\bar{\nu}$ rates.

On the other hand, enhancements of $b \to s\tau^+\tau^-$ rates that are independent of $b \to s\nu\bar{\nu}$ are possible in models with right-handed lepton currents. In such scenarios the current experimental bounds can in principle be saturated.

### 2.5.2 TAU DECAYS

From Table 2.4 we see that at the tera-$Z$ factory of CEPC we can expect approximately $3 \times 10^{10}$ $\tau^+\tau^-$ pairs produced from $Z$ decays. This is comparable to the expected number of taus produced at Belle II, i.e., roughly $5 \times 10^{10}$. This suggests that the sensitivities to lepton-flavor violating decays of taus at CEPC can be similar to the sensitivities expected at Belle II. The large boost of taus from the $Z$ decays might allow CEPC to measure the standard leptonic branching ratios of the tau and to test lepton universality in $\tau \to \ell\nu\bar{\nu}$ with unprecedented precision.

#### 2.5.2.1 LEPTON UNIVERSALITY IN $\tau \to \ell\nu\bar{\nu}$

The best measurements of the leptonic branching ratios of the tau, BR$(\tau \to \mu\nu_\tau\bar{\nu}_\mu)$ and BR$(\tau \to e\nu_\tau\bar{\nu}_e)$, still come from LEP [291]. The most precise individual results are from ALEPH [461] and read BR$(\tau \to \mu\nu_\tau\bar{\nu}_\mu) = (17.319 \pm 0.070 \pm 0.032)\%$ and BR$(\tau \to e\nu_\tau\bar{\nu}_e) = (17.837 \pm 0.072 \pm 0.036)\%$, where the first uncertainty is due to statistics and the second due to systematics. One can see that the measurements were statistics limited with relative systematic uncertainties at the level of approximately 2 permille. This implies that the larger statistics of a tera-$Z$ program at the CEPC will result in the world best measurement of these branching ratios with uncertainties at the permille level or even much better.

Indeed, it is likely that the much larger number of $\tau$ pairs will also allow the experiments to gain a much better control of systematic uncertainties. Assuming that systematics can be reduced by an order of magnitude (which requires exquisite control of the electron and muon efficiencies), the leptonic tau branching ratios could be measured at CEPC with a relative uncertainty of $10^{-4}$. Dedicated studies are required to establish the precise sensitivity of CEPC.

The leptonic branching ratios of the tau can in principle be predicted with very high precision in the SM [462]. The SM precision is limited by the uncertainty in the measured tau lifetime, $\tau_\tau$. The most precise tau lifetime determination comes currently from Belle [463] and has an uncertainty of approximately 2 permille. Given the much higher statistics expected at Belle II, future measurements may be able to improve the precision of $\tau_\tau$ by up to an order of magnitude. We expect that CEPC could reach a precision for $\tau_\tau$ similar to Belle II. The precise relation between the $\tau$ lifetime and the leptonic branching ratios in the SM, combined with future precise determinations of $\text{BR}(\tau \to \mu\nu_\tau\bar{\nu}_\mu)$ and $\text{BR}(\tau \to e\nu_\tau\bar{\nu}_e)$ at CEPC would allow to scrutinize the weak interactions in tau decays with an unprecedented precision.

Additional information can be extracted from measurements of kinematic distributions in tau decays and the determination of the tau decay parameters (also known as Michel parameters) [291], which are highly sensitive to the structure (spin and chirality) of the current that mediates tau decays. CEPC might substantially improve the existing (LEP) and expected (Belle II) measurements of tau decay parameters.

In addition to measurements of the absolute leptonic branching ratios and their kinematic distributions, it is of particular interest to look at the lepton-flavor universality ratio

$$R_\tau = \frac{\text{BR}(\tau \to \mu\nu_\tau\bar{\nu}_\mu)}{\text{BR}(\tau \to e\nu_\tau\bar{\nu}_e)} . \tag{2.47}$$

This ratio is independent of the tau lifetime and can be predicted with extremely high precision in the SM, $R_\tau^{\text{SM}} = 0.972559 \pm 0.000005$ [462]. The currently most precise direct measurement of this ratio comes from BaBar and has an uncertainty of approximately 4 permille, $R_\tau^{\text{BaBar}} = 0.9796 \pm 0.0016 \pm 0.0036$ [464]. A measurement of $R_\tau$ with a precision of $10^{-4}$ may be possible at CEPC if systematic uncertainties can be controlled (cf. discussion above about the expected precision in the absolute branching ratios).

Most new-physics models that explain the current hints for lepton-flavor universality violation in $B$ decays, $R_{K^{(*)}}$ [443, 444] and $R_{D^{(*)}}$ [455] also lead to lepton-flavor universality violation in $\tau \to \mu\nu_\tau\bar{\nu}_\mu$ vs. $\tau \to e\nu_\tau\bar{\nu}_e$ [465, 466]. Typical new-physics effects are at the level of a few permille and should be well within the reach of CEPC. Therefore, more accurate measurements of $R_\tau$ would be invaluable to scrutinize many motivated new-physics scenarios.

#### 2.5.2.2 LEPTON-FLAVOR VIOLATING $\tau$ DECAYS

In the SM the rate of lepton-flavor violating tau decays is tiny because it is controlled by the small neutrino masses; branching ratios like $\tau \to \mu\gamma$ are predicted to be at the level of $10^{-45}$. However, in models of new physics such branching ratios could be enhanced by many orders of magnitude and could be in reach of experimental searches. In this sense, any observation of lepton-flavor violating tau decays would be an unambiguous sign of physics beyond the SM.

Lepton-flavor violating tau decays have been searched for in a multitude of channels at the $B$ factories BaBar and Belle. Among them are the radiative modes $\tau \to \mu\gamma$ and $\tau \to e\gamma$, purely leptonic modes like $\tau \to 3\mu$, $\tau \to 3e$, $\tau \to \mu ee$, etc., as well as many hadronic modes like $\tau \to \mu\pi^0$, $\tau \to e\pi^0$, $\tau \to \mu K$, etc. Most of these decays have been constrained at the level of $10^{-8}$ [431]. Thanks to its increase in statistics, Belle II is expected to improve the sensitivities to the lepton-flavor violating tau decays by at least one order of magnitude or even more in very clean modes like $\tau \to 3\mu$.

The clean signature of three muons also allows LHCb to search for the decay $\tau \to 3\mu$ with high sensitivity. The current limit, which has been obtained with $3\,\text{fb}^{-1}$ of the combined 7 TeV and 8 TeV data, reads $\text{BR}(\tau \to 3\mu)_{\text{LHCb}} < 4.6 \times 10^{-8}$ [467] and is competitive with the existing bounds from BaBar and Belle. In the high-luminosity phase of LHC, LHCb will likely improve this bound by one order of magnitude down to few times $10^{-9}$.

Given the comparable numbers of taus that will be produced at Belle II and that could be expected from the tera-$Z$ factory at CEPC, we expect similar sensitivities to these decays at both machines. While dedicated studies would need to be performed to ascertain that backgrounds would be under control at CEPC, we expect CEPC's sensitivities to lepton-flavor violating tau decays across the board at the level of $10^{-9}$ or better.

### 2.5.3 FLAVOR VIOLATING $Z$ DECAYS

Rare decays of the $Z$ boson that violate quark flavor, $Z \to qq'$, are absent in the SM at tree level and therefore strongly suppressed. The largest branching ratio in the SM is expected to be $Z \to bs$ and can be estimated as $\text{BR}(Z \to bs) \sim \left|\frac{g^2}{16\pi^2} V_{tb} V_{ts}^*\right|^2 \times \text{BR}(Z \to bb) \sim 10^{-9}$. Even with the statistics expected from $10^{12}$ $Z$ bosons, a measurement of the SM rate would be extremely challenging as the $Z \to bs$ events will be buried under an enormous background from $Z \to q\bar{q}$ and $Z \to b\bar{b}$ decays. New physics can induce effective quark-flavor violating $Z$ couplings, but such effects are typically constrained by rare meson decays and meson-mixing observables. Rates of $Z \to qq'$ that are far above SM expectations are therefore unlikely.

Lepton-flavor violating decays are completely absent in the SM without neutrino masses. Including neutrino masses, $Z \to \ell\ell'$ decays can in principle arise but the branching ratios are suppressed by the tiny neutrino masses and predicted to be in the ballpark of $10^{-50} - 10^{-60}$. However, new physics could enhance these branching ratios by many orders of magnitude.

Searches at LEP established the following upper bounds using few$\times 10^6$ $Z$ bosons [468–470]: $\text{BR}(Z \to \mu e) < 1.7 \times 10^{-6}$, $\text{BR}(Z \to \tau e) < 9.8 \times 10^{-6}$, and $\text{BR}(Z \to \tau\mu) < 1.2 \times 10^{-5}$. Due to the huge numbers of $Z$ bosons produced at the LHC, searches at ATLAS and CMS for the clean $Z \to \mu e$ decay have recently set limits at the level of few$\times 10^{-7}$ [471, 472]. Searches for the final states with taus are more challenging at the LHC. The current ATLAS limits for $Z \to \tau e$ and $Z \to \tau\mu$ are at the level of few$\times 10^{-5}$ [473]. With the high statistics that are be expected from the future LHC runs, it is conceivable that the bounds on lepton-flavor violating $Z$ decays will improve by an order of magnitude or more.

Assuming that the sensitivities at the tera-$Z$ factory of CEPC can be scaled from the LEP bounds. We present here the rescaling assuming the production of $10^{12}$ $Z$ bosons. We show two projections, the first one assumes a background dominated measurement, i.e., rescaling with the square root of number of events, and the second assumes zero backgrounds, i.e., rescaling with number of events.

$$\text{BR}(Z \to \mu e)_{\text{CEPC}} \lesssim 3 \times 10^{-9} \;\; [1/\sqrt{N} \text{ scaling}]\,, \; 7 \times 10^{-12} \;\; [1/N \text{ scaling}]\,, \tag{2.48}$$

$$\text{BR}(Z \to \tau e)_{\text{CEPC}} \lesssim 2 \times 10^{-8} \;\; [1/\sqrt{N} \text{ scaling}]\,, \; 4 \times 10^{-11} \;\; [1/N \text{ scaling}]\,, \tag{2.49}$$

$$\text{BR}(Z \to \tau\mu)_{\text{CEPC}} \lesssim 2 \times 10^{-8} \;\; [1/\sqrt{N} \text{ scaling}]\,, \; 5 \times 10^{-11} \;\; [1/N \text{ scaling}]\,, . \tag{2.50}$$

The LEP measurements are to a large extend background free, so we expect CEPC's sensitivity to be within the range of the two values above. This is a substantial improvement compared to existing and expected bounds. A more realistic analysis, including explicit background studies from e.g. $Z \to \tau\tau$ would need to be performed to provide a more precise estimate of the sensitivities [474]. Nevertheless, the above estimates indicate promising sensitivities to new-physics models that induce lepton-flavor violating $Z$ decays, as for example extensions of the SM with heavy sterile neutrinos [286].

### 2.5.4 SUMMARY

A CEPC that produces $10^{12}$ $Z$ bosons provides large statistics samples of $b$ and $c$ hadrons as well as tau leptons in a clean experimental environment. This results in unique opportunities for various flavor measurements that in several cases should be unparalleled in current or any other future machine. For example, the observation of the rare tauonic decays $B \to K^*\tau^+\tau^-$ and $B_s \to \phi\tau^+\tau^-$ at the SM rate might be achieved at CEPC, whereas the SM rates of such tauonic decays are not in reach of neither LHCb nor Belle II. It appears that sufficient statistics could be accumulated such that even a partial angular analysis of $B \to K^{(*)}\tau^+\tau^-$ may be possible. It is also conceivable that CEPC could achieve the world's best sensitivity to the related tauonic decay modes $B_s \to \tau^+\tau^-$ and $B \to \tau^+\tau^-$ at a level of $10^{-5}$. New physics in the rare tauonic decays is particularly well motivated given the current hints for lepton-flavor universality violation in $R_{K^{(*)}}$ and $R_{D^{(*)}}$.

A future circular electron–positron collider is also the only machine where measurements of the rare FCNC decays of $B_s$ mesons and $\Lambda_b$ baryons to neutrinos, i.e., $B_s \to \phi\nu\bar{\nu}$ and $\Lambda_b \to \Lambda\nu\bar{\nu}$, might be possible. The corresponding sensitivities could be at the level of $\sim 10^{-6}$, thus complementing the sensitivity of Belle II to $B \to K^{(*)}\nu\bar{\nu}$.

A tera-$Z$ factory of CEPC will also likely reach sensitivities to lepton-flavor violation in tau decays at a level of $10^{-9}$, which is comparable to the sensitivities expected at Belle II. The leptonic decays of taus, $\tau \to \mu\nu\nu$ and $\tau \to e\nu\nu$ could be measured at CEPC with unprecedented precision, providing extremely sensitive tests of the weak interaction in tau decays. Furthermore, it might be possible to test lepton universality in $\tau \to \ell\nu\nu$ at the level of $10^{-4}$. Many new-physics explanations of the observed anomalies in $R_{K^{(*)}}$ and $R_{D^{(*)}}$ predict violation of lepton-flavor universality in tau decays at the permille level and could, therefore, be scrutinized at CEPC.

Finally, the CEPC measurements would improve the bounds on lepton-flavor violating $Z$ decays by orders of magnitude compared to the current best bounds from LEP, down to a level of $10^{-8}$ and better.

| Observable | Current sensitivity | Future sensitivity | Tera-$Z$ sensitivity |
|---|---|---|---|
| $\mathrm{BR}(B_s \to ee)$ | $2.8 \times 10^{-7}$ (CDF) [438] | $\sim 7 \times 10^{-10}$ (LHCb) [435] | $\sim$ few $\times 10^{-10}$ |
| $\mathrm{BR}(B_s \to \mu\mu)$ | $0.7 \times 10^{-9}$ (LHCb) [437] | $\sim 1.6 \times 10^{-10}$ (LHCb) [435] | $\sim$ few $\times 10^{-10}$ |
| $\mathrm{BR}(B_s \to \tau\tau)$ | $5.2 \times 10^{-3}$ (LHCb) [441] | $\sim 5 \times 10^{-4}$ (LHCb) [435] | $\sim 10^{-5}$ |
| $R_K, R_{K^*}$ | $\sim 10\%$ (LHCb) [443, 444] | $\sim$few% (LHCb/Belle II) [435, 442] | $\sim$few % |
| $\mathrm{BR}(B \to K^*\tau\tau)$ | – | $\sim 10^{-5}$ (Belle II) [442] | $\sim 10^{-8}$ |
| $\mathrm{BR}(B \to K^*\nu\nu)$ | $4.0 \times 10^{-5}$ (Belle) [449] | $\sim 10^{-6}$ (Belle II) [442] | $\sim 10^{-6}$ |
| $\mathrm{BR}(B_s \to \phi\nu\bar{\nu})$ | $1.0 \times 10^{-3}$ (LEP) [452] | – | $\sim 10^{-6}$ |
| $\mathrm{BR}(\Lambda_b \to \Lambda\nu\bar{\nu})$ | – | – | $\sim 10^{-6}$ |
| $\mathrm{BR}(\tau \to \mu\gamma)$ | $4.4 \times 10^{-8}$ (BaBar) [475] | $\sim 10^{-9}$ (Belle II) [442] | $\sim 10^{-9}$ |
| $\mathrm{BR}(\tau \to 3\mu)$ | $2.1 \times 10^{-8}$ (Belle) [476] | $\sim$ few $\times 10^{-10}$ (Belle II) [442] | $\sim$ few $\times 10^{-10}$ |
| $\frac{\mathrm{BR}(\tau\to\mu\nu\bar{\nu})}{\mathrm{BR}(\tau\to e\nu\bar{\nu})}$ | $3.9 \times 10^{-3}$ (BaBar) [464] | $\sim 10^{-3}$ (Belle II) [442] | $\sim 10^{-4}$ |
| $\mathrm{BR}(Z \to \mu e)$ | $7.5 \times 10^{-7}$ (ATLAS) [471] | $\sim 10^{-8}$ (ATLAS/CMS) | $\sim 10^{-9} - 10^{-11}$ |
| $\mathrm{BR}(Z \to \tau e)$ | $9.8 \times 10^{-6}$ (LEP) [469] | $\sim 10^{-6}$ (ATLAS/CMS) | $\sim 10^{-8} - 10^{-11}$ |
| $\mathrm{BR}(Z \to \tau\mu)$ | $1.2 \times 10^{-5}$ (LEP) [470] | $\sim 10^{-6}$ (ATLAS/CMS) | $\sim 10^{-8} - 10^{-10}$ |

**Table 2.5:** Order of magnitude estimates of the sensitivity to a number of key observables for which the tera-$Z$ factory at CEPC might have interesting capabilities. The expected future sensitivities assume luminosities of $50\,\mathrm{fb}^{-1}$ at LHCb, $50\,\mathrm{ab}^{-1}$ at Belle II, and $3\,\mathrm{ab}^{-1}$ at ATLAS and CMS. For the tera-$Z$ factory of CEPC we have assumed the production of $10^{12}$ $Z$ bosons.

Table 2.5 summarizes a number of key observables. We stress that all sensitivities listed for CEPC are rough estimates only and are mainly based on rescaling by the number of expected events. They need to be followed up by dedicated sensitivity studies that carefully take into account detection efficiencies, background systematics, etc.

**CHAPTER 3**

# EXPERIMENTAL CONDITIONS, PHYSICS REQUIREMENTS AND DETECTOR CONCEPTS

The CEPC physics program spans a wide range of center-of-mass energies and beam luminosities to achieve the highest yields of Higgs, $W$, and $Z$ bosons produced in the exceptionally clean environment of an $e^+e^-$ collider. As described in Chapter 2, the CEPC data will provide new levels of high precision tests of the Standard Model (SM) and in the search for physics beyond the SM (BSM). This chapter describes the design requirements for the CEPC detectors to achieve these physics goals, taking into account the CEPC collision environment and the related backgrounds. The CEPC precision physics program places stringent requirements on the detector performance. These include large and precisely defined solid angle coverage, excellent particle identification, precise particle energy/momentum measurements, efficient vertex reconstruction, superb jet reconstruction and flavor tagging.

Two CEPC detector concepts are introduced in this chapter. They derive from detector concepts proposed for the International Linear Collider project [1], benefiting from a long period of prior development, and incorporate modifications motivated by the circular collider experimental environment and by the higher luminosity. Although the overall design and main building blocks of the these concepts are similar, the particular technology choices are different. The CEPC baseline detector concept follows closely the International Large Detector (ILD) design [2, 3]. It is guided by particle flow principles and includes a silicon vertex detector, a silicon tracker, a Time Projection Chamber (TPC), an ultra high granularity calorimetry system, a 3 Tesla solenoid, and a muon detector embedded in the flux return yoke. A variant of the baseline concept substitutes the TPC and the silicon tracking sytem with a Full-Silicon Tracker (FST). The alternative detector concept is based on a lower magnetic field of 2 Tesla, a drift chamber, and dual readout calorimetry. While the baseline concept detector with the TPC option is used for the physics performance studies in this Conceptual Design Report, the other options are considered fully valid alternatives. The final two CEPC detectors are likely to be composed of the detector technologies included in these concepts and possibly beyond.

## 3.1 CEPC EXPERIMENTAL CONDITIONS

The CEPC is a circular electron-positron collider with 100 km circumference and two interaction points (IP). The details of the full CEPC accelerator complex are described in the CDR Volume I [4]. The final stage of the CEPC complex is a double-ring collider. Electron and positron beams circulate in opposite directions in separate beam pipes. They collide at two IPs which house large detectors.

The detectors must operate in three primary sets of conditions, corresponding to three different center-of-mass energies ($\sqrt{s}$): Higgs factory at $\sqrt{s} = 240$ GeV for the $e^+e^- \to ZH$ production, $Z$ factory at $\sqrt{s} \sim 91.2$ GeV for the $e^+e^- \to Z$ production, and $WW$ threshold scan at $\sqrt{s} \sim 160$ GeV for the $e^+e^- \to W^+W^-$ production. The instantaneous luminosities are expected to reach $3 \times 10^{34}$, $32 \times 10^{34}$ and $10 \times 10^{34}$ cm$^{-2}$s$^{-1}$, respectively, as shown in Table 3.1, and will remain relatively constant throughout the operation through a process of full-energy top-up injection [5, 6] by the CEPC accelerator complex. The current tentative operation plan will allow the detectors to collect one million Higgs bosons or more, close to one trillion $Z$ boson events, and over one hundred million $W^+W^-$ events.

The detector designs must comprehensively meet the requirements imposed by the CEPC experimental conditions and the physics program. Each of the beam conditions and corresponding detector implications are presented below.

### 3.1.1 THE CEPC BEAM

The detectors will record collisions in beam conditions presented in Table 3.1. Several of these parameters impose important constraints on the detectors. The bunch spacing of the colliding beams differ greatly in the three operational modes (680 ns, 25 ns, and 210 ns, respectively for the Higgs, $Z$ and $W$ operations) as does the power dissipated into synchrotron radiation (16.5 MW for the $Z$ factory and 30 MW for the $WW$ threshold scan and the Higgs factory). Other important differences are also present in the expected beam backgrounds, described in more detail below, and, most importantly, in the event rates and types of events to be recorded, according to the cross sections shown in Figure 3.1 for different center-of-mass energies.

### 3.1.2 BEAM BACKGROUNDS

Three most important sources of radiation backgrounds are evaluated for the CEPC:

1. synchrotron radiation photons from the last bending dipole magnet;
2. $e^+e^-$ pair production following the beamstrahlung process;
3. off-energy beam particles lost in the interaction region.

#### 3.1.2.1 SYNCHROTRON RADIATION

Synchrotron Radiation (SR) photons are prevalent at circular machines. A large flux of SR photons are generated in the last bending dipole magnets. They can hit the central beam pipe, either directly or after scattered by the beam pipe in the forward region. SR photons can also be generated in the final focusing magnets but contribute little to the detector backgrounds because they are produced with extremely small polar angles and can leave the interaction region without interacting in the beam pipe. To suppress the SR photons, three sets of mask tips made with high-$Z$ material are introduced at $|z| = 1.51$, $1.93$ and $4.2$ m away from the interaction point. The studies using the BDSim software [8] show that the masks can reduce the number of SR photons hitting the central beam pipe effectively, from almost 40,000 to below 80 from one of the two beams per bunch crossing. Further optimization may suppress SR photons even more and make this particular background well controlled.

| | Higgs | W | Z (3T) | Z (2T) |
|---|---|---|---|---|
| Number of IPs | 2 | | | |
| Beam energy (GeV) | 120 | 80 | 45.5 | |
| Circumference (km) | 100 | | | |
| Synchrotron radiation loss/turn (GeV) | 1.73 | 0.34 | 0.036 | |
| Crossing angle at IP (mrad) | $16.5 \times 2$ | | | |
| Piwinski angle | 3.48 | 7 | 23.8 | |
| Bunch number | 242 | 1524 | 12000 (10% gap) | |
| Bunch spacing (ns) | 680 | 210 | 25 | |
| No. of particles/bunch $N_e (10^{10})$ | 15 | 12 | 8 | |
| Beam current (mA) | 17.4 | 87.9 | 461 | |
| Synch. radiation power (MW) | 30 | 30 | 16.5 | |
| Bending radius (km) | 10.7 | | | |
| $\beta$ function at IP: $\beta_x^*$ (m) | 0.36 | 0.36 | 0.2 | 0.2 |
| $\beta_y^*$ (m) | 0.0015 | 0.0015 | 0.0015 | 0.001 |
| Emittance: $x$ (nm) | 1.21 | 0.54 | 0.18 | 0.18 |
| $y$ (nm) | 0.0024 | 0.0016 | 0.004 | 0.0016 |
| Beam size at IP: $\sigma_x$ ( µm) | 20.9 | 13.9 | 6.0 | 6.0 |
| $\sigma_y$ ( µm) | 0.06 | 0.049 | 0.078 | 0.04 |
| Beam-beam parameters: $\xi_x$ | 0.018 | 0.013 | 0.004 | 0.004 |
| $\xi_y$ | 0.109 | 0.123 | 0.06 | 0.079 |
| RF voltage $V_{RF}$ (GV) | 2.17 | 0.47 | 0.1 | |
| RF frequency $f_{RF}$ (MHz) | 650 | | | |
| Natural bunch length $\sigma_z$ (mm) | 2.72 | 2.98 | 2.42 | |
| Bunch length $\sigma_z$ (mm) | 4.4 | 5.9 | 8.5 | |
| Natural energy spread (%) | 0.1 | 0.066 | 0.038 | |
| Energy spread (%) | 0.134 | 0.098 | 0.08 | |
| Photon number due to beamstrahlung | 0.082 | 0.05 | 0.023 | |
| Lifetime (hour) | 0.43 | 1.4 | 4.6 | 2.5 |
| F (hour glass) | 0.89 | 0.94 | 0.99 | |
| Luminosity/IP ($10^{34}$ cm$^{-2}$s$^{-1}$) | 3 | 10 | 17 | 32 |

**Table 3.1:** Main beam parameters for the CEPC operation at three center-of-mass energies. The detector solenoid magnetic field affects the beam quality in the $Z$-factory operation mode. The last two columns compare the beam parameters for the case of a 2 or 3 Tesla detector solenoid.

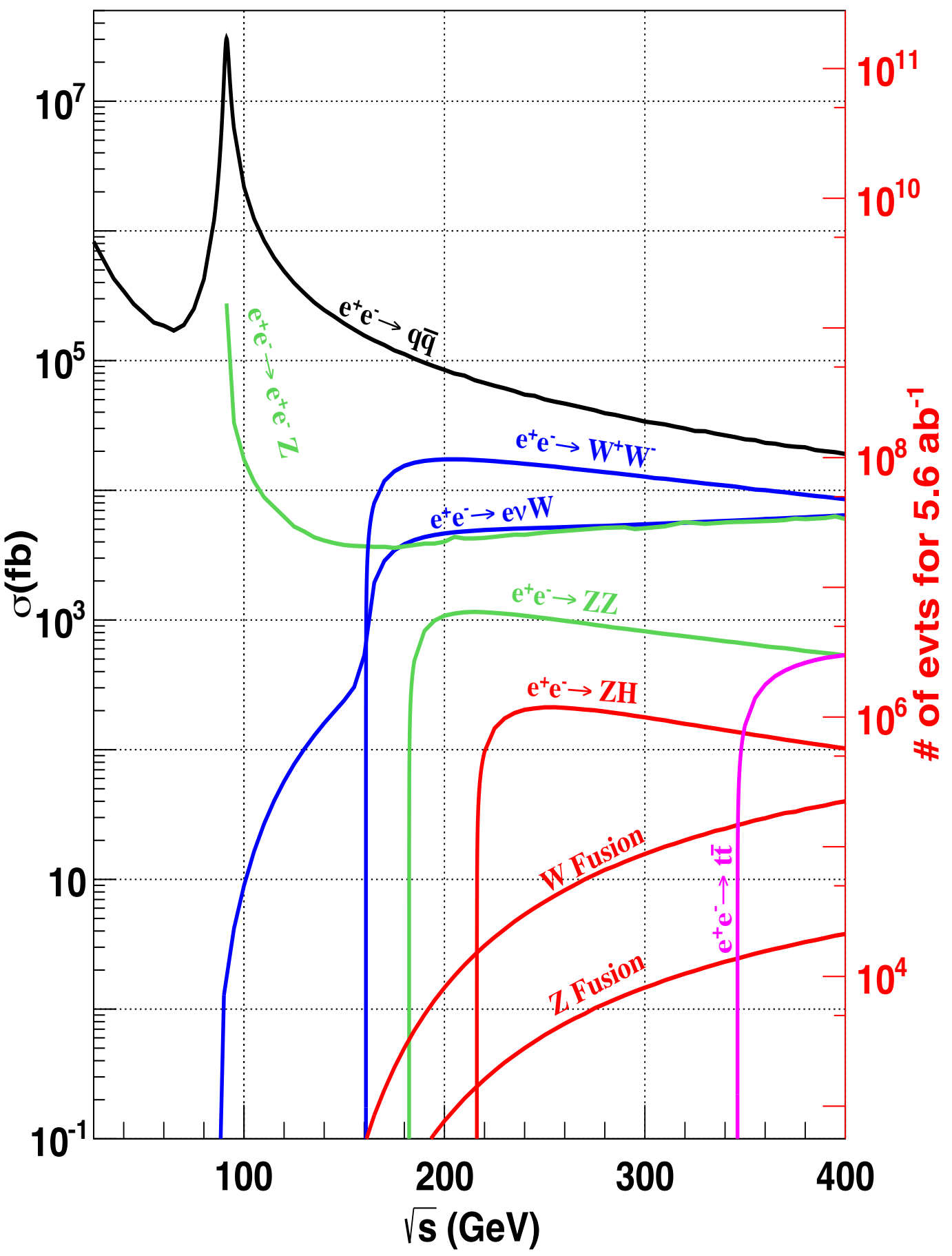

**Figure 3.1:** Cross sections of the leading Standard Model processes for unpolarized electron-positron collisions and the numbers of events expected in a dataset corresponding to an integrated luminosity of 5.6 ab$^{-1}$ as functions of the center-of-mass energy. The $W$ and $Z$ fusion processes refer to $e^+e^- \to \nu\bar{\nu}H$ and $e^+e^- \to e^+e^-H$ production, respectively. The $e^+e^- \to q\bar{q}$ curve includes contributions from $u$, $d$, $s$, $c$, and $b$ quarks, and the $e^+e^- \to t\bar{t}$ production is shown separately. These cross sections are obtained from the Whizard program [7].

| Operation mode | $Z$ factory | $WW$ threshold scan | Higgs factory |
|---|---|---|---|
| $\sqrt{s}$ (GeV) | 91.2 | 158 – 172 | 240 |
| Running time (years) | 2 | 1 | 7 |
| L ($10^{34}$ cm$^{-2}$s$^{-1}$) | 17 – 32 | 10 | 3 |
| Integrated Luminosity (ab$^{-1}$) | 8 – 16 | 2.6 | 5.6 |
| Higgs yield | – | – | $10^6$ |
| $W$ yield | – | $10^7$ | $10^8$ |
| $Z$ yield | $10^{11-12}$ | $10^8$ | $10^8$ |

**Table 3.2:** Running time, instantaneous and integrated luminosities at different values of the center-of-mass energy and anticipated corresponding boson yields at the CEPC. The $Z$ boson yields of the Higgs factory and $WW$ threshold scan operation are from the initial-state radiative return $e^+e^- \to \gamma Z$ process. The ranges of luminosities for the $Z$ factory correspond to the two possible solenoidal magnetic fields, 3 or 2 Tesla.

### 3.1.2.2 PAIR PRODUCTION

Electron-positron pairs are produced via the interaction of beamstrahlung photons with the strong electromagnetic fields of the colliding bunches. Pair production, in particular the incoherent pair production, represents the most important detector background at the CEPC. The process is simulated with GUINEAPIG [9] interfaced to GEANT4 [10–12] for the detector simulation. Despite of the magnitude of beam squeezing being different in the $x$ and $y$ directions, the hit distribution is almost uniform in the azimuthal ($\phi$) direction. The resulting hit density at the first vertex detector layer ($r = 1.6$ cm) is about 2.2 hits/cm$^2$ per bunch crossing when running at $\sqrt{s} = 240$ GeV. The total ionizing dose (TID) and non-ionizing energy loss (NIEL) are 620 kRad/year and $1.2 \times 10^{12}$ 1 MeV $n_{eq}$/cm$^2 \cdot$ year, respectively. For the background estimation, safety factors of ten are applied to cope with the uncertainties on the event generation and the detector simulation.

### 3.1.2.3 OFF-ENERGY BEAM PARTICLES

Beam particles after losing a certain amount of energy, *i.e.* 1.5% of the nominal beam energy, can be kicked off their nominal orbits. Such off-energy beam particles may hit machine and/or detector elements close to the interaction region and give rise to important backgrounds. The three main scattering processes are radiative Bhabha scattering, beamstrahlung and beam-gas interaction. After the introduction of two sets of collimators upstream of the IPs, backgrounds due to beamstrahlung and beam-gas interaction become negligible. The residual backgrounds due to radiative Bhabha scattering yields hit densities of about 0.22 hits/cm$^2$ per bunch crossing when operating at $\sqrt{s} = 240$ GeV. The corresponding TID and NIEL are 310 kRad/year and $9.3 \times 10^{11}$ 1 MeV $n_{eq}$/cm$^2 \cdot$ year, respectively.

#### 3.1.2.4 BACKGROUNDS AT DIFFERENT ENERGIES

When operating the machine at the center-of-mass energy of $\sqrt{s} = 240$ GeV, the main detector backgrounds come from the pair-production and off-energy beam particles. At lower operational energies, *i.e.* $\sqrt{s} = 160$ GeV for $WW$ and $\sqrt{s} = 91$ GeV for $Z$, the background particles are usually produced with lower energies but with higher rates given the higher machine luminosities. The pair-production becomes dominant, while contributions from other sources tend to be negligible.

## 3.2 PHYSICS REQUIREMENTS

As factories for the Higgs, $W$, and $Z$ bosons, the CEPC should be equipped with detectors that can reconstruct and identify their decay products with high efficiency and high purity and measure them with high precision. The CEPC physics program also requires precise determination of luminosities and beam energies. The integrated luminosity should be measured to a relative accuracy of 0.1% for the Higgs factory operation, and $\mathcal{O}(0.01\%)$ for the $Z$ factory operation. The beam energy needs to be controllable and known with an accuracy of the order of 1 MeV for the Higgs factory operation and 100 keV for the $Z$ factory operation and the $WW$ threshold scan.

The requirements on the physics object reconstruction and identification are briefly quantified below drawing on past experiences and using benchmark physics processes wherever possible. It should be noted, however, there are no yardsticks to define the requirements precisely. In most cases, the requirements are of the nature of *the more the better*. They are limited by practicalities and technological feasibility.

### 3.2.1 PARTICLE MULTIPLICITY

In physics events, visible particles include electrons, muons, photons, charged and neutral hadrons. The Monte Carlo (MC) truth-level multiplicities of these basic ingredients, with the charged particles collectively referred to as tracks, are shown in Figure 3.2 for the leading SM processes at the CEPC Higgs factory operation. The majority of the visible particles are charged particles whose multiplicity can be as high as $\sim 100$. These visible particles can have very small angles in between, particularly for those produced in high energy jets. An efficient separation of these particles provides a solid basis for the reconstruction and identification of physics objects, the high-level objects such as leptons, photons and jets that are input to physics analyses.

### 3.2.2 TRACKING

The CEPC detector should have excellent track finding efficiency and track momentum resolution. Figure 3.3 shows the expected energy and polar angle distributions of charged particles from the leading SM processes at the Higgs factory operation. The energy spectra cover a wide range from a very small energy to as high as the beam energy. For tracks within the detector acceptance and transverse momenta larger than 1 GeV, a track finding efficiency better than 99% is required. In order to measure the $H \to \mu^+\mu^-$ signal and to reconstruct precisely the Higgs boson mass from the recoil mass distribution at $ZH \to \ell^+\ell^- H$ events, a per mille level relative momentum resolution is required.

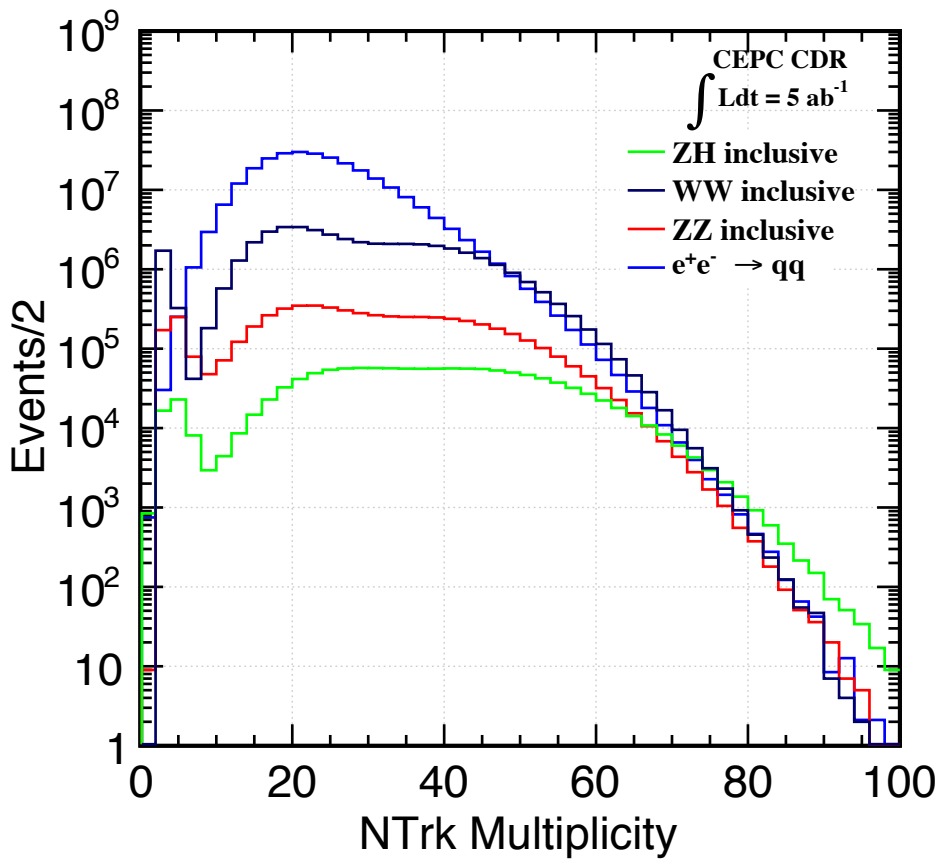

(a)

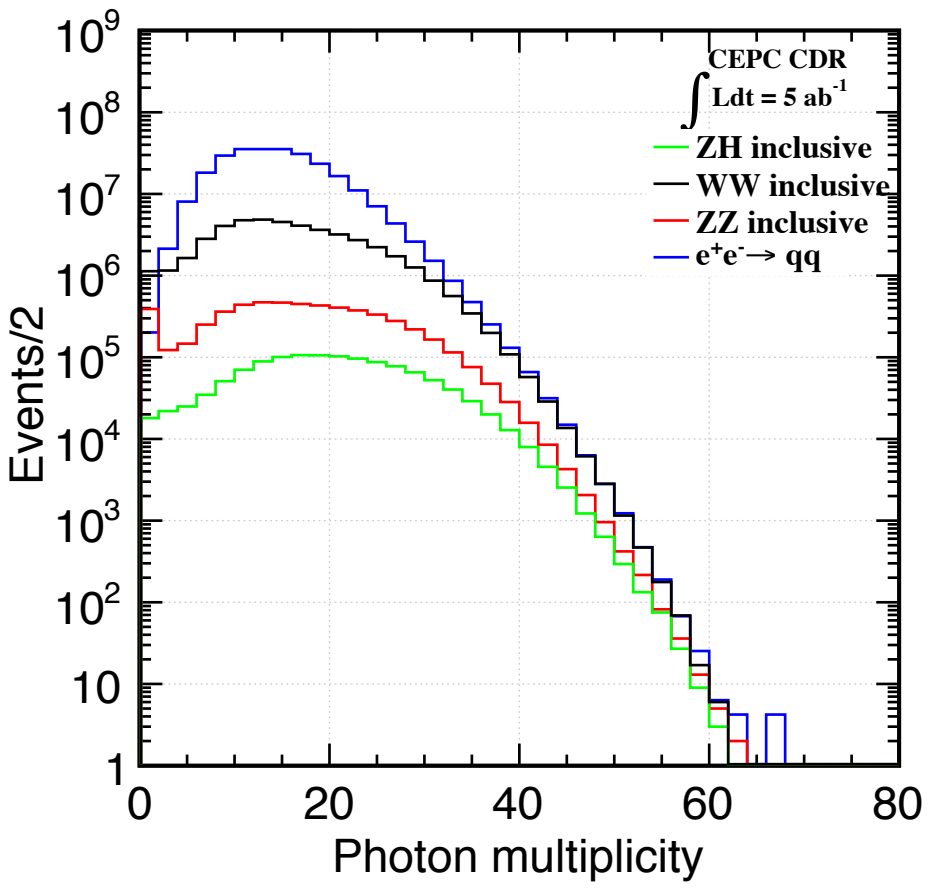

(b)

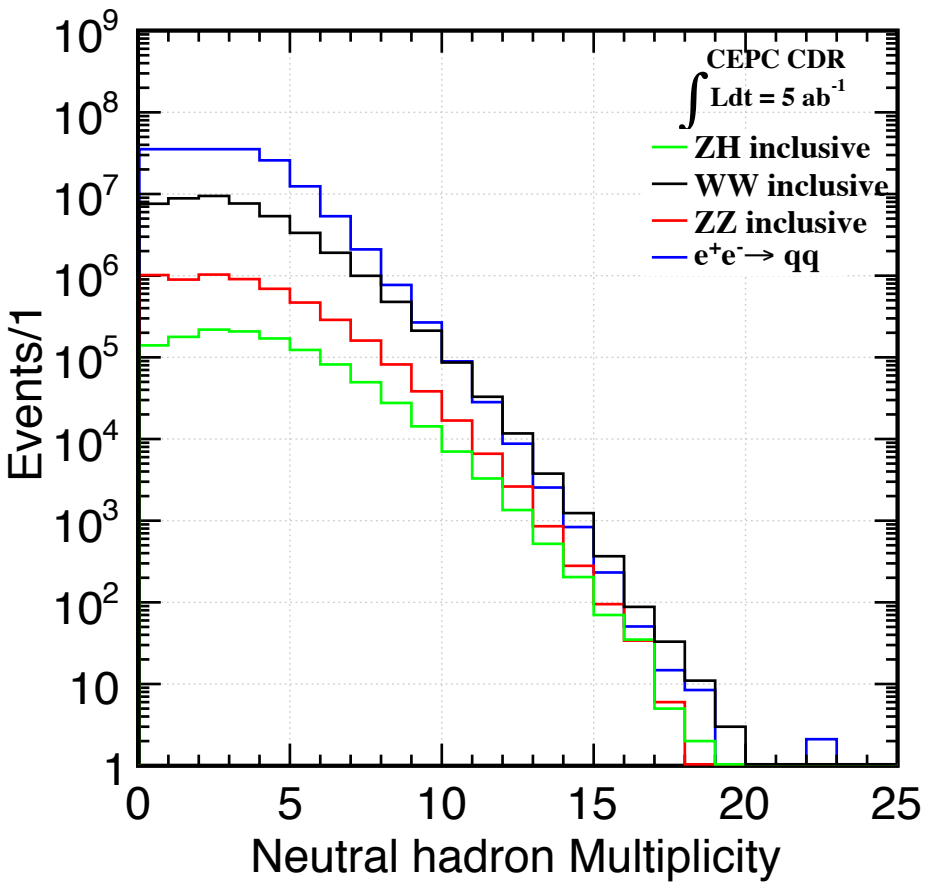

(c)

**Figure 3.2:** The MC truth-level multiplicity distributions of (a) charged particles, (b) photons, and (c) neutral hadrons with energies above 0.5 GeV from the leading SM processes at the CEPC Higgs operation, normalized to an integrated luminosity of 5 ab$^{-1}$. The $e^+e^- \to q\bar{q}$ distributions include contributions from both the $e^+e^- \to Z^{(*)}/\gamma^* \to q\bar{q}$ and $e^+e^- \to \gamma Z^{(*)} \to \gamma q\bar{q}$ processes.

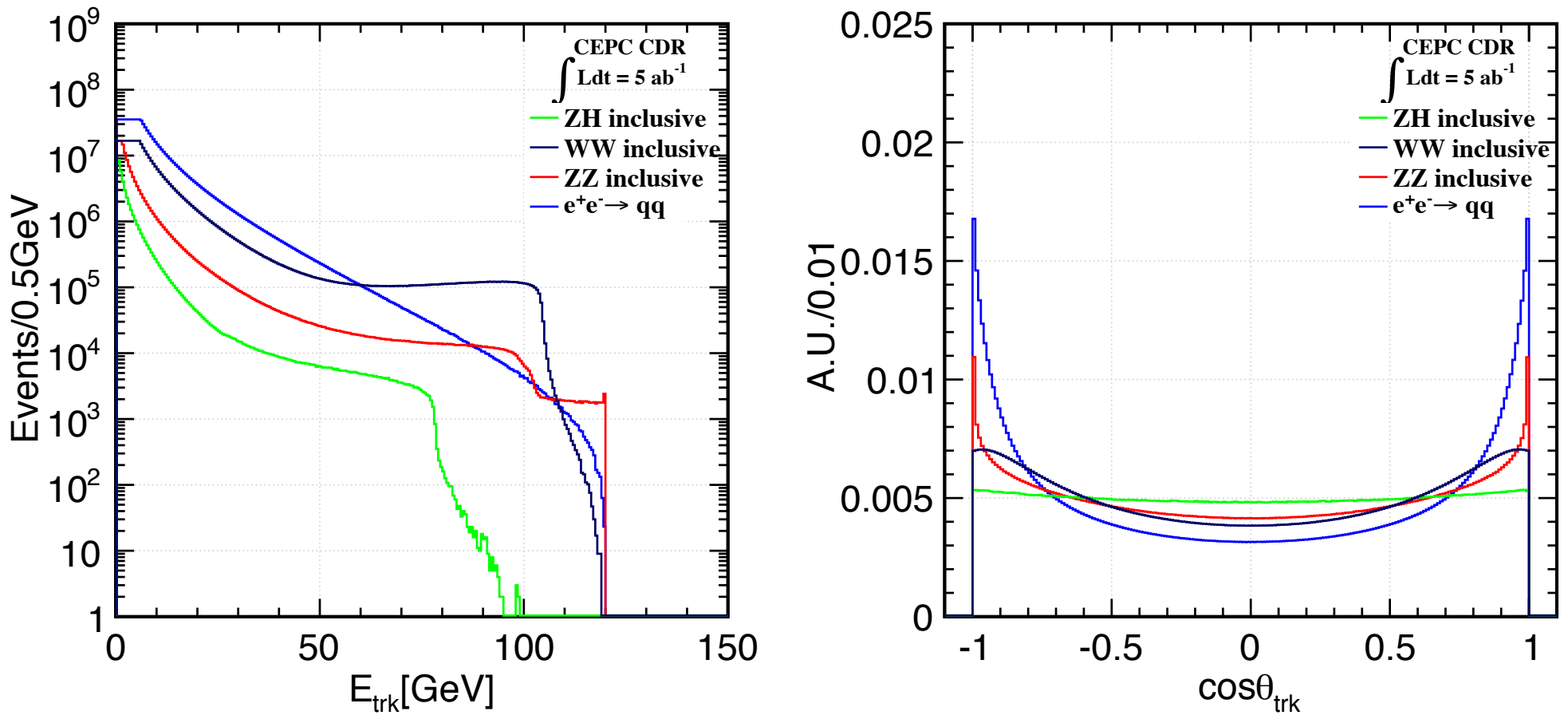

(a) (b)

**Figure 3.3:** The MC truth-level distributions of (a) energy and (b) cosine of the polar angle of charged particles with energies above 0.5 GeV from the leading SM processes at the CEPC Higgs operation. The energy distributions are normalized to an integrated luminosity of 5 ab$^{-1}$ whereas the $\cos\theta$ distributions are normalized to unity. The $e^+e^- \to q\bar{q}$ distributions include contributions from both the $e^+e^- \to Z^{(*)}/\gamma^* \to q\bar{q}$ and $e^+e^- \to \gamma Z^{(*)} \to \gamma q\bar{q}$ processes.

The $ZH$ process has a flat $\cos\theta$ distribution, whereas the background processes are more forward region dominated. A large solid angle coverage is essential for large acceptance and for the separation of different processes. Thus a coverage of up to $|\cos(\theta)| = 0.99$ is benchmarked.

### 3.2.3 LEPTONS

Leptons (electrons and muons, collectively denoted as $\ell$) are one of the most important physics signatures and play a crucial role in the classification of different physics events. An efficient lepton identification with high purity is fundamental for the CEPC physics program.

For example, approximately 7% of the Higgs bosons are produced with a pair of leptons. These $e^+e^- \to \ell^+\ell^- H$ events are golden events for the recoil mass analysis. Figure 3.4 shows the energy distributions of these leptons along with those produced in the Higgs boson decay cascades. The basic requirements on the lepton identification for the CEPC detector is to identify these prompt leptons with high efficiency and high purity. Therefore, a lepton identification with efficiency higher than 99% and misidentification rate smaller than 2% is required for isolated leptons with energies above 5 GeV.

Leptons from heavy quark decays are important for the jet flavor tagging and charge measurement. Therefore, a good identification of leptons in jets will be highly benefi-cial. However, more study is needed to quantify the requirements on the lepton-in-jet identification.

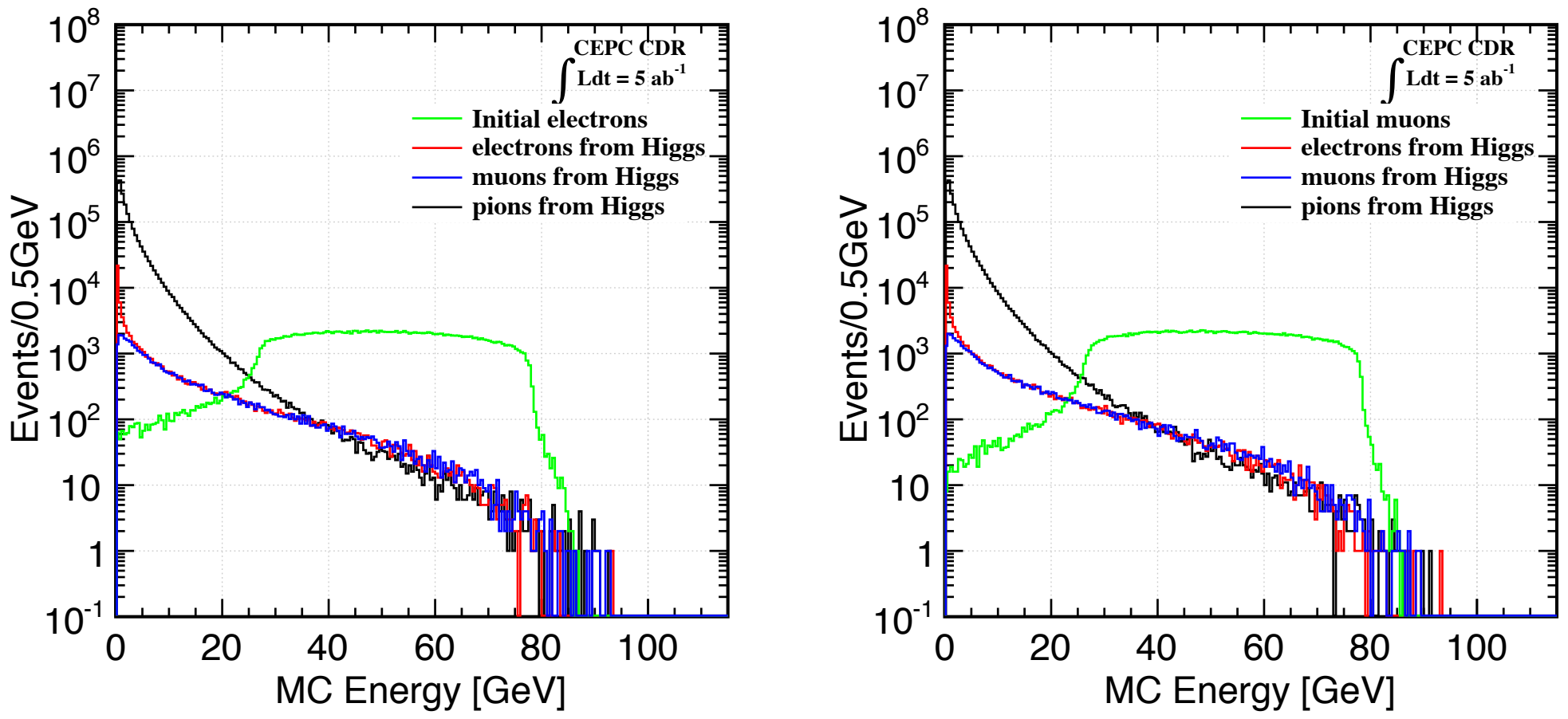

(a) (b)

**Figure 3.4:** Comparisons of the MC truth-level energy distributions of leptons and charged hadrons in Higgs boson events: (a) electrons and charged hadrons in the $e^+e^-H$ events and (b) muons and charged hadrons in the $\mu^+\mu^-H$ events. The energy spectra of leptons produced in association with the Higgs boson ("initial lepton") shown in green exhibit flat plateaus between 20–90 GeV as expected from the $ZH \to \ell^+\ell^-H$ production.

### 3.2.4 CHARGED HADRON IDENTIFICATION

Particle identification, especially the identification of charged kaons, is important for the flavor physics program. Similar to the leptons in jets, the identification of charged kaons is highly valuable for the jet flavor tagging and charge measurement. For an inclusive $Z$ boson sample, both the efficiency and purity of the kaon identification are required to be better than 90%.

### 3.2.5 PHOTONS

Photons are crucial for the jet energy resolution, the $H \to \gamma\gamma$ measurement, studies of radiative processes and the final states with $\tau$-leptons. They are prevalent in $e^+e^-$ collisions. As an example, Figure 3.5 shows the energy and polar angle distributions of photons expected from the leading SM processes at $\sqrt{s} = 240$ GeV.

For the reconstruction of unconverted and isolated photons with energies above 1 GeV, the identification efficiency higher than 99% and a misidentification rate smaller than 5% are required. To observe at least 50% of a diphoton resonance with a pair of unconverted photons, the material budget in front of the calorimeter must be less than $0.35X_0$ averaged over all solid angle. To identify the $\tau$-leptons in the different decay modes, the photons should be distinguishable from $\pi^0$'s with an efficiency and purity higher than 95% measured in the $Z \to \tau^+\tau^-$ event sample at the CEPC $Z$ factory operation.

To fully explore the hadronic decays of the Higgs, $W$, and $Z$ bosons, the requirements on the jet energy resolution, described in the next section, impose a photon energy resolu-tion requirement of better than $20\%/\sqrt{E} \oplus 1\%$. This requirement is sufficient to meet the needs of the precision Higgs physics program at the CEPC. However, the photon energy resolution requirement for the precision electroweak measurements need to be evaluated.

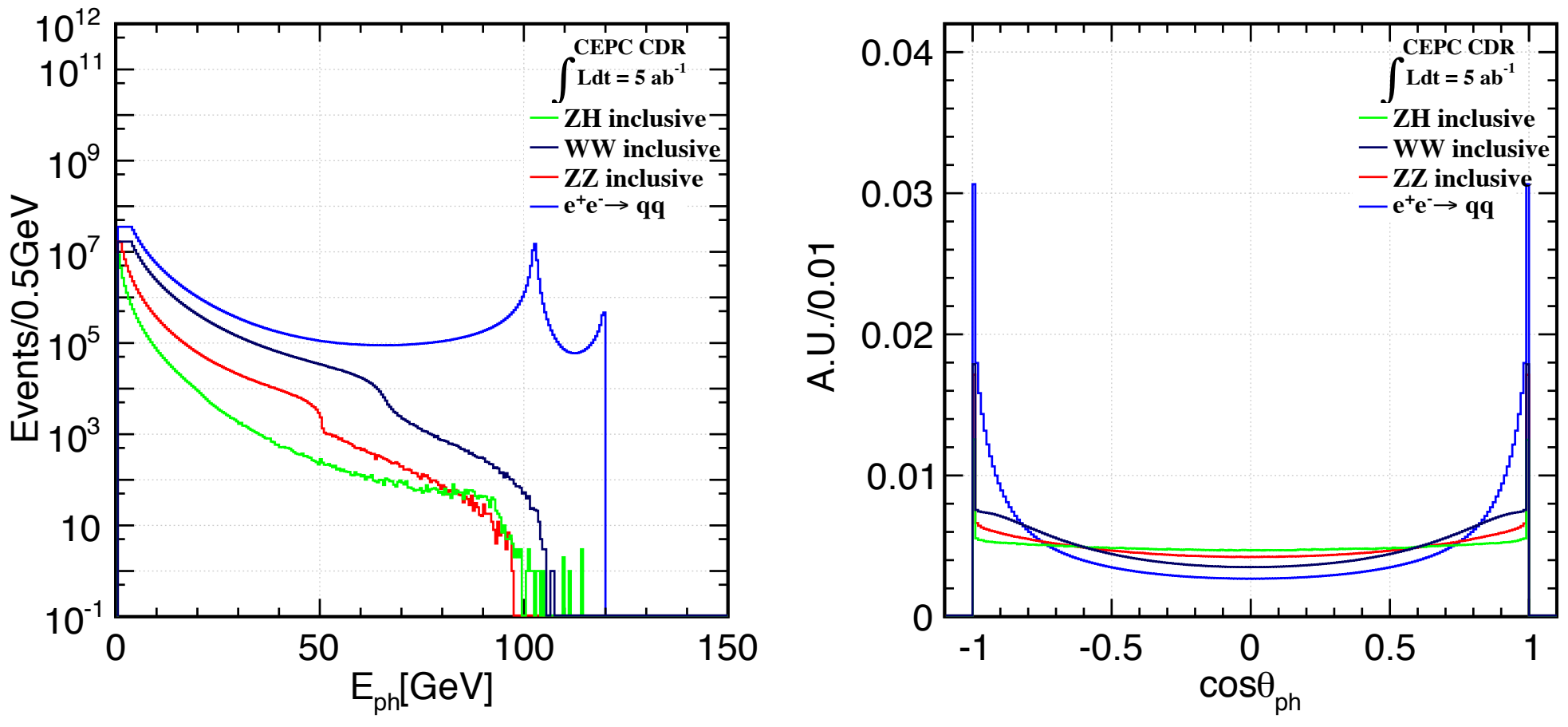

(a) (b)

**Figure 3.5:** The MC truth-level distributions of the (a) energy and (b) cosine of the polar angle of photons from the leading SM processes at the CEPC Higgs operation. The energy distributions are normalized to an integrated luminosity of 5 ab$^{-1}$, whereas the $\cos\theta$ distributions are normalized to unity. The $e^+e^- \to q\bar{q}$ distributions include contributions from both the $e^+e^- \to Z^{(*)}/\gamma^* \to q\bar{q}$ and $e^+e^- \to \gamma Z^{(*)} \to \gamma q\bar{q}$ processes.

### 3.2.6 HADRONIC FINAL STATES, JETS AND MISSING ENERGY

The majority of the Higgs, $W$, and $Z$ bosons decay into quarks or gluons which fragment into hadronic final states. Traditionally these final states are reconstructed using jet algorithms. However, due to color connections between quarks and gluons, energy deposited in the detector cannot always be cleanly assigned to jets. Therefore, jets may not be the best tools for analyzing these events.

Taking for example the $ZZ \to \nu\bar{\nu}q\bar{q}$ and $ZH \to \nu\bar{\nu}(b\bar{b}, c\bar{c}, gg)$ events, the mass of the hadronic system is the main discriminant variable for these two event types. Conventionally the mass is calculated from jet pairs after the jet reconstruction. This procedure is dependent on the jet reconstruction performance and suffers from radiation effects such as hard gluon radiation. These issues can be avoided by determining the mass of the hadronic system directly without going through the intermediate step of the jet reconstruction.

A quantity called Boson Mass Resolution (BMR) defined as the mass resolution of the hadronic system is introduced to quantify the detector performance. The hadronic system can also be defined in events with leptons from the $W$ and $Z$ boson decays and with high energy photons from radiation as these leptons and photons can be cleanly identified and removed. Thus the BMR concept has a broad application. Figure 3.6 illustrates the mass distributions of the hadronic decays of the Higgs, $W$, and $Z$ bosons for four different values of BMR. To achieve approximately a $2\sigma$ separation of the $W$ and $Z$ bosons in their hadronic decays, a BMR of 4% or better will be needed.

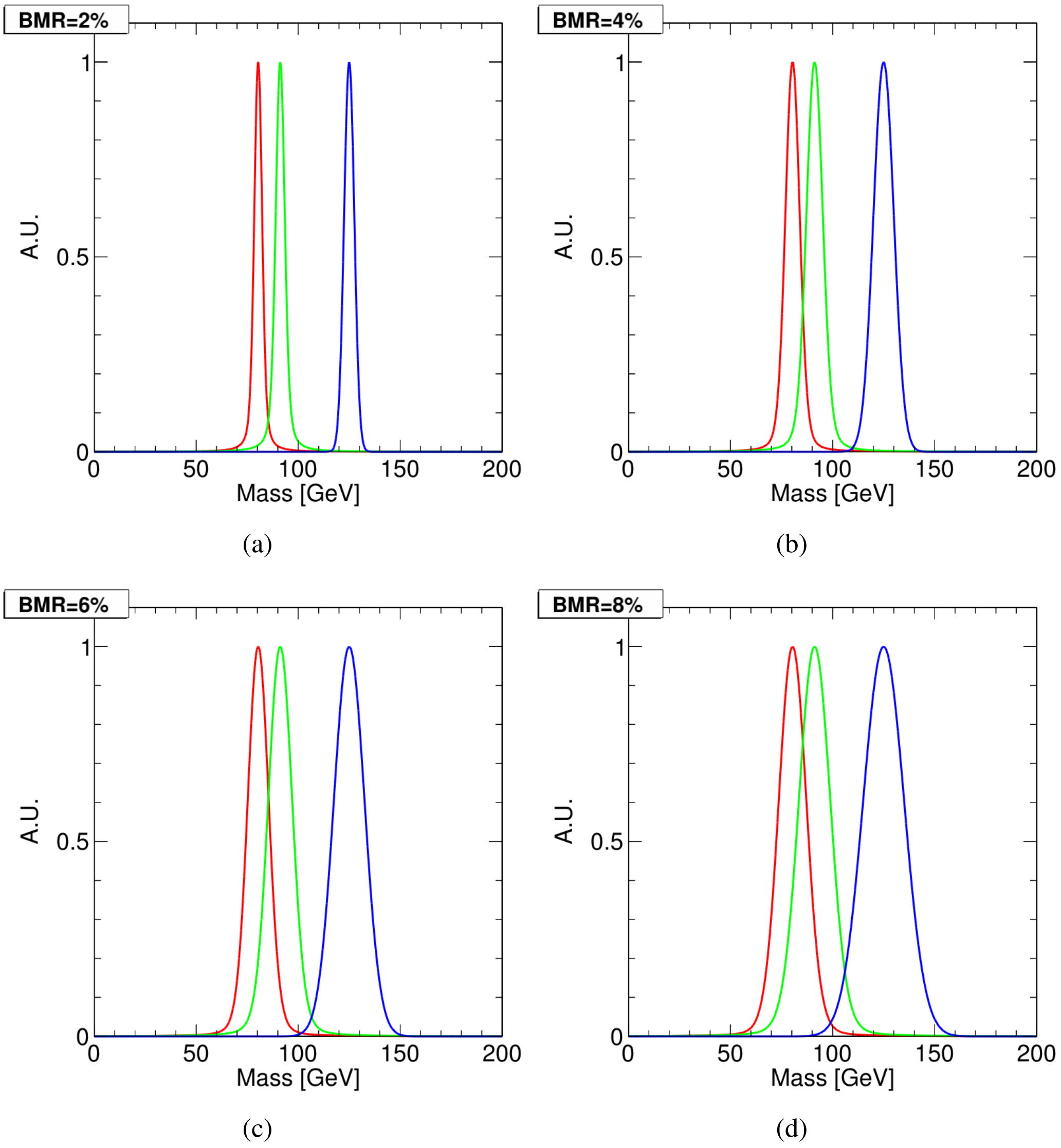

**Figure 3.6:** The invariant mass distributions of the hadronic decays of the $W$ (red), $Z$ (green) and Higgs (blue) bosons for different BMR values. All distributions are normalized to unit height. The distributions of the $W$ and $Z$ bosons are modeled with Breit-Wigner distributions of widths of 2 and 2.5 GeV, respectively, convoluted with detector resolutions of different BMR values. The distributions of the Higgs bosons are obtained similarly, but from an intrinsic narrow width distribution.

Despite its issues, jets are indispensable in studying hadronic final states. For example, the $H \to WW^* \to 4j$ and $H \to ZZ^* \to 4j$ decays are expected to have the same hadronic mass and therefore cannot be separated using this information. However, they can, in principle, be distinguished using the dijet mass information after the jet reconstruction. This will place stringent requirements on the dijet mass resolution. Though studies are ongoing, a resolution of about 4%, similar to that for BMR, is likely needed.

BMR is also a good yardstick to quantify the missing energy performance in $e^+e^-$ collisions as the missing energy resolution is largely determined by the hadronic energies. For example, the most sensitive final state for searching for the invisible decays of the Higgs boson is the $ZH$ production with $Z \to q\bar{q}$. The dominant background is from the $ZZ \to \nu\bar{\nu}q\bar{q}$ production. The signal and background events are expected to have the same hadronic mass but differ in the missing mass (or the dijet recoil mass). BMR will also impact the missing mass resolution as demonstrated in Figure 3.7. If the BMR is worse than 4%, the $ZZ$ background starts to have significant overlap with the signal. Thus, a BMR of 4% or better is also a good benchmark for the missing energy performance. In general, since the majority of the Higgs bosons are produced in association with a pair of jets from the Z boson decays, a BMR better than 4% and the corresponding missing mass measurement are crucial for the identification of Higgs bosons in their non-hadronic decay modes.

### 3.2.7 FLAVOR TAGGING

The identification of jet flavors (jet flavor tagging) allows for the separate measurements of processes with heavy ($b$ or $c$) quarks, light quarks or gluons. For example, the tagging of $c$-quark jets ($c$-jets) from the large backgrounds of $b$-quark jets ($b$-jets) and light-quark jets (light-jets, including gluon jets) is a prerequisite for the measurement of the Higgs boson coupling to the $c$-quark, an important physics goal of the CEPC.

The classification of different kinds of jets depends strongly on the reconstruction of secondary vertex, where the performance of the vertex system is crucial. The clean collision environment of the CEPC allows for a much aggressive vertex system design as described in Section 4.1.

Benchmarked using the $Z \to q\bar{q}$ sample at $\sqrt{s} = 91.2$ GeV, the efficiency and purity are both required to be greater than 80% for the $b$-jet identification and greater than 60% for the $c$-jet identification. To achieve these performance metrics, an impact parameter resolution of about 5 μm is required.

### 3.2.8 SUMMARY

The discussion above quantifies the physics requirements on the physics object reconstruction, identification and measurements. Below is a brief summary.

**Tracking performance:** For tracks with transverse momenta greater than $1\,\mathrm{GeV}$ and within the detector acceptance, a reconstruction efficiency of better than 99% is required. The track momentum resolution should achieve per mille level.

**Lepton identification:** For isolated leptons with momenta greater than 5 GeV, an identification efficiency of 99% and a misidentification rate smaller than 2% are required. Leptons inside jets also need to be identified well, as they provide information on the jet flavor and jet charge.

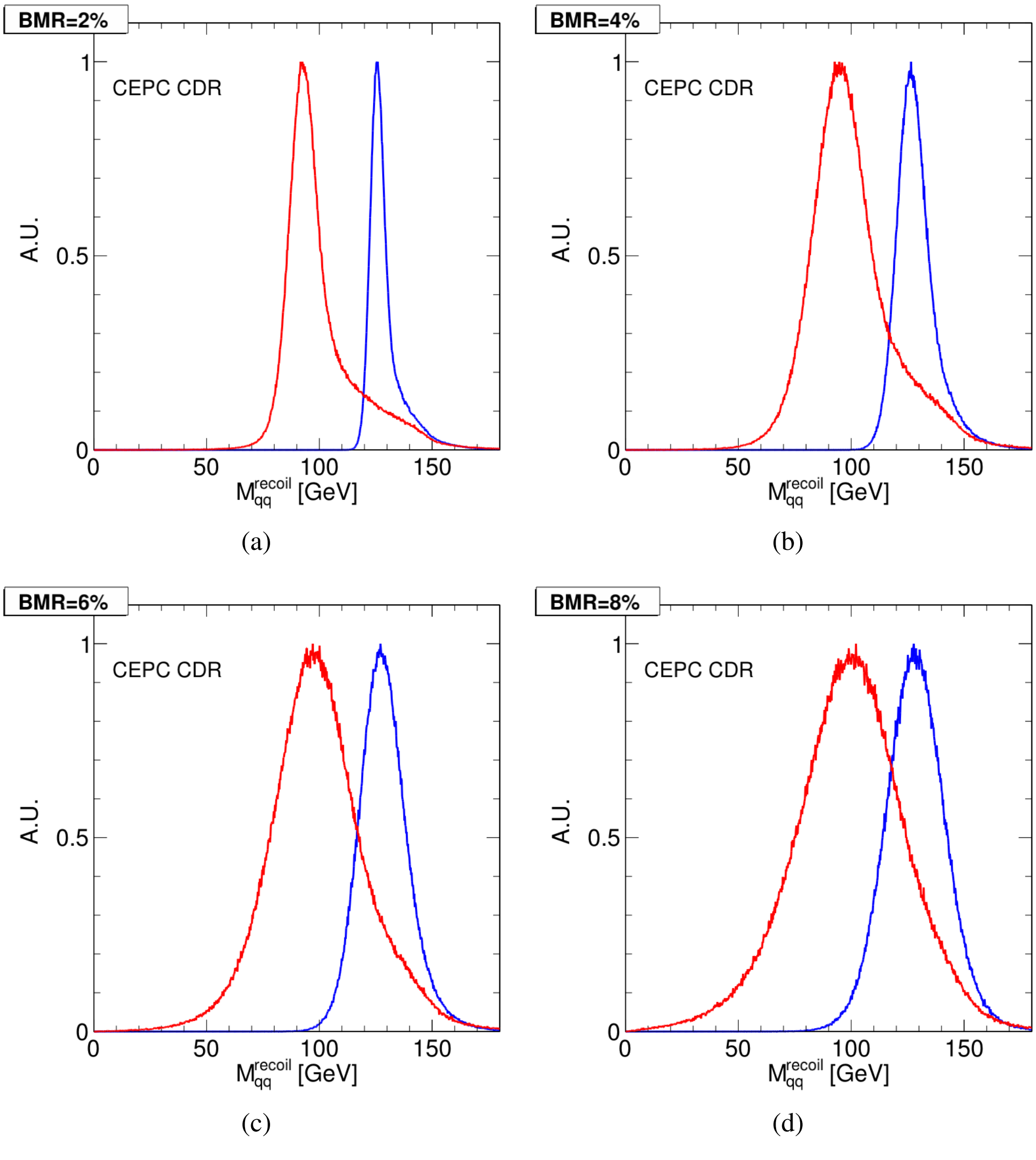

**Figure 3.7:** The dijet recoil mass distributions of the $ZZ \to \nu\bar{\nu}q\bar{q}$ (red) and $ZH \to q\bar{q}H$ (blue) events for different BMR values. The invisible Higgs boson decays are considered. All distributions are normalized to unit height.

| Physics process | Measurands | Detector subsystem | Performance requirement |
|---|---|---|---|
| $ZH, Z \to e^+e^-, \mu^+\mu^-$<br>$H \to \mu^+\mu^-$ | $m_H, \sigma(ZH)$<br>$\mathrm{BR}(H \to \mu^+\mu^-)$ | Tracker | $\Delta(1/p_T) =$<br>$2 \times 10^{-5} \oplus \frac{0.001}{p(\mathrm{GeV}) \sin^{3/2}\theta}$ |
| $H \to b\bar{b}/c\bar{c}/gg$ | $\mathrm{BR}(H \to b\bar{b}/c\bar{c}/gg)$ | Vertex | $\sigma_{r\phi} =$<br>$5 \oplus \frac{10}{p(\mathrm{GeV}) \times \sin^{3/2}\theta}(\mu\mathrm{m})$ |
| $H \to q\bar{q}, WW^*, ZZ^*$ | $\mathrm{BR}(H \to q\bar{q}, WW^*, ZZ^*)$ | ECAL<br>HCAL | $\sigma_E^{\mathrm{jet}}/E =$<br>$3 \sim 4\%$ at 100 GeV |
| $H \to \gamma\gamma$ | $\mathrm{BR}(H \to \gamma\gamma)$ | ECAL | $\Delta E/E =$<br>$\frac{0.20}{\sqrt{E(\mathrm{GeV})}} \oplus 0.01$ |

**Table 3.3:** Physics processes and key observables used as benchmarks for setting the requirements and the optimization of the CEPC detector.

**Charged kaon identification:** For the inclusive $Z \to q\bar{q}$ sample at $\sqrt{s} = 91.2$ GeV, the charged kaon identification should have both the efficiency and purity higher than $90\%$.

**Photon identification and energy measurement:** The photon energy should be measured to a precision better than $20\%/\sqrt{E} \oplus 1\%$. Photons should be identified from $\pi^0$'s with an efficiency and purity higher than 95% in the $Z \to \tau^+\tau^-$ event sample at the CEPC $Z$ factory operation.

**Jet and missing energy:** Benchmarked with the separation of massive SM bosons ($W$, $Z$, and Higgs boson) and the $\mathrm{BR}(H \to \mathrm{invisible})$ measurements, a BMR better than $4\%$ is identified.

**Flavor tagging:** Benchmarked with the $Z \to q\bar{q}$ sample at $\sqrt{s} = 91.2$ GeV, the efficiency and purity are both required to be above $80\%$ for the $b$-jet tagging and above $60\%$ for the $c$-jet tagging.

Most of the above-mentioned requirements are driven by the precision Higgs physics program. Some examples are shown in Table 3.3. However, these requirements also apply to the precise EW measurements as the $W$ and $Z$ bosons decay into similar physics objects.

## 3.3 DETECTOR CONCEPTS

To address the physics requirements of the CEPC, a baseline and an alternative detector concepts are introduced. A variant baseline option with a different tracker is also proposed.

The baseline concept was developed from the ILD concept [2, 3], optimized for the CEPC collision environment. It employs an ultra high granular calorimetry system to efficiently separate the final state particle showers, a low material tracking system to minimize the interaction of the final state particles in the tracking material, and a large volume

| Concept | ILD | CEPC baseline | IDEA |
|---|---|---|---|
| Tracker | TPC/Silicon | TPC/Silicon or FST | Drift Chamber/Silicon |
| Solenoid B-Field (T) | 3.5 | 3 | 2 |
| Solenoid Inner Radius (m) | 3.4 | 3.2 | 2.1 |
| Solenoid Length (m) | 8.0 | 7.8 | 6.0 |
| L* (m) | 3.5 | 2.2 | 2.2 |
| VTX Inner Radius (mm) | 16 | 16 | 16 |
| Tracker Outer Radius (m) | 1.81 | 1.81 | 2.05 |
| Calorimeter | PFA | PFA | Dual readout |
| Calorimeter $\lambda_I$ | 6.6 | 5.6 | 7.5 |
| ECAL Cell Size (mm) | 5 | 10 | - |
| ECAL Time resolution (ps) | - | 200 | - |
| ECAL $X_0$ | 24 | 24 | - |
| HCAL Layer Number | 48 | 40 | - |
| HCAL Absorber | Fe | Fe | - |
| HCAL $\lambda_I$ | 5.9 | 4.9 | - |
| DRCAL Cell Size (mm) | - | - | 6.0 |
| DRCAL Time resolution (ps) | - | - | 100 |
| DRCAL Absorber | - | - | Pb or Cu or Fe |
| Overall Height (m) | 14.0 | 14.5 | 11.0 |
| Overall Length (m) | 13.2 | 14.0 | 13.0 |

**Table 3.4:** Comparison of parameters of the ILD detector and the CEPC detector concepts. L* is the distance between the IP and the final focusing quadrupole magnet.

3 Tesla solenoid that encloses the entire calorimetry system. Two options for its tracking system are being considered. The default option is a combination of a silicon tracker and a Time Projection Chamber (TPC). The other one is a full silicon tracker.

An alternative detector concept, IDEA, uses a dual readout calorimeter to achieve the excellent energy resolution for both the electromagnetic and hadronic showers. Comparing to the baseline detector, IDEA has a lower solenoidal field of 2 Tesla, but compensates with a large tracking volume. The IDEA is also been proposed as a reference detector for FCC-ee studies.

The main detector parameters of both concepts are summarized in Table 3.4.

### 3.3.1 BASELINE DETECTOR CONCEPT

The baseline detector concept is guided by the particle flow principle of measuring final state particles in the most suited detector subsystem. The Particle Flow Algorithm (PFA) reconstructs a list of low-level particles (called PFA particles) and associates every detec-tor hit with these particles. For each physics event, all physics objects that are input to physics analyses are then identified or reconstructed from this unique list of PFA particles. Single particle physics objects, such as leptons and photons, are identified directly from the list. Compound physics objects like converted photons, neutral kaons $K^0_S$, and the jets are reconstructed from the PFA particles using dedicated algorithms based on their unique characteristics. This global interpretation of the final state particles leads to high efficiency and high purity reconstruction of all physics objects.

From innermost to outermost, as shown in Figure 3.8, the baseline concept is composed of a silicon pixel vertex detector, a silicon inner tracker, a TPC surrounded by a silicon external tracker, a silicon-tungsten sampling Electromagnetic Calorimeter (ECAL), a steel-Glass Resistive Plate Chambers (GRPC) sampling Hadronic CALorimeter (HCAL), a 3 Tesla superconducting solenoid, and a flux return yoke embedded with a muon detector. In addition, five pairs of silicon tracking disks are placed in the forward regions at either side of the IP.

The vertex detector consists of six concentric cylindrical pixel layers at radii between 1.6 cm and 6.0 cm, providing impact parameter measurements with a resolution of $\sim$ 5 $\mu$m. Outside the vertex detector is the silicon inner tracker consisting of two microstrip layers at radii of 15.3 cm and 30 cm before the TPC. The silicon external tracker consists of one microstrip layer at radius of 181 cm after the TPC. The forward silicon tracker is composed of five pairs of silicon disks located at either side of the IP in between $z$ positions of 20–100 cm. The two disks closest to the IP on either side will use the pixel technology due to their expected high occupancy while the rest will use the microstrip technology. The silicon tracking system is shown in detail in Figure 4.1 and is expected to have a single hit spatial resolution better than 7 $\mu$m.

The TPC, located in between the silicon inner and external tracker, has an inner radius of 0.3 m, an outer radius of 1.8 m, and a length of 4.7 m. It is divided into 220 radial layers in steps of 6 mm. Along the azimuth, each layer is segmented into 1 mm wide cells. In total, the TPC has one million readout channels in each endcap. Operating in a 3 Tesla solenoidal magnetic field, the TPC has a single-hit spatial resolution of 100 $\mu$m in the $r$–$\phi$ plane and 500 $\mu$m in the $z$ direction. The TPC can also measure the ionization energy loss ($dE/dx$), allowing the identification of low momentum charged particles.

Combining the measurements from the silicon tracking system and the TPC, the track momentum resolution reaches $\Delta(1/p_T) \sim 2\times10^{-5}$ GeV$^{-1}$. In comparison, the resolution from the TPC alone is $\Delta(1/p_T) \sim 10^{-4}$ GeV$^{-1}$. While the silicon tracking system dominates the precision of the momentum measurement, the TPC complements with excellent pattern recognition and track finding capability.

The baseline concept employs high granular sampling ECAL and HCAL, providing 3-dimensional spatial and the energy information. The ECAL is composed of 30 longitudinal layers of alternating silicon sensors and tungsten absorbers, splitting into two sections of different absorber thickness. The total absorber length at $\theta = 90°$ is 8.4 cm, or equivalently 24 radiation lengths ($X_0$). Transversely, each sensor layer is segmented into $10 \times 10\,\mathrm{mm}^2$ cells. These geometric parameters are optimized through a dedicated study [13]. A scintillator and tungsten sampling ECAL is also being investigated. The HCAL uses steel as absorbers and scintillator titles or gaseous detectors such as RPCs as sensors. It has 40 longitudinal layers, each consists of a 2.5 cm steel absorber. The total steel thickness is about 100 cm at $\theta = 90°$, corresponding to 5.6 interaction lengths. Like the ECAL, the HCAL is segmented into $10 \times 10\,\mathrm{mm}^2$ cells transversely.

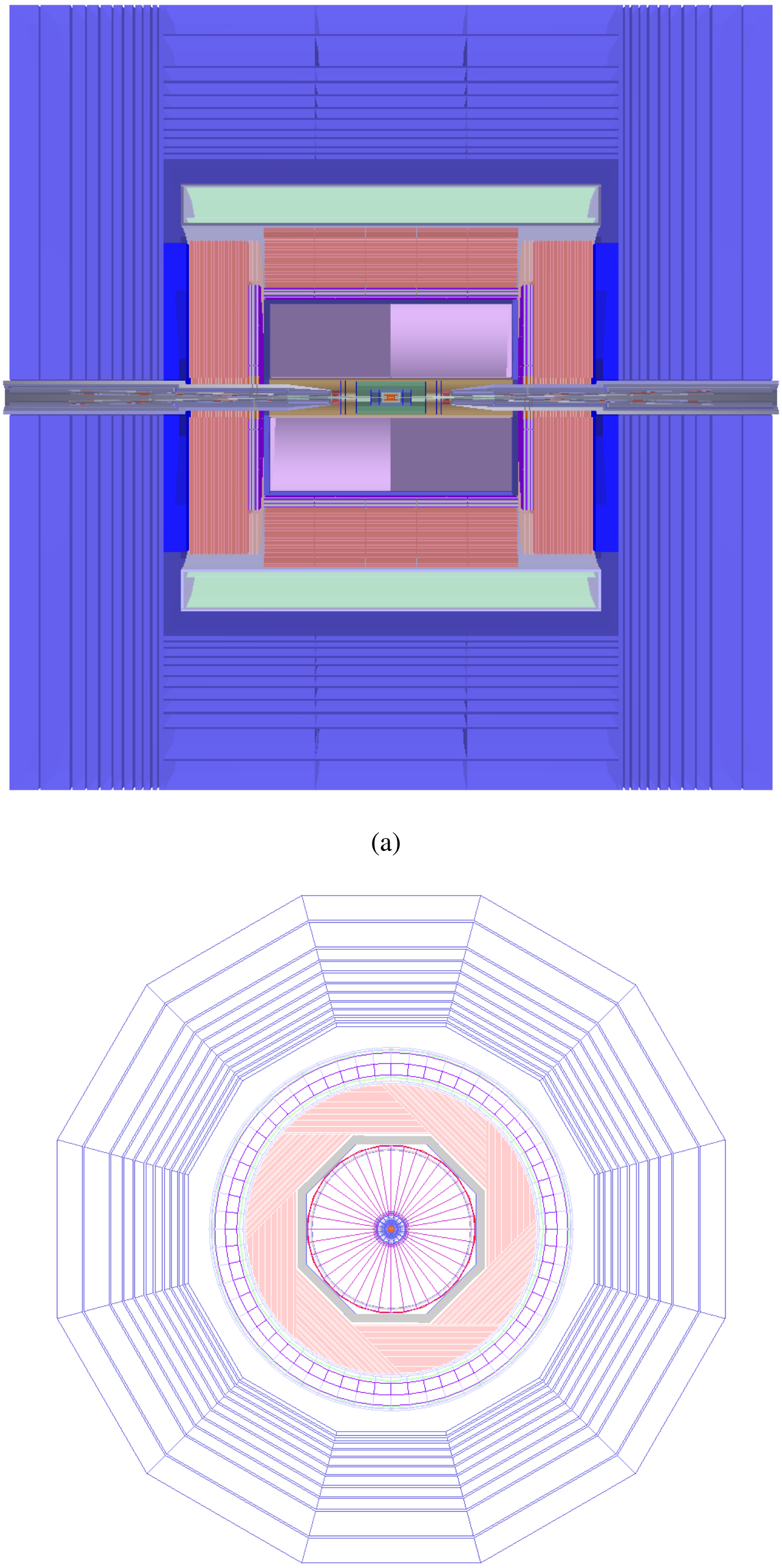

(a)

(b)

**Figure 3.8:** The (a) $r$–$z$ and (b) $r$–$\phi$ view of the baseline detector concept. In the barrel from inner to outer, the detector is composed of a silicon pixel vertex detector, a silicon inner tracker, a TPC, a silicon external tracker, an ECAL, an HCAL, a solenoid of 3 Tesla and a return yoke with embedded a muon detector. In the forward regions, five pairs of silicon tracking disks are installed to enlarge the tracking acceptance (from $|\cos(\theta)| < 0.99$ to $|\cos(\theta)| < 0.996$).

The calorimetry system provides good energy measurements for neutral particles: approximately $16\%/\sqrt{E/\text{GeV}}$ for photons and $60\%/\sqrt{E/\text{GeV}}$ for neutral hadrons. A jet energy resolution of 3–5% are expected for jet energies between 20 and 100 GeV. In addition, the finely segmented calorimeters record enormous amount of information of both electromagnetic and hadronic shower spatial development, ensuring efficient separation between nearby showers and providing essential information for lepton identification, see Section 10.2.1. The silicon-tungsten ECAL could also be instrumented to provide precise time measurements with a 50 ps or better resolution. Such Time-of-Flight (ToF) measurements will aid the charged particle identification complementary to that of the TPC $dE/dx$ measurements, see Section 10.2.7.

Surrounding the HCAL is a cylindrical superconducting solenoid with a room temperature bore of 6.8 m in diameter and 8.3 m in length. The solenoid will provide a 3 Tesla axial magnetic field. Outside the solenoid is an iron yoke spanning radii from 4.4 m to 7.24 m serving as the magnetic flux return. The yoke will be instrumented with the muon detector of either Resistive Plate Chambers (RPC) or $\mu$-RWELL [14] detector technology.

The CEPC detector requires a dedicated Machine-Detector Interface (MDI). The accelerator uses a double ring configuration, with a crossing angle of 33 mrad at the IP. The beam emittance (particularly the vertical emittance) increases via the coupling to the detector solenoid magnetic field as the beam passes through the IP. This magnetic field needs to be compensated locally. Therefore, a compensating solenoid will be installed in the forward regions in $|z|$ positions between 110–600 cm. A Luminosity Calorimeter (LumiCal) will be mounted at the end of the nose structure of this solenoid. For the $Z$ factory operation, the detector solenoid field may need to be lowered to 2 Tesla to achieve high luminosities.

**Full silicon tracker** Silicon detectors are at present the most precise tracking devices for charged particles in high energy physics experiments. They have an excellent spatial resolution and granularity to separate tracks in the environment of dense jets and high hit occupancy from beam related backgrounds at high luminosities. A variant of the CEPC baseline detector concept with a full-silicon tracker without the TPC is, therefore, an attractive option. The FST will provide excellent tracking efficiency, momenta resolution, and vertexing capability for charged particles from the IP as well as from the secondary vertices. The challenge is to build such a tracker with minimal amount of material to preserve the momentum resolution and coverage hermiticity down to a polar angle of $|\cos\theta| < 0.992$.

The FST replaces the TPC and the silicon tracking system with a full-silicon tracker while keeping other detector subsystems unchanged. The number of layers, single-sided versus double-sided layers, and the layout geometry are optimized through a toy Monte Carlo simulation. An alternative approach based on the ILC-SID design [3] is also being considered. The FST detector geometry, as shown in Figure 3.9, has been implemented in the simulation and track reconstruction software. Initial studies show promising tracking performance. However, many improvements in the simulation and reconstruction are needed to demonstrate the full potential of the FST. Detailed design is described in Section 4.3.

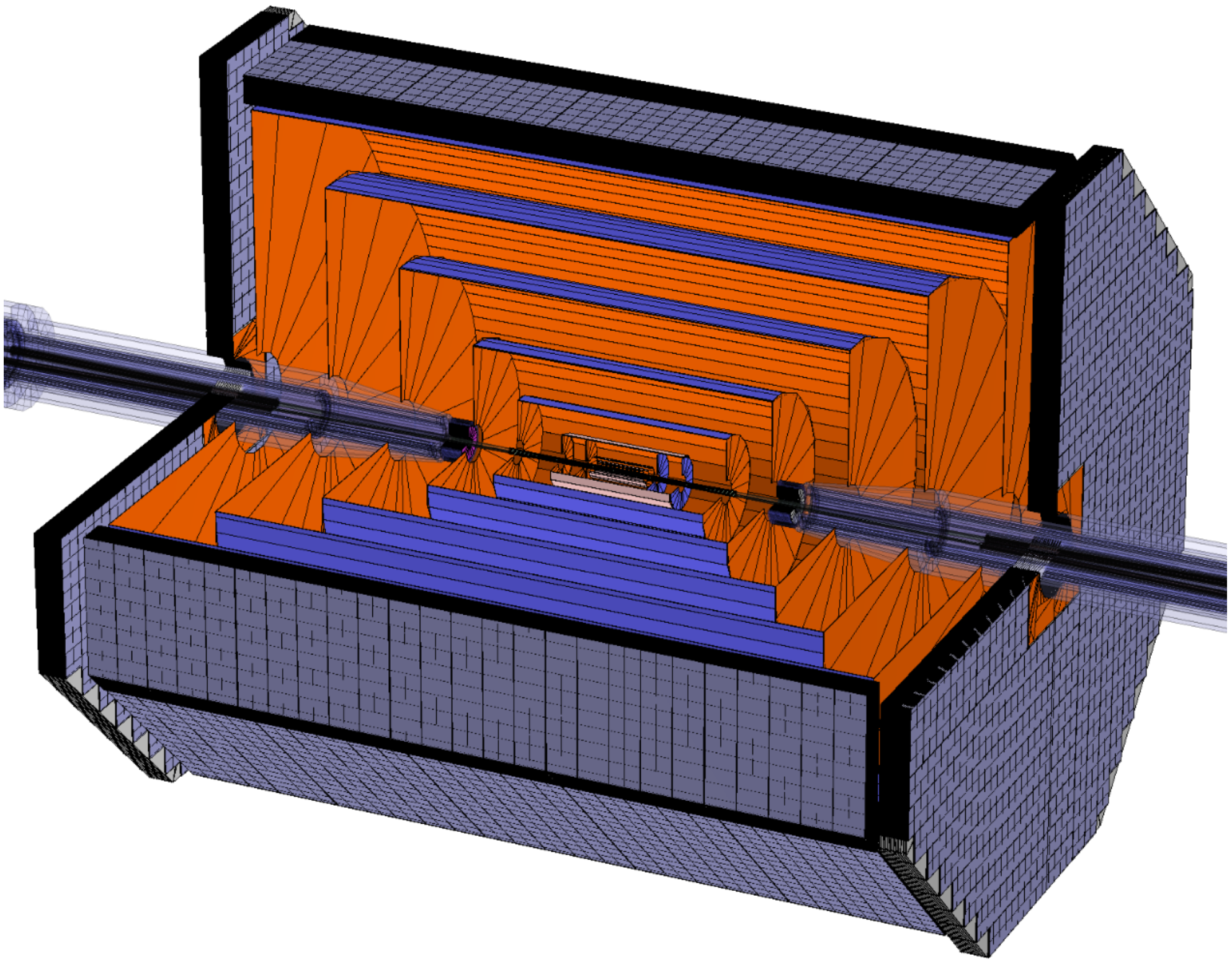

**Figure 3.9:** The cutaway view of the full silicon tracker proposed as an option for the CEPC baseline detector concept.

### 3.3.2 ALTERNATIVE DETECTOR CONCEPT

An alternative detector concept, Innovative Detector for Electron-positron Accelerator (IDEA), has been designed for a circular electron-positron collider and it is also being adopted as a reference detector for FCC-ee studies. The concept design attempts to economize on the overall cost of the detector and proposes different technologies than the baseline concept for some of the main detector subsystems. It provides therefore an opportunities to leverage challenges and advances in detector development prior to the CEPC detector constructions.

The detector requirements at CEPC are tied to the operational parameters of the storage ring at each energy point. For example, the typical luminosity at the $Z$ pole ($\sqrt{s} = 91.2\,\mathrm{GeV}$) is expected to be up to two orders of magnitude higher than at $ZH$ threshold ($\sqrt{s} = 240\,\mathrm{GeV}$). Bunch spacing will be significantly smaller. One would therefore prefer an intrinsically fast main tracker to fully exploit the cleanliness of the $e^+e^-$ environment while integrating as little background as possible. Additional issues of emittance preservation, typical of circular machines, set limits on the maximum magnetic field usable for the tracker solenoid, especially when running at lower center-of-mass energies.

Additional specific requirements on a detector for CEPC come from precision physics at the $Z$ pole, where the statistical accuracy on various electroweak parameters is expected to be over an order of magnitude better than at LEP. This calls for a very tight control of the systematic error on the acceptance, with a definition of the acceptance boundaries at the level of a few $\mu\mathrm{m}$, and a very good $e - \gamma - \pi_0$ discrimination to identify $\tau$ leptons efficiently and measure their polarization. A layer of silicon microstrip detectors around the main tracker can provide the needed acceptance control for charged tracks, while also improving the tracking resolution. Similarly, the acceptance accuracy and improved identification efficiency of $\gamma$'s can be obtained with a preshower based on micro-pattern gas detectors (MPGD) located just outside the detector magnet, which serves as a radiator.

Particle flow calorimeters, like the ones proposed in the baseline concept, have been extensively studied. They are expected to deliver excellent energy measurement and unmatched particle identification capabilities but feature an extremely large number of readout channels and require significant data processing to obtain the optimal performance. The IDEA concept adopts an alternative calorimeter based on the dual readout technique [15], which has been extensively studied and demonstrated in over ten years of R&D by the DREAM/RD52 collaboration [16, 17]. With this technology the electromagnetic and hadronic calorimeters come in a single package that plays both functions and allows an excellent discrimination between hadronic and electromagnetic showers [18]. Since all the readout electronics is located in the back of the calorimeter, its cooling can be easily implemented.

Finally recent developments in micro-pattern gas detector technology, such as $\mu$Rwell [19], can significantly reduce the cost of large area tracking chambers to be used for tracking muons outside the calorimeter volume.

The structure of the IDEA detector is outlined in Figure 3.10, which also shows its overall dimensions. A key element of IDEA is a thin, ~30 cm, and low mass, $\sim 0.8\,X_0$, solenoid with a magnetic field of 2 Tesla. The low mass and thickness of the solenoid allows it to be located between the tracking volume and the calorimeter without a significant performance loss. The low-magnetic field is optimal, according to studies done for FCC-ee, as it minimizes the impact on emittance growth and allows for manageable fields in the compensating solenoids [20]. On the other hand, it puts stringents constraints on the tracker design required to achieve the necessary momentum resolution. IDEA has consequently adopted a large low-mass cylindrical drift chamber has its main tracker.

The innermost detector, surrounding the 1.5 cm radius beam pipe, is a silicon pixel detector for the precise determination of the impact parameter of charged particle tracks. Recent test beam results on the detectors planned for the ALICE inner tracker system (ITS) upgrade, based on the ALPIDE readout chip [21], indicate an excellent resolution, ~5 $\mu\text{m}$, and high efficiency at low power and dark noise rate [22]. This looks like a good starting point for the IDEA vertex detector and a similar approach is proposed for the CEPC baseline detector (see Section 4.1). The two detector concepts could then share the same pixel technology as well as profit from the electronic and mechanical work of the ALICE ITS.

Outside the vertex detector is a 4 m long cylindrical drift chamber starting from a radius of ~35 cm and extending until 2 m. The chamber can be made extremely light, with low mass wires and operation using 90% helium gas; less than 1% $X_0$ is considered feasible for 90° tracks. Additional features of this chamber, which is described in detail in Section 4.4, are a good spatial resolution, <100 $\mu\text{m}$, $dE/dx$ resolution at the 2% level and a maximum drift time of only 400 ns. A layer of silicon microstrip detectors surrounds the drift chamber in both barrel and forward/backward regions. Track momentum resolution of less then 0.5% for 100 GeV tracks is expected when vertex detector and silicon wrapper information is included in the track fit. It is worth noting that the design of this chamber is the evolution of work done over many years on two existing chambers, that of the KLOE detector [23] and that of the recent MEG experiment upgrade [24]; major R&D work was done also for the 4th concept detector at ILC [25] and then for the Mu2E tracker [26].

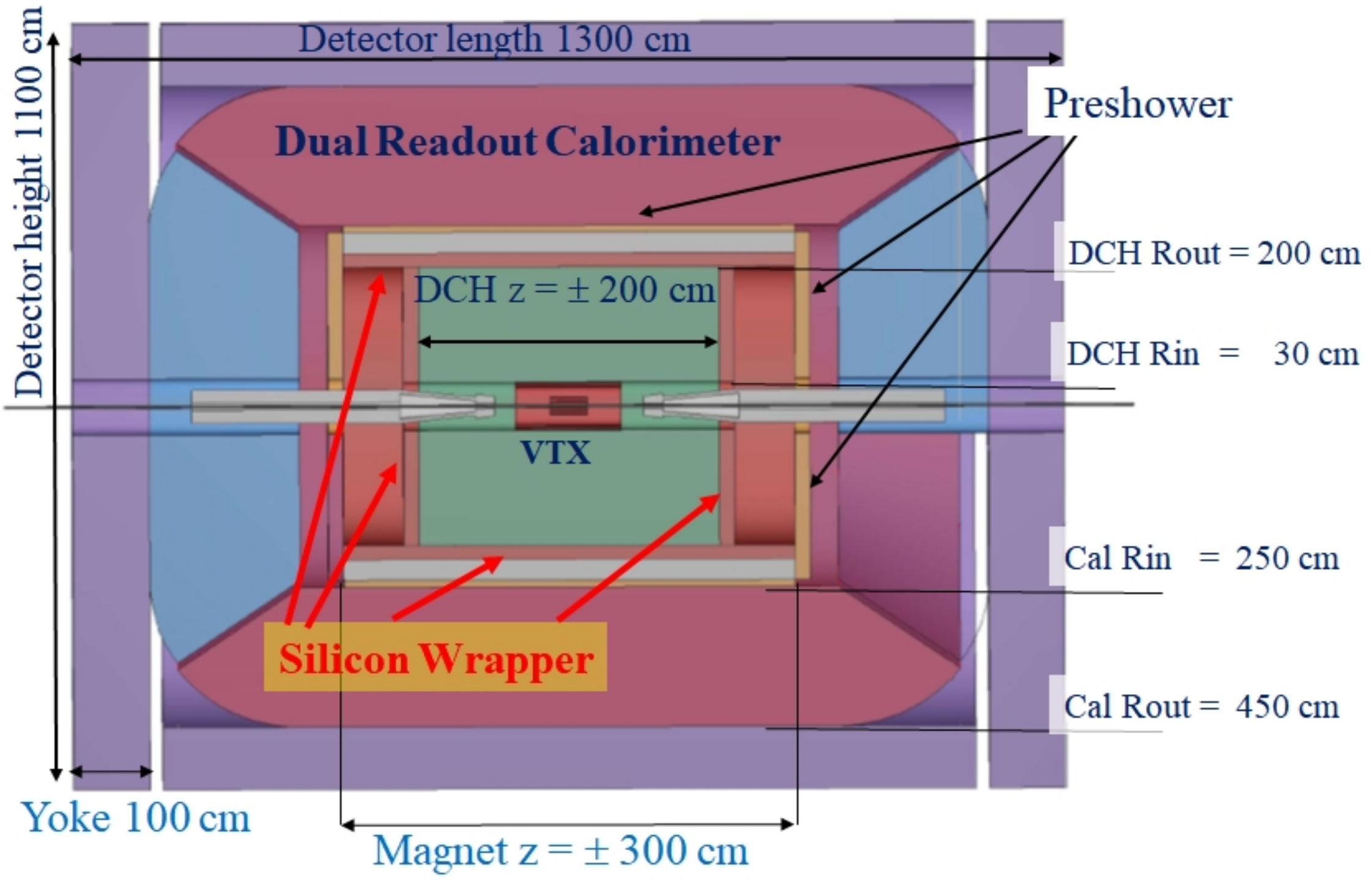

**Figure 3.10:** Schematic layout of the IDEA detector.

A preshower is located between the solenoid magnet and the calorimeter in the barrel region and between the drift chamber and the endcap calorimeter in the forward region. This detector consists of two passive material radiators each followed by a layer of MPGD detectors. In the barrel region the solenoidal magnet plays the role of the first radiator, while in all other cases the radiators are made of lead. The actual thickness of the radiators are still being optimized based on test beams currently in progress. In the extreme case of using a total of two radiation lengths about 75% of the $\pi^0$'s can be tagged by having both $\gamma$'s from their decay identified by the preshower. Additional $\pi^0$ identification power comes from the high granularity of the calorimeter.

A solenoidal magnet surrounds the tracking system. The currently planned dimensions are 6 m of length and 4.2 m inner diameter. The relatively low two Tesla field and the small dimensions have important implications on the overall magnet package thickness, that can be kept at the 30–40 cm level, and on the size of the flux return yoke, which scales linearly with the field and the square of the coil diameter. With the given dimensions a yoke thickness of less than 100 cm of iron is sufficient to completely contain the magnetic flux and provide adquate muon filtering and support for the muon chambers.

A dual readout fiber calorimeter (see Section 5.5) is located behind the second preshower layer. We assume a total calorimeter depth of 2 m, corresponding to approximately seven pion interaction lengths. The detector resolution is expected to be about $10.5\%/\sqrt{E}$ for electrons and $35\%/\sqrt{E}$ for isolated pions with negligible constant terms, as obtained from extrapolations from test beam data using GEANT4 without including the preshower. This detector has very good intrinsic discrimination between muons, electrons/photons and hadrons for isolated particles [18]. This discrimination power is further enhanced when the information of the preshower and the muon chambers is added, extending the separation power also into hadronic jets and making it suitable for the application of particle flow algorithms. The intrinsic high transverse granularity provides a good matching of showers to tracks and preshower signals.

The muon system consists of layers of muon chambers embedded in the magnet yoke. The area to be covered is substantial, several hundreds of square meters, requiring an inexpensive chamber technology. Recent developments in the industrialization of $\mu$Rwell based large area chambers, as planned for the CMS Phase II upgrade, are very promising (see Section 7.3).

CHAPTER 4

# TRACKING SYSTEM

The CEPC physics program demands a robust and high performance charged particle tracking system. Charged particles are used directly in physics analyses; they are input to determine primary and secondary vertices; and they are crucial input to particle flow calorimetry.

The tracking system has two major components. The vertex detector has excellent spatial resolution and is optimized for vertex reconstruction. The main tracker is optimized for tracking efficiency and resolution required for the CEPC physics program.

This Chapter introduces all tracking system options of the detector concepts discussed in this report. Section 4.1 describes the CEPC baseline vertex detector and the inner tracker. An outer tracking system, composed of a Time Projection Chamber (TPC), a silicon external tracker and a forward tracking detector, is discussed in Section 4.2. This system, together with the vertex detector and the inner tracker from Section 4.1, composes the tracking system of the baseline detector concept which is represented in Figure 4.1. Section 4.3 discusses in some detail the option of a full-silicon tracker that could substitute the tracking system of the CEPC baseline detector concept. Finally, in Section 4.4 a Drift Chamber Tracker is proposed as an option for the CEPC main outer tracker. This chamber, together with a layer of silicon microstrip detectors that surrounds it in both barrel and forward/backward regions, and the inner vertex detector, constitute the tracking system of the CEPC alternative detector concept.

## 4.1 VERTEX DETECTOR

The identification of heavy-flavor ($b$- and $c$-) quarks and $\tau$ leptons is essential for the CEPC physics program. It requires precise determination of the track parameters of charged particles in the vicinity of the Interaction Point (IP), permitting reconstruction of the displaced decay vertices of short-lived particles. This drives the need for a vertex detector with low material budget and high spatial resolution. The baseline design of the CEPC vertex detector is a cylindrical barrel with six silicon pixel sensor layers. It is optimized for the CEPC energy regime and utilizes modern sensors.

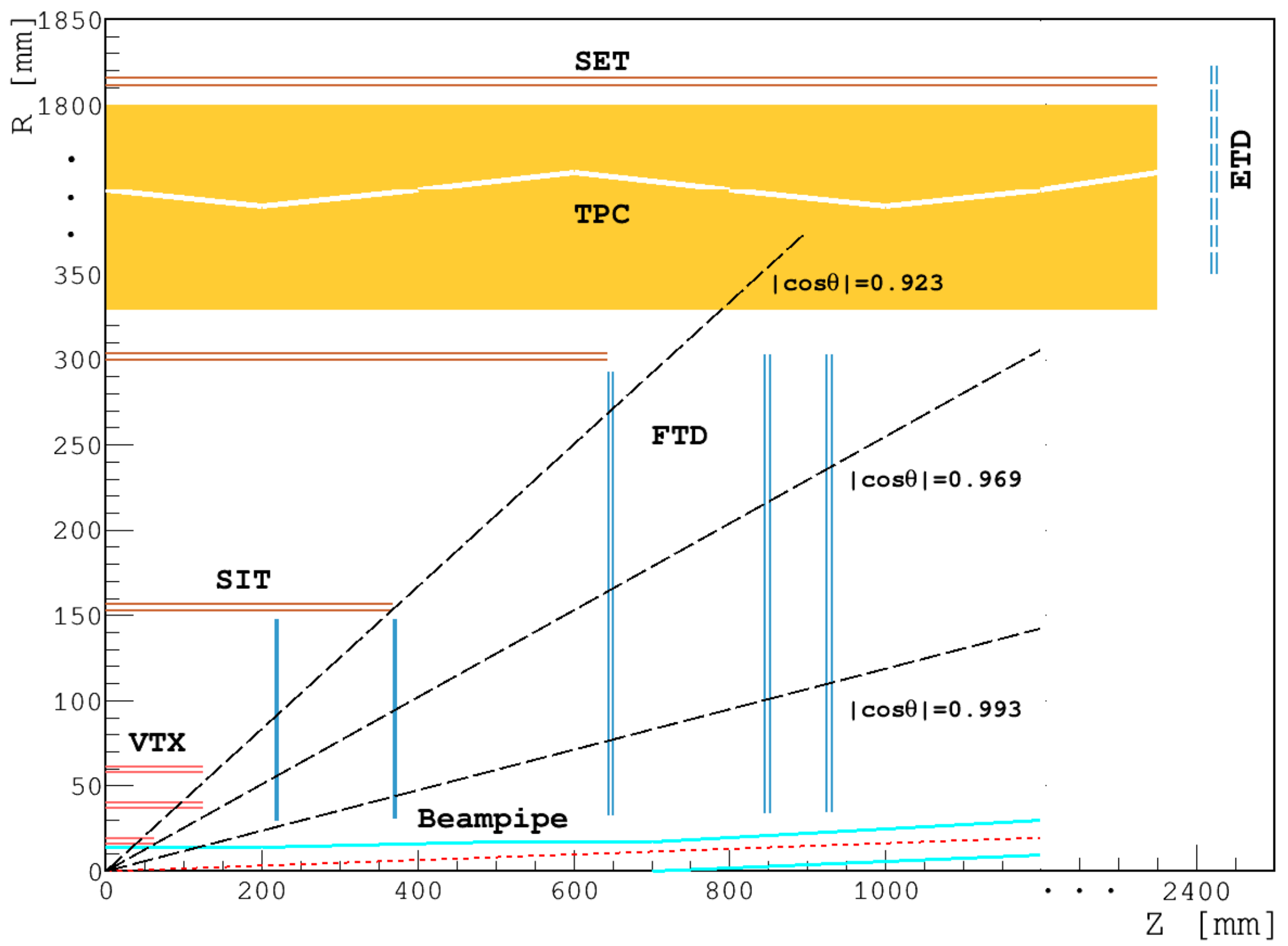

**Figure 4.1:** Preliminary layout of the tracking system of the CEPC baseline detector concept. The Time Projection Chamber (TPC) is embedded in a Silicon Tracker. Colored lines represent the positions of the silicon detector layers: red lines for the Vertex Detector (VTX) layers; orange lines for the Silicon Inner Tracker (SIT) and Silicon External Tracker (SET) components of the silicon tracker; gray-blue lines for the Forward Tracking Detector (FTD) and Endcap Tracking Detector (ETD) components of the silicon tracker. The cyan lines represent the beam pipe, and the dashed red line shows the beam line position with the beam crossing angle of 16.5 mrad. The ETD line is a dashed line because it is not currently in the full simulation. The radial dimension scale is broken above 350 mm for display convenience.

### 4.1.1 PERFORMANCE REQUIREMENTS AND DETECTOR CHALLENGES

As required for the precision physics program, the CEPC vertex detector is designed to achieve excellent impact parameter resolution, which in the $r\phi$ plane can be parametrized by:

$$\sigma_{r\phi} = a \oplus \frac{b}{p(\mathrm{GeV}) \sin^{3/2}\theta} \tag{4.1}$$

where $\sigma_{r\phi}$ denotes the impact parameter resolution, $p$ the track momentum, and $\theta$ the polar track angle. The first term describes the intrinsic resolution of the vertex detector in the absence of multiple scattering and is independent of the track parameters, while the second term reflects the effects of multiple scattering. The parameters $a = 5$ µm and $b = 10$ µm $\cdot$ GeV are taken as the design values for the CEPC vertex detector. The main physics performance goals can be achieved with a three concentric cylinders of double-layer pixelated vertex detector with the following characteristics:

- Single-point resolution of the first layer better than 3 µm;
- Material budget below 0.15% $X_0$ per layer;
- First layer located close to the beam pipe at a radius of 16 mm, with a material budget of 0.15% $X_0$ for the beam pipe;
- Detector occupancy not exceeding 1%.

The power consumption of the sensors and readout electronics should be kept below 50 mW/cm$^2$, if the detector is air cooled. The readout time of the pixel sensor needs to be shorter than 10 µs, to minimize event accumulation from consecutive bunch crossings. The radiation tolerance requirements, which are critical for the innermost detector layer, are driven by the beam-related backgrounds as described in Chapter 9.

### 4.1.2 BASELINE DESIGN

The baseline layout of the CEPC vertex detector consists of six concentric cylindrical layers of high spatial resolution silicon pixel sensors located between 16 and 60 mm in radii from the beam line (see Figure 4.2), providing six precise space-points for charged particles traversing the detector. The main mechanical structure is called a ladder. Each ladder supports sensors on both sides; thus, there are three sets of ladders for the vertex detector. The material budget of each detector layer amounts to ~0.15% $X_0$, including their corresponding supporting material. Extensive simulation studies (see Section 4.1.3) show that this configuration with the single-point resolutions listed in Table 4.1 achieves the required impact parameter resolution.

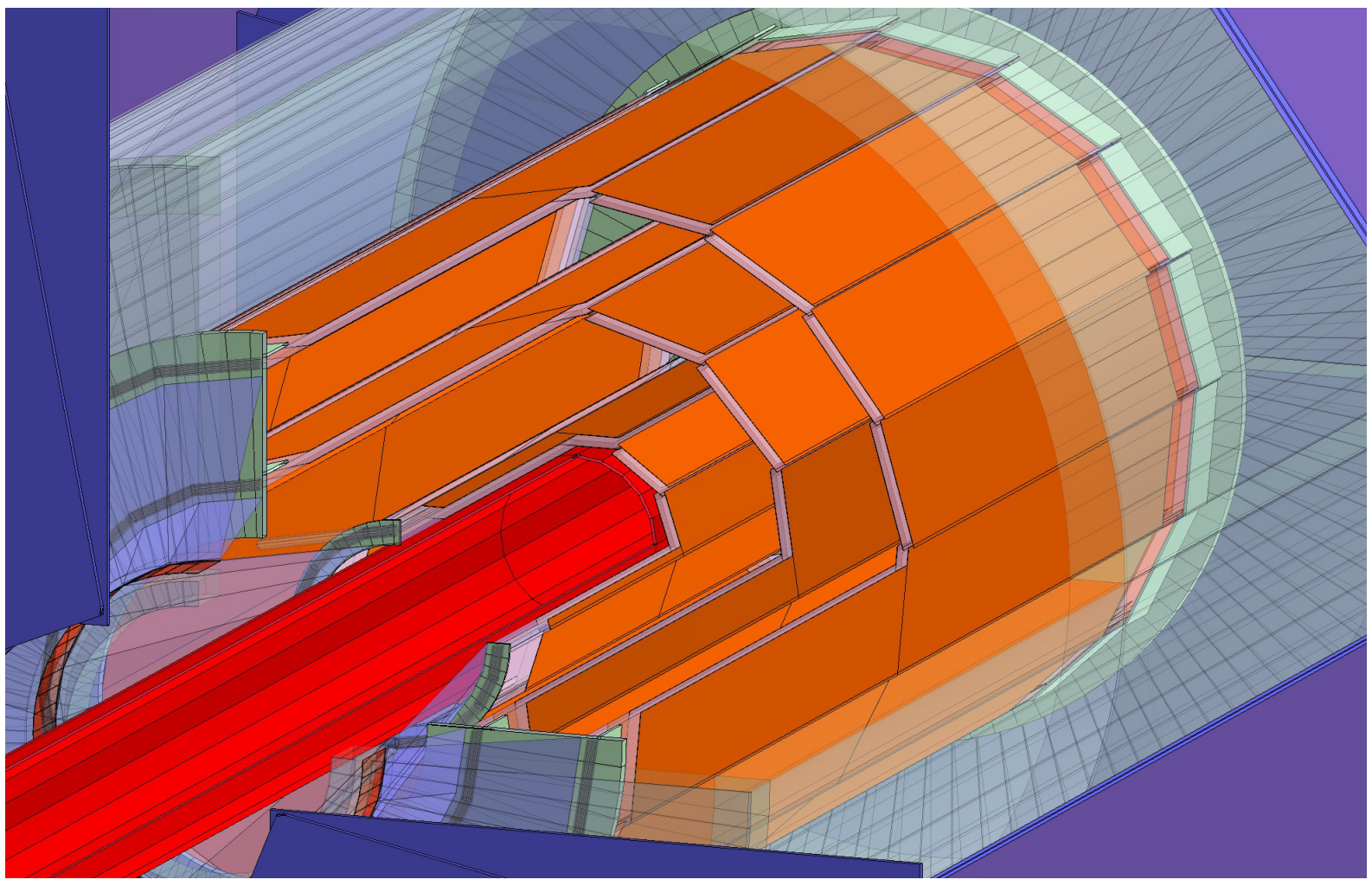

**Figure 4.2:** Schematic view of pixel detector. Two layers of silicon pixel sensors are mounted on both sides of each of three ladders to provide six space points. Only the silicon sensor sensitive region (in orange) is depicted. The vertex detector surrounds the beam pipe (red).

| | $R$ (mm) | $\|z\|$ (mm) | $\|\cos\theta\|$ | $\sigma$( μm) |
|---|---|---|---|---|
| Layer 1 | 16 | 62.5 | 0.97 | 2.8 |
| Layer 2 | 18 | 62.5 | 0.96 | 6 |
| Layer 3 | 37 | 125.0 | 0.96 | 4 |
| Layer 4 | 39 | 125.0 | 0.95 | 4 |
| Layer 5 | 58 | 125.0 | 0.91 | 4 |
| Layer 6 | 60 | 125.0 | 0.90 | 4 |

**Table 4.1:** The baseline design parameters of CEPC vertex detector including position and single-point resolution. The values of single-point resolution for layer 1 and layer 2 consider a double-sided ladder concept based on a high resolution sensor on one side, and a faster sensor on the other side to provide necessary time-stamp for tracking.

### 4.1.3 DETECTOR PERFORMANCE STUDIES

The identification of $b/c$-quark jets (called "flavor-tagging") is essential in physics analysis where signal events with $b/c$-quark jets in the final state have to be separated from one another and from light-quark jets. Flavor tagging requires the precise determination of the trajectory of charged tracks embedded in the jets. For CEPC operation at the center-of mass energy of 240 GeV, those tracks are often of low momentum, for which the multiple scattering effect dominates the tracking performance as illustrated by Eq. 4.1.

The CEPC vertex detector layout has been fully implemented in the GEANT4-based simulation framework, MOKKA [1]. In addition, the LiC Detector TOY fast simulation and reconstruction framework (LDT) [2] have been used for detector performance evaluation and layout optimization. Preliminary optimization studies have been done to evaluate the sensitivity of the flavor-tagging performance to the detector geometry and material budget, resulting in the chosen parameters. The detector simulations include the full tracker: vertex detector, silicon tracker and TPC, however, beam-induced background have not yet been included.

#### 4.1.3.1 PERFORMANCE OF THE BASELINE CONFIGURATIONS

The impact parameter resolution, following from the single-point resolutions provided in Table 4.1, is displayed in Figure 4.3 as a function of the particle momentum, showing that the ambitious impact parameter resolution is achievable.

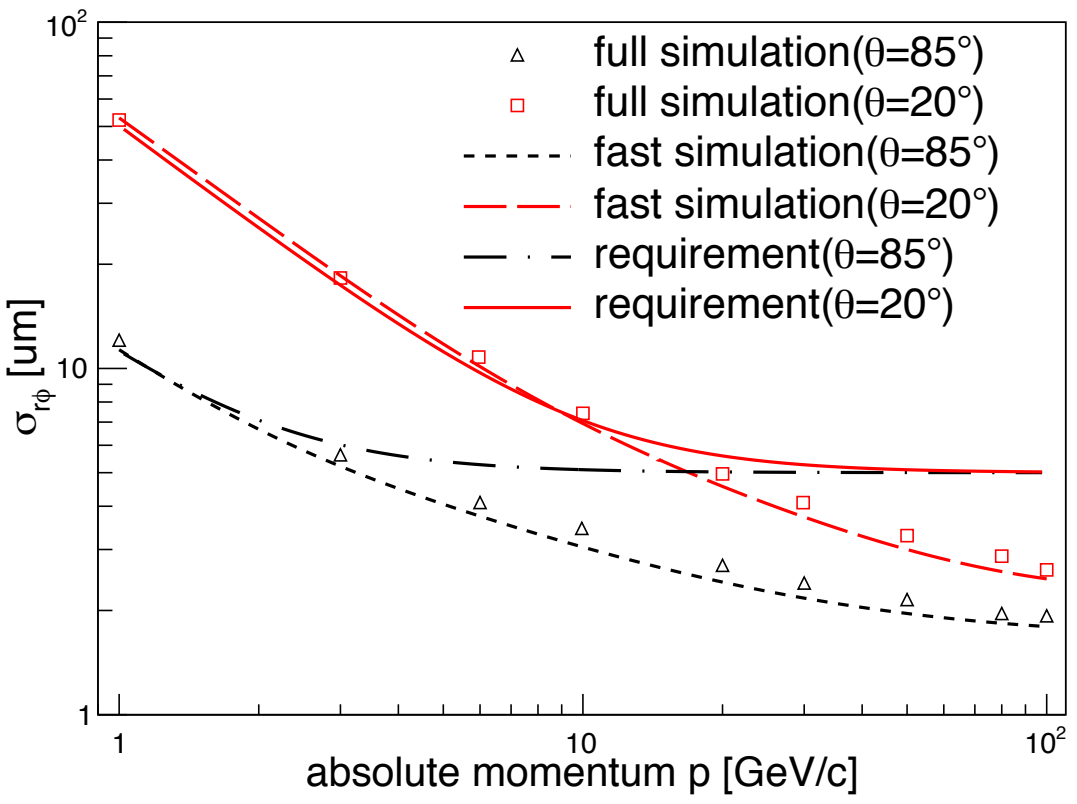

**Figure 4.3:** Transverse impact-parameter resolutions for single muon events as a function of momentum for two polar angles $20°$ and $85°$. The results are shown for both fast simulation and full simulation method.

#### 4.1.3.2 MATERIAL BUDGET

The baseline design includes very small material budget for the beam pipe as well as for the sensor layers and their support. To assess the sensitivity of the performance on the amount of material, the material budget of the beam pipe and the vertex detector layers has been varied. The resulting transverse impact-parameter resolutions for low-momentum tracks are shown in Figure 4.4. When increasing the material of all detector layers by a factor of two, the resolution degrades by approximately 20%.

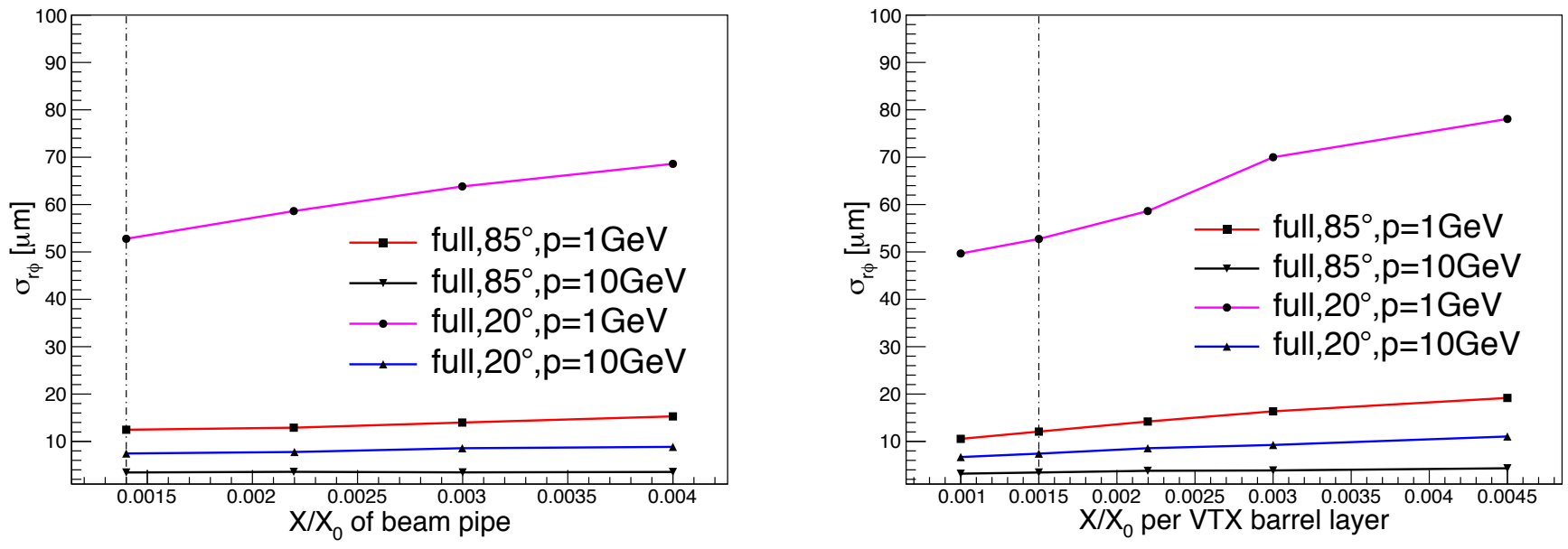

(a) (b)

**Figure 4.4:** Transverse impact-parameter resolution (a) as function of the amount of material in the beam pipe and, (b) in each vertex detector barrel layer, as obtained from full simulation. The results are shown for 1 GeV and 10 GeV muon tracks and for polar angles of $\theta = 20$ degrees and of $\theta = 85$ degrees. The material budget corresponding to the baseline configuration is indicated by dashed lines.

#### 4.1.3.3 DEPENDENCE ON SINGLE-POINT RESOLUTION

The dependence of the transverse impact-parameter resolution on the pixel size was studied by worsening the single-point resolution of the vertex detector layers by 50% w.r.t. the baseline values. The resulting impact parameter resolutions for high and low momentum tracks as functions of the polar angle $\theta$ are shown in Figure 4.5. The impact parameter resolution for track momenta of 100 GeV is found to degrade by approximately 50% in the barrel region, which is expected. They are better than the target value for the high-momentum limit of $a \sim 5$ µm in both cases, as expected from the corresponding single-point resolutions. For 1 GeV, where multiple-scattering effects dominate, the transverse impact-parameter resolution is only 10% worse. The target value for the multiple-scattering term of $b \sim 10$ µm $\cdot$ GeV is approximately reached in both cases. It should be noted, however, that the pixel size is also constrained by the background occupancies (see Section 4.1.4) and the ability to separate adjacent tracks in very dense jets in the presence of such backgrounds.

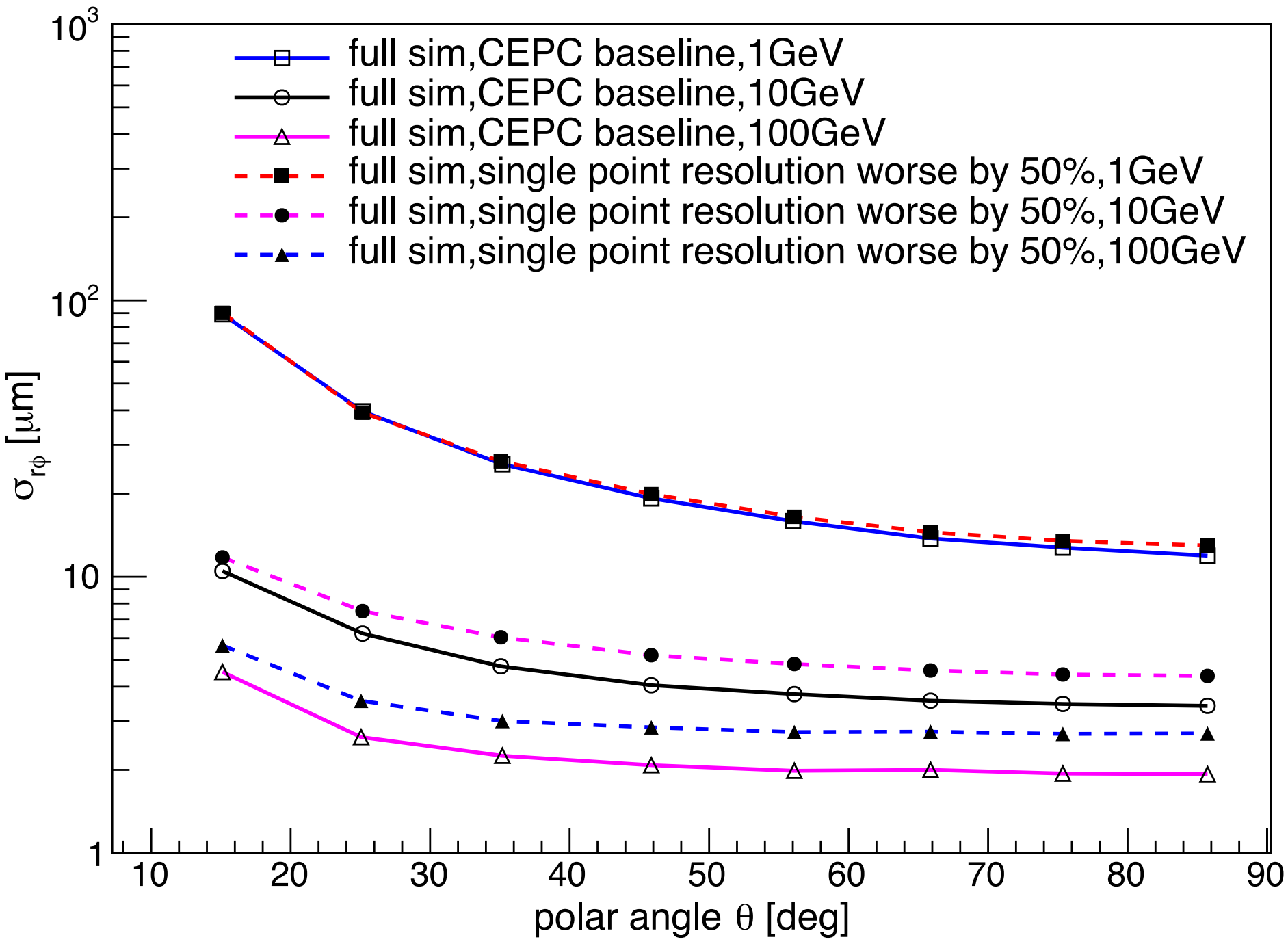

**Figure 4.5:** Transverse impact-parameter resolutions as function of the polar angle $\theta$ for different values of the single-point resolution of the CEPC vertex detector. Shown are the resolutions for 1 GeV, 10 GeV and 100 GeV tracks.

#### 4.1.3.4 DISTANCE TO BEAM LINE

The radius of the vertex detector innermost layer has a strong effect on the impact parameter resolution, and should be kept as small as possible. Figure 4.6 shows the transverse impact-parameter resolution at $\theta = 85°$= as a function of the momentum and for different radius of the innermost barrel vertex layer from the IP. The radius from the IP of the first ladder, that supports the first two vertex layers, was varied by $\pm 4$ mm relative to baseline geometry of the CEPC vertex detector ($R_{VTX1} = 16$ mm). For low momentum tracks, the transverse impact-parameter resolution is observed to be proportional to the radius, as expected from the parameter formula. When the radius of the innermost layer was consider to be 12 mm, the radius of the beam pipe was reduced to 10.5 mm. Such beam pipe radius size is smaller than the current baseline beam pipe, and it poses challenges regarding beam backgrounds and mechanical assembly that would require further studies to demonstrate its feasibility. We conclude that the current radius of the innermost layer is adequate to achieve the impact parameter resolution requirements.

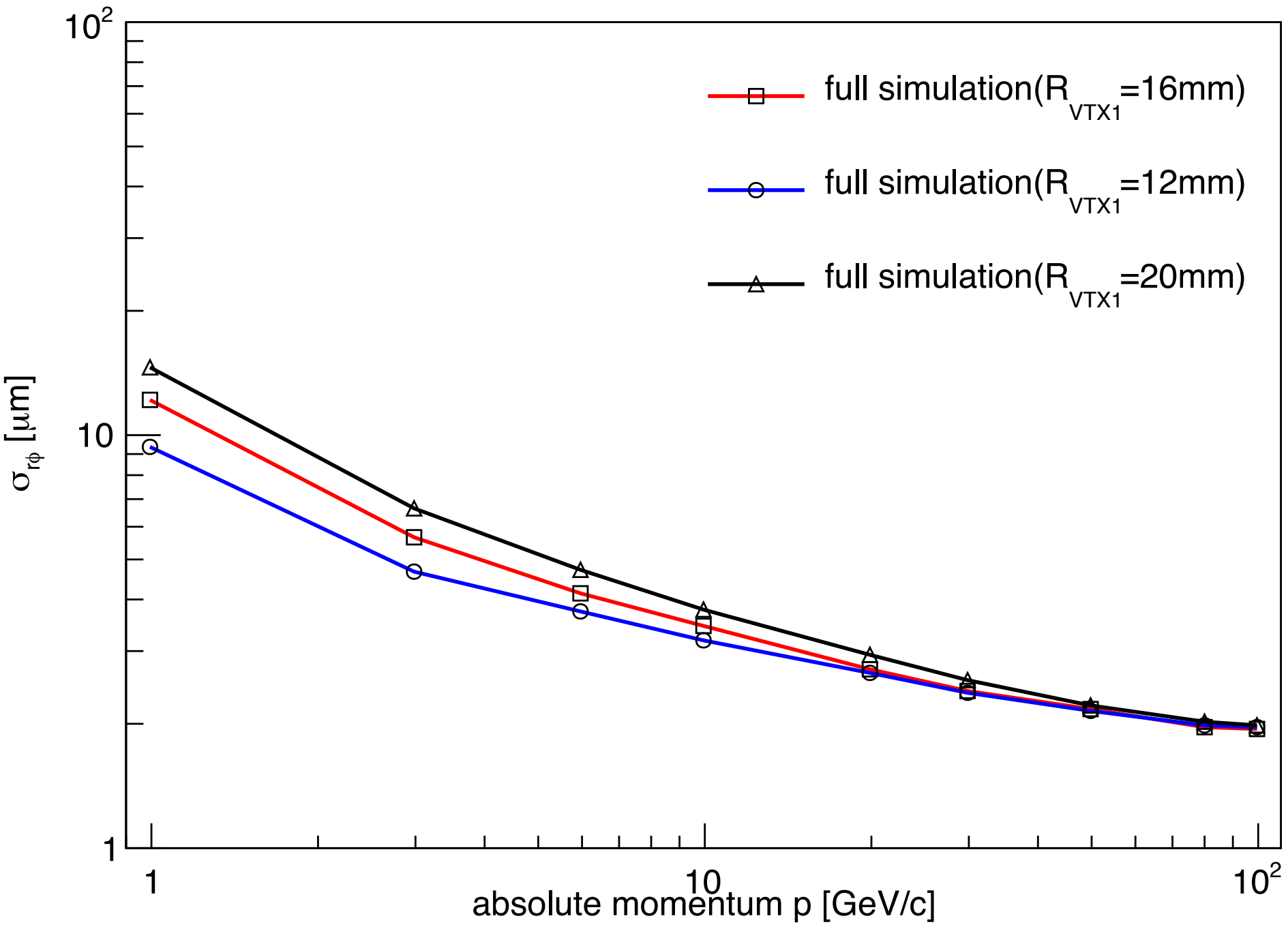

**Figure 4.6:** Transverse impact-parameter resolution at $\theta = 85$ degrees as function of the momentum for different values of innermost layer radius $R_{VTX1}$. The red curve indicates the baseline configuration of $R_{VTX1} = 16\,\mathrm{mm}$.

### 4.1.4 BEAM-INDUCED BACKGROUND IN THE VERTEX DETECTOR

Pair-production and off-energy beam particles are expected to be the dominating source of detector backgrounds originating from the interaction region. These processes have been studied with detailed MC simulation described in Chapter 9. For the first vertex detector layer, the maximum annual values of the Total Ionizing Dose (TID) and Non-Ionizing Energy Loss (NIEL) are estimated to be 3.4 MRad and $6.2 \times 10^{12}$ 1 MeV $\mathrm{n_{eq}/cm^2}$ per year, respectively, with a safety factor of 10 included (see Table 9.4 in Chapter 9). This happens when the machine is operating at the $Z$-pole energy, and imposes radiation tolerance requirements on the silicon pixel sensor and associated readout electronics.

The beam-induced background will have impacts on vertex detector occupancy, which is critical for the innermost detector layer. Table 4.2 shows the expected hit density and occupancies of the first vertex detector layer at different machine operation energies. The result of occupancies depends on assumptions of detector readout time and average cluster size. Here we assume 10 μs of readout time for the silicon pixel sensor and an average cluster size of 9 pixels per hit, where a pixel is taken to be $16{\times}16\ \mu\text{m}^2$. The resulting maximal occupancy at each machine operation mode is below $1\%$.

| Operation mode | H (240) | W(160) | Z (91) |
|---|---|---|---|
| Hit density ($\text{hits} \cdot \text{cm}^{-2} \cdot \text{BX}^{-1}$) | 2.4 | 2.3 | 0.25 |
| Bunching spacing ( µs) | 0.68 | 0.21 | 0.025 |
| Occupancy (%) | 0.08 | 0.25 | 0.23 |

**Table 4.2:** Occupancies of the first vertex detector layer at different machine operation energies: 240 GeV for $ZH$ production, 160 GeV near $W$-pair threshold and 91 GeV for $Z$-pole.

### 4.1.5 SENSOR TECHNOLOGY OPTIONS

Significant progress has been made since the first silicon pixel detector was used in high-energy physics experiments, and considerable R&D efforts have taken place to develop pixel sensors for vertex tracking at future high-energy physics experiments [3], driven by track density, single-point resolution and radiation level.

As outlined in Section 4.1.1, the detector challenges for the CEPC include high impact-parameter resolution, low material budget, low occupancy and sufficient radiation tolerance (mild comparing to the LHC but not necessarily easy to achieve). To fulfill these requirements at system level, sensor technologies which achieve fine pitch, low power and fast readout must be selected. These considerations present unique challenges to the CEPC vertex detector. CEPC has a bunch spacing of 0.68 µs for the Higgs factory oepration and 25 ns for the $Z$ factory operation, and power pulsing cannot be utilized to reduce average power as is planned at the ILC. Experiments such as the STAR [4], BELLE II [5] and ALICE upgrade [6] readout continuously as the CEPC. They have, however, less stringent requirements in terms of impact parameter resolution and material budget.

The monolithic pixel sensor has the potential to satisfy the low-material and high-resolution requirements of the CEPC vertex detector. This technology has been developing fast. The first generation MAPS-based vertex detector for the STAR HFT upgrade [4, 7] just completed 3-year physics run successfully, while the new generation HR Complementary Metal-Oxide-Semiconductor (CMOS) Pixel Sensor (CPS) for ALICE-ITS upgrade [6] is in mass production. In the previous 0.35 µm double-well process, only N-MOS transistors can be used in the pixel design. This constraint is removed in the new 0.18 $\mu\text{m}$ quadruple-well process [8]. Both N- and P-MOS transistors could be used in the pixel design. Combining with the smaller feature size, it becomes a very appealing technology. A good start point for the CEPC vertex would be the ALPIDE design [9], which is developed for the aforementioned ALICE-ITS upgrade and has achieved performance very close to the requirements of the CEPC. Further R&Ds are needed to shrink the pixel pitch to 16 $\mu\text{m}$ (binary readout) in order to accomplish the required 2.8 $\mu\text{m}$ single-point resolution. Another monolithic option is the Silicon On Insulator (SOI) pixel sensor. After more than 10 years of evolution, SOI has entered a new stage of maturity. Fundamental issues, including the transistor shielding [10] and the TID tolerance [11], have been addressed and wafer thinning [12] has been demonstrated. In the meanwhile, R&Ds for the ILC and CLIC [13, 14] are exploring time stamping and analog readout schemes. The SOI has a unique feature of a fully-depleted substrate as the active silicon. And its 0.2 μm CMOS process provides the necessary density of transistors as the 0.18 μm CMOS in HR CMOS does. Therefore it is envisaged that the readout design for the CEPC vertex may be adapted for both processes and to exploit each ones potentials.

Depleted P-channel Field Effect Transistor (DEPFET) [15] is referred to as semi-monolithic because the first amplification stage can be integrated into the pixel combined with subsequent processing circuit in separate readout ASICs. The BELLE II is anticipating its full detector operation with a DEPFET-based vertex detector [5] installed at the end of 2018. It is very helpful to have the readout ASICs, as the major heat sources, located outside the detector acceptance area, while keeping the sensors exceptionally low power and low material. The challenge is to periodically sample the modulated current over a large pixel array within required intervals, 20 μs/frame or even less.

Hybrid pixel has been used at hadron colliders for the past decades, and now CLIC R&D is pushing for 50 μm thinned sensors, bump-bonded on 25 μm pitch to 50 μm thinned ASICs [16]. The hybrid approach evolves constantly and profits from industrial technology developments. Apart from the Very Deep Sub-Micro (VDSM) ASIC technology that enables complex functionalities and superior performance, a close watch on industrial developments of the vertical and lateral inter-connection technologies will also be very helpful to meet the material budget.

### 4.1.6 MECHANICS AND INTEGRATION

The design of the vertex detector is conceived as a barrel structure with three concentric cylinders of double-sided layers. Each double-sided layer is equipped with pixel sensors on both sides, and has a common support frame. In the azimuthal direction, each layer is segmented in elements called ladders. The ladder, which extends over the whole length of the layer, is the basic building block of the detector. It contains all structural and functional components, such as chips, flex cable, support frame and cold plate if it is necessary. Pixel chips in a row are connected to flex cable by wire bonding or other bonding techniques, and then glued to the support frame, which is composed of low-$Z$ materials, such as carbon fiber and silicon carbide, providing stable mechanical support. The other side of the support frame is equipped with another layer of pixel sensors.

The design of the ladders should take into account the specifications of the vertex detector. In order to reduce a small multiple Coulomb scattering contribution to the charged-track vertex resolution and control deformations from gravity and cooling forces for the sensor position stability, the ladder mechanical support must fulfill stringent requirements in terms of minimum material budget and highest stiffness. Ladder designs similar to the STAR pixel detector, the ALICE ITS, the BELLE II PXD, and the ILD double-sided ladder are under consideration.

The ladder mechanical support is inherently linked to the layout of the cooling system that will be adopted to remove the heat dissipated by the pixel sensors since the cooling system is integrated in the mechanical structure. The cooling system of the CEPC vertex detector must balance the conflicting demands of efficient heat dissipation with a minimal material budget. Therefore a suitable, high thermal conductivity and low material budget, cold plate coupled with pixel sensors should be implemented in the ladder design. There are two main types of cooling methods in particle physics experiments, air cooling and active cooling. Table 4.3 gives a list of cooling methods and the corresponding material of each layer of the aforementioned experiments. The upgrade of ALICE ITS [6] adopts water cooling with respect to a chips power dissipation value of 300 mW/cm$^2$. Polyimide cooling pipes fully filled with water are embedded in the cold plate. STAR-PXL [17] uses air cooling according to its chips power consumption of 170 mW/cm$^2$. For ILD [18] vertex system, two different cooling options are considered, depending on the sensor technology. The sensors and SWITCHER chips of BELLE II PXD [19] require air cooling, while active cooling will be used for readout chips on each end of the detector, which is out of the sensitive region of the detector. For the CEPC vertex detector, the suitable cooling method will be determined according to the sensor option and the power consumption.

<table>
<tr><th>Vertex detector</th><th>Power dissipation</th><th>Cooling method</th><th>Material budget requirement/layer</th></tr>
<tr><td>Alice ITS</td><td>300 mW/cm$^2$</td><td>water</td><td>0.3%</td></tr>
<tr><td>STAR PXL</td><td>170 mW/cm$^2$</td><td>air</td><td>0.39%</td></tr>
<tr><td rowspan="2">ILD vertex</td><td>< 120 mW/cm$^2$ (CPS and DEPFET)</td><td>air or $N_2$</td><td rowspan="2">0.15%</td></tr>
<tr><td>35 W inside cryostat (FPCCD)</td><td>two-phase $CO_2$</td></tr>
<tr><td rowspan="2">BELLE-II PXD</td><td>20 W for sensor and SWITCHER</td><td>Air</td><td rowspan="2">0.2%</td></tr>
<tr><td>180 W on each end</td><td>$CO_2$</td></tr>
</table>

**Table 4.3:** Cooling methods for several vertex detector designs. The chip power dissipation, coolant type and corresponding material budget requirement per sensor layer are indicated. The active $CO_2$ cooling adds additional material in the forward region, outside the sensitive area. For the ILD FPCCD option, this additional material budget is $0.3\%\,X_0$ averaged over the end-plate region, while for the BELLE-II PXD, it is $\sim 0.1 - 0.2\%\,X_0$ per layer.

Simulation and module prototype studies will be carried out to find suitable designs that can meet requirements of stability, cooling and the performance of the vertex detector. For the design of the whole mechanical structure of the vertex detector, some criteria must be taken into account. Firstly, minimum material has to be used in the sensitive region to reduce multiple Coulomb scattering. Secondly, to ensure high accuracy in the relative position of the detector sensors and provide an accurate position of the detector with respect to the central tracker of TPC and the beam pipe, a mechanical connector or locating pin at each end of the ladder should be considered to allow the fixation and alignment of the ladder itself on the end rings. Thirdly, the cooling system should be arranged reasonably to ensure stable heat dissipation. Lastly, to reduce the dead region caused by the boundary of each ladder, neighboring ladders should be partially superimposed.

In addition, the main mechanical support structures of the vertex should also meet the requirements of the integration with the other detectors, such as time-projection chamber (TPC) and forward tracking disks.

### 4.1.7 CRITICAL R&D

The inner most layers have to fulfill the most demanding requirements imposed by the physics program. In addition, the system is bounded by stringent running constraints. The technology options in Section 4.1.5 are able to meet each individual requirement, including single-point resolution, low material budget, fast readout, low power consumption and radiation tolerance, but R&D is needed to select the specific design which can achieve the combination of all these criteria. Due to the limited manpower and availability of process, presently R&D efforts have been put into CMOS and SOI pixel sensor development to address the challenges concerning single-point resolution and low power consumption. Further developments are foreseen to follow in the future, including enhancement of density, radiation hardness and ultra-light module assembly.

The current R&D activities have access to two advanced processes. The TowerJazz 0.18 μm quadruple-well process enables the full CMOS pixel circuit, while Lapis 0.2 μm double-SOI process has properly solved the crosstalk between sensor and digital part, and improved TID tolerance significantly.

In order to exploit the potential of these new developments, two design teams have started chip designs using HR CMOS and SOI technologies, respectively. Two designs have been submitted to the TowerJazz foundry. The first one uses simple three transistor (3T) analog amplification circuit to carry out the optimization of sensing diode and evaluate the influence of radiation damage [20]. The second one implements a well-proved rolling shutter readout as well as an innovative data-driven readout [21, 22]. Another two designs that adopt the SOI technology have also been submitted [23]. With the amplifier and discriminator integrated into each pixel, the pixel size has been shrunk to 16 μm pitch. The chip has been thinned to 75 μm successfully and an infrared laser test has shown that a single-point resolution of 2.8 μm is achievable with that pitch [12], but it needs to be certified further at the beam test. All the designs for current R&D are in line with the same principle of in-pixel discrimination even though each one has its own implementation. An in-pixel discriminator can reduce analog current therefore lead to reduced power consumption.

Enhancements of the TowerJazz 0.18 μm process or Lapis 0.2 μm process are possible by migrating to a smaller feature size, 0.13 μm for example, or combining with a micro-bump 3D integration process. The latter is able to attach a second layer of pixel circuit on top of the existing layer of the sensing diode and front-end circuit. The upper tier can be the fully digital part that implements data-driven readout architecture, while the lower tier can be HR CMOS or SOI pixel matrix. A promising result has been demonstrated by the successful formation of 2.5 μm Au cone bump with NpD (Nano-particle deposition) technique [24]. However, the throughput needs further improvement and the thinning of sensors has to be compatible with micro-bump 3D integration.

The TowerJazz process is expected to be sufficiently radiation hard for the expected TID. An N-type plain implant has recently been added to improve the charge collection efficiency [25], which therefore will benefit the non-ionization radiation damage. In terms of the SOI process, the weak point is the BOX layer of $SiO_2$. Although the TID tolerance of the SOI process has been improved dramatically by the introduction of Double-SOI and the optimization of transistor doping recipe (LDD, lightly doped drain) [11], SOI needs carefully study on the irradiation of large scale chips and of low power designs.

Sensor thinning and ultra-low material construction of modules are subject to the constraint of 0.15% $X_0$/layer. HR CMOS wafer thinned to 50 μm is routine in semiconductor industry nowadays. SOI wafers thinned to 75 μm with backside implant have also been demonstrated by current R&D. However, low material detector modules need to integrate mechanical support, power and signal connections, and have sufficient stiffness to avoid vibration.

A pixel detector prototype will be built with full-size pixel sensors to develop and test some of these critical aspects, including the mechanical design of low-mass support structures, cooling, fast readout and radiation tolerance.

### 4.1.8 SUMMARY

The basic concept of the CEPC Vertex detector, including the detector layout, the material budget per layer, and the pixel sensors specifications required by the impact parameter resolution are implemented in the baseline detector simulation. This is an essential requirement for the detailed mechanical design. Small pixel sensor prototypes that can satisfy some of the CEPC requirements have already been produced.

It will be crucial to continue the pixel sensor R&D program and develop pixel sensors with radiation tolerance, lower power consumption and fast readout electronics because of continuous colliding mode and strong beam-related background. Detailed designs for low-mass mechanical supports, cooling, cabling, and power conversion are also necessary. Most of these issues will be addressed by R&D for the CEPC and by exploring synergies with experiments which have similar requirements.

## 4.2 TIME PROJECTION CHAMBER AND SILICON TRACKER

The TPC is the default option of the outer tracker of the CEPC baseline detector concept. The high density of space points provides unparalleled pattern recognition capability. The TPC is complemented by an envelope of silicon detectors to improve the track momentum resolution. The silicon detectors are also useful for monitoring possible field distortions in the TPC and for alignment.

### 4.2.1 TIME PROJECTION CHAMBER

Time Projection Chambers have been extensively studied and used in different fields, especially in particle physics experiments such as ALEPH [26], DELPHI [27], STAR [28] and ALICE [29]. Furthermore, the LC-TPC collaboration [30] has made extensive studies, from which this report benefits, of a large TPC for a linear $e^+e^-$ collider detector. The technology directly provides three-dimensional space points; the gaseous detector volume gives a low material budget; and the high density of such space points enables excellent pattern recognition capability. However, care must be taken to address space charge distortion resulting from the accumulation of positive ions in the drift volume [31]. This issue is especially important in high rate conditions.

There have been extensive R&D on readout modules to optimize position resolution and to control ion backflow. These studies will continue for the next few years in order to understand and resolve several critical technology challenges.

#### 4.2.1.1 CEPC TIME PROJECTION CHAMBER

The CEPC TPC consists of a field cage, which is made with advanced composite materials, and two readout end-plates that are self-contained including the gas amplification, readout electronics, supply voltage, and cooling. The TPC has a cylindrical drift volume with an inner radius of 0.3 m an outer radius of 1.8 m, and a full length of 4.7 m. The central cathode plane is held at a potential of 50 kV, and the two anodes at the two end-plates are at ground potential. The cylindrical walls of the volume form the field cage, which ensures a highly homogeneous electrical field of 300 V/cm between the electrodes. The drift volume is filled with $Ar/CF_4/iC_4H_{10}$ in the ratio of 95%/3%/2%. Ionization electrons released by charged particle tracks drift along the electric field to the anodes where they are amplified in an electron avalanche and read out using a Micro-Pattern Gas Detector (MPGD).

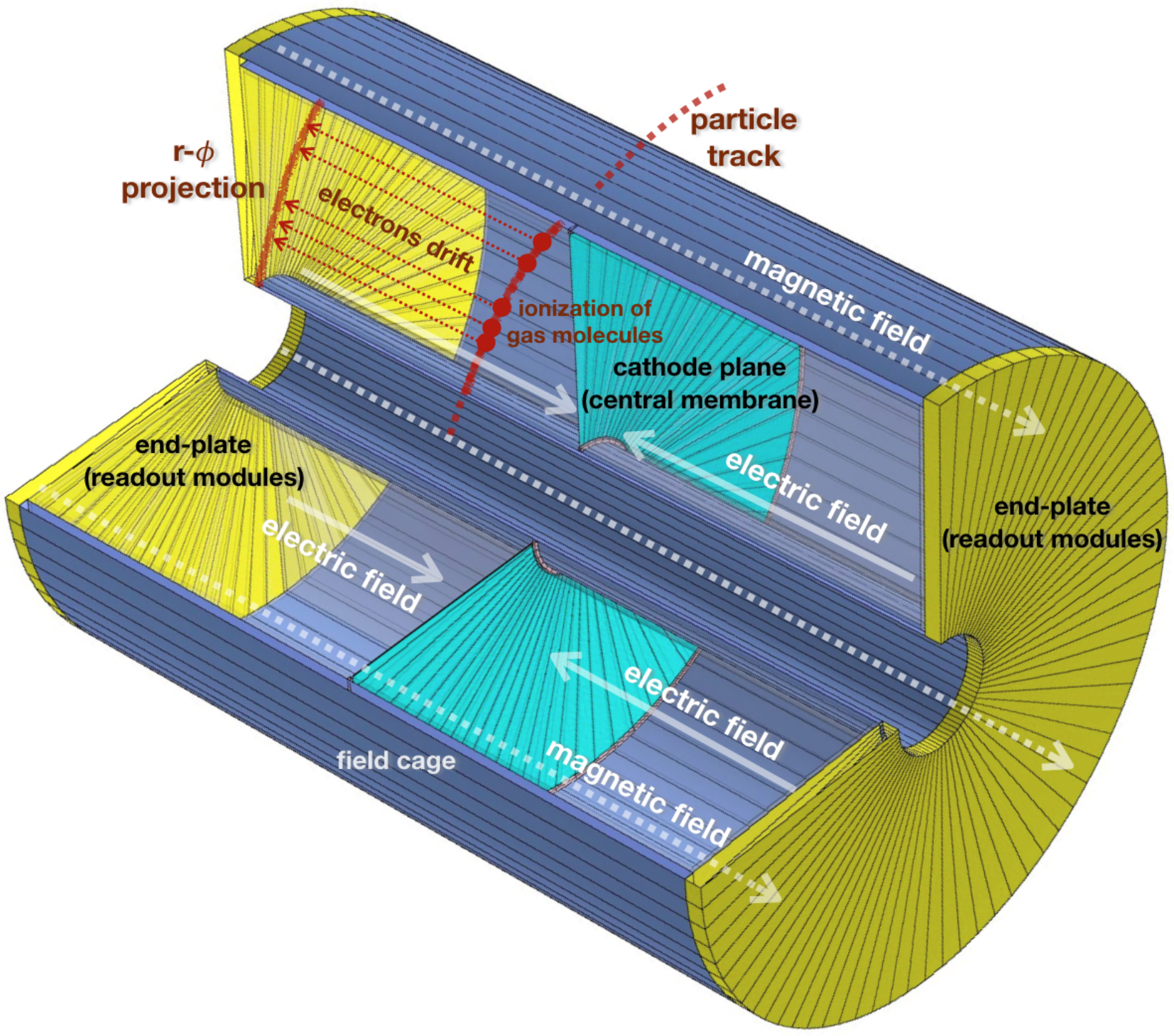

**Figure 4.7:** Sketch of the TPC detector. The TPC is a cylindrical gas detector with an axial electric field formed between the end-plates (yellow) and a central cathode plane/membrane (light blue). The cylindrical walls of the volume form the electric field cage (dark blue). Gas ionization electrons due to charged particles drift to the end-plates where they are collected by readout modules (yellow).

The CEPC TPC will be operated at the atmospheric pressure resulting in a material budget of less than $1\% X_0$ in the central region. The 3-Tesla solenoidal magnetic field suppresses transverse diffusion and improves position resolution. It also curls up low-momentum tracks resulting in higher occupancy near the beam line. The readout modules are attached to the end-plates from the inside to minimize the dead area between adjacent readout modules. Thus, a particular mounting technique is required to enable rotation and tilting of the readout modules during the installation.

The chamber's cylindrical inner and outer walls serve multiple functions. They hold the field forming strips, which are attached to a divider chain of non-magnetic resistors. Since the central cathode will be held at approximately 50 kV, the walls must withstand this enormous potential. The field cage will be designed to maintain the electric field uniform over the whole active TPC volume. Advanced composite material will be used for the cylindrical walls because of its low mass.

The MPGD detector on each end-plate is divided into many independent readout modules to facilitate construction and maintenance. The modules are mounted closely together on the end-plate to provide nearly full coverage. Power cables, electronic connectors, cooling pipes, PCB boards and support brackets wall are also mounted on the end-plate. The end-plate needs to constructed from a lightweight material in order to minimize the amount of material in the forward region but should also be sufficiently rigid to maintain stable positioning of the detector modules with a position accuracy better than 50 $\mu m$. The endcap structure has a thickness of $8\% X_0$, $7\%$ of which originate from the material for the readout planes, front-end electronics and cooling. Adding power cables and connectors, the total thickness increases from $8\% X_0$ up to $10\% X_0$.

The CEPC TPC provides 220 space point measurements per track with a single-point resolution of 100 $\mu m$ in $r - \phi$. In addition to position information, the TPC measures the ionization energy loss ($dE/dx$) on each readout pad. This can be combined with the measurement of momentum in the magnetic field to provide particle identification.

### 4.2.1.2 BASELINE DESIGN AND TECHNOLOGY CHALLENGES

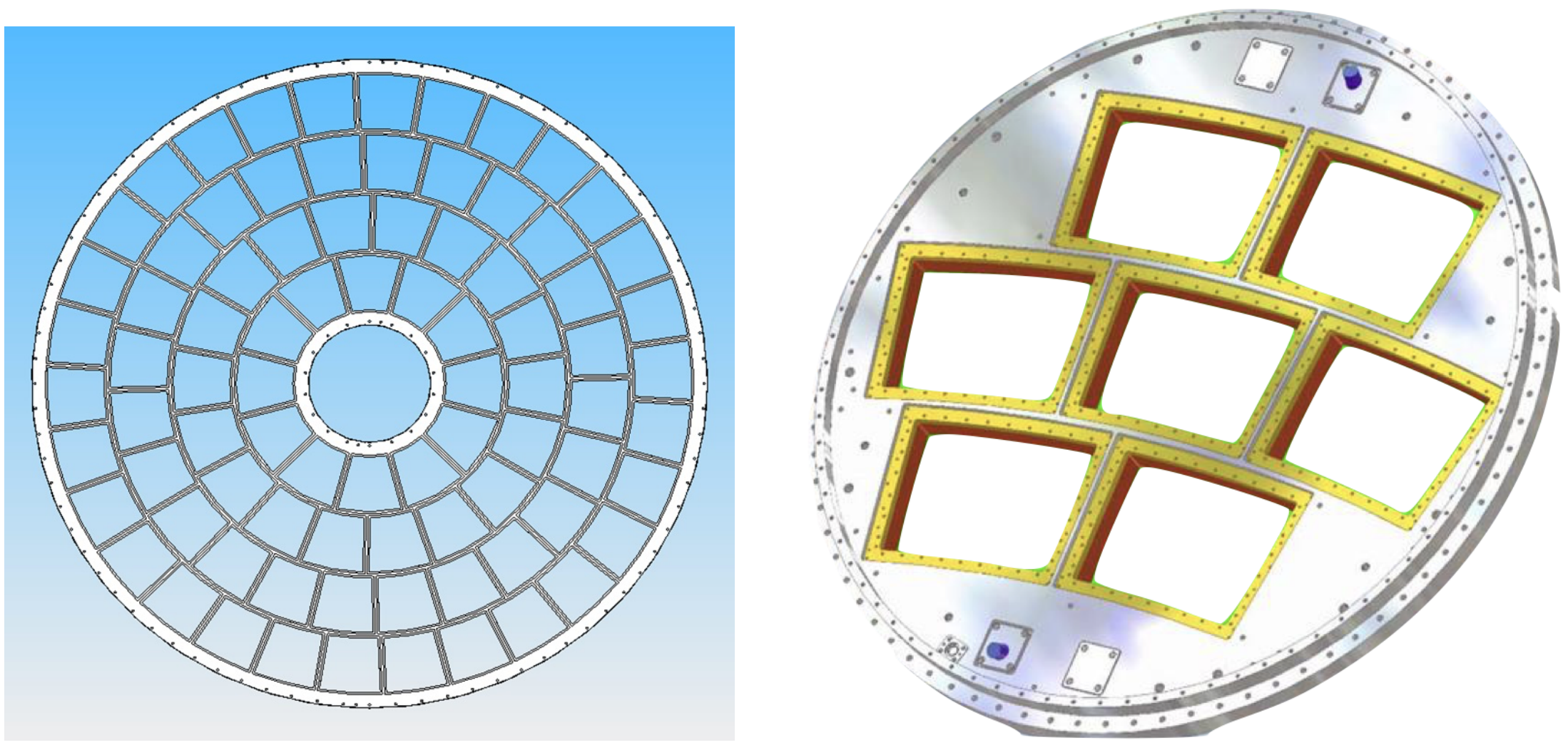

(a) (b)

**Figure 4.8:** Mechanical support (TPC endwall) for mounting the readout modules. Each opening corresponds to a readout module of roughly 170 mm×210 mm in size. (a) Full size design of TPC endwall; (b) Small prototype of the endwall showing details of the openings for the module insertion. The readout modules will be inserted, and installed on the inside of the endwall to minimize dead space [32].

The readout structure is designed to be modular to facilitate construction and maintenance. Each module will consist of a gas amplification system, a readout plane and the associated front-end electronics. An MPGD-based gas amplification system will be necessary to achieve the required performance, and the charge from the amplification system will be collected on the readout board. The readout module will also have to provide all necessary power and cooling. Each module will be approximately 170 mm in width and 200 mm in height [32].

Figure 4.8 shows the design of the mechanical support for the mounting of the readout modules on the inside of the TPC endwall developed by the LC-TPC collaboration. Figure 4.8(b) shows the details of the mounting openings from a small endwall prototype. Readout modules are inserted at an angle through openings in the endwall (white in Figure 4.8(b)) into the drift volume. They are then rotated to the proper orientation, pulled back against the mounting frame (yellow in the figure), and bolted into place from the outside. This approach allows the active areas of adjacent readout modules to be very close to one another and therefore minimizes dead space between them.

#### Gas amplification detector module

The required physics performance can be achieved with amplification technologies with gain in the range of $10^3 - 10^4$ combined with a spatial granularity of approximately 1 mm$^2$. Gas Electron Multiplier (GEM) [33] and Micro-Mesh Gaseous Structure (MicroMegas) detector [34] of the MPGD family of detectors [35] are both viable technologies for the large-area application in the CEPC TPC readout module. They can generate the very high fields necessary for gas amplification with modest voltages (300–400 V) across 50–100 μm structures. In the case of GEM, two or three will be stacked together to achieve sufficient charge amplification while MicroMegas have enough amplification in a single stage. Micro-pattern gaseous devices such as GEM and MicroMegas provide:

1. High gain;
2. High rate capability: MPGDs provide a rate capability over $10^5$ Hz/mm$^2$ without discharges that can damage electronics;
3. Intrinsic ion backflow suppression: Most of the ions produced in the amplification region will be neutralized on the mesh or GEM foil and do not go back to the drift volume;
4. A direct electron signal, which gives good time resolution ($< 100$ ps) and spatial resolution (100 μm).

The baseline gas amplification and readout module for the CEPC TPC is a novel configuration detector module: a combination of GEM and a MicroMegas. This detector will be called a GEM-MM detector for short throughout this report. This device aims to deliver gains around 5000 with ion backflow at 0.1% level.

#### Optimization readout strip size

Signals are read out in two orthogonal sets of strips. The readout strips in the X direction are 193 μm wide with a pitch of 752 μm. The readout strips in the Y direction are 356 μm wide with a pitch of 457 μm. The difference in strip widths is to improve signal sharing between adjacent strips. Strips are approximately 6 mm long, and each strip is connected to one electronic channel to process the signal. Each readout unit contains 267 channels for the X direction and 437 channels for the Y direction.

Figure 4.9 is a typical layout of the X and Y readout strips, and two representative electron clusters are also superimposed. Each X-Y strip crossing has an area about 1 mm$^2$. Thus each cluster spans a large number of such crossings, allowing the use of the center-of-gravity method to reach a position resolution finer than the strip pitches.

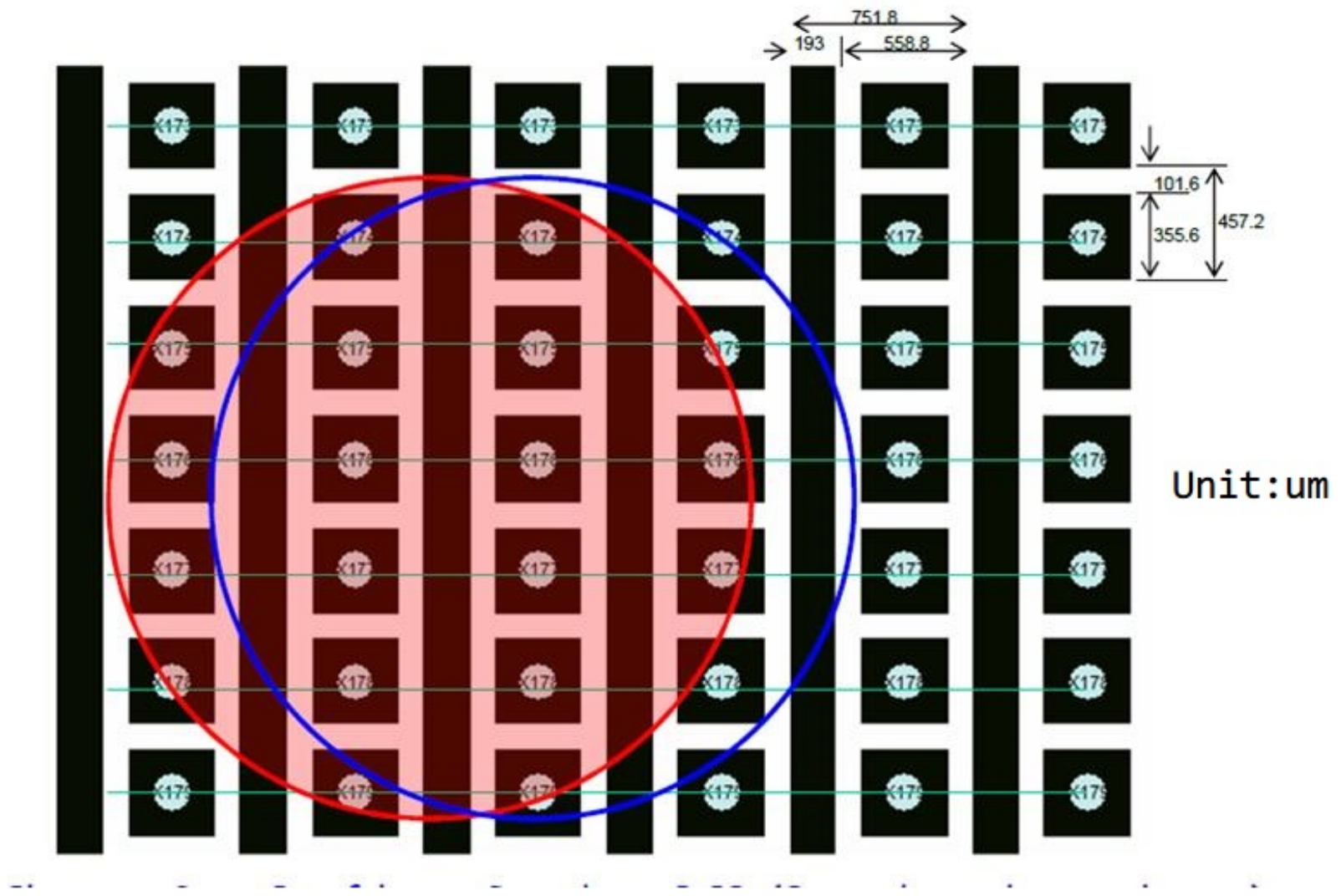

**Figure 4.9:** The profile of an electron cluster in triple GEMs. Vertical black lines are the strips in the X direction with a pitch of 752 μm. The light blue horizontal lines connect electrodes to form strips in the Y direction. The red and blue circles are two representative clusters.

### Operation gas

The choice of chamber gas strongly affects the properties and eventually the performance of a TPC. Desirable characteristics are:

1. High drift velocity (to avoid accumulation of too many events inside the chamber);
2. Low electron capture probability (to preserve signal size);
3. Low transverse and longitudinal diffusion coefficients (to prevent deterioration of the spatial resolution);
4. Large specific energy loss $dE/dx$ (to improve particle identification);
5. Stability against electrical breakthroughs (to allow reliable operation of the amplification device);
6. Nonhazardous chemical properties (to address safety concerns like inflammability and damages to the hardware).

Due to the long drift distance of ∼2.5 m and the fact that ions are more massive and much slower than electrons, a large number of ions can accumulate in the chamber. This effect can lead to electric field distortions and should be minimized. To decrease this effect, the structure of the readout chambers is generally designed to avoid ions from escaping into the gas volume. A gas with a large drift velocity is also chosen in experiments with large interaction rate.

In a given working gas, the drift velocity of the electron is a function of $E/P$ where $E$ denotes the electric field and $P$ the gas pressure. Figure 4.10 shows the drift velocity obtained in two different gas mixtures. The mixture of Ar/$CF_4$/i$C_4H_{10}$ (95%/3%/2%) is widely used in a number of experiments and is the default for CEPC. At an atmospheric pressure, this mixture has a saturated drift velocity of approximately $8\,\mathrm{cm}/\,\mu\mathrm{s}$ in a drift field of ∼300 V/cm. In addition, the gas has a transverse diffusion coefficient of $30\,\mu\mathrm{m}/\sqrt{\mathrm{cm}}$.

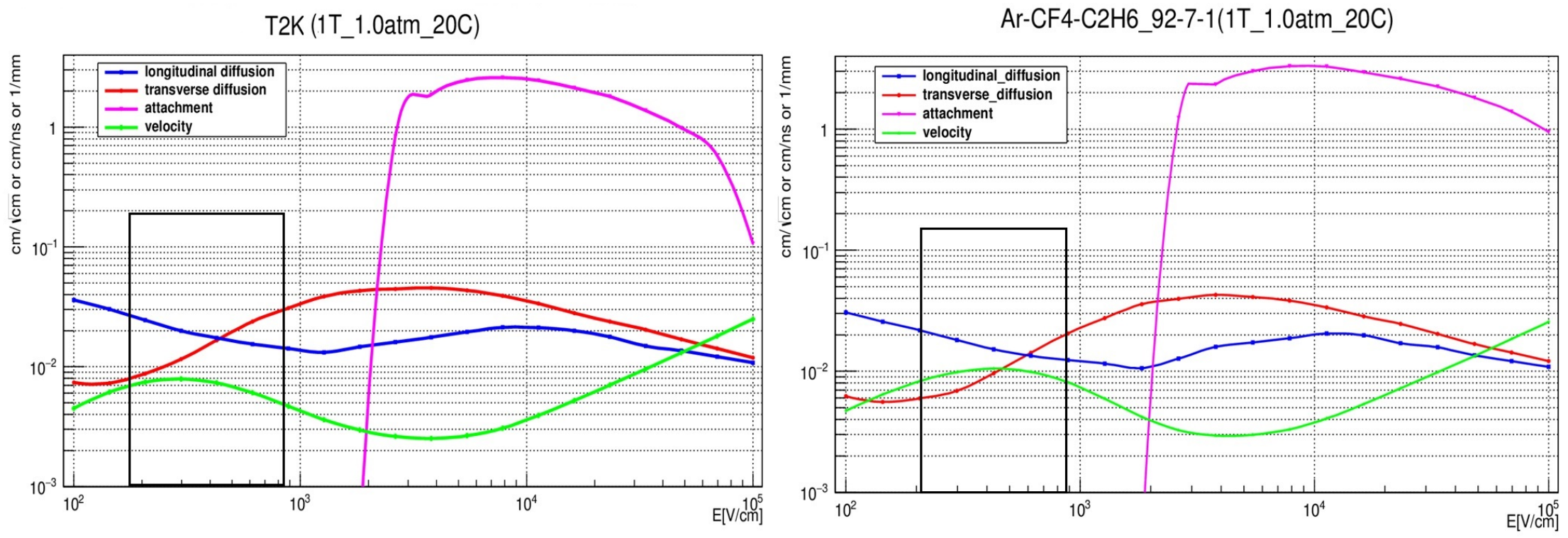

**Figure 4.10:** Drift velocity study of gas mixtures. Left: Ar/$CF_4$/i$C_4H_{10}$ (95%/3%/2%) - the T2K gas and the default option for the CEPC TPC; Right: Ar/$CF_4$/$C_2H_6$ (92%/7%/1%) - the gas mixture with the highest drift velocity. The plots show the drift velocity (green line), transverse diffusion coefficient (red), longitudinal diffusion coefficient (blue) and attachment coefficient (purple) as functions of the electric field. The black rectangle indicates the possible operation range.

The bunch spacing is 0.68 μs for the CEPC Higgs factory operation and 25 ns for the $Z$ factory operation. Since the Ion Backflow (IBF) problem scales with the number of collisions within the maximum drift time, a working gas with a higher saturated drift velocity would be beneficial and should be considered. The mixture Ar/$CF_4$/$C_2H_6$ (92%/7%/1%) is a candidate: its saturated drift velocity is roughly 20% higher than the default gas mixture and the diffusion coefficients are lower. Further R&D is needed to confirm that its other properties are compatible with CEPC needs.

#### Low-power consumption readout electronics

Small readout pads of a few square millimeters (e.g. $1\,\mathrm{mm}\times 6\,\mathrm{mm}$, there are 3 pads in the electron cluster size, see Figure 4.9) are needed to achieve high spatial and momentum resolution in the TPC, leading to about 1 million readout channels per end-cap. The total power consumption of the front-end electronics is limited by the cooling system to be several kilo-watts in practice. The architecture of the TPC readout electronics is shown in Figure 4.11, selected from a broad range of survey on current electronics installed or under development during past decades, including ALTRO/S-ALTRO and more recently SAMPA for ALICE, AFTER/GET for T2K and Timepix for ILC. It consists of the front-end electronics on the detector panel and the data acquisition system several meters away from the detector.

The waveform sampling front-end is the preferred option, including a preamplifier and shaper as the Analog Front-End (AFE), a waveform sampling ADC operating at $\geq$ 20 MSPS, a dedicated Digital Signal Processors (DSP) and zero-suppression unit, and a de-randomize event buffer for each channel. To satisfy the stringent requirements on the integration and the power consumption, a front-end Application Specific Integrated Circuits (ASIC) will be developed using the 65 nm CMOS process. The key specifications of the front-end ASIC are summarized in Table 4.4.

The power consumption of CMOS digital circuits decreases with the reduction of the feature size, while the density of the circuitry increases. Digital circuits usually use minimum sized transistors, hence part of the ADC, digital filter, control logics and data buffer dimensions will be reduced by 1/4 when migrating the same design from a 130 nm to a 65 nm process. The power consumption is also reduced since it is proportional to the gate capacitance of the transistor. Therefore, the 4 mW/ch power consumption of the DSP circuits in the 130 nm process reported in Ref. [33] can be reduced by a factor of at least two by migrating the same design to 65 nm. For the analog circuitry, the transistor size is determined by performance factors, including noise, dynamic range, and bandwidth, hence the power consumption does not benefit from the smaller feature size of the 65 nm process. In this case, the development of a low-power analog front-end, with a design strategy to keep it as simple as possible, is essential. The block diagram of the analog front-end and the Successive Approximation Register (SAR) ADC are shown in Figures 4.12 and 4.13, respectively. The CR-RC shaper and the SAR ADC instead of a pipeline ADC will be used for their simplicity in analog circuits and hence the higher power-efficiency.

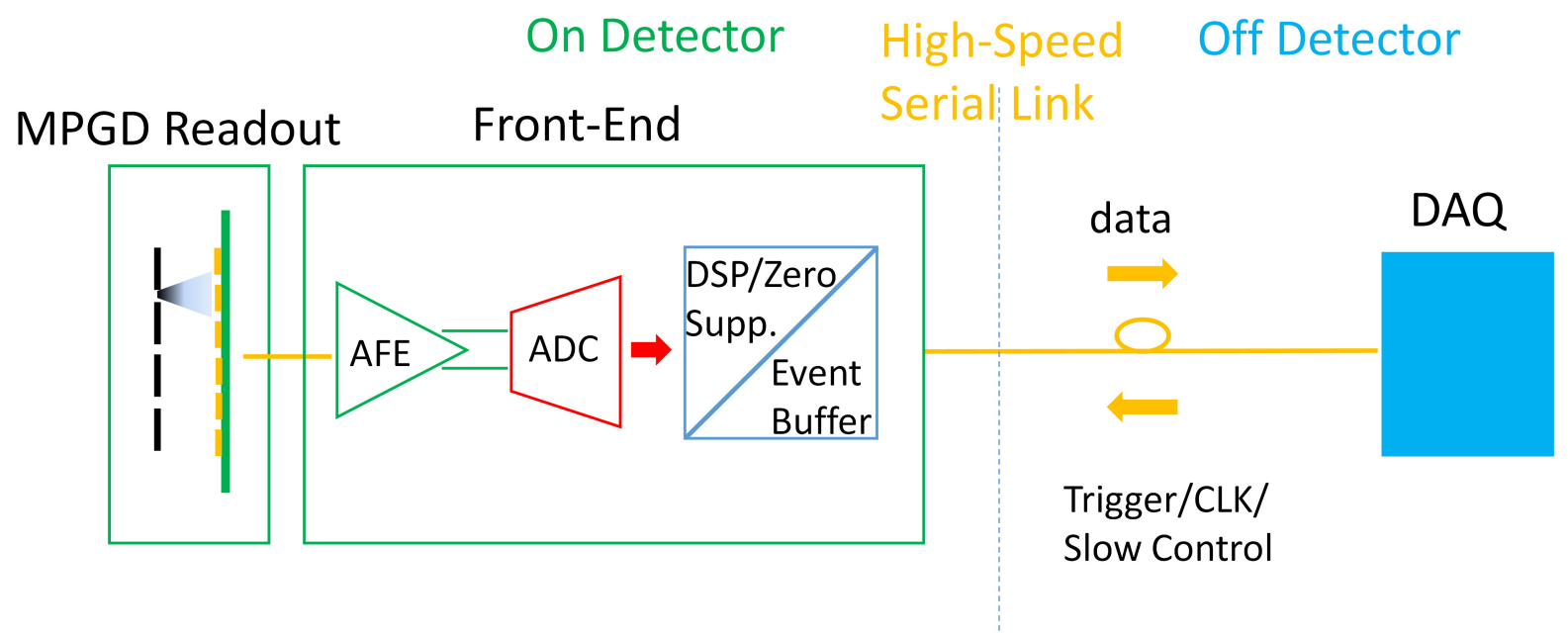

**Figure 4.11:** Architecture of the TPC readout electronics. The front-end electronics will be mounted directly on the MPGD readout board. It includes the analog front-end (AFE), with a preamplifier and shaper, followed by an Analog-to-Digital Convert (ADC), and a dedicated signal processing (DSP) unit. Signals are sent through high-speed serial links to the off-detector DAQ system.

| | | |
|---|---|---|
| Total number of channels | | 1 million per endcap |
| AFE (Analog Front-End) | ENC (Equivalent Noise Charge) | 500e @ 10pF input capacitance |
| | Gain | 10 mV/fC |
| | Shaper | CR-RC |
| | Peaking time | 100 ns |
| ADC | Sampling rate | $\geq$ 20 MSPS |
| | Resolution | 10 bit |
| Power consumption | | $\leq$ 5 mW per channel |
| Output data bandwidth | | 300–500 MB/s |
| Channel number | | 32 |
| Process | | TSMC 65 nm LP |

**Table 4.4:** Key specifications of the front-end readout ASIC for the CEPC TPC. Each endcap has a total of about 1 million channels that will be readout by over 30,000 ASICs with 32 channels each.

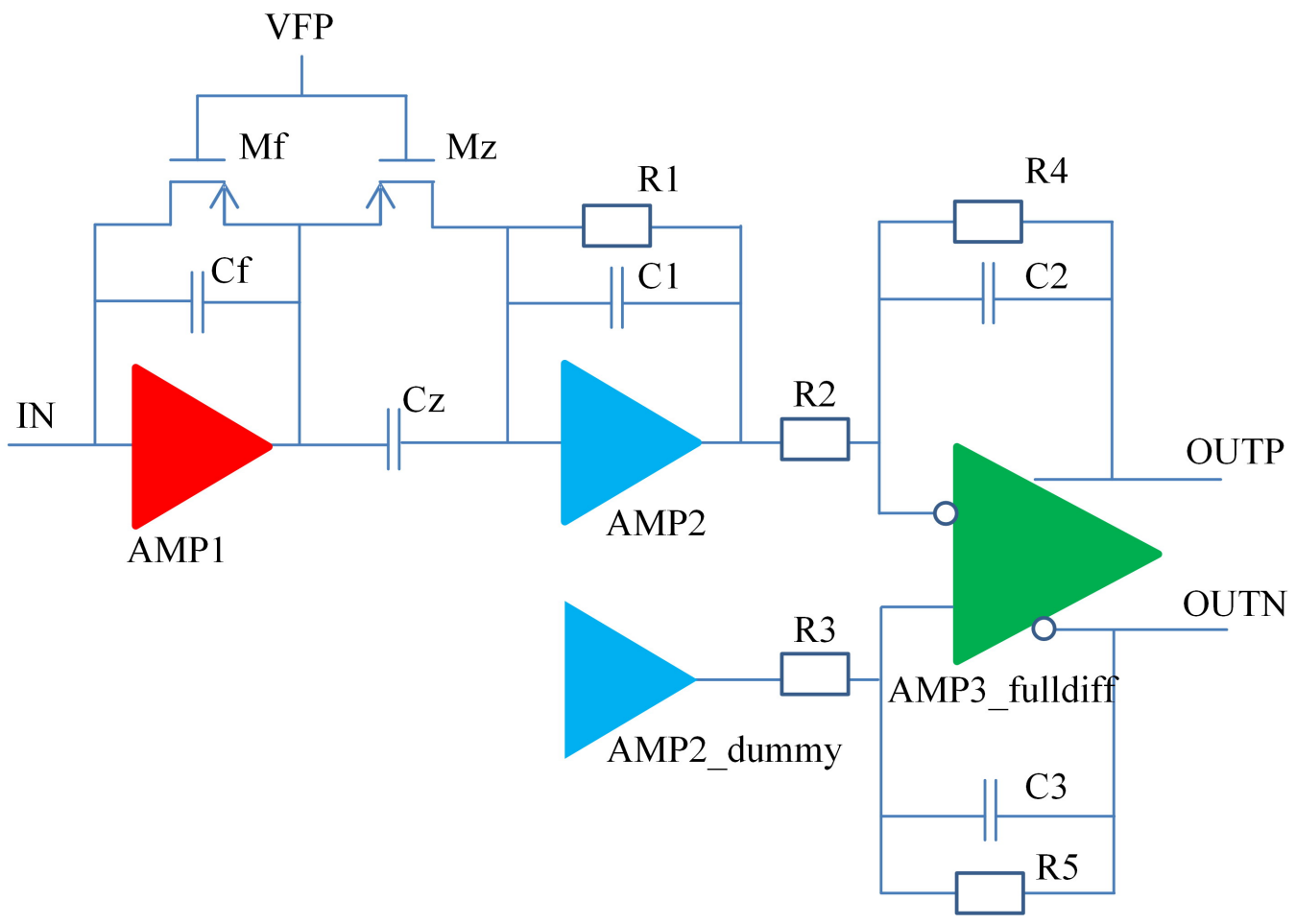

**Figure 4.12:** Block diagram of the analog front-end component of the TPC on-detector readout electronics. AMP1 is the core amplifier of the preamplifier; AMP2/AMP_dummy amplifiers compose the first stage shaper; AMP3_fulldiff is the fully differential amplifier of the second stage shaper; Feedback for the preamplifier is adjusted by the bias voltage (VFP).

Dedicated digital filters will be applied to the continuously digitized input signals to suppress the pedestal perturbations caused by the non-ideal effects such as temperature variation and environmental disturbance. The data will then be compressed by only storing the data packets above a programmable threshold with a specified number of pre- and post-samples. A data head will be added to each packet with its timestamp and other information for reconstruction afterward. The buffered data are readout through high-speed serial links to the DAQ system. The front-end electronics can support both external trigger and self-trigger mode.

Even with state of the art technology, the TPC front-end electronics on the end-plate needs a cooling system to keep the temperature stable. Two-phase $CO_2$ cooling [36] is a well-developed technology and can be used as a baseline solution to extract the heat generated by the front-end electronics and to keep the temperature of the TPC chamber stable at 20 °C. Micro-channel $CO_2$ cooling has lower mass and may be studied further as an alternative technique to copper pipes [37].

The TPC readout electronics are a few meters away from the collision point, and the radiation dose is rather low ($< 1$ krad) at the CEPC, which allows the use of standard, radiation soft technologies. On the other hand, energetic particles can always produce instantaneous failures (SEU or SEL) from time to time. Hence, sensitivity to radiation still need to be considered in the electronics design.

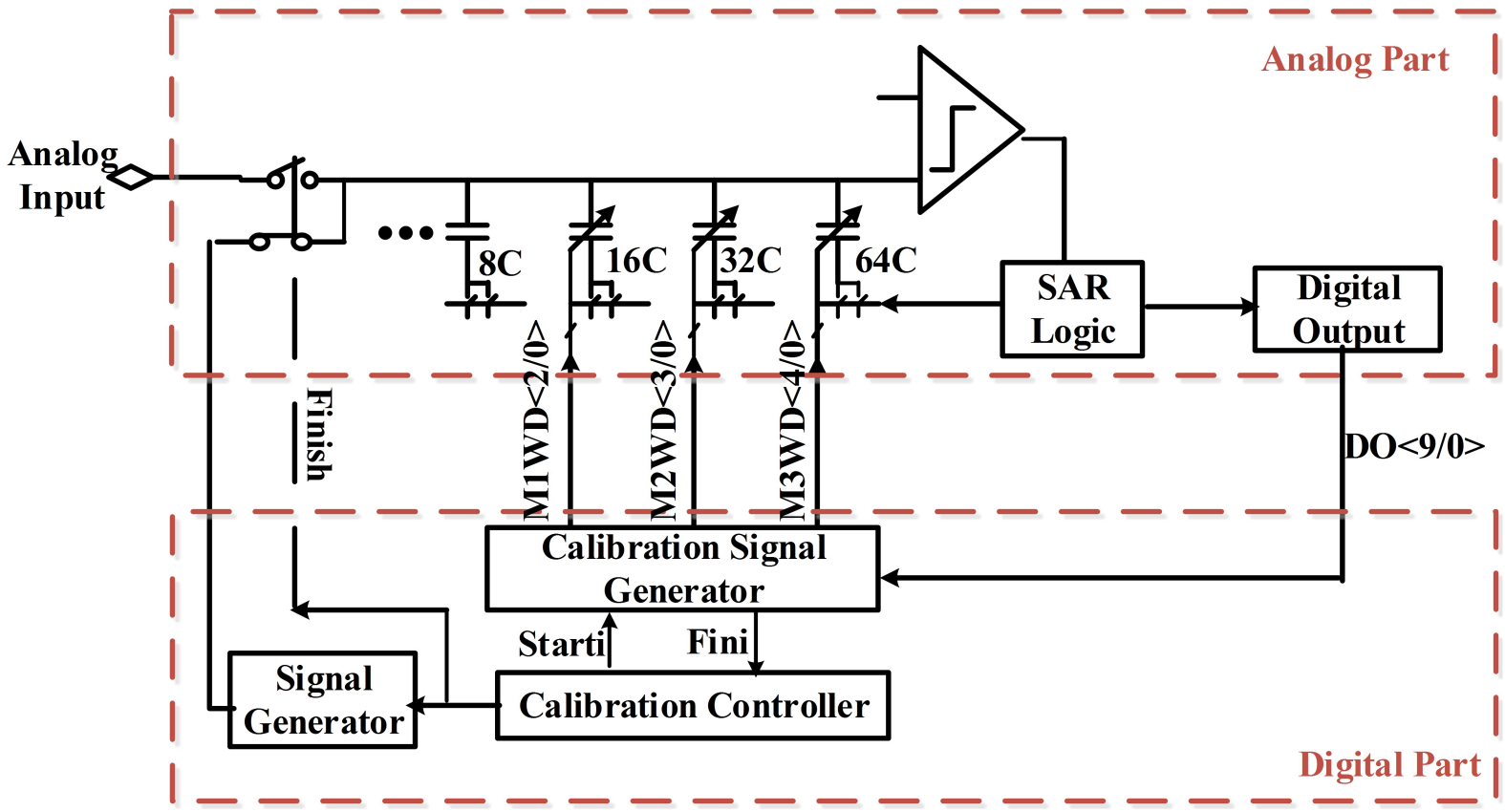

**Figure 4.13:** Block diagram of the analog-to-digital converter (ADC) using Successive Approximation Registers (SAR). Both analog and digital parts are shown.

### Critical technology challenges of the TPC detector

It will be challenging to design and manufacture the support structure with a relatively light material, and at the same time very rigid. It is required to maintain accuracy, robustness in all directions, and stability over long time periods. As the field cage is not strong enough due to the limited material budget, the end-plates become the only choice, where the support structure connects to. In the current stage of design, the TPC end-plate support scheme has not yet been finalized. A promising solution is to suspend it from the solenoid, in which a number of spokes run radically along the faces of the calorimeter to the TPC end-plates.

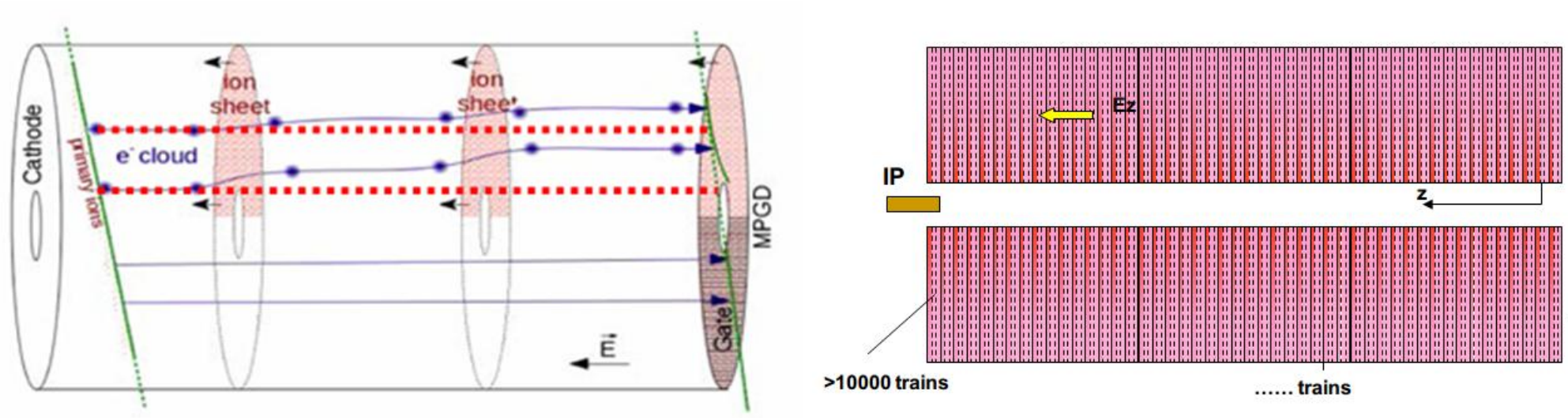

**Figure 4.14:** Ion backflow effects on the TPC tracking for the CEPC beam conditions. Left: Diagram of the distortion effects on TPC tracks caused by the ion backflow disks. The electrons from gas ionization originated by a track crossing the TPC (green line), in the absence of ion backflow, would drift directly towards the end planes following the red dotted lines. The ion disk clouds cause distortions in their path (blue lines) degrading the track measurement. (The lower part of the diagram shows the operation in case of the usage of a gating grid [38], a solution adopted for ILC but that is not applicable to the CEPC due to the short bunch space.) Right: The profile of the ions disks under the beam structure of a high-luminosity circular machine such as the CEPC.

Ions in the drift volume of the TPC move towards the cathode at a much lower velocity than electrons, and they can accumulate in this volume to build up a significant space charge in the form of ‘ion discs’ that distort the trajectory of electrons moving towards the anodes.[1] In the CEPC TPC, the majority of ions inside the drift volume are created in the amplification region and backflow to the drift region. It is therefore important to suppress this ion backflow in order to minimize the deteriorating influence on spatial resolution.

Figure 4.14 shows a diagram of the distortion due to ion backflow and ion disks in the CEPC TPC. An often used method of backflow suppression is a so-called gating grid;[2] however, it is not applicable here because the bunch spacing of 25 – 680 ns is short compared with the maximum electron drift time ($\sim$ 40 µs). Another promising option is to exploit the 'built-in' ion backflow suppression of GEMs or MicroMegas. In next section, the R&D study of an hybrid detector module that has been proposed to control ions continuously, and its updated results will be described.

### 4.2.1.3 SIMULATION AND ESTIMATION OF KEY ISSUES

#### Occupancy and charge distortion

The maximum event rate of the CEPC occurs at its $Z$-pole operation. At the nominal instantaneous luminosity of $16 \times 10^{34}\,\mathrm{cm}^{-2}\mathrm{s}^{-1}$ for operation with the 3 Tesla solenoid magnet, the $Z \to q\bar{q}$ process reaches an event rate close to 5 kHz. Using a fully simulated $Z \to q\bar{q}$ sample at $\sqrt{s} = 91.2$ GeV, we estimated the number of 3D readout cells (voxels) occupancy and the local ion charge density [39]. The voxel occupancy is the number of voxels that have a signal divided by all voxels in the TPC. A voxel has the size of a pad in the $x - y$ direction and the length of one time sample multiplied by the drift velocity in the $z$ direction. With a typical electron drift velocity of 5 cm/ µs and a readout frequency of 40 MHz (25 ns), the voxel size is $\sim 1.2$ mm in the $z$ direction. The signal from a single primary ionization can occupy several voxels, due to the shaping of the electronics, even if the signal is recorded in just one pad.

The CEPC TPC has $2 \times 10^6$ independent channels that are read out at a frequency of 40 MHz. Therefore, the TPC produces voxels at a very high rate. As a result, at the nominal CEPC $Z$ pole operation, the average voxel occupancy is extremely low ($1 \times 10^{-5}$ for the innermost layer and $3 \times 10^{-6}$ on average, assuming each hit occupies 10 voxels along the timing direction) and poses no problem to the TPC operation.

During the TPC operation, the primary and backflow ions form stable currents from the end-plates to the central cathode plane. These ions, accumulated in the gaseous volume, induce secondary electric fields and distort the TPC hit position resolution. Figure 4.15 shows the distortion as a function of the hit (primary ionization) initial radial position $r$ at the nominal luminosity for the CEPC $Z$ pole operation (and a reference luminosity of $3 \times 10^{34}\,\mathrm{cm}^{-2}\mathrm{s}^{-1}$). If the ion backflow could be controlled to Gain×IBF = 5, the maximal distortion would be about $40\,\mu\mathrm{m}$, which is roughly half of the intrinsic TPC position resolution in the $r - \phi$ direction. In other word, in the nominal parameters of the CEPC, the ion distortion would not prohibit the TPC usage, but it start to limit its performance. A few options could be applied to mitigate the ion charge distortion effects, and require further studies:

1. Better ion backflow control technology;
2. Dedicated distortion correction algorithms;
3. Global optimization of the TPC parameters.

[1] With the electron drift velocity of 5 cm/ µs, it takes $\sim$ 40 µs for all the electrons to drift 2 m to reach the end-plate. On the other hand, ions drift with a velocity of only 5 µm/ µs. This leads to ions from hundreds of thousands of events overlapping in the TPC volume.

[2] Early TPCs were equipped with Multi-Wire Proportional Chambers (MWPCs) as gas amplification devices. The IBF ratio in a standard MWPC is 30–40%, so a gating grid is essential to prevent ions from reaching the drift volume [38]. In the presence of a trigger, the gating grid switches to the open state to allow ionization electrons to travel into the gas amplification region. After a maximum drift time of about 100 µs the gating grid is closed to prevent positive ions from drifting back into the drift volume. Since it must remain closed until the ions have been collected on the grid wires, the ionization electrons are also blocked during this time and dead time is consequently generated.

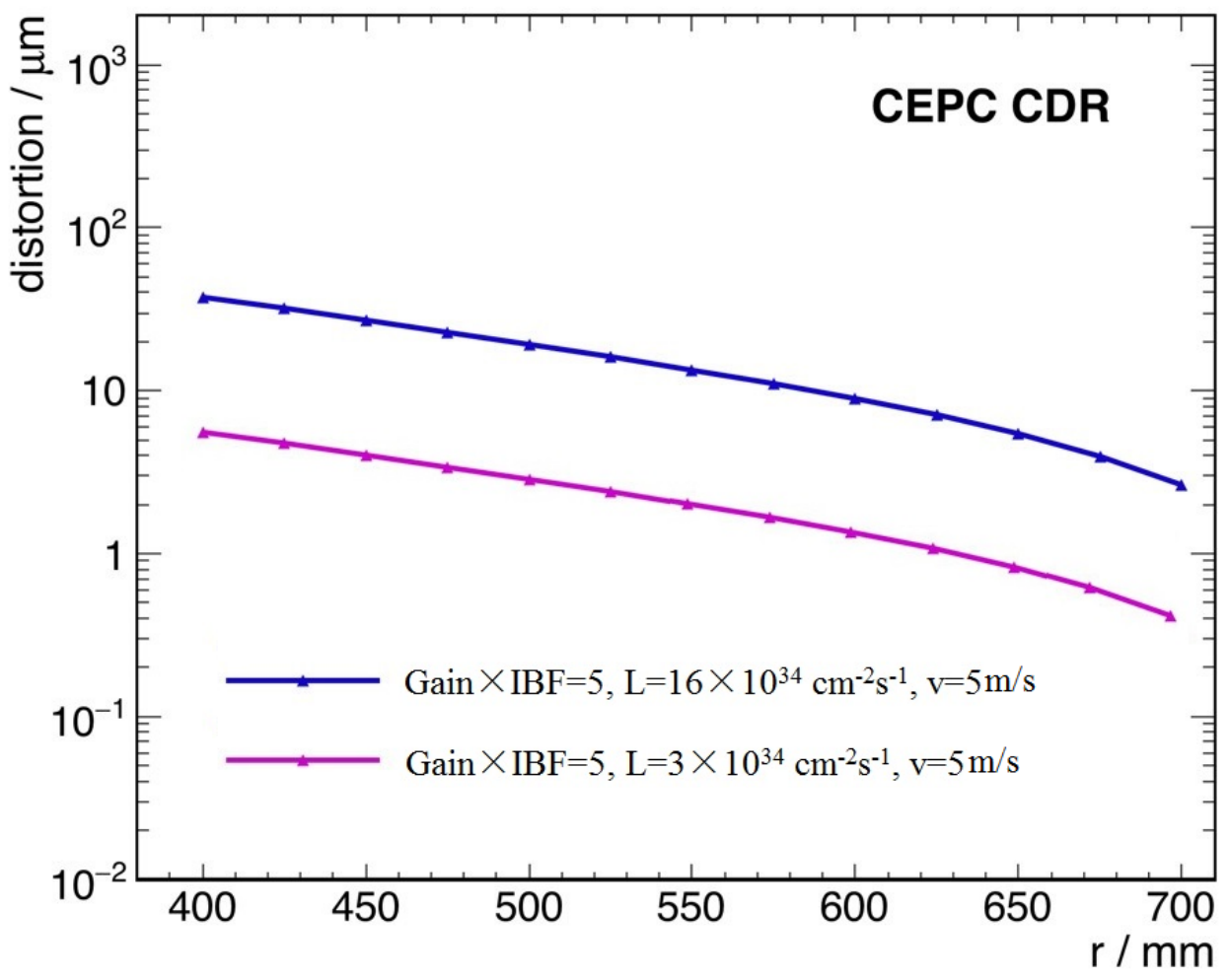

**Figure 4.15:** Distortion on the hit position reconstruction, as a function of the hit (primary ionization) radial position $r$ for different TPC parameters. Gain×IBF refers to the number of ions that will escape the end-plate readout modules per primary ionization, obtained by the multiplication of the readout modules gain and the ion backflow rate (IBF). The ion drift velocity is $v = 5$ m/s.

To conclude, the pad occupancy and distortion posses little pressure on the TPC operation if the Gain×IBF can be controlled to a value smaller than 5.

### 4.2.1.4 FEASIBILITY STUDY OF TPC DETECTOR MODULE AND FUTURE WORK

#### Hybrid structure TPC detector module

TPC readout with MPGDs, especially GEM and MicroMegas, is very attractive, because the IBF of those detectors is intrinsically low, usually around a few percent. GEM detectors have been extensively investigated in the last decade and are considered to be the prime candidate, as they offer excellent results for spatial resolution and low IBF. Numerous GEM foils can be cascaded, allowing multilayer GEM detectors to be operated at an overall gas gain above $10^4$ in the presence of highly ionized particles. MicroMegas is another kind of MPGD that is likely to be used as endcap detectors for the TPC readout. It is a parallel plate device, composed of a very thin metallic micromesh which separates the detector region into a drift and amplification volumes. The IBF of this detector is equal to the inverse of the field ratio between the amplification and the drift electric fields. Low IBF, therefore, favors high gain. However, the high gain will make it particularly vulnerable to sparking. The idea of combining GEM with MicroMegas was first proposed with the goal of reducing the spark rate of MicroMegas detectors. Pre-amplification using GEMs also extends the maximum achievable gain.

This hybrid configuration, GEM-MM is currently the baseline readout module for the CEPC TPC. Figure 4.16 shows a small prototype of $100 \times 100$ mm$^2$ that has been produced and tested. The device has a 4 mm drift region GEM, followed by a 1.4 mm transfer region and a MicroMegas with an avalanche region of 0.128 mm. The preliminary results of this detector module are described next.

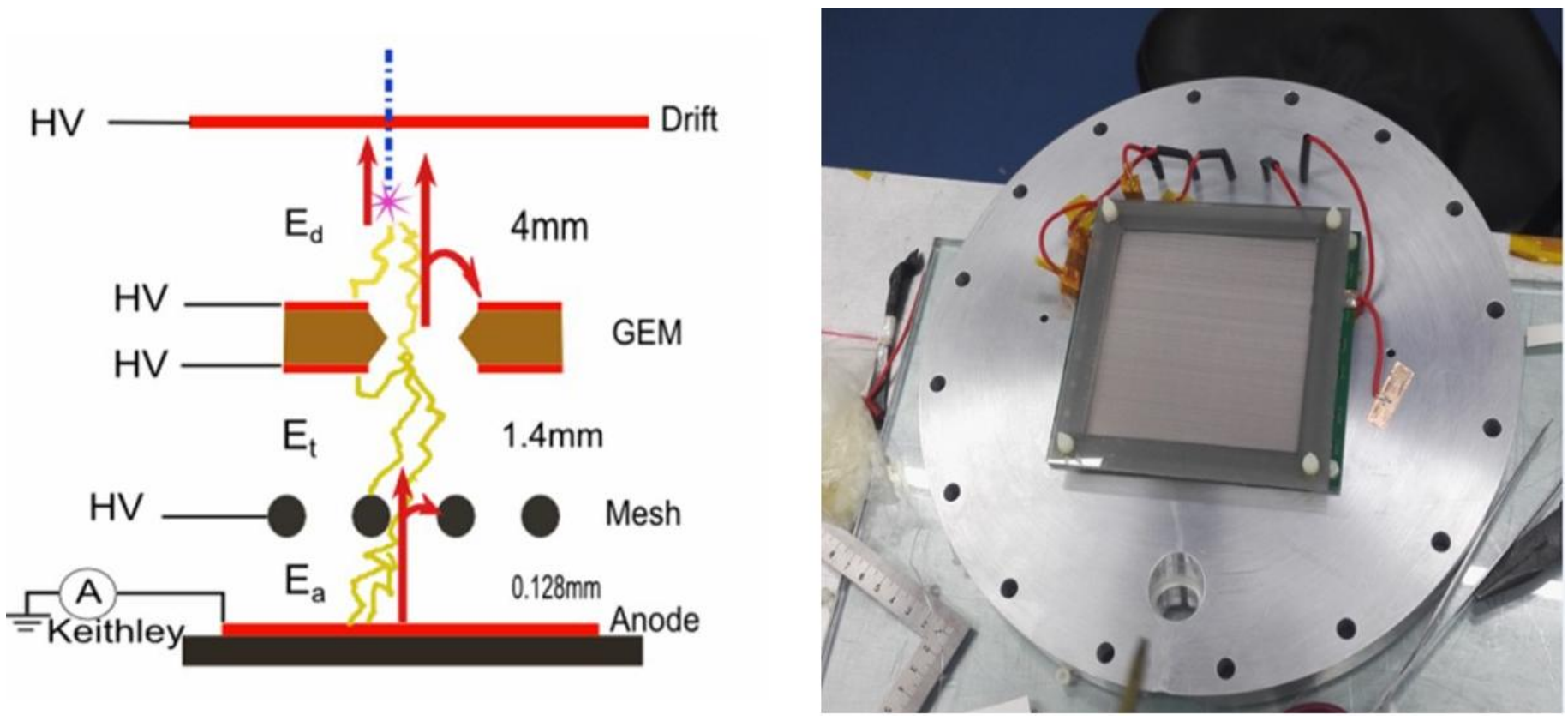

**Figure 4.16:** Left: Schematic diagram of the GEM-MM hybrid detector module. The device has a GEM with a 4 mm drift region, followed by a 1.4 mm transfer region and a MicroMegas with an avalanche region of 0.128 mm. Right: Photo of the detector prototype with 100 cm$^2$ active area designed based on this concept.

The IBF tests have been carried out with a $^{55}$Fe X-ray source and several gas mixtures: Ar/$CO_2$ (90%/10%), Ar/i$C_4H_{10}$ (95%/5%) and Ar/$CF_4$/i$C_4H_{10}$ (95%/3%/2%) - T2K gas. The currents on the anode and drift cathode were measured precisely with an electrometer.

The $^{55}$Fe X-ray source has a characteristic energy of 5.9 keV. In the Ar/$CO_2$ gas, a typical pulse height spectrum for a GEM or MicroMegas detector contains one major peak corresponding to the 5.9 keV X-rays and an escape peak at lower pulse height corresponding to the ionization energy of an electron from the argon K-shell.

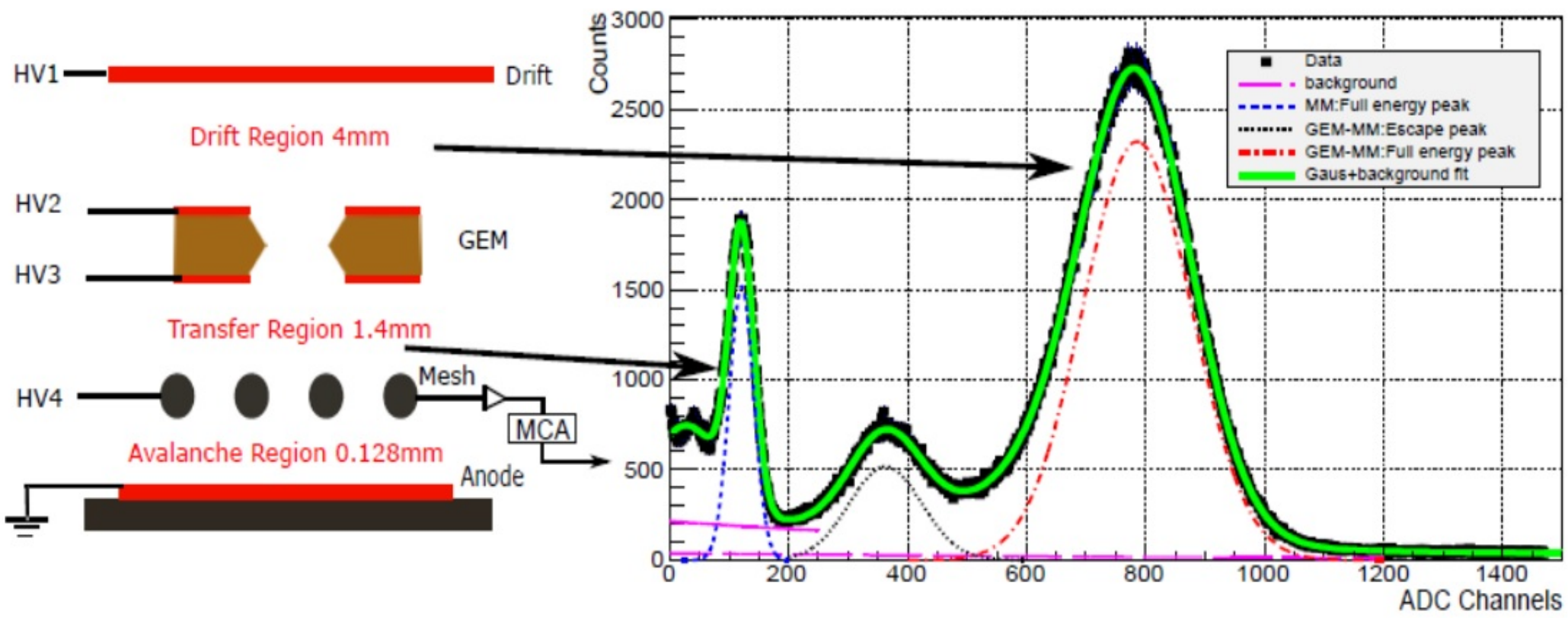

**Figure 4.17:** Energy spectrum of the $^{55}$Fe radioactive source in Ar/$CO_2$ (90%/10%) as measured by the GEM-MM hybrid module. The green curve is the whole energy spectrum from the module. The last two peaks correspond to the GEM and MicroMegas amplification in tandem. The first two peaks are from MicroMegas amplification only.

In the GEM-MM detector, the situation is different. There are two amplification stages inside the detector. The primary ionization created by photon absorption can be in the drift region or in the transfer region (Figure 4.17). Photoelectrons starting from the drift region get amplified by both the GEM detector and the MicroMegas detector before they are collected in the anode. If the photons are absorbed in the transfer region, the primary electrons will be amplified only once (by the MicroMegas).

Figure 4.17 depicts a typical $^{55}$Fe pulse height spectrum obtained by the GEM-MM detector. Four peaks are seen in the spectrum. From the left, the first peak and the second peak are the escape peak and the full energy peak of the stand alone MicroMegas. The last two peaks are created by photons with their energy deposited in the drift region. These primary electrons show combination amplification. The pre-amplification effect of GEM foil has been thus demonstrated in this energy spectrum measurement. The effective gain of the GEM can be measured even when it is relatively low. The energy resolution of the GEM-MM detector was measured to be 27% (FWHM). The gain properties of the device were also measured. A gain up to about 5000 can be achieved without any apparent discharge behavior. Finally, the IBF was reduced and measured to be $\sim 0.1\%$ at this gain.

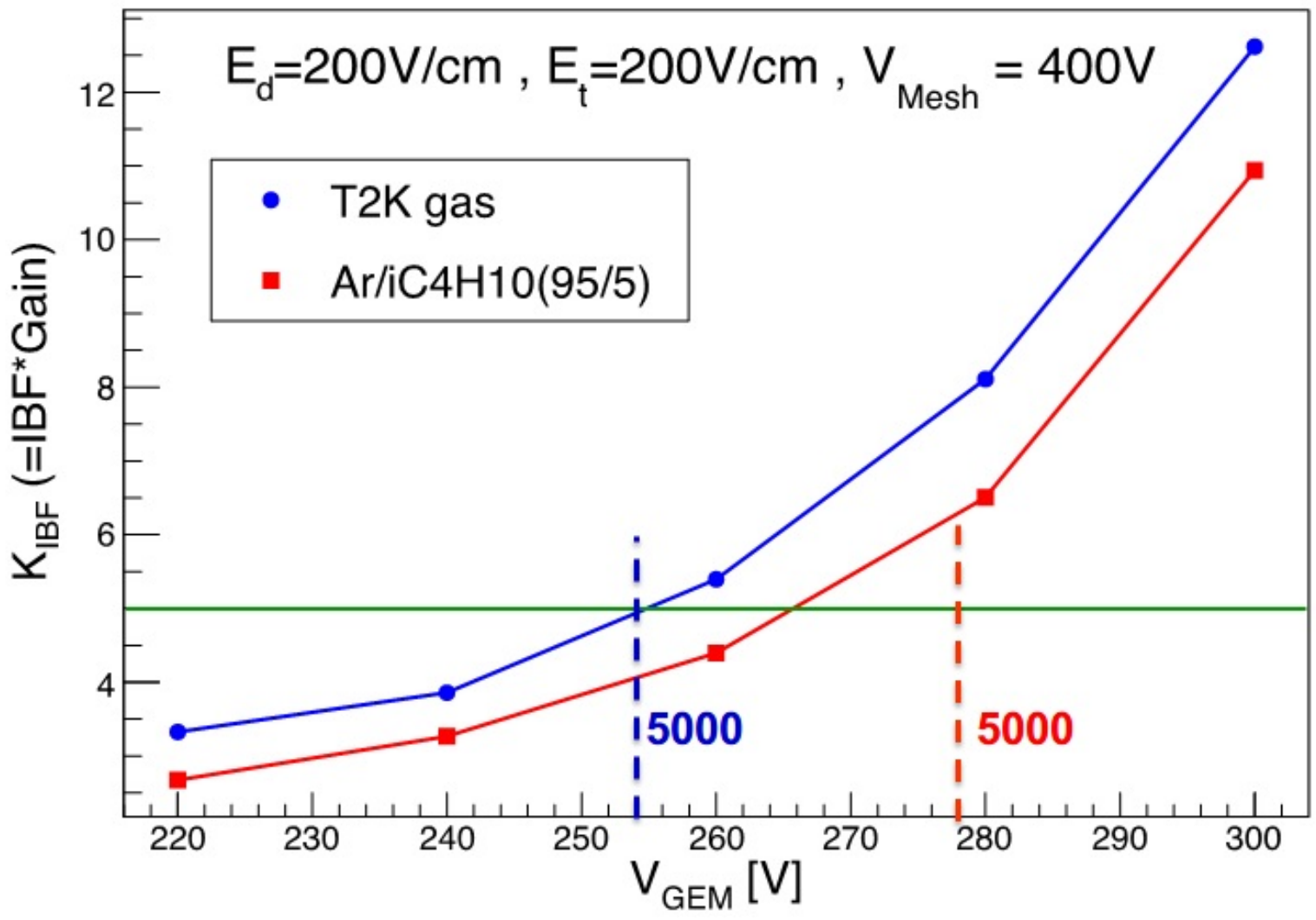

**Figure 4.18:** The IBF result of the GEM-MM module operating in two different gas mixtures: T2K gas and Ar/i$C_4H_{10}$. The nominal gain of 5000 is reached at different GEM operating voltage for the two gas mixtures. At that nominal gain, the Gain$\times$IBF is about 5 for the T2K gas, and about 7 for the Ar/i$C_4H_{10}$ gas. Thus the T2K gas mixture has lower ion backflow for the same operational gain.

Different operational parameters are being investigated, including different gas mixtures, with the goal of optimizing the TPC functionality. The Gain $\times$ IBF for the T2K and Ar/i$C_4H_{10}$ (95%/5%) gas mixtures are shown in Figure 4.18. The T2K gas reaches the nominal gain of 5000 at a lower GEM amplification voltage, corresponding also to a lower Gain $\times$ IBF value, when compared with the Ar/i$C_4H_{10}$ gas mixture. Thus, the T2K gas results in lower ion backflow and is therefore considered the baseline gas for the hybrid module.

Space charge effects could disturb the IBF measurement and result in reduced measured IBF values. To quantify this, the IBF value was studied as a function of the space-charge density by varying the X-ray's voltage and current as shown in Figure 4.19. The IBF results reported here were obtained in the green rectangle area. There is no obvious discharge or spark, and there is no large number of electrons to lead the high space charge to reduce the value of IBF. No indication of space charge affecting the IBF measurement is found.

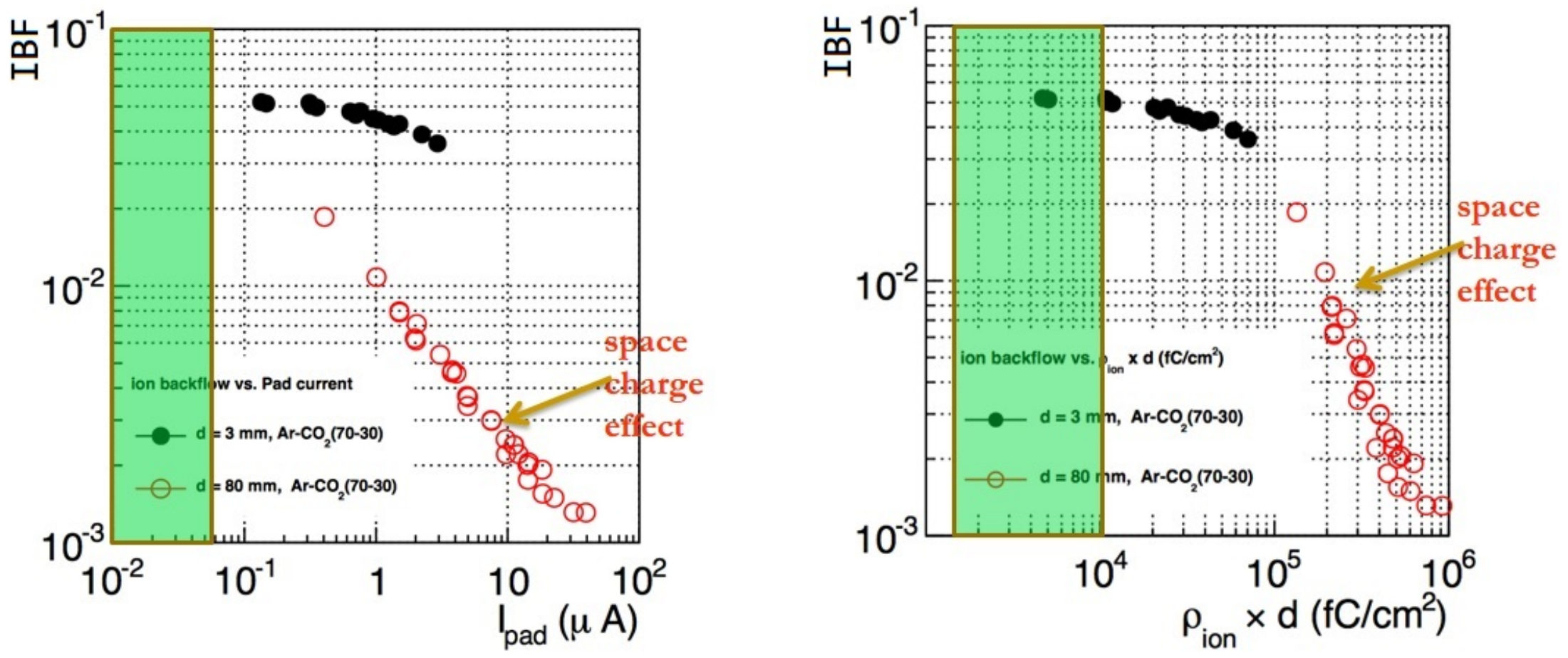

**Figure 4.19:** Comparison of the IBF with the different X-ray's voltage and current. The test results of the GEM-MM detector prototype appear in the green area where there is no space charge effect.

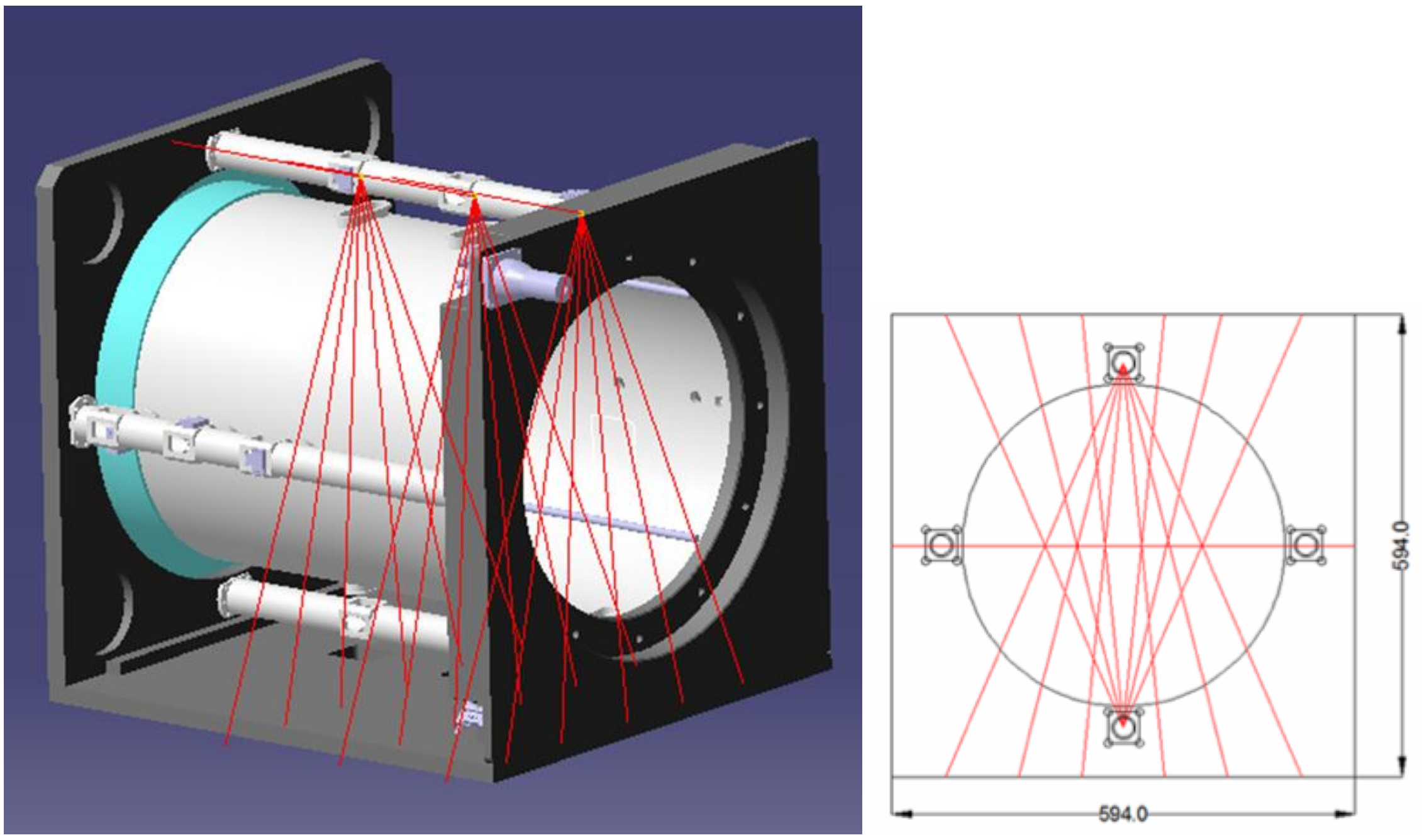

(a) (b)

**Figure 4.20:** (a) Schematic diagram of the detector module with the 266 nm laser system. The red lines show the split laser beam injection into drift chamber. (b) There are several laser beam planes, each composed of 6 downward beams, and 6 upwards beams. Several single horizontal laser beams also transverse the chamber.

#### Laser calibration and alignment system

A laser calibration system with narrow beams inside the drift volume to simulate ioniz-ing tracks at predefined locations will be used for calibration and distortion measurements. The goal is to map the uniformity of the TPC drift field within a reasonable relative uncertainty corresponding to a spacial resolution of $\sigma_{r\phi} = 100\ \mu\text{m}$. The system can be used for tests and calibration either outside or during normal data taking with the aim of understanding the chamber performance and control track distortions. Of particular interest is the testing of electronics, alignment of the readout modules, and monitoring of the drift velocity variations due to mechanical imperfections and non-uniformities in the gas, temperature and electric and magnetic fields.

The laser system will be composed of a UV laser beam, such as a Nd:YAG laser with a wavelength of 266 nm, split by transmission and reflection mirrors into several laser planes, as shown in Figure 4.20. The laser entrance window into the chamber will likely be made of fused silica as it has $\sim$99% transmission efficiency at 266 nm. There are several laser beam planes distributed longitudinally throughout the chamber. Each laser plane is composed of 6 downward beams, and 6 upwards beams. Several single-laser beams also transverse the chamber horizontally. Ionization in the gas volume occurs along the laser path via two-photon absorption by organic impurities.

A prototype of this laser calibration system has been built with one readout module and a Nd:YAG laser device. The complete optical path and the laser power is split into planes of 6 laser beams. The laser power reaches over $\sim 10\mu$ J/mm$^2$ that is equivalent to $\sim$10 MIP. Tests of the laser system and prototype have been performed at gain of 3000 and 5000. Figure 4.21 reports the response of the prototype readout module as a function of the laser beam size. At the nominal gain of 5000, the desired output signal range of 300–500 mV can be obtained with a small beam diameter. Ultimately, the beam incident spot area will be in the range of 0.8 mm$^2$ to 1.0 mm$^2$. Larger beam diameters eventually result in the saturation of the signal response from the readout module, but no damage has been observed.

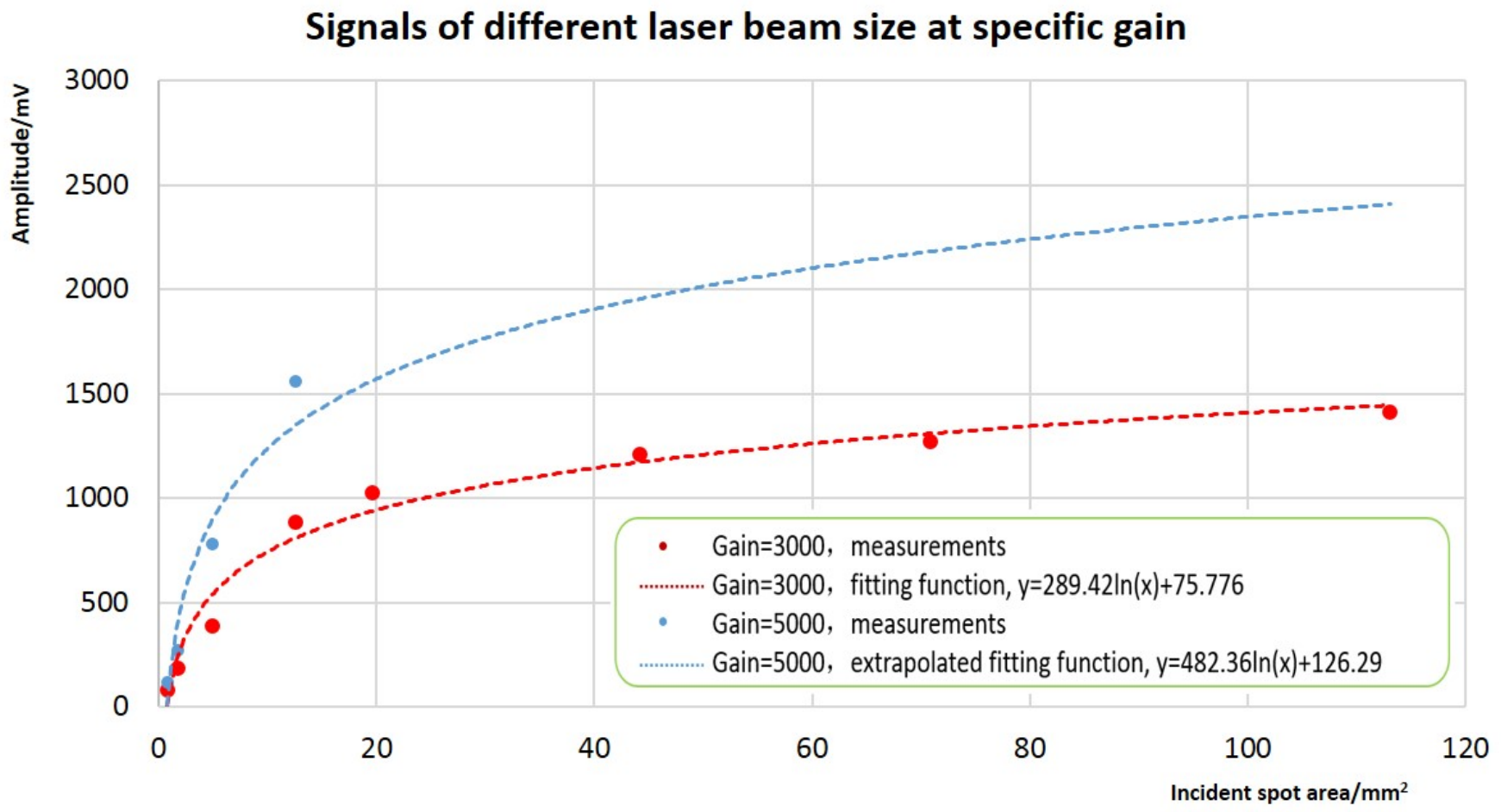

**Figure 4.21:** Signal amplitude collected by the readout module as a function of the incident area of the laser beam. Measurements have been done for gain of 3000 and 5000. At the nominal gain of 5000, the desired output signal range of 300–500 mV can be obtained with a laser beam diameter of 0.8 mm. The detector response saturates for larger laser beam sizes without damaging the readout module.

An additional UV-lamp can generate additional ions in the prototype volume via photoelectric effect. A Deuterium lamp with wavelengths of 160–400 nm illuminates a smooth Aluminum film cathode, and produces photoelectrons which can be used to study and monitor the distortions. To mimic the bunch structure and the ion distortions, the UV lamp can be controlled with a specific time structure and produce more than 10000 electrons/$\mathrm{s} \cdot \mathrm{mm}^2$.

#### 4.2.1.5 CONCLUSION

The Time Projection Chamber presented here provides an excellent starting point for the research and development in the context of the CEPC beam environment. Several studies have already been performed and many more are foreseen. Modifications are still expected due to the performance requirements and severe experimental conditions at the CEPC. Several critical R&D issues have been identified in previous studies. Possible solutions to these issues have been suggested and will continue to be investigated with the TPC prototype. A hybrid-structure MPGD detector module has been developed, and preliminary results have been obtained and analyzed. Further studies will be done from this combination detector module; another small TPC prototype with the 266 nm laser calibration system and UV photoelectric function has been designed and assembled. This calibration experiment will continue to be further studied for the CEPC.

### 4.2.2 SILICON TRACKER

The silicon tracker and the TPC together with the vertex detector form the complete tracking system of the CEPC baseline detector concept. With sufficiently low material budget to minimize the multiple-scattering effect, the silicon tracker provides additional high-precision hit points along the trajectories of charged particles, greatly improving tracking efficiency and precision. In addition, the silicon tracker also provides the following functionalities:

- monitoring possible field distortion in the TPC,
- contributing detector alignment,
- separating events between bunch crossings with relative time-stamping, and
- potentially the $dE/dx$ measurement.

The track momentum resolution can be parametrized in terms of the resolution on $1/p_T$ as

$$\sigma_{1/p_\mathrm{T}} = a \oplus \frac{b}{p \sin^{3/2}\theta} \quad [\,\mathrm{GeV}^{-1}] \tag{4.2}$$

where $p$ ($p_T$) is the (transverse) momentum of the track and $\theta$ is the polar angle. The constant term $a$ represents the intrinsic resolution of the tracker and the term with $b$ parametrizes the multiple-scattering effect. The CEPC physics program requires

$$a \sim 2 \times 10^{-5}\,\mathrm{GeV}^{-1} \quad \text{and} \quad b \sim 1 \times 10^{-3}. \tag{4.3}$$

At $\theta = 90°$, the resolution is dominated by the multiple-scattering effect for tracks with momenta below 50 GeV and by the single-point resolution for tracks with momenta above 50 GeV.

#### 4.2.2.1 BASELINE DESIGN

The silicon tracker of the baseline design consists of four components: the Silicon Inner Tracker (SIT), the Silicon External Tracker (SET), the Forward Tracking Detector (FTD), and the Endcap Tracking Detector (ETD). The overall layout is shown in Figure 4.1 and the main parameters are summarized in Table 4.5.

| **Detector** | | **Radius $R$ [mm]** | | **$\pm z$ [mm]** | **Material budget [$X_0$]** |
|---|---|---|---|---|---|
| **SIT** | Layer 1 | 153 | | 371.3 | 0.65% |
| | Layer 2 | 300 | | 664.9 | 0.65% |
| **SET** | Layer 3 | 1811 | | 2350 | 0.65% |
| | | $R_{in}$ | $R_{out}$ | | |
| **FTD** | Disk 1 | 39 | 151.9 | 220 | 0.50% |
| | Disk 2 | 49.6 | 151.9 | 371.3 | 0.50% |
| | Disk 3 | 70.1 | 298.9 | 644.9 | 0.65% |
| | Disk 4 | 79.3 | 309 | 846 | 0.65% |
| | Disk 5 | 92.7 | 309 | 1057.5 | 0.65% |
| **ETD** | Disk | 419.3 | 1822.7 | 2420 | 0.65% |

**Table 4.5:** Main parameters of the CEPC silicon tracker. Silicon pixel sensors are planned for the two inner disks of the FTD whereas silicon microstrip sensors are envisioned for the rest. The column labelled $\pm z$ shows the length of the SIT and SET layers, and the $z$ position of the FTD and ETD disks.

The barrel components SIT and SET provide precise hit points before and after the TPC, improving the overall tracking performance in the central region. The SIT helps the link between the vertex detector and the TPC, enhancing the reconstruction efficiency, particularly for low-momentum charged particles. The SET sits between the TPC and the calorimeter and helps in extrapolating from the TPC to the calorimeter. In addition, the good timing resolution of silicon sensors provides hit time-stamps for the separation of bunch crossings.

The ETD is positioned in the gap between the endplate of the TPC and the endcap calorimeter. It improves the reconstruction of charged particles with a reduced path in the TPC. The SIT, SET and ETD covers the central tracking region. They form the complete silicon envelope and help in calibrating the tracking system. However, the ETD has not been included in the current version of full simulation.

The FTD is installed between the beam pipe and the inner cage of the TPC, covering the very forward region. It consists of five silicon disks on each side. The FTD is essential for precise and efficient tracking down to very small (or large) polar angles, where a number of challenges exist: the net magnetic field approaching zero along the beam pipe due to the compensating MDI solenoid, significantly larger occupancies due to forward going particles and high backgrounds from the interaction region. To achieve the best tracking performance, the FTD needs to provide precise space point measurements with maximal possible lever arms, but with low material budget.

#### 4.2.2.2 SENSOR TECHNOLOGIES

The basic sensor technology is silicon microstrip sensors for all tracker components except the two innermost FTD disks where silicon pixel sensors are foreseen. Requirements of the single point resolution vary with positions of tracker components, but a general condition of $\sigma_{\text{SP}} < 7\,\mu\text{m}$ is required for high precision tracking. The microstrip sensors have been proven to be capable of reaching this resolution, taking into account material budget and power consumption. The baseline design of the microstrip sensors calls for a large detection area of $10 \times 10\ \text{cm}^2$, a fine pitch of $50\,\mu\text{m}$, and a thickness $< 200\,\mu\text{m}$ to minimize the multiple-scattering effect. All microstrip layers or disks will consist of two back-to-back mounted single-sided microstrips with a stereo angle of $7°$ for SIT and SET and $5°$ for FTD. Double-sided microstrips are also being considered.

An alternative design being investigated is a fully pixelated silicon tracker. Although the choice of pixel technologies is open, the CMOS Pixel Sensors (CPS) have gained particular interest because of two main performance advantages compared to the microstrip sensors:

- Granularity. The CPS provides better single-point spatial resolution and significantly reduces the ambiguity caused by multiple hits in a single strip.
- Material budget. The CPS can be thinned to be less than $50\,\mu\text{m}$, whereas the strip sensor is usually a few hundred microns.

In addition, production cost could be significantly reduced for fabricating large area pixelated sensors because CPS is based on the standard industrial CMOS procedure. Furthermore, it is possible to embed circuits in the pixel to simplify the tracker readout circuitry. Initial R&D on large area CPS has been carried out.

The pixelated silicon tracker alternative is used to set data acquisition requirements because it is more demanding. Table 4.6 shows the estimated occupancies of the first layers of SIT and FTD based on the following assumptions:

1. The pixel dimension is assumed to be $50\,\mu\text{m} \times 350\,\mu\text{m}$, with which at least in one dimension spatial resolution can reach $7\,\mu\text{m}$ by implementing in-pixel ADC with multiple bits.
2. The track multiplicities in different operation modes are inferred from hit densities in Table 9.4.
3. Readout time of pixel sensors is set as $10\,\mu\text{s}$, the same as that of VTX.
4. Cluster size is set as 9 hits per track.

#### 4.2.2.3 FRONT-END ELECTRONICS

The Front-End (FE) electronics will depend on the choice of the sensors, namely microstrips or pixels.

For the microstrips, custom designed ASICs with deep sub-micron CMOS technology will be used. The chips will provide functions of the analogue to digital conversion (ADC), zero suppression, sparsification and possibly time stamping, together with necessary control circuitry. The high degree digitization is for relaxing the data processing pressure on downstream electronics.

| Operation mode | **H (240)** | **W (160)** | **Z (91)** |
|---|---|---|---|
| Track multiplicity ($\mathrm{BX}^{-1}$) | 310 | 300 | 32 |
| Bunching spacing (ns) | 680 | 210 | 25 |
| SIT-L1 occupancy (%) | 0.19 | 0.58 | 0.52 |
| FTD-D1 occupancy (%) | 0.17 | 0.54 | 0.48 |

**Table 4.6:** Estimated occupancies of the first layers of the SIT (SIT-L1) and the FTD (FTD-D1). See context for more details.

As for the pixels, all FE functions can be realized in a pixel chip, even with some functions, e.g., ADC on pixels themselves. Particular concerns are readout time and the number of electronic channels.

The FE chip will be developed in mind with low noise, low power consumption and high radiation tolerance. New developments, such as in the SiLC collaboration and for the LHC experiment upgrades, will be good references.

#### 4.2.2.4 POWERING AND COOLING

Powering and cooling are challenges for the CEPC silicon tracker. It is important to investigate the novel powering scheme based on the DC-DC converter, which has already been actively pursued by the ATLAS and CMS experiments for their silicon detector upgrades [40–42]. It allows significant reduction in material budget for the low-voltage power cables and gives less power dissipation in the delivery system. Cooling is another critical issue. Although cooling based on forced cooled gas flow might be still feasible to efficiently conduct away the heat generated by the sensors, ASICs and other electronics, it is important to look into other cooling techniques, such as silicon micro-channel cooling [43], which are being investigated by several other experiments. The technique chosen will have to provide sufficient cooling without compromising the detector performance.

#### 4.2.2.5 MECHANICS AND INTEGRATION

There will always be challenging aspects of the mechanical design for a large area silicon tracker. A lightweight but stiff support structure can be built based on Carbon fiber Reinforced Plastic material [44]. The support structure, cable routing and electronics common to other sub-detectors need to be carefully designed to minimize the overall quantity of material and facilitate the construction and integration. Precise and fast system alignment might be achieved with dedicated laser monitoring systems, while the final alignment will be accomplished using tracks from well-understood physics events [45].

#### 4.2.2.6 CRITICAL R&D

Silicon technology for large-area tracking detectors will continue to evolve over the next few years [46]. There are ongoing R&D activities conducted by the ATLAS and CMS experiments to develop advanced silicon detectors for the High Luminosity LHC as well as several pioneering R&D projects by the Silicon tracking collaboration for the Linear Collider (SiLC). Despite the rather different operation conditions and requirements, it is always important to exploit synergies with existing R&D from other experiments to share expertise. During the preliminary studies, several critical R&D items have been identified for the CEPC silicon tracker. All of them, as listed below, will be pursued in the R&D phase of the CEPC project and made available for engineering construction.

- Alternative pixelated strip sensors with CMOS technologies;
- $p^+$-on-n silicon microstrip sensors with slim-edge structure;
- Front-end electronics with low power consumption and low noise, fabricated with CMOS technologies of small feature size;
- Efficient powering and cooling techniques with low material budget;
- Lightweight but robust support structure and related mechanics;
- Detector layout optimization, particularly in the end regions.

It will be vital to develop necessary instrumentation for the module assembly and to verify the detector module performance with beam tests. Prototypes of support structures, including cooling solutions, shall be also built for mechanical and thermal tests.

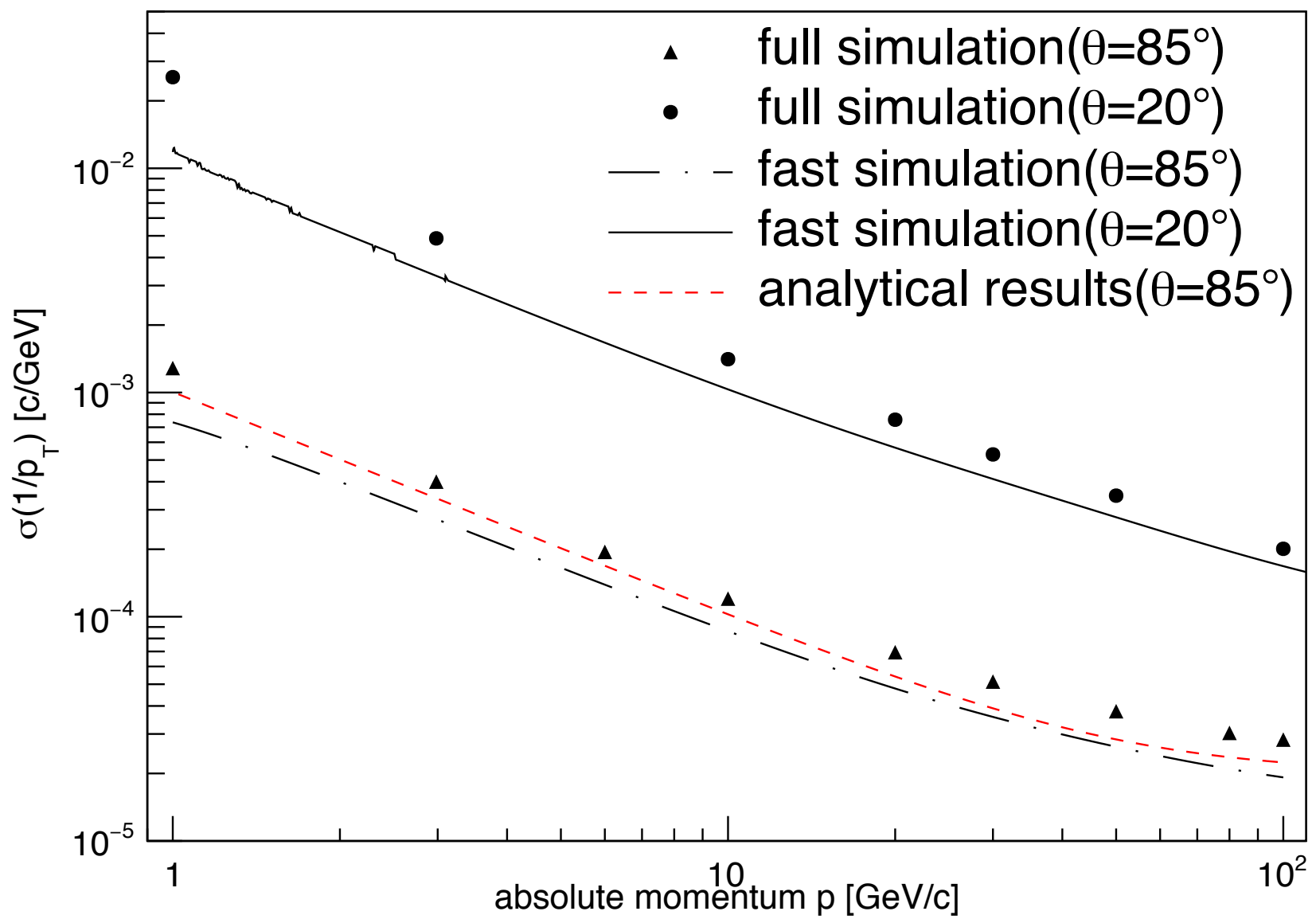

**Figure 4.22:** Transverse momentum resolution for single muon tracks as a function of the track momentum estimated for the CEPC baseline design with full simulation (dots) and fast simulation (black lines) compared to the analytical results obtained with Eqs. 4.2 and 4.3 (red line).

### 4.2.3 TPC AND SILICON TRACKER PERFORMANCE

The performance study described in this section is based on the full tracking system: vertex detector, silicon tracker and TPC. While the tracking performance in the central region has been studied [47], the performance in the forward region, which has been designed to cope with the rather short $L^*$, requires additional careful evaluation. Figure 4.22 shows the estimated transverse momentum resolution for single muon tracks at two polar angles $\theta = 20°$ and $85°$, and the analytical results from Equation (4.2) and Equation (4.3). Due to the reduced lever arm of the tracks and fewer FTD disks in the forward region ($\theta = 20°$), the resolution is worse.

## 4.3 FULL SILICON TRACKER

A full-silicon tracker is also an option for the CEPC baseline detector concept. The tracker will utilize a well known technology that provides excellent space point resolution and granularity to cope with track separation in dense jets and in high occupancy environment from beam related backgrounds at high luminosities. Potential drawbacks include more material within the tracking volume, fewer space points available for pattern recognition, and limited $dE/dx$ measurements.

This section will demonstrate that the full-silicon tracker concept is a viable option for the CEPC. The tracker replaces the entire baseline tracking system while leaving all other detector subsystems unchanged. While the boundary conditions are fixed in this study, the layout including the number of silicon layers, single- or double-strip layers, and support material has been optimized. The parameters used in this simplified simulation study are summarized in the following:

- the solenoidal magnetic field is kept at 3 Tesla,
- the cylindrical tracking volume has a radius of 1.83 m and a length of 4.6 m,
- the tracking coverage extends down to 7.25 degree from the beam pipe,
- the beam pipe has a radius of 1.45 cm and is 14 cm long.

### 4.3.1 FULL SILICON TRACKER LAYOUT

Two layouts of the full-silicon tracker concept have been investigated. The first, referred to as the FST, has six double-strip layers in the barrel and five double-strip disks in the endcap. They are labeled as Silicon Outer Tracker (SOT) and Endcap Outer Tracker (EOT), respectively. A double-strip layer or disk consists of two single-sided microstrips mounted back-to-back or one double-sided microstrip, both with a stereo angle. In addition, two pixel disks called Endcap Inner Tracker (EIT) are placed at either side of the IP inside the EOT. Like the baseline, the FST has a vertex detector with six silicon pixel layers, but with reoptimized geometry. The FST layout is shown in Figure 4.23. The FST provides 12 precise space point measurements (6 from the vertex detector and 6 from the silicon trackers) in the transverse plan for tracks in the central region, and at least 7 points down to an angle of about 7.25 degree from the beam line, as shown in Figure 4.24.

An alternative approach based on the ILC-SiD design [48], referred to as the FST2, five single-strip layers in the barrel and four double-strip disks in the endcaps. Moreover, the FST has five pixel layers for VTX and seven pixel disks for EIT. The complete FST2 layout is shown in Figure 4.23 with the expected of number of hits shown in Figure 4.24. The FST2 provides 10 space point measurements for tracks in the central region.

Tables 4.7 and 4.8 summarize the geometric parameters of the layers and disks of the FST and FST2 configurations. The total radiation length, including dead material such as the readout, for the entire tracking system is about 5–7$\%X_0$ for the FST and 7–10$\%X_0$ for the FST2.

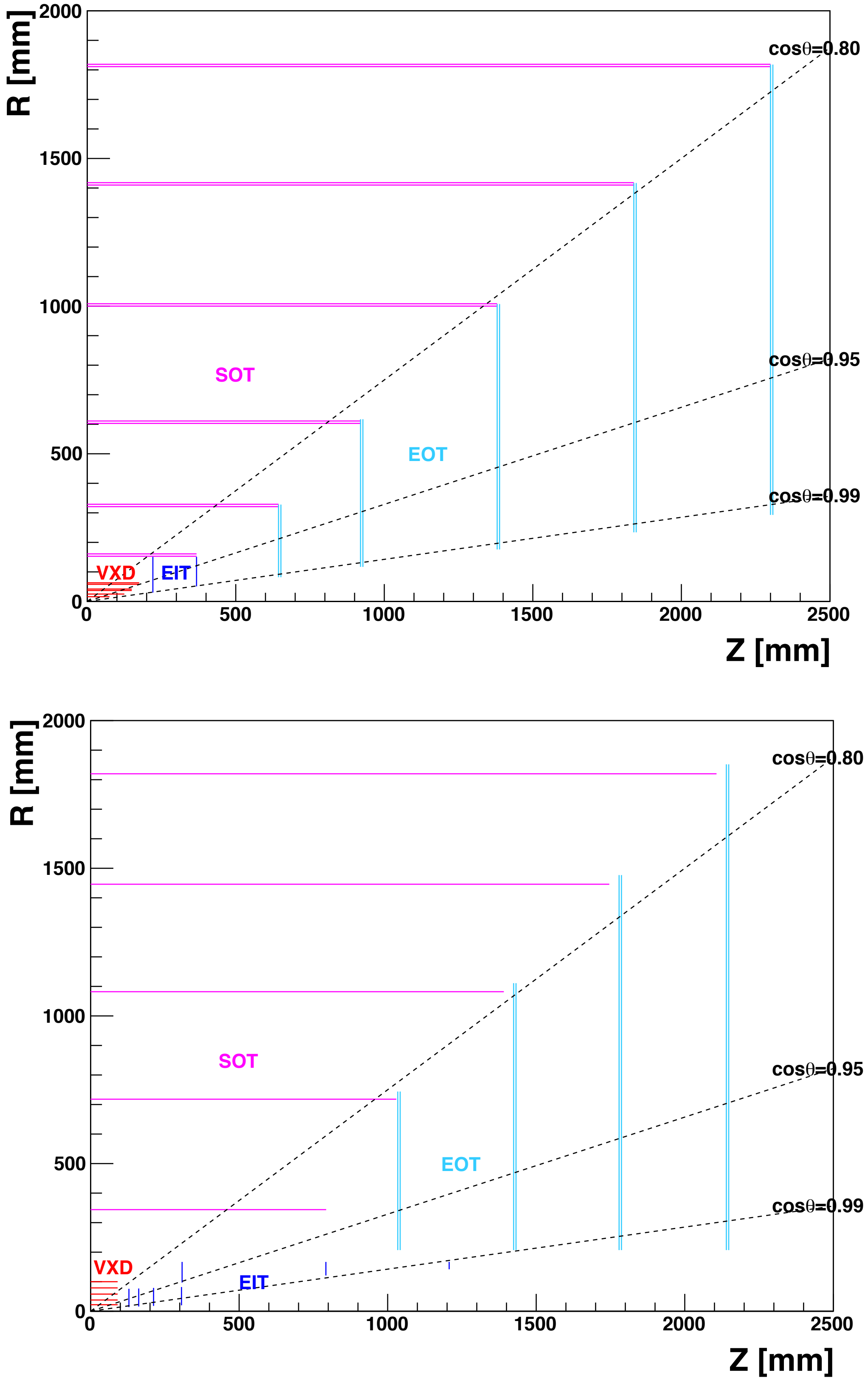

**Figure 4.23:** $R-Z$ views of the full-silicon tracker options, FST (top) and FST2 (bottom). In the FST layout, the full strip detector (SOT and EOT) is composed of double silicon strip layers. In the FST2 layout, the SOT consists of single layers, while the EOT consists of double-strip layers.

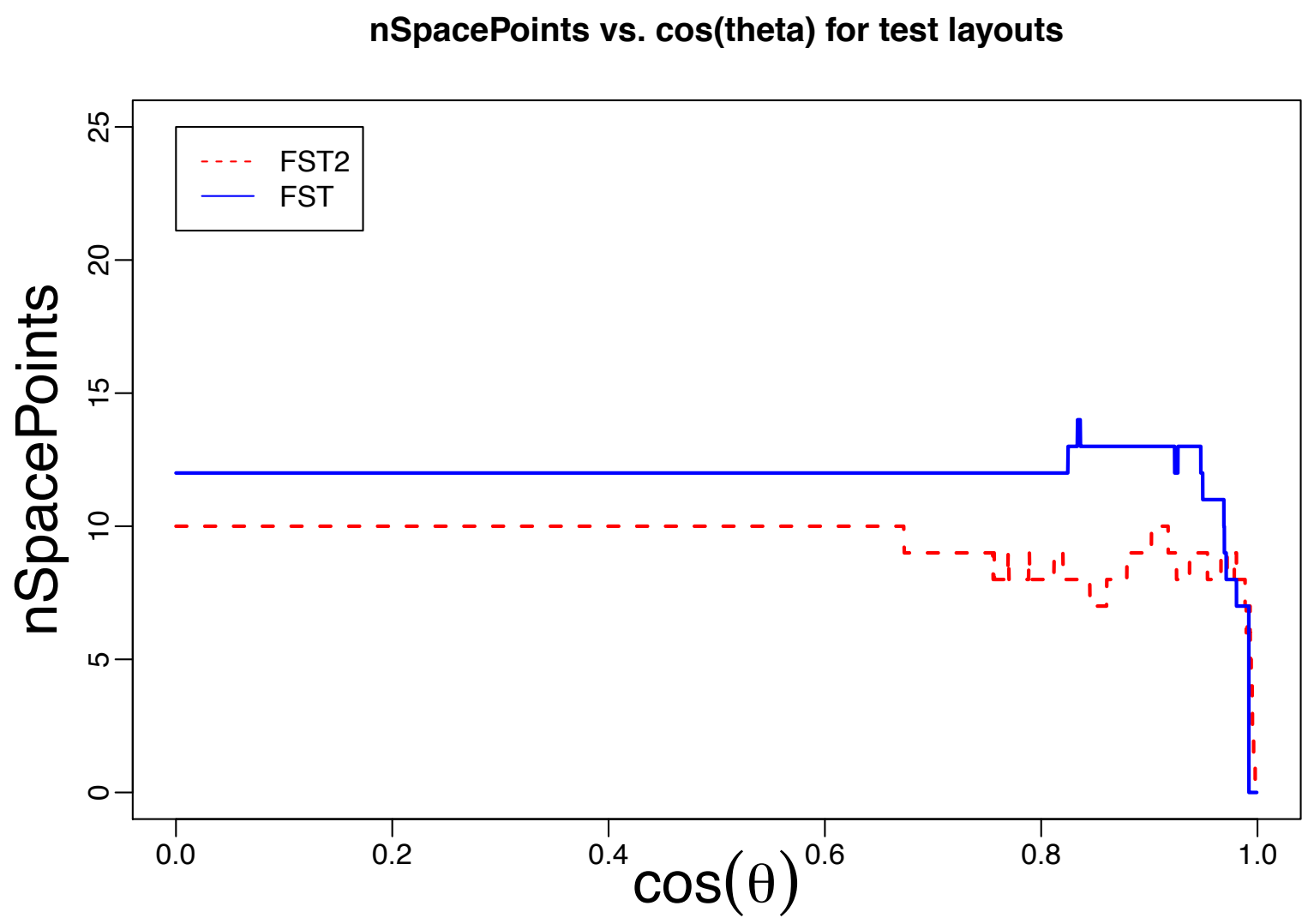

**Figure 4.24:** The expected numbers of hits as functions of the cosine of the track polar angle.

| | FST | | | FST2 | | |
|---|---|---|---|---|---|---|
| **VXD** | R (m) | | $\pm z$ (m) | R (m) | | $\pm z$ (m) |
| Layer 1 | 0.016 | | 0.078 | 0.022 | | 0.091 |
| Layer 2 | 0.025 | | 0.125 | 0.038 | | 0.091 |
| Layer 3 | 0.037 | | 0.150 | 0.058 | | 0.091 |
| Layer 4 | 0.038 | | 0.150 | 0.079 | | 0.091 |
| Layer 5 | 0.058 | | 0.175 | 0.100 | | 0.091 |
| Layer 6 | 0.059 | | 0.175 | | | |
| **EIT** | $R_{in}$ (m) | $R_{out}$ (m) | $\pm z$ (m) | $R_{in}$ (m) | $R_{out}$ (m) | $\pm z$ (m) |
| Disk 1 | 0.030 | 0.151 | 0.221 | 0.014 | 0.076 | 0.129 |
| Disk 2 | 0.051 | 0.151 | 0.368 | 0.016 | 0.077 | 0.162 |
| Disk 3 | | | | 0.018 | 0.079 | 0.212 |
| Disk 4 | | | | 0.020 | 0.082 | 0.306 |
| Disk 5 | | | | 0.097 | 0.167 | 0.308 |
| Disk 6 | | | | 0.121 | 0.167 | 0.792 |
| Disk 7 | | | | 0.142 | 0.167 | 1.207 |

**Table 4.7:** Geometric parameters of the silicon pixel detectors of the FST and FST2. The vertex detector has six layers for the FST option and five layers for the FST2 option. The EIT has two disks in the FST case, and seven disks in the FST2 case.

| | FST | | | | FST2 | | | |
|---|---|---|---|---|---|---|---|---|
| **SOT** | R (m) | | $\pm z$ (m) | Type | R (m) | | $\pm z$ (m) | Type |
| Layer 1 | 0.153 | | 0.368 | D | 0.344 | | 0.793 | S |
| Layer 2 | 0.321 | | 0.644 | D | 0.718 | | 1.029 | S |
| Layer 3 | 0.603 | | 0.920 | D | 1.082 | | 1.391 | S |
| Layer 4 | 1.000 | | 1.380 | D | 1.446 | | 1.746 | S |
| Layer 5 | 1.410 | | 1.840 | D | 1.820 | | 2.107 | S |
| Layer 6 | 1.811 | | 2.300 | D | | | | |
| **EOT** | $R_{in}$ (m) | $R_{out}$ (m) | $\pm z$ (m) | Type | $R_{in}$ (m) | $R_{out}$ (m) | $\pm z$ (m) | Type |
| Disk 1 | 0.082 | 0.321 | 0.644 | D | 0.207 | 0.744 | 1.034 | D |
| Disk 2 | 0.117 | 0.610 | 0.920 | D | 0.207 | 1.111 | 1.424 | D |
| Disk 3 | 0.176 | 1.000 | 1.380 | D | 0.207 | 1.477 | 1.779 | D |
| Disk 4 | 0.234 | 1.410 | 1.840 | D | 0.207 | 1.852 | 2.140 | D |
| Disk 5 | 0.293 | 1.811 | 2.300 | D | | | | |

**Table 4.8:** Geometric parameters of the silicon strip detectors of the FST and FST2. Types S and D stand for single- and double-strip layer, respectively. The FST design has six double-strip layers for the SOT and five double-strip disks for the EOT, whereas the FST2 design has five single-strip layers for the SOT and four double-strip disks for the EOT.

### 4.3.2 DETECTOR SIMULATION AND RECONSTRUCTION

To optimize the detector design, benchmark processes of $e^+e^- \to ZH \to \nu\bar{\nu}\mu^+\mu^-$ and $e^+e^- \to ZH \to \nu\bar{\nu}gg$ (two gluon jets) as well as single muon events were generated. These events were simulated with different tracking geometries and reconstructed accordingly. They were then used for the tracking performance studies.

#### 4.3.2.1 FST TRACKER

The performance of the FST tracker was studied using the same Mokka simulation tool as for the study of the CEPC baseline detector by substituting the baseline tracker with the FST tracker while keeping all other detector elements unchanged. In the simulation, the silicon tracker was represented by planar structures with each plane consisting of a silicon layer of 150 $\mu$m thick with a pitch size of 50 $\mu$m. Each layer was composed of several ladders which were further divided into multiple sensors. The stereo angles are $7^\circ$ for the SOT layers and $5^\circ$ for the EOT layers.

The amount of material of the whole tracker is about 5% in the barrel and about 8% in the endcap as shown in Figure 4.25, including breakdowns from individual components of the tracker. The zigzag structures in the endcap are caused by the alternation and overlap of layers.

A conformal tracking algorithm developed for CLIC [49] has been adapted for the pattern recognition of the FST. Through the conformal transformation of $u = \frac{x}{x^2+y^2}$ and $v = \frac{y}{x^2+y^2}$, where $x$ and $y$ are hit coordinates in the transverse plane, a track in a uniform magnetic field becomes a straight line in the $(u, v)$ conformal space. Thus the track finding is reduced to the search of straight lines, significantly simplifying the pattern recognition. Currently, a cellular automaton is used as pattern recognition for the straight line searching.

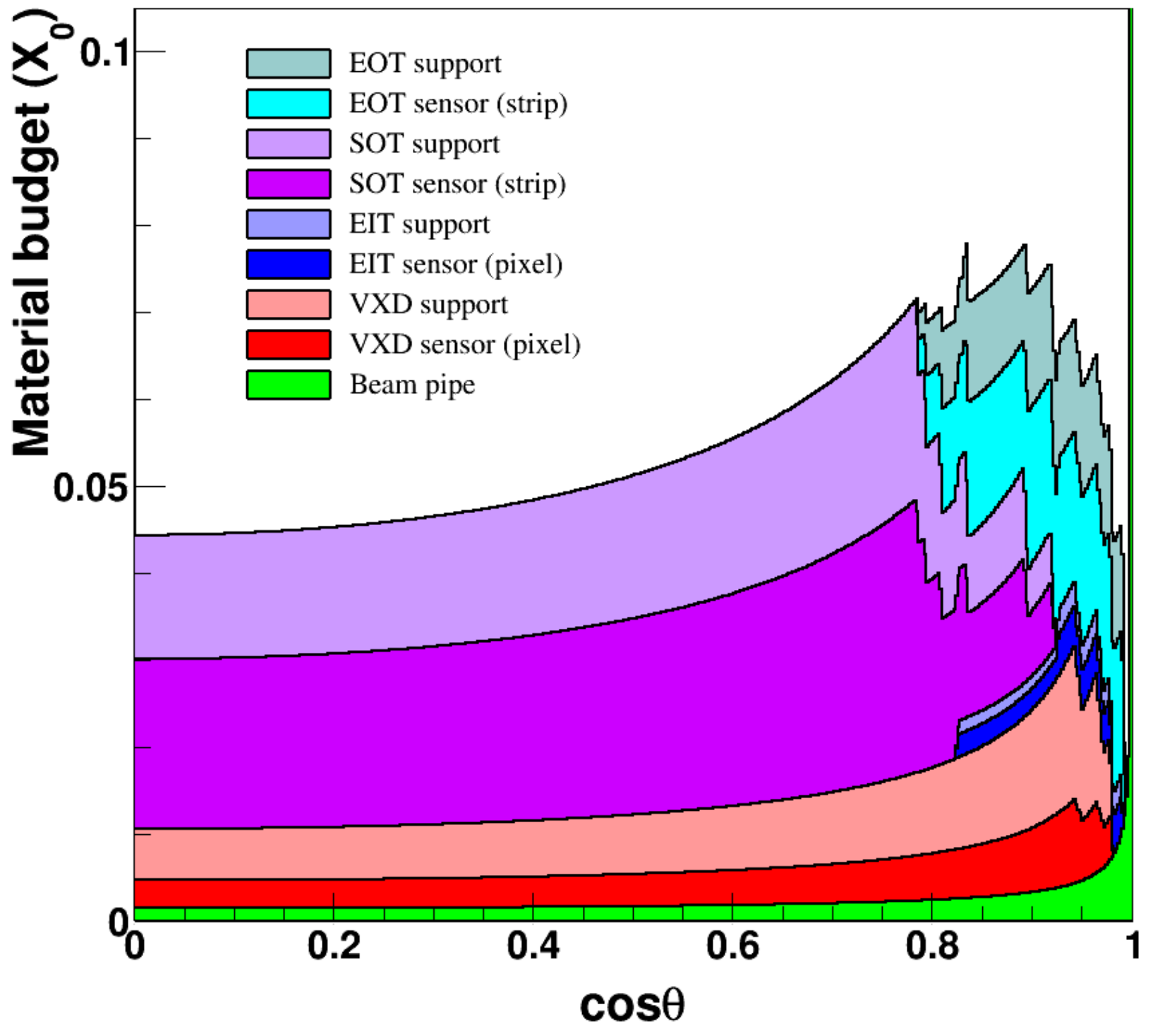

**Figure 4.25:** The amount of material of the full tracker with the FST option highlighting contributions from the VXD and SOT in the barrel, and the EIT and EOT in the endcap.

#### 4.3.2.2 FST2 TRACKER

For the FST2 option, event simulation and reconstruction were performed using the software developed for the ILC [48, 50]. Tracks were reconstructed with the LCSIM 4.0 package [49] using the "seed tracker" algorithm developed for the SiD detector simulation. Track candidates with at least six hits in the pixel and strip layers were considered. Only tracks with $p_T$ greater than 100 MeV were accepted. The particle-flow algorithms implemented in the PANDORA package [51, 52] were used to reconstruct PFA objects.

### 4.3.3 TRACKING PERFORMANCE

**Expected resolutions:** The semi-analytical program IdRes [53], developed by the ATLAS collaboration, was used to calculate the expected tracking resolution as a function of the track momentum for a given incident angle $\theta$, taking into account the effect of multiple scatterings. The results were cross checked using the LDT program [54] and consistent results were obtained from the two programs. Figure 4.26 shows the expected resolutions for the transverse momentum, transverse and longitudinal impact parameters ($d_0$ and $z_0$) as functions of track $p_{\mathrm{T}}$ for two incident angles of $\theta = 20°$ and $\theta = 80°$, comparing the performance of the FST and FST2 options. Slight better resolutions are obtained for the FST option, largely because the FST option has more silicon layers.

Tracking performance were also characterized in terms of efficiencies, momentum resolution, and the impact parameter resolutions using single muons or $e^+e^- \to ZH$ events. The tracking efficiency is defined as the fraction of stable charged particles that can be matched to the reconstructed tracks. The stable charged particles are charged particles with $p_T$ >1 GeV in the detector fiducial region ( $9° < \theta < 170°$), originated from the IP, and lived long enough to reach the calorimeter. Since the CEPC baseline tracker and the FST tracker share the common software, the performance comparisons are done for these two designs to demonstrate that the full-silicon tracking concept is a viable option for the CEPC.

**Single muon particles:** Figure 4.27 shows the tracking efficiency of the FST for single muons as a function of $p_T$. The efficiency is close to 100% at $p_T > 1$ GeV. The CEPC baseline has a similar performance, suggesting that both trackers are capable of finding tracks efficiently in the detector fiducial region.

The resolutions of track momentum, impact parameters of $d_0$, and $z_0$ as functions of track $p_{\mathrm{T}}$ in the barrel and endcap are shown in Figure 4.28. The performance of the FST is again comparable to that of the CEPC baseline. However, the latter has a slightly better momentum resolution at low momentum because it has less material.

**Dimuon mass resolution:** Figure 4.29 compares the dimuon invariant mass distributions from the $ZH \to \nu\bar{\nu}\mu^+\mu^-$ events. The FST has a mass resolution of $\sigma = 0.21$ GeV, approximately 16% better than that of the CEPC baseline.

**Tracking inside the jets:** The $ZH \to \nu\bar{\nu}gg$ events were used to study tracking performance inside jets. Figure 4.30 shows the tracking efficiency inside jets as function of the track momentum. The efficiency of finding tracks inside jets is close to 97% for both the CEPC baseline and the FST.

### 4.3.4 CONCLUSION

A preliminary study of a full-silicon tracker for the CEPC is presented. Two design approaches are considered: the first keeps the vertex detector of the baseline design and replaces the TPC and the silicon tracker with a full-silicon tracker, the second replaces the entire tracking system of the baseline with an ILC-SiD inspired tracker to achieve the excellent momentum resolution using 3 Tesla magnetic field. Initial studies of the tracking performance are promising. Many improvements, however, are needed in the simulation and reconstruction in order to understand the full potential of the full-silicon tracker.

## 4.4 DRIFT CHAMBER TRACKER

The Drift CHamber (DCH) is another option for the CEPC main outer tracker, and it is being purposed in conjunction with the alternative detector concept using a 2-Tesla magnet. It is designed to provide good tracking, high precision momentum measurement and excellent particle identification by cluster counting. In addition, a layer of silicon microstrip detectors surrounds the drift chamber in both barrel and forward/backward regions, and together with the vertex detector, improve the momentum resolution to better then 0.5% for 100 GeV charged particles.

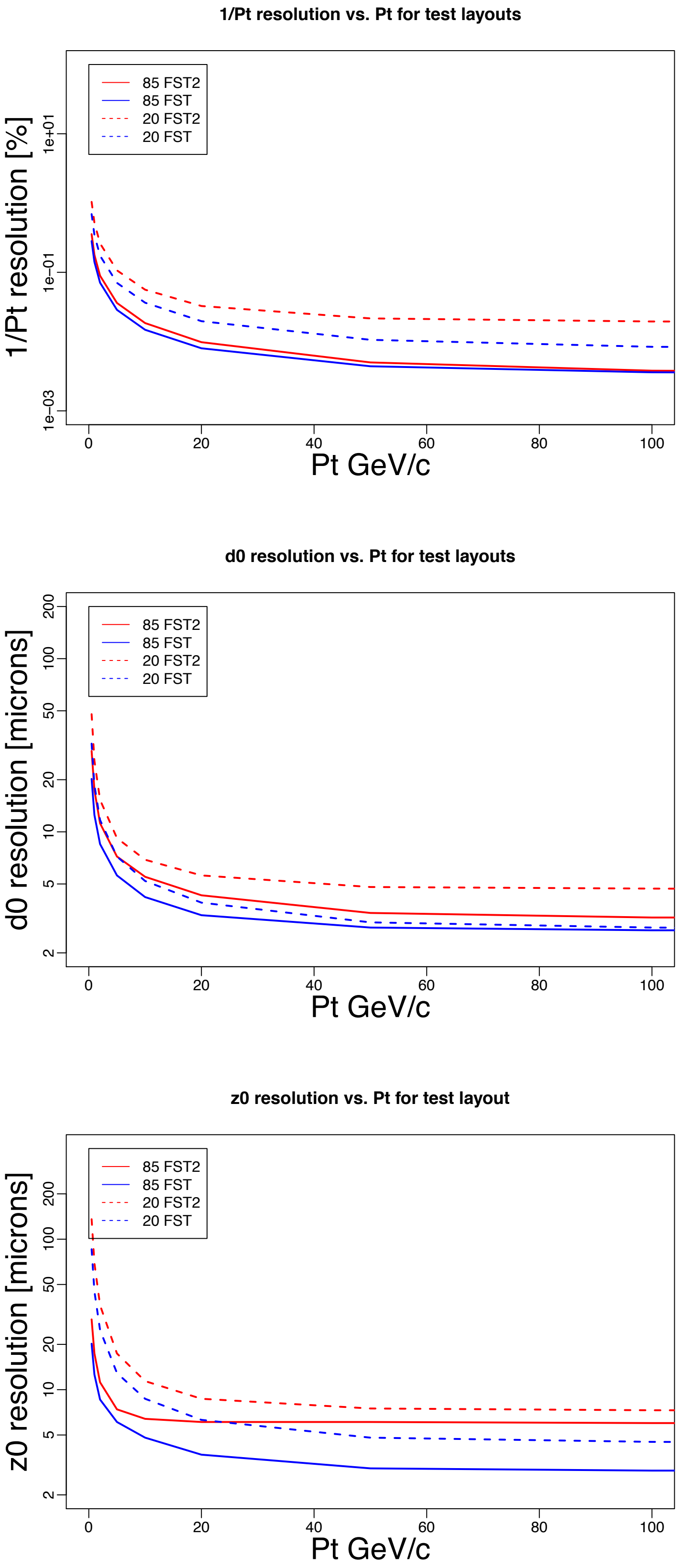

**Figure 4.26:** The expected resolutions of $1/p_T$, $d_0$, and $z_0$ from the IdRes simulation as functions of track $p_T$ for tracks with incident angles of $\theta = 85°$ and $20°$, respectively, comparing the options of the FST and FST2.

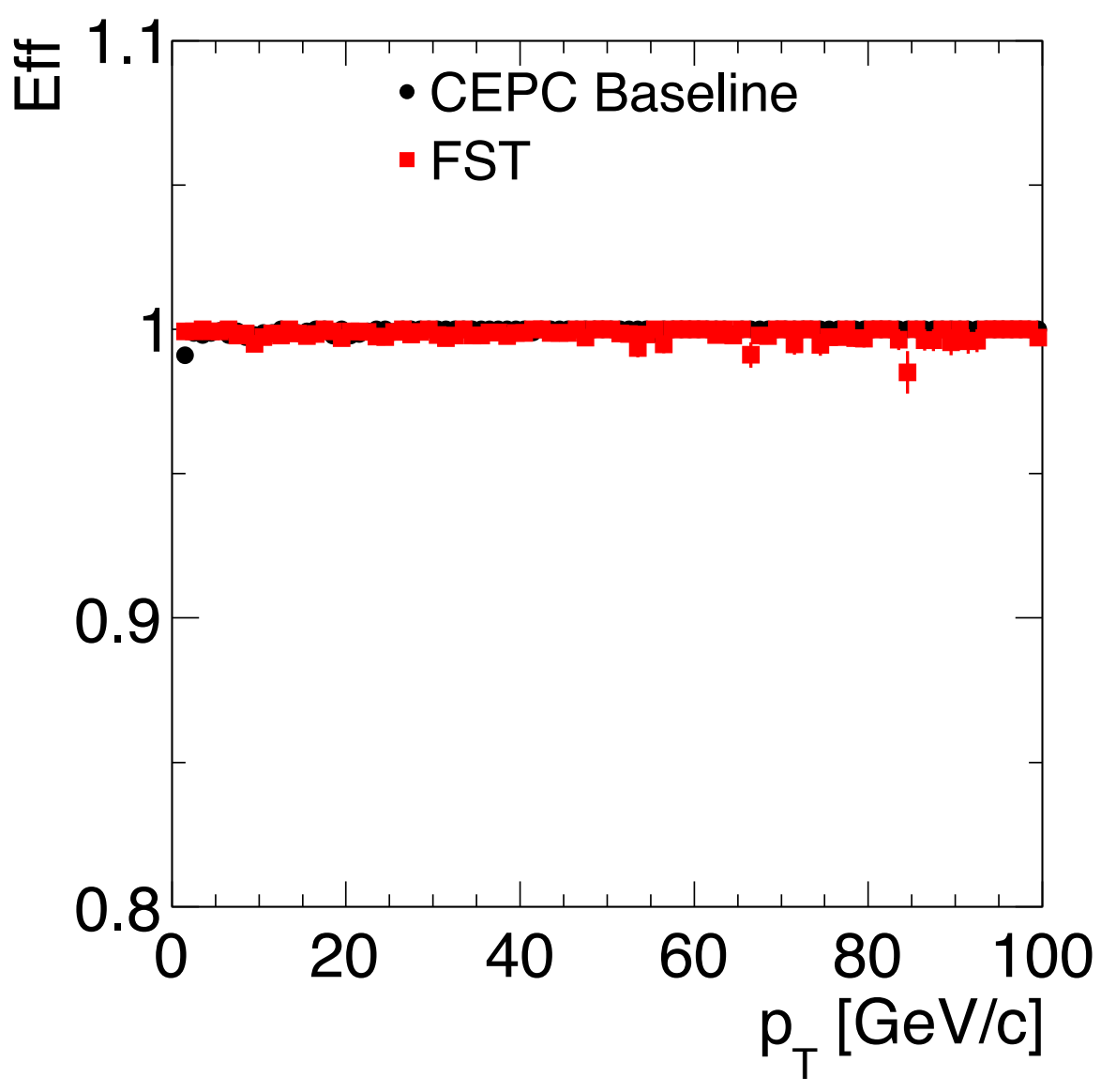

**Figure 4.27:** The tracking efficiencies as functions of $p_T$ measured using single muons for the CEPC baseline and the FST option.

### 4.4.1 INTRODUCTION

The special feature of this drift chamber is its high transparency, in terms of radiation lengths, obtained thanks to the novel approach adopted for the wiring and assembly procedures. The design concept originated with the KLOE experiment [55], and more recently culminated in the realization of the MEG2 [56] drift chamber. As implemented here for the CEPC main tracker, the total amount of material in radial direction, towards the barrel calorimeter, is of the order of $1.6\%\,X_0$, whereas, in the forward and backward directions, this is equivalent to about $5.0\%\,X_0$, including the endplates instrumented with front end electronics. The high transparency is particularly relevant for precision electroweak physics at the $Z$ pole and for flavor physics, where the average charged particles momenta are in a range over which the multiple scattering contribution to the momentum measurement is significant.

The DCH is a unique volume, high granularity, all stereo, low mass cylindrical drift chamber, co-axial to the 2 T solenoid field. It extends from an inner radius $R_{in}$ = 0.35 m to an outer radius $R_{out}$ = 2 m, for a length $L$ = 4 m and consists of 112 co-axial layers, at alternating sign stereo angles (in the range from 50 mrad to 250 mrad), arranged in 24 identical azimuthal sectors. The square cell size (5 field wires per sense wire) varies between 12.0 and 14.5 mm for a total of 56,448 drift cells. Thanks to the peculiar design of the wiring procedures, successfully applied to the recent construction of the MEG2 drift chamber, such a large number of wires poses no particular concern.

A system of tie-rods directs the wire tension stress to the outer endplate rim, where a cylindrical carbon fiber support structure bearing the total load is attached. Two thin carbon fiber domes, suitably shaped to minimize the stress on the inner cylinder and free to deform under the gas pressure without affecting the wire tension, enclose the gas volume. The angular coverage, for infinite momentum tracks originated at the interaction point and efficiently reconstructed in space, extends down to approximately $13°$. In order to facilitate track finding, the sense wires are read out from both ends to allow for charge division and time propagation difference measurements. The chamber is operated with a very light gas mixture, 90%He-10%i$C_4H_{10}$, corresponding to about 400 ns maximum drift time for the largest cell size. The number of ionization clusters generated by a Minimum Ionizing Particle (MIP) in this gas mixture is about 12.5 $cm^{-1}$, allowing for the exploitation of the cluster counting/timing techniques for improving both spatial resolution ($\sigma_x < 100\,\mu$m) and particle identification ($\sigma(dN_{cluster}/dx)/(dN_{cluster}/dx) \approx 2\%$).

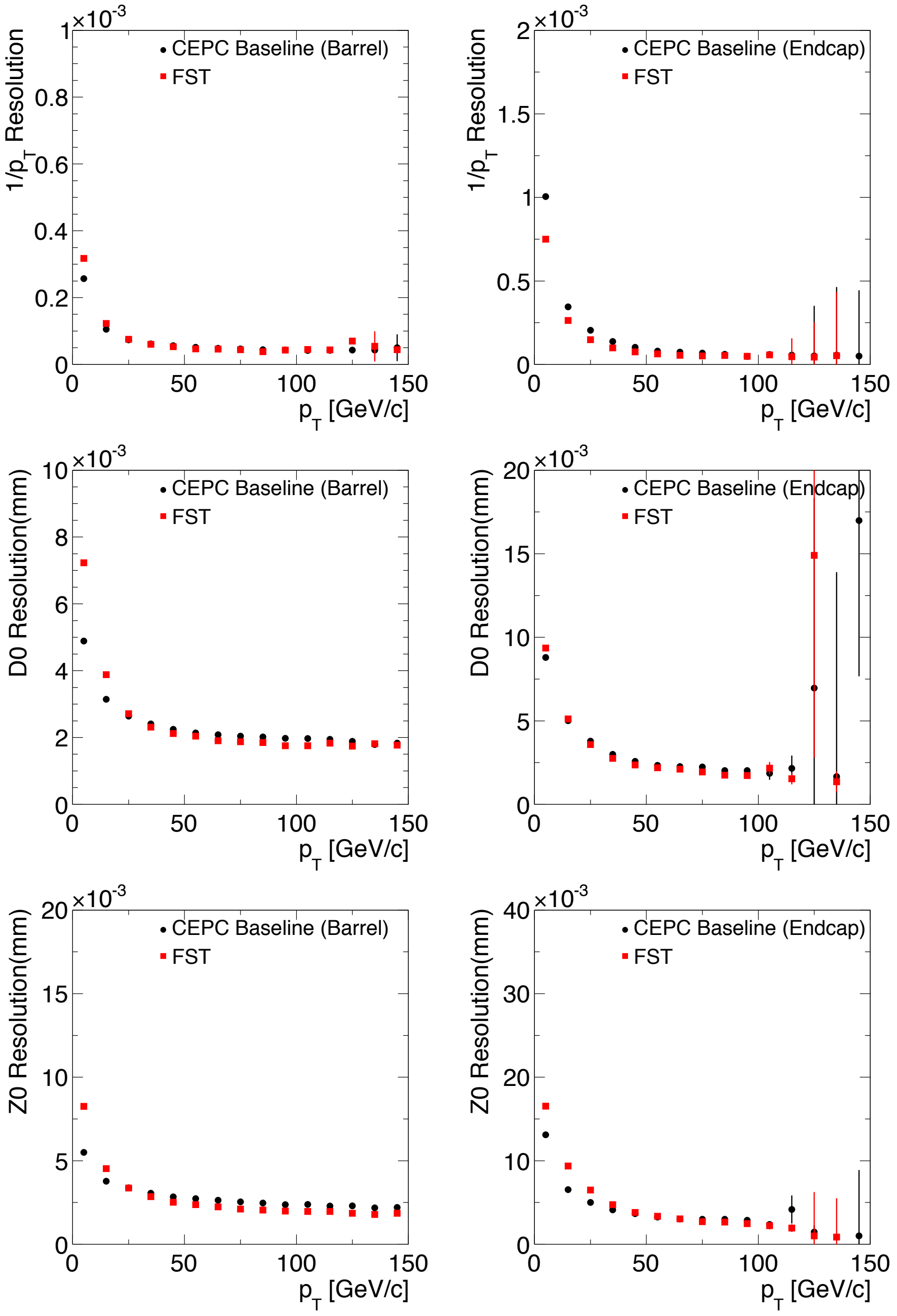

**Figure 4.28:** The track $p_T$, $d_0$, and $z_0$ resolutions are as functions of $p_T$ measured using single muons in the barrel region (left) and endcap region (right), comparing the performance of the CEPC baseline and the FST options. The FST option shows slightly worse resolutions at low momentum due to its extra material.

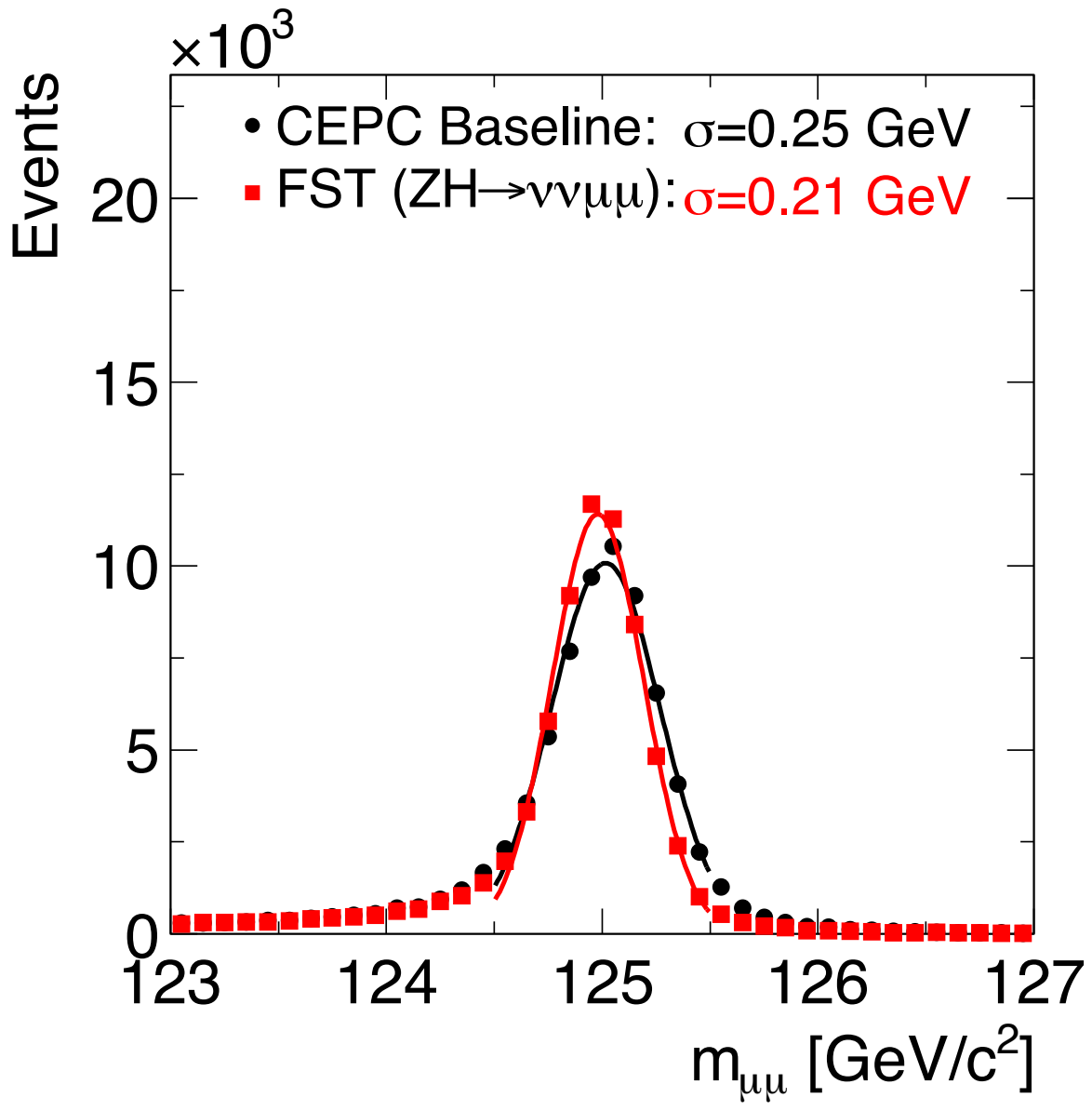

**Figure 4.29:** Comparison of the dimuon mass distributions from the $ZH \to \nu\bar{\nu}\mu^+\mu^-$ events of the CEPC baseline and the FST option.

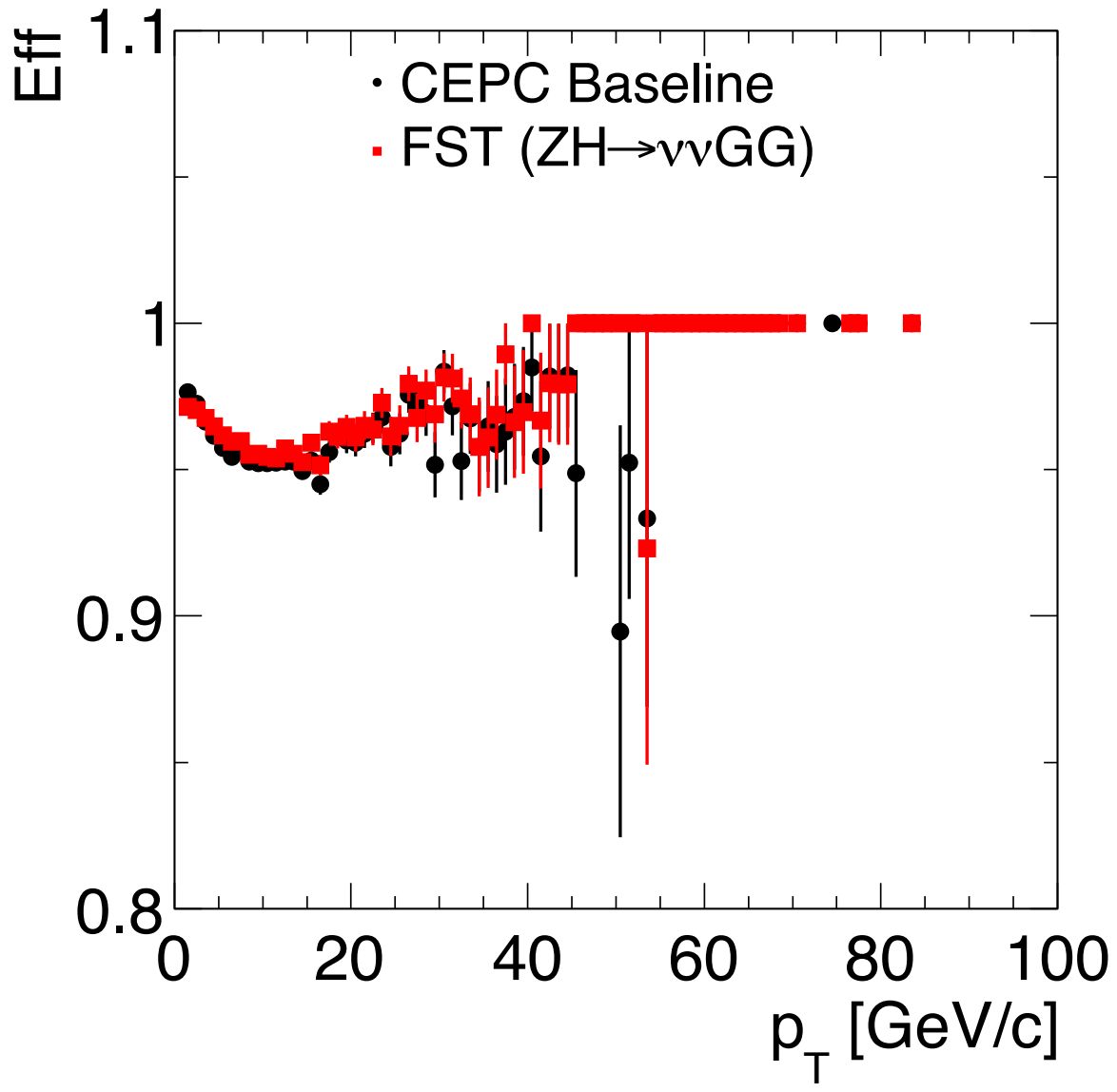

**Figure 4.30:** The tracking efficiencies for the stable charged particles inside the gluon jets from the $ZH \to \nu\bar{\nu}gg$ events as functions of track $p_T$ for the CEPC baseline and the FST. The loss of efficiency for track momentum around 10 GeV is under investigation.

### 4.4.2 EXPECTED PERFORMANCE

The expected performance of the drift chamber has been studied with a MEG2 drift chamber prototype with 7 mm cell size and very similar electrostatic configuration and gas mixture [57]. Figure 4.31 indicates a 100 $\mu$m drift distance resolution, averaged over all drift times. A better resolution is expected for the drift chamber proposed here because of the longer drift distances. Cluster timing technique may further improve it. Analytical calculations for the expected transverse momentum and angular resolutions are plotted in Figure 4.32.

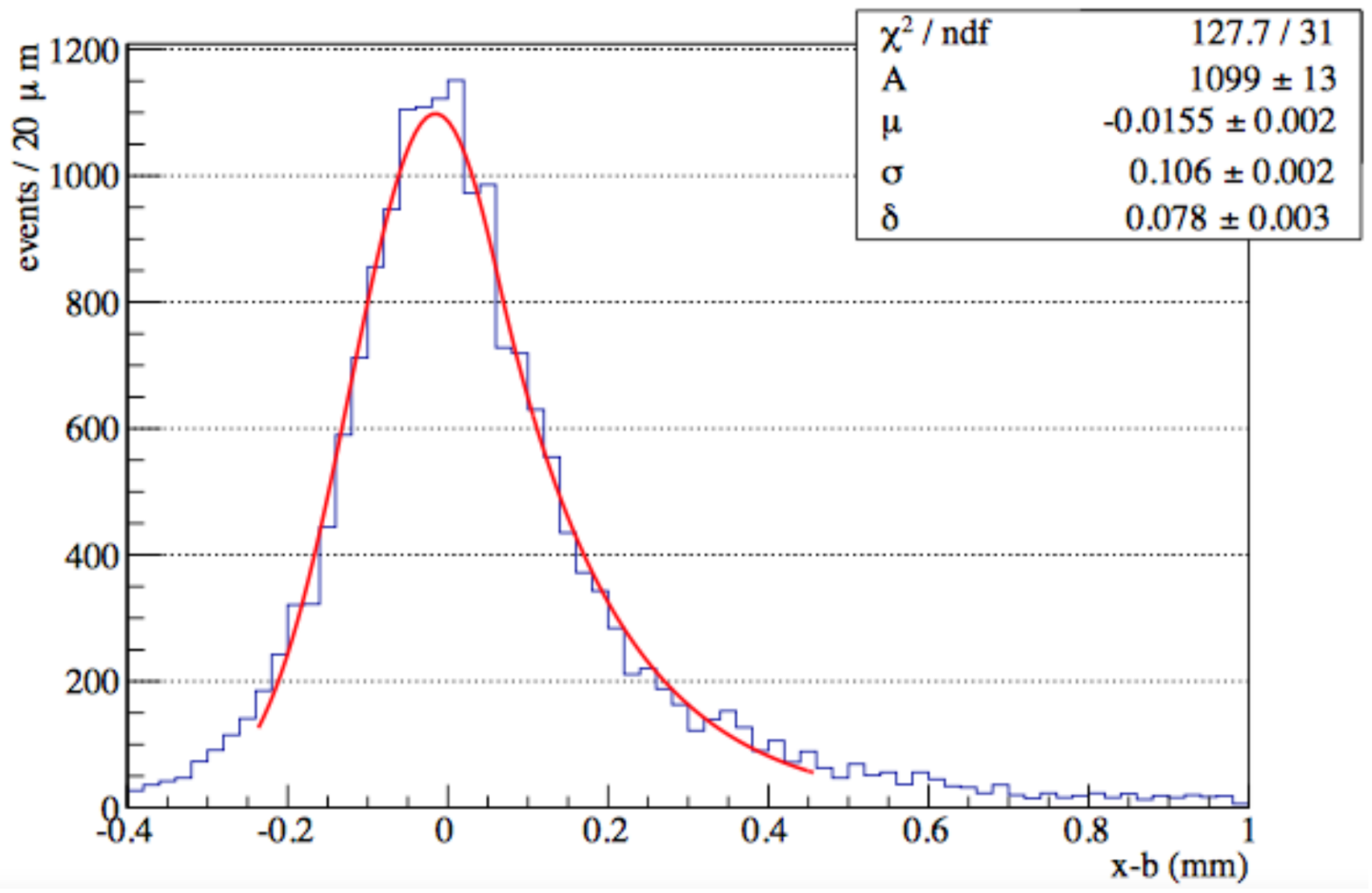

**Figure 4.31:** Measured drift distance residual distribution in the MEG2 drift chamber prototype using a 85%He-15%i$C_4H_{10}$ gas mixture. Cosmic rays tracks indicate a resolution better than 110 $\mu$m, averaged over all drift times and in a wide range of track angles.

Based on the assumption that one can, in principle, reach a relative resolution on the measurement of the number of primary ionization clusters, $N_{cl}$, equal to $1/\sqrt{N_{cl}}$, the expected performance relative to particle separation in number of units of standard deviations is presented in Figure 4.33 as a function of the particle momentum. Solid curves refer to cluster counting technique applied to a 2 m track length with 80% cluster identification efficiency and negligible (a few percent) fake clusters contamination. Dashed curves refer to the best theoretical prediction attainable with the $dE/dx$ technique for the same track length and same number of samples. For the whole range of momenta, particle separation with cluster counting outperforms $dE/dx$ technique by more than a factor of two, estimating an expected pion/kaon separation better than three standard deviations for all momenta below 850 MeV and slightly above 1.0 GeV.

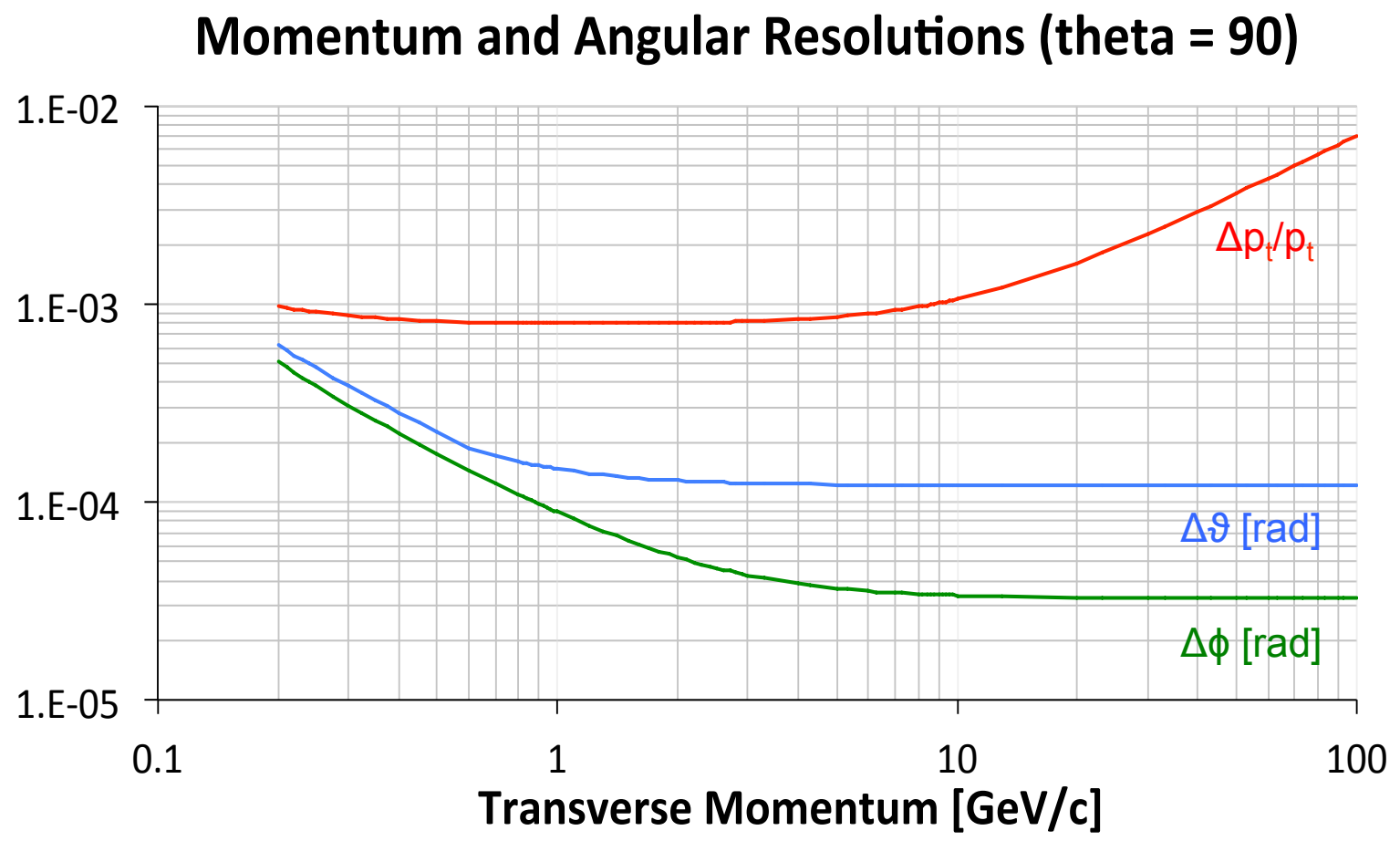

**Figure 4.32:** Analytical calculations for the expected transverse momenta (red) and angular resolutions (blue/green) as a function of the particle momentum for $\theta = 90°$.

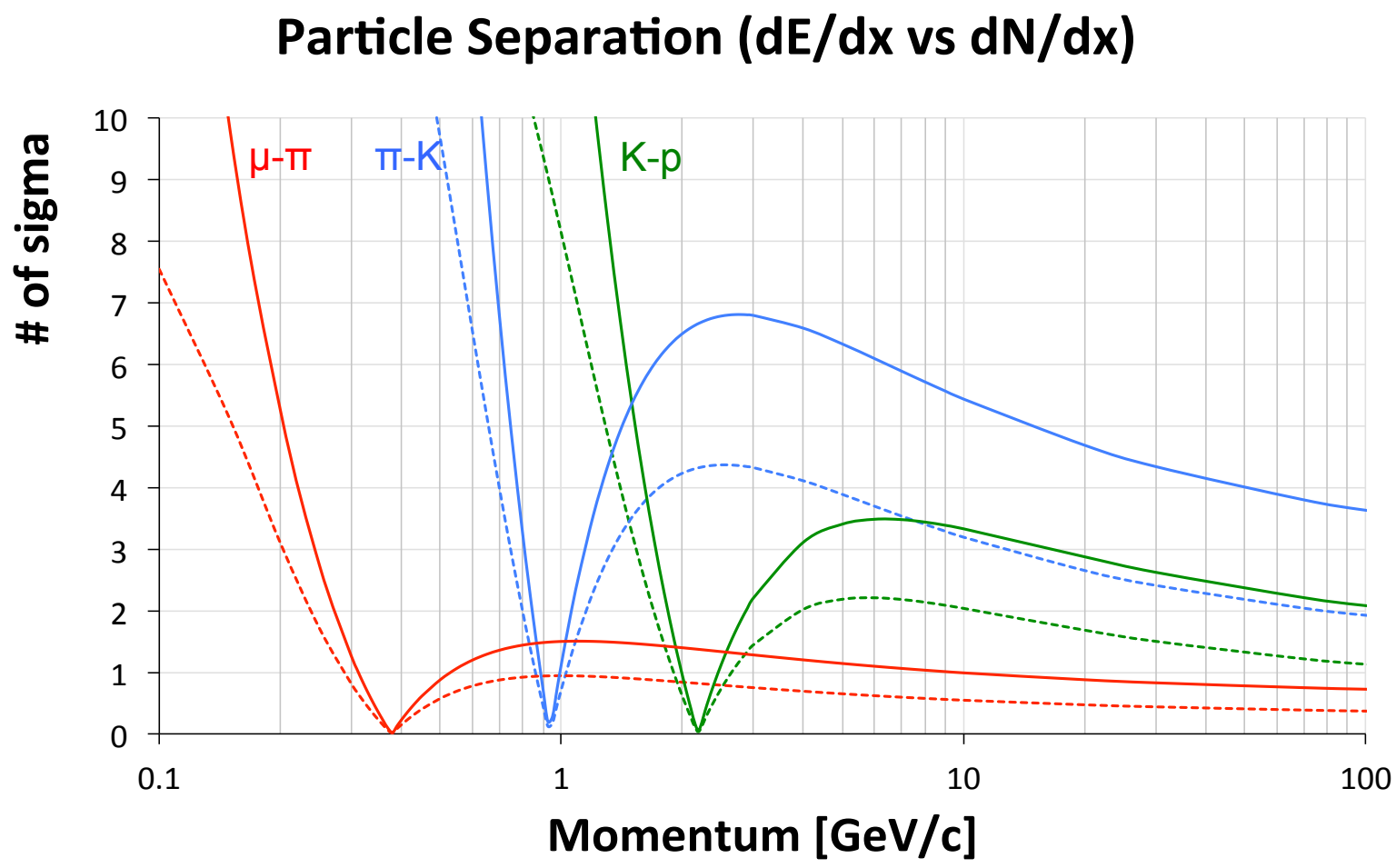

**Figure 4.33:** Particle type separation in units of standard deviations, with cluster counting (solid lines) and with $dE/dx$ (dotted lines) as a function of the particle momentum. A cluster counting efficiency of 80% and a $dE/dx$ resolution of 4.2% have been assumed.

## 4.4.3 TRACKING SYSTEM SIMULATION RESULTS

For the purpose of optimizing the track reconstruction performance, a vertex detector (different from the baseline choice) made of seven cylindrical layers, inside the drift chamber inner radius, and of five forward disks, has been simulated together with a layer of silicon microstrip detectors surrounding the drift chamber both in the barrel and in the forward regions, followed by a preshower detector system within a homogeneous 2 T longitudinal magnetic field. Details of ionization clustering for cluster counting/timing analysis have not been included in the simulations, limiting the drift chamber performance both in spatial resolution (a 100 $\mu$m Gaussian smeared point resolution has been assumed) and in particle separation (no $dN_{cl}/dx$ analysis has been simulated). A simplified track finding algorithm in its preliminary stage of development has been used to feed the space points to the fitter interface for the ultimate track fit. Figure 4.34 shows the momentum and angle resolutions as a function of the track momentum obtained by the tracking system simulation. No optimization has been tried yet. Momentum resolutions $\Delta p/p = 4 \times 10^{-3}$ at $p = 100$ GeV, for $\theta = 65°$, and angular resolutions $\leq 0.1$ mrad for $p \geq 10$ GeV, are within reach. Lastly, Figure 4.34 shows the $z_0$ and $d_0$ impact parameter resolutions obtained from the same simulation. A fit to the right plot in Figure 4.35 gives a $d_0$ impact parameter resolution:

$$\sigma_{d_0} = a \oplus \frac{b}{p \sin^{3/2} \theta}$$

with $a = 3$ µm and $b = 15$ µm $\cdot$ GeV$/c$.

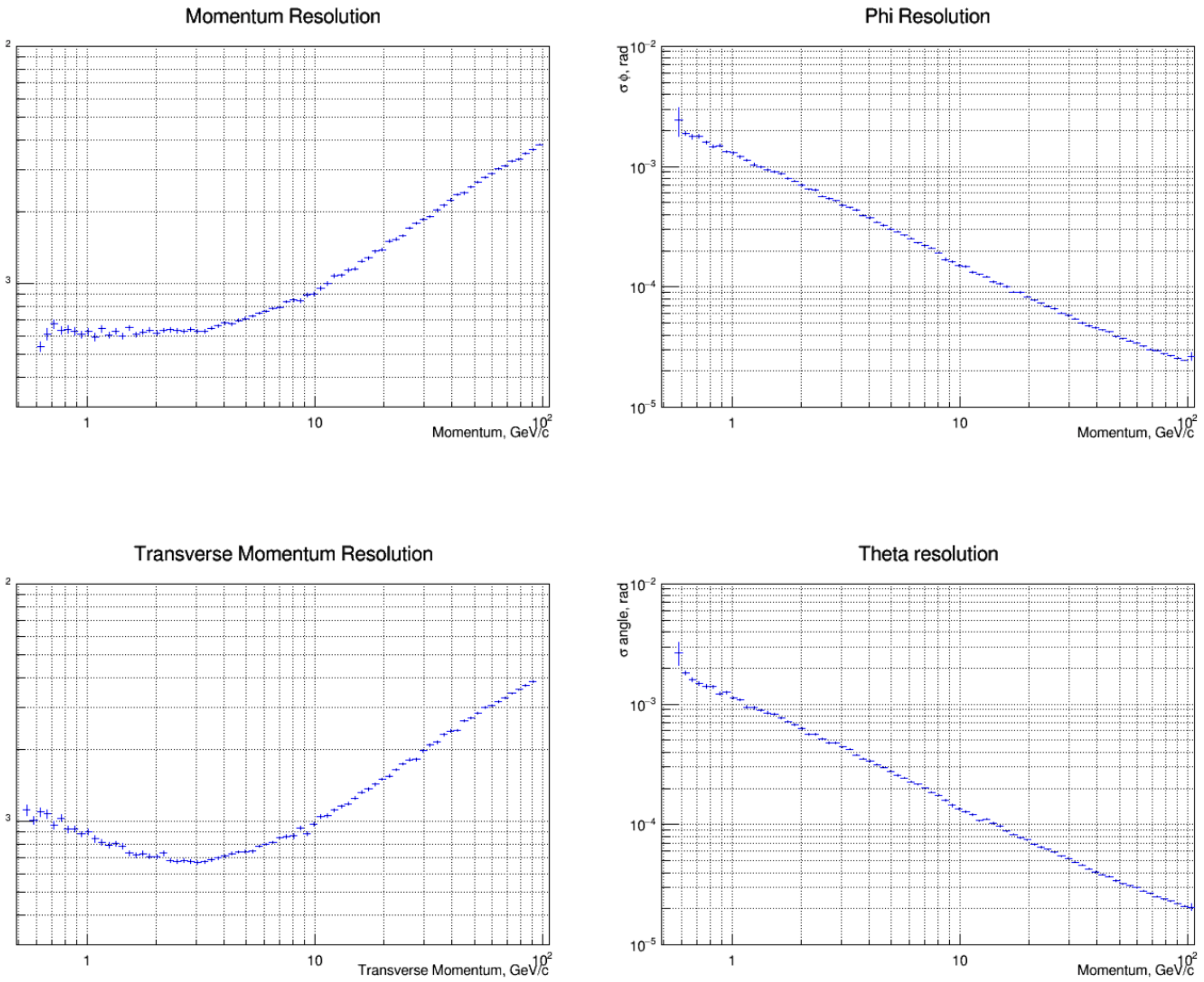

**Figure 4.34:** Momentum and transverse momentum resolutions (left) and angular $\phi$ and $\theta$ resolutions (right) (in µm) versus track momentum, from simulation of isolated tracks and a preliminary track finding algorithm.

### 4.4.4 BACKGROUNDS IN THE TRACKING SYSTEM

The main sources of backgrounds in the tracking system come from incoherent pair production, synchrotron radiation and $\gamma\gamma$ to hadrons. The incoherent pair production background is dominant among these, however, only very few of the primary $e^{\pm}$ particles will have a transverse momentum large enough to reach the inner radius of the drift cham-ber. The majority of the hits will be generated by secondary particles (mainly photons of energy below 1 MeV) produced by scattering off the material at low radii. Based on experience from the very similar MEG2 drift chamber, which has a smaller number of hits per track and a much more complicated event topology, occupancies of up to several percent will not affect tracking efficiency and single track momentum resolution. The level of occupancy here is expected to be even smaller with the use of the drift chamber timing measurement. As opposed to charged particles that leave a string of ionization in the drift cells they traverse, photons are characterized by a localized energy deposition. Signals from photons can therefore be effectively suppressed at the data acquisition level by requiring that a threshold be reached by the number of ionization clusters within a reasonable time window. In addition, charge strings with holes longer than the average cluster separation can be interpreted as due to separate signals, thus avoiding piling up of any remaining photon induced background. With this effective suppression of photon induced signals, the background from incoherent pair production is expected to remain low and is unlikely to cause adverse issues for the track reconstruction.

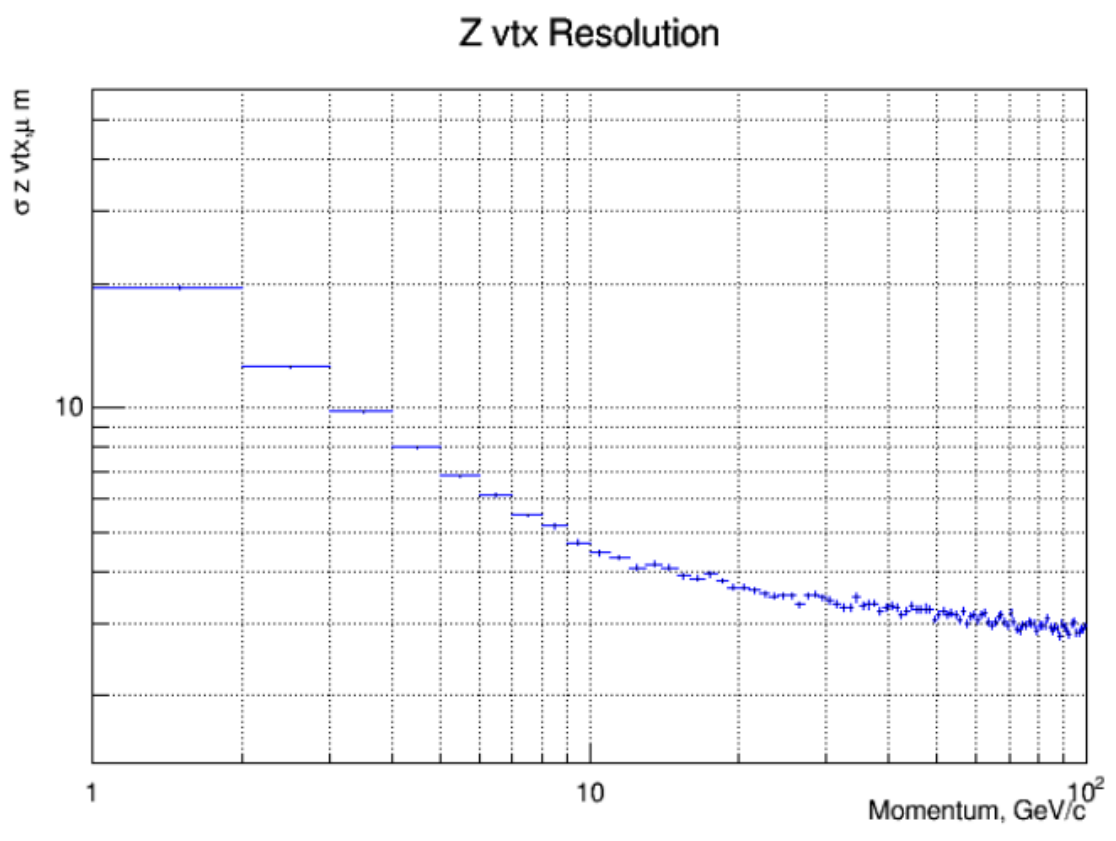

(a)

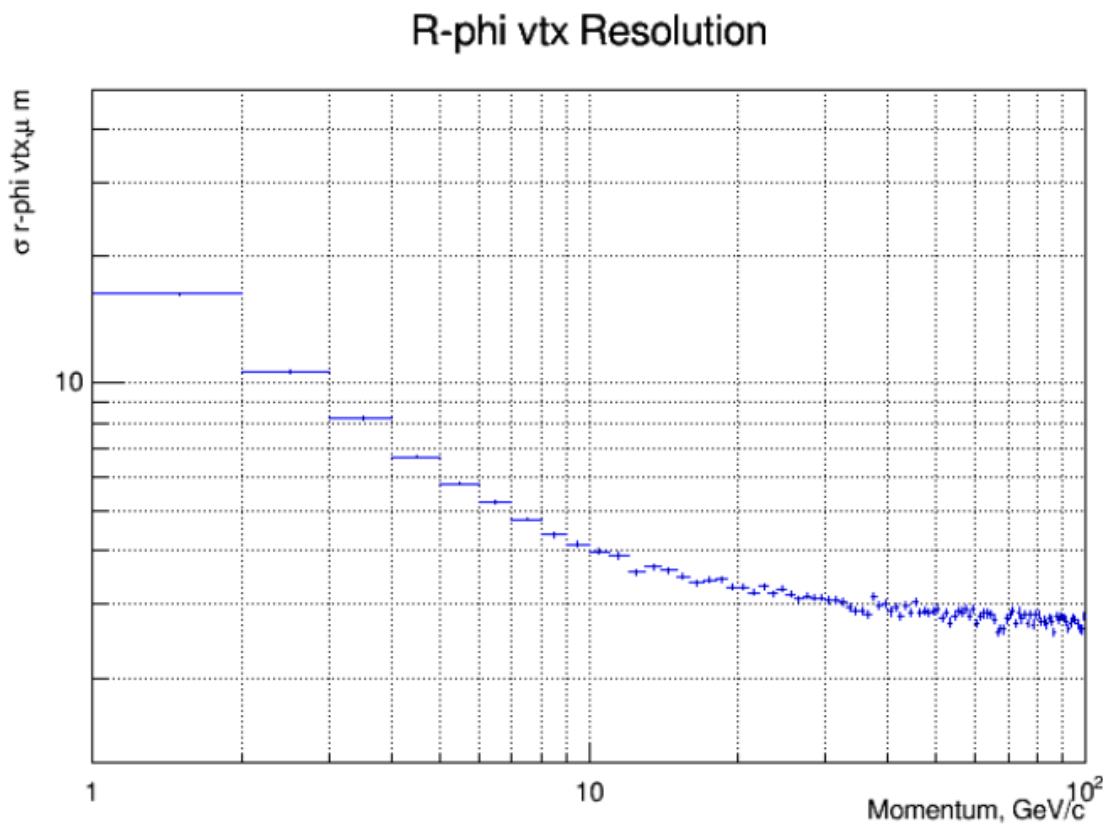

(b)

**Figure 4.35:** Impact parameters, $z_0$ and $d_0$, resolution (in $\mu$m) versus track momentum, from simulation of isolated tracks and a preliminary track finding algorithm.

### 4.4.5 CONSTRAINTS ON THE READOUT SYSTEM

With a drift chamber, all digitized hits generated at the occurrence of a trigger are usually transferred to data storage. The IDEA drift chamber transfers 2 B/ns from both ends of all wires hit, over a maximum drift time of 400 ns. With 20 tracks/event and 130 cells hit for each track, the size of a hadronic $Z$ decay in the DCH is therefore about 4 MB, corresponding to a bandwidth of 40 GB/s at the $Z$ pole (at a trigger rate of approximately 10 kHz). The contribution from $\gamma\gamma$ to hadrons amounts to 6 GB/s. As mentioned in the previous paragraph, the incoherent pair production background causes the read-out of additional 1400 wires on average for every trigger, which translates into a bandwidth of 25 GB/s. A similar bandwidth is taken by the noise induced by the low single electron detection threshold necessary for an efficient cluster counting. Altogether, the various contributions sum up to a data rate of about 0.1 TB/s. Reading out these data and sending them into an "event builder" would not be a challenge, but the data storage requires a large reduction. Such a reduction can be achieved by transferring, for each hit drift cell, the minimal information needed by the cluster timing/counting, i.e., the amplitude and the arrival time of each peak associated with each individual ionization electron, each encoded in 1 Byte, instead of the full signal spectrum. The data generated by the drift chamber, subsequently digitized by an ADC, can be analyzed in real time by a fast read-out algorithm implemented in a FPGA [58]. This algorithm identifies, in the digitized signal, the peaks corresponding to the different ionization electrons, stores the amplitude and the time for each peak in an internal memory, filters out spurious and isolated hits and sends these reduced data to the acquisition system at the occurrence of a trigger. Each hit cell integrates the signal of up to 30 ionization electrons, which can thus be encoded within 60 B per wire end instead of the aforementioned 800 B. Because the noise and background hits are filtered out by the FPGA algorithm, the data rate induced by $Z$ hadronic decays is reduced to 3 GB/s, for a total bandwidth of about 3.6 GB/s, roughly a factor 30 reduction.

CHAPTER 5

# CALORIMETRY

A calorimetry system is employed in the CEPC detectors to provide hermetic coverage for high-resolution energy measurements of electrons, photons, taus and hadronic jets. Section 5.1 provides an overview of the calorimetry systems being considered. Two distinct approaches are being pursued: particle-flow and dual-readout calorimeters. The current baseline detector concept adopts the particle flow approach. Section 5.2 introduces design considerations for these calorimeters. It is followed by the description of the corresponding particle flow oriented electromagnetic calorimeter (ECAL) in Section 5.3, and hadronic calorimeter (HCAL) in Section 5.4. Different technology options are described in both cases. The technology options that have been integrated into the full detector simulation are silicon-tungsten for the ECAL and steel-GRPC for the HCAL. The dual-readout calorimeter concept, described in Section 5.5, is an integral part of IDEA, the alternative detector concept for the CEPC.

## 5.1 OVERVIEW

To fully exploit the potential of the CEPC physics program for Higgs and electroweak physics, all possible final states from decays of the intermediate vector bosons, $W$ and $Z$, and the Higgs boson need to be separately identified and reconstructed with high sensitivity. In particular, to clearly discriminate the $H \to ZZ^* \to 4j$ and $H \to WW^* \to 4j$ final states, the energy resolution of the CEPC calorimetry system for hadronic jets needs to be pushed quite beyond today's limits. Indeed, in order to distinguish the hadronic decays of $W$ and $Z$ bosons, a 3–4% invariant mass resolution for two-jet systems is required. Such a performance needs a jet energy resolution of $\sim 30\%/\sqrt{E}$, at energies below 100 GeV. This would be about a factor of two better than that of the LEP detectors and the currently operating calorimeters at the LHC, and would significantly improve the separation of the $W$ and $Z$ bosons in their hadronic decays, as shown in Figure 5.1. Two different technology approaches are pursued for the CEPC calorimetry system, the first one aiming to measure individual particles in a jet using a calorimetry system with very high granularity based on the particle flow concept, while the second aiming at a homogeneous and integrated solution based on the dual-readout concept. Both approaches will be described in this chapter, while the first approach is the current baseline for the design of the CEPC calorimetry system in that it is integrated in the full CEPC detector simulation.

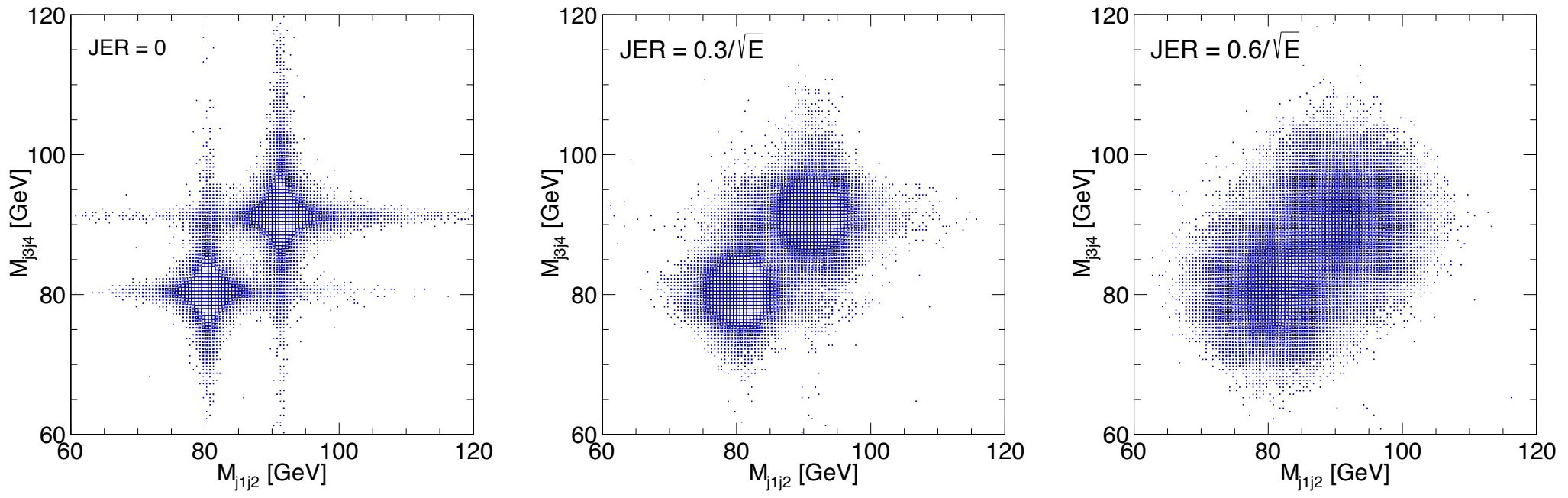

(a) (b) (c)

**Figure 5.1:** Separation of $W$ and $Z$ bosons in their hadronic decays with different jet energy resolutions: (a) $0/\sqrt{E}$, (b) $30\%/\sqrt{E}$, and (c) $60\%/\sqrt{E}$. A jet energy resolution of $30\%/\sqrt{E}$ is required to separate the hadronic decays of $W$ and $Z$ bosons.

The particle flow algorithm (PFA [1]) is a very promising approach to achieve the unprecedented jet energy resolution of 3–4%. The basic idea of the PFA is to make use of the optimal detector subsystem to determine the energy/momentum of each particle in a jet. An essential prerequisite for realization of this idea is to distinguish among energy deposits of individual particles from a jet in the calorimetry system. High, three-dimensional spatial granularity is required for the calorimetry system to achieve this. Therefore, PFA calorimeters feature finely segmented, three-dimensional granularity and compact, spatially separated, particle showers to facilitate the reconstruction and identification of every single particle shower in a jet. It is for this feature PFA calorimeters are usually also called imaging calorimeters. A PFA calorimetry system generally consists of an electromagnetic calorimeter (ECAL), optimized for measurements of photons and electrons, and a hadronic calorimeter (HCAL) to measure hadronic showers.

In a typical jet, 65% of its energy is carried by charged particles, 25% by photons, and 10% by neutral hadrons. The charged particles in a jet can be precisely measured with a tracking system, especially for low momentum particles where the relatively small, multiple scattering term dominates in the resolution, and their tracks can be matched to their energy deposits in a PFA calorimetry system. This combination maximizes the overall resolution of the jet energy measurement by compensating for the worsening of calorimeter-only energy resolution for low energy particles by leveraging the improved resolution from the tracking system. Energy deposits in the PFA calorimetry system without matched tracks are considered to originate from the neutral particles of photons and neutral hadrons in the jet. Among these neutral particles, photons are measured using the ECAL with good energy resolution, while only the neutral hadrons are primarily measured using a combination of the ECAL and HCAL with a limited energy resolution. Therefore, in the PFA, the jet energy is determined by combining the best measurement in a detector system of each single particle in the jet: the track momenta of charged particles measured using the tracking system, the energies of photons measured using the ECAL and the energies of neutral hadrons measured primarily using the HCAL.

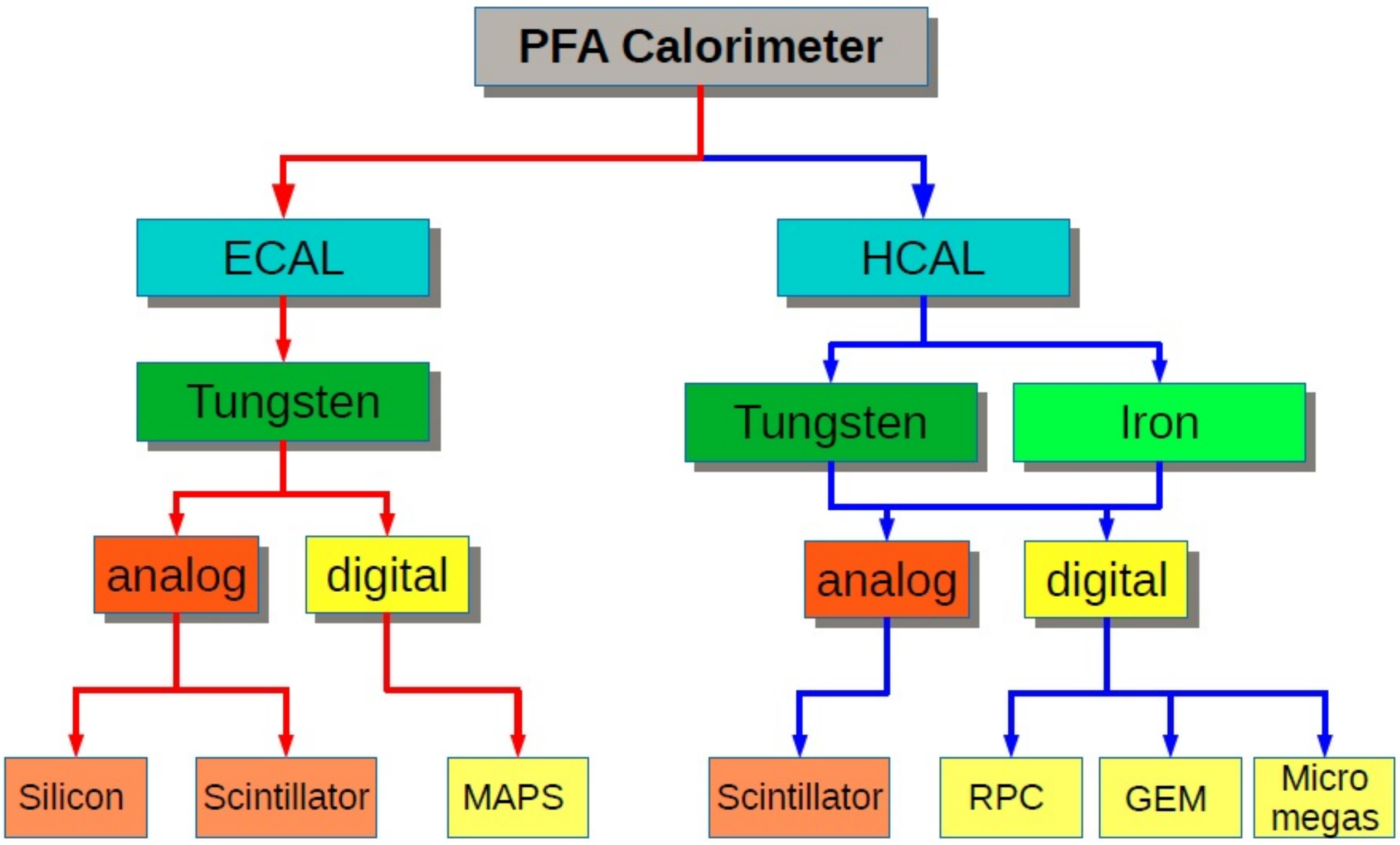

**Figure 5.2:** : Overview of world-wide development of imaging calorimeters. Various technology options have been explored in aspects including absorber material, active medium and readout scheme.

Extensive studies have been carried out within the CALICE collaboration and in the world-wide detector R&D efforts for the ILC [2, 3] to develop compact PFA calorimeters. Various detector technology options have been explored to address challenges from stringent performance requirements as shown in Figure 5.2. Prototypes with high granularity using several technological options have been developed and exposed to particle beams, which have demonstrated the in-depth understanding of the PFA calorimetry performance.

An alternative approach for a combined, high-performance, electromagnetic and hadronic calorimeter aims at reaching an even better (standalone) hadronic resolution, without spoiling the electromagnetic one, by exploiting the dual-readout (DR) technique. Indeed the main limiting factor to the energy resolution in hadron calorimetry arises from the fluctuations of the electromagnetic component ($f_{em}$) that each hadronic shower develops as consequence of $\pi^0$ and $\eta$ production. Since typically the detector response to the hadronic and $em$ components is very different ($h/e \neq 1$), the reconstructed signal heavily depends on the actual value of $f_{em}$. By using two independent processes (namely, scintillation and Čerenkov light production) that have a very different sensitivity to the hadronic and $em$ components, it is possible to reconstruct $f_{em}$, event by event, and minimize the effects of its fluctuations.

Among the possible DR implementations, a fiber-sampling DR calorimeter, based on either copper or lead as absorber material, looks the most suitable to provide the required performance in a cost-effective way. Preliminary results of GEANT4 simulations point to possible resolutions better than $15\%$ and around 30–40%/$\sqrt{E}$, for electromagnetic and hadronic showers, respectively (see Section 5.5.6).

Moreover, if the fibers are readout with Silicon PhotoMultipliers (SiPMs), the high detector granularity and the possibility of longitudinal segmentation will make this solution easily compatible with particle flow algorithms.

In the following sections, several possible concrete implementations of a calorimeter system are discussed in sufficient detail to describe the readiness of the technologies and the performance of these systems in current test beams and prototypes and their corresponding general implementation in the simulation performance studies of the physics objects and benchmarks presented in subsequent chapters.

## 5.2 DESIGN CONSIDERATIONS FOR THE PFA CALORIMETRY SYSTEM

The CEPC PFA calorimetry system is composed of two separate sampling calorimeters: ECAL and HCAL, installed inside the bore of the solenoid to minimize the amount of inactive material in front of the calorimetry system and to enable the reliable association of tracks with energy deposits. Both ECAL and HCAL are organized into one cylindrical barrel and two disk-like endcap sections.

The ECAL consists of sensitive layers of either silicon pads or scintillator tiles interleaved with tungsten absorber plates. It will be read out using analog signals. The HCAL has steel plates as the absorber. Both digital and analog readout are being considered. The Digital HCAL (DHCAL) uses either Glass Resistive Plate Chambers (GRPC) or Thick Gas Electron Multiplier detectors (THGEM) as the active medium, whereas the Analog HCAL (AHCAL) uses scintillator tiles as the active medium.

Driven by the PFA requirement of the excellent particle shower separation capability, the calorimeters will need to be finely segmented, in both transverse and longitude, regardless the design options. The baseline technology options that have been integrated into the full CEPC detector simulation are silicon-tungsten for the ECAL and steel-GRPC for the HCAL.

In the baseline design, the ECAL is segmented into 30 longitudinal layers with a total tungsten thickness of 84 mm, or equivalently $24X_0$. It is split longitudinally into 2 sections with different absorber layer thickness. The first section contains 20 layers of 2.1 mm thick tungsten plates while the second has 10 layers of 4.2 mm tungsten plates (see section 5.3.1 for details). The silicon pads sandwiched in tungsten plates are each 0.5 mm thick and have a size of $10 \times 10\ \mathrm{mm}^2$. The HCAL consists of 40 longitudinal layers each containing 2 cm thick steel with a thin layer of GRPC readout in a cell size of also $10 \times 10\ \mathrm{mm}^2$. These design parameters are the result of dedicated optimization studies described below.

## 5.3 PARTICLE FLOW ORIENTED ELECTROMAGNETIC CALORIMETER

The particle flow paradigm has tremendous impact on the design of the ECAL. With excellent capability for pattern recognition, the ECAL is expected to identify photons from close-by showers, reconstruct detailed properties of a shower (i.e. shower shape, starting point and energy distribution), and distinguish electromagnetic showers from hadronic ones. For a PFA-oriented ECAL, the shower imaging capability is more important than its intrinsic energy resolution, although the latter is still important to the particle flow performance for electrons, photons and jets. Because about half of the hadronic showers start inside the ECAL, a high three-dimensional granularity is of primary importance to the ECAL performance. In order to separate close-by showers in the calorimeter, absorber material with small Moliere radius is required. Moreover, a large ratio of the interaction length over the radiation length of the absorber material is advantageous for the separation of the electromagnetic and hadronic showers. This is because a short radiation length makes an electromagnetic shower starting early in the ECAL, while a long interaction length reduces the fraction of a hadronic shower in the ECAL. A short radiation length also makes a compact ECAL, reducing the overall cost of the detector.

The requirements for a high granularity and compact ECAL with an excellent capability for shower separation lead to the choice of tungsten (the radiation length $X_0 = 3.5$ mm, the Molière radius = $9$ mm, and the interaction length $\lambda_{\mathrm{I}} = 99$ mm) as the absorber material. Two options for the active medium are considered for the ECAL: silicon and scintillator. The silicon option is taken as the baseline, while the scintillator option is being investigated as an alternative. Both options are discussed in this section.

### 5.3.1 DESIGN OPTIMIZATION

The ECAL design parameters including the total thickness of the absorber, the thickness and transverse size of the silicon sensors, the number of sampling layers have been optimized based on simulation studies with a simplified and standalone ECAL geometry. There is no material in front of the ECAL and no gaps or dead area between detector modules in this geometry. In addition, all absorber layers have the same thickness, so do all sensitive layers.

The $H \to \gamma\gamma$ mass resolution for the $\nu\bar{\nu}H$ ($\to \gamma\gamma$) process was used as a figure of merit in optimizing the total thickness of the absorber. The total thickness was scanned by coherently varying the thickness of each absorber layer at a very fine step. The number of sampling layers was fixed to 30 and the thickness of silicon sensors was fixed to 0.5 mm in the scan. The result from the scan is shown in Figure 5.3. The best mass resolution is achieved when the total thickness is 84 mm. The number of sampling layers was then varied fixing the total tungsten thickness to 84 mm, and the photon energy resolution was examined as a function of energy for different numbers of sampling layers as shown in Figure 5.4(a). The absorber has a uniform thickness in all layers in this case. The energy resolution degrades by 11% and 26% when the number of sampling layers is reduced from 30 to 25 and 20, respectively. However, such degradation can be compensated by increasing the thickness of the silicon sensors. As shown in Figure 5.4(b), the energy resolution for 20 (25) layers with 1.5 mm (1 mm) thick silicon sensors is about the same as that for 30 layers with 0.5 mm thick sensors. Considering potential challenges of thick silicon sensors, 0.5 mm was chosen as the thickness of silicon sensors. To further improve the energy resolution of the ECAL, a configuration with varying absorber thicknesses for the 30 sampling layers was examined. The absorber is 2.1 mm thick for the first 20 layers and 4.2 mm for the last 10 layers in this configuration, and the photon energy resolution with this configuration is shown in Figure 5.4 (black curves) as a function of the photon energy. For energies smaller than 10 GeV, a relative improvement of 10% in the resolution can be observed using this absorber configuration, which is adopted for the ECAL baseline design.

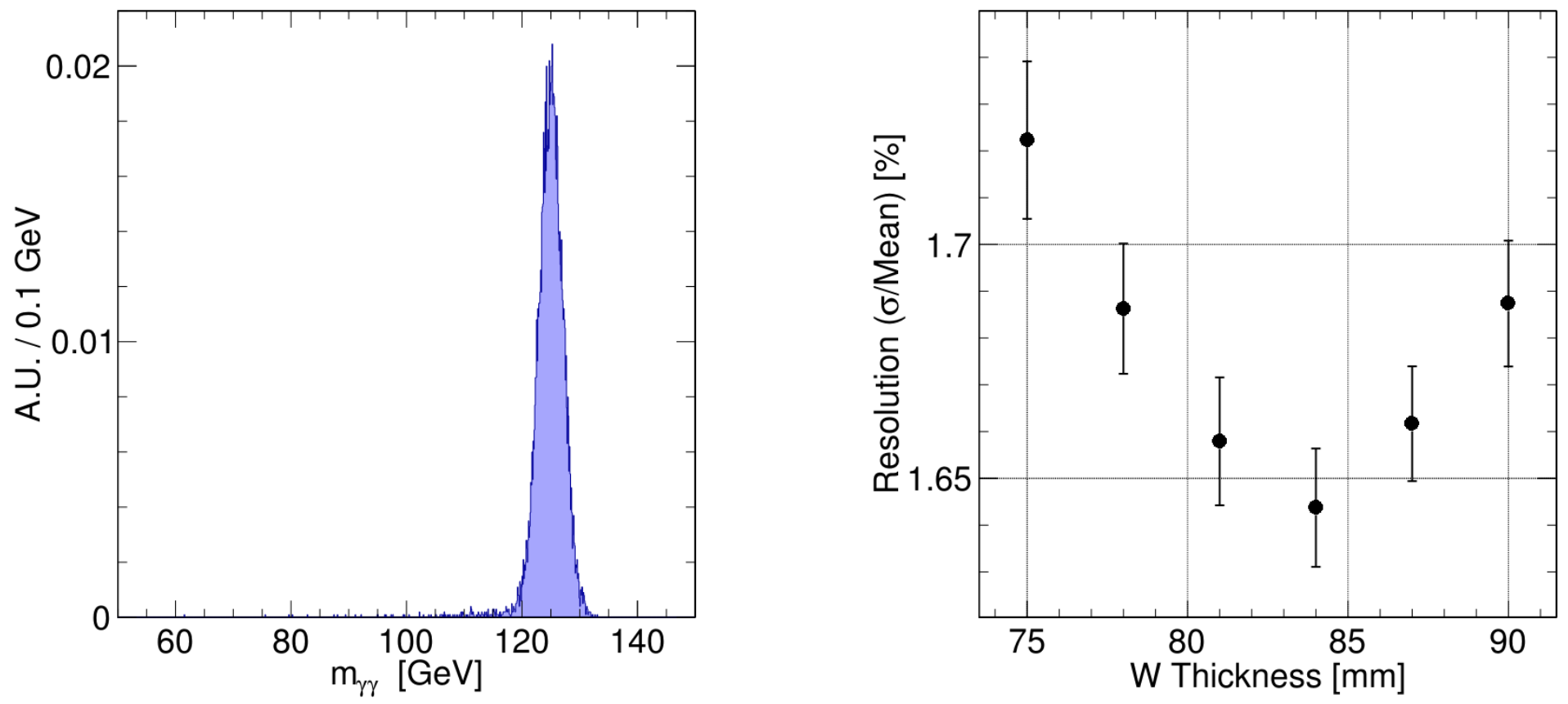

(a) (b)

**Figure 5.3:** (a) The diphoton invariant mass distribution of the $\nu\bar{\nu}H \to \nu\bar{\nu}\gamma\gamma$ events and (b) the relative mass resolution for different total thicknesses of the tungsten absorber of the ECAL. The best resolution is achieved when the total thickness is 84 mm.

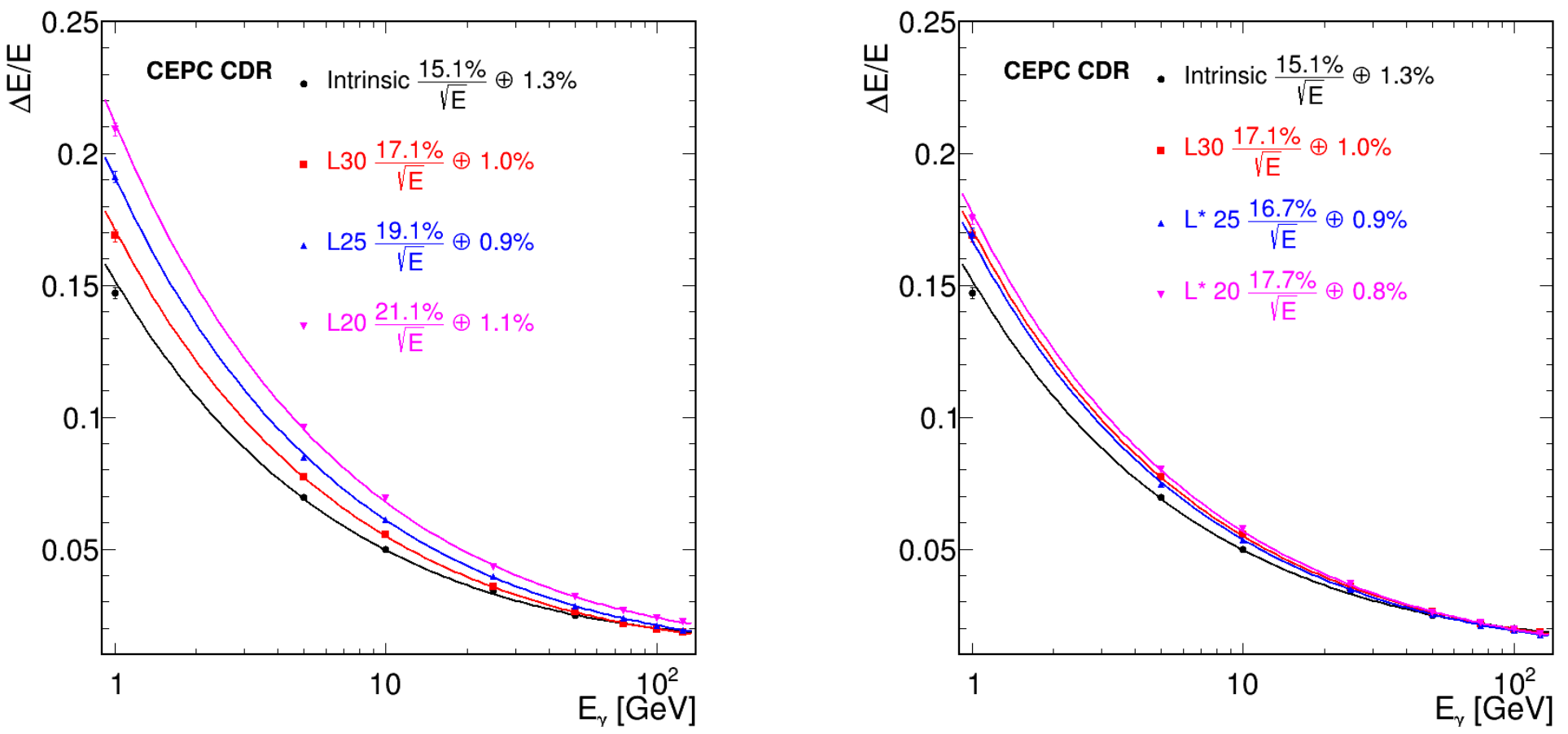

(a) (b)

**Figure 5.4:** Photon energy resolution of the silicon-tungsten ECAL as a function of photon energy for different numbers of sampling layers of the ECAL: 30 layers (black, red), 25 layers (blue), and 20 layers (magenta). In (a), a silicon sensor thickness of 0.5 mm is used for all cases. In (b), different silicon sensor thicknesses are used: 0.5 mm, 1.0 mm and 1.5 mm for the 30, 25 and 20 layers, respectively. The absorber has a uniform thickness in all layers for the magenta, blue and red curves, while the absorber is 2.1 mm thick for the first 20 layers and 4.2 mm for the last 10 layers of the baseline design for the black curves. Note that the resolutions are obtained for a standalone ECAL without the tracking material in front and dead areas between ECAL modules. Thus the resolutions are labeled as "intrinsic". Photons used in these studies are uniformly distributed over the $4\pi$ solid angle.

In the ECAL baseline design, silicon sensors have square shapes, and their transverse sizes directly determine the number of readout channels and have significant impact on the shower separation power. The former has strong implications on both the cost and power consumption of the ECAL. Therefore, it is attractive to have a large sensor size to reduce the number of readout channels as long as the physics performance is not significantly compromised. The sensor size is not expected to have a significant effect on the electromagnetic energy resolution, but could potentially impact the separation of photons inside jets and therefore affect jet energy measurement and the reconstruction of $\tau$ decay final states. The sensor size was optimized in two performance metrics: the mass resolution of the hadronic decays of the Higgs bosons and the capability of identifying photons from close-by showers.

The dijet mass distribution of the $\nu\bar{\nu}H \to \nu\bar{\nu}gg$ events was used to investigate the effect of the sensor size on the jet energy measurement. Showers induced by the photons in the gluon jets need to be well separated from the rest of the jet activities in the ECAL for the precise measurement of the jet energy and the dijet mass. The expected dijet mass resolutions for different sensor sizes are shown in Table 5.1. No significant degradation in the resolution is expected when the sensor size is varied from $5 \times 5\,\mathrm{mm}^2$ to $10 \times 10\,\mathrm{mm}^2$. A 5% degradation is seen when the size is increased to $20 \times 20\,\mathrm{mm}^2$.

| Silicon sensor size (mm$^2$) | Relative dijet mass resolution (%) (with statistic uncertainty) |
|---|---|
| $5 \times 5$ | 3.74 $\pm$ 0.02 |
| $10 \times 10$ | 3.75 $\pm$ 0.02 |
| $20 \times 20$ | 3.93 $\pm$ 0.02 |

**Table 5.1:** The relative dijet mass resolution for the $H \to gg$ decays of the $\nu\bar{\nu}H \to \nu\bar{\nu}gg$ events for different ECAL sensor sizes. No significant impact on the resolution is found.

The impact of the sensor size on the identification of close-by photons was evaluated using simulation of two parallel photons separated with different distances. Figure 5.5 shows the reconstruction efficiency of the two photons as a function of their separation for different sensor sizes. The efficiency reaches a plateau when the separation is approximately twice of the sensor size or larger. Thus a small ECAL sensor size is needed to separate close-by photon showers. This is particularly relevant for the identification of $\tau$-leptons which often have multiple photons in their decays. Table 5.2 presents the fractions of photons in the decays of the $\tau$-leptons of the $Z \to \tau\tau$ events that failed to be reconstructed due to overlapping showers in the ECAL for different sensor sizes. The fraction remains low for sensor sizes up to $10 \times 10\,\mathrm{mm}^2$, but jumps to 20% for a sensor size of $20 \times 20\,\mathrm{mm}^2$.

These studies suggest that a good ECAL performance can be maintained with a sensor size as large as $10 \times 10\,\mathrm{mm}^2$. The above studies were performed for the silicon-tungsten ECAL option. The same set of studies was also carried out for the scintillator-tungsten option and consistent results were obtained. However, much thick scintillator sensors are preferred in this case.

### 5.3.2 SILICON-TUNGSTEN SANDWICH ECAL

#### 5.3.2.1 SILICON SENSORS

Among several sensor technologies, silicon PIN diodes with high resistivity offer several unique intrinsic advantages:

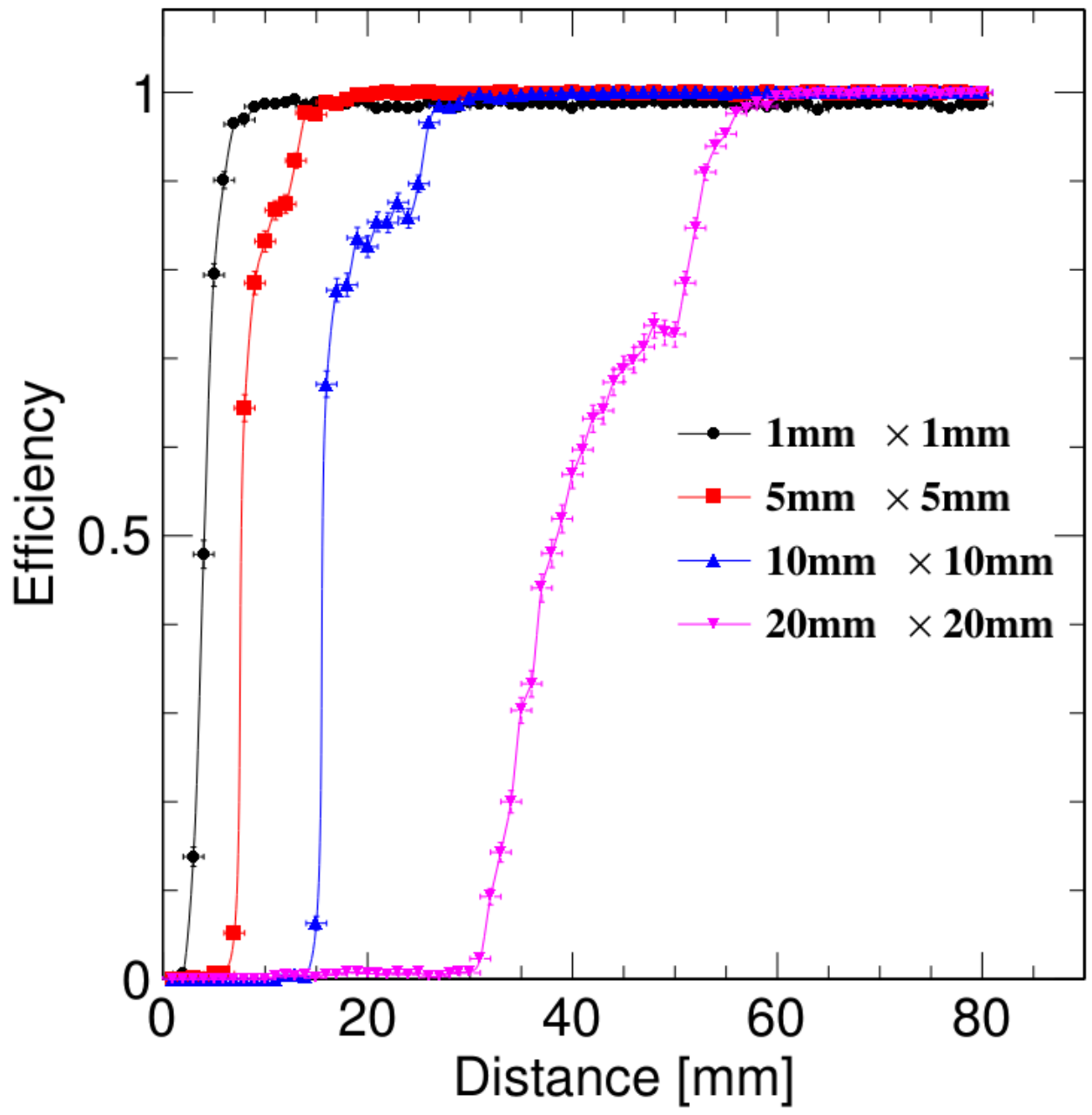

**Figure 5.5:** Reconstruction efficiency of two parallel 5 GeV photons as a function of their separation distance for different ECAL sensor sizes. The reconstruction is fully efficient once the separation is approximately twice of the sensor size. The secondary step structures of the efficiency curves are the edge effects of the finite sensor sizes.

| Cell size ($mm^2$) | Percentage of inseparable photons |
|---|---|
| $1 \times 1$ | 0.07% |
| $5 \times 5$ | 0.30% |
| $10 \times 10$ | 1.70% |
| $20 \times 20$ | 19.6% |

**Table 5.2:** Percentage of photons in $\tau$ decays of the $Z \to \tau\tau$ events that cannot be separated from neighboring particles in the ECAL for different ECAL sensor sizes. Almost all photons in the $\tau$ decays can be well reconstructed for a ECAL sensor size up to $10 \times 10\,\mathrm{mm}^2$. A significant portion of the photons fail to be separated from other particles in the ECAL when the sensor size exceeds $10 \times 10\,\mathrm{mm}^2$.

- Stability: under a reasonable bias voltage, a completely depleted silicon PIN diode has unity gain, and a signal response to a MIP mostly defined by the sensor thickness, with a relatively low dependence on the operating environment including temperature, humidity, etc.

- Uniformity: the control of the sensor thickness within large production batches (typically to less than a percent) ensures uniform responses within a wafer and between different wafers. The non-sensitive area between wafers has recently been reduced by the use of laser cutting, thinned guard-ring design [4], and would benefit from the use of larger ingot size (8″ becoming the standard).

- Flexibility: the dimension and geometry of the cells can be flexibly defined. The readout pads on the PCB need to be compatible.
- High Signal-to-Noise (S/N) ratio: for a MIP, the most probable number of electron-hole pairs generated in $1\ \mu\mathrm{m}$ thick silicon is around 76 (while the average number is 108), which yields an excellent S/N ratio of silicon sensors. Thus, MIP tracks can be easily tracked in the calorimeter, which is critical to the good performance of the ECAL.
- Timing capability: A completely depleted silicon PIN diode has great potential for fast timing. This has been demonstrated in the detector R&D for the CMS high granularity calorimeter upgrade project [5]. And a good timing resolution of 20–30 ps has been achieved with a silicon diode sensor for a signal amplitude corresponding to 5 MIPs. Time measurements of energy depositions in the ECAL can be useful to particle flow algorithms to help disambiguate particle contributions. For the CEPC as a lepton collider, normally with a single primary vertex, precision timing of individual ECAL cells - or group of ECAL cells - could still be useful to reduce the shower confusion in the calorimeter and improve the energy resolution, which however needs further studies to quantitatively explore this potential.

One concern with the silicon sensors is the price, which could be very high. By integrating the silicon sensors with tungsten plates and carbon fiber structures, the SiW-ECAL offers an excellent option for the PFA optimized calorimetry.

#### 5.3.2.2 GEOMETRY AND MECHANICAL DESIGN

A key requirement for the calorimeter system is to ensure the best possible hermeticity. Three regions need to be considered, including the boundaries of mechanical modules, the overlap region between the barrel and endcap sections, and very forward regions near forward detectors. A design with large ECAL modules is preferred to minimize crack regions in the barrel section, and the inter-module boundaries should not point back to the interaction point (IP). As shown in Figure 5.6, an octagonal shape is adopted to approximate the cylindrical symmetry and the modules are designed in such a way that the cracks are at a large angle with respect to the radial direction. One eighth of the barrel ECAL is called a stave. Each stave is fastened to the HCAL front face with a precise system of rails. Some space is left between the ECAL and HCAL to accommodate services including cooling, power and signal cabling. Along the beam direction, a stave is divided into five modules. The two ECAL endcap sections are fastened to the front face of HCAL endcap sections using a similar rail system.

**Longitudinal arrangement:** The ECAL is composed of 30 layers in the longitudinal direction. To optimize the ECAL energy resolution, the 30 layers are split into 2 sections with different thickness of absorber layers. The first section contains 20 layers of $0.6X_0$ (or 2.1 mm) thick tungsten plates corresponding to 12 radiation lengths. It is followed by the second section with 10 layers of $1.2X_0$ thick (4.2 mm) tungsten plates corresponding to another 12 radiation lengths. (see section 5.3.1 for details). The ECAL starts with an sensitive layer. Each sensitive layer is equipped with $0.5\ \mathrm{mm}$ thick silicon sensors. The granularity of sensitive layers is determined by the silicon sensor size which is chosen to be $10 \times 10\ \mathrm{mm}^2$ for all layers. The two longitudinal sections are both held on a 20 mm thick base plate made of carbon-fiber.

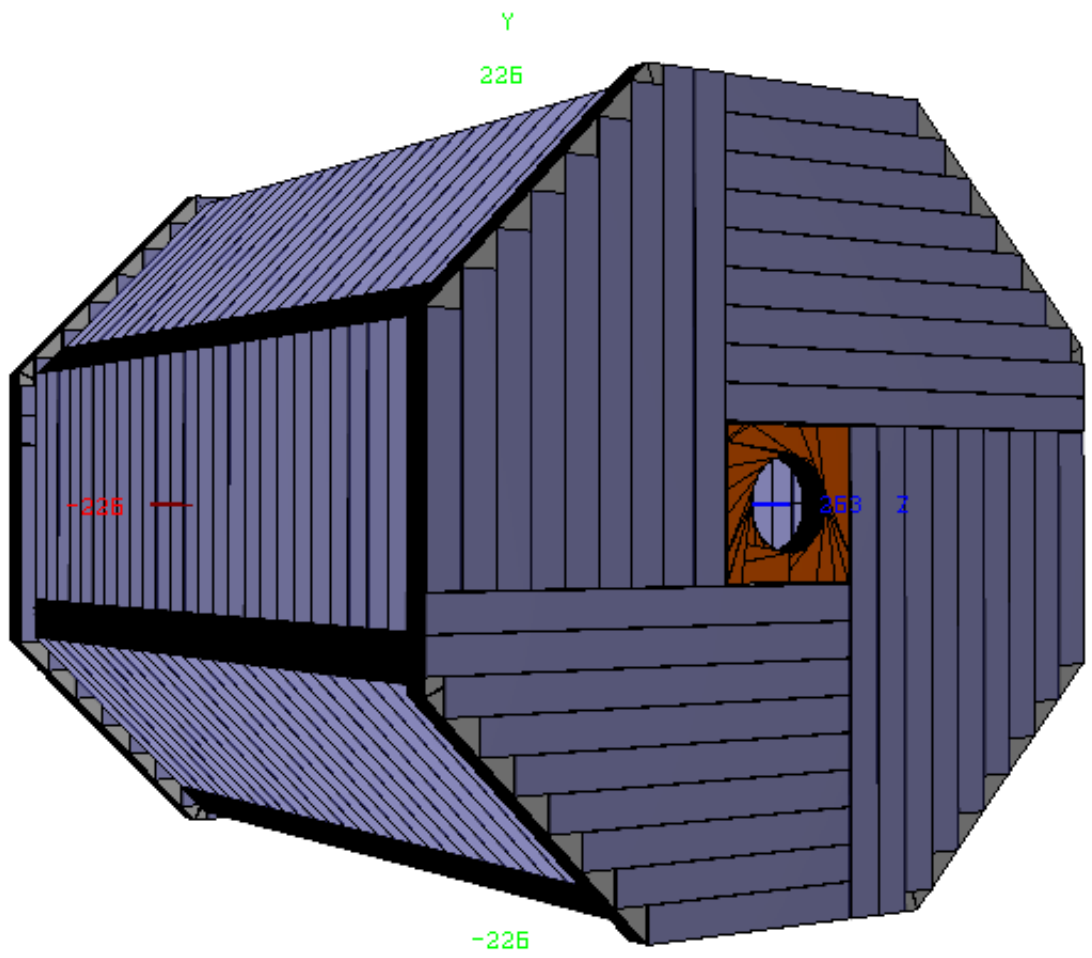

**Figure 5.6:** Schematic of the CEPC ECAL layout in its baseline design. The ECAL is organized into one cylindrical barrel and two disk-like endcap sections, with 30 layers in each section. The barrel section is arranged into 8 staves, each consisting of 5 trapezoidal modules. Each of the two endcap sections is made of four quadrants, each consisting of 2 modules. The ECAL barrel overall radius is 2028 mm in X-Y plane, the two endcaps are located at ±2635 mm.

**Structures:** The ECAL barrel section consists of 8 staves, each composed of 5 trapezoidal modules as shown in Figure 5.7. A barrel module contains 5 columns. The numbers of modules and columns are positioned along the beam line and chosen to be odd in number and symmetrically placed in order to avoid any pointing-like dead regions at the azimuthal plane perpendicular to the beam direction at the IP. The column size is 186 mm by mechanical limits and by cost optimization considerations, in order to contain exactly two 6-inch wafers or one and a half 8-inch wafers. Integrating the column size, walls of modules and contingencies, the barrel length adds up to 4700 mm in the beam direction. A gap of typically 100 mm is left between the barrel sides and endcap front faces. The precise dimension will depend on the amount of services for the ECAL, the HCAL and the tracker system (including power and DAQ cabling, cooling pipes, patch panels, etc.).

The two endcap sections are made up of 4 quadrants, each of which is then made up of 2 modules with one of the modules containing 4 columns and other other 3 columns. The endcap inner radius is fixed by the ECAL ring at 400 mm. With 7 columns, the endcap outer radius is 2088 mm. An overshoot of 32 mm is kept between the outer radii of the barrel and of the endcaps, in order to contain the EM shower impinging the overlap region. This fixes the inner radius of the barrel section to 1843 mm. For the above structures, summing up all barrel and endcap sections, 256 ECAL columns are needed.

**Active Sensor Unit:** Each ECAL column is made up of several ECAL slabs. Each slab consists of two symmetric sensitive layers and one tungsten plate. Each sensitive layer contains a layer of silicon sensors glued on a PCB, equipped with readout ASICs, a high voltage distribution by a Kapton foil and copper layers for passive cooling. The components are attached on both sides of an H-shaped carbon fiber cradle, with a tungsten core, and shielded by an aluminum cover. To insure scalability and industrial production, the design has been made as modular as possible: each basic unit is an Active Sensor Unit (ASU), which currently has a $18 \times 18\ \mathrm{mm}^2$ PCB glued with 4 pieces of $90 \times 90\ \mathrm{mm}^2$ silicon wafers. Each ASU will handle 256 silicon pads with 4 ASIC chips, for the cell size of $10 \times 10\ \mathrm{mm}^2$. The ASUs are chained together for the clock and configuration distributions and data collection.

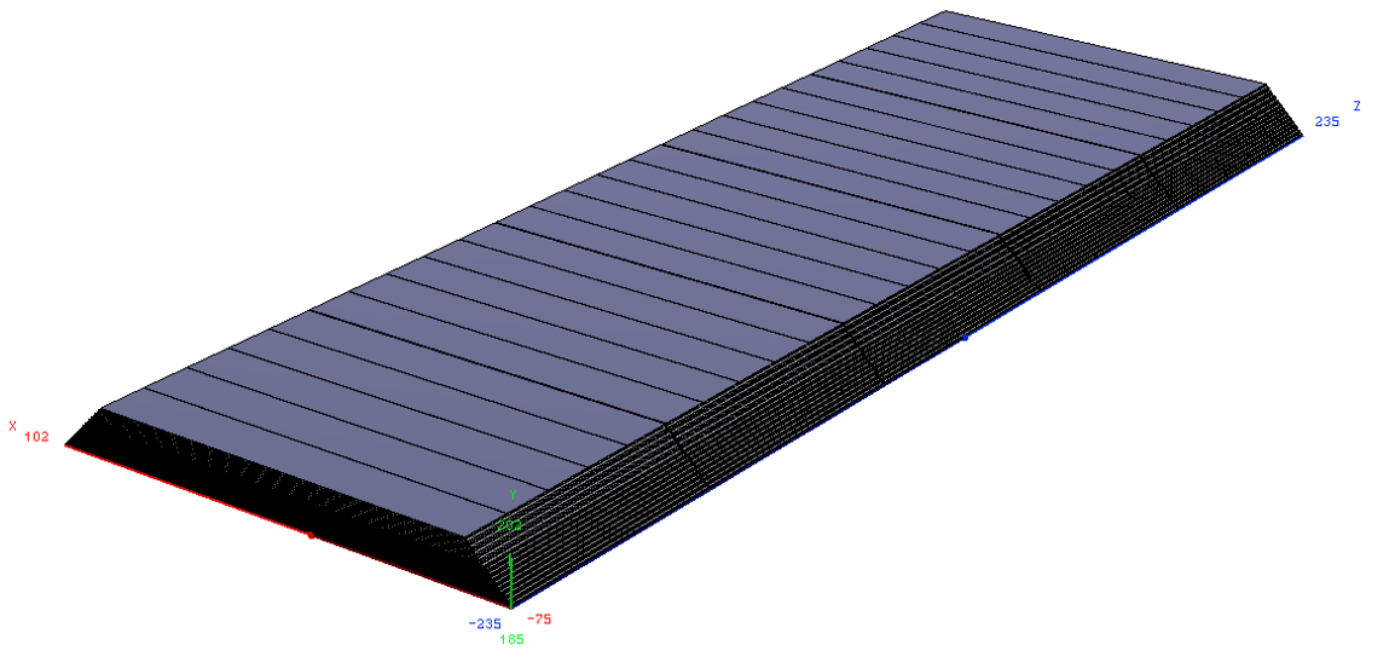

**Figure 5.7:** Schematic of the structure of one ECAL stave. Each stave is made up of 5 trapezoidal modules, and each module contains up to 5 columns.

#### 5.3.2.3 SIW-ECAL ELECTRONICS

One of the most critical elements of the CEPC calorimeters is the readout electronics which is defined by the dynamic range, the effective digitization, mode of trigger, the rate of working and power consumption per channel.

**Dynamic range:** A MIP going through a $500\ \mu\mathrm{m}$ silicon diode will produce around 60000 electron-hole pairs (or a charge of 9.6 fC) as the Most Probable Value (MPV). To record MIPs with an efficiency higher than 95%, this determines the lower limit of the dynamic range to a 1/3 of the MPV. The higher limit is given by the number of MIP equivalents at the core of the high-energy EM showers, which can reach up to 10000 MIPs (or 96 pC) within a $10 \times 10\ \mathrm{mm}^2$ cell.

**Timing:** As described earlier, time measurements of energy depositions in the calorimeter can be useful to particle flow algorithms. A recent version of SiW-ECAL ASIC (SKIROC2A) has been tested on a test board and reached a timing resolution close to 1.1 ns for a signal amplitude corresponding to 5 MIPs [6]. And a much better timing resolution of 20–30 ps has even been achieved for the same signal amplitude with a dedicated readout ASIC chip (HGCROC) developed for the CMS high granularity calorimeter upgrade project [5].

**Power consumption:** The running conditions of a circular collider exclude pulsed operation as is planned for the linear colliders. As a point of reference, the current power consumption for the SKIROC2 chip is around 5 mW per channel in the continuous operating mode.

**Occupancy:** The occupancy of the calorimeters is expected to be very low. This offers room for an ultra-low power electronics design when there is no signal.

#### 5.3.2.4 SIW-ECAL POWER CONSUMPTION AND COOLING

To the first order, the amount of the power dissipation scales up with the number of electronics channels. One critical issue for the calorimeters is the cooling scheme. As for now there are two options. The CEPC ECAL is at the boundary of both options, with a limit for the purely passive option of the order of $20 \times 20\ \mathrm{mm}^2$ cells for a increase of temperature limited to $\Delta\mathrm{T} \sim 10\ ^\circ\mathrm{C}$ at the far end of the slab.

- Passive cooling: this option requires a reduced number of channels in order to use only passive cooling at the rear of the detector. As an example, a $400\ \mu\mathrm{m}$ thick copper sheet will drain the heat to the end of an ECAL slab, where it is then removed by an active cooling system installed near boundaries between barrel and endcap parts. A leak-less water cooling system can also be an option to extract the heat at the end of each slab from the copper. Details of implementation can be found in Ref. [7]. Full simulation studies based on PFA should be performed to provide the quantitative impact from a reduced granularity and the corresponding calorimeter performance.

- Active cooling: this option is the baseline high granularity design and requires the cooling system to provide cooling near the sensors and front-end electronics throughout the entire calorimeter system. A two-phase, low mass $\mathrm{CO_2}$ cooling system is a promising option, which can be embedded in the absorber plates. There are already some simulation studies on a similar system adapted to the SiW-ECAL [8], where 3 mm thick copper plates, equipped with 1.6 mm inner diameter pipes for $\mathrm{CO_2}$ circulation, with the ASICs glued on both sides of the slab. The study assumed a fully transversely isolated system, with ASICs as the primary heat source at equilibrium dissipating 0.64 W (10 mW per channel times 64 channels), and a fixed working point of $20\ ^\circ\mathrm{C}$ for $\mathrm{CO_2}$ (i.e. assuming perfect heat absorption). A doubled-sided module of $252 \times 252\ \mathrm{mm}^2$ holding 32 chips cooled by $2 \times 2$ pipes was simulated. Preliminary simulations in "ideal conditions" show a difference of $\Delta\mathrm{T} \sim 2\ ^\circ\mathrm{C}$ mostly centered on the ASICs (and only $0.3\ ^\circ\mathrm{C}$ in the heat exchanger).

#### 5.3.2.5 SIW-ECAL R&D STATUS

The performance of a SiW-ECAL have been explored using the “physical prototype“ developed within the CALICE collaboration, with extensive beam tests during the years 2005 – 2011 [9–11]. Some ASUs have been operated in beam test campaigns: first at CERN in 2015, where 3 ASUs were mounted on test boards which behaved as expected [12]; a Signal to Noise Ratio (SNR) (defined as the Most Probable Value of a Landau fit on data, divided by the Gaussian width of the noise) reached typical values of 15–18, with a very limited number of masked channels.

In a recent beam test at DESY with 1–5 GeV electrons, "short slabs" (featuring all the elements as required but limited to a single ASU on a single side) could reach a SNR of around 20 on average [13].

The collected data is still under analysis, but they are expected to be similar to the SiW-ECAL physical prototype. The construction of a “long slab“ is being actively pursued, and should be completed toward the end of year 2019; the R&D involves all the power, cooling and frond-end electronics issues. The results and design will have to be optimized for a circular collider, where the power-pulsing operation is not an option.

## 5.3.3 SCINTILLATOR-TUNGSTEN SANDWICH ECAL

### 5.3.3.1 INTRODUCTION

Alternatively, a sampling calorimeter with scintillator-tungsten structure is proposed. It can be built in a compact and cost effective way. The layout and structure of the scintillator-tungsten ECAL is very similar to that of the silicon-tungsten ECAL. Major design parameters for the scintillator-tungsten ECAL were also studied and optimized, with an outcome quite similar to that of the silicon-tungsten ECAL. The primary difference is in the thickness of the active layers, and another difference being in the sensor shape of the active layers. The active layers of the scintillator-tungsten ECAL consists of 2 mm thick and $5 \times 45\ \mathrm{mm}^2$ large scintillator strips. The scintillator strips in adjacent layers are perpendicular to each other to achieve a small effective transverse readout cell size. However, the performance of a ECAL with this configuration may be subject to degradation due to ambiguity in pattern recognition of showers, and therefore the effectiveness of this configuration of scintillator strips still needs to be demonstrated. Each strip is covered by a reflector film to increase light collection efficiency and improve the uniformity of scintillation light yield w.r.t. incident position by a particle on the strip. Photons from each scintillator strip are read out by a very compact photo-sensor, SiPM, attached to the strip. The SiPM and highly integrated readout electronics make the dead area in the scintillator-tungsten ECAL almost negligible. Figure 5.8 shows the schematic structure of a scintillator-tungsten ECAL module in the above configuration. Although a SiPM is coupled to a scintillator strip by side in this schematic, it should be pointed out that various schemes for coupling the SiPM to the scintillator strip are considered for optimum performance.

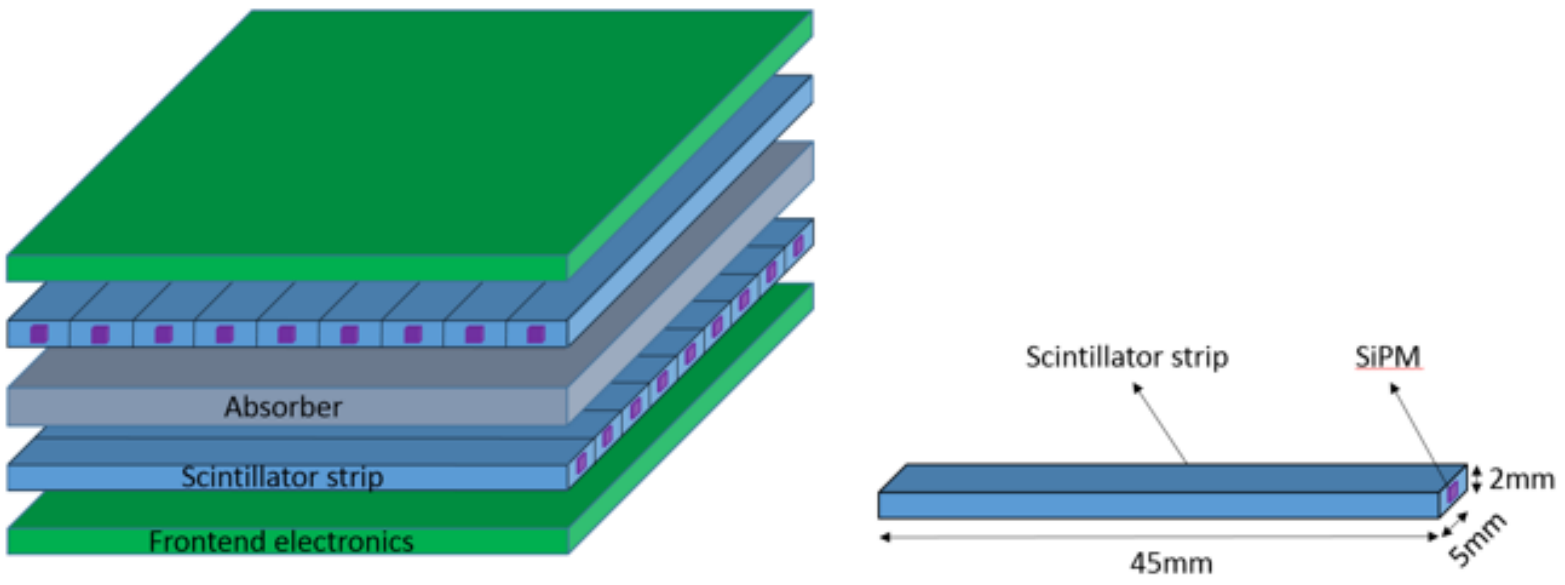

**Figure 5.8:** Layout of a scintillator-tungsten ECAL module and dimensions of a scintillator strip. The scintillator strips in adjacent layers are perpendicular to each other to achieve a small effective transverse readout cell size.

Plastic scintillator is a robust material which has been used in many high energy physics experiments. Production of scintillator strips can be made at low cost by the extrusion method. And prices for SiPMs on the market have also been falling constantly with the rapid development of the SiPM technology. Moreover, the number of readout channels can also be significantly reduced due to the strip readout configuration. So the total construction cost of the scintillator-tungsten ECAL is expected to be lower than that of the silicon-tungsten ECAL. Some key aspects of the scintillator-based ECAL technology were studied and optimized.

### 5.3.3.2 SIPM DYNAMIC RANGE

Because each pixel on a SiPM can only detect one photon at a time and a few nanoseconds are needed before it is recovered, a SiPM is not a linear photon detection device, particularly when illuminated with high intensity light. Therefore, the dynamic range and linearity of SiPM needs to be addressed for its application in the scintillator-tungsten ECAL.

For a very short light pulse , the response of a SiPM can be formulated as

$$N_{fired} = N_{pixel}(1 - e^{-N_{pe}/N_{pixel}}), \tag{5.1}$$

where $N_{fired}$ is the number of fired pixels of a SiPM and $N_{pixel}$ is the number of total pixels. However, light pulses produced in plastic scintillator last long enough for some pixels of a SiPM to detect more than one photon in one event of light generation. The response function of a SiPM is then modified in this case as

$$N_{fired} = N_{eff}(1 - e^{-N_{pe}/N_{eff}}), \tag{5.2}$$

where $N_{eff}$ stands for the effective number of pixels on a SiPM, which is a function of the width of incident light pulse. The response curve of a 10000-pixel (10 $\mu\mathrm{m}$ pitch size) and a 1600-pixel (25 $\mu\mathrm{m}$ pitch size) SiPMs with an active area of $1 \times 1\ \mathrm{mm}^2$ were measured for light pulses with different widths, as shown in Figure 5.9. The dynamic range of the 10000-pixel SiPM is much larger than that of the 1600-pixel one, as expected, and can reach 4000 photon-electrons with very good linearity for short light pulses and much beyond if saturation correction is made. Also the linearity of response of SiPMs is improved with increasing of the width of incident light pulses. So care has to be taken if operation of SiPMs reaches saturated regions and correction is required. Rough estimation suggests a SiPM dynamic range of at least up to 10000 photon-electrons is needed for a scintillator-tungsten ECAL at the CEPC experiment for $H \to \gamma\gamma$ measurement. So large-area SiPMs with a large number of pixels are favored for use in the CEPC scintillator-tungsten ECAL because of high dynamic ranges. This is also becoming increasingly practical as the SiPM price per $\mathrm{cm}^2$ has been dropping significantly.

### 5.3.3.3 PERFORMANCE OF SCINTILLATOR SENSITIVE UNIT

A scintillator sensitive unit is a scintillator strip coupled with a SiPM. When a SiPM is coupled to a scintillator at one position, the light output is expected to depend on the scintillation light propagation distance along the strip to the SiPM coupling point. This dependence would introduce non-uniformity of light output, hence affecting the ECAL performance. Three configurations of a SiPM coupling to a scintillator strip were explored as shown in Figure 5.10, to minimize the non-uniformity of light output. And the uniformity of light output was measured with a Sr90 source for the three SiPM coupling configurations, respectively, where the pitch size of the SiPM is 10 $\mu\mathrm{m}$. Figure 5.11 shows the measured uniformities. The side and bottom-end configurations give a similar uniformity of 30%, while the bottom-center configuration presents a very good uniformity of 10% without reduction of light output. Furthermore, such a configuration has additional advantages of completely eliminating dead areas between scintillator strips due to mounting of SiPMs and allowing to use SiPMs with very large areas which is essential for meeting the requirement on dynamic range of SiPMs. For these attractive features, the bottom-center SiPM coupling configuration is adopted for the scintillator-tungsten ECAL.

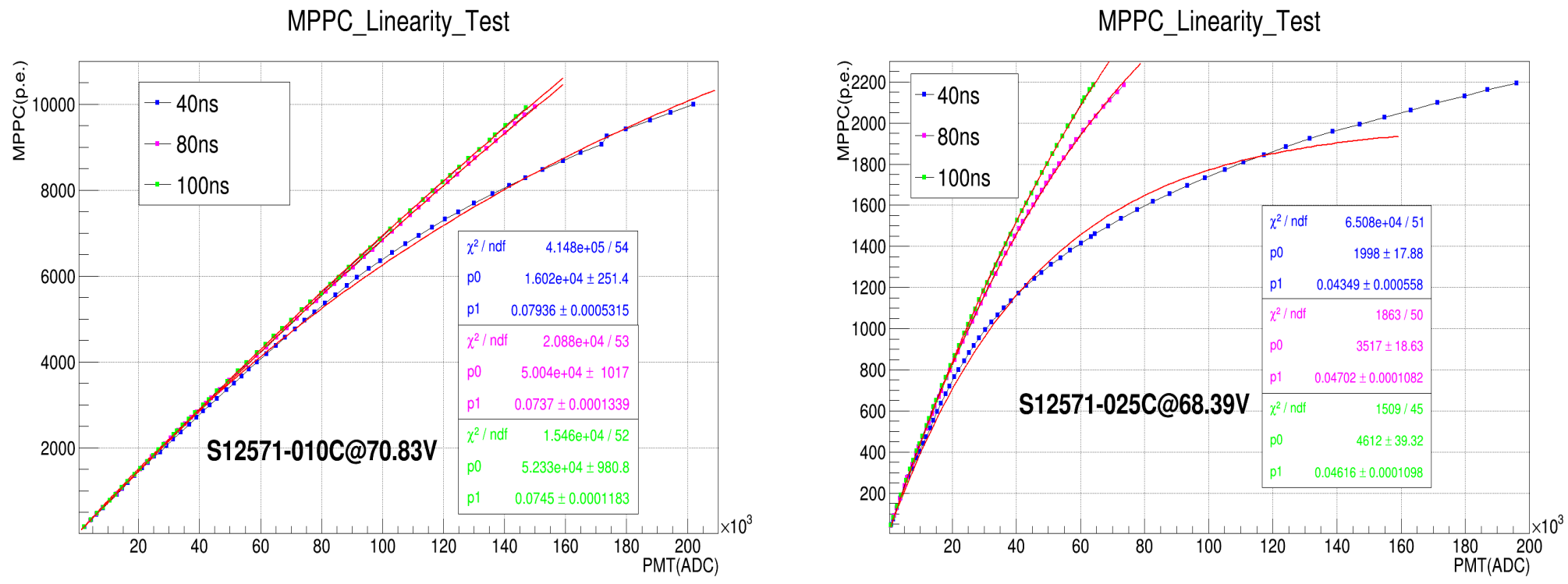

(a) (b)

**Figure 5.9:** Response linearity ( the number of photo-electrons detected with a SiPM as a function of the number of incident photons) of SiPMs with different numbers of pixels [(a): 10000-pixel SiPM, (b): 1600-pixel SiPM] for light pulses with different widths (blue: 40 ns, red: 80 ns and green: 100 ns). The linearity of SiPM response with 10000 pixels is better than that with 1600 pixels. And the range for linear response of SiPM gets larger for a wider light pulse.

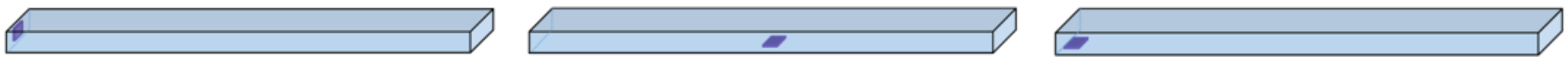

(a) (b) (c)

**Figure 5.10:** Three configurations of the SiPM-scintillator coupling explored for the design of scintillator-tungsten ECAL: (a) a SiPM is embedded in a scintillator strip on one side (side), (b) a SiPM is embedded in a scintillator strip at the center of the bottom face (bottom-center), (c) a SiPM is embedded in a scintillator strip at one end of the bottom face (bottom-end).

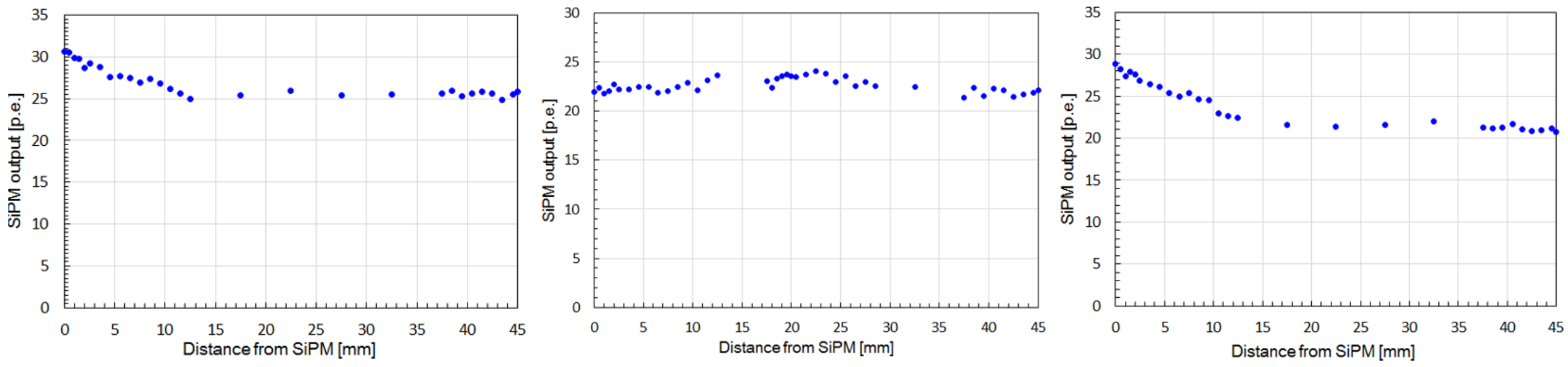

(a) (b) (c)

**Figure 5.11:** Light output of a scintillator sensitive unit with three SiPM coupling configurations: (a) side, (b) bottom-center, (c) bottom-end. The bottom-center configuration gives the best uniformity of 10% without reduction of light output.

Light output of scintillator sensitive unit was also studied with the scintillator strip wrapped with different reflectors as shown in Figure 5.12. ESR reflector gives much higher light out than Tyvek reflector.

Light output of scintillator sensitive unit would depend on the pitch size of the SiPM due to different photon detection efficiency. Figure 5.13 shows the light output of scintillator sensitive units with SiPMs that have the same sensitive area ($1 \times 1\ \mathrm{mm}^2$) but with different pitch sizes ( 25 μm vs. 10 μm). The light output with the 10 μm SiPM is only about 1/3 of that with the 25 μm SiPM due to its much lower photodetection efficiency.

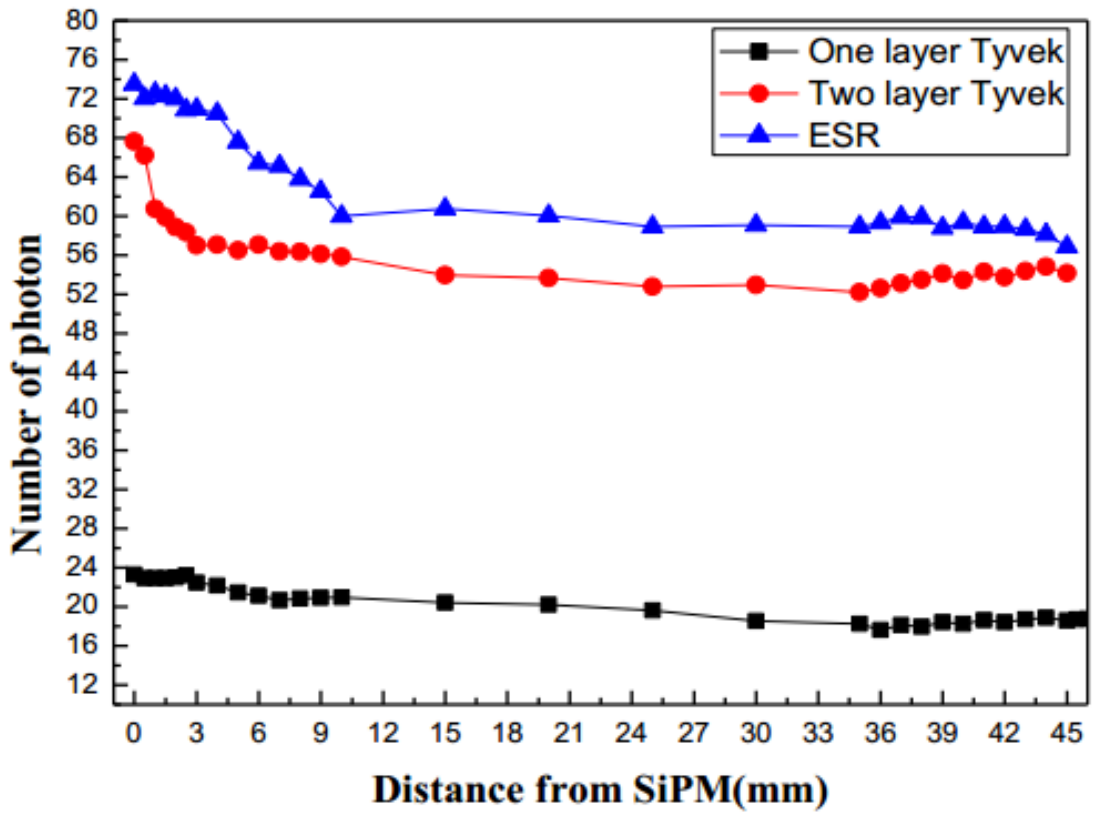

**Figure 5.12:** Light yields of scintillator strips with different reflectors versus the distance of an incident particle from the SiPM . The scintillator with ESR gives the highest light yield.

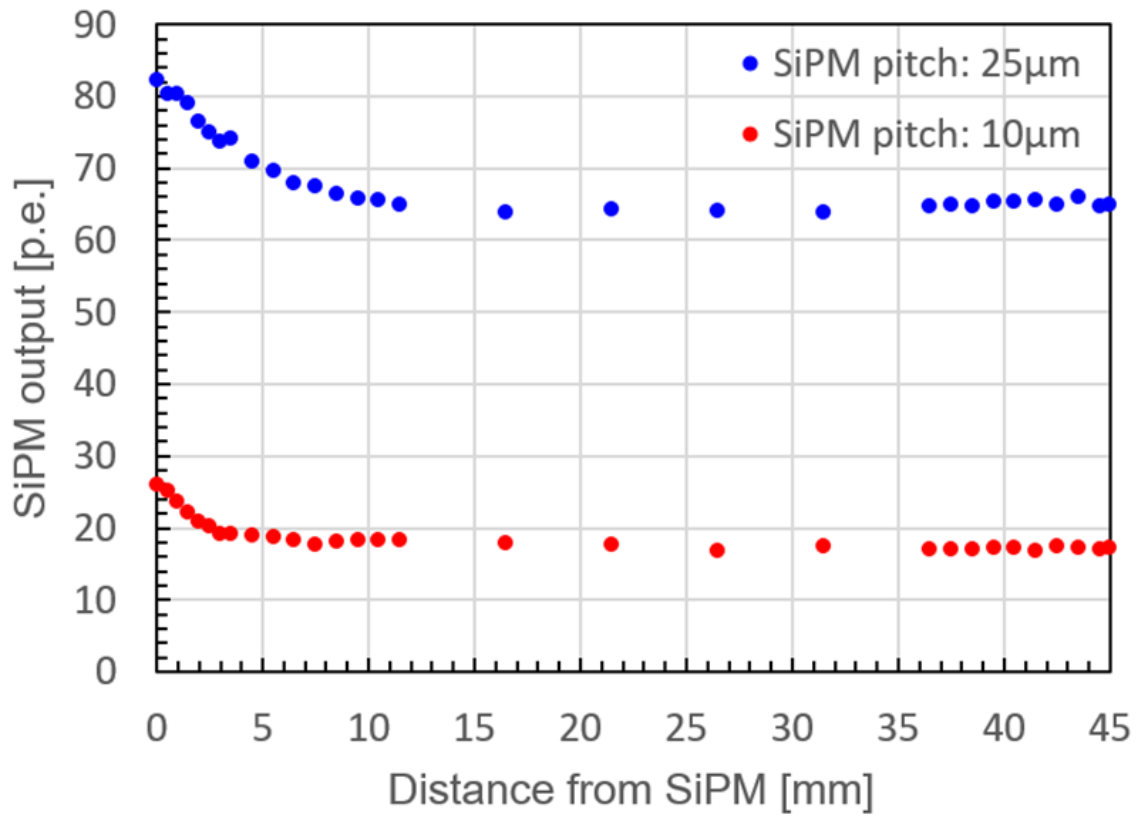

**Figure 5.13:** Light yields of scintillator strips coupled with SiPMs with different pitch sizes ( red: 10 μm, blue: 25 μm). The SiPM with the larger pitch size of 25 μm has a significant higher light yield (about 65 p.e.) than that with the smaller pitch size of 10 μm (about 18 p.e.) due to a higher photon detection efficiency.

So light output should be taken into account when choosing small-pitch SiPMs for a high dynamic range. It has to be ensured the scintillator sensitive unit is sensitive to MIPs. Figure 5.14 shows the pulse height distribution for cosmic-ray signals of the scintillator sensitive unit with a 10 μm SiPM using the readout electronics described in 5.3.3.4. Cosmic-ray signals are seen well separated from noise demonstrating the sensitivity to MIPs.

### 5.3.3.4 SIPM READOUT ELECTRONICS

The readout electronics of the ECAL has to provide high dynamic range for energy measurements. A 100 GeV photon shower may leave an energy deposit of 1∼800 MIP-equivalent in a single cell. A high spatial granularity of the ECAL readout, typically 10 mm, is required for the PFA. This, in turn, results in a large total channel count and a high density of channels. For this reason, multi-channel readout chips are considered.

The full readout chain of the electronics consists of two parts: Front-End and Back-End. The Front-End (FE) electronic is embedded into the layers of ECAL. It performs amplification, auto-triggering, digitization and zero-suppression, with local storage of data between the working phases. The Back-end Electronics (BE) collects data and configures the readout chips before system running.

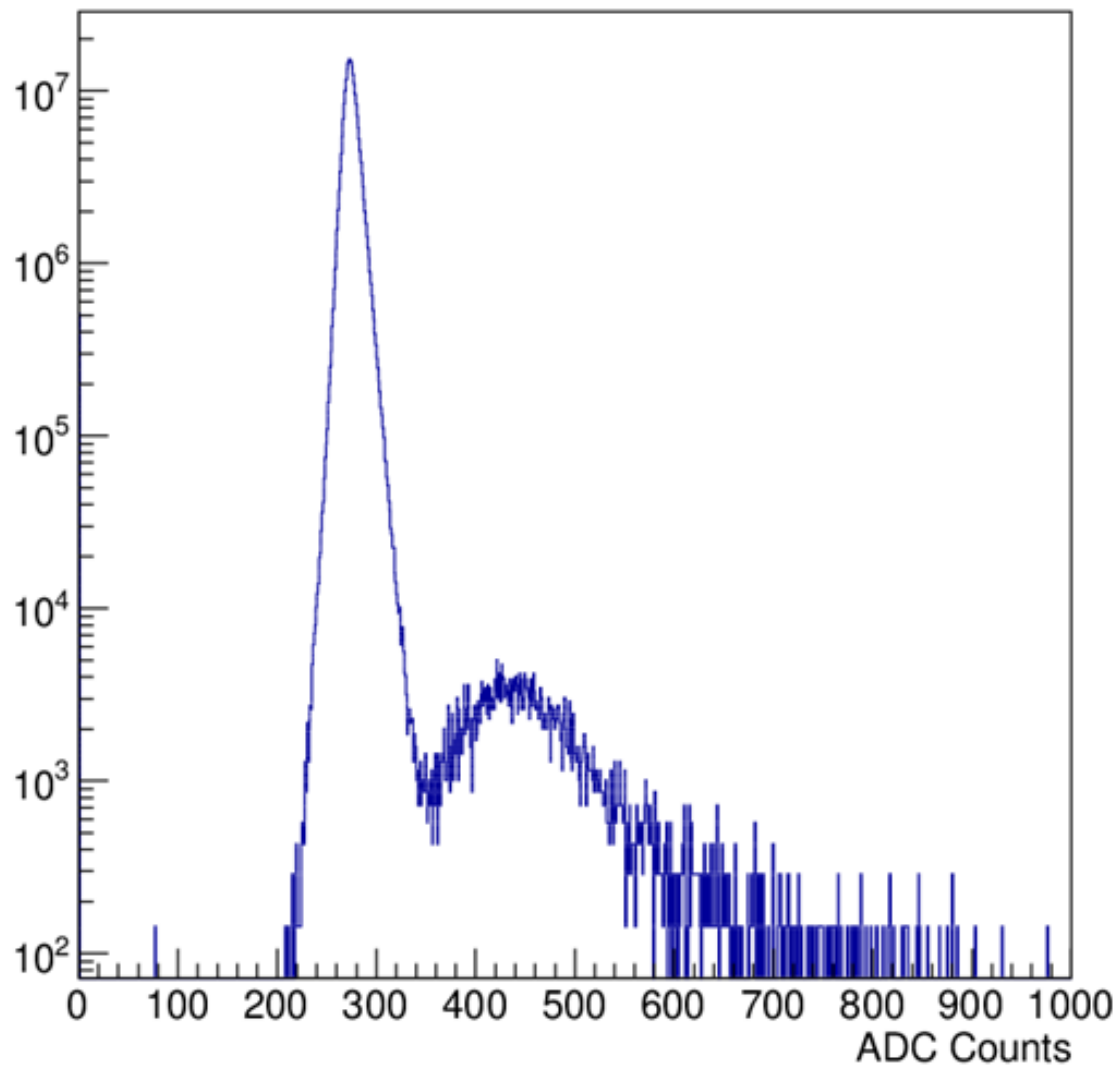

**Figure 5.14:** The pulse height distribution for cosmic-ray signals of the scintillator sensitive unit with a 10 μm SiPM. Cosmic-ray signals can be found well separated from noise demonstrating the sensitivity of the scintillator sensitive unit to MIPs.

Several studies and existing calorimeter readout electronics have shown that one can obtain optimized energy resolutions using a preamplifier-shaper and digitizing the pulse at the peak amplitude. For instance, a preamp-shaper-Switched-Capacitor-Array (SCA) structure of analog circuit was applied on an ILC HCAL and implemented in an ASIC named SPIROC [14]. The basic principle consists of a readout chain with an amplifier-shaper using a RCn-CRp filter delivering a pulse length of about 50–200 ns duration for a SiPM pulse signal. This signal is also shaped by a fast shaper in parallel to generate fast and narrow pulse for pulse discrimination. Then, the discriminator sends the trigger to a Switched-Capacitor-Array (SCA) for locking the peak value of the slowly shaped signal. The locked voltage value corresponds to the charge that the circuit received. A ADC is used to digitize the analog voltage in the SCA. A similar approach can be applied for the CEPC ECAL readout electronics. But the continuous operation mode of the CEPC has to be fully taken into account in the design and implementation of the ECAL front-end electronics.

### 5.3.3.5 CALIBRATION SYSTEM

The Scintillator-Tungsten ECAL consists of about ten million channels of scintillator strip units. The stability of the light output has to be monitored. A light distribution system is under study to monitor possible gain drifts of the SiPMs by monitoring photon-electron peaks. The system consists of pulse generator circuit and chip LEDs. A schematic diagram of the system is shown in Figure 5.15. The pulse generator circuit and chip LEDs are integrated into the front-end electronics board. The LEDs are placed in the holes under scintillator strips. So each scintillator strip can be illuminated by fast light pulses from the LED under it which is driven by the pulse generator circuit.

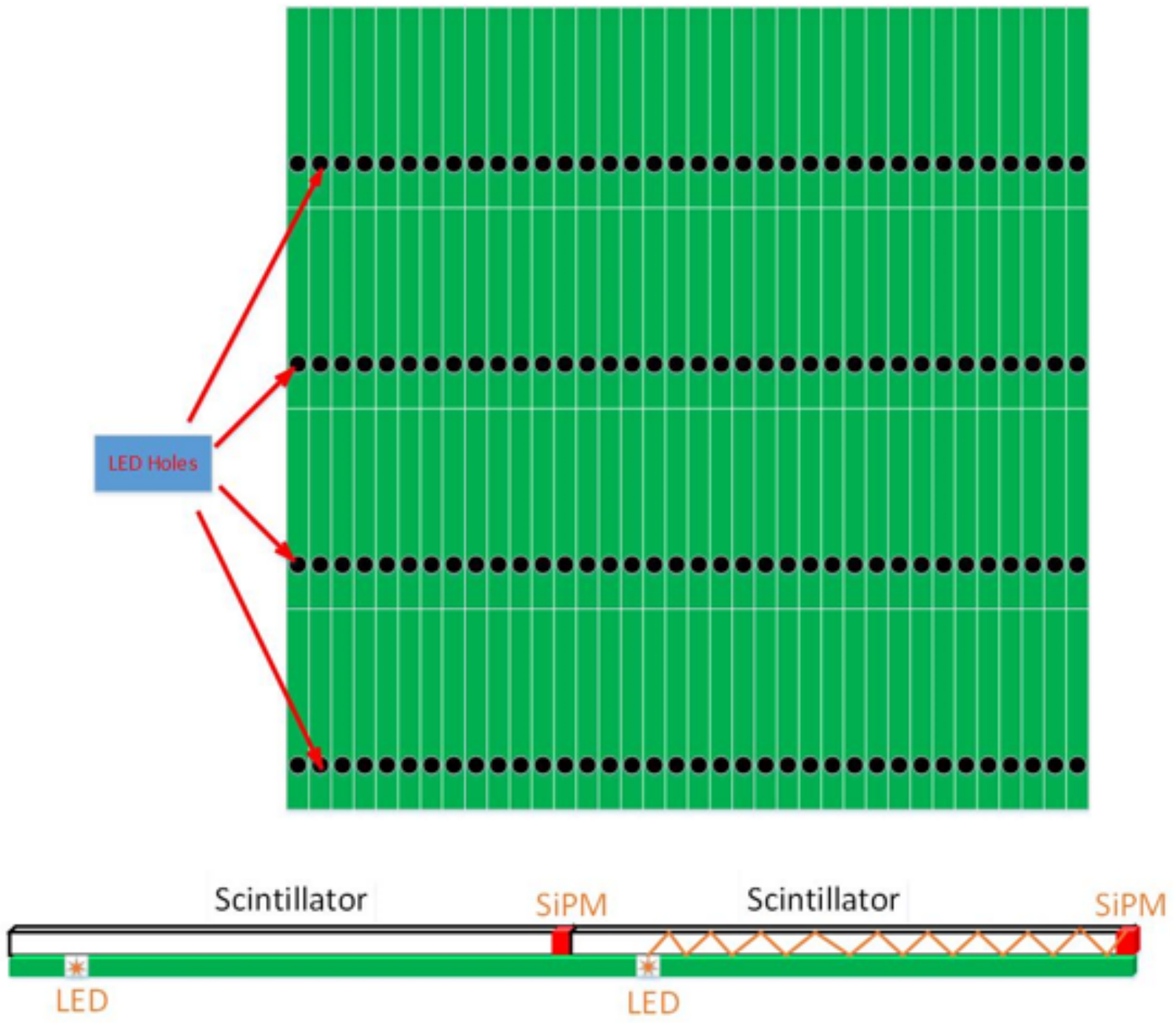

**Figure 5.15:** Schematic of the SiPM gain monitoring system with LEDs. Each scintillator strip is illuminated by fast light pulses from a LED placed under it.

## 5.4 PARTICLE FLOW ORIENTED HADRONIC CALORIMETER

### 5.4.1 INTRODUCTION

High-granularity HCAL is an essential component of PFA-based detectors such as the baseline detector proposed for the CEPC. The high spatial granularity provides means to separate the energy deposits of charged and neutral hadrons and therefore allow the precise measurement of the energy of the neutral particles. The contribution of the neutral particles to the jet energy, around 10% on average, fluctuates in a wide range from event-to-event. This uncertainty is the dominant contribution to the PFA jet energy resolution for energies up to about 100 GeV. At higher energies, the performance is dominated by a term in the PFA resolution called confusion term, which arises from both the failure of the topological pattern recognition and the incorrect assignment of charged and neutral energies. A high-granularity hadronic calorimeter can minimize both of these effects and thus can achieve excellent jet energy resolution.

The HCAL systems considered here are sampling calorimeters with steel as the absorber and scintillator tiles or gaseous devices with embedded electronics active the active medium. The steel was chosen due to its rigidity allowing for the self-support without auxiliary supports (dead regions). Moreover, the moderate ratio of hadronic interaction length ($\lambda_I = 17$ cm) to electromagnetic radiation length ($X_0 = 1.8$ cm), allows a fine longitudinal sampling in terms of $X_0$ with a reasonable number of layers in $\lambda_I$, thus keeping the detector volume and readout channel count small. This fine sampling is beneficial both for the measurement of the sizable electromagnetic energy in hadronic showers and for the topological resolution of shower substructure, needed for particle separation.

The active detector element has finely segmented readout pads, with $1 \times 1$ cm$^2$ or $3 \times 3$ cm$^2$ size, for the entire HCAL volume. Each readout pad is readout individually, so the readout channel density is approximately $4 \times 10^{4-5}$/m$^3$. For the entire HCAL with a volume of $\sim$100 m$^3$, the total number of channels will be $4 \times 10^{6-7}$ which is one of the biggest challenges for the HCAL system. On the other hand, simulation suggests that, for a calorimeter with cell sizes as small as $1 \times 1$ cm$^2$, a simple hit counting provides already a good energy measurement for hadrons. As a result, the readout can be greatly simplified to record 'hit' or 'no hit' according to a single threshold (equivalent to a '1-bit' ADC). A hadron calorimeter with this simplified readout is called a Digital Hadron Calorimeter (DHCAL). In a DHCAL, each readout channel is used to register a 'hit', instead of measuring energy deposition as in the traditional HCAL systems. In this context, gas detectors (such as RPC, GEM) become excellent candidates for the active element of a DHCAL. Another technology option is the AHCAL which is based on scintillator coupled with SiPMs as the active sensor.

A drawing of the HCAL structure is shown in Figure 5.16, the barrel part is made of 5 independent and self-supporting wheels along the beam axis. The segmentation of each wheel in 8 identical modules is directly linked with the segmentation of the ECAL barrel. A module is made of 40 stainless steel absorber plates with independent readout cassettes inserted between the plates. The absorber plates consist of a total of 20 mm stainless steel: 10 mm absorber from the welded structure and 10 mm from the mechanical support of the detector layer. Each wheel is independently supported by two rails on the inner wall of the cryostat of the magnet coil. The cables as well the cooling pipes will be routed outside the HCAL in the space left between the outer side of the barrel HCAL and the inner side of the cryostat.

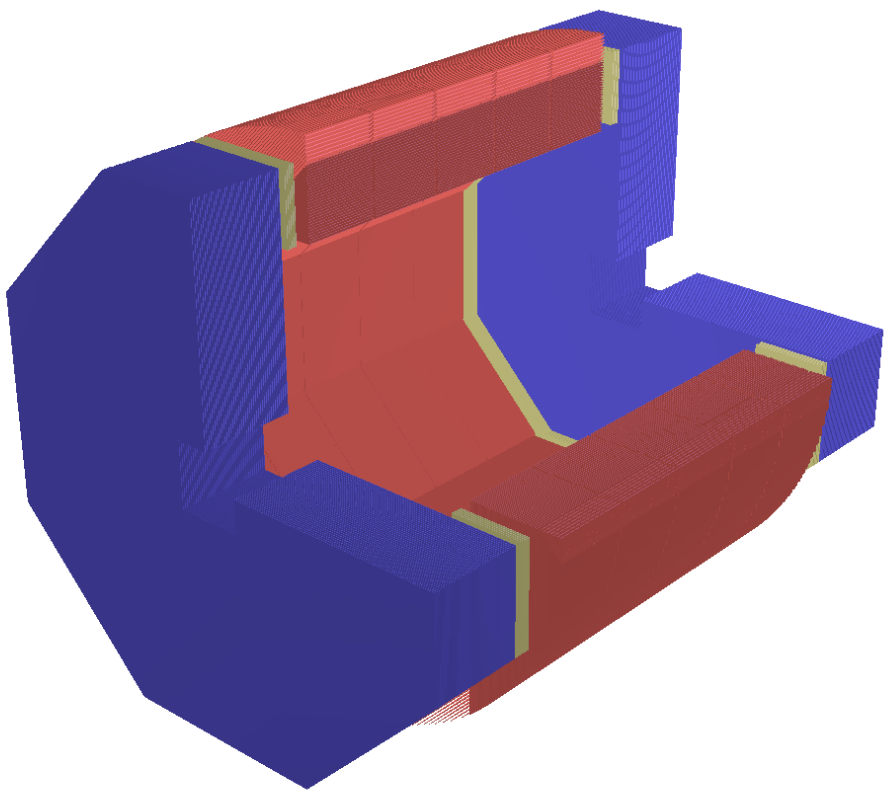

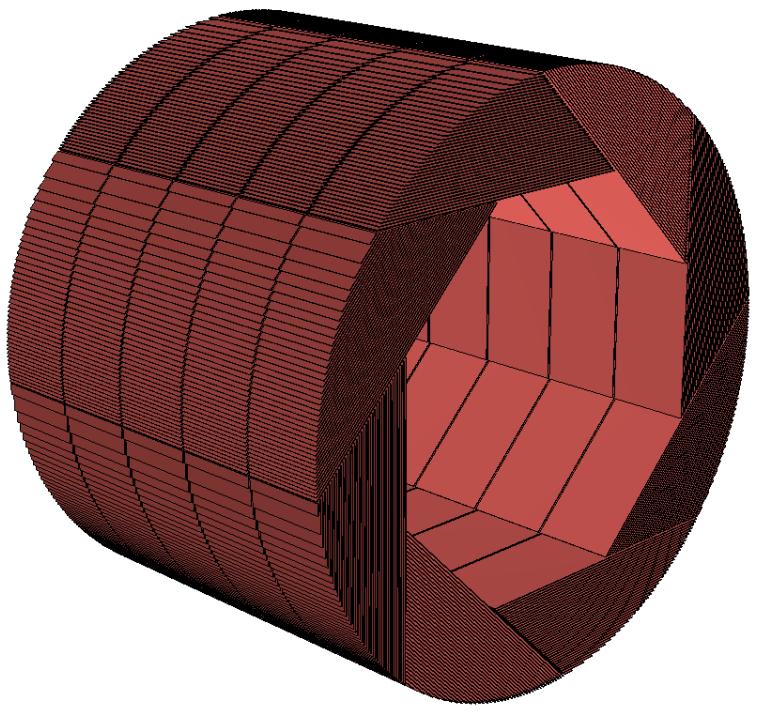

**Figure 5.16:** Schematic of the CEPC HCAL layout in its baseline design (left) consisting of one cylindrical barrel (red) spanning from 2058 mm to 3144 mm radially and two endcaps (blue) between 2650 mm and 3736 mm in $|z|$. An isometric view of the barrel HCAL is shown on the right.

## 5.4.2 SEMI-DIGITAL HADRONIC CALORIMETER

### 5.4.2.1 INTRODUCTION

For the CEPC, a Semi-Digital Hadronic CALorimeter (SDHCAL) based on gaseous detector is proposed. This is motivated by the excellent efficiency and very good homogeneity the gaseous detectors could provide. Another important advantage of gaseous detectors is the possibility to have very fine lateral segmentation. Indeed, in contrast to scintillator tiles, the lateral segmentation of gaseous devices is determined by the readout electronics and not by the detector segmentation itself. The thickness of active layers is also importance as the CEPC HCAL is to be placed inside the solenoid. Highly efficient gaseous detectors can indeed be built with a thickness of less than 3 mm. While other detectors could achieve such performance, gaseous detectors have the advantage of being cost-effective and discharge free. They are also known for their fast timing performance which could be used to perform 4-dimensional (4D)[1] reconstruction of hadronic showers. Such a reconstruction can better resolve hadronic showers from different particles and identify delayed neutrons and ultimately improve the energy measurement.

To achieve excellent resolution in the hadronic shower energy measurement, a binary readout of the gaseous detector is the simplest and most effective approach. However, a lateral segmentation of a few millimeters is needed to ensure good linearity and to resolve energy deposits. Such a lateral segmentation leads to a huge number of electronic channels resulting in a complicated readout system design and excessive power consumption. A cell size of $1 \times 1\ \mathrm{cm}^2$ is found to be a good compromise that still provides a good resolution at moderate energies. However, simulation studies show that saturation effects are expected to show up at higher energies ($> 40\,\mathrm{GeV}$). This happens when many particles cross one cell in the center of a hadronic shower. To reduce these effects, multi-threshold electronics (Semi-Digital) readout is chosen to improve the energy resolution by utilizing the particle density information. These elements were behind the development of a SDHCAL that is proposed for one of the CEPC detectors.

Even with a $1\times1$ cm$^2$ lateral granularity of the readout system, the number of electronic channels is still large. This has two important consequences. The first is the high power consumption and the resulting increase in temperature which affects the performance of the active layers. The other consequence is the number of service cables needed to power and readout these channels. These two aspects can degrade the performance of the HCAL and destroy the principle of PFA if they are not addressed properly.

The R&D pursued by the CALICE SDHCAL groups has succeeded to pass almost all technical hurdles of a PFA-based HCAL. The SDHCAL groups have successful built the first technological prototype [15] of these new-generation calorimeters with 48 active layers of GRPC, each with a size of $1\,\mathrm{m}^2$. The prototype validates the concept of the high-granularity gaseous detector and enables the study of the energy resolution of hadrons achievable with such a calorimeter. Figure 5.17 shows the energy resolutions of SDHCALs with different number of layers from simulated pion samples. A SDHCAL with 40 layers has a decent performance for pions with energies up to 80 GeV, sufficient for a CEPC detector.

A baseline SDHCAL detector has been designed with 40 layers in total. Each layer is composed of a 20 mm thick stainless steel layer, a 3 mm thick GRPC and 3 mm thick readout electronics with $1 \times 1$ cm$^2$ readout pads.

In order to investigate appropriate options for the active detector of the SDHCAL, two parallel detector schemes, the GRPC and the Thick Gas Electron Multiplier detectors (THGEM) are proposed for the active layers of the SDHCAL.

[1]Time is the fourth dimension.

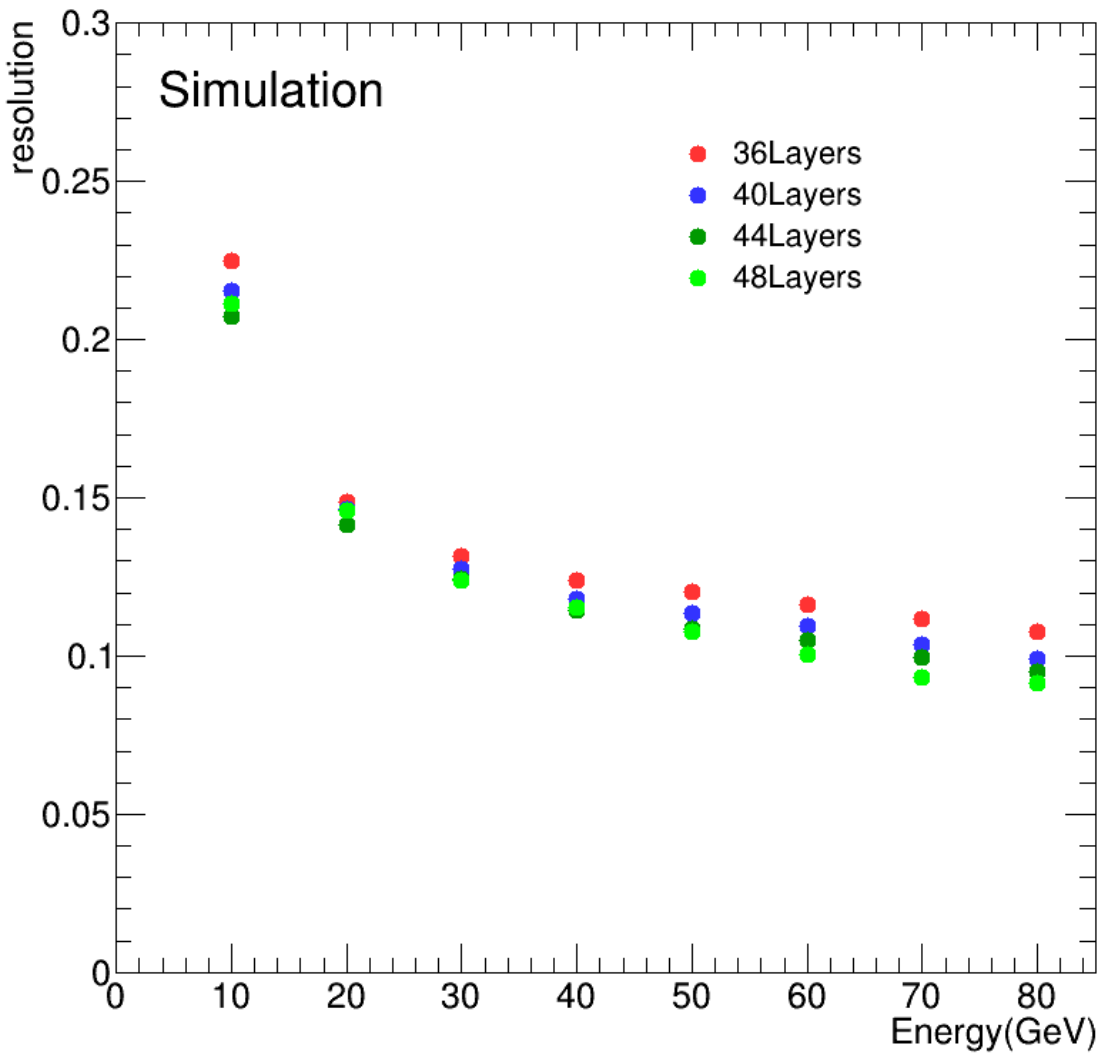

**Figure 5.17:** Energy resolutions from simulation of SDHCALs with different number of layers for pions with energies ranging from 10 GeV to 80 GeV. Energy resolutions of the SDHCALs with 48-layer, 44-layer, 40-layer, 36-layer are shown in light green, green, blue and red dots, respectively. A SDHCAL with 40-layer has decent energy resolutione, about 15% and 10% for pion energies of 20 GeV and 80 GeV, respectively.

#### 5.4.2.2 GRPC BASED SDHCAL

**The GRPC scheme** The structure of the HCAL with GRPC as an active layer for the CEPC is shown in Figure 5.18. It is made out of two glass plates of 0.7 mm and 1.1 mm thickness. The thinner plate is used to form the anode while the thicker one forms the cathode. Ceramic balls of 1.2 mm diameter are used as spacers between the glass plates. The balls are glued on only one of the glass plates. In addition to those balls, 13 cylindrical fiber-glass buttons of 4 mm diameter are also used. Contrary to the ceramic balls the buttons are glued to both plates ensuring a robust structure. Special spacers (ceramic balls) were used to maintain uniform gas gap of 1.2 mm. Their number and distribution were optimized to reduce the noise and dead zones $(0.1\%)$.

The distance between the spacers (10 cm) was fixed so that the deviation of the gap distance between the two plates under the glass weight and the electric force does not exceed 45 $\mu\mathrm{m}$. The choice of these spacers rather than fishing lines was intended to reduce the dead zones $(0.1\%)$. It was also aimed at reducing the noise contribution observed along the fishing lines in the standard GRPC chambers. The gas volume is closed by a 1.2 mm thick and 3 mm wide glass-fiber frame glued on both glass plates. The glue used for both the frame and the spacers was chosen for its chemical passivity and long term performance. The resistive coating on the glass plates which is used to apply the high voltage and thus to create the electric field in the gas volume was found to play important role in the pad multiplicity associated to a MIP [16]. A product based on colloids containing graphite was developed. It is applied on the outer faces of the two electrodes using the silk screen print method, which ensures very uniform surface quality. The measured surface resistivity at various points over a $1\mathrm{m}^2$ glass coated with the previous paint showed a mean value of 1.2 $\mathrm{M}\Omega/\mathrm{cm}^2$ and a ratio of the maximum to minimum values of less than 2 ensuring a good homogeneity of the detector.

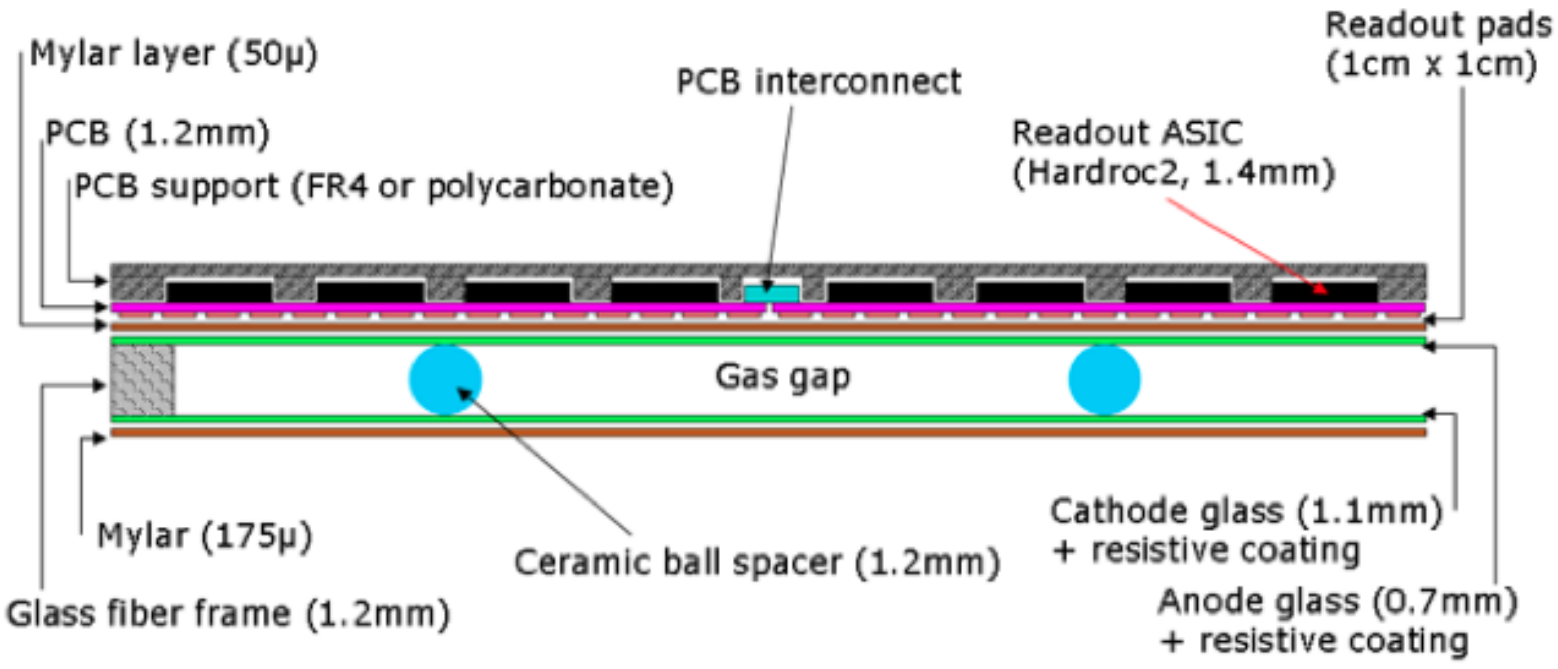

**Figure 5.18:** Cross-sectional view of an active layer with GRPC and readout layer. The GRPC gas gap is 1.2 mm, with two glass plates of 1.1 mm (cathode plate) and 0.7 mm (anode plate) thickness. The thickness of PCB is 1.2 mm and that of readout ASIC is 1.4 mm. (Note: Figure taken from [17].)

Another important aspect of this development concerns the gas circulation within the GRPC taking into account that for the CEPC SDHCAL, gas outlets should all be on one side. A realization of this system was developed. It is based on channeling the gas along one side of the chamber and releasing it into the main gas volume at regular intervals. A similar system is used to collect the gas on the opposite side. A finite element model has been established to check the gas distribution. The simulation confirms that the gas speed is reasonably uniform over most of the chamber area. The GRPC and its associated electronics are housed in a special cassette which protects the chamber and ensures that the readout board is in close contact with the anode glass. The cassette is a thin box consisting of 2.5 mm thick stainless steel plates separated by 6 mm wide stainless steel spacers. The plates are also a part of the absorber.

The electronics board is assembled with a polycarbonate spacer which is also used to fill the gaps between the readout chips and to improve the overall rigidity of the detector. The electronics board is fixed on the small plate of the cassette with tiny screws. The assembled set is fixed on the other plate which hosts the detector and the spacers. The total thickness of the cassette is 11 mm with 6 mm of which due to the sensitive medium including the GRPC detector and the readout electronics.

**GRPC technological prototype** An SDHCAL prototype fulfilling the efficiency, robustness and the compactness requirements of the future PFA-based leptonic collider experiments [15] was built. A total of 48 cassettes as the one described above were built. They fulfilled a stringent quality control. It is worth mentioning that 10500 HR ASICs were produced and tested using a dedicated robot for this purpose. The yield was found to be higher than 92%. The ASICs were then fixed on the PCBs over a surface area of $1m^2$ and then subsequently fixed on the cassette cover once successfully tested. The cassettes were inserted in a self-supporting mechanical structure that was conceived and built in collaboration with the Spanish group of CIEMAT. The structure is made of Stainless Steel plates of 1.5 cm each. The plates were machined to have an excellent flatness and well controlled thickness. The flatness of the plates was measured using a laser-based interferometer system. It was found that the flatness of the plates are less than 500 μm. In April 2012 the prototype was exposed to pion, muon, electron beams of both the PS and the SPS of CERN as shown in Figure 5.19. The data were collected continuously in a triggerless mode. Figure 5.20 shows the efficiency (a) and pad multiplicity (b) of the pro-totype GRPC chambers measured using the muon beam. Figure 5.21 shows a display of two events collected in the SDHCAL. One is a produced by a pion interaction (a) and the other by an electron interaction (b).

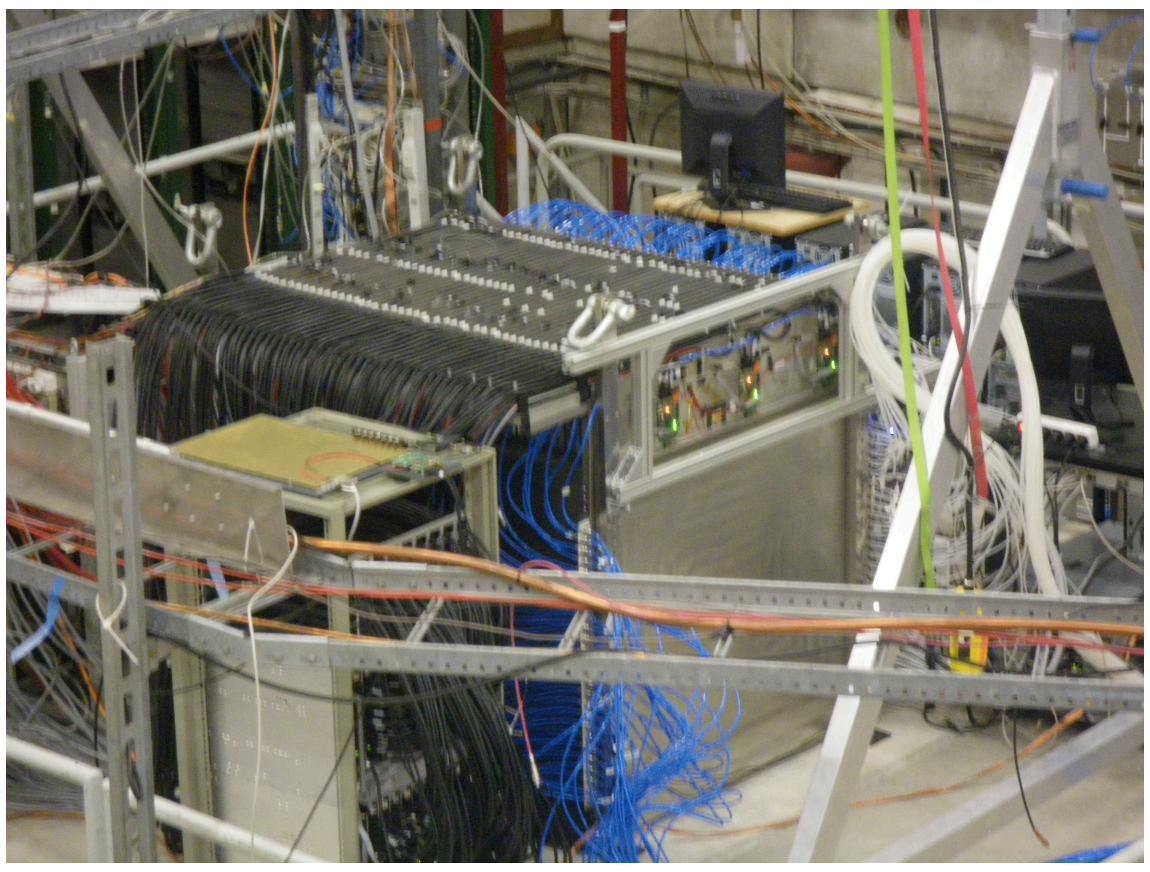

**Figure 5.19:** The SDHCAL prototype in beam test at CERN. (Note: Figure taken from [17].)

The SDHCAL prototype results obtained with a minimum data treatment (no gain correction) show clearly that excellent linearity and good resolution [17] could be achieved on large energy scale as can be shown in Figure 5.22 where results obtained in two different beam lines are obtained using the same detector configurations. As is clearly demonstrated from this data, the high granularity of the SDHCAL allows one to study thoroughly the hadronic showers topology and to improve on the energy resolution by, among others, separating the electromagnetic and the hadronic contribution. The separation between close-by showers will also benefit from the high granularity on the one hand and the very clean detector response ( $< 1$ Hz/cm$^2$) on the other. The results obtained with the SDHCAL [18] confirm the excellent efficiency of such separation due to the SDHCAL performance.

The quality of data obtained during several campaigns of data taking at the CERN PS and SPS beam lines validates completely the SDHCAL concept. This is especially encouraging since no gain correction was applied to the electronics channels to equalize their response. Still, improvement was further achieved by applying gain and threshold correction schemes in terms of the calorimeter response homogeneity.

A digitizer describing the response of the GRPC within the SDHCAL was developed [19]. It allows to study the SDHCAL behavior in a realistic manner in the future experiments.

In parallel to the prototype construction, a single cassette was tested in a magnetic field of 3 Tesla (H2 line at CERN) applying the power-pulsed mode. The test beam results [20] indicated clearly that the use of the power-pulsed mode in such a magnetic field is possible. The behavior of the detector in terms of efficiency, multiplicity, and other factors was found to be similar to those obtained in the absence of both the magnetic field and the power-pulsed mode.

**Current SDHCAL R&D** Large GRPC of 1m$^2$ were developed and built for the technological prototype. However, larger GRPC are needed in the SDHCAL proposed for future leptonic collider experiments. These large chambers with gas inlet and outlet on one side need a dedicated study to guarantee a uniform gas gap everywhere notwithstanding the angle of the plate. It is necessary also to ensure an efficient gas distribution as it was done for the 1 m$^2$ chambers. To obtain this different gas distribution systems were studied. A new scheme with two gas inlets and one outlet was found to ensure an excellent homogeneity of the gas distribution. This system will be used in the near future to build large detectors exceeding 2 m$^2$.

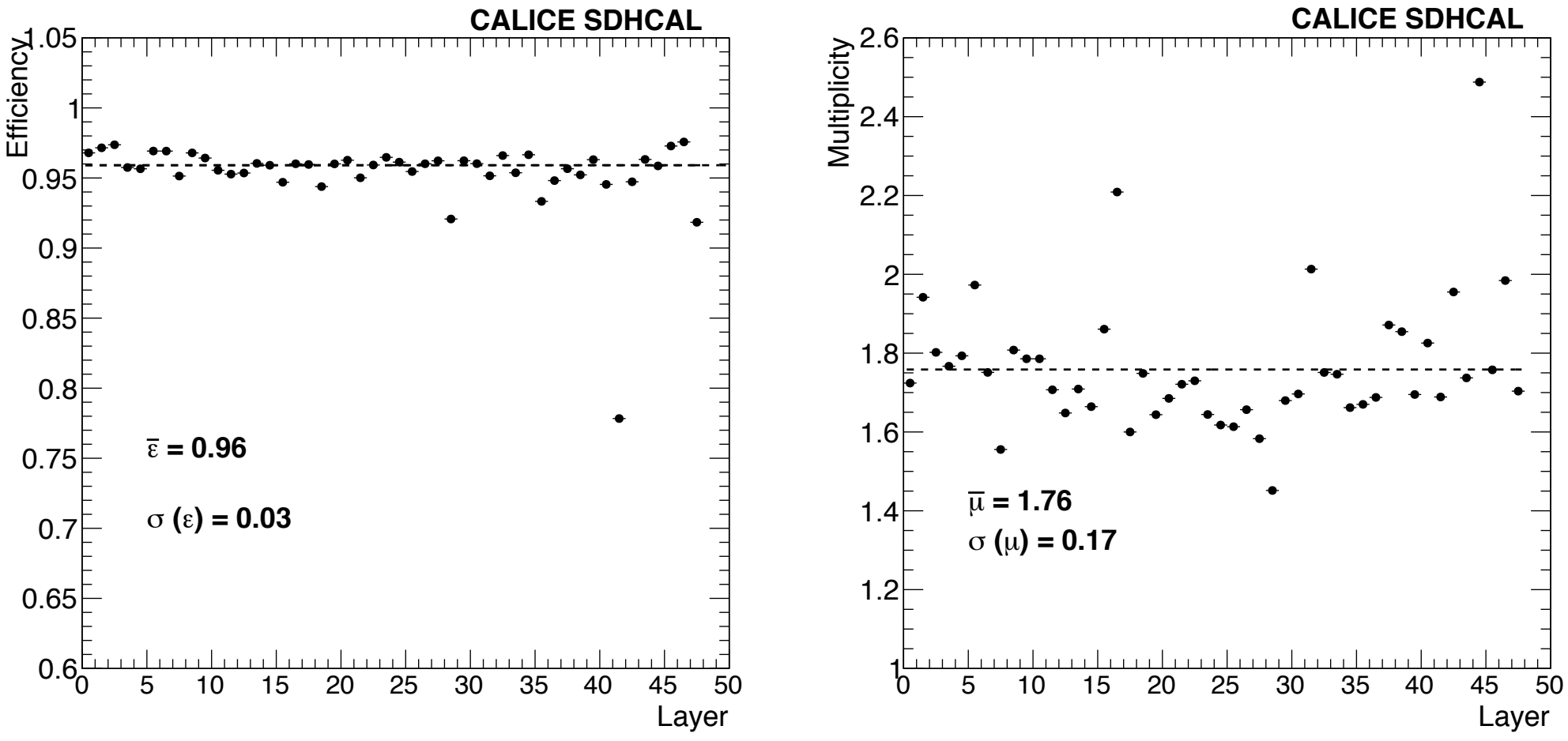

(a) (b)

**Figure 5.20:** (a) Efficiency of the GRPC detectors of the SDHCAL, the average efficiency is $0.96 \pm 0.03$. (b) The pad multiplicity of the GRPCs, it is $1.76 \pm 0.17$. (Note: Figures taken from [17].)

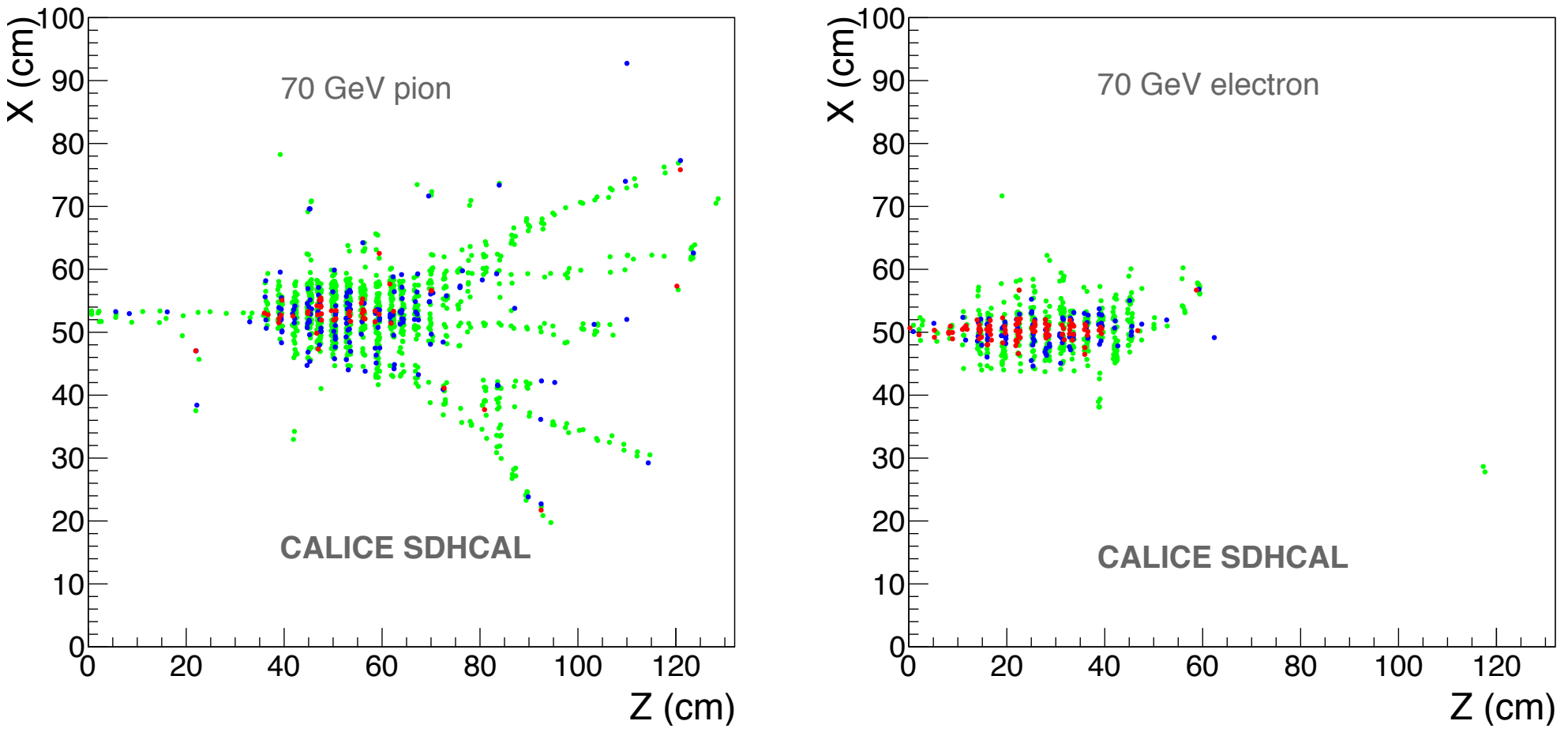

(a) (b)

**Figure 5.21:** (a) Event display of an 70 GeV pion interaction in the SDHCAL prototype. (b) Event display of a 70 GeV electron interaction in the SDHCAL prototype. (Note: Figures taken from [17].)

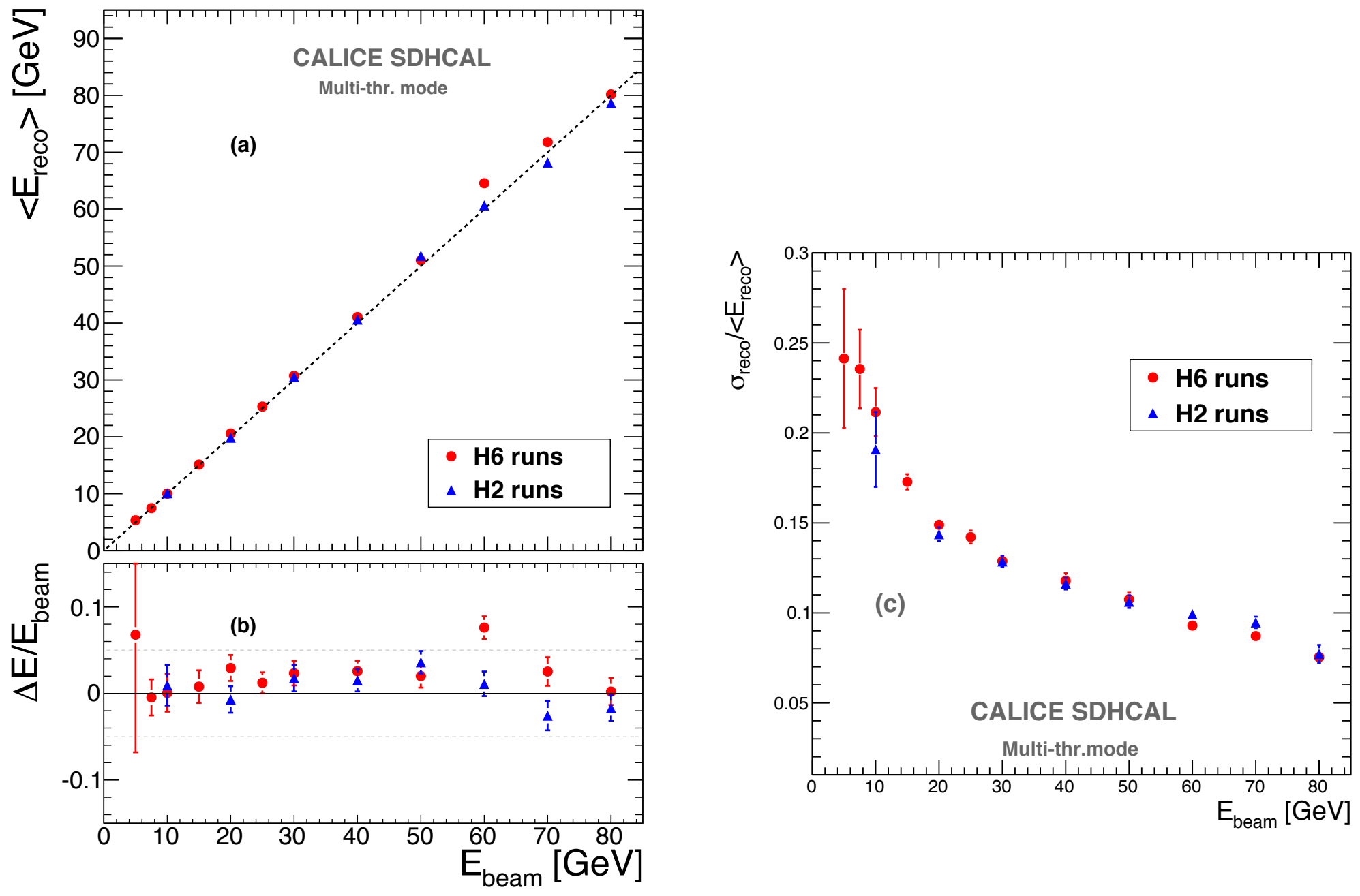

**Figure 5.22:** (a) Reconstructed energy of the hadronic showers collected in both H2 and H6 SPS beamlines. (b) The relative deviation of the reconstructed energy with respect to the beam energy. Right: Relative energy resolution of the reconstructed hadronic shower. The pion beam of H6 beamline is largely contaminated by protons at high energy ($> 50\,\mathrm{GeV}$). (Note: Figures taken from [17].)

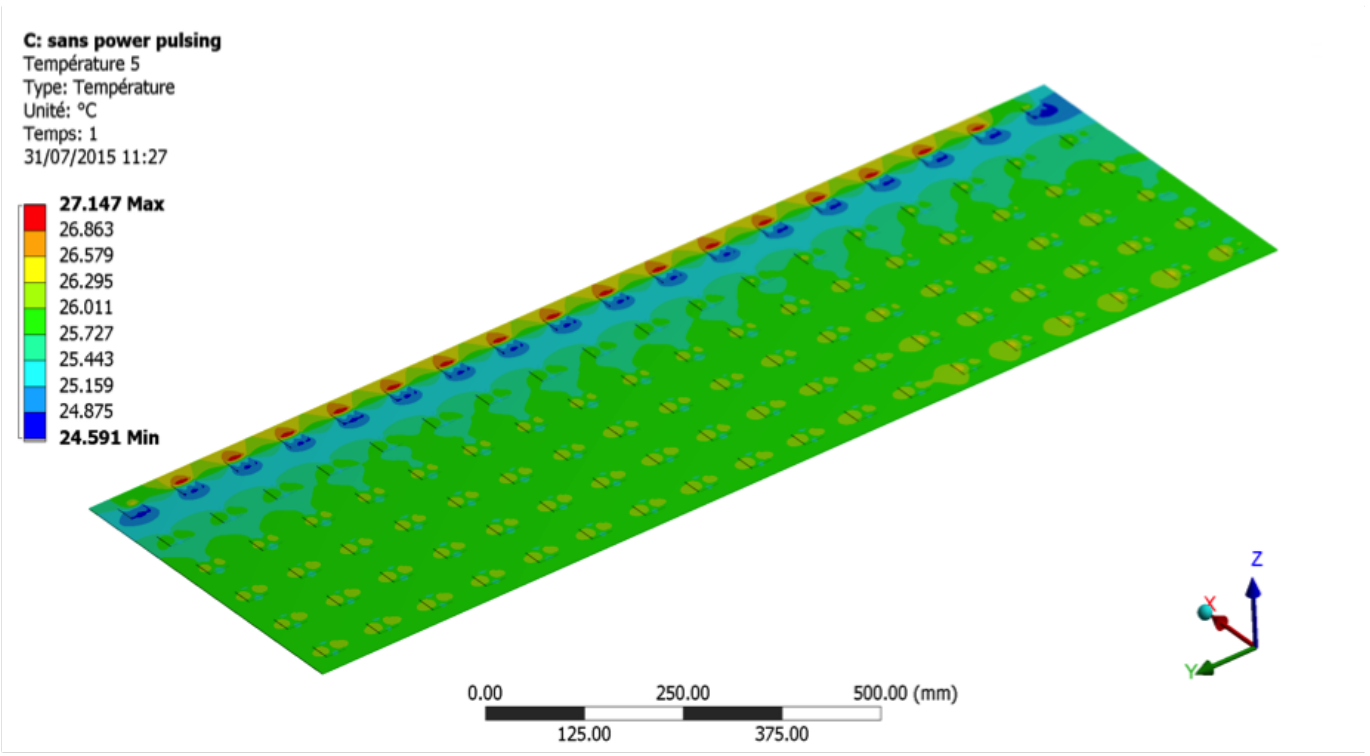

**Figure 5.23:** Temperature distribution in an active layer of the SDHCAL operated at continuous mode. The cooling system is based on circulating water inside copper tubes in contact with the ASICs.

To cope with the heating produced by the embedded readout system, a new active cooling system is being studied. Figure 5.23 shows a study of a water-based cooling system to absorb the excess of heat in the SDHCAL. The cooling system is very simple but very effective as well. It allows to keep the average temperature as well as the temperature dispersion of the GRPC well under control.

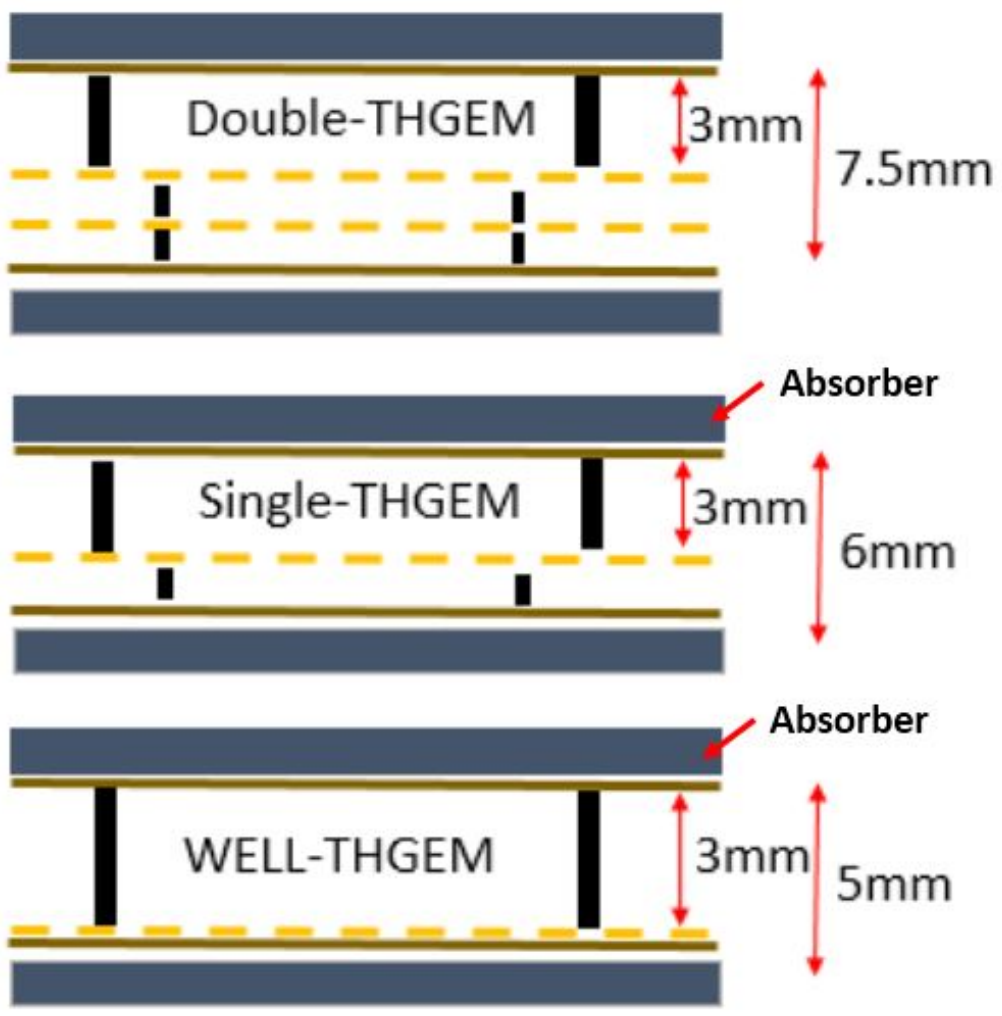

**Figure 5.24:** Schematic of three different types of THGEMs, eg. doubly-THGEM with thickness of 7.5 mm (top plot), single-THGEM with thickness of 6 mm (middle plot) and well-THGEM with thickness of 5 mm (bottom plot).

#### 5.4.2.3 THGEM-BASED DHCAL

**The THGEM scheme** The THGEM can be built in large quantities at low cost, which might make them suitable for the large CEPC HCAL. THGEM detectors can provide flexible configurations, which allow small anode pads for high granularity. They are robust and fast, with only a few nano-seconds rise time, and have a short recovery time which allows a higher rate capability compared to other detectors. They are operated at a relatively low voltage across the amplification layer with stable high gain. The ionization signal from charged tracks passing through the drift section of the active layer is amplified using a single layer or WELL-type THGEM structure. The amplified charge is collected at the anode layer with pads at zero volts. As the HCAL is located within the coil, WELL-THGEM, a single layer structure with thinner thickness, as shown in Figure 5.24, can be considered as the sensitive medium, to keep the HCAL compact.

Digital readout has been proposed to limit the total amount of data, which simplifies the data treatment without comprising the energy resolution performance. The readout electronics of the DHCAL will be integrated into the sensitive layer of the system, thus minimizing dead areas. Large electronics boards are assembled together to form extra large boards before being attached to the THGEM. The board assembly will utilize a mechanical structure made of 4 mm stainless steel plates. In addition, to keep the HCAL as compact as possible, the fully equipped electronic boards are designed to be less than 2 mm thick in total.

A THGEM based detector for DHCAL has been designed with 40 layers in total. Each layer contains 2.0 cm thick stainless steel, 0.8 cm thick THGEM and readout electronics with $1 \times 1$ cm$^2$ readout pads. As THGEM production technology matures, the maximum area of THGEM is limited only by the size of the CNC drilling area. The low cost of materials and fabrication, robustness against occasional discharges, high gain and count rate capability of up to 10 MHz/cm$^2$ make THGEM very attractive for building the DHCAL. As illustrated in Figure 5.24, the total thickness of the sensitive medium is 5 mm, which consists of 3 mm drift gap, 1 mm transfer gap and 1 mm induction gap. The absorber between the active layers is made of 20 mm thick stainless steel. The thickness of the readout electronics board is about 3 mm, and the total thickness of a single sensitive layer is less than 10 mm. Each layer corresponds to about 1.2 radiation lengths and 0.65 nuclear interaction lengths. The whole DHCAL detector is evenly divided into 40 layers, with a total stainless steel absorber thickness of 4.7 nuclear interaction lengths.

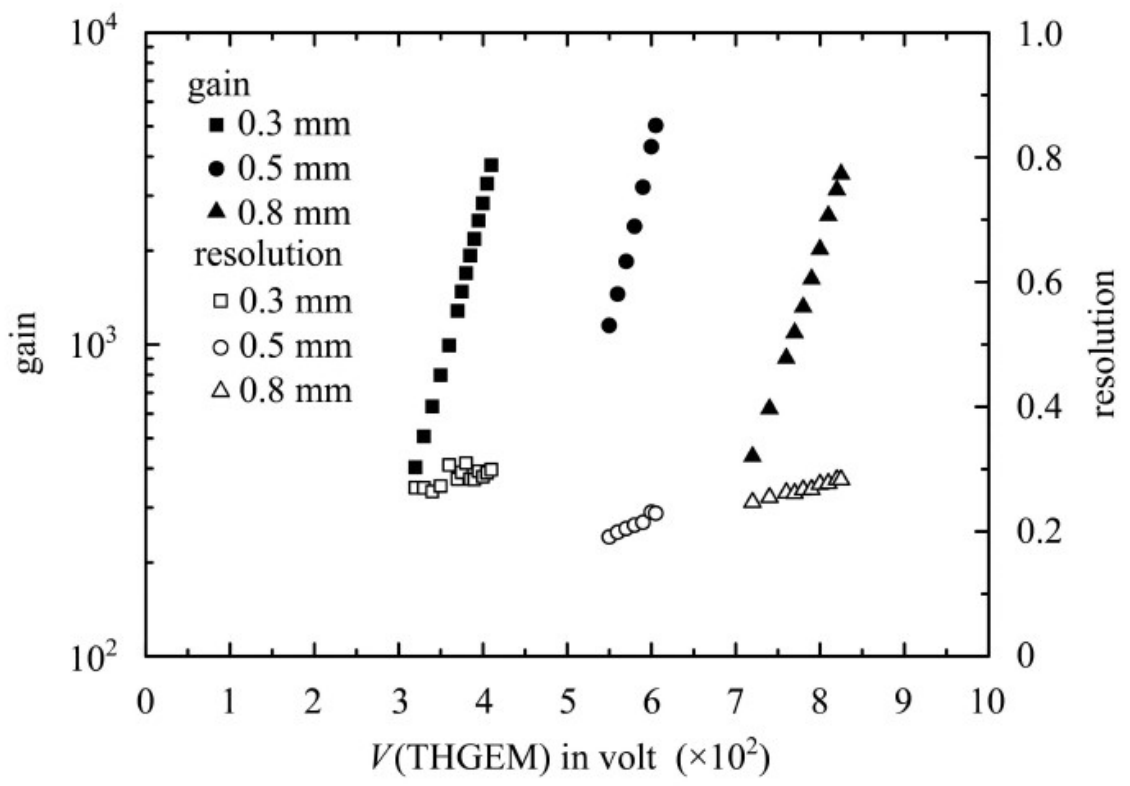

**Figure 5.25:** Gain and energy resolution of THGEM detector obtained with $^{55}$Fe source. Black boxes (0.3 mm thick), dots (0.5 mm thick) and triangles (0.8 mm thick) represent THGEM gain versus voltage, gain is achieved up to 8000. Hollow boxes, dots and triangles show energy resolution, typically around 20% to 25%.

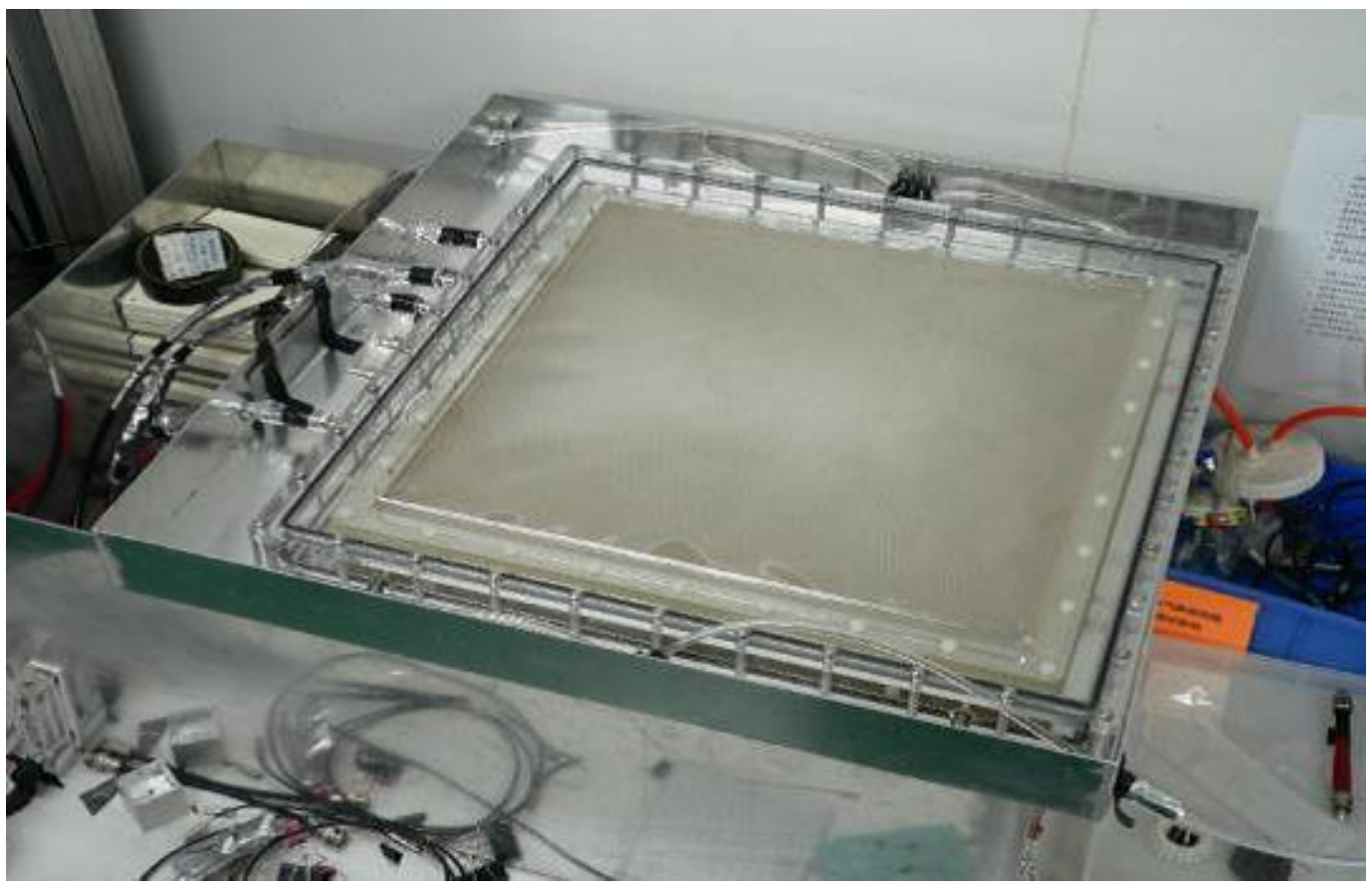

**Figure 5.26:** A double THGEM was produced with a size of $40 \times 40$ cm$^2$.

**THGEM prototype** A THGEM with an area of $40 \times 40$ cm$^2$ has been successfully fabricated, as shown in Figure 5.26, and a gain of $2 \times 10^5$ has been achieved with a double THGEM, with an energy resolution of about 20% for an $^{55}$Fe source. The THGEM produced has the following features:

1. Standard PCB processes are used, which keeps the cost low;
2. Excellent performance in terms of energy resolution, gas gain and stability (as shown in Figure 5.25);
3. Rim around the hole formed by full-etching process, the size of which can be varied between 10 $\mu$m and 90 $\mu$m, as depicted in Figure 5.25 - this allows adjustment according to gas requirements.

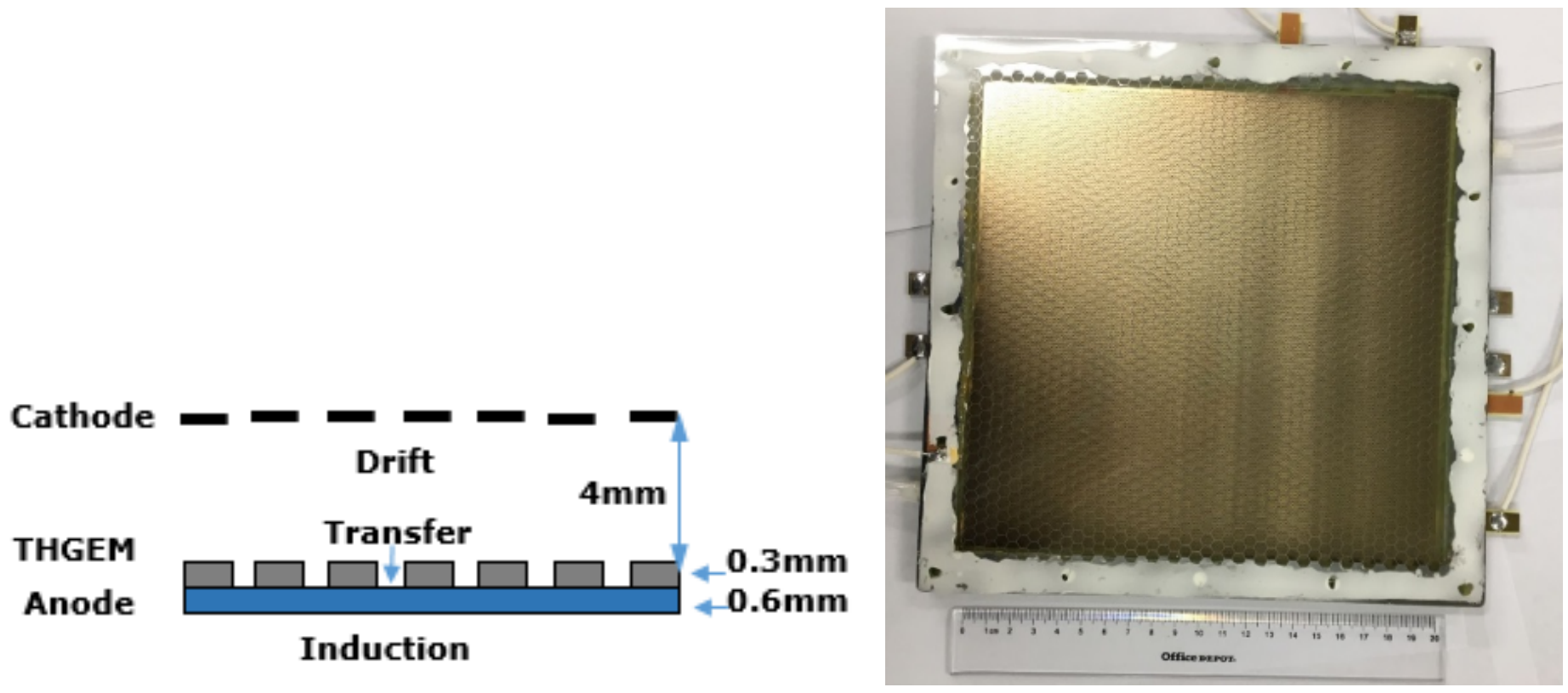

(a) (b)

**Figure 5.27:** The schematic diagram of the WELL-THGEM (a), the thickness is 5 mm. A $20\ \mathrm{cm} \times 20\ \mathrm{cm}$ WELL-THGEM detector was produced (b).

Figure 5.27 shows the schematic diagram of a new THGEM detector, where a micro-plate directly attached to the readout plate. Since the micro-porous structure is similar in shape to a well, these detectors are known as well-type THGEM (WELL-THGEM). This structure contains a single-layer THGEM, so that the thickness of detector can be reduced to $4 \sim 5$ mm, and the total thickness of the detector including ASIC electronics could be lowered to about 6 mm. A $20\ \mathrm{cm} \times 20\ \mathrm{cm}$ WELL-THGEM detector using thin-type THGEM has been developed as shown in the right plot of Figure 5.27.

In addition, large THGEM detectors have been studied. Single THGEM detectors and WELL-THGEM detectors are being developed to reduce detector instability and inefficiency. Gas recycling systems are built to lower gas consumption and pollution. The achieved THGEM detection rates of 1 MHz/cm$^2$ with efficiencies greater than 95% already meet the CEPC requirements.

**THGEM digital readout system** A MICRO-mesh gaseous structure Read-Out Chip (MICROROC), which is developed at IN2P3 by OMEGA/LAL and LAPP microelectronics groups was used to readout the THGEM-based SDHCAL. The MICROROC is a 64-channel mixed-signal integrated circuit based on 350 nm SiGe technology. Each channel of the MICROROC chip contains a very low noise fixed gain charge preamplifier which is optimized to cover a dynamic range from 1 fC to 500 fC and allow an input detector capacitance of up to 80 pF, two gain-adjustable shapers, three comparators for triple-threshold readout and a random access memory used as a digital buffer. In addition, the chip has a 10-bit DAC, a configuration register, a bandgap voltage reference, a LVDS receiver shared by 64 channels and other features. A 1.4 mm total thickness is achieved by using the Thin Quad-Flat Packaging (TQFP) technology.

### 5.4.3 AHCAL BASED ON SCINTILLATOR AND SIPM

The AHCAL (Analog Hadron CALorimeter) is a sampling calorimeter with steel as the absorber and scintillator tiles with embedded electronics. Within the CALICE collaboration, a large technological prototype [21] using scintillator tiles and SiPMs has been built to demonstrate the scalability to construct a final detector via automated mass as-sembly. The outcome of CALICE-AHCAL R&D activities can be an essential input for the conceptual design of the hadron calorimeter system at the CEPC.

### 5.4.3.1 AHCAL GEOMETRY AND SIMULATION

The AHCAL will consist of 40 sensitive and absorber layers, and the total thickness is about 100 cm. The AHCAL barrel consists of 32 super modules, each super module consists of 40 layers (Figure 5.28 shows the AHCAL structure). Figure 5.29 shows the single layer structure of AHCAL. The scintillator tiles wrapped by reflective foil are used as sensitive medium, interleaved with stainless steel absorber. The thickness of active layer including the scintillator and electronics is about $4 \sim 5$ mm. Assuming the scintillator cell size of $3 \times 3\ cm^2$, the total readout channels for AHCAL is about $4 \times 10^6$.

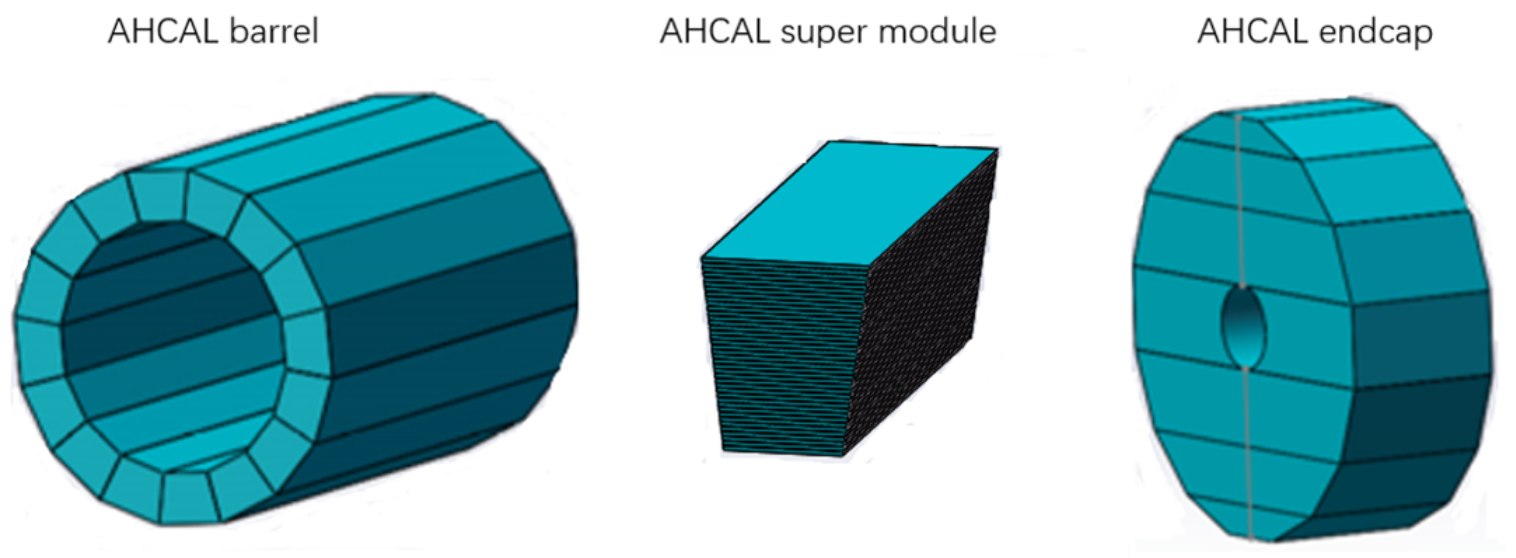

**Figure 5.28:** The layout of AHCAL barrel (left plot) and endcap regions (right plot), the middle plot shows a super module of AHCAL. The total thickness of AHCAL is about 100 cm. The AHCAL barrel consists of 32 super modules, each with 40 layers.

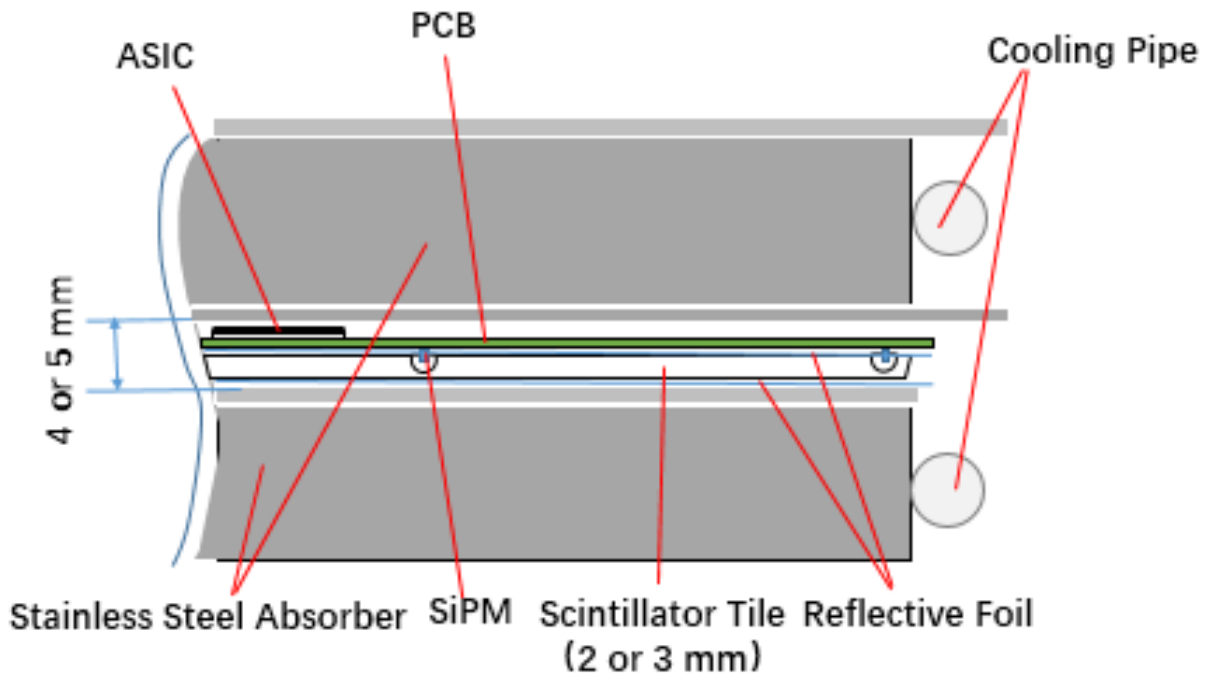

**Figure 5.29:** Cross-sectional view of a single layer of AHCAL with stainless steel absorber. The thickness of active layer including scintillator and readout electronics is about 5 mm.

The structure of scintillator tiles is shown in Figure 5.30. A dome-shaped cavity was processed in the center of the bottom surface of each tile by injection molding technology. The diameter and height of the cavity [22] are 6 mm, 1.5 mm, respectively, as shown in Figure 5.30 (right). The "SiPM-on-Tile" design has advantage to mount SiPMs on PCB so that automated mass assembly of all components can be achieved. Good response uniformity and low dead area will be achieved by the design of the cavity. More optimizations of the cavity structure will be done by GEANT4 simulation.

The AHCAL prototype detector was simulated by GEANT4 to show the expected performance of combined ECAL and HCAL using single hadrons. An earlier version of the detector model was used here. The geometry information was extracted by Mokka at runtime and the generated events were stored in Slcio, which contains primary information regarding the energy deposition, hit position, time and Monte Carlo particle causing the energy deposition. The ECAL was simulated with 30 layers, and the HCAL has 40 active layers interleaved with 20 mm stainless steel as absorber plates. Each active layer consists of plastic scintillator (3 mm) and readout layer (2 mm PCB). The detector cell size is 30×30×3 mm$^3$, as shown in Figure 5.31.

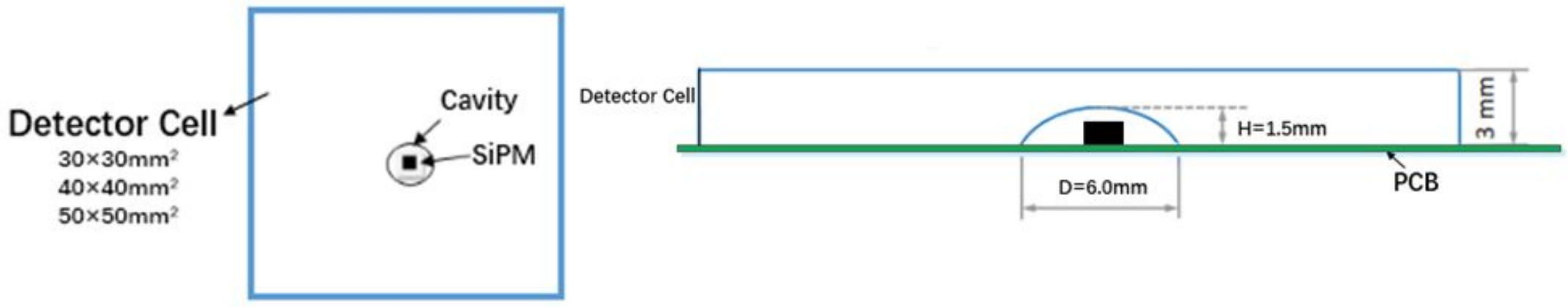

**Figure 5.30:** Top view of a detector cell (left plot) and cross-sectional view of a detector cell with a dome-shaped cavity (right plot). The detector cell size varies from 30×30 mm$^2$, 40×40 mm$^2$ to 50×50 mm$^2$. The height of dome-shaped cavity in the center of detector cell is 1.5 mm with diameter of 6 mm.

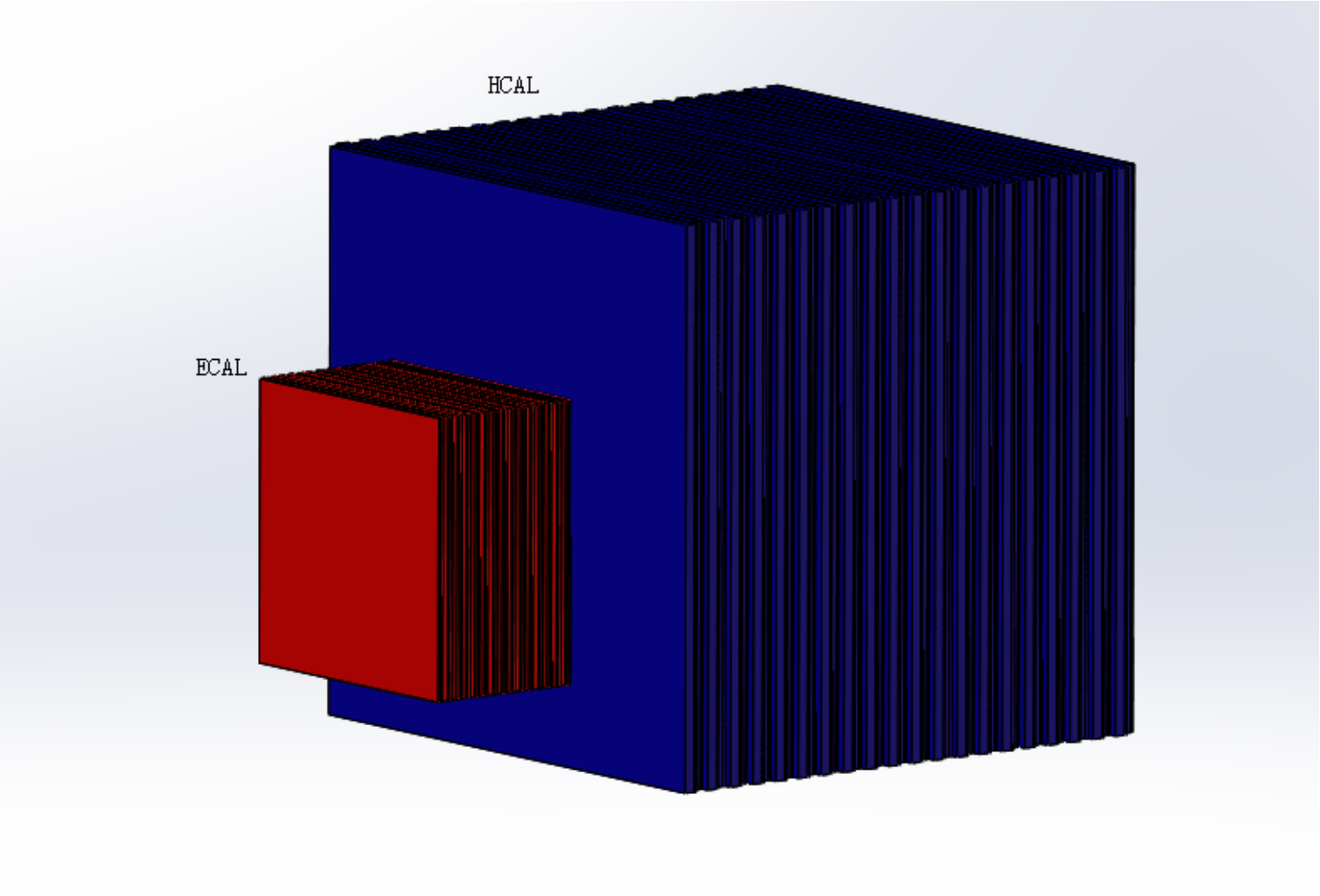

**Figure 5.31:** The structure of simulated calorimeters which is a part of the simplified geometry. Red part is the Silicon ECAL (30-layer), blue part is the scintillator AHCAL (40-layer).

In order to obtain the resolution of calorimeters (ECAL and AHCAL) as shown in Figure 5.31, the energy reconstruction formula 5.3 is employed [23], the coefficients a and b in this formula represent ECAL and HCAL calibration constants, respectively. After optimization, the calibration constants are $a = 44.4$ and $b = 44.2$, respectively, which were corrected to the energy scale of 60 GeV pions. The calibration constants compensate for the energy leakage from the calorimeters. Equation 5.4 [23] is used to fit for the energy resolution, as shown in Figure 5.32.

$$E_{REC} = a \times E_{ECAL} + b \times E_{HCAl} \tag{5.3}$$

$$\frac{\sigma}{E} = \frac{p_0}{\sqrt{E}} + p_1 \tag{5.4}$$

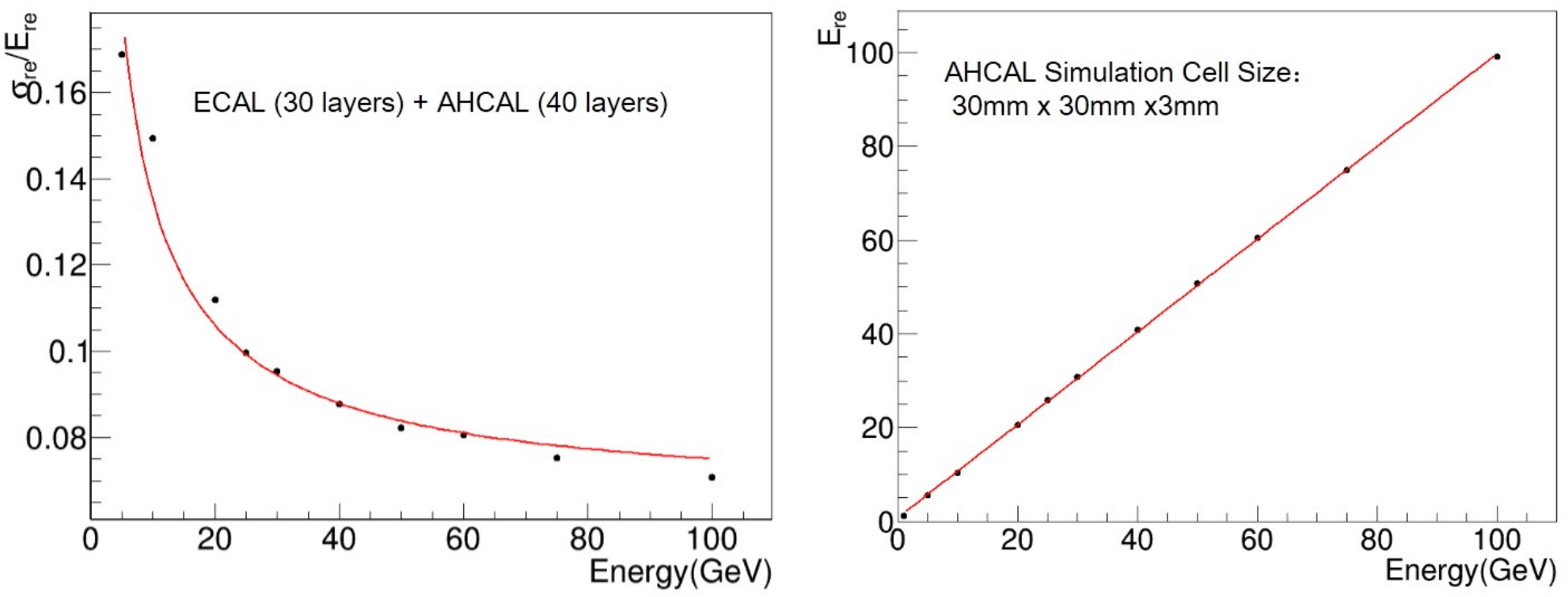

**Figure 5.32:** The left plot is the energy resolution from the SiW-ECAL and AHCAL for pions. The right plot is the corresponding results of reconstruction energy linearity. The energy resolution is 11% and 8% for energy at 20 GeV and 80 GeV, respectively.

### 5.4.3.2 PLASTIC SCINTILLATOR DETECTOR CELL DESIGN AND TEST

According to studies by the CALICE collaboration, a scintillator detector cell size of 30×30 mm$^2$ is an optimal size. The simulation results of the CALICE collaboration [24] also suggest that it is possible to use the detector cells of larger sizes. A large detector cell size of 40×40 mm$^2$ would reduce by nearly half the number of electronics channels compared to the 30×30 mm$^2$ size. Therefore, the construction costs can be greatly reduced if the larger detector cells can meet the physics requirements. Two larger sizes of detector cells were considered. Four kinds of scintillator tiles with different sizes were fabricated and tested.

The SiPM is soldered onto a readout Printed Circuit Board (PCB) and the scintillator tile wrapped by ESR reflective foil is directly glued onto the PCB. A cavity design provides enough space for the SiPM package and improves collection efficiency of the light produced by incident particles penetrating the tile at different positions.

A strongly non-uniform tile response can lead to a distortion of the energy reconstruction in a complete calorimeter, and also compromises the calibration of the detector cells based on single particle signals. Three different sizes tiles (30×30×3 mm$^3$, 30×30×2 mm$^3$ and 50×50×3 mm$^3$) were tested with the Hamamatsu MPPC S12571-025P and S13360-025PE. The spatial distribution of photon equivalents number (p.e.) with different detector cell areas are shown in Figure 5.33. The result shows that the number of p.e. in the center area is slightly larger than that of the surrounding area. The three detector cells show good response uniformity, within 10% deviation from their mean response.

Seven detector cells of different sizes, polishing methods and wrapping foil types were measured. The larger the area of the cell is, the less p.e. are detected, and the results of same size cells varied greatly because of the polishing methods.

The detection efficiency of 30×30×3 mm$^3$ and 50×50×3 mm$^3$ were measured with cosmic rays. The detection efficiency of 30×30×3 mm$^3$ and 50×50×3 mm$^3$ cells are 99% and 98.2%, respectively. According the cosmic-ray test result, the detection efficiency of 30×30×2 mm$^3$ with S13360-025PE MPPC also can reach 98%.

The good response uniformity and high detection efficiency results indicate that scintillator detector cells are acceptable for AHCAL. The size of 30×30×3 mm$^3$ detector cell is the baseline of AHCAL and more optimization of the detector cell size will be done by the simulation and test beam measurements.

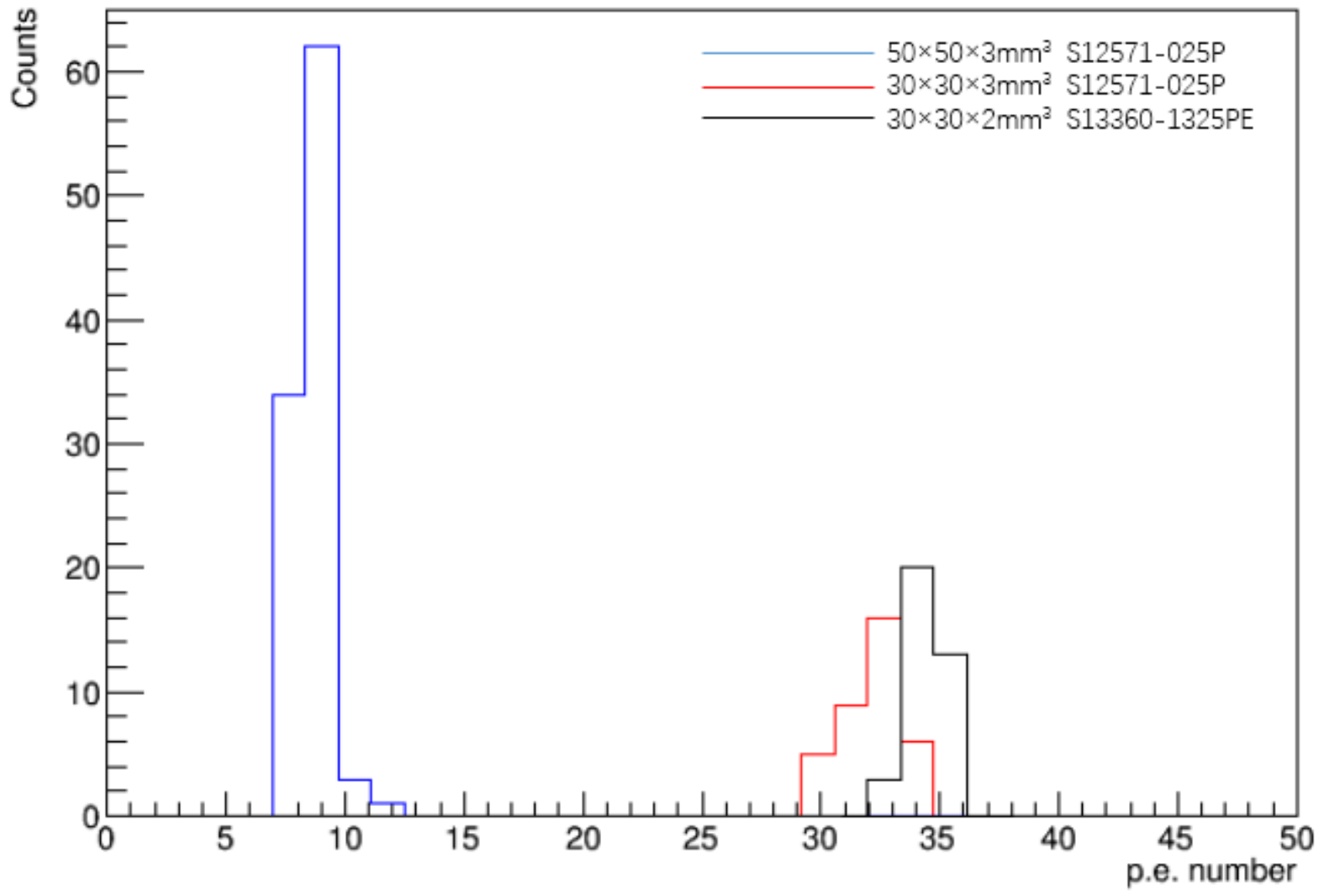

**Figure 5.33:** The uniformity measurement result of $30\times30\times3$ mm$^3$ (red histogram, p.e. is about 30–34), $50\times50\times3$ mm$^3$ (blue histogram, p.e. is about 8-10) and $30\times30\times2$ mm$^3$ (black histogram, p.e. is about 34–36) detector cell.

#### 5.4.3.3 DEVELOPMENT OF SIPM

Several kinds of SiPM were developed by Hamamatsu and other companies, they have been used for scintillator ECAL systems. The SiPM with Epitaxial Quenching Resistors (EQR SiPM) is one of the main SiPM technologies under development in China [25]. As shown in Figure 5.34, each APD cell (pixel) forms a high electric field, composing an enriched region between N-type epitaxial silicon substrate and P++ cap layer, and it employs the un-depleted region in the epitaxial silicon layer below P/N junction as the quenching resistor. Compared to conventional SiPM configurations that employ poly-silicon quenching resistors on the device surface, it is easier to achieve high density and small micro-APD cells, thus obtaining a small junction capacitor; the EQR SiPMs are expected to have short recovery time and high counting rate capability.

#### 5.4.3.4 ELECTRONICS AND DAQ

**Front-end electronics ASIC:** High-density electronics is indispensable to the instrumentation of high-granularity calorimetry. An ASIC chip named SPIROC, developed by the OMEGA group, is capable of handling 36 SiPMs. For each channel, it can be operated in an auto-trigger mode and has a dual-gain charge preamplifier with high dynamic range. It allows to measure the charge from 1 to 2000 photo-electron and the time to within 1 ns using a 12-bit digitizing circuit. With one 8-bit 5V input DAC per channel, the bias voltage for each SiPM can be adjusted to reach its optimum. In each channel, there are 16 analog memory cells that can buffer both charge and timing signals to be digitized afterwards consecutively. The digitization circuit is shared for both charge and timing measurements to minimize the power consumption, which needs to be as low as 25 $\mu$W per channel.

DAQ system is also required to be compatible to the final detector layout, where two hardware parts are essential. One part is so-called LDA (Link to Data Aggregator), which collects all the data via DIFs from active layers in an HCAL segment and transmit them to a back-end PC for further processing or storage. Smart units like Field-Programmable Gate Arrays (FPGAs) are equipped on this board for data packaging and transmission. Modern FPGAs integrated with RAMs are an ideal option to have a capability of data buffering and some advanced feature like system on chip.

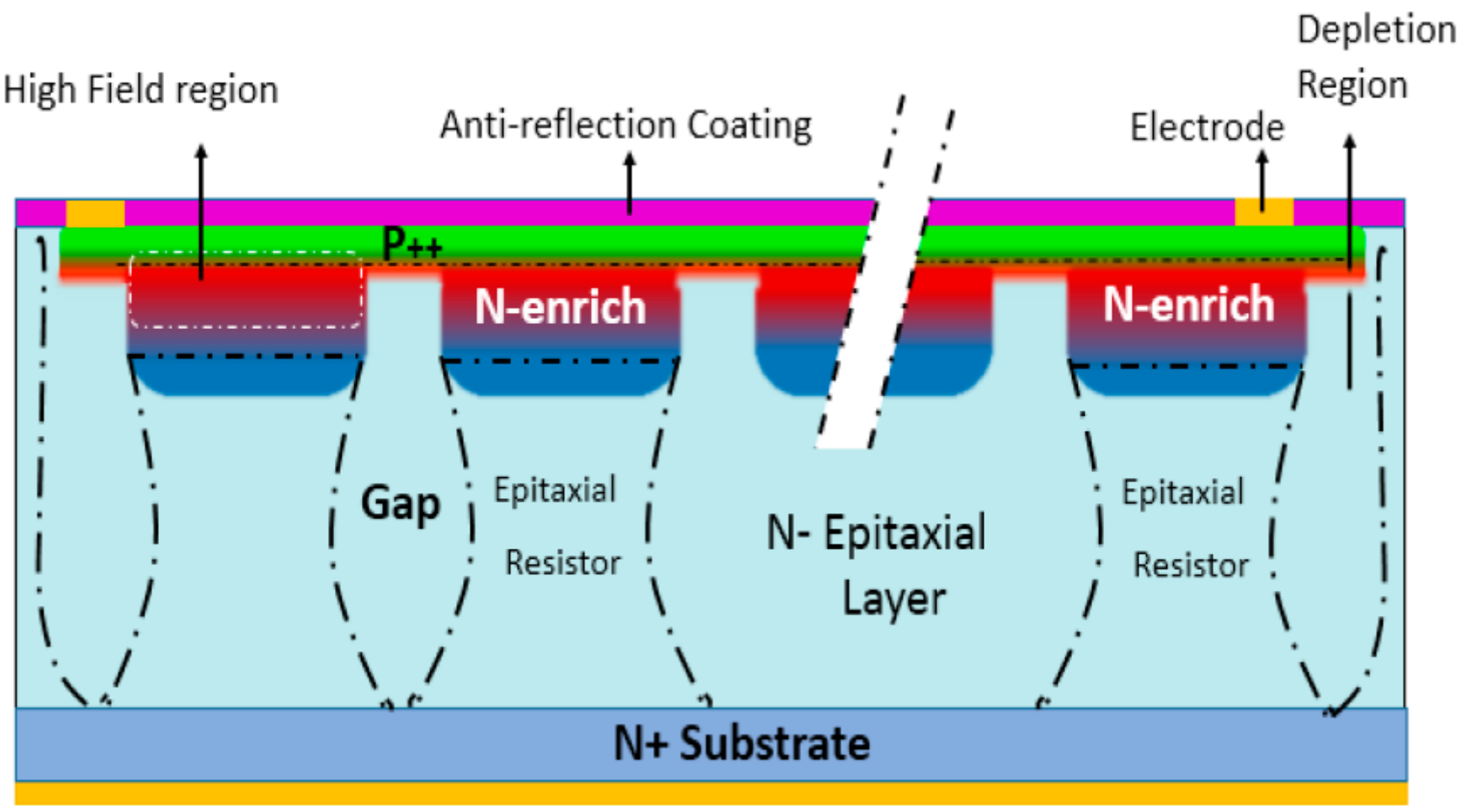

**Figure 5.34:** Schematic structure of EQR SiPM. APD cell consists of N-enriched regions forming high electric fields between the N-type epitaxial silicon wafer and the P++ surface layer, the un-depleted region in the epitaxial silicon layer below the P/N junction as the quenching resistor. The APD cells are isolated from each other by the Gap depletion region. Figure taken from Ref. [25].

#### 5.4.3.5 COOLING SYSTEM

Inside the active layers of the calorimeter, the total power consumption of SPIROC ASIC chip and SiPM is about 5 mW/channel [26]. The scintillator detector cell size is $30 \times 30$ mm$^2$, and the total channel number is about 5 million. For whole AHCAL, the total power consumption from ASIC chips is about 30 kW. The copper cooling water pipes are expected to be embedded in the stainless steel absorber. The cooling pipes are in the layer structure, as shown in Figure 5.29. Detailed design and optimization of a cooling system is needed.

## 5.5 DUAL-READOUT CALORIMETER

### 5.5.1 INTRODUCTION

The dual-readout approach envisages designing a combined, homogeneous, detector with excellent performance for both electromagnetic and hadronic particle showers.

With conventional calorimeters, the performance obtained in hadronic energy measurements is by far worse than for the electromagnetic ones. The origin of this disparity is in the showers from single hadrons or jets of hadrons. Hadronic showers develop an electromagnetic component, from $\pi^0$ and $\eta$ production, that exhibits large event-by-event fluctuations and dependence on the particle type and energy [27]. The $em$ and non-$em$ components of a hadronic shower are normally sampled with very different sensitivity, producing large differences in the measured signals, heavily affecting the energy resolution capability.

The variation of the $em$ fraction is intrinsic to hadronic showers. As a matter of fact, the $em$ fraction depends on the kind of particle initiating the shower (e.g., $\pi$, $K$, $p$) since, for example, impinging $\pi^{\pm}$ mesons can undergo a charge-exchange reaction with a nucleon as first interaction and generate a pure $em$ shower, while a $p$ cannot do the same due to baryon number conservation. Moreover, since $\pi^0$ production happens at any stage of shower development, the average $em$ fraction $\langle f_{em} \rangle$ increases with the energy as well as with the depth ("age") of the shower.

To overcome the problem two methods have been exploited: compensation and Dual Readout (DR). The first relies on equalizing the detector response to electromagnetic ($e$) and non-electromagnetic ($h$) shower particles (i.e. $h/e = 1$), but this requires a fixed ratio of absorber-to-sensor volumes, which limits the electromagnetic energy resolution, and the integration of the signals over large volumes and long times, to increase the response to the $h$ component. The dual-readout method avoids these limitations by directly measuring $f_{em}$ on an event-by-event basis. The showers are sampled through two independent processes, namely scintillation ($S$) and Čerenkov ($C$) light emissions. The former is sensitive to all ionizing particles, while the latter is produced by highly relativistic particles only, almost exclusively found inside the $em$ shower component. By combining the two measurements, energy and $f_{em}$ of each shower can be simultaneously reconstructed. The performance in hadronic calorimetry may be boosted toward its ultimate limit.

The results obtained so far with prototypes, support the statement that fiber-sampling DR calorimeters may reach resolutions of the order of $10\%/\sqrt{E}$ or better for $em$ showers and around $30 - 40\%/\sqrt{E}$ for hadronic showers and jets, coupled with strong standalone Particle IDentification (PID) capabilities. One of the strengths of a DR calorimeter is that it achieves excellent jet energy resolution while not sacrificing performance in electromagnetic energy measurements. This would allow $W \to jj$ separation from $Z \to jj$ by invariant mass, high-precision missing three-momentum reconstruction by subtraction, $e$-$\mu$-$\pi$ separation and particle tagging.

While the dual-readout concept has been extensively demonstrated and experimentally validated in a series of beam tests, the use of standard PhotoMultiplier (PM) tubes to readout the $S$ and $C$ light has so far limited its development towards a full-scale system compliant with the integration in a particle detector at a colliding beam machine. These limitations should be overcome using SiPMs, low-cost solid-state sensors of light with single photon sensitivity, magnetic field insensitivity and design flexibility.

As it will be shown in the following, the high readout granularity in the plane perpendicular to the shower development and few other signal properties will probably make redundant or even inessential the need of a longitudinal segmentation into $em$ and hadronic compartments (that is anyway possible). In case of a segmented calorimeter, both compartments need to provide dual-readout signals, in order to allow for the measurement of $\langle f_{em} \rangle$.

### 5.5.2 PRINCIPLE OF DUAL-READOUT CALORIMETRY

The independent sampling of hadronic showers, through scintillation and Čerenkov light emission, allows one to fully reconstruct, at the same time, energy and $f_{em}$ of hadronic showers. In fact, the total detected signals, measured with respect to the electromagnetic energy scale, can be expressed as:

$$S \ = \ E \, [ \, f_{em} \ + \ \eta_S \cdot (1 - f_{em}) \, ] \tag{5.5}$$

$$C \ = \ E \, [ \, f_{em} \ + \ \eta_C \cdot (1 - f_{em}) \, ] \tag{5.6}$$

where $\eta_S = (h/e)_S$ is the ratio of the average $S$ response for the non-$em$ component to the $em$ component in hadronic showers. The response being defined as the average signal per unit of deposited particle energy. The parameter $\eta_C = (h/e)_C$ is the response ratio for the $C$ signal. In a typical dual-readout calorimeter, $\eta_S \approx 0.7$ and $\eta_C \approx 0.2$. These two equations are easily solved giving:

$$\frac{C}{S} \ = \ \frac{[\, f_{em} \ + \ \eta_C \cdot (1 - f_{em}) \,]}{[\, f_{em} \ + \ \eta_S \cdot (1 - f_{em}) \,]} \tag{5.7}$$

$$E \ = \ \frac{S \ - \ \chi C}{1 \ - \ \chi} \tag{5.8}$$

where:

$$\chi \ = \ \frac{1 - \eta_S}{1 - \eta_C} \ = \ \cot \ \theta \tag{5.9}$$

This is the simplest formulation of hadronic calorimeter response: an $em$ part with relative response of unity, and a non-$em$ part with relative response $\eta$.

There are two unknowns for each shower, $E$ and $f_{em}$, and two measurements, $S$ and $C$. The electromagnetic fraction, $f_{em}$, is determined entirely by the ratio $C/S$, and the shower energy calculated as in Eq. 5.8. Both, $S$ and $C$, $\eta = (h/e)$ ratios have event-by-event fluctuations and should be considered stochastic variables, nevertheless the average $<h/e>$ values are essentially independent of hadron energy and species [28–30]. The global parameter $\chi$ can be extracted with a fit to calibration data:

$$\chi \ = \ \frac{E_0 - S}{E_0 - C} \tag{5.10}$$

$$S \ = \ (1 - \chi)E_0 + \chi C \tag{5.11}$$

where $E_0$ is the beam energy.

The geometrical meaning of the $\theta$ angle in Eq. 5.9 can be understood by looking at the scatter plot of $C$ versus $S$ signals in Figure 5.35. An illustration of the prediction for the scatter plot for protons and pions is shown in Figure 5.35(a) and the scatter plot for 60 GeV pions measured in the RD52 lead-fiber calorimeter is shown in Figure 5.35(b).

The plot in Figure 5.35(b) shows that the data points are located on a locus, clustered around a line that intersects the $C/S = 1$ line at the beam energy of 60 GeV. In first approximation, the signal generated in the Čerenkov fibers is produced only by the $em$ components of the hadron showers. The smaller the $em$ fraction $f_{em}$, the smaller the $C/S$ signal ratio. All signals are relative to the $em$ scale meaning that both the Čerenkov and the scintillation responses are calibrated with beam electrons only, i.e. no hadronic calibration is required. This is one of the most qualifying and important points of dual-readout calorimetry.

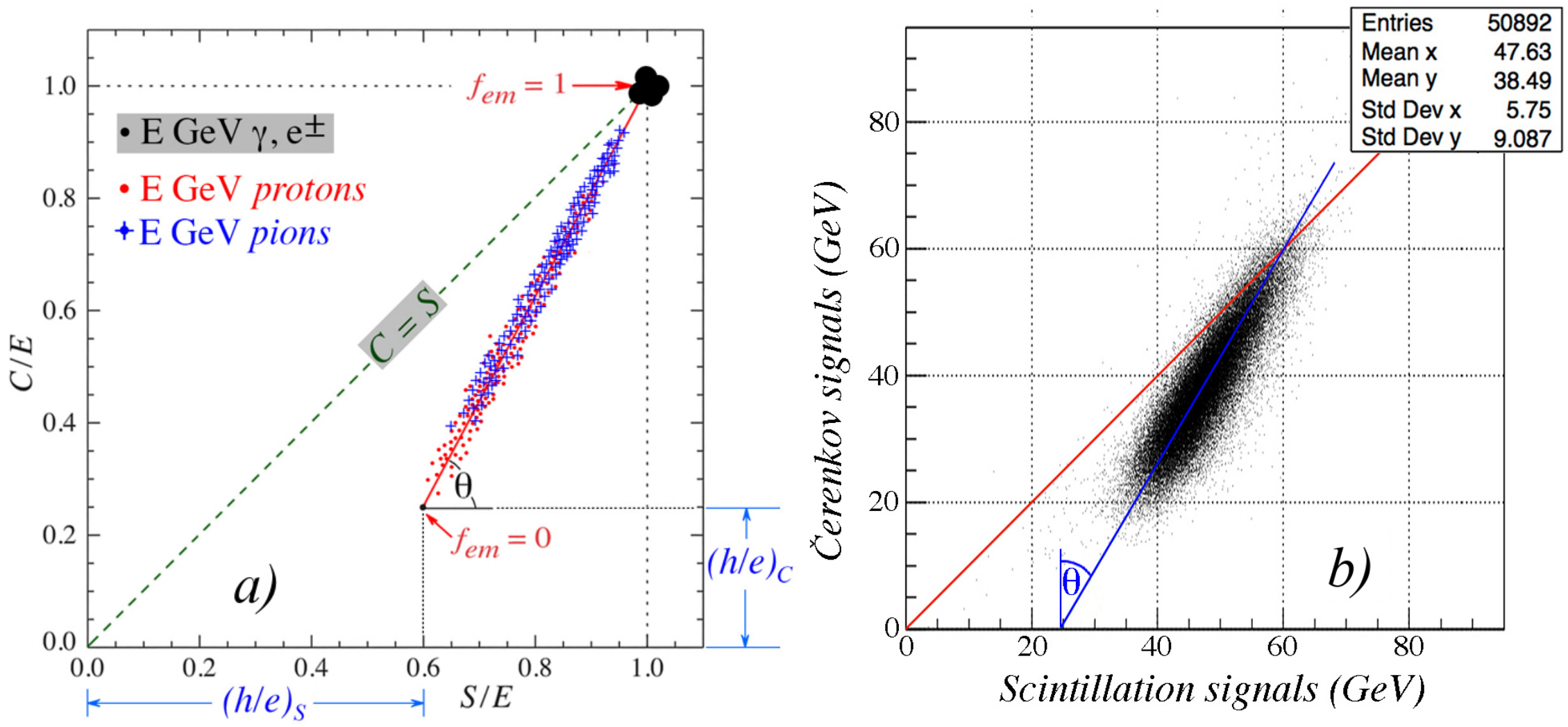

(a) (b)

**Figure 5.35:** (a) Scatter plot of $C/E$ versus $S/E$ in a dual-readout calorimeter for $p$ and $\pi$ [31]; (b) scatter plot of $C$ and $S$ signals for 60 GeV pions in the RD52 lead-fiber calorimeter [32].

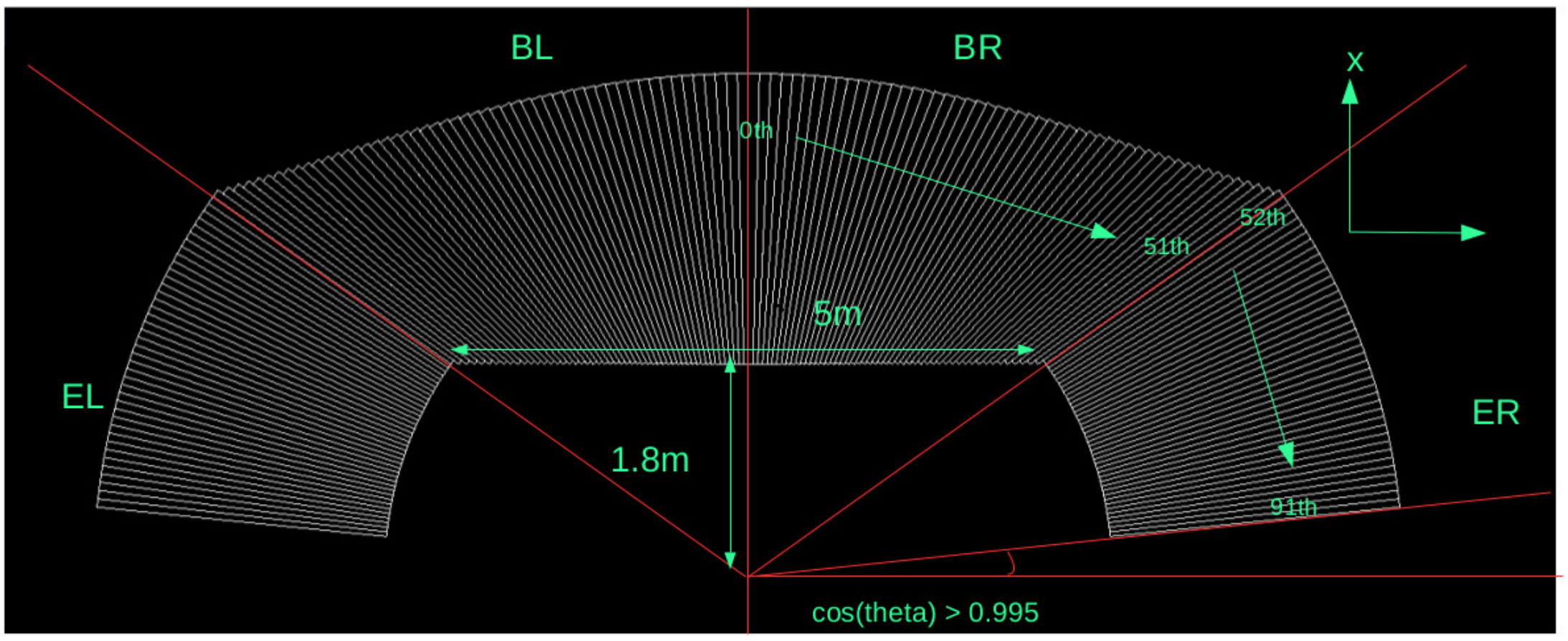

**Figure 5.36:** A possible $4\pi$ solution (called "wedge" geometry).

The effectiveness of this approach has been demonstrated by the DREAM/RD52 collaboration over a 15-year research program with a variety of detector solutions. Results and simulations [32–37] provide, so far, confidence that a fiber-sampling calorimeter, even without longitudinal segmentation, may meet the requirements of the CEPC physics program in a cost-effective way. Linearity and energy resolution, for both $em$ and hadronic showers, $e/\pi/\mu$ separation, spatial resolution, all show adequate performance.

## 5.5.3 LAYOUT AND MECHANICS

### 5.5.3.1 LAYOUT

A possible projective layout ("wedge" geometry, Figure 5.36) has been implemented in the simulations. Based on the work done for the 4th Detector Collaboration (described in its Letter of Intent [38]), it covers, with no cracks, the full volume up to $|cos(\theta)| = 0.995$, with 92 different types of towers (wedges). A typical one in the barrel region is shown in Figure 5.37(b), together with the fiber arrangement (Figure 5.37(a)): it has a granularity of $\Delta\theta \times \Delta\phi = 1.27° \times 1.27°$, a depth of about 250 cm ($\sim$ 10 $\lambda_{\text{Int}}$), and contains a total of about 4000 fibers. The sampling fraction is kept constant by fibers starting at different depths inside each tower. This layout has been already imported in the simulations for the CEPC detector. Preliminary results on performance are shown in the next chapters.

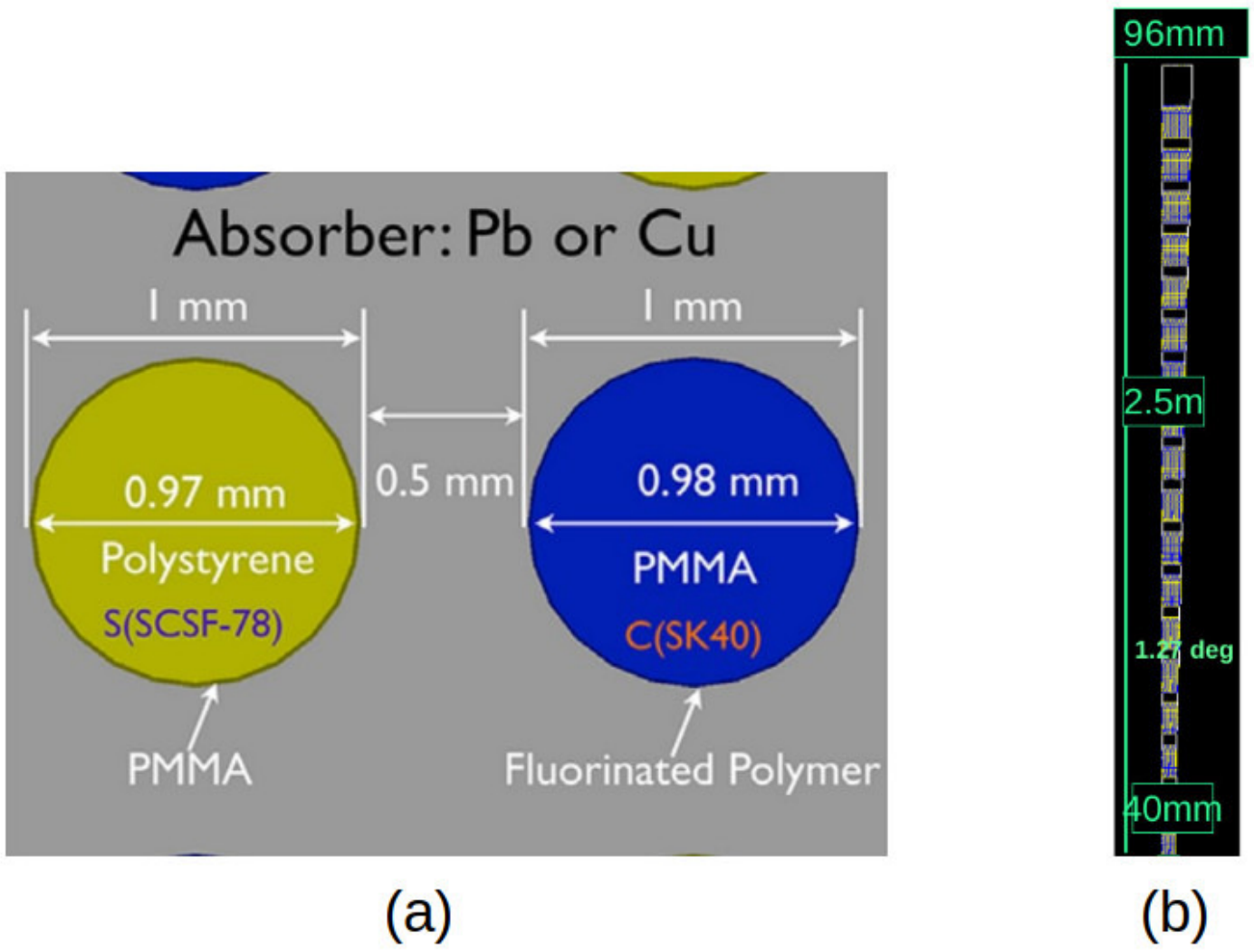

**Figure 5.37:** (a) Fiber arrangement inside the modules [35]. (b) Dimensions of a module in the barrel region (at $\eta = 0$): from inside to outside the number of fibers more than doubles.

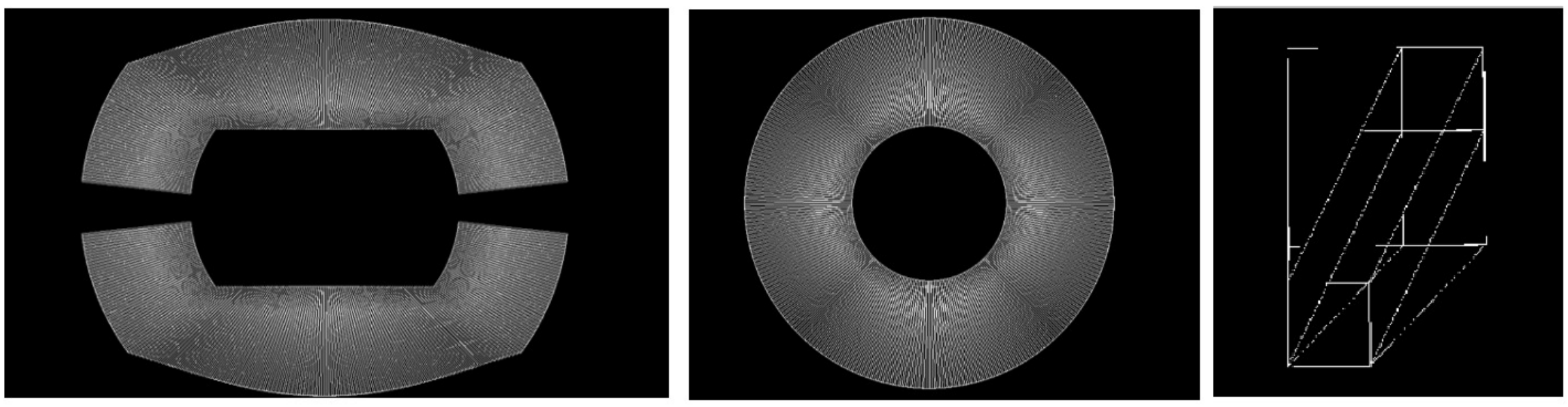

**Figure 5.38:** An alternative $4\pi$ solution (called "wing" geometry).

A different layout implementing the "wing" geometry (see Figure 5.38) is also under study and preliminary results on the $em$ performance will also been shown in the next chapters. In this case, the calorimeter is made of rectangular towers coupled with triangular ones.

In both cases, the total number of fibers is of the order of $10^8$ for a complete $4\pi$ calorimeter.

### 5.5.3.2 MECHANICS (MATERIAL CHOICE AND MACHINING)

Copper, lead and brass (Cu260) have been used as absorber materials by the DREAM/RD52 collaboration. Their main properties are shown in the Table 5.3, that also reports the calculation for the RD52 lead-prototype geometry. The values for iron are also shown, for comparison. From the table it can be seen that, for hadronic showers, a full-coverage solution with lead will give 6% broader and longer showers and a total mass 56% heavier than using brass. A full-containment $3 \times 3 \times 10\ \lambda^3$ prototype will need $\sim 5$ tons of material with lead and $\sim 3.2$ tons with brass.

| Material | $\rho$ (g/cm$^3$) | $X_0$ (cm) | $R_{\text{Molière}}$ (cm) | $\lambda_{\text{Int}}$ (cm) | $\rho \times \lambda^3_{\text{Int}}$ (kg) |
|---|---|---|---|---|---|
| Copper (Cu) | 8.96 | 1.44 | 1.57 | 15.3 | 32.2 |
| Brass (Cu260) | 8.53 | 1.49 | 1.64 | 16.4 | 37.8 |
| Lead (Pb) | 11.35 | 0.56 | 1.60 | 17.6 | 61.8 |
| Iron (Fe) | 7.874 | 1.76 | 1.72 | 16.8 | 37.1 |
| Fibers:Copper (38:62) | 5.98 | 2.26 | 2.28 | 21.9 | 62.8 |
| Fibers:Brass (38:62) | 5.72 | 2.35 | 2.38 | 23.3 | 72.1 |
| Fibers:Lead (38:62) | 7.46 | 0.90 | 2.33 | 24.7 | 112.8 |
| Fibers:Iron (38:62) | 5.31 | 2.75 | 2.48 | 23.7 | 70.8 |

**Table 5.3:** Main properties of lead, copper, brass and iron absorber material and of fiber sampling matrices (RD52 lead-fiber prototype geometry).

A possibly stronger reason in favor of copper/brass is the fact that, since the $e/\text{MIP}$ ratio is 50% higher for copper than for lead, the Čerenkov light (almost exclusively produced by the $em$ component of the shower) has a larger yield for copper, resulting in a better hadronic resolution [27]. However this statement needs to be quantified since it depends on the absolute level of the Čerenkov light yield(s).

On the other hand, lead is easily and accurately extruded, whereas forming copper into the desired shape, either by extrusion, molding, or machining, with the required tolerances in planarity and groove parallelism, is not yet an established industrial process. A variety of techniques (extrusion, rolling, scraping, and milling) for forming the converter layers have been tested. None has been qualified for a large-scale production and identifying an industrial and cost-effective process, including moulding, is a key point.

Alternative copper alloys (e.g. bronze) and/or materials (e.g. iron) may be investigated as well, both for addressing the production process issues and for optimizing the detector performance.

### 5.5.4 SENSORS AND READOUT ELECTRONICS

To separately read out the signals from the $S$ and $C$ fiber forest and avoid oversampling of late developing showers is an issue that may be successfully addressed through the use of Silicon PhotoMultipliers (SiPMs). They would allow the separate reading of each fiber and provide magnetic field insensitivity. In principle, assuming powering and cooling do not pose issues, the transverse segmentation could be made as small as a fiber spacing, or 1.5 mm. Less aggressive options, such as adding/grouping together the analog signals of a handful of fibers prior to digitization, will be studied, both with simulations and testbeam measurements, in order to find the optimal solution in all respects.

SiPMs are low-cost solid state light sensors with single photon sensitivity that underwent an impressive development over the last few years. Tests done in the last two years by the RD52 collaboration indicate that effective solutions for small-scale prototypes are very close already now. Thanks to their higher photon detection efficiency with respect to a standard PM, the higher number of Čerenkov ($pe$) should result in an improved resolu-tion for both $em$ and hadronic showers. On the other hand, the scintillation light spans a very large dynamic range and saturation and non-linearity effects were observed already for low-energy $em$ showers.

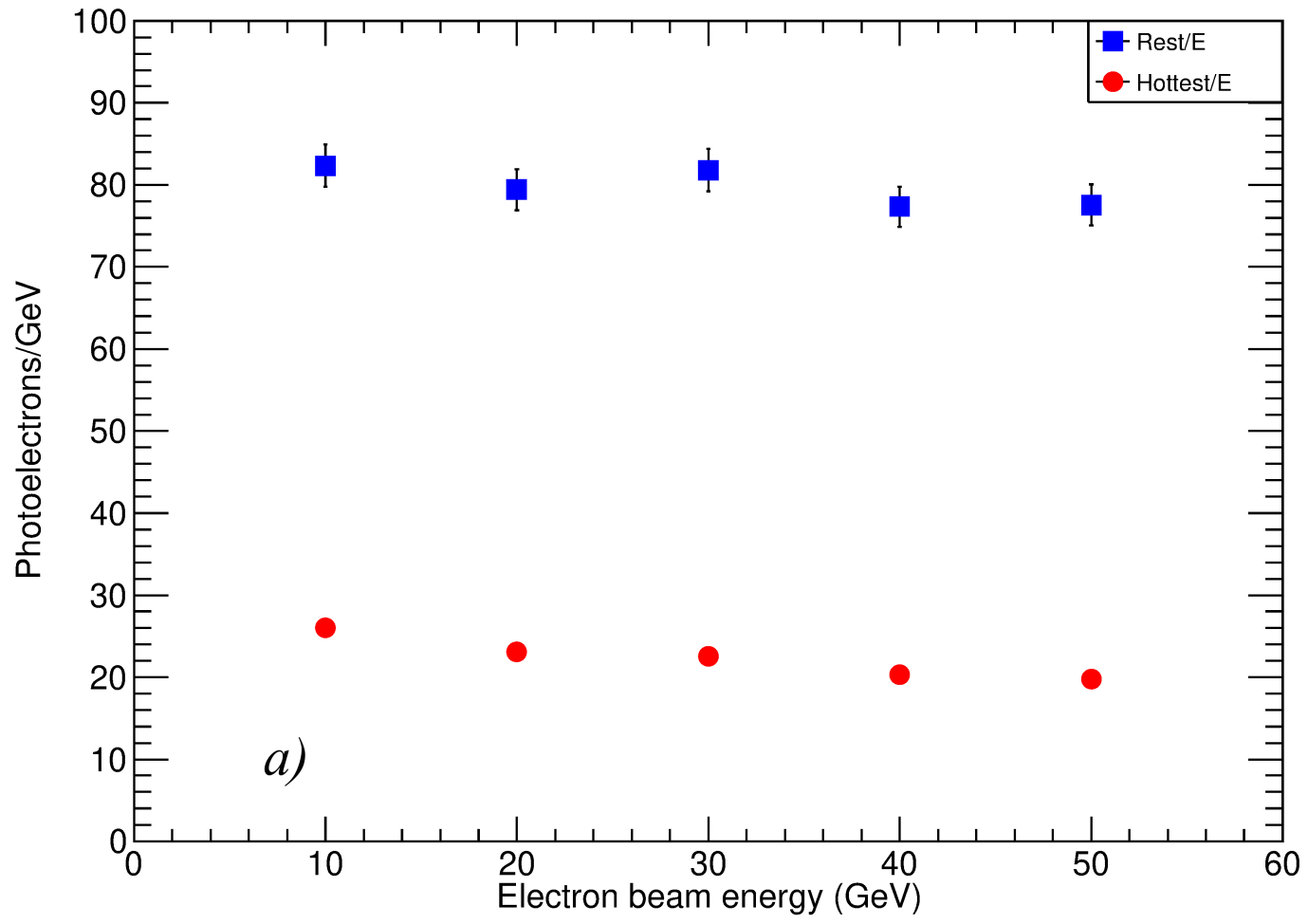

(a)

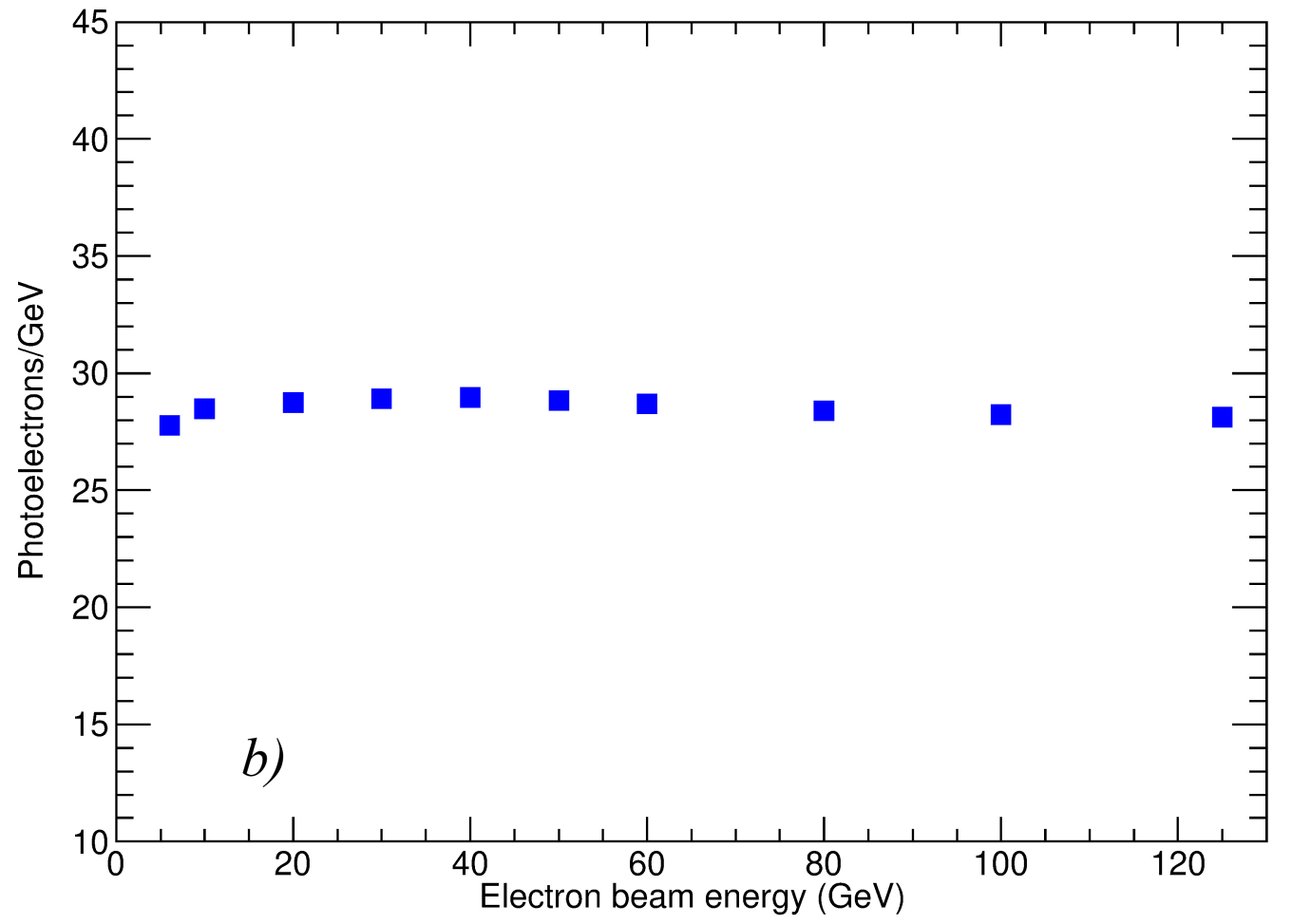

(b)

**Figure 5.39:** Number of photoelectrons per GeV $(\mathrm{pe/GeV})$ for (a) $S$ and (b) $C$ signals, as a function of the electron energy, from 10 to 50 GeV, in a small 64-fiber brass module. In (a), the results are shown separately for the hottest fiber and for the sum of the signals measured by the other 31 scintillating fibers obtained at the (ultra low) PDE of $\sim 2\%$. The main sensor specifications were: $1600$, $25 \times 25$ $\mu\mathrm{m}^2$, cells, and a $25\%$ nominal PDE.

In Figure 5.39, the number of photoelectrons per GeV $(\mathrm{pe/GeV})$ measured, in July 2017, with a very small module ($\sim 1.44$ $\mathrm{cm}^2$ cross section, 32 + 32 fibers), is shown. The most relevant sensor characteristics are $1600$, $25 \times 25$ $\mu\mathrm{m}^2$, cells, and a $25\%$ nominal PDE. Due to the large $S$ light yield, the data for the $S$ signal were obtained at an (ultra low) PDE of $\sim 2\%$, and corrected for non-linearity. Rescaled to a $25\%$ efficiency, the yield of $S$ photoelectrons results in $\sim 108 \times 12.5 = 1350$ $\mathrm{pe/GeV}$. By removing from the sum the hottest fiber, more heavily affected by non-linearity effects, the estimate grows to $\sim 1530\ \mathrm{pe/GeV}$.

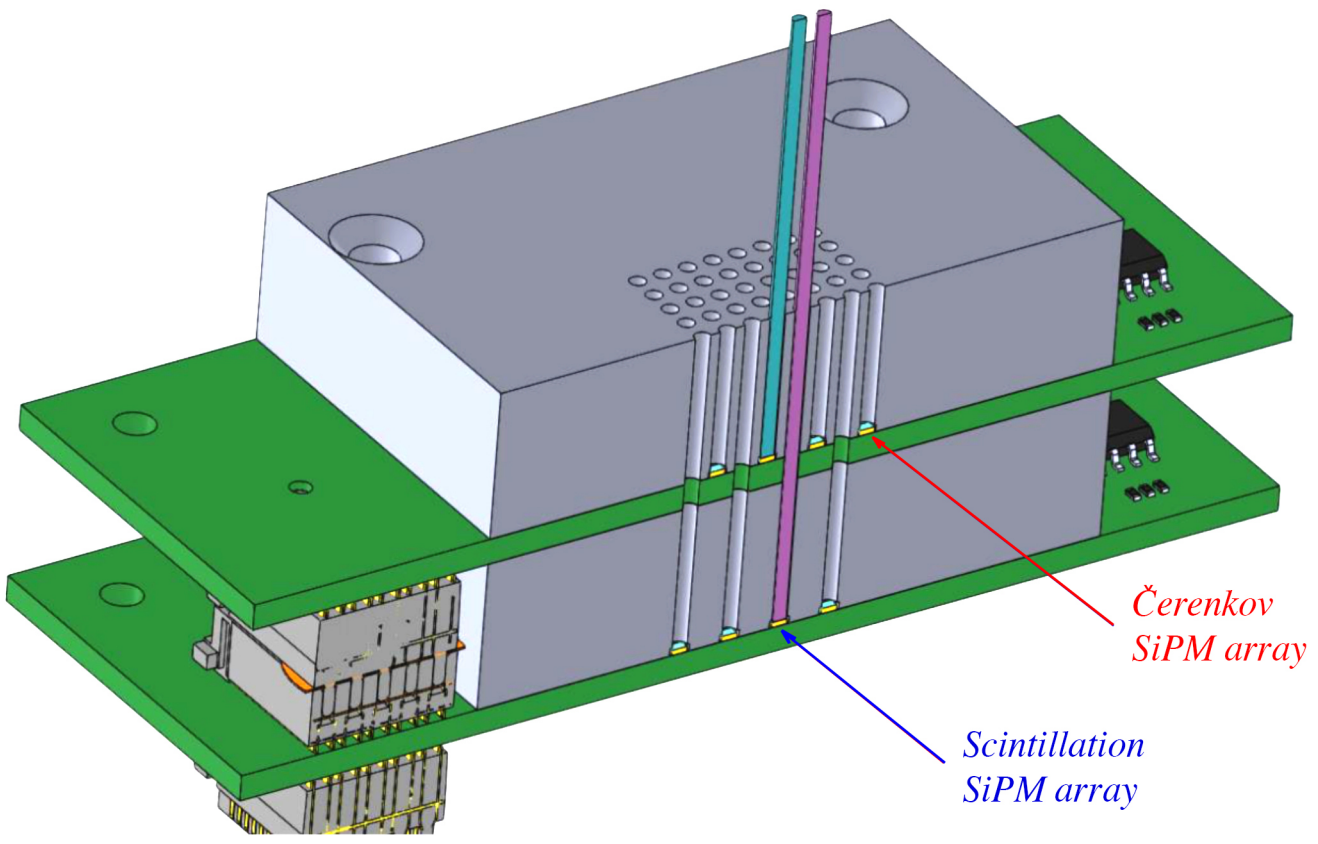

**Figure 5.40:** Staggered readout scheme: the scintillation and Čerenkov fibers are readout at different planes to minimize light leakage into neighboring channels [39].

The $C$ signals show a linear response at $\sim\ 30\ \mathrm{pe/GeV}$. It should be mentioned that the shower containment was estimated from GEANT4 simulations to be $\sim\ 45\%$. In addition, the problem of large light leaks from the $S$ fibers into the neighboring $C$ SiPM channels, observed in the 2016 tests, seems to be largely but not completely solved by a staggered readout of the $S$ and $C$ fibers (Figure 5.40). The contamination of the $C$ signal was estimated to be $\sim 16\% \pm 6\%$.

### 5.5.4.1 SENSOR CHOICE

As far as the scintillation light detection is concerned, saturation and non-linearity should be solvable using higher density devices (e.g. with 10000, $10 \times 10\ \mu\mathrm{m}^2$, cells) in combination with some light filtering. The definition of the optimal dynamic range and the qualification of existing SiPMs in that regard, will be likely addressed in a short-term R&D phase.

For the Čerenkov light, improvements of the photon collection are possible with the use of an aluminized mirror on the upstream end of the fibers. The acceptance cone may also be enlarged with the use of cladding with a different refractive index. Over a longer term, it could be possible that the R&D on new devices, such as Silicon Carbide (SiC) sensors, expected to provide exclusive UV sensitivity (i.e. visible-light blindness), will allow us to obtain significantly larger $pe$ yields.

### 5.5.4.2 FRONT-END ELECTRONICS AND READOUT

Concerning the front-end, the development shall certainly evaluate the use of Application Specific Integrated Circuits (ASIC) to handle and reduce the information to be transferred to the DAQ system. A major question is finding the optimal way for summing signals from a plurality of sensors into a single output channel. A dedicated feature-extracting processor, capable of extracting timing information such as time-over-threshold, peaking, leading and/or falling times, may allow to disentangle overlapping $em$ and hadronic showers without the need for longitudinal segmentation. With the present fibers, a resolution of the order of 100 ps corresponds to a spatial resolution of about $\sim$ 6 cm along the fiber axis (relativistic particles take 200 ps to cover 6 cm while light needs 300 ps).

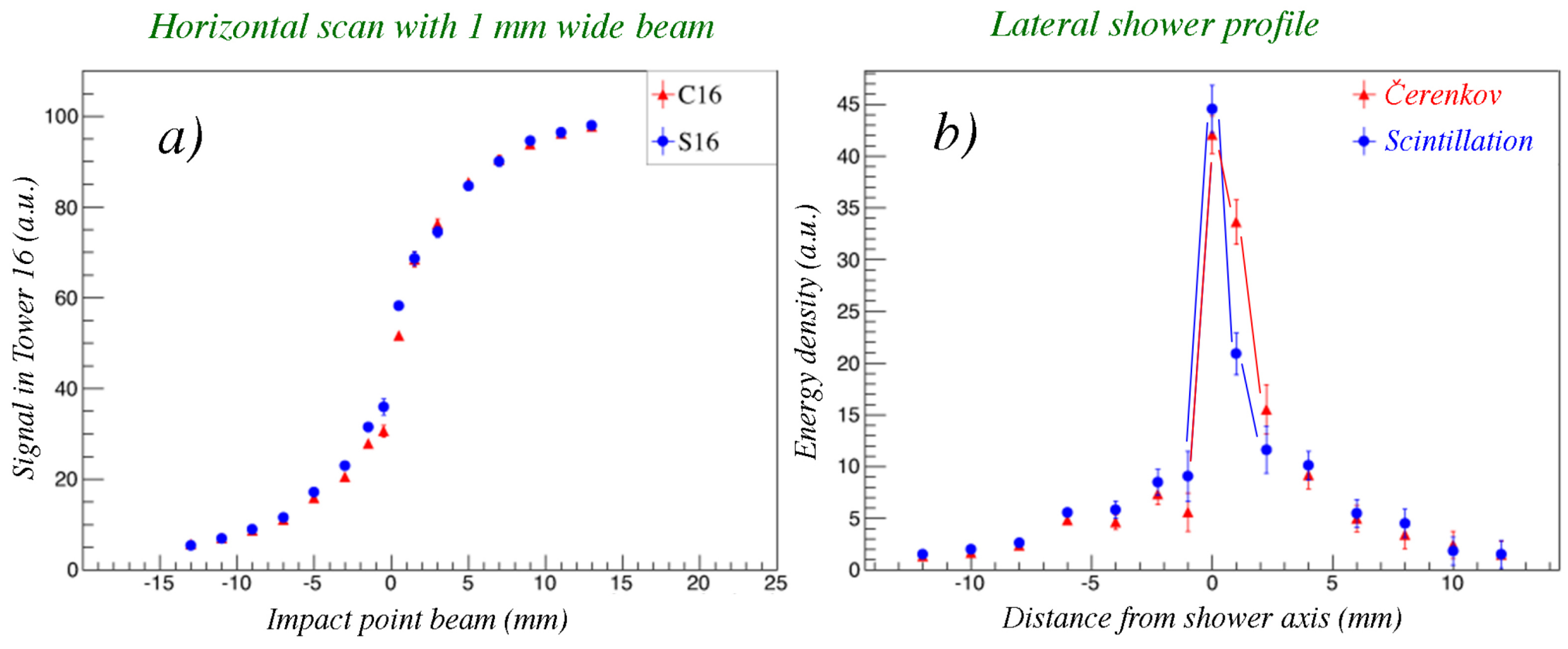

(a)

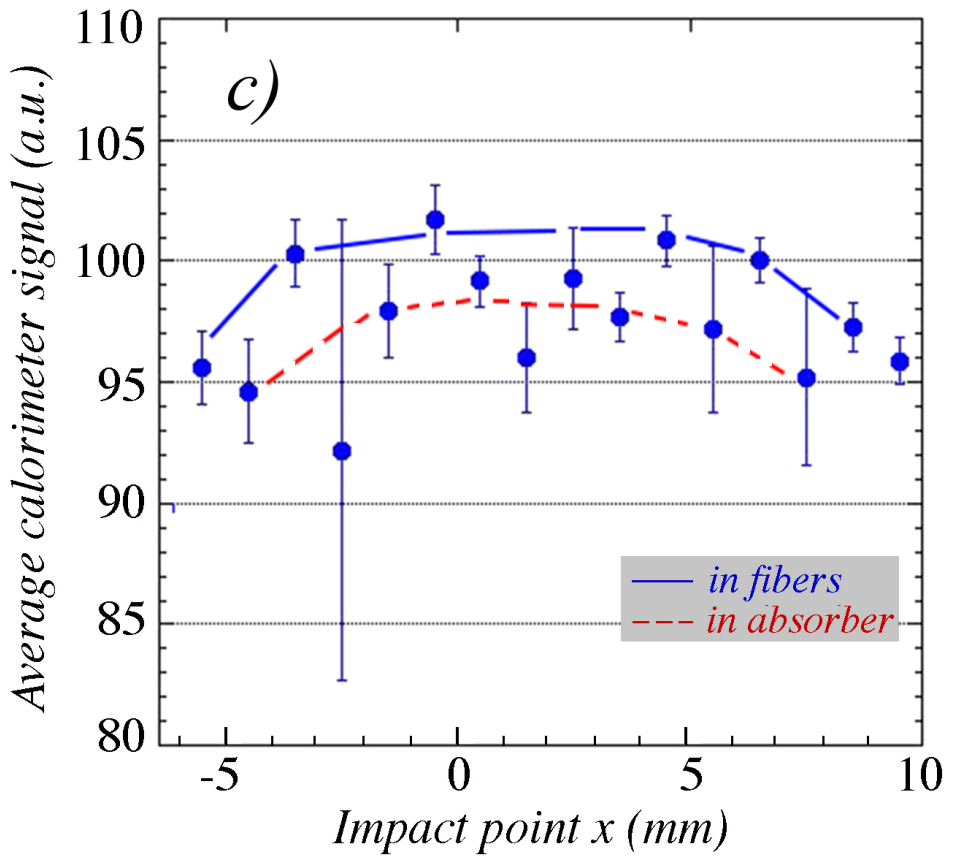

(b)

**Figure 5.41:** (a) The signal from a 1 mm wide beam of 100 GeV electrons, in the RD52 lead-fiber prototype, as a function of the impact point; (b) the lateral shower profiles derived from this measurement; (c) the dependence of the scintillation signal on impact point for a beam impinging parallel to the fibers. All plots from [34].

### 5.5.5 PERFORMANCE STUDIES WITH FIBER-SAMPLING PROTOTYPES

Different prototypes were built and studied by the DREAM/RD52 collaboration, with copper or lead as absorber and photomultipliers as light sensors [32–37]. With electrons and pions, in the range of ~10-150 GeV, the response linearity was found at the level of $1\%$ for both the $em$ and the hadronic energy reconstruction (having applied the dual-readout formula, equation 5.8, for hadronic showers). The $em$ resolution was estimated to be close to $\sim 10\%/\sqrt{E}$, while the hadronic resolution was found to be at the level of $60\text{-}70\%/\sqrt{E}$, to be corrected for the fluctuations introduced by lateral leakage and light attenuation in the fibers. None of the prototypes built thus far were large enough to substantially contain hadronic showers and an R&D program to assess the hadronic performance of a real detector, is under way. Preliminary simulations of standalone modules indicate a possible ultimate resolution of $\sim 30-40\%/\sqrt{E}$. More details can be found in the next paragraphs.

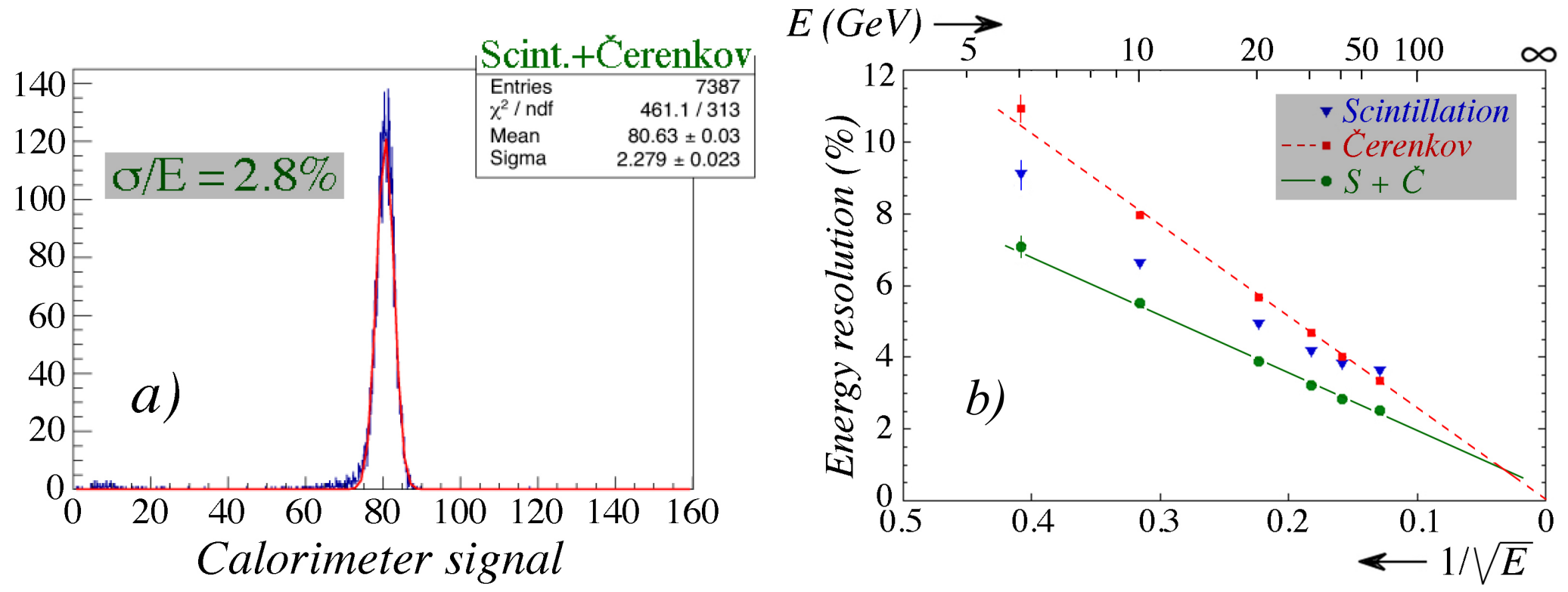

(a) (b)

**Figure 5.42:** In the RD52 copper-fiber module: (a) signal distribution of the sum of all fibers for 40 GeV electrons; (b) the $em$ energy resolution as a function of the beam energy. Shown are the results for the two types of fibers, and for the average combined signal. All plots from [34].

### 5.5.5.1 ELECTROMAGNETIC PERFORMANCE

Figure 5.41(a) and 5.41(b) show the radial shower profile and the sensitivity to the impact point: the core of the signal spans just a few mm. Figure 5.41(c) shows the dependence of the $S$ signal on the impact point for particles entering parallel to the fibers. This introduces a constant term in the resolution that can be avoided with a small tilt of the fiber axis. In the $C$ fibers, the problem does not show up since the early (collimated) part of the shower produces photons outside the fiber numerical aperture.

For the reconstruction of the energy of $em$ showers, $S$ and $C$ signals provide independent uncorrelated measurements, with different sensitivity of the response. They are affected by different problems: $S$ signals have photoelectron statistics one or two orders of magnitude higher than $C$ signals, and their fluctuations are largely dominated by the sampling fluctuation of the energy deposits. $C$ signal fluctuations are generally dominated by the limited photoelectron statistics, especially at low energies. Nevertheless, at high energies, the constant term for $C$ signals is negligible, giving a better resolution. Averaging the two measurements improves the resolution up to a factor of $\sqrt{2}$. For the copper matrix, in Figure 5.42(a) the sum of $S$ and $C$ signals for 40 GeV electrons is plotted, while Figure 5.42(b) shows the $em$ resolution, for $S$, $C$ and the (average) combined signal.

### 5.5.5.2 HADRONIC PERFORMANCE

The response of a lead-fiber matrix was studied with pion and proton beams [32]. The energy was reconstructed with the dual-readout relation (Eq. 5.8) and shows a restored Gaussian response function (Figure 5.43) and linearity of the mean response.

The comparison of $p$ and $\pi^+$ signals confirms that the dual-readout method largely compensates for the differences in shower composition, i.e., differences in the electromagnetic fraction, $f_{em}$, and between baryon-initiated and pion-initiated hadronic showers.

Due to the limited lateral size of the matrix (the effective diameter was $\sim 1\lambda_{\text{Int}}$), the containment for hadronic showers was $\sim 90\%$ so that leakage fluctuations dominated the energy resolution. Selecting contained showers improved the resolution by a factor of $\sim 2$. Although that selection was introducing a bias in favor of high $f_{em}$ showers, a significant improvement is expected for a realistic-size module.

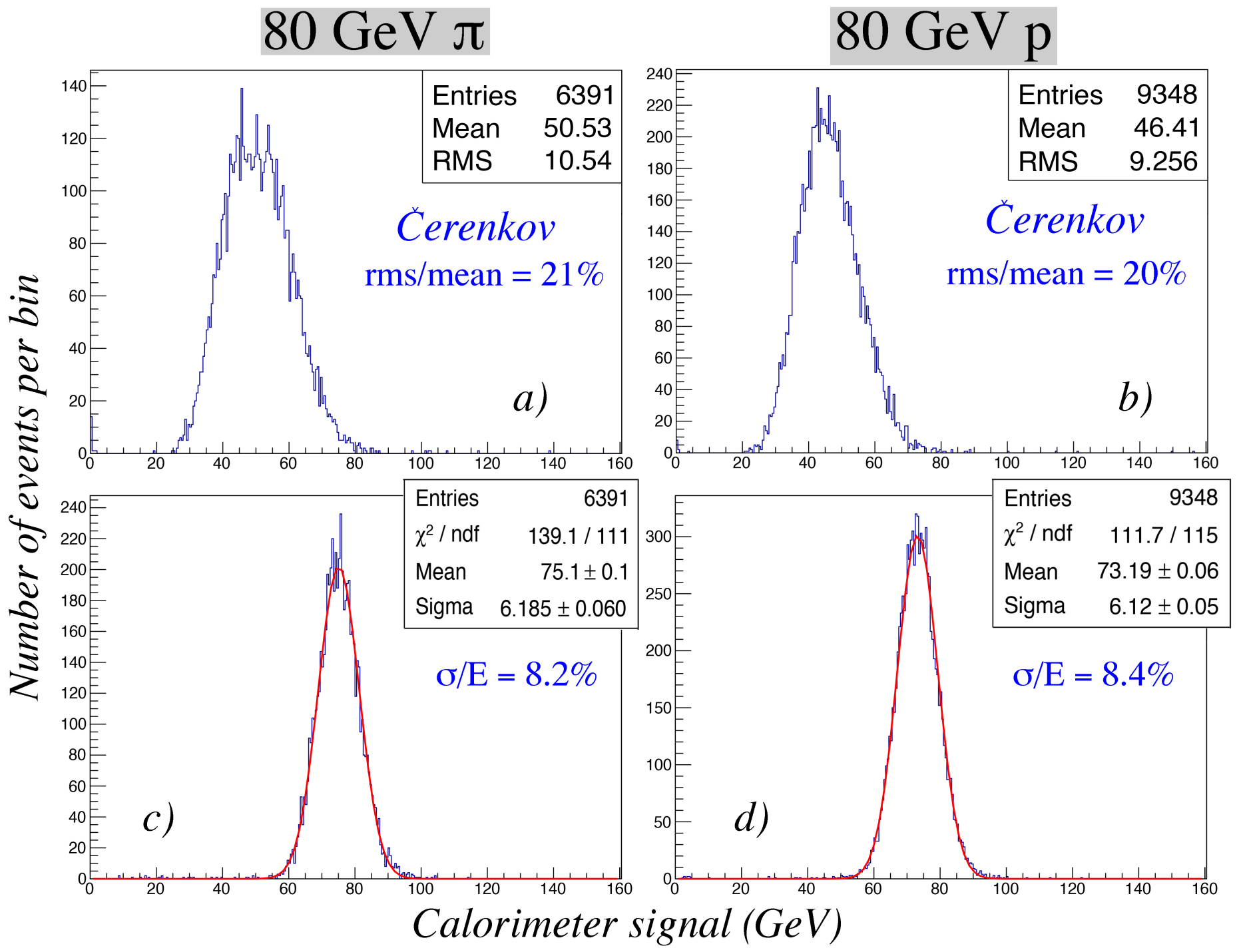

**Figure 5.43:** Signal distributions for 80 GeV pions and protons measured with the RD52 lead-fiber calorimeter. Shown are the distributions for the Čerenkov signals from 80 GeV (a) $\pi^+$ and (b) protons, as well as the dual-readout total signals for 80 GeV(c) $\pi^+$ and (d) protons. The dual-readout signals were obtained by applying Equation 5.8 with $\chi = 0.45$. Data from [32].

The resolution was also affected by the finite light attenuation length of the fibers, causing early starting showers to be observed at lower signal values. The hadronic resolution, yet to be corrected for both effects, was reconstructed to be $\sim 70\%/\sqrt{E}$.

#### 5.5.5.3 $E/\pi$ SEPARATION

Four discriminating variables were identified for implementing $e/\pi$ separation: the fraction of energy in the central tower, the $C/S$ signal ratio, the signal starting time and the total charge/amplitude ratio, shown in Figure 5.44. The plots are relative to test beam data taken with the RD52 lead-fiber prototype [33].

A multivariate neural network analysis showed that the best $e/\pi$ separation achievable for 60 GeV beams was $99.8\%$ electron identification efficiency with $0.2\%$ pion misidentification. Further improvements may be expected by including the full time structure information of the pulses, especially if the upstream ends of the fibers are made reflective.

### 5.5.6 MONTE CARLO SIMULATIONS

GEANT4 simulations[2] are under development and analysis for understanding the performance of both test beam modules and a $4\pi$ calorimeter integrated in a detector, with magnetic field, tracking and preshower elements.

[2]version 10.02.p01-10.03.p01, with FTFP_BERT_HP physics list

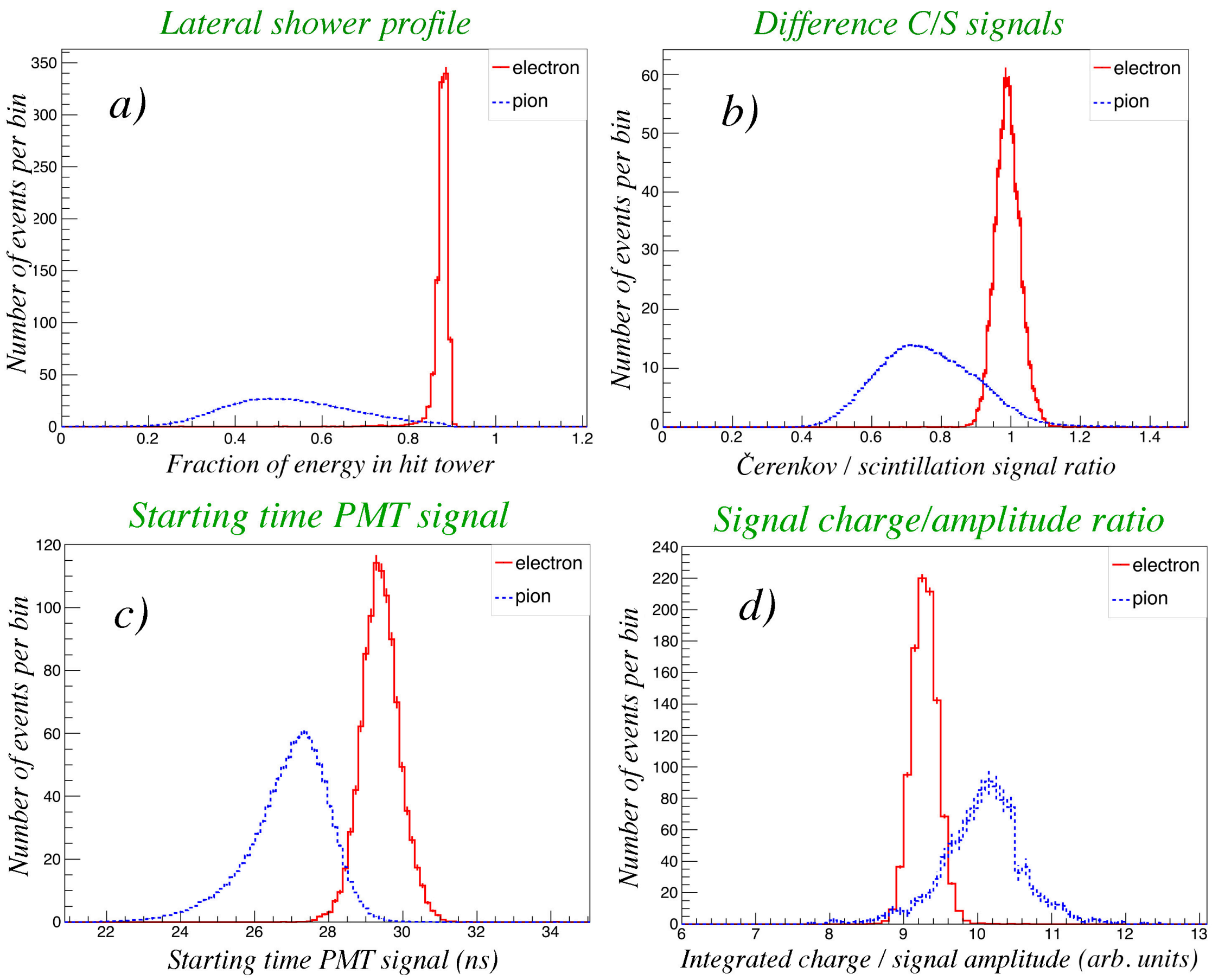

**Figure 5.44:** Distribution of four discriminating variables for 60 or 80 GeV electrons and pions, as measured with the RD52 lead-fiber prototype [33]: (a) energy fraction deposited in the hit tower; (b) $C/S$ signal ratio in the hit tower; (c) starting time of the PM signal; (d) ratio of the integrated charge and the amplitude of the signals.

| *fibers used* | Fitted Gaussian $em$ energy resolution |
|---|---|
| S-fibers only | $\sigma/E = 10.1\%/\sqrt{E} \oplus 1.1\%$ |
| C-fibers only | $\sigma/E = 17.3\%/\sqrt{E} \oplus 0.1\%$ |
| S-fibers and C-fibers | $\sigma/E = 10.1\%/\sqrt{E} \oplus 0.4\%$ |

**Table 5.4:** Fit to the $em$ resolution (MC simulations)

#### 5.5.6.1 ELECTROMAGNETIC PERFORMANCE

A Cu matrix of dimensions $\sim 31 \times 31 \times 100 \text{ cm}^3$ , with 1 mm fibers at 1.4 mm distance, compatible with the RD52 prototypes, has been simulated for the evaluation of the electromagnetic performance. PMMA clear fibers and Polystyrene scintillating fibers, with a $3\%$ thick cladding ($C_2F_2$ Fluorinated Polymer for clear and PMMA for scintillating fibers), were the sensitive elements.

A small ($\lesssim 1°$) tilt angle was introduced to avoid large non-Gaussian tails in the scintillation signal due to channeling.

The energy containment for 20 GeV electrons was estimated to be $\geq 99\%$, with sampling fractions of 5.3% and 6.0% for scintillating and clear fibers, respectively.

Given the integral sampling fraction of $11.3\%$ and the 1 mm diameter fibers, the contribution to the energy resolution due to sampling fluctuations can be estimated to be $\sim 9\%/\sqrt{E}$, ultimate limit on the $em$ resolution for this detector.

The scintillation light yield is so large ($\sim 5500$ pe/GeV) that the fluctuations of the $S$ signals are dominated by the energy sampling process (Figure 5.45(a)). This is not true for the Čerenkov signals (Figure 5.45(b)), whose sensitivity is estimated to be $\sim 100$ pe/GeV.

In the simulations, the process of generation and propagation of the scintillation light was switched off and the energy deposited in the fibers was taken as signal since this does not introduce any bias to the detector performance. This statement does not apply to the Čerenkov photons for which a parameterization that convolutes the effect of light attenuation, angular acceptance and PDE, was introduced.

In Figure 5.46 the resolutions are shown for both $C$ and $S$ signals, separately, and for the unweighted average value of the two. The variable on the horizontal axis and in the formulae for the fitted resolutions is the beam energy. The results of the fit to the data points are shown in Table 5.4. A slightly better result may be obtained with a weighted average.

#### 5.5.6.2 HADRONIC PERFORMANCE

A simulation of larger ($\sim 72 \times 72 \times 250$ $cm^3$) matrices was implemented in order to get a hadronic shower containment of $\sim 99\%$. Calibration was done with 40 GeV electron beams.

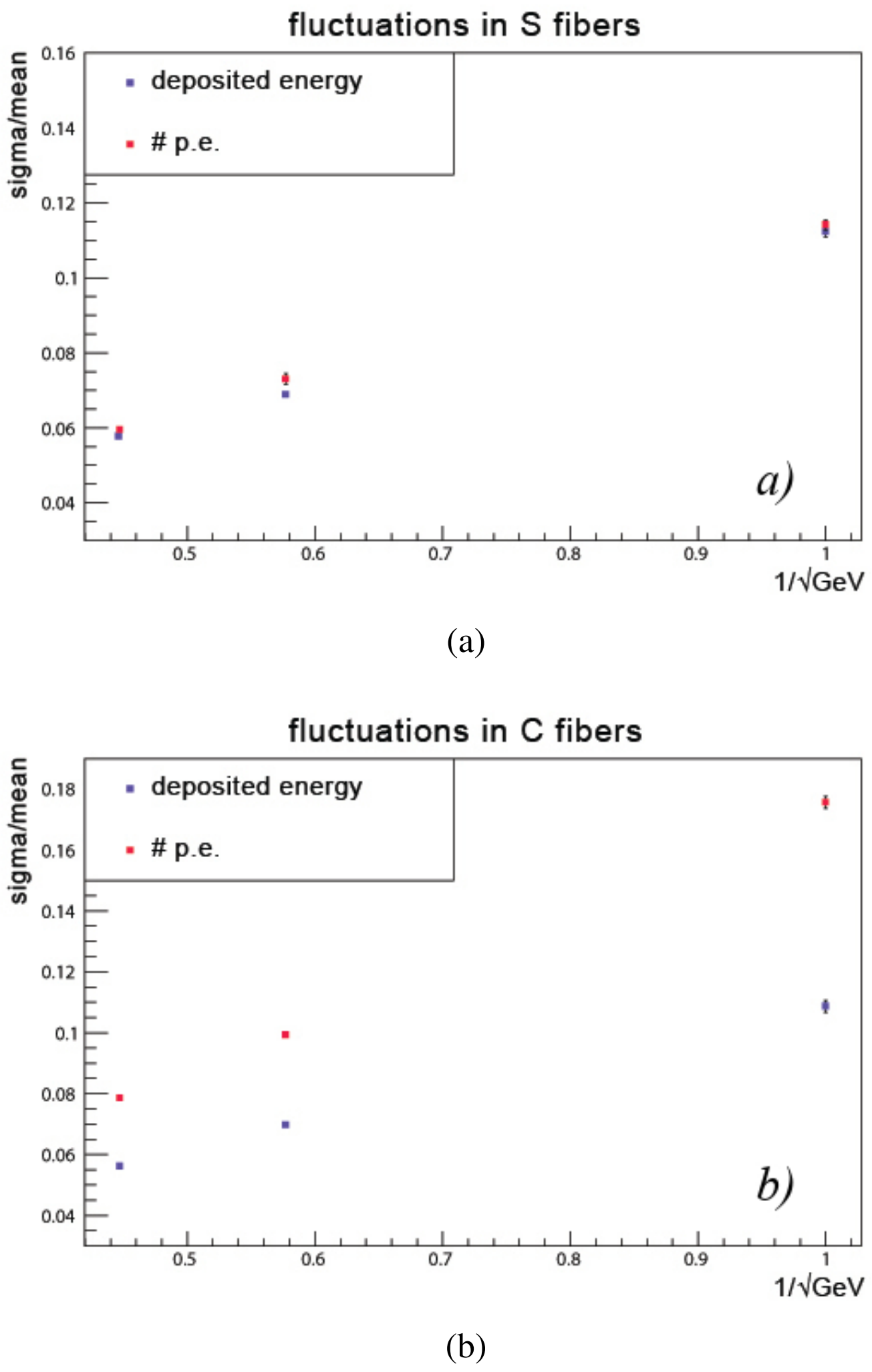

**Figure 5.45:** Relative fluctuation of the total signal detected in the (a) scintillating and (b) Čerenkov fibers, for both the energy deposit and the number of photoelectrons (MC simulations).

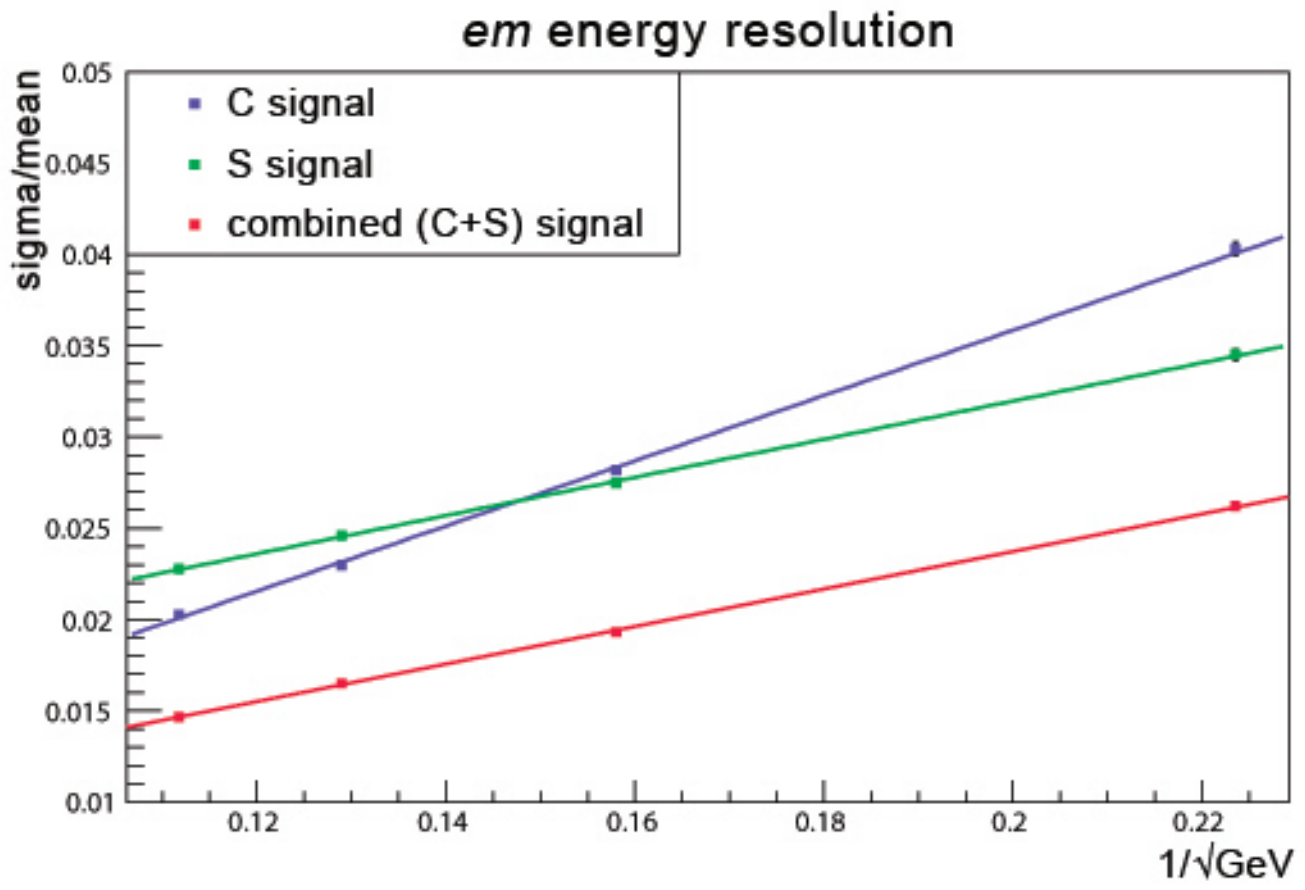

**Figure 5.46:** Relative resolution for $em$ showers for the $C$ and $S$ signals, independently, and for the average of the two (MC simulations).

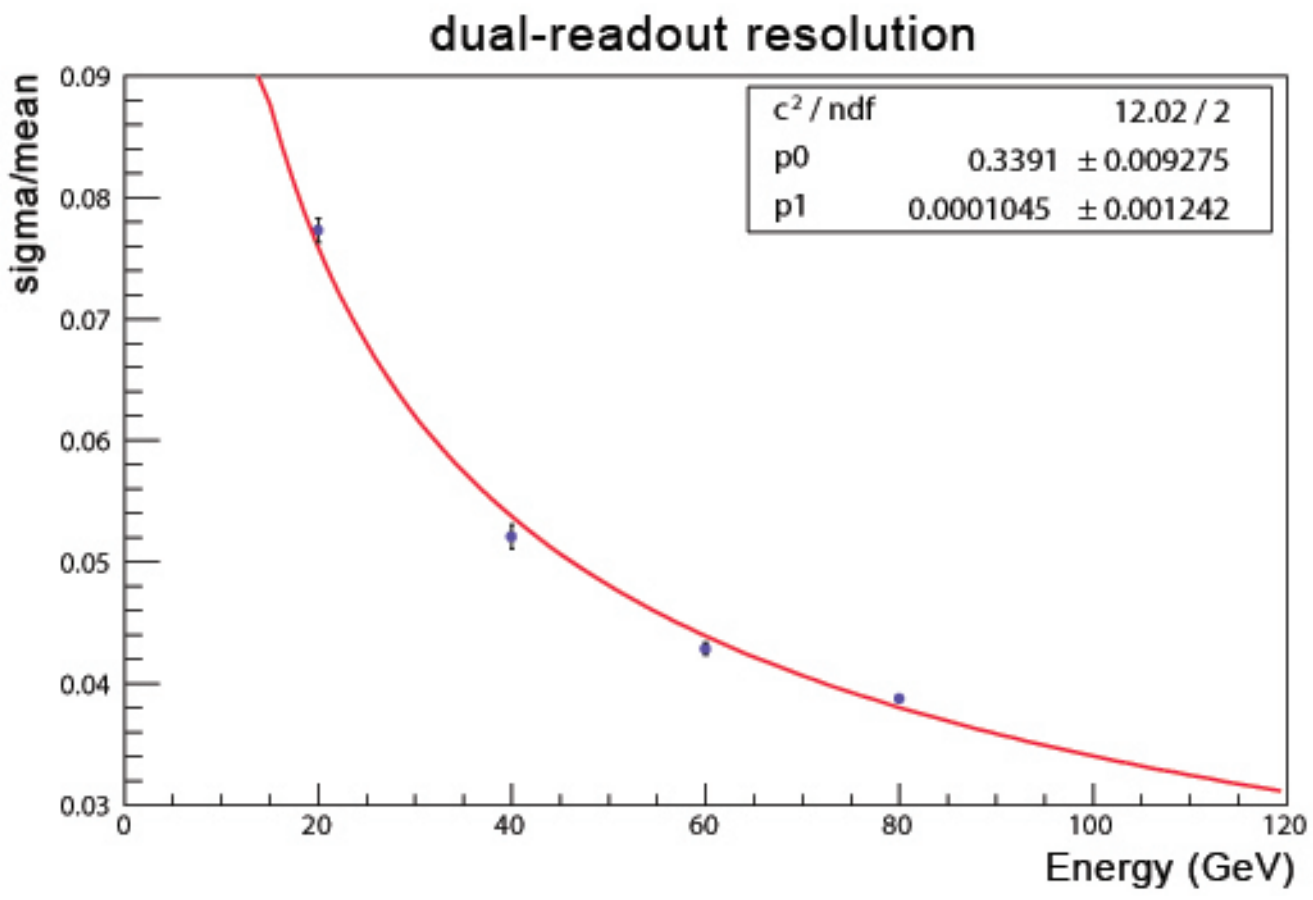

(a)

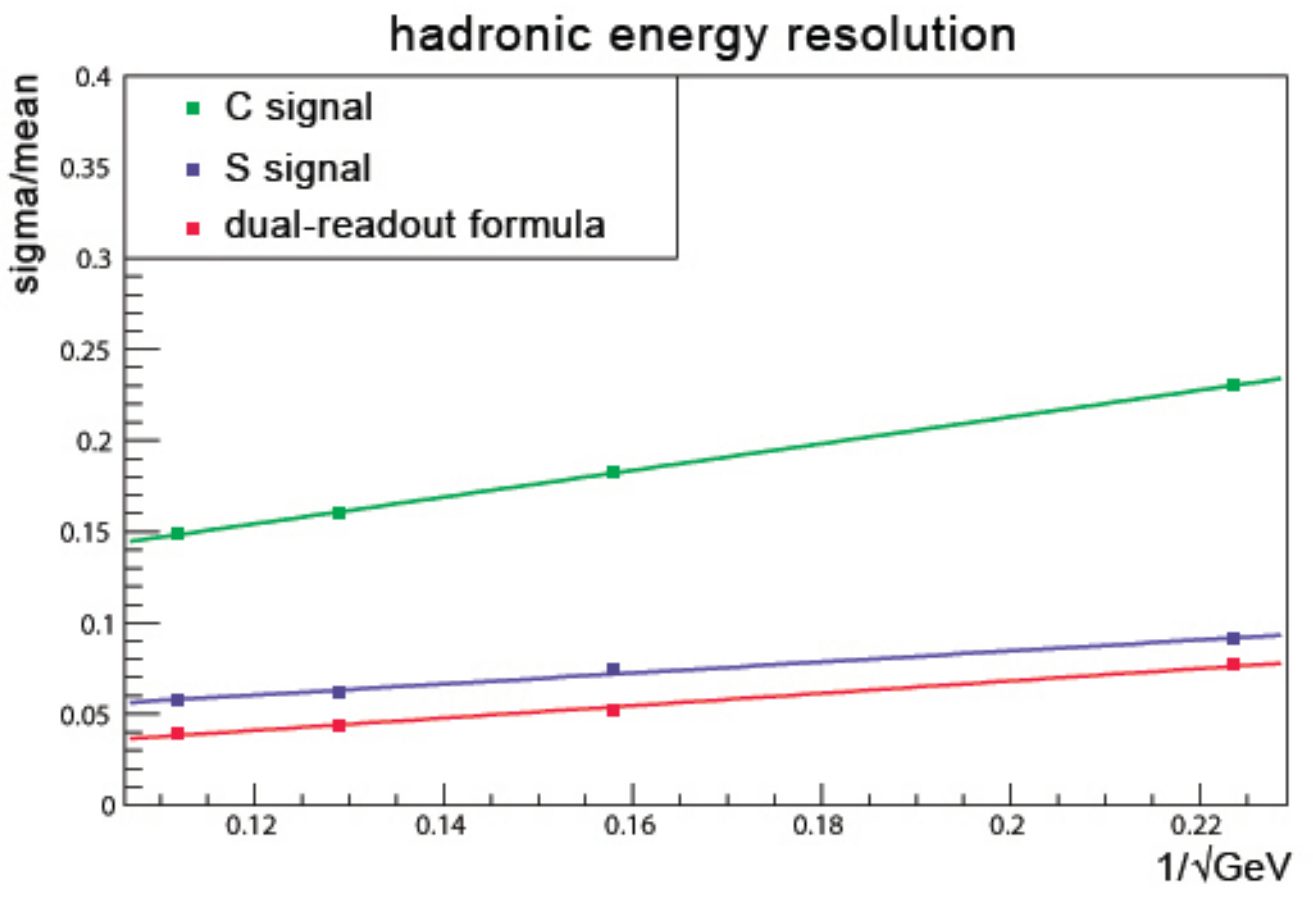

(b)

**Figure 5.47:** Monte Carlo simulations showing: (a) the relative hadronic resolution as reconstructed with the dual-readout formula; (b) the relative hadronic resolution independently for the $C$ and $S$ signals and for the dual-readout combination of the two.

| *fibers used* | Fitted Gaussian hadronic energy resolution |
|---|---|
| S-fibers only | $\sigma/E = 30\%/\sqrt{E} \oplus 2.4\%$ |
| C-fibers only | $\sigma/E = 73\%/\sqrt{E} \oplus 6.6\%$ |
| Dual-readout S-fibers and C-fibers | $\sigma/E = 34\%/\sqrt{E} \oplus \text{(negligible)}\%$ |

**Table 5.5:** Fit to the hadronic resolution (MC simulations)

In Figure 5.47 GEANT4 predictions for the hadronic energy resolution, with copper absorber, are shown. Table 5.5 lists the results of the fit to the curves.

The large constant terms, for both $S$ and $C$ signals, are generated by the $f_{em}$ correlated fluctuations. Simulations with lead absorber give equivalent but even slightly better results. The energy E in the plot (and in the expressions for the fitted resolutions) is the beam energy, corresponding in average to the energy reconstructed with the Equation 5.8 when the containment is properly accounted for (i.e., the reconstructed energy corresponds, in average, to the beam energy times the average containment). The fact that the experimental resolution was, so far, about a factor of two worse than simulations, is in our understanding, largely due to the small lateral size of the prototypes. In order to fully validate the MC predictions, an R&D program is being pursued.

The correlation of the invisible energy with all the other components of hadronic showers was also analyzed. Preliminary results seem to indicate that the most appropriate variable to account for the fluctuations of the invisible energy component is, by far, the $f_{em}$, with correlation coefficients of $90\%$, $92\%$, $94\%$, for copper, iron and lead respectively. The kinetic energy of the neutrons is predicted to be, at best, correlated at the $76\%$ level. If confirmed, this would prove that compensation through neutron signal pickup or amplification will anyway give worse results than the dual-readout method [31].

In terms of particle ID capabilities, in Figure 5.48 the $C/S$ ratio is shown for 80 GeV $e^-$ and protons in copper (a) and lead (b). For an electron efficiency of $\sim 98\%$, the rejection factor for protons is $\sim 50$ in copper and $\sim 600$ in lead. Of course, this is an ideal detector and in reality it is likely that the numbers will be worse. On the other hand, there are more variables that can be easily used in order to enhance the particle ID performance (namely the lateral shower profile, the starting time of the signal, the charge-to-amplitude ratio).

### 5.5.6.3 PROJECTIVE GEOMETRY

Each tower, in the wedge geometry implementation, was exposed to 20 GeV electron beams, with an incident angle of $(1^\circ, 1.5^\circ)$, and the calibration constants calculated as the average deposit energy (in each tower) divided by the average $C$ or $S$ signal (of each tower). The response to an electron beam of the same energy is plotted in Figure 5.49. In the barrel region the response of all towers is within $0.2\%$, while in the forward the systematics are within $2\%$. All results were obtained with the quantum efficiency for the Čerenkov channel of each tower tuned to a light yield of $\sim$ 30 $\mathrm{pe/GeV}$, as estimated in the RD52 beam tests.

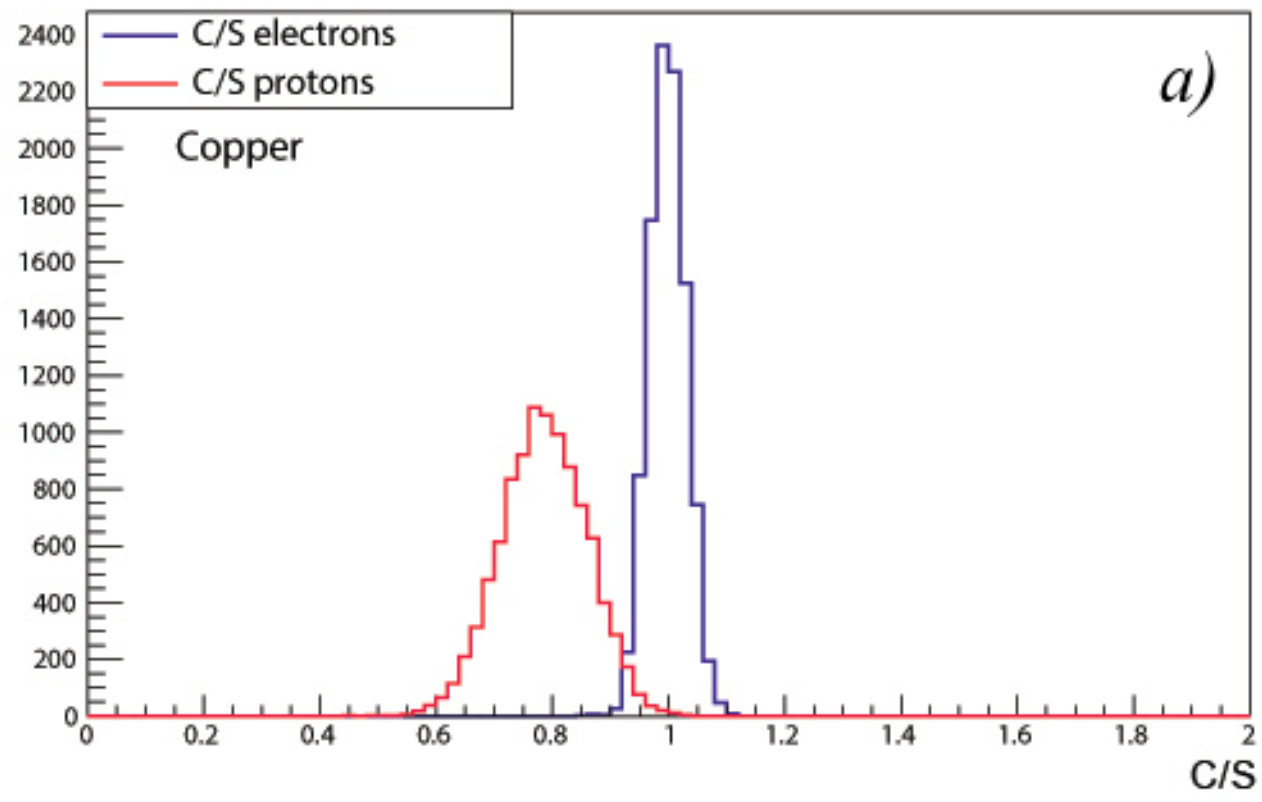

(a)

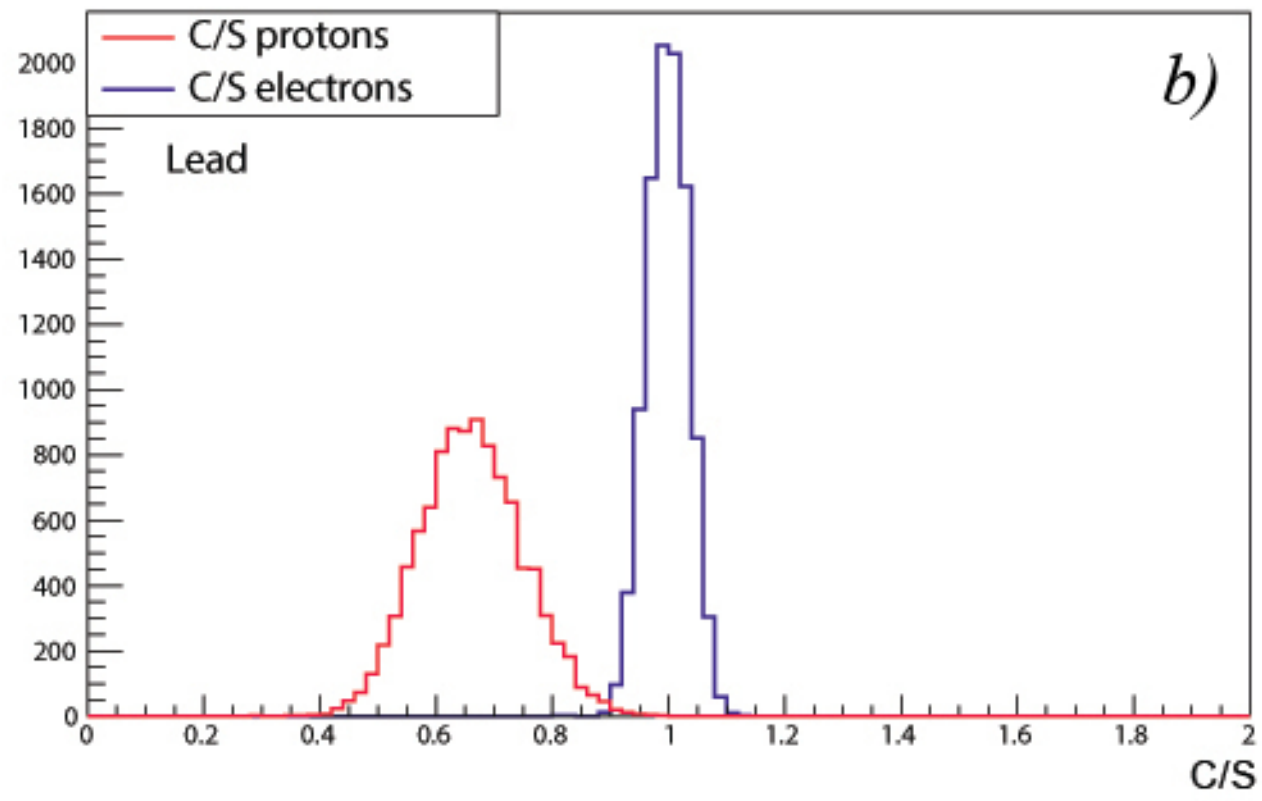

(b)

**Figure 5.48:** $C/S$ ratio (MC simulations) for 80 GeV $e^-$ and protons in (a) copper and (b) lead.

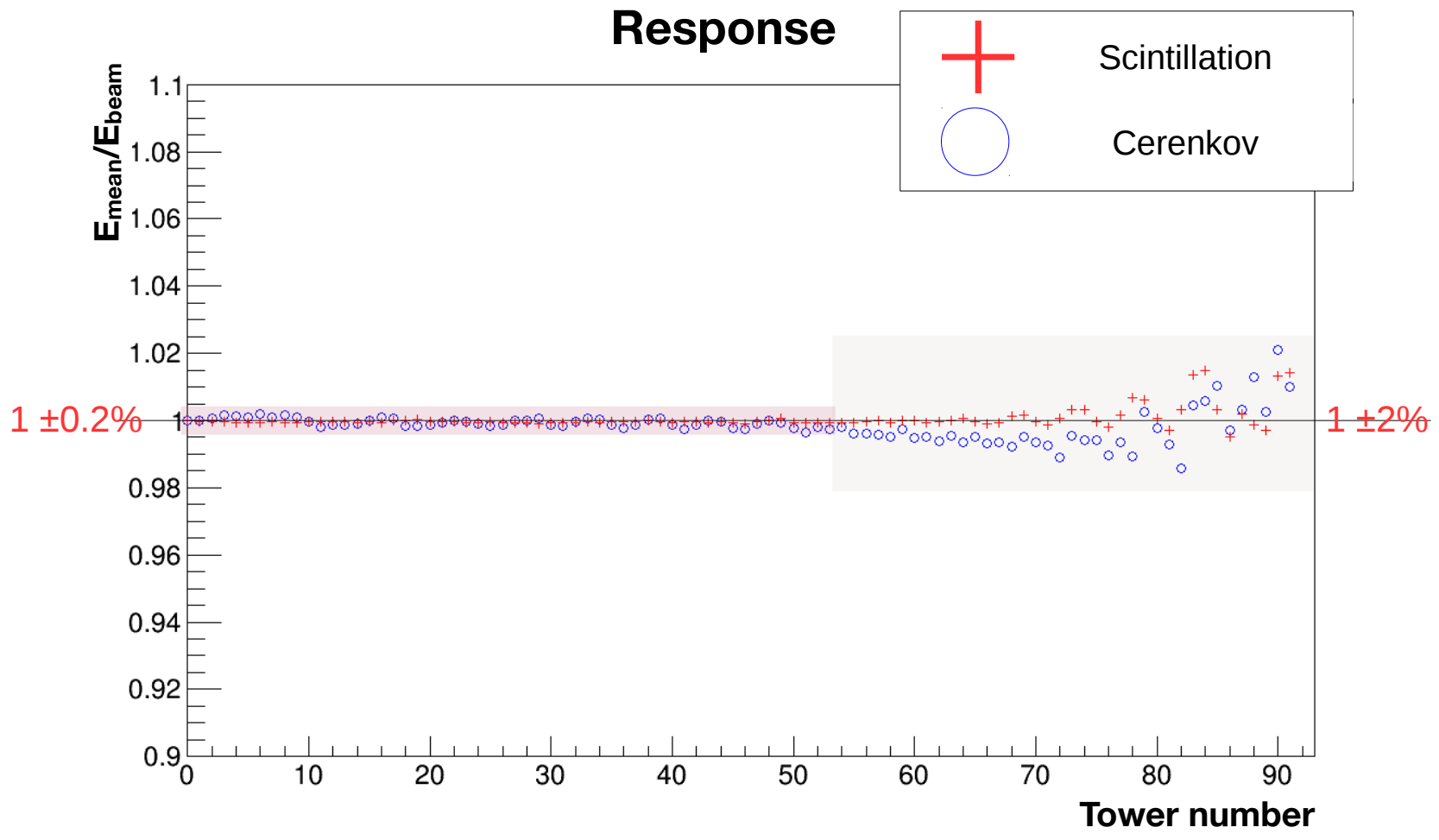

**Figure 5.49:** Ratio of reconstructed energy to the beam energy for 20 GeV $e^-$, as a function of the tower number, in the wedge geometry (MC simulations).

The performance of a few towers was studied with electron beams in the range of 10–100 GeV. Figure 5.50 shows the linearity and $em$ energy resolutions for towers #0 and #45. In both cases, the combined $S$ and $C$ signal shows a resolution of $\sim 14\%/\sqrt{E}$ with a constant term of $\sim 0.1\%$ while the average response is constant within $0.4\%$.

The hadronic resolution was studied with pions in the same energy range. A $\chi$ value of 0.29, the value measured for the DREAM calorimeter [40], was used to reconstruct the shower energy with Eq. 5.8. In the linearity plots for both tower #0 and #45 in Figure 5.51, the $C$ and $S$ responses to single pions increase non-linearly as the pion beam energy increases. On the other hand, the value reconstructed with the dual-readout formula shows a constant response to single pions $\sim 8\%$ lower than that to electrons (the reason being the shower containment). This effect in the GEANT4 simulations is described in reference [41]. In addition, the energy resolution after the correction (shown in Figure 5.51 for towers #0 and #45) is $\sim 26\%/\sqrt{E}$, with a constant term of less than $1\%$. These results support the statement that the hadronic energy resolution and the response to single hadrons should be constant (and appropriate) over the full barrel region. We may reasonably expect to obtain good performance over the entire $4\pi$ detector.

For the wing geometry, the results, at present, are limited to the $em$ performance of few towers and the results (linearity and $em$ resolution) substantially reproduce the wedge geometry ones.

#### 5.5.6.4 SHORT TERM PLANNING AND OPEN ISSUES

The performance for single hadrons, jets and $\tau$ leptons has to be understood and the work has just started. For validation, the comparison with a prototype with a non-marginal hadronic shower containment, like the RD52 lead matrix, will be pursued.

For the $em$ simulations, a program for the comparison with the 2017 RD52 data is ongoing. Some initial understanding of the absolute photoelectron scale for the Čerenkov light should be available in a very short time.

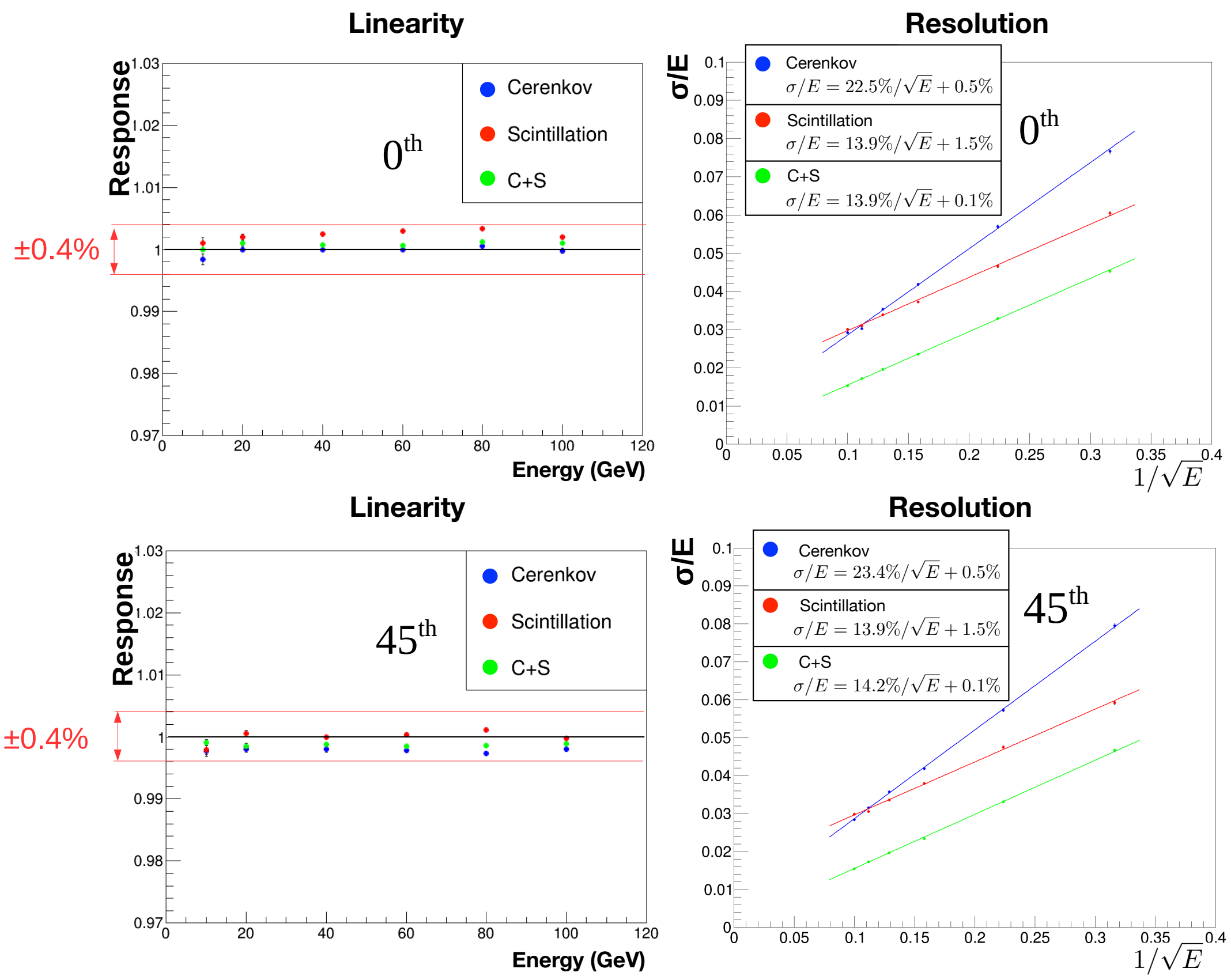

**Figure 5.50:** Linearity and $em$ energy resolution for towers #0 (top) and #45 (bottom), in the wedge geometry (MC simulations).

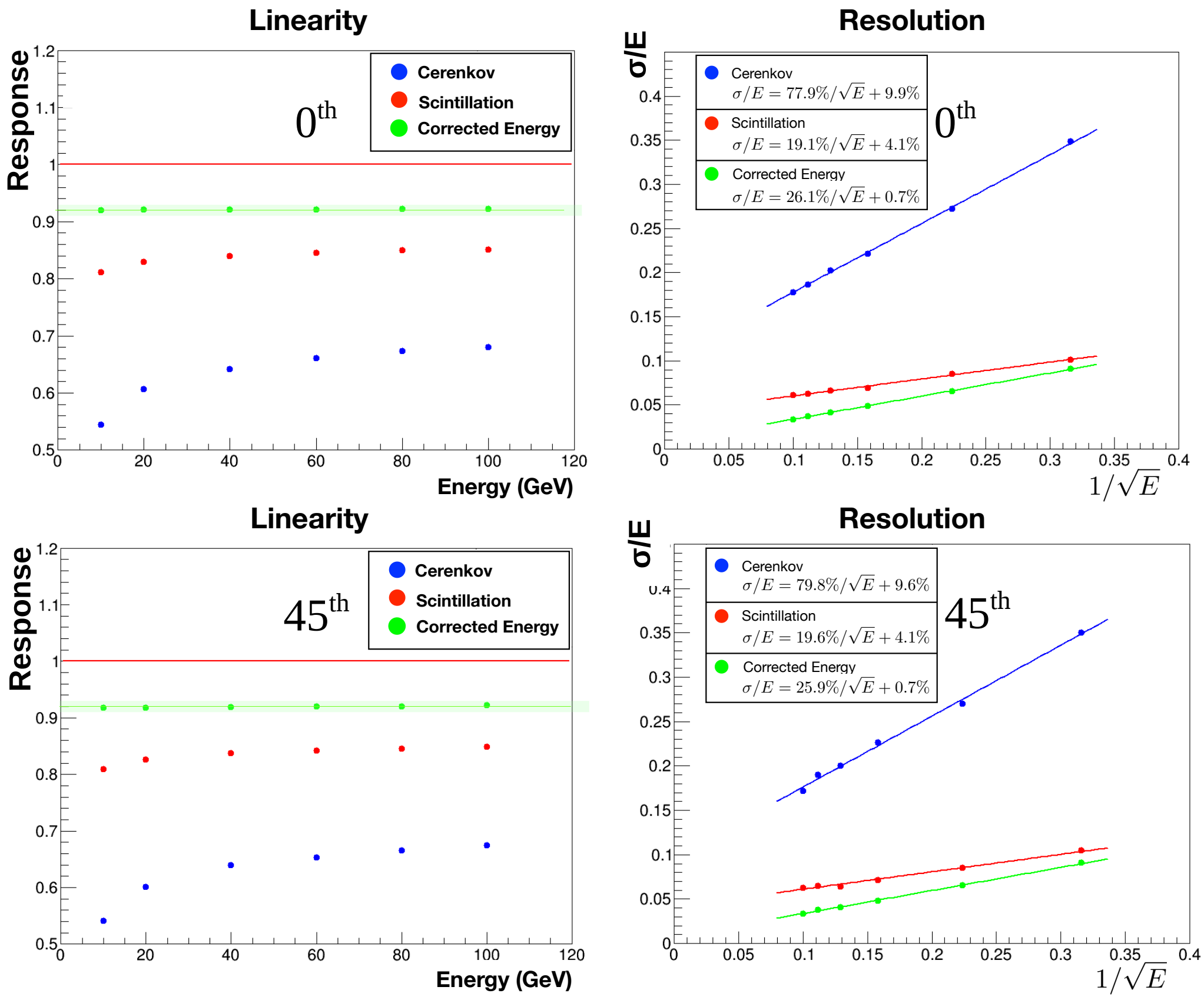

**Figure 5.51:** Linearity and energy resolution with pions, for towers #0 (top) and #45 (bottom), in the wedge geometry (MC simulations).

In general, light attenuation effects need also to be considered, for a $\sim 2 - 2.5\ m$ long fiber detector, since they may introduce a constant term in the hadronic resolution as a function of the shower development point (late starting showers will give bigger and earlier signals).

The evaluation of advantages and disadvantages of filters (to dump the short attenuation-length components) and mirrors (to increase the number of photons that reach the photodetectors) may be relevant in this context.

The effects of the integration of a preshower detector have to be evaluated and the $e/\pi$ separation capability assessed and quantified, for both isolated particles and particles within jets.

Possible longitudinally segmented solutions must be studied and understood. For example, one could imagine the calorimeter made of long and short fibers, where the short ones start after one interaction length, in order to mimic a hadronic compartment. Together with timing information, this should allow to improve the capability to discriminate $em$ and hadronic showers, at the price of a worse $em$ resolution.

### 5.5.7 FINAL REMARKS ON DUAL-READOUT CALORIMETRY

The 15-year-long experimental research program on dual-readout calorimetry of the DREAM/RD52 collaboration has yielded a technology that is mature for application at CEPC. The results show that the parallel, independent, readout of scintillation and Čerenkov light, makes it possible to cancel the effects of the fluctuations of the electromagnetic fraction in hadronic showers, dominating the energy resolution of most (if not all) the calorimeters built so far. In conjunction with high-resolution $em$ and hadronic energy measurements, excellent standalone particle-ID capability has been demonstrated as well.

Those results give increasing support to the conviction that a matrix of alternating scintillating and clear fibers, inserted in copper or lead strips and readout by Silicon PhotoMultipliers (SiPMs), will be able to provide performance more than adequate for the physics programs at the CEPC collider.

Nevertheless, there is a series of technical and physics issues that need to be solved, within the next 2-3 years in order to arrive up to the design of a realistic $4\pi$ detector. A non-exhaustive list must include:

1. The industrial machining of foils of copper, lead or some other material, with the required precision.
2. The development of a mechanical integration design.
3. The readout of the high granularity matrices of SiPM that, in order to be effective, will require the development of a dedicated Application Specific Integrated Circuit (ASIC). Possible aggregations of more fiber outputs into a single channel have also to be implemented and studied.
4. The need and, in case, the way for a longitudinally segmented calorimeter system and the performance of Particle Flow Algorithms to further boost the performance of dual-readout.
5. The development of a modular solution and the assessment, at all levels, of its performance, through beam tests of small modules and simulations. An intensive program of simulations is already ongoing for a dual-readout calorimeter system at CEPC. The response to single particles and jets is under study, in standalone configurations. The work for understanding the behavior of a $4\pi$ calorimeter integrated in a full detector, with a tracking and a magnetic system, has also started. This will include, as well, the evaluation of the combined performance with a preshower detector in front.

CHAPTER 6

# DETECTOR MAGNET SYSTEM

The CEPC detector magnet is an iron-yoke-based solenoid to provide an axial magnetic field of 3 Tesla at the interaction point. A room temperature bore is required with 6.8 m in diameter and 8.3 m in length. This chapter describes the conceptual design of the magnet, including the design of field distribution, solenoid superconducting coil, cryogenics, quench protection, power supply and the yoke. In the end of this chapter, the R&D Section 6.5 brings up other concept options and some reach projects. The compensating magnets designed to minimize the disturbance from the detector solenoid on the incoming and outgoing beams are briefly discussed in Section 9, and in more detail in the CDR accelerator volume, Chapter 9.2 [1].

## 6.1 MAGNETIC FIELD DESIGN

The CEPC detector magnet follows the same design concepts of the BES III magnet [2], as well as CMS [3] and ILD [4]. The aluminum superconductor stabilization with indirect Liquid Helium (LHe) cooling will be adopted. The self-supporting winding turn with aluminum alloy reinforcement is suitable for the CEPC magnet. The conductor with reinforcement wrapped around the pure aluminum as a box configuration is chosen for the magnet design and is manufactured in the R&D project.

The magnetic stray fields outside the iron return yoke of the detector need to meet the requirements of the electronics, the accelerator components and the maintenance interventions. Furthermore, the booster is located in a tunnel 25 m away from the interaction region [1], and the magnetic field there needs to be kept below two Gauss. Additional shielding may therefore be needed.

### 6.1.1 MAIN PARAMETERS

The CEPC magnet system requires a 3 T central field. The superconducting coil is designed with 5 modules wound with 4 layers. The three middle coil modules and the two end coil modules are wound with 78 and 44 turns, respectively. The operating current is 15,779 A for each turn. The geometrical layout of magnet are shown in Figure 6.1. The main magnetic and geometrical design parameters are given in Table 6.1.

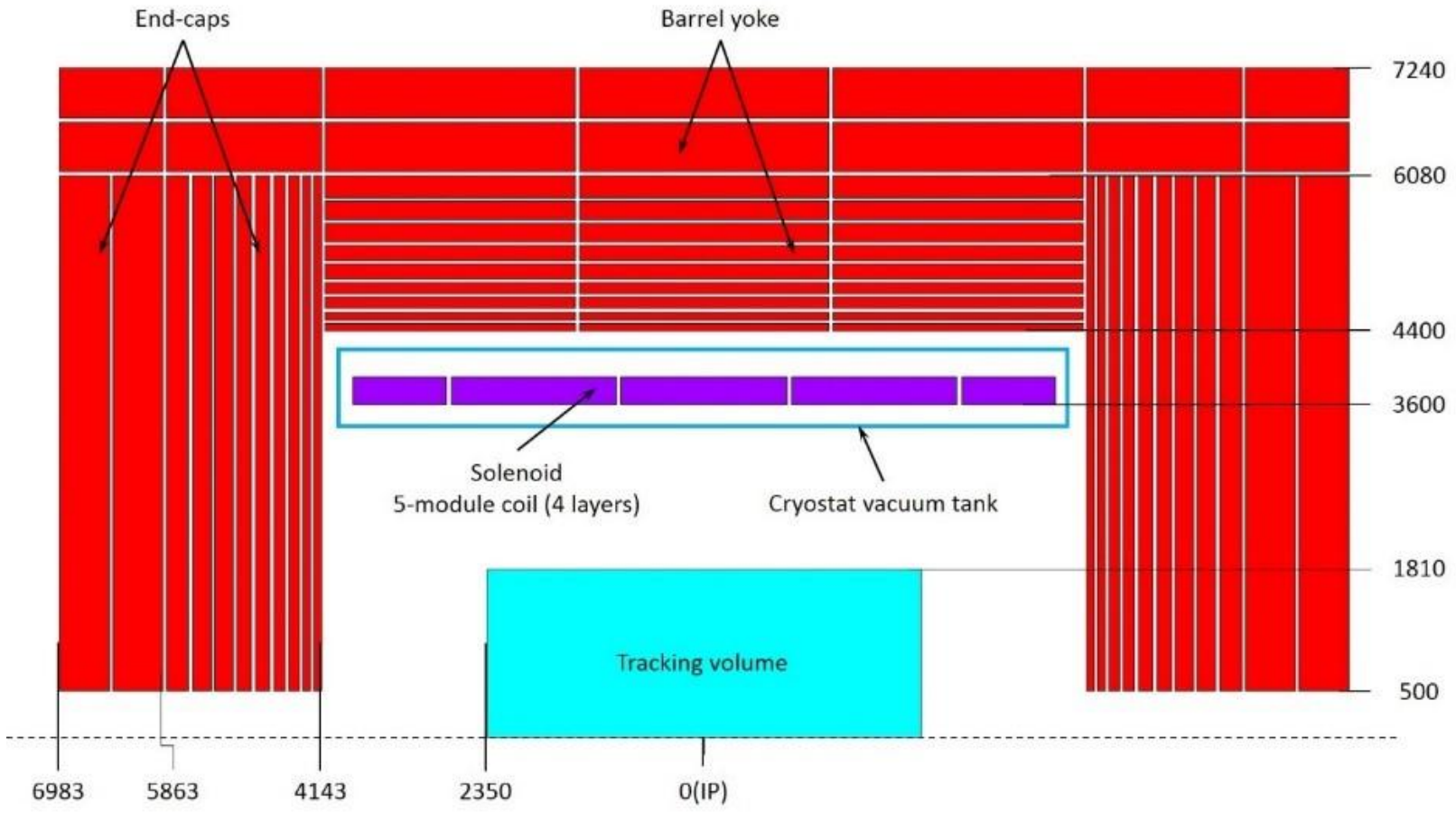

**Figure 6.1:** Geometrical layout of CEPC detector magnet. The simulation model consists of the superconducting coil and the iron yoke with a barrel yoke and two endcap yokes. The superconducting coil is designed with 5 modules wound with 4 layers. Eleven iron plates with 4 cm gaps form both the barrel and endcap yokes.

| Parameter | Value | Parameter | Value |
|---|---|---|---|
| Solenoid central field | 3 T | Working current | 15779 A |
| Maximum field on conductor | 3.5 T | Total ampere-turns of the solenoid | 20.3 MAt |
| Coil inner radius | 3600 mm | Inductance | 10.5 H |
| Coil outer radius | 3828 mm | Stored energy | 1.3 GJ |
| Coil length | 7445 mm | Cable length | 30.1 km |

**Table 6.1:** Main parameters of the CEPC superconducting coil.

### 6.1.2 MAGNETIC FIELD CALCULATION

The magnetic field simulation has been calculated in 2D Finite Element Analysis (FEA) model, with fine structure of the barrel yokes and endcap yokes. The axial magnetic force is maximum at full current; there is no iron saturation effect. The magnetic stray fields outside the iron return yokes of the detectors need to meet the requirements of the electronics, the accelerator components and the interventions for maintenance. Figure 6.2 shows the magnetic field contour of the magnet. The maximum field on NbTi cable is 3.5 Tesla. Figure 6.3 shows the stray field map of the magnet, going respectively from 50 to 250 Gauss on the center plane (beam orbit plane). The edge of 50 Gauss stray field is located relative to the IP at 13.6 m along the beam axis, and 15.8 m in the axial direction. The Booster tunnel located 25 m away is indicated by the red line on the map. The field is about 12 Gauss in the center of the booster tunnel.

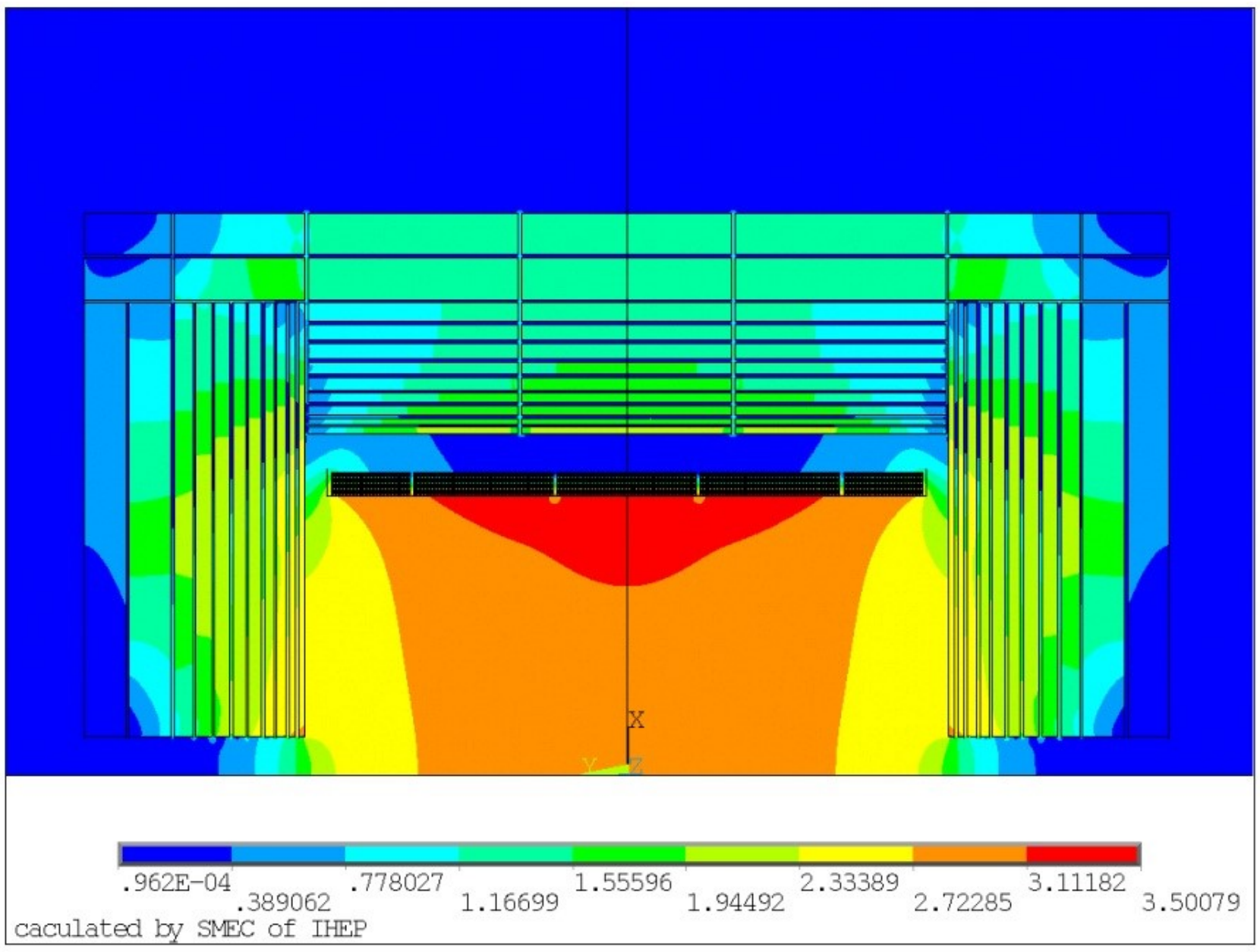

**Figure 6.2:** Field map of magnet in 2D FEA model. The magnetic field at IP is 3.0 T. The maximum field on superconducting cable is 3.5 T. The color scale is in Tesla.

## 6.2 SUPERCONDUCTING COIL SYSTEM

The CEPC solenoid conductor baseline design is the box configuration, based on the self-supporting conductor design of the CMS detector magnet, composed of NbTi Rutherford cable, pure aluminum stabilizer and aluminum alloy reinforcement.

The CMS conductor is fabricated by ebeam welding aluminum alloy to the coextruded high purity Al/superconducting cable insert, whereas the CEPC conductor is fabricated by coextrusion of all components. The configuration is shown in Figure 6.4. The Rutherford cable contains 32 NbTi strands. All magnet finite element analysis has been with this conductor with overall dimensions of 22 mm $\times$ 56 mm.

The coil is wound by inner winding technique on the support aluminum-alloy cylinder, as an external supporting mandrel. The support cylinder also takes away the heat energy induced by quench. The superconducting coil in the cryostat requires cooling at liquid helium temperature. The total weight of cold mass is about 120t. The energy over mass density is about 10.8kJ/kg. The cold mass will be indirectly cooled by a network of LHe tubes. These tubes are welded to the support cylinder. The indirect cooling method is designed in a thermosiphon process. The siphon cooling circulation loop operates under a suitable filling amount 50% to 85% with high efficient heat transfer properties. In addition, it is optimized to minimize the temperature difference throughout the whole magnet.

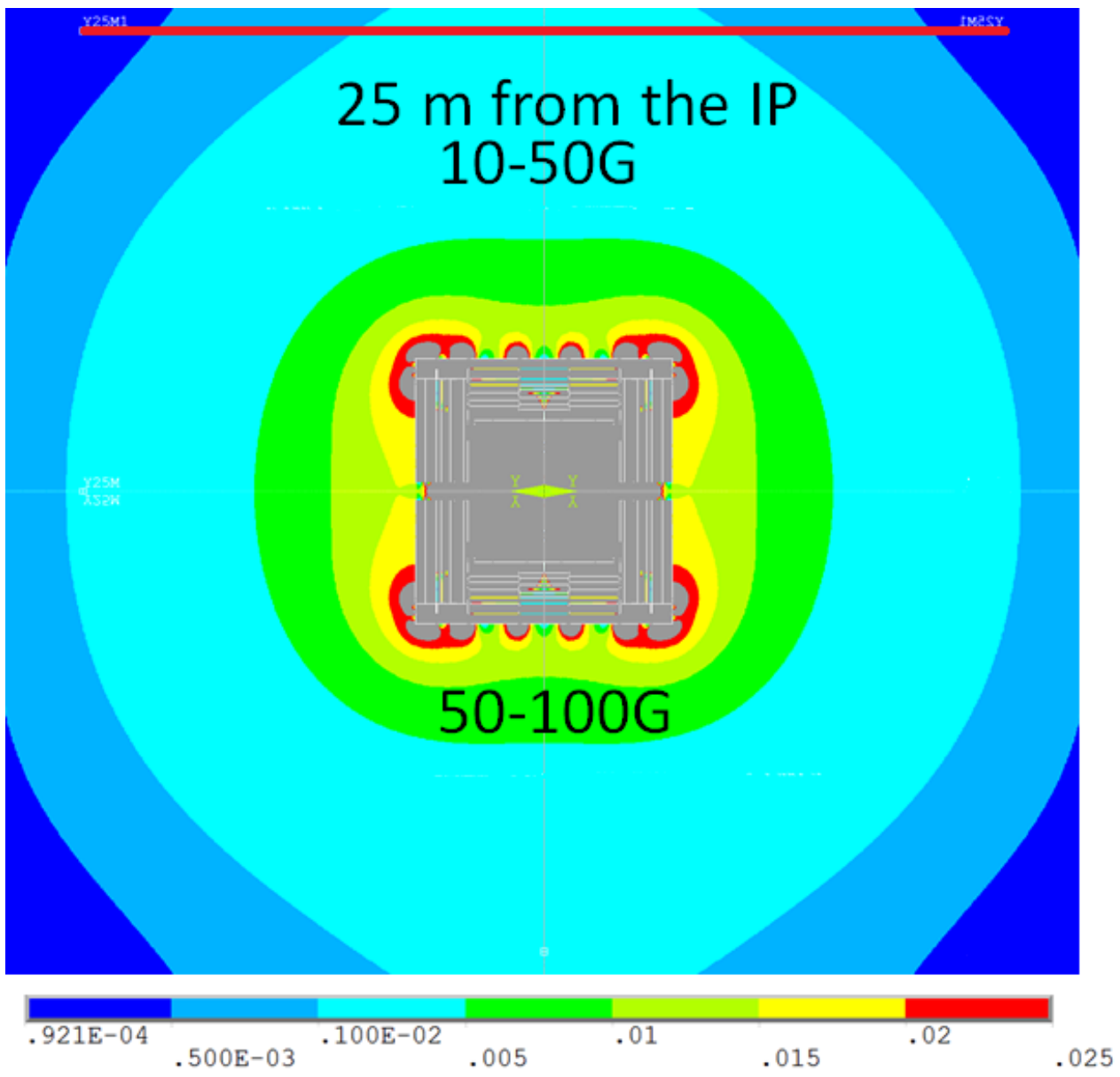

**Figure 6.3:** Stray field map of the magnet, going respectively from 50 to 250 Gauss on the center plane (beam orbit plane). The edge of 50 Gauss stray field is located relative to the IP at 13.6 m along the beam axis, and 15.8 m in the axial direction. The Booster tunnel located 25 m away is indicated by the red line on the map. The field is about 12 Gauss in the center of the booster tunnel. The color scale is in Tesla.

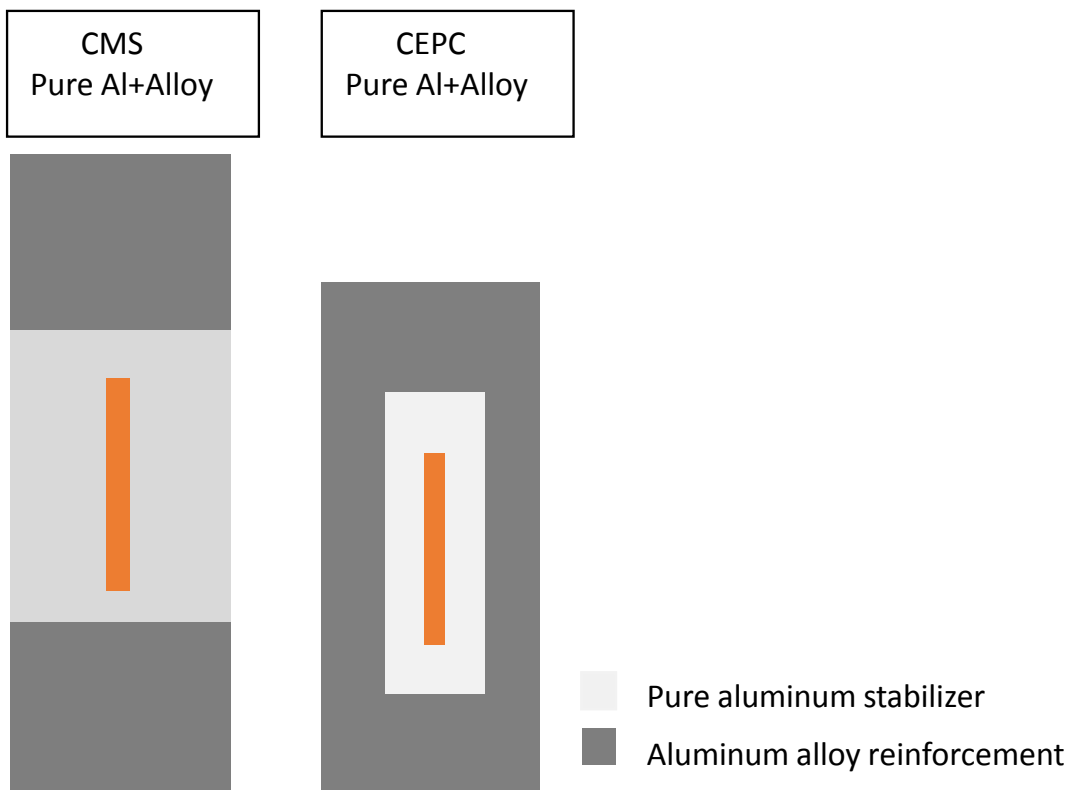

**Figure 6.4:** CMS conductor [3] and the baseline design of the CEPC conductor. The CEPC NbTi/Cu Rutherford cable is wrapped by purity aluminum stabilizer and aluminum alloy reinforcement with the box configuration. The Rutherford cable contains 32 NbTi strands. The overall dimensions of this conductor are 22 mm $\times$ 56 mm.

## 6.3 ANCILLARY SYSTEMS

### 6.3.1 CRYOGENICS SYSTEM

A cryoplant with a capacity of 750 W @ 4.5 K is under design for the operation of the superconducting facility. The cryogenic system provides the liquefaction and refrigeration at 4.5 K in varying proportions depending on the operating modes, which include cooling down from 300 K to 4.5 K, normal operation, energy dump and warming up. It is also designed to extract the dynamic losses during the various magnet ramps or discharges.

Helium liquefier is in a position close to the magnet compatible with the fringing field and the maintenance activities. It supplies the liquid helium to the coils and taking the helium gas return back from the coils and the power lines. The CEPC detector magnet cryogenic system, shown in Figure 6.5, is composed of two sub-system: the external system and the inner system. The external system includes cold box and intermediate Dewar; the inner system includes coil cooling circuit and phase separator.

The refrigerator is dimensioned for acceptable cool-down time. Moreover, the refrigeration plant design will also take into account the losses occurred in the current leads and during the magnet ramping, plus an additional safety margin.

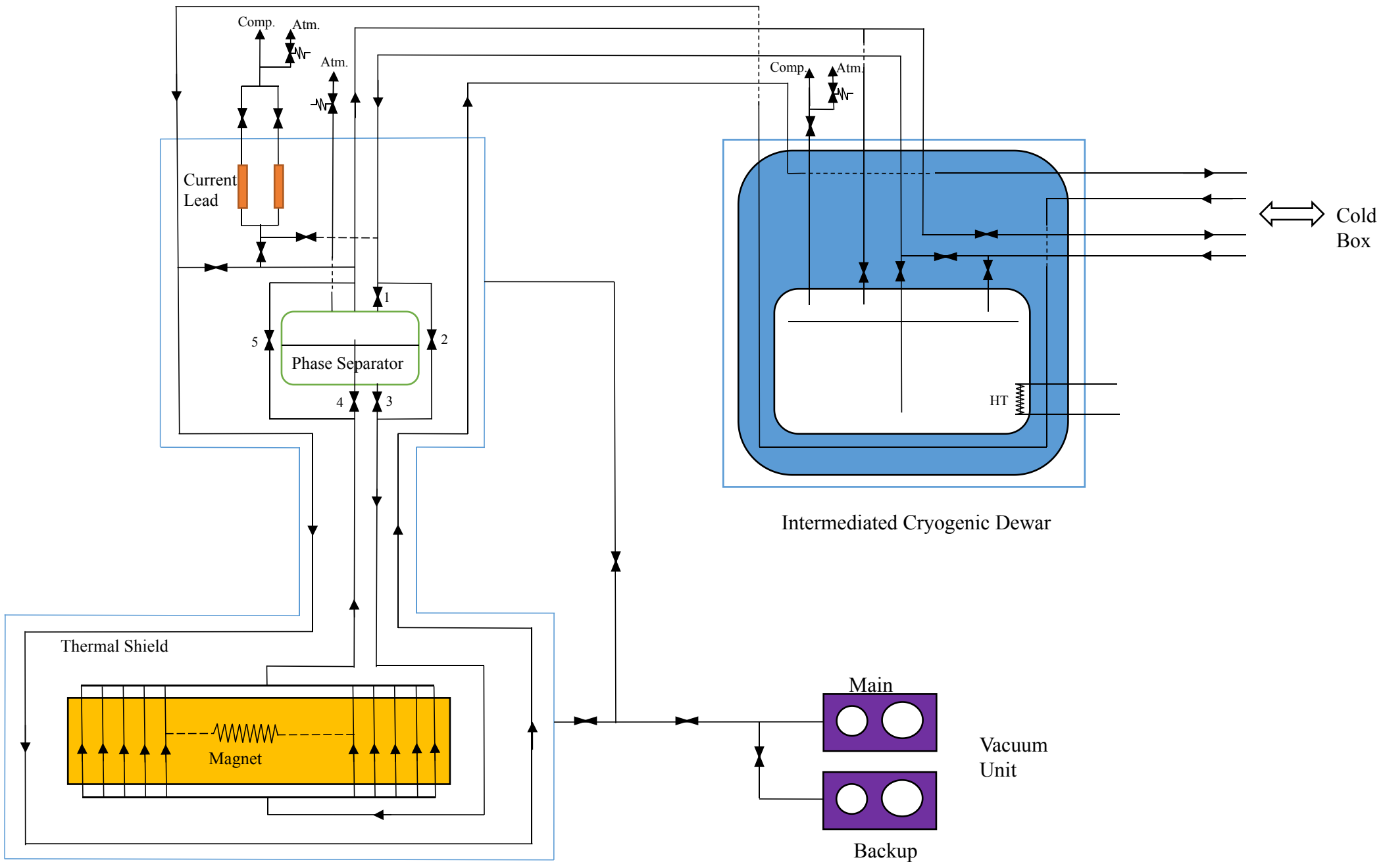

**Figure 6.5:** Thermosiphon cooling circuit. The CEPC detector magnet cryogenic system is composed of two sub-system: the external system and the inner system. The external system includes cold box and intermediate Dewar; the inner system includes coil cooling circuit and phase separator. The cryogenic sequences include cooling down, normal operation, energy dump and warming up. The first operation mode is the cooling down process by forced flow helium from 300 K to 100 K and to 4.5 K. The second operation mode is the normal operation process in thermosiphon flow condition, which is the main working mode for the magnet. The third one is the energy dump process, which is divided into fast discharge, slow discharge and post quench re-cooling. The last operation mode is the warming up.

### 6.3.2 POWER SUPPLY

A low ripple DC current-stabilized power supply, with low output voltage and high output current, will be used for CEPC detector magnet. The power supply is expected to have a free-wheel diode system and to be cooled with demineralized water. The main circuit of a standard power supply includes 12 pulse diode rectifiers and 4 Insulated Gate Bipolar Transistor (IGBT) chopper units with a switching frequency of 10 kHz.

### 6.3.3 QUENCH PROTECTION AND INSTRUMENTATION

Selected voltage signals from the CEPC detector magnet coil and current leads are monitored by an FPGA board for quench detection. If a quench happens, the power supply will be switched off and a dump resistor will be switched into the electrical circuit, the huge stored energy will be extracted mainly by the dump resistor and partially by the coil itself.

## 6.4 IRON YOKE DESIGN

The iron yoke serves as the magnetic flux return and the main mechanical structure of the sub-detectors. Therefore high permeability material with high mechanical strength is required for the yoke material. The gaps between yokes provide room for the muon detector, data cables, cooling pipes, gas pipes and etc. through the yoke. The yoke is divided into two main components, one cylindrical barrel yoke and two endcap yokes. The total weight of the yoke assembly is about 10,000 tons. We are studying the possibility of reducing the yoke weight due to cost concerns.

The barrel yoke is a dodecagonal shape structure with a length of 8,206 mm (Figure 6.6). The outer diameter of the dodecagon and the inner diameter are 14,480 mm and 8,800 mm respectively. The barrel yoke is subdivided along the beam axis into 3 rings, with 11 radial layers in each ring. Each ring of the barrel yoke is composed of 12 azimuthal segments. 40 mm gap is designed between the rings and the layers for placing the muon detector and the electronics cables and services. From the inner to the outer, the layer thicknesses are 80 mm, 80 mm, 120 mm, 120 mm, 160 mm, 160 mm, 200 mm, 200 mm, 240 mm, 540 mm, 540 mm, respectively.

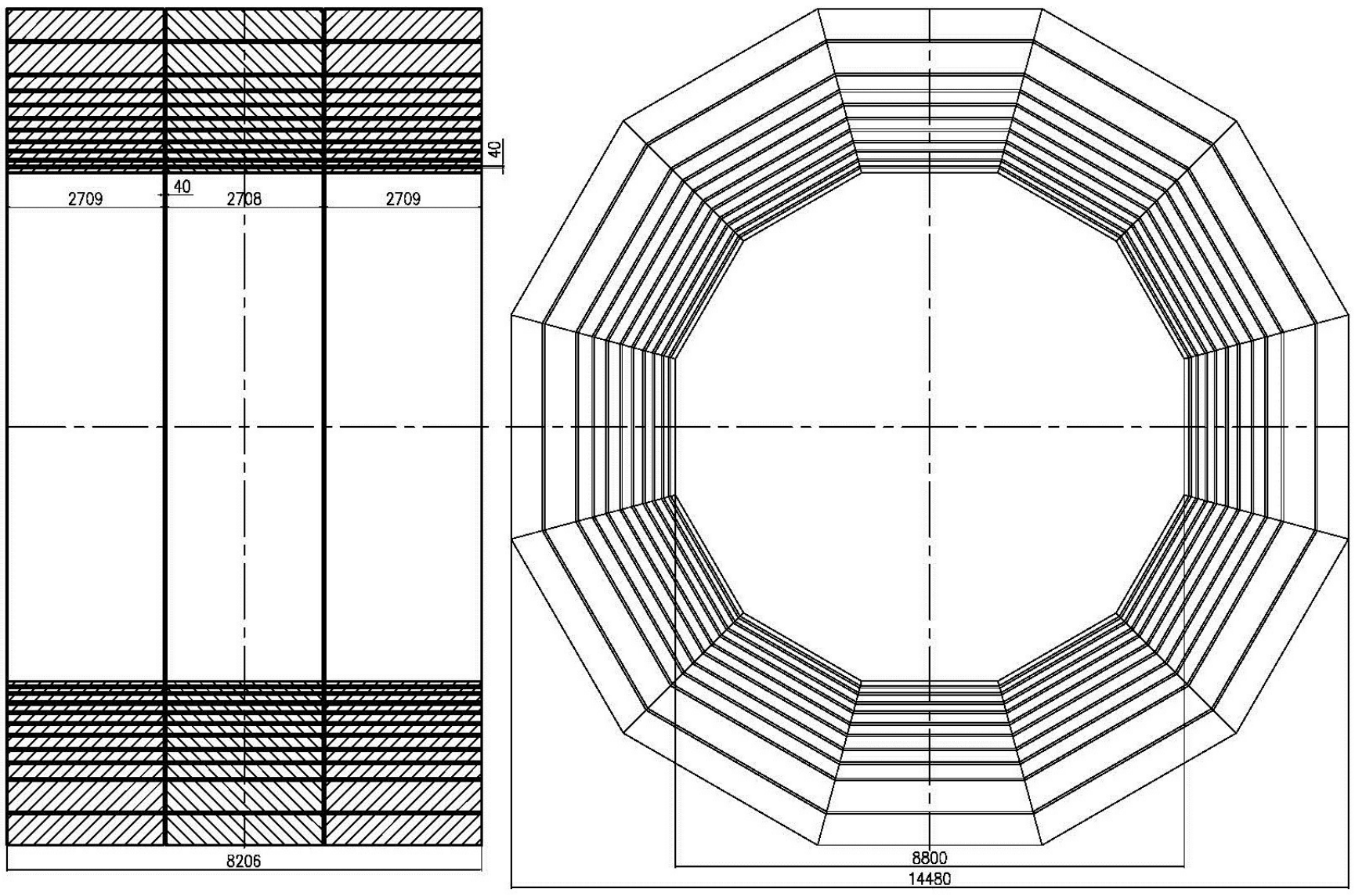

**Figure 6.6:** Barrel yoke of CEPC detector magnet. The barrel yoke is a dodecagonal shape structure with a length of 8,206 mm. The outer diameter of the dodecagon and the inner diameter are 14,480 mm and 8,800 mm respectively. The barrel yoke is subdivided along the beam axis into 3 rings, with 11 radial layers in each ring. Each ring of the barrel yoke is composed of 12 azimuthal segments. 40 mm gap is designed between the rings and the layers for placing the muon detector and the electronics cables and services. From the inner to the outer, the layer thicknesses are 80 mm, 80 mm, 120 mm, 120 mm, 160 mm, 160 mm, 200 mm, 200 mm, 240 mm, 540 mm, 540 mm, respectively.

The endcap yokes are designed to dodecagonal structure with the out diameter of 14,480 mm. Each endcap yoke will consist of 11 radial layers (Figure 6.7). Each endcap yoke is composed of 12 azimuthal segments. The layer thicknesses are 80 mm, 80 mm, 120 mm, 120 mm, 160 mm, 160 mm, 200 mm, 200 mm, 240 mm, 540 mm, 540 mm, respectively.

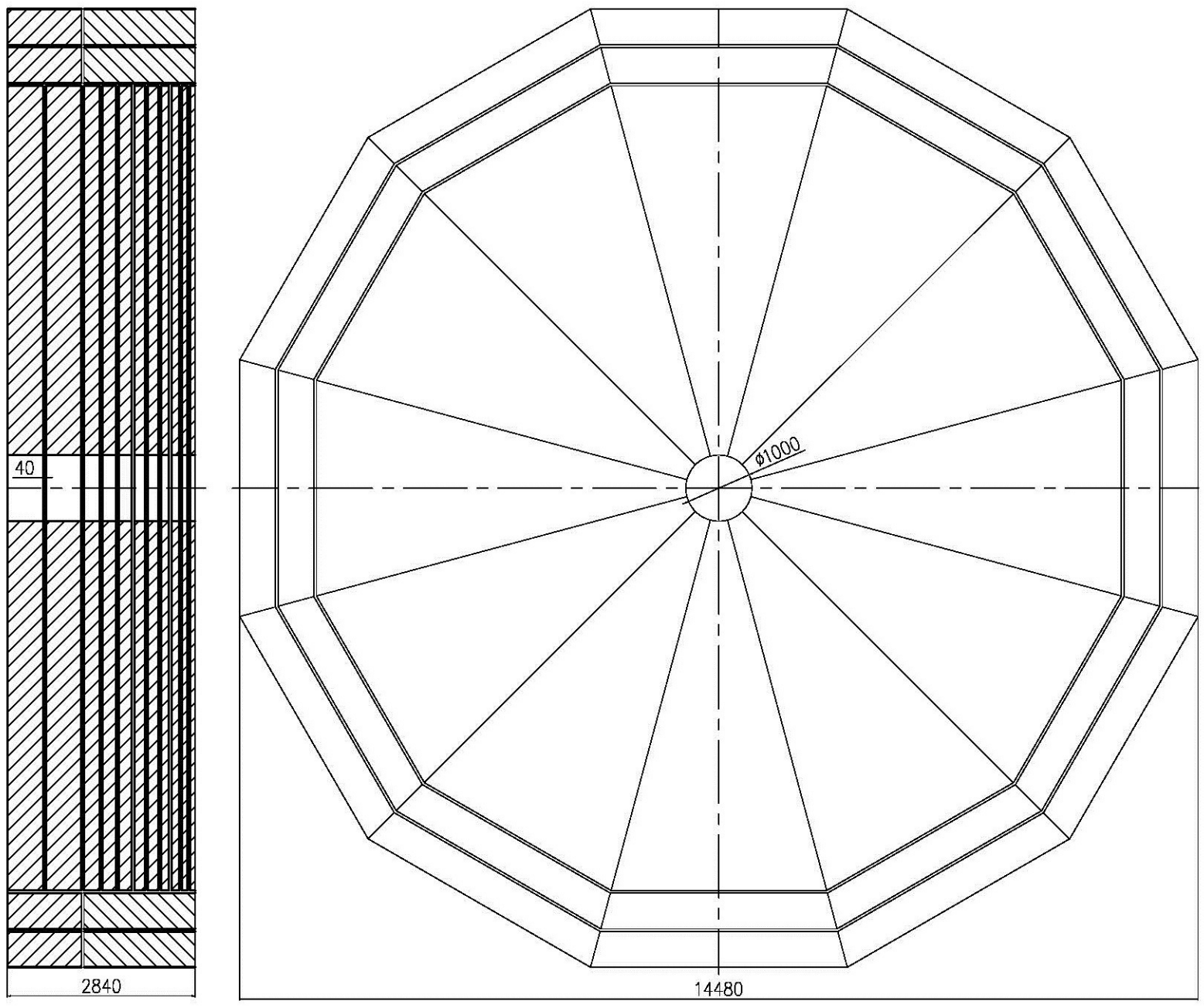

**Figure 6.7:** Endcap yokes of CEPC detector magnet. The endcap yokes are designed to dodecagonal structure with the out diameter of 14,480 mm. Each endcap yoke will consist of 11 radial layers (Figure 6.4). Each endcap yoke is composed of 12 azimuthal segments. The layer thicknesses are 80 mm, 80 mm, 120 mm, 120 mm, 160 mm, 160 mm, 200 mm, 200 mm, 240 mm, 540 mm, 540 mm, respectively.

## 6.5 ALTERNATIVE DESIGNS AND R&D

### 6.5.1 LTS SOLENOID FOR THE IDEA DETECTOR

A "thin" 6 m long solenoid with an inner bore of 2.1 m radius and a field of 2 Tesla is a key element of the IDEA detector, where the calorimeter is located outside of the solenoid, allowing the maximum possible volume for tracking, but requiring maximum transparency. By "thin" we mean that the magnet should give minimal perturbation to the calorimetric measurements. The current design is mostly derived by scaling the present 2 Tesla solenoid of the ATLAS detector and uses a self-supporting aluminum stabilized NbTi conductor. Preliminary engineering studies [5, 6] indicate that the coil and the cryostat can fit in a total thickness of 30 cm using current technology; for 0.46 radiation lengths of the coil and 0.28 radiation lengths of the cryostat or 0.16 interaction lengths at normal incidence. Up to 20% additional reduction in the overall thickness may be achieved with more R&D and engineering.

### 6.5.2 HTS SOLENOID FOR IDEA DETECTOR

A High Temperature Superconductor (HTS) solenoid is being studied for the IDEA detector. The HTS solenoid is designed to use Yttrium Barium Copper Oxide, $YBa_2Cu_3O_7$ (YBCO) stacked-tape cable as the conductor. The radiation length of single YBCO tape coated with 10 $\mu$m copper is about 0.004 $X_0$. Each tape carries 700 A at 20 K. The 35 YBCO tapes stacked together allow 24.5 kA. These tapes are embedded in 5 mm pure aluminum. The radiation length of this YBCO stacked-tape cable is estimated to be 0.2 $X_0$. The radiation length of HTS coil will be less than half of the current "thin" Low Temperature Superconductor (LTS) coil design. If the operation temperature of the cold mass is raised to 20 K, the heat conductivity parameters of all components are improved. In addition, the electricity consumption of cooling station is much lower than that at 4.2 K. Therefore, the YBCO stacked-tape cable and the cryogenics are brought into R&D.

### 6.5.3 DUAL SOLENOID DESIGN

The dual solenoid design is presented for a conceptual option for CEPC detector magnet, which contains two series connected superconducting solenoids carrying the current in opposite directions, based on FCC twin solenoid [7]. The main solenoid provides central field within the room temperature bore. The outer solenoid provides the stray field shielding and a magnetic field between the two solenoids to facilitate muon tracking. The main advantage of this dual solenoid is that the system becomes comparatively light-weight and cost saving without iron yoke. The sketch is shown in Figure 6.8.

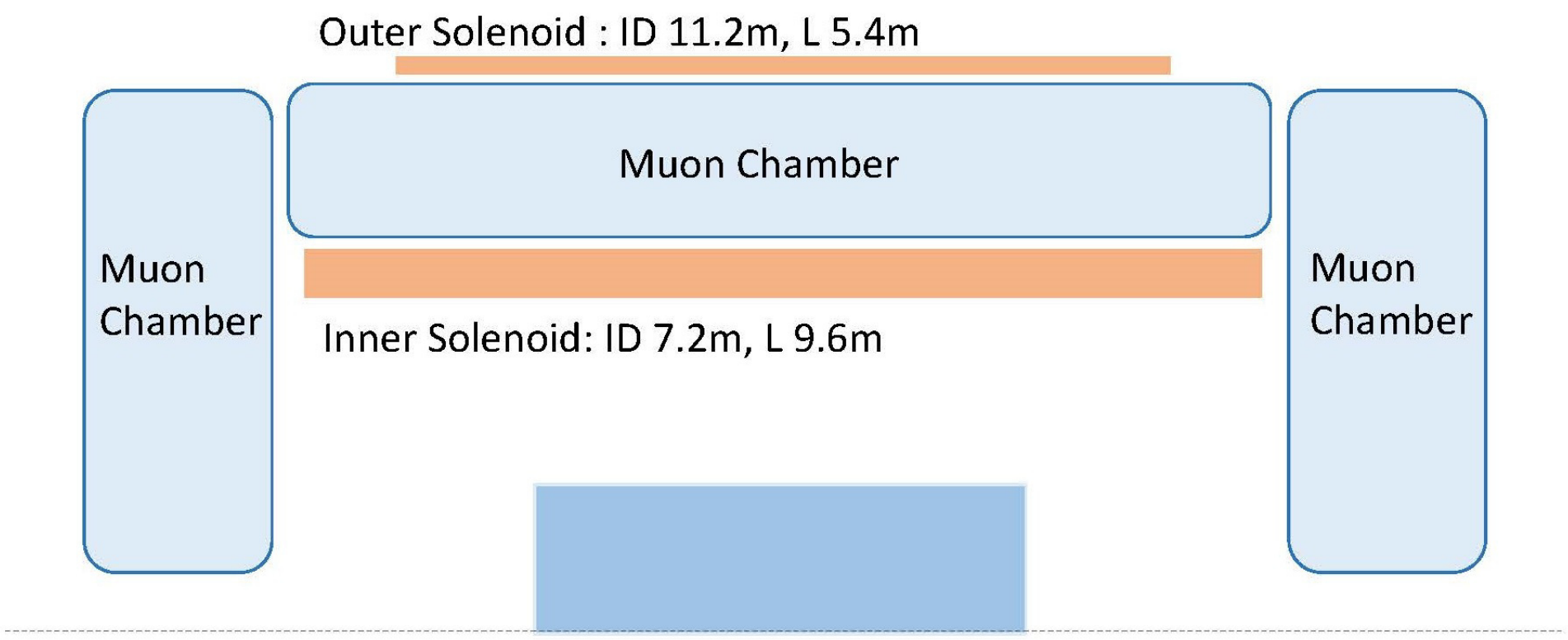

**Figure 6.8:** Sketch of dual solenoid design. The main solenoid provides central field within the room temperature bore. The outer solenoid provides the stray field shielding and a magnetic field between the two solenoids to facilitate muon tracking. The inner diameter of the main solenoid is 7.2 m. The inner diameter of the outer solenoid is 11.2 m.

### 6.5.4 SUPERCONDUCTING CONDUCTOR

The coil is simulated with an elasto-plastic 2D FEA model. Mechanical analysis requires the experimental material properties of all conductor components. We have developed a 10 m long NbTi Rutherford cable embedded inside stabilizer which provides Ic 5 kA at 4 T background magnetic field. Meanwhile we measured the material properties and the tensile stress of 10 m cable. The R&D for longer conductor with higher Ic 15 kA at 4 T background is ongoing.

### 6.5.5 THERMOSYPHON CIRCUIT

Thermosyphon principle is used to cool CEPC detector superconducting magnet by the U-shaped circuit configuration (more detail in the thermal shield helium circuit of CMS [3]), carrying LHe on the outer surfaces of the coil supporting cylinders. The thermosyphon circuit consists of helium phase separator located in an elevated position and the cooling tubes. In order to study the phase transition process of helium in the circuit, the changes of the temperature distribution and the density distribution over the time, a 1:10 scale thermosyphon circuit will be established for simulation and experiment.

CHAPTER 7

# MUON DETECTOR SYSTEM

For the baseline detector concept, muons are identified primarily in the the PFA-oriented calorimeters and their momenta are measured in the tracking system. An outermost muon detector system is envisioned to provide redundancy, aid muon identification in busy environments and reduce backgrounds. Embedded in the solenoid flux return yoke, the detector is designed to identify muons with high standalone efficiency ($\geq 95\%$) and purity for muon $p_T$ down to $\sim 3\,\mathrm{GeV}$ over the largest possible solid angle. While the design is driven by the identification, the muon detector could provide standalone measurements of the muon momenta as well.

The muon detector will significantly improve the identification of muons produced inside jets such as those from $B$ hadron decays. Moreover, the detector can compensate for leaking energetic showers and late showering pions from the calorimeters, which could help to improve the jet energy resolution [1]. It can also aid in searches for long-lived particles that decay far from the IP, but still inside the detector.

This Chapter presents design considerations and technology options for the CEPC muon detector. Section 7.1 introduces the conceptual design. Both the Resistive Plate Chamber (RPC) and an innovative type of Micro Pattern Gas detector (MPGD), the $\mu$-RWELL detector, are being considered. The main difference between the two technologies lies in the position resolution and the cost. Details are presented in Section 7.2 for the RPC technology and in Section 7.3 for the $\mu$-RWELL technology. Though not described here, other gas detectors such as Gas Electron Multiplier (GEM), MicroMegas and Monitored Drift Tubes (MDT) are also possible options. Section 7.4 briefly describes the future R&D program.

## 7.1 BASELINE DESIGN

The baseline design of the CEPC muon detector is divided into one barrel and two endcaps, as shown in Figure 7.1, consisting of azimuthal segmented dodecagon modules. The design parameters are summarized in Table 7.1. These parameters will be further optimized together with other detector subsystems, in particular the ECAL and the HCAL. The number of sensitive layers and the thickness of the absorber (iron in this case) are two critical parameters for the muon system. The baseline design consists of 8 sensitive layers alternating with iron absorber layers. The total iron thickness is 6.7 interaction lengths, sufficient to reduce punch-through backgrounds. The total sensitive area amounts to 8600 $m^2$.

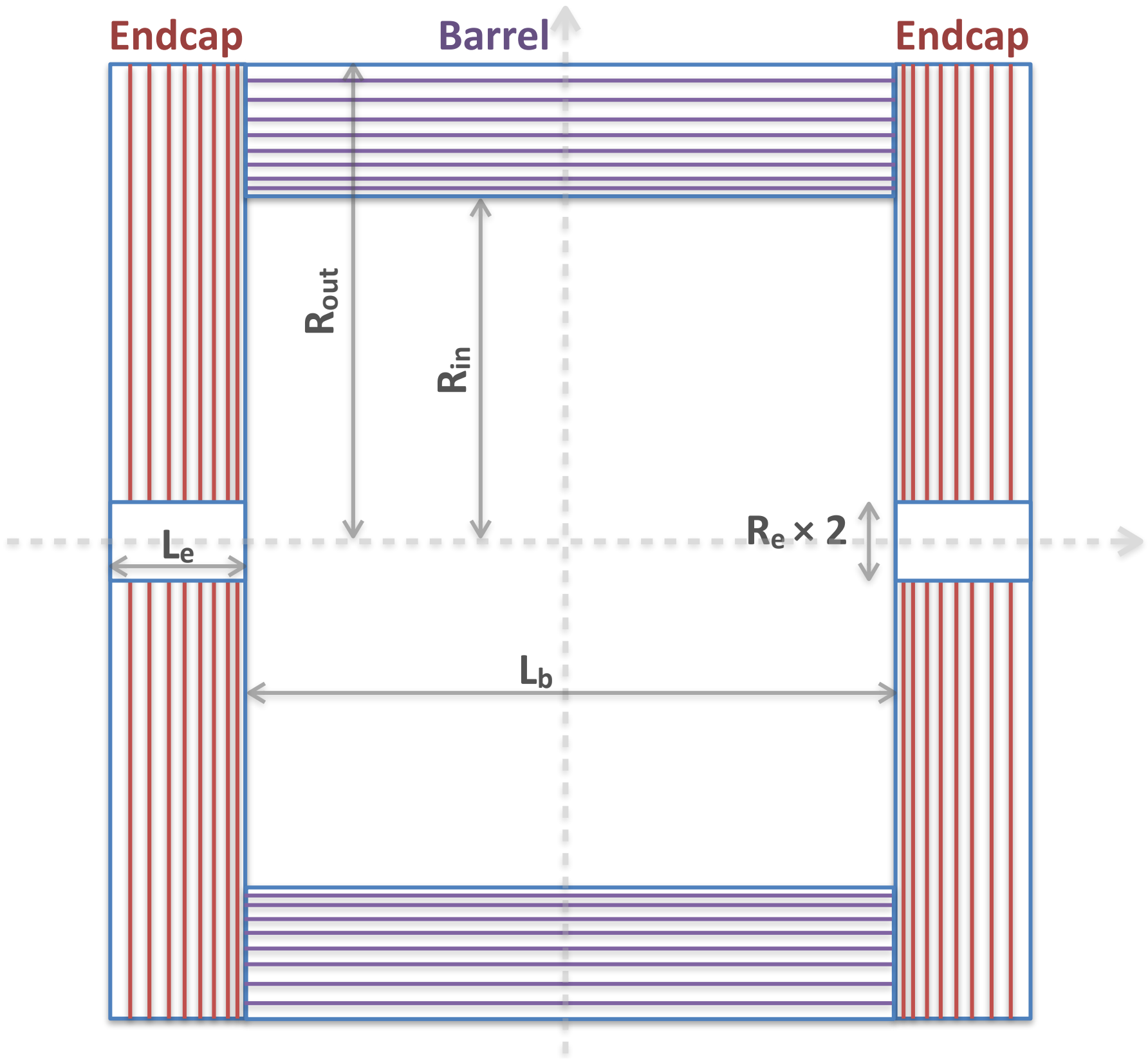

**Figure 7.1:** The $R - Z$ view of the basic layout of the muon system, subdivided into a barrel and two endcaps. $L_b$ is the length of the barrel and $L_e$ is the length of each endcap. $R_{out}$ ($R_{in}$) is the outer (inner) radius of the barrel. $R_e$ is the inner radius of each endcap. The extra iron yoke that exists past the instrumented region is not depicted here.

The solid angle coverage of the CEPC muon system should be up to $0.98 \times 4\pi$ in accordance with the tracking system. Minimum position resolutions of $\sigma_{r\phi} = 2.0$ cm and $\sigma_z = 1.5$ cm are required. The muon system should provide several space point measurements, a time resolution of a few nanosecond and a rate capability of 50 – 100 Hz/$cm^2$. The position measurements should provide information on muon momenta which can be used independently or combined with the measurements in the tracking system.

The performance of the baseline muon detector has been studied using simplified simulations. On average, muons need a momentum larger than 2 GeV to reach the first layer of the muon detector, and a momentum larger than 4 GeV to penetrate all 8 layers. A standalone muon identification efficiency greater than 95% with a pion fake rate smaller than 1% can be achieved for muons above 5 GeV. These preliminary simulation studies are performed with an earlier detector geometry and simulation configuration. Further studies as described in Section 7.4 are planned.

| **Parameter** | **Baseline** |
|---|---|
| $L_b/2$ [m] | 4.14 |
| $R_{in}$ [m] | 4.40 |
| $R_{out}$ [m] | 6.08 |
| $L_e$ [m] | 1.72 |
| $R_e$ [m] | 0.50 |
| Segmentation in $\phi$ | 12 |
| Number of layers | 8 |
| Total thickness of iron ($\lambda = 16.77$ cm) | $6.7\lambda$ (112 cm) (8/8/12/12/16/16/20/20) cm |
| Solid angle coverage | 0.98 $\times 4\pi$ |
| Position resolution [cm] | $\sigma_{r\phi}$: 2<br>$\sigma_z$ : 1.5 |
| Time resolution [ns] | 1 – 2 |
| Detection efficiency ($P_\mu > 5$ GeV) | $> 95\%$ |
| Fake($\pi \rightarrow \mu$)@30GeV | $< 1\%$ |
| Rate capability [Hz/cm$^2$] | $\sim$60 |
| Technology | RPC (super module, 1 layer readout, 2 layers of RPC ) |
| Total area [m$^2$] | Barrel: $\sim$4450<br>Endcap: $\sim$4150<br>Total: $\sim$8600 |

**Table 7.1:** Design parameters of the CEPC baseline muon system. $L_b$ is the length of the barrel and $L_e$ is the length of each endcap. $R_{out}$ ($R_{in}$) is the outer (inner) radius of the barrel. $R_e$ is the inner radius of each endcap. Further optimizations are expected in the near future and different technologies are being considered.

It is particularly interesting to study the muon identification performance for muons produced in a non-isolated environment, e.g. inside jets, where the PFA muon identification using the calorimeter information is expected to be challenging. Preliminary studies using $e^+e^- \to ZH \to \nu\bar{\nu}b\bar{b}$ samples where muons are produced from $b$-quark decays have been performed. For non-isolated muons with momenta above 10 GeV, the average identification efficiency is found to be around 80% using the LICH algorithm [2] with the full calorimetry information, but without the input from the muon detector. The corresponding purity is estimated to be around 95%. This performance will significantly deteriorate for lower momentum muons. Combining both the muon detector and calorimeter information in the object reconstruction and identification procedure is expected to deliver a significant improvement in the performance for non-isolated muons (Section 7.4).

## 7.2 RESISTIVE PLATE CHAMBER TECHNOLOGY

Resistive Plate Chamber (RPC) is suitable for building large area detectors with centimeter spatial resolution. It has been applied in muon systems for experiments including BaBar [3], Belle [4], CMS [5] , ATLAS [6], BESIII [7], and Daya Bay [8]. It provides a common solution with the following advantages: low cost, robustness, easy construction of large areas, large signal, simple front-end electronics, good time and spatial resolution. It is chosen as the baseline design of the CEPC muon system.

RPCs can be built with glass or Bakelite, and run in avalanche or streamer mode. Bakelite RPCs of about 1200 m$^2$ and 3200 m$^2$ were produced for the BESIII and Daya Bay muon systems, respectively. Compared with glass RPC, Bakelite RPC has the advantages of easier construction, lower density, larger chamber size and lower cost, especially if the event rate is below 100 Hz/cm$^2$ as expected at the CEPC. On the other hand, glass RPC have a better long-term stability, detecting efficiency and lower noise level. The characteristics of Bakelite and glass RPCs are compared in Table 7.2.

## 7.3 $\mu$-RWELL TECHNOLOGY

The $\mu$-RWELL is a compact, spark-protected and single amplification stage Micro-Pattern Gas Detector (MPGD). A $\mu$-RWELL detector [9] is composed of two PCBs: a standard GEM Drift PCB acting as the cathode and a $\mu$-RWELL PCB that couples in a unique structure the electron amplification (a WELL patterned matrix) and the readout stages. The layout is shown in Figure 7.2(a). A standard GEM 50 µm polyimide foil is copper clad on one side and Diamond Like Carbon (DLC) sputtered on the opposite side. The thickness of the DLC layer is adjusted according to the desired surface resistivity value (50–200 MΩ/□) and represents the bottom of the WELL matrix providing discharge suppression as well as current evacuation. The foil is then coupled to a readout board.(as shown in Figure 7.2(b). A chemical etching process is then performed on the top surface of the overall structure in order to create the WELL pattern (conical channels 70 µm (50 µm) top (bottom) in diameter and 140 µm pitch) that constitutes the amplification stage.

| **Parameters** | | **Bakelite** | **Glass** |
|---|---|---|---|
| Bulk resistivity [$\Omega \cdot$ cm] | Normal | $10^{10} \sim 10^{12}$ | $> 10^{12}$ |
| | Developing | $10^8 \sim 10^9$ | |
| Max unit size (2 mm thick) [m] | | 1.2×2.4 | 1.0×1.2 |
| Surface flatness [nm] | | $< 500$ | $< 100$ |
| Density [g/cm$^3$] | | 1.36 | 2.4∼2.8 |
| Min board thickness [mm] | | 1.0 | 0.2 |
| Mechanical performance | | Tough | Fragile |
| Rate capability [Hz/cm$^2$] | Streamer | 100@92% | |
| | Avalanche | 10K | 100@95% |
| Noise rate [Hz/cm$^2$] | Streamer | $< 0.8$ | 0.05 |

**Table 7.2:** Comparison of the main parameters of Bakelite and glass RPCs. Both technologies would satisfy the CEPC detector requirements.

The process is shown in Figure 7.3. The high voltage applied between the copper and the resistive DLC layers produces the required electric field within the WELLs that is necessary to develop charge amplification. The signal is capacitively collected at the readout strips/pads. Two main schemes for the resistive layer can be envisaged: a *low-rate* scheme ( for particles fluxes lower than 100 kHz/cm$^2$) based on a simple resistive layer of suitable resistivity; and an *high-rate* scheme (for a particle flux up to 1 MHz/cm$^2$) based on two resistive layers intra-connected by vias and connected to ground through the readout electrodes. Finally, a drift thickness of 3-4 mm allows for reaching a full efficiency while maintaining a versatile detector compactness.

The $\mu$-RWELL technology, especially in its *low-rate* version, is a mature solution, with whom single detectors of a 0.5 m$^2$ have been realized and successfully operated in the laboratory as well as in test beams. They can withstand particle rates up to a few tens of kHz/cm$^2$, providing a position resolution as good as $\sim$ 60 $\mu$m with a time resolution of 5–6 ns. The requirements of a muon detector for CEPC are not as stringent and therefore can be easily and cost-effectively achieved with the $\mu$-RWELL technology. Moreover the $\mu$-RWELL technology is a robust solution, intrinsically safer against sparks than, for example, the widely used GEM detectors. The muon system could be realized by using tiles of $\mu$-RWELL detectors of a size $50 \times 50$ cm$^2$. This would make the whole muon detector very modular with components bought directly from industry. A CEPC muon detector made of $\mu$-RWELL tiles could consist of three or four detector stations, each equipped with a couple of layers of $\mu$-RWELL detectors in order to provide a very precise, of the order of 200–300 $\mu$m, position resolution on the coordinates of a muon track. The expected parameters of a $\mu$-RWELL based muon detector would perfectly match the specifications required for a CEPC muon detection system.

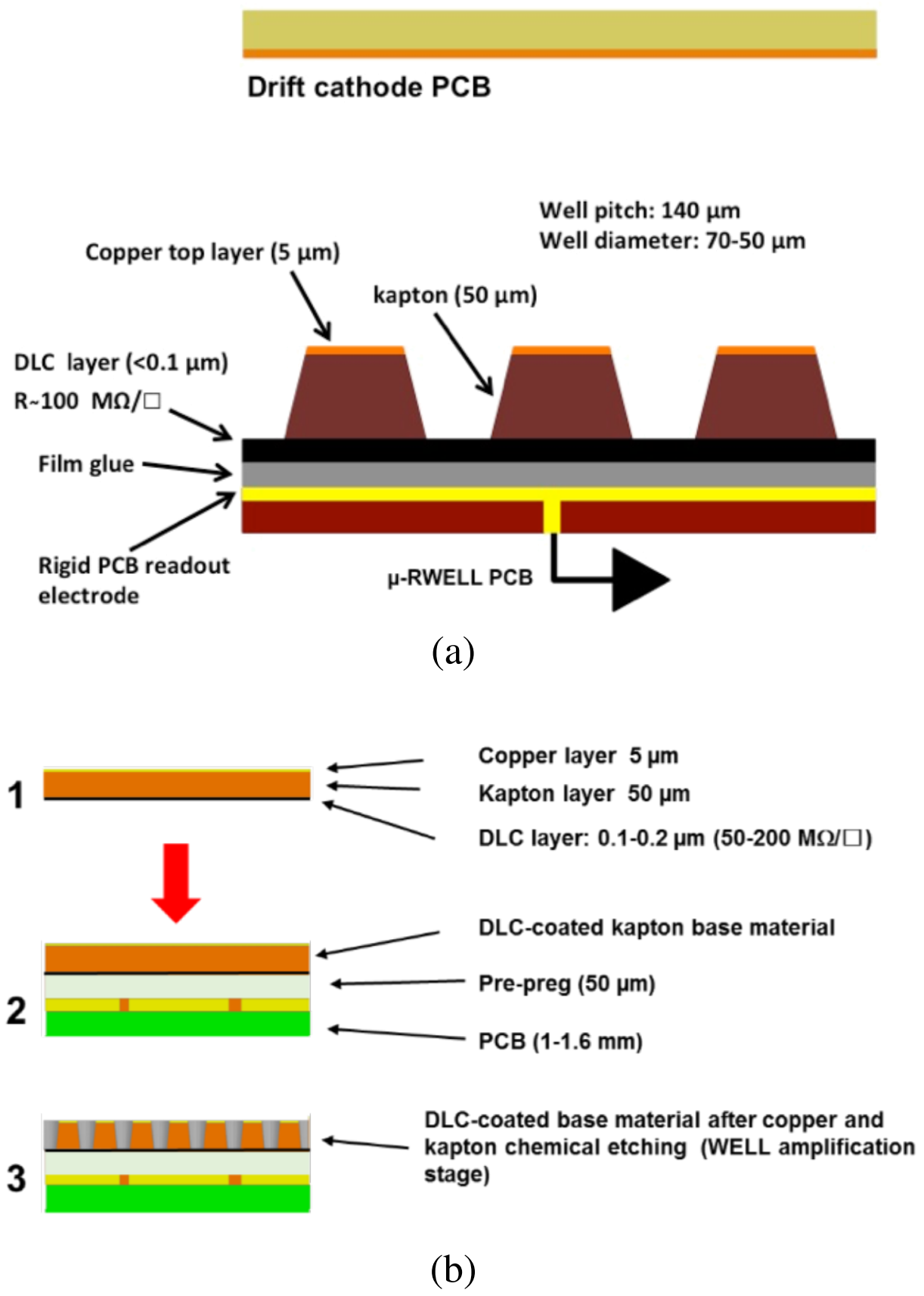

**Figure 7.2:** (a) Layout of a $\mu$-RWELL detector module with the cathode and the $\mu$-RWELL PCBs; (b) Coupling steps of the $\mu$-RWELL PCB.

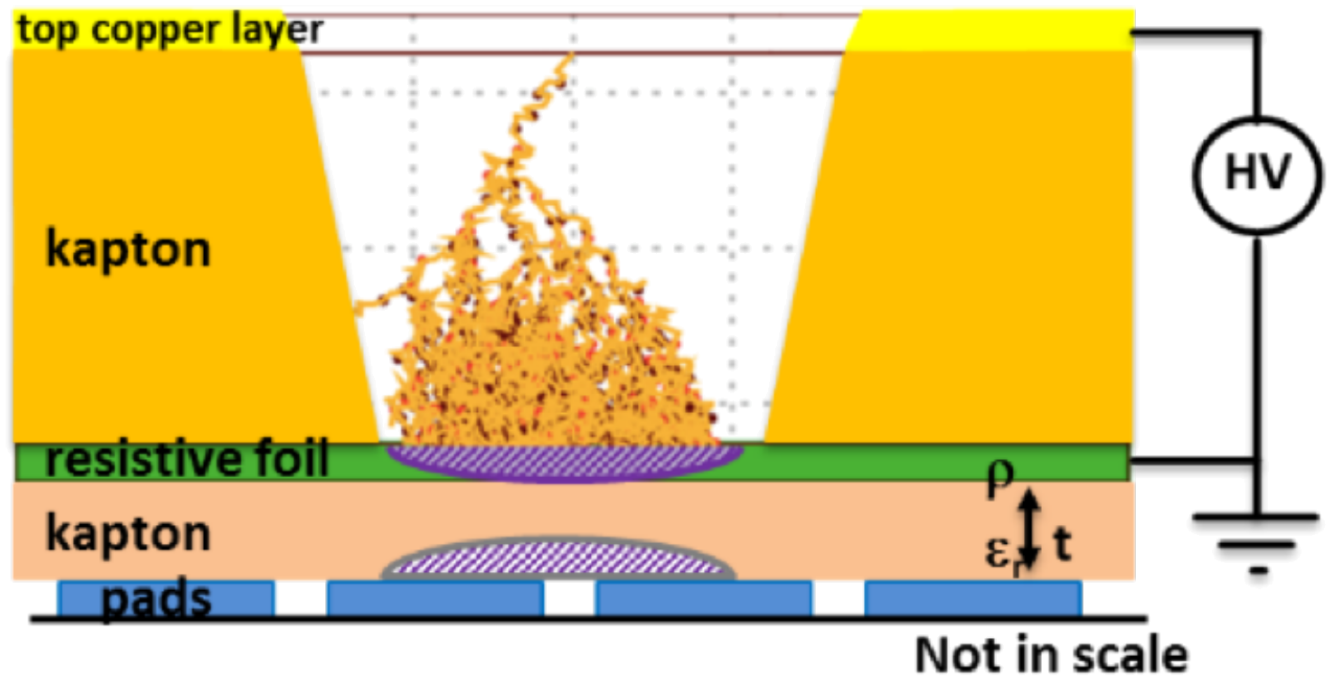

**Figure 7.3:** Amplification stage of a $\mu$-RWELL detector directly coupled to the readout PCB.

## 7.4 R&D PROGRAM

The baseline conceptual design and most promising technologies for the CEPC muon system have been discussed. Future R&D requires detailed studies of different technologies and further optimization of baseline design parameters. Several critical R&D items have been identified, including:

- **Combined optimization with ECAL and HCAL**: Study and improve muon identification performance when combining the muon detector and calorimeter information.
- **Long-lived particles optimization**: Explore new physics scenario of long-lived particles and exotic decays. Optimize detector parameters and technologies.
- **Layout and geometry optimization**: Detailed studies on the structure of the segments and modules need to be carried out to minimize the dead area and to optimize the interface for routing, support and assembly. The geometry and dimensions need to be optimized together with the inner detectors, in particular the ECAL and the HCAL.
- **Detector optimization**: Study aging effects, improve long-term reliability and stability, readout technologies.
- **Detector industrialization**: Improve massive and large area production procedures for all technologies. One example is the engineering and the following industrialization of the $\mu$-RWELL technology. The engineering of the detector essentially coincides with the technological transfer of the manufacturing process of the anode PCB and the etching of the kapton foil to suitable industrial partners.

CHAPTER 8

# READOUT ELECTRONICS, TRIGGER AND DATA ACQUISITION

The readout electronics and Data AcQuisition (DAQ) systems for the detectors at the CEPC need to operate synchronously with the circular collider time structure for beam collisions and deliver high efficiency for recording all events without compromising on rare or yet unknown physics processes. The beam conditions and time structure are adjusted to operate in three different modes, corresponding to three different center-of-mass energies ($\sqrt{s}$): Higgs factory ($e^+e^- \to ZH$) at $\sqrt{s} = 240$ GeV, $Z$ boson factory ($e^+e^- \to Z$) at $\sqrt{s} = 91.2$ GeV and $W$ threshold scan ($e^+e^- \to W^+W^-$) at $\sqrt{s} \sim 160$ GeV. The instantaneous luminosities are expected to reach $3 \times 10^{34}$, $32 \times 10^{34}$ and $10 \times 10^{34}$ cm$^{-2}$s$^{-1}$, respectively, as shown in Table 1.1. The current tentative operation plan will allow the CEPC to collect one million Higgs particles or more, close to one trillion Z boson events, and ten million $W^+W^-$ event pairs.

In conjunction with the recording of central events, the forward luminosity monitors are required to measure Bhabha scattering events to determine the delivered integrated luminosity to a relative accuracy of $0.1\%$ for the Higgs factory operation, and $10^{-4}$ for the $Z$ line shape scan. This imposes dedicated readout capabilities to maintain the high rates of the luminosity calorimeter (LumiCal) discussed in Section 9.5.

The following sections detail specifications for the on-detector front-end electronics and off-detector back-end electronics for the detector subsystems and their interface to the central DAQ, trigger, clock and control systems. The event builders provide data to the event filters to determine the final event selection and data storage.

## 8.1 TRIGGER STRATEGY

The specifics of the trigger system remain to be defined. At the conceptual stage, two approaches are being considered. The first is a standard approach where trigger primitives will be based on sub-detector data and correlations will be used. The back-end electronics of the sub-detectors will include trigger hardware that will be designed in a board frame compliant to the xTCA standard providing a high level of interconnectivity. The alternative approach is a trigger-less scheme that would provide a continuous steam of data processing, assuming a high fraction of the data is signal. Further study on the hardware trigger will be done before the technical design report and incorporated into the readout structure.

## 8.2 READOUT ELECTRONICS

The readout electronics of each detector subsystem consists of on-detector front-end electronics and off-detector back-end electronics. The front-end electronics directly receive the analog signals from the sensors. These signals are fed into Application Specific Integrated Circuits (ASIC) to produce digital signals that are further processed by configurable front-end chips, such as digital signal processors (DSP) and field-programmable gate arrays (FPGA), that format and buffer the data to be sent on the data link to the back-end electronics.

The specific details of how the analog signals are processed to yield buffered digital data varies depending on the each detector subsystem, as described in the Chapter 4 for the vertex detector and tracking systems, Chapter 5 for the electromagnetic and hadronic calorimeters, Chapter 7 for the muon systems, and Chapter 9 for the LumiCal. A common set of specifications for the readout electronics parameters are needed to ensure that the detector data from a CEPC collision collected across all subsystems can be fully assembled into a single event containing all measurements above threshold of all final state particles produced within the detector acceptance.

The most fundamental step in the readout is to provide synchronized data. This is achieved foremost by distributing phase-locked copies of the machine clock from the accelerator to the front-end systems that digitize and buffer the data. The specifications on the clock jitter depend on the level of precision required for timing measurements, where for reference the LHC clock distribution for the HL-LHC upgrade is expected to have an RMS jitter of less than 10 ps. The requirements for the digital transmission and event building are to keep the data aligned on the same clock boundary and depend on the speed of the data transmission, where transfer rates of 25 Gbps have been achieved. The standardization of the clock distribution and data links across the detector subsystems is advantageous to provide uniform performance and robust synchronization of the data links.

The second parameter common to all readout systems is the maximum latency for receiving a trigger decision to initiate the transfer of data from the front-end buffers to the back-end systems or from the back-end systems to the central DAQ, depending on where the data are buffered. The latency is set by the total transit time for the collision data used by the trigger to provide the data to trigger processors, to process a trigger decision, and to receive the trigger decision by the data buffers. Depending on the complexity of the triggers and internal response times of the detectors providing the data, the latency will be set. With the maximum latency, the required data buffering per sub-detector will depend on the occupancy of the channels served by the same readout segment. The occupancy, channel capacity and amount of data per channel, for channels above threshold, will set the average data volume per DAQ link per trigger. The buffer needs to account for fluctuations with respect to the average to avoid buffer overrun. Control signals to monitor the data buffers and back pressure are used to throttle the trigger, as needed, to avoid data loss. Most notably, the detector data occupancies per readout segment need to model well the beam background contributions in addition to the expected occupancies from collisions.

The off-detector back-end readout systems will provide the data links to the trigger processors and the central DAQ system. Current back-end designs using Advanced Telecommunications Computing Architecture (ATCA) readout platforms are able to provide a common framework for configuring the data management from different front-end systems [1, 2]. Mezzanine boards are typically implemented to allow customization of the number and types of front-end links to optimize the resources of the back-end readout boards. The ATCA readout crates support high-speed commercial data links that can directly feed commercial network switches in the central event builder.

## 8.3 DATA ACQUISITION SYSTEM

The main task of the central DAQ system is to read out data from the electronics with the level-1 trigger decision given by trigger system, then build into a full event with data fragments from different sub-detectors and process data, such as data compression and event filter. Finally, the data are sent to permanent storage.

### 8.3.1 READOUT DATA RATE ESTIMATION

The CEPC conceptual detectors proposed in this CDR include eight major types of sub-detectors: vertex detector, silicon tracker, TPC, drift chamber, ECAL, HCAL, dual-readout calorimeter and LumiCal.

Table 8.1 shows the estimated data rate of sub-detectors of CEPC. The event rate reaches $\sim$32 kHz for $Z$ factory operation from $Z$ boson decays and Bhabha events with the 2 Tesla solenoid option ($\mathcal{L}= 3.2 \times 10^{35}$ cm$^2$/s). We apply a safety factor and assume a maximum event rate of 100 kHz. TPC and drift chamber are two options for the outer tracker. AHCAL and DHCAL are two options for the PFA hadronic calorimeter, while the Dual Readout Calorimeter is a calorimeter option to cover both the ECAL and HCAL functionality. We assume a 10 $\mu$s readout window to calculate vertex and silicon tracker occupancy. The estimated rates for Bhabha events in the LumiCal detector are within the nominal event rate, however, a dedicated high-rate LumiCal data stream is envisioned to study the beam backgrounds and deliver the required luminosity uncertainty. With the level-1 trigger operating at 100 kHz, the total raw data rate is about 2 TBytes/s.

### 8.3.2 CONCEPTUAL DESIGN SCHEMA

The current LHC experiments have up to $10^8$ front-end readout channels and a maximum event building rate of 100 kHz, moving data at speeds of up to 300 GBytes/s (with an average throughput of < 200 GBytes/s required). The HL-LHC Phase-2 Upgrades reach 6000 GBytes/s and average event sizes of 7.4 MBytes [1]. The proposed CEPC DAQ system has the similar requirement in terms of data throughput. Upon the reception of the data, the computing requirements for event processing at the CEPC, in terms of storage and CPU, depend on the reconstruction times and trigger algorithms. As these algorithms are evolving, the current approach is to remain as compatible as possible with the rapidly developing technologies in the computing and network markets.

| | **Total # channels** [M($10^6$)] | **Occupancy** [%] | **Nbit /channel** | **# Channels readout/evt** [k($10^3$)] | **Volume /evt** [MBytes] | **Data rate @100 kHz** [GBytes/s] |
|---|---|---|---|---|---|---|
| Vertex | 690 | 0.3 | 32 | 2070 | 8.3 | 830 |
| Silicon Tracker | | | | | | |
| Barrel | 3238 | $0.01 \sim 1.6$ | 32 | 1508 | 3.15 | 315 |
| Endcap | 1238 | $0.01 \sim 0.8$ | 32 | 232 | 0.4 | 40 |
| TPC | 2 | 0.1-8 | 30 | 1375 | 5 | 500 |
| Drift Chamber | 0.056 | 5-10 | 480 | | 3 | 300 |
| ECAL | | | | | | |
| Barrel | 17/7.7 | 0.17 | 32 | 28.8/13.1 | 0.117/0.053 | 11.7/5.3 |
| Endcap | 7.3/3.3 | 0.31 | 32 | 22.4/10.2 | 0.090/0.041 | 9.0/4.1 |
| AHCAL | | | | | | |
| Barrel | 3.6 | 0.02 | 32 | 0.72 | 0.0029 | 0.3 |
| Endcap | 3.1 | 0.12 | 32 | 3.72 | 0.015 | 1.5 |
| DHCAL | | | | | | |
| Barrel | 32 | 0.004 | 8 | 1.28 | 0.00128 | 0.13 |
| Endcap | 32 | 0.01 | 8 | 3.2 | 0.0032 | 0.32 |
| Dual Readout | | | | | | |
| Calorimeter | 22 | 0.4-1.6 | 64 | 88-352 | 0.704-2.8 | 70-280 |
| Muon | | | | | | |
| Barrel | 4.9 | 0.0002 | 24 | 0.01 | < 0.0001 | < 0.01 |
| Endcap | 4.6 | 0.0002 | 24 | 0.01 | < 0.0001 | < 0.01 |
| LumiCal | 0.5 | 0.2 | 12 | 0.5 | 0.0007 | 0.07 |

**Table 8.1:** CEPC DAQ data rate estimation. TPC and drift chamber are options for the outer tracker. AHCAL and DHCAL are two options for the PFA hadronic calorimeter, while the Dual Readout Calorimeter is a calorimeter option to cover both the ECAL and HCAL functionality. With the level-1 trigger operating at 100 kHz, the total raw data rate is 2 TBytes/s.

Figure 8.1 shows the conceptual software architecture design of the CEPC DAQ based on the experience gained from BES III and Daya Bay experiments. The DAQ system is connected with the sub-detector back-end electronics in counting rooms through commercial network switches with the TCP/IP protocol. For 2 TBytes/s data readout requirement, it need about 1600 10 Gbit or 640 25 Gbit network links. All other DAQ devices are deployed in a dedicated machine room. Event building will be performed on the online farm connected to the back-end electronics via network switches. An event filter will also run on an online farm. Each node of the online farm will process the data of one complete event at a time. The purpose of the online event processing will mainly be event classification, data quality monitoring and online filtering to reduce background events. The DAQ system will provide other common functions including run control, run monitoring, information sharing, distributed process manager, software configure, Elog, data quality monitoring, remote monitoring and so on.

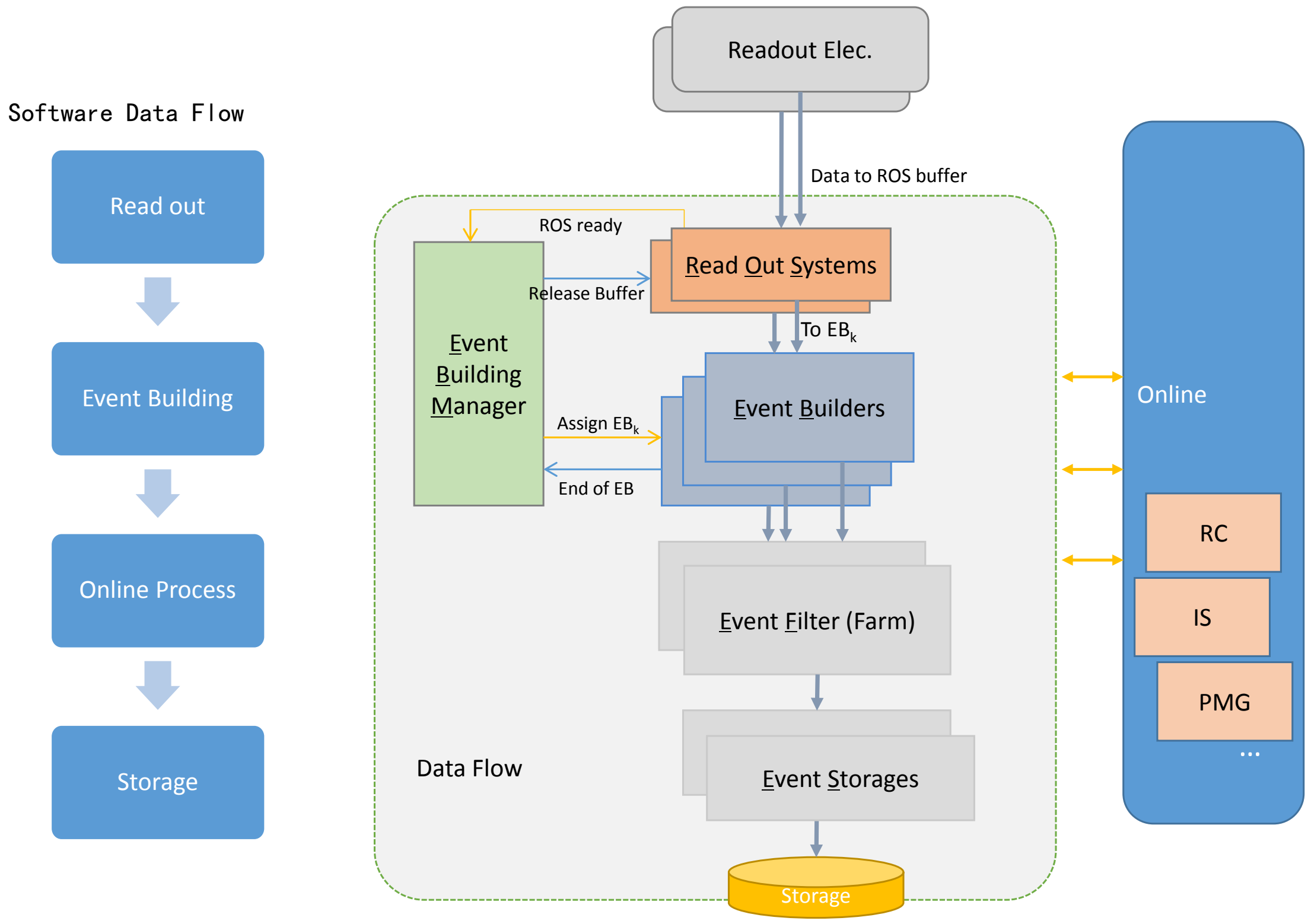

**Figure 8.1:** DAQ Conceptual Software Architecture Design Diagram. ROS readout data from electronics and send data to EB with EBM by data driven. After event building the event filter or software trigger will be processed at EF farm. Then the passed events will be stored at storage nodes.

There are two levels of event building in the conceptual design data flow. The first level is implemented in the readout farm which reads out the data from the back-end electronics and builds into a data fragment. The second level is implemented in the online farm which reads out the data from readout farms and builds into a full event. The two levels of event building could be implemented as follows:

1. Electronics boards send data to the Read Out System (ROS) through the network.
2. ROS receives all data slices of one event and sends event ID to the Event Building Manager (EBM).
3. EBM assigns the event ID to a free Event Builder (EB) node when EBM gets all the data fragments corresponding to a particular event ID from the ROSs.
4. EB sends data request to each ROS.
5. ROSs send requested data to EB.
6. EB receives all ROSs data fragments of one event and finishes full event building, then sends event ID back to EBM.
7. EBM send event ID to ROSs to clear data buffer.

A software trigger can be deployed in the event filter farm. Each event filter node requests a full event from EBs, then sends the event data to process tasks to analyze for software trigger and data quality monitoring, and then in the last step sends triggered event to event storage nodes.

## CHAPTER 9

# MACHINE DETECTOR INTERFACE AND LUMINOSITY DETECTORS

The MDI represents one of the most challenging tasks for the CEPC project. In general, it will have to address all common issues relevant to both the machine and detector. Topics summarized in this chapter include the interaction region, the final focusing magnets, the beam pipe, the detector radiation backgrounds and the luminosity instrumentation. Integration of all the machine and detector components in the interaction region is also briefly discussed. It is critical to achieve comprehensive understanding of MDI issues to assure the optimal performance of the machine and detector.

## 9.1 INTERACTION REGION

The Interaction Region (IR) is where both electron and positron beams are focused to small spot sizes at the interaction point (IP) to maximize the machine luminosity. The two storage rings are merged and subsequently separated again. The IR layout, as illustrated in Figure 9.1, has received several updates with respect to the preliminary CDR [1], to cope with the latest double-ring design and a beam-crossing angle of 33 mrad. The two final focusing magnets, QD0 and QF1, sit inside the detector. The focal length $(L^{*})$, defined as the distance from the final focusing magnet (*i.e.,* QD0) to the IP, has increased from 1.5 m to 2.2 m. This allows increased separation between the two single apertures of the QD0. Compensating magnets are positioned in front of the QD0 and surrounding both the QD0 and QF1 magnets. They are introduced to cancel out the detector solenoid field and minimize the coupling between horizontal and vertical betatron motion. Furthermore, the outer radius of the compensating magnets defines the detector acceptance to be $|\cos\theta| \leq 0.993$. The luminosity calorimeter (so called "LumiCal"), located right in front of the compensating magnets, is designed to measure the integrated luminosity to a precision of $10^{-3}$ or better. Tracking disks, labeled as FTD, are designed to measure charged particle trajectories in the forward region.

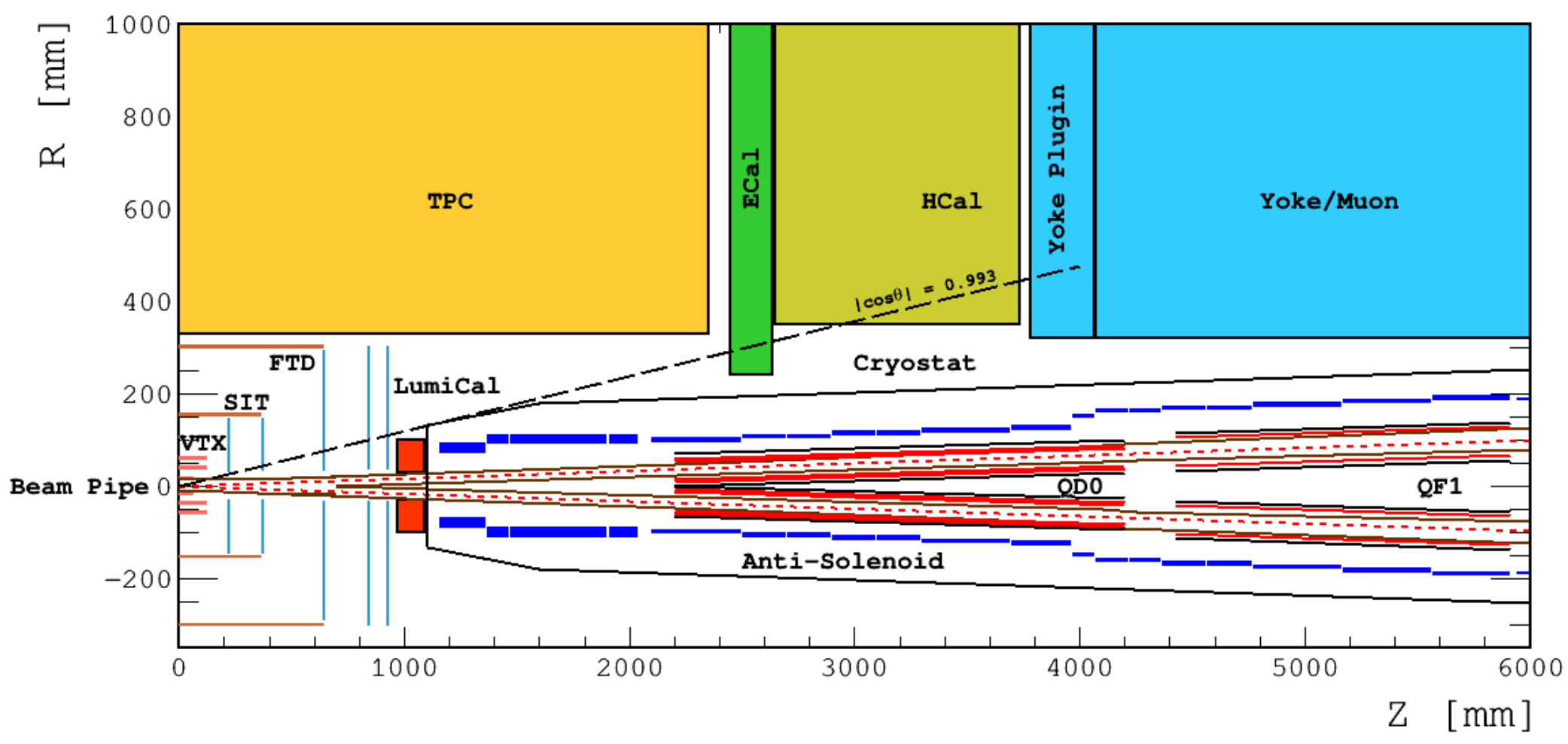

**Figure 9.1:** Layout of the CEPC interaction region. The two beam pipes merge into one at $|z| = 70$ cm, with the central part between $z = \pm 7$ cm made with Beryllium. The two final focusing magnets (QD0 and QD1) are surrounded with the anti-solenoid magnets ssegmented into 22 sections. The magnets are placed inside the cryostat. The LumiCal (red) sitting in front of the cryostat provides precise luminosity measurement. Silicon tracking detectors, VTX and SIT, are in the barrel region, while FTD disks are covering the forward region.

## 9.2 FINAL FOCUSING MAGNETS

In the interaction region, compact high gradient quadrupole magnets are designed to focus the electron and positron beams. The two final focusing quadrupoles (QD0 and QF1), are placed inside the CEPC detector and must operate in the background field of the detector solenoid. QD0 is the quadrupole magnet close to the interaction point, with a distance of 2.2 m to the IP. It is designed as a double aperture superconducting magnet and can be realized with two layers of Cos-Theta quadrupole coil using NbTi Rutherford cables without iron yoke. It is designed to deliver a gradient field of 136 T/m and control the field harmonics in the sensitive area to be below $3 \times 10^{-4}$. Design parameters are summarized in Table 9.1. The QF1 magnet is similar to the QD0, except that there is an iron yoke around the quadrupole coil for the QF1.

Additional anti-solenoid magnets are introduced to minimize the disturbance from the detector solenoid on the incoming and outgoing beams. The anti-solenoid sections in front of the QD0 are designed to achieve an almost zero integral longitudinal field before entering the QD0. And the sections right surrounding the QD0 and QF1 are necessary to screen the detector field. The magnets are based on wound of rectangular NbTi-Cu conductors. To minimize the magnet size and better field quality, the anti-solenoid magnets are segmented into 22 sections with different inner coil diameters. Inside the first section, the central field reaches the peak value of 7.2 Tesla. More detailed design of the final focusing magnets and the compensating magnets can be found in [2].

| Magnet | QD0 | QF1 |
|---|---|---|
| Field gradient [T/m] | 136 | 110 |
| Magnetic length [m] | 2.0 | 1.48 |
| Coil turns per pole | 23 | 29 |
| Excitation current [A] | 2510 | 2250 |
| Coil layers | 2 | 2 |
| Stored energy [kJ] | 25.0 | 30.5 |
| Inductance [H] | 0.008 | 0.012 |
| Peak field in coil [T] | 3.3 | 3.8 |
| Coil inner diameter [mm] | 40 | 56 |
| Coil outer diameter [mm] | 53 | 69 |
| X direction Lorentz force/octant [kN] in [0, 45°] | +68 | +110 |
| Y direction Lorentz force/octant [kN] in [0, 45°] | -140 | -120 |

**Table 9.1:** Main design parameters of the two final focusing magnets, QD0 and QF1.

## 9.3 BEAM PIPE

The beam pipe design foresees several constraints. In the central region ($z = \pm 7$ cm or longer), its radius should be small enough ($r = 1.4$ cm) to allow precise measurement of track impact parameters, but still large enough not to interfere with the beam backgrounds. It shall be made with Beryllium to reduce photon conversions and hadronic interactions, but has to be rigid enough (sufficient wall thickness of $\sim$500 $\mu$m) to withstand the high vacuum pressure. In the forward region, the beam pipe opens conically away from the IP to allow enough space for the beam-induced backgrounds and splits into two pipes at $|z| = 70$ cm. Beam pipe in the forward region can be built with stainless steel or copper. Bellows required for installation are located at about $|z| = 70$ cm. Preliminary estimation shows there will be non-negligible High Order Mode (HOM) heat generated in the central region. Additional HOM absorbers and active cooling need to be deployed to control the HOM heat. For the beam pipe extending into the final quadrupoles, a room temperature beam pipe has been chosen because of the 4 mm gap between the beam pipe and the Helium vessel.

## 9.4 DETECTOR BACKGROUNDS

Beam and machine induced radiation backgrounds can be the primary concern for the detector design [3–6]. They can cause various radiation damages to the detectors and electronic components, and degrade the detection performance or even kill the detector completely in the extreme case. During data-taking, high rate radiation backgrounds may significantly increase the detector occupancy and impair the data-taking capability of the detector. Therefore it is always desirable to characterize the potential backgrounds at the machine and detector design stage and mitigate their impacts with effective measures. Detailed Monte Carlo simulation, along with lessons and experience learned from other experiments, can serve as the basis for such studies.

The deleterious effects of the radiation backgrounds can be represented with hit density, total ionizing dose (TID), and Non-Ionizing Energy Loss (NIEL). The expected hit density can be used to evaluate the detector occupancy. TID is an important quantity for understanding surface damage effects in electronics. NIEL, represented in the 1 MeV neutron equivalent fluence, is important for understanding the bulk damage to silicon devices. The background simulation starts with either generating background particles directly in the IR (e.g., pair production) or propagating them to the region close enough to the IR (e.g., SR photons and off-energy beam particles). Particle interactions with detector components are simulated with GEANT4 [7–9]. The characterization methodology for the ATLAS detector background estimation [10] has been adopted. In the following, main radiation backgrounds originating from synchrotron radiation, beam-beam interactions, and off-energy beam particles, are discussed and their contributions are carefully evaluated. Safety factors of ten are always applied to cope with the uncertainties on event generation and detector simulation. More complicated radiation backgrounds, e.g., offset during beam injection and beam particles significantly off the design orbit, are subject to future studies together with the evolving machine design.

### 9.4.1 SYNCHROTRON RADIATION

Synchrotron radiation (SR) photons are prevalent at circular machines. At the CEPC, they are mostly produced in the last bending dipole magnets and in the focusing quadrupoles inside the interaction region. The innermost tracking detectors can be sensitive to photons above 10 keV and vulnerable to high levels of soft photon radiation. [1] In order to reduce the energy and flux of SR photons that enter the straight sections, the field strength of the last bending dipole magnet has been reduced and becomes much weaker than the normal arc dipole fields. This controls the critical energy of SR photons to be below 100 keV and makes the collimation design less difficult.

The BDSim [11] software based on GEANT4 has been deployed for the detailed studies. It generates SR photons from the last bending dipole magnet and transports both beam particles and synchrotron radiation photons to the central detector region. Both the accelerator and detector geometries are implemented in simulation. Particular care has been taken for a realistic simulation in the tails of the beam density distributions (up to 10 $\sigma_{x/y}$) and for both beam core and halo, as particles form the tails are most effective in producing background particles and interacting with the beam pipe. Additional sampling techniques are deployed to improve the computation efficiency.

As shown in Figure 9.3, SR photons from the last dipole magnet form the light yellow band and can reach the beam pipe in the interaction region. A considerable amount of them are scattered and can then hit the central Beryllium beam pipe between $z = \pm 7$ cm as shown in Figure 9.3(a). They will enter the central detector and increase significantly the detector occupancy. It is necessary to introduce collimators made with high-$Z$ materials

[1] It should be noted that the SR photon energy increases rapidly with the beam energy and additional measures might have to be introduced to allow detector operation at higher operation energies.

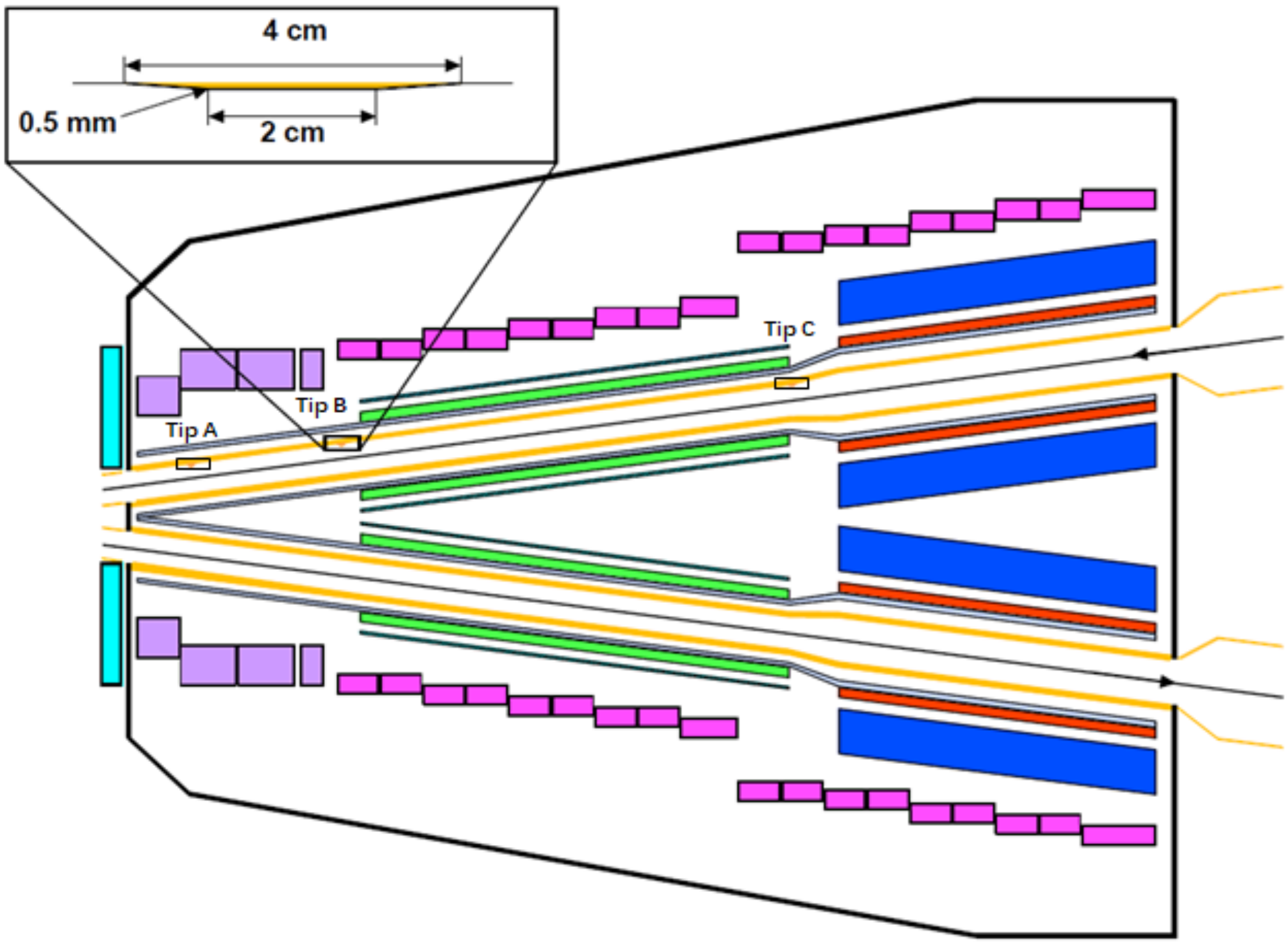

**Figure 9.2:** Three sets of mask tips (Tip A, B and C) located at $|z| = 1.51$, 1.93 and 4.2 m along the incoming beam pipe to the interaction point, are introduced to suppress the scattered SR photons. The mask tip, made of high-$Z$ material, is 2 cm and 4 cm wide at its inner and outer radii, respectively and only 0.5 mm thick as depicted for Tip B at $|z|$=1.51 m.

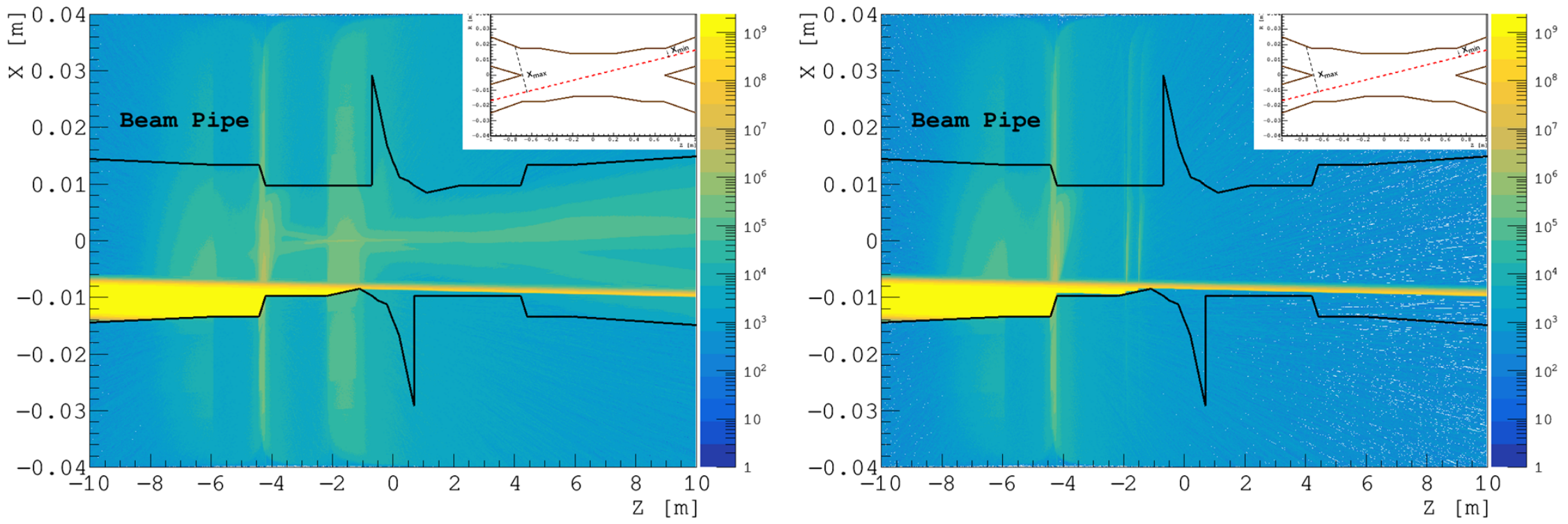

(a) (b)

**Figure 9.3:** Illustration of the synchrotron photon flux formed by the upstream bending magnet on the left side and the photons scattered by the beam pipe before (a) and after (b) introducing the mask tips. The black lines indicate the beam pipe wall seen from the central beam orbit and reaches the maximum at close to the two positions ($z = \pm 70$ cm), where the single beam pipe splits into two beam pipes.

(e.g., Tungsten), with particular shapes to block those direct and scattered photons. As shown in Figure 9.2, three sets of mask tips, located at $|z| = 1.51$, 1.93 and 4.2 m along the beam pipe to the IP and as thin as 0.5 mm, are designed to suppress such SR photons. They can reduce the number of SR photons hitting the central beam pipe from nearly $40,000$ to below 80. This significant reduction leads to a much lower power deposition in the beam pipe and shall ease the cooling design for the beam pipe. The resulting photon flux distribution after collimation is shown in Figure 9.3(b). SR photons generated in the final focusing magnets are also carefully evaluated. According to the simulation results, they travel in the highly forward region and do not strike directly the central beam pipe unless the particles are 40 $\sigma_x$ off the central orbit. It should be pointed out that beam particles may be off orbit before entering the last bending dipole magnets. This can lead to additional SR photons hitting the central beam pipe and needs to be carefully evaluated with better understanding of the machine performance.

| **Machine Parameters** | **H** (240 GeV) | **W** (160 GeV) | **Z** (91 GeV) |
|---|---|---|---|
| Beam energy [GeV] | 120 | 80 | 45.5 |
| Particles per bunch [$10^{10}$] | 15 | 12 | 8 |
| Transverse size $\sigma_x/\sigma_y$ [ µm] | 20.9/0.068 | 13.9/0.049 | 6.0/0.078 |
| Bunch length $\sigma_z$ [ µm] | 3260 | 5900 | 8500 |
| Emittance $\varepsilon_x/\varepsilon_y$ [nm] | 12.1/0.0031 | 0.54/0.0016 | 0.18/0.004 |

**Table 9.2:** The input parameters to the GUINEA-PIG for the pair production simulation for the machine operations at $\sqrt{s}$ = 240, 160 and 91 GeV.

### 9.4.2 BEAM-BEAM INTERACTIONS

Beamstrahlung and its subsequent process of pair production ($\gamma\gamma \to e^+e^-$) give rise to important radiation backgrounds at the CEPC. Due to the pinch effect in the beam-beam interaction, the trajectories of beam particles in the bunches are bent, which causes the emission of beamstrahlung photons. This process can be studied with the Monte Carlo simulation program GUINEA-PIG [12], which takes into account dynamically changing bunch effects, reduced particle energies and their impacts on the electric and magnetic fields. The simulation program has been customized to implement the external detector solenoid field for the charged particle tracking. This allows improved determination of the positions and momenta of the out-going charged particles before interfacing to the GEANT4 detector simulation. Machine parameters for operation at different energies are listed in Table 9.2, and they serve as the input to the GUINEA-PIG simulation. It should be noted that compared to other processes, electron-positron pair production generates most significant detector backgrounds. The processes can be categorized as:

- *Coherent Production*: $e^+e^-$ pairs are produced via the interaction of virtual or real photons (e.g., beamstrahlung photons) with the coherent field of the oncoming bunch. Particles can be highly energetic but are dominantly produced with small angle and confined in the beam pipe. Its contribution to the detector backgrounds is negligible.

- *Incoherent Production*: $e^+e^-$ pairs are produced through interactions involving two real and/or virtual photons. Most of the particles are confined in the beam pipe by the strong detector solenoid field. However, a small fraction of them are produced with high transverse momentum and large polar angle. The incoherent production dominates the contribution to the detector backgrounds.

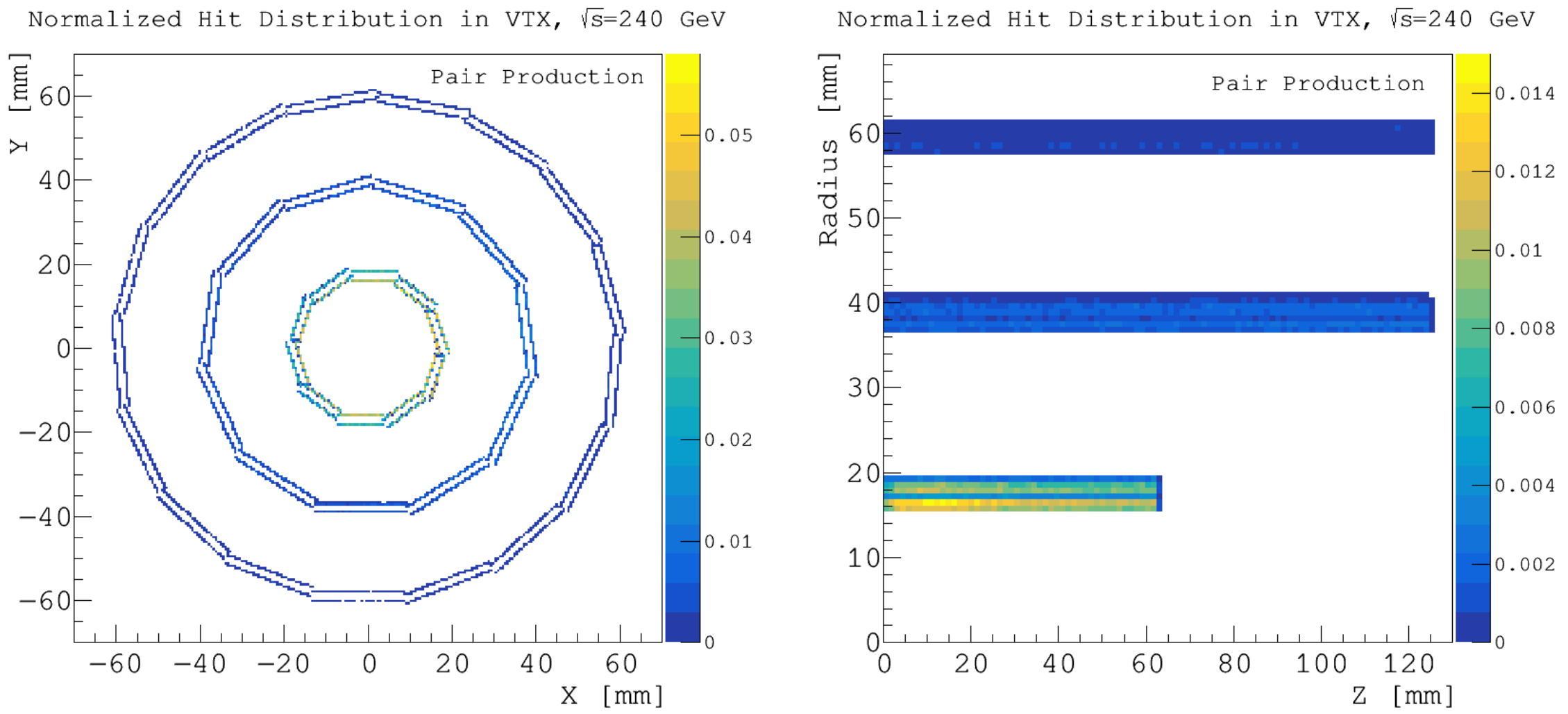

(a) (b)

**Figure 9.4:** Hit distributions due to pair production alone in the $x-y$ and $r-z$ planes of the vertex detector for the machine operation at $\sqrt{s}=240$ GeV. The incoherent production dominates detector backgrounds.

Particles interacting with the detector components are simulated with GEANT4. As shown in Figure 9.4(a), the resulting hit distribution is nearly uniform in the $\phi-$ direction, even though the beam squeezing is different in the $x$ and $y$ directions. On the other hand, the hit distribution is more dense in the central region as shown in Figure 9.4(b), but decreases rapidly with the increased radius, as shown in Table 9.3. At the inner most vertex detector layer ($r$ =1.6 cm), the hit density reaches the maximum of 2.2 hits/cm$^2$·BX when operating at $\sqrt{s}=240$ GeV. The corresponding TID and NIEL are about 620 kRad/year and 1.2×10$^{12}$ 1 MeV $n_{eq}$/cm$^2$·year, respectively.

| | **Hit Density** [hits/cm$^2$·BX] | **TID** [kRad/year] | **NIEL** [1 MeV $n_{eq}$/ cm$^2$·year] |
|---|---|---|---|
| Layer 1 ($r=1.6$ cm) | 2.2 | 620 | $1.2\times10^{12}$ |
| Layer 2 ($r=1.8$ cm) | 1.5 | 480 | $9.1\times10^{11}$ |
| Layer 3 ($r=3.7$ cm) | 0.18 | 60 | $1.2\times10^{11}$ |
| Layer 4 ($r=3.9$ cm) | 0.15 | 45 | $1.0\times10^{11}$ |
| Layer 5 ($r=5.8$ cm) | 0.03 | 9.7 | $3.3\times10^{10}$ |
| Layer 6 ($r=6.0$ cm) | 0.02 | 6.8 | $3.0\times10^{10}$ |

**Table 9.3:** Maximum hit density, total ionizing dose (TID) and non-ionizing energy loss (NIEL) due to pair production ($\gamma\gamma\rightarrow e^+e^-$) alone at each vertex detector layer for the operation at $\sqrt{s}=240$ GeV.

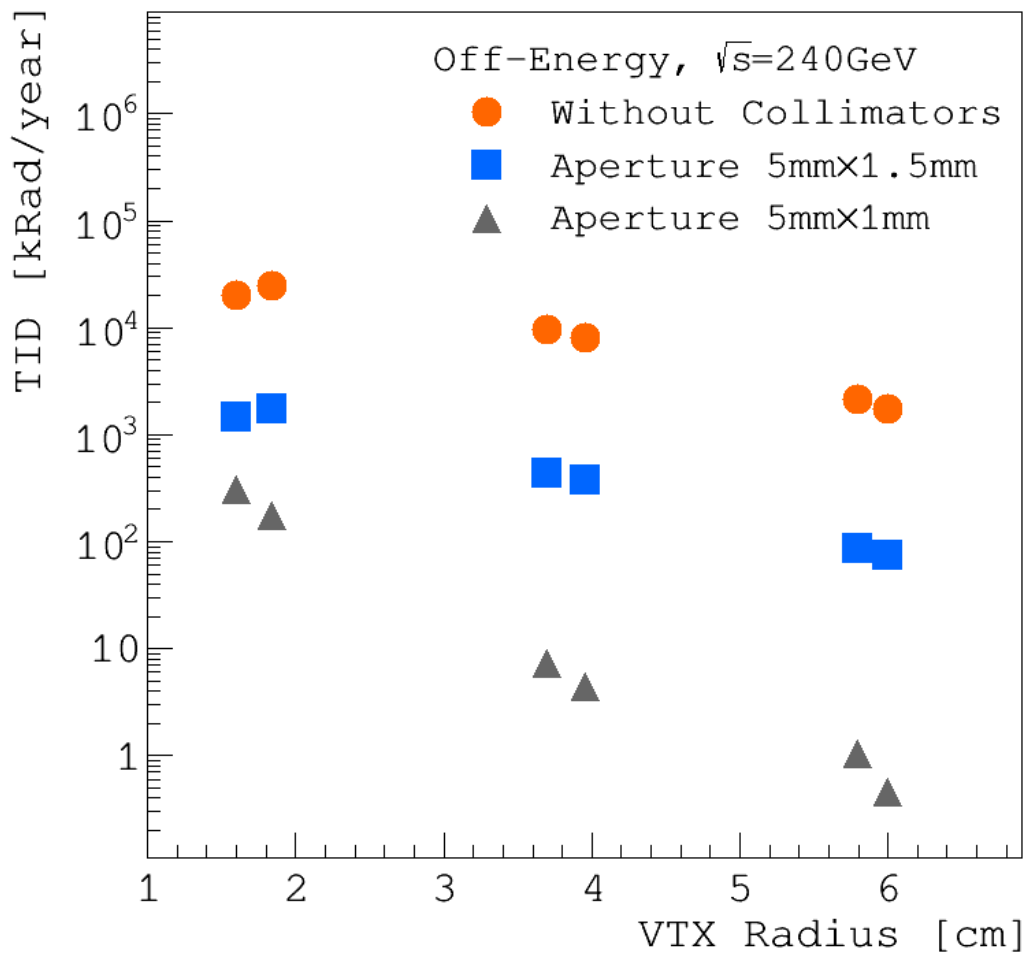

(a)

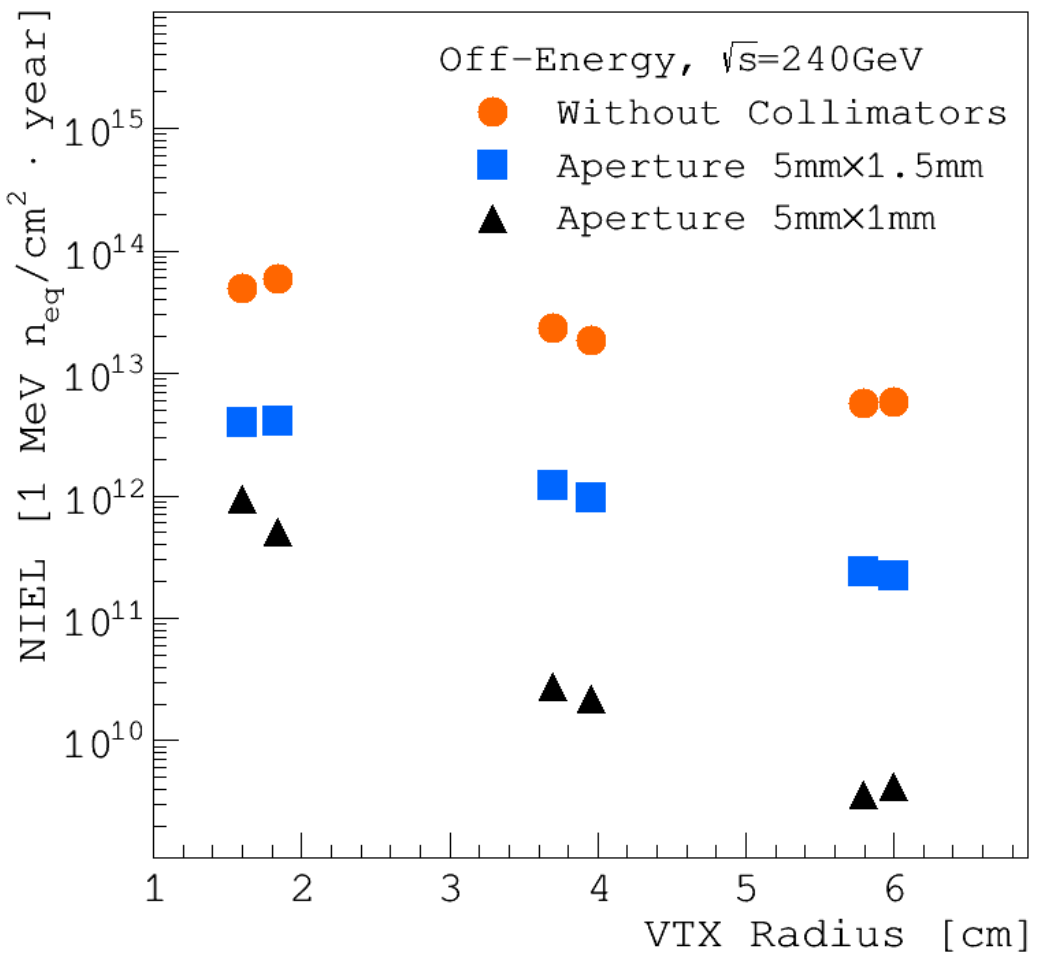

(b)

**Figure 9.5:** Total ionizing dose (TID) (a) and and non-ionizing energy loss (NIEL) (b) caused by off-energy beam particles at each vertex detector layer are effectively reduced after introducing the two sets of collimators.

### 9.4.3 OFF-ENERGY BEAM PARTICLES

Circulating beam particles can lose significant amounts of energy in scattering processes. If exceeding 1.5% of the nominal energy (defined as the machine energy acceptance), scattered particles can be kicked off their orbit. A fraction of them will get lost close to or in the interaction region. They can interact with machine and/or detector components and contribute to the radiation backgrounds. There are three main scattering processes that are almost entirely responsible for the losses of beam particles, including beamstrahlung, radiative Bhabha scattering and beam-gas interaction.

Beamstrahlung events out of beam-beam interactions are generated with GUINEA-PIG. Radiative Bhabha events with small angles are generated with the BBBREM program [13]. Interactions between the beam particles and the residual gas in the beam pipe are simulated with custom code, assuming the gas pressure to be $10^{-7}$ mbar. The backgrounds originating from the beam-gas interaction is much smaller compared to that from the Radiative Bhabha scattering. Beam particles after interactions are tracked with SAD [14] and transported to the interaction region. Particles lost close to the interaction region, either right after the bunch crossing or after traveling multiple turns (simulated up to 40 turns), are interfaced to detector simulation.

Backgrounds introduced by the off-energy beam particles can be effectively suppressed with proper collimation. The collimator aperture has to be small enough to stop as much as possible the off-energy beam particles, but must be sufficiently large without disturbing the beam. Four collimators are deployed in the design. APTX1 and APTX2 , with an aperture size of 5 mm, are placed in the horizontal plane, and APTY1 and APTY2, with an aperture size of 1 mm, are placed in the vertical plane. All the four collimators are located in the upstream of the IP, in the range between 1700 m and 2300 m. The aperture sizes are chosen to be equivalent to 14 $\sigma_x$ and 39 $\sigma_y$, for being sufficiently away from the beam clearance region. Figure 9.5 shows detector backgrounds from the off-energy beam particles are reduced significantly after introducing the collimation system. As shown in Figure 9.6(a), the resulting hit distribution is maximized towards the $-x$ direction due to the nature of the off-energy beam particles that are swept away by the magnets. But along the $z$ direction, the hit distribution is more or less uniform with the additional contribution of the back-scattered particles by the LumiCal downstream. Although the exact locations and shapes of the collimators are subject to further optimization as the machine design evolves, the current design demonstrates the feasibility to control such specific background to a sufficiently low level. For background estimation, the maximum values in the $-x$ direction are taken. At the first vertex detector layer ($r = 1.6$ cm), the hit density is about 0.22 hits/cm$^2$ per bunch crossing from radiative Bhabha scattering events alone. The TID and NIEL are 310 kRad per year and $9.3 \times 10^{11}$ 1 MeV $n_{eq}$/cm$^2$ per year, respectively. Although the hit density due to radiative Bhabha scattering is 1/10 of that from pair production, the corresponding TID and NIEL become comparable to those introduced by pair production. Additional detector shielding might have to been considered to suppress further this particular type of radiation background.

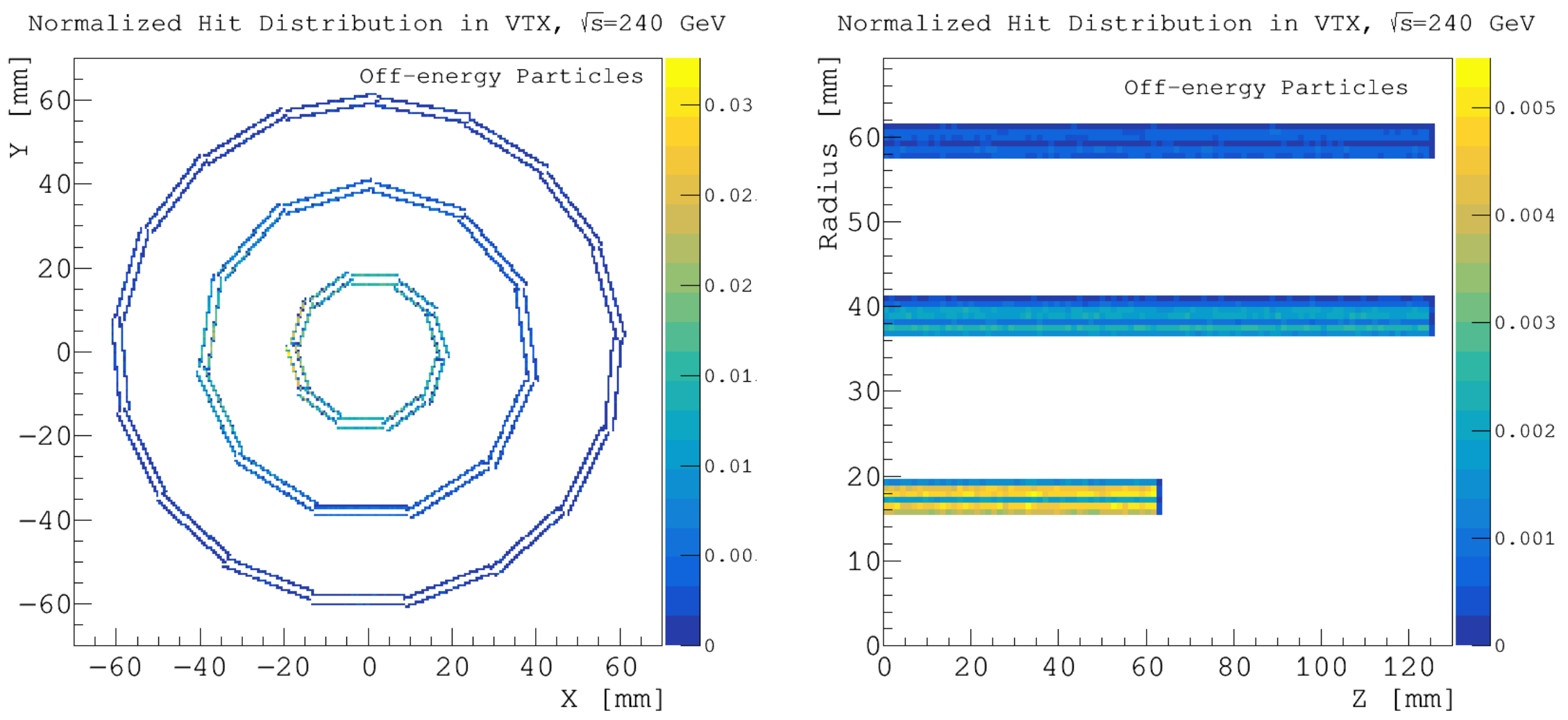

(a) (b)

**Figure 9.6:** Hit distributions due to the radiative Bhabha scattering process in the $x - y$ and $r - z$ planes of the vertex detector for the machine operation at $\sqrt{s} = 240$ GeV.

### 9.4.4 SUMMARY OF RADIATION BACKGROUNDS

When operating the machine at the center-of-mass energy of $\sqrt{s} = 240$ GeV, the main detector backgrounds come from pair-production. The contribution from the off-energy beam particles is nearly an order of magnitude lower. Figure 9.7 shows the hit density, TID and NIEL at different vertex detector layers, originating from pair production, off-energy beam particles and the two combined. In addition, TID and NIEL distributions covering the silicon detectors in $r - z$ are shown in Figure 9.8.

At lower operation energies, *i.e.*, $\sqrt{s} = 160$ GeV for $W$ and $\sqrt{s} = 91$ GeV for $Z$, the background particles are usually produced with lower energies but with higher rates given the higher machine luminosities. In addition, the pair-production dominates the radiation backgrounds and contributions from other sources become negligible. The resulting radiation backgrounds at the first vertex detector layer at different operation energies are summarized in Table 9.4.

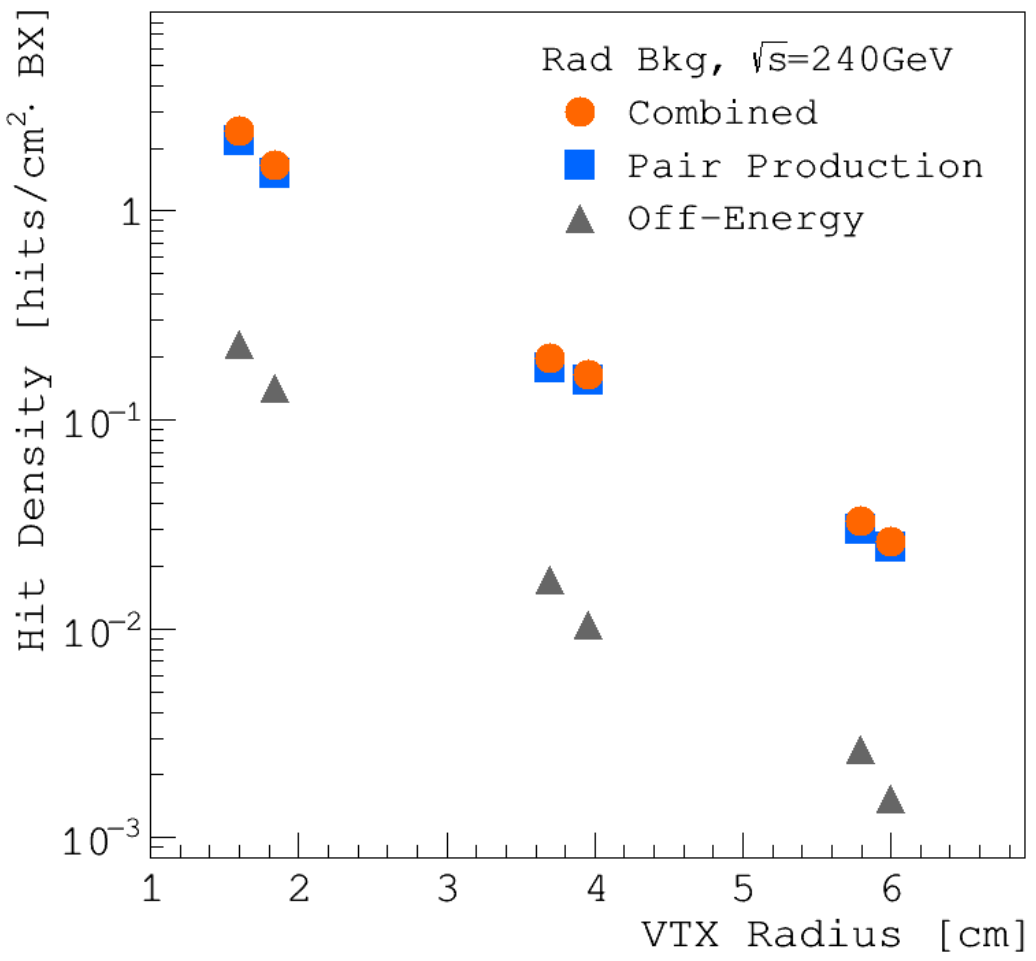

(a)

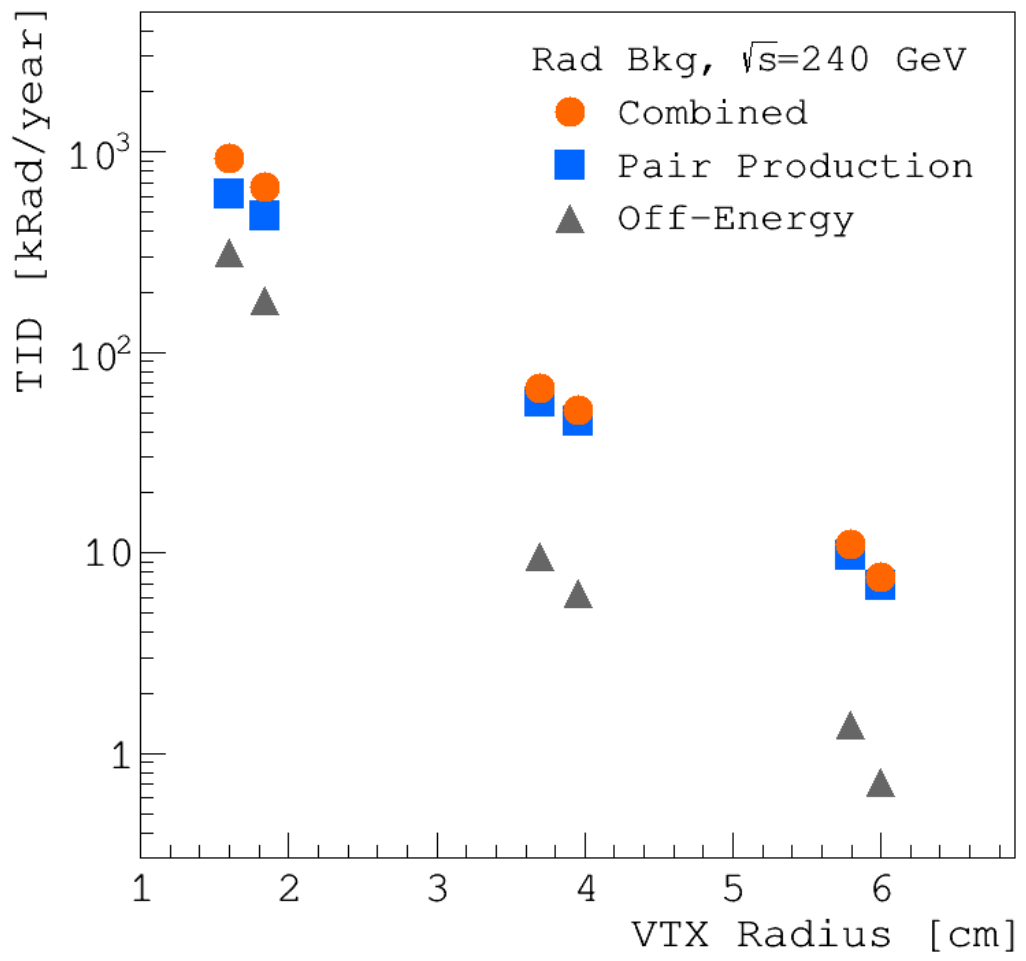

(b)

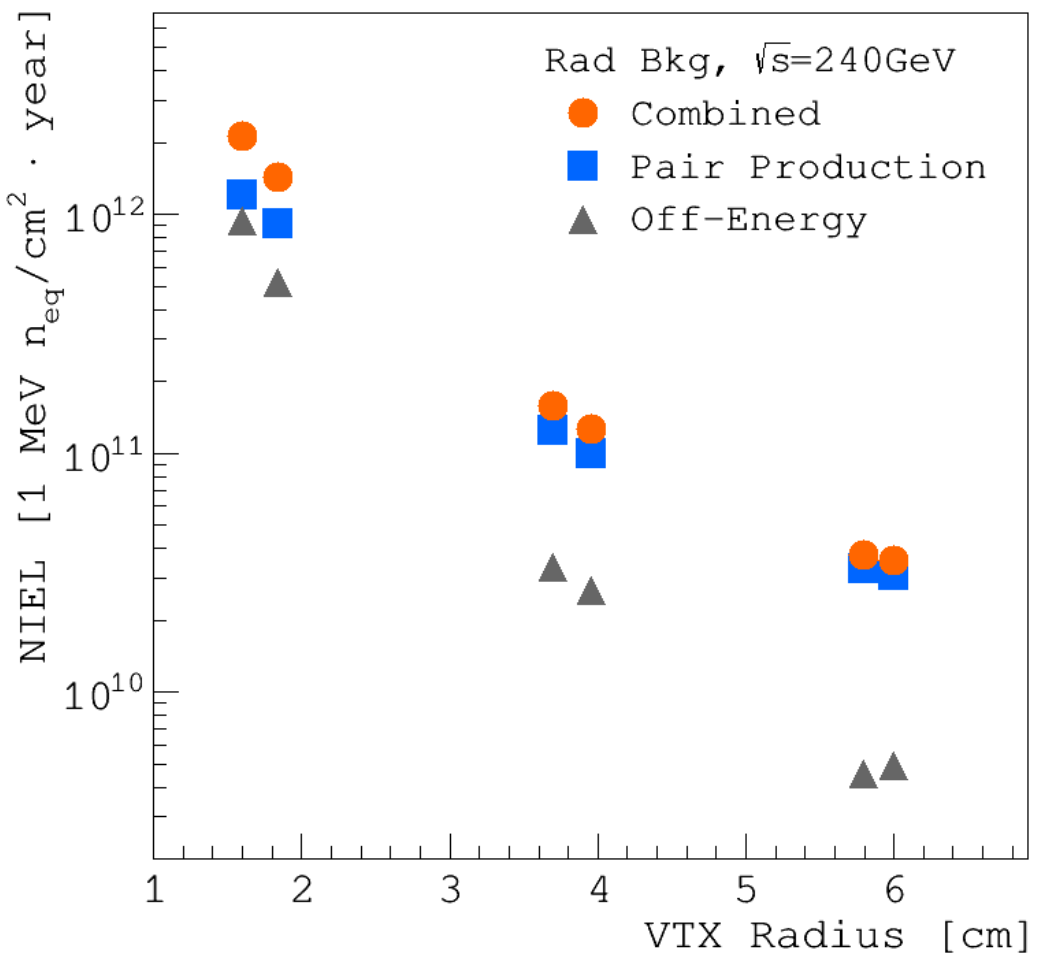

(c)

**Figure 9.7:** Hit density, total ionizing dose (TID) and non-ionizing energy loss (NIEL) at different vertex detector layers due to pair production, off-energy beam particles and the two combined for the machine operation at $\sqrt{s} = 240$ GeV.

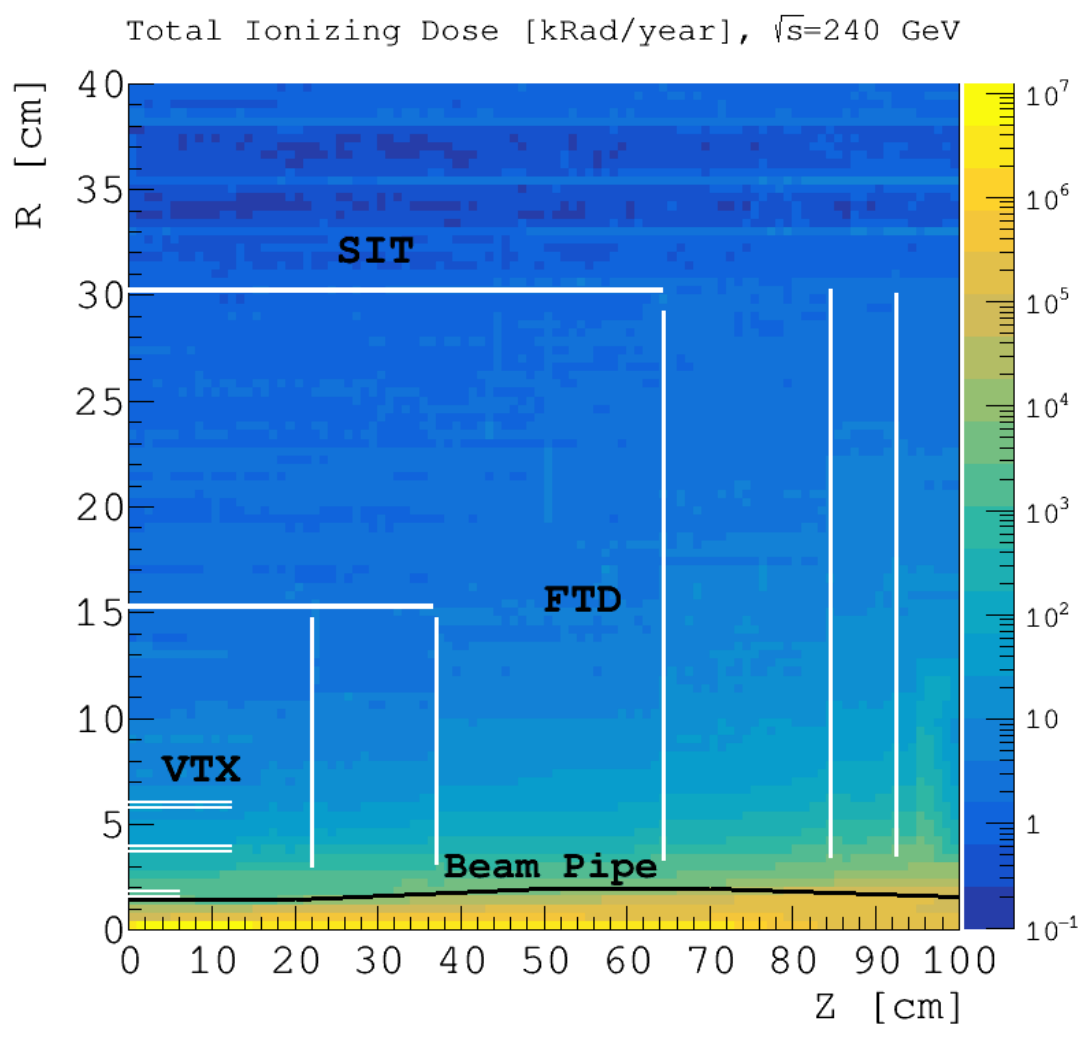

(a)

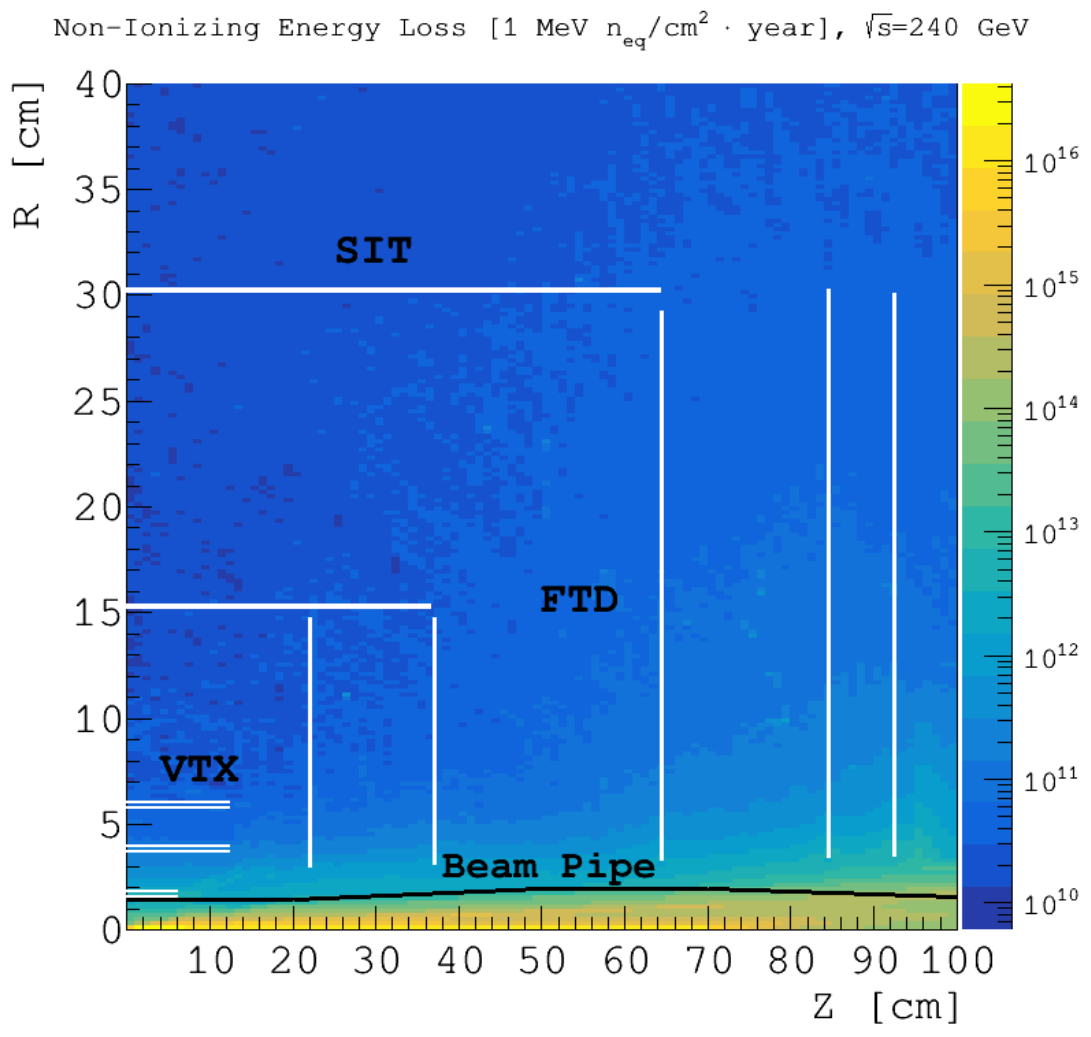

(b)

**Figure 9.8:** Total ionizing dose (TID) and non-ionizing energy loss (NIEL) distribution in $r - z$ for the machine operation at $\sqrt{s} = 240$ GeV. The white lines indicate the locations of the vertex detector (VTX), the forward tracking disks (FTD) and the silicon inner tracker (SIT).

| | **H** (240) | **W** (160) | **Z** (91) |
|---|---|---|---|
| Hit Density [hits/cm$^2$·BX] | 2.4 | 2.3 | 0.25 |
| TID [MRad/year] | 0.93 | 2.9 | 3.4 |
| NIEL [$10^{12}$ 1 MeV $n_{eq}$/cm$^2$·year] | 2.1 | 5.5 | 6.2 |

**Table 9.4:** Summary of hit density, total ionizing dose (TID) and non-ionizing energy loss (NIEL) with combined contributions from pair production and off-energy beam particles, at the first vertex detector layer ($r = 1.6$ cm) at different machine operation energies of $\sqrt{s} = 240$, 160 and 91 GeV, respectively.

## 9.5 LUMINOSITY INSTRUMENTATION

The forward region of the CEPC detector will be instrumented with a luminometer (LumiCal), aiming to measure integrated luminosity with a precision of $10^{-3}$ and $10^{-4}$ in $e^+e^-$collisions for Higgs production at the center-of-mass energy of 240 GeV, and for studies of the Standard Model at the $Z$ pole and $WW$ production energies, respectively. The precision requirements on the integrated luminosity measurement are motivated by the CEPC physics program, intended to test the validity scale of the Standard Model through precision measurements in the Higgs and the electroweak sectors with $10^6$ Higgs and $10^{10-12}$ $Z$ bosons. Many sensitive observables such as $\Gamma_Z$ and $\sin\theta_{eff}$ depend on the uncertainty of the integrated luminosity.

Luminosity at an $e^+e^-$ collider is best measured by counting Bhabha events of elastic $e^+e^-$ scattering. Its theoretical uncertainty is better than 0.05% at the $Z$ pole [15]. The scattered electrons are distributed in the forward direction with a $1/\theta^3$ dependence. The cross section of the BHLUMI [16] simulation is illustrated in Figure 9.9(a).

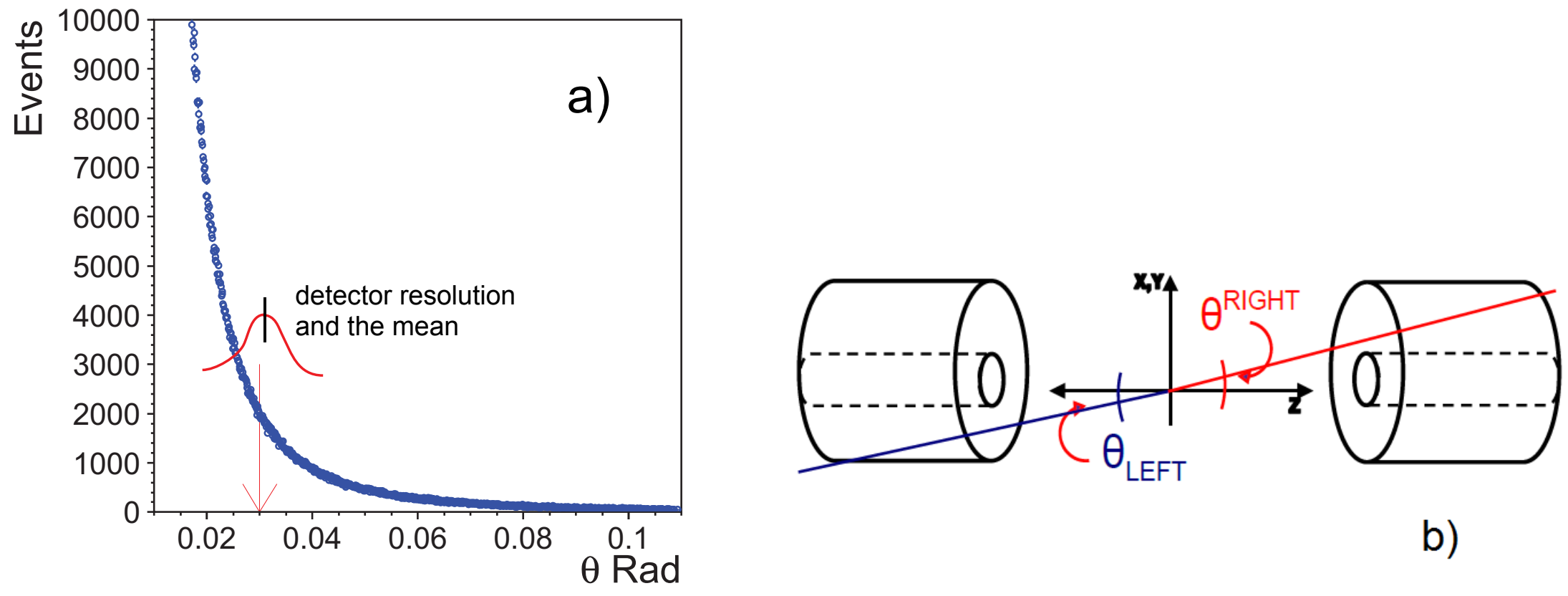

**Figure 9.9:** a) Distribution of scattered electrons in polar angle of the BHLUMI simulation. The Gaussian curve illustrates the detector resolution to $\theta$ measured at a given fiducial edge. The offset of the mean in measurement contributes to the systematic uncertainties. b) Bhabha events is measured preferably in the forward direction of the $e^+e^-$collision characterized by the back-to-back of elastic scattering and the electromagnetic shower of the electrons.

A Bhabha event is detected with a pair of scattered electrons back-to-back in direction, and the momenta of beam energy. Therefore the luminosity detector is consisted of a pair of forward calorimeters with high precision on detecting electron impact positions. The configuration is sketched in Figure 9.9(b). Bhabha events are detected in the angular coverage ($\theta_{min} < \theta < \theta_{max}$) of the forward calorimeters. The integrated luminosity ($L$) of the leading order calculation is

$$\sigma^{vis} = \frac{16\pi\alpha^2}{s}\left(\frac{1}{\theta^2_{min}} - \frac{1}{\theta^2_{max}}\right), \quad \mathcal{L} = \frac{1}{\epsilon}\frac{N_{acc}}{\sigma^{vis}}, \quad \frac{\Delta\mathcal{L}}{\mathcal{L}} \sim \frac{2\Delta\theta}{\theta_{min}}, \tag{9.1}$$

where $\epsilon$ is the detection efficiency. The systematic uncertainties are mostly from the precision on $\theta_{min}$, mainly due to mechanical alignment and the detector resolution. The uncertainty propagates to the luminosity calculation is about twice in magnitude.

The dimension of the detector is favorable to have the $\theta_{min}$ as low as possible to optimize coverage of the Bhabha cross section. The luminosity detector is planned to be mounted in front of the quadruple magnets at $z = \pm 100$ cm. With the $\theta_{min}$ of $\sim$30 mrad, corresponding to a radius of 30 mm to the beam pipe at $z = 100$ cm, the corresponding Bhabha cross-section, $\sigma^{vis}$, after event selection will reach $\sim$ 50 nb. A large detector coverage of $\sigma^{vis}$ is necessary for statistics required for the $Z$ line-shape study, where the $Z \to q\bar{q}$ cross section is 41 nb. The precision for $Z$-pole studies reaching $10^{-4}$ makes a strong demand on the detector resolution. At $\theta = 30$ mrad, it corresponds to an offset of $\Delta\theta \sim 1.5$ $\mu$rad, which is equivalent to 1.5 μm in radius at $z = 100$ cm.

Several technological options for LumiCal design under study, are described in Section 9.5.1, with emphases on the precision of polar angle and energy reconstruction of Bhabha particles scattered in the $t$-channel $V(V = \gamma, Z)$ exchange. The dual beam-pipe configuration with the beam-crossing at 33 mrad results to a boost to particles of $e^+e^-$ collisions. The back-to-back characteristics of Bhabha electrons is shifted by approximately a horizontal offset of 33 mm. The impact to LumiCal design is discussed. The LumiCal together with the quadruple magnet are inserted into the tracking volume that extended to $z = \pm 200$ cm. Shower leakage of electrons off the LumiCal to central tracker is studied by simulation, which is also discussed.

The LumiCal is a precision device with challenging requirements on the integrated luminosity measurement depending on the opening aperture and positioning of the LumiCal. Various sources of luminosity uncertainty in this respect are reviewed in Section 9.5.2. Encouraging estimations on feasibility reaching the goals on the luminosity precision are presented. Detailed studies are ongoing, to include the full simulation of physics and machine induced processes and of the detector itself, for various LumiCal positioning and technology choices.

### 9.5.1 TECHNOLOGICAL AND DESIGN OPTIONS

In the current design of the forward region, LumiCal is foreseen to cover the polar angle region between 26 mrad and 105 mrad corresponding to the detector aperture of 25 mm for the inner radius and 100 mm for the outer, at $z = \pm 100$ cm of the LumiCal front plane from the IP. The detector options shall be considered for

1. precision of the electron impact position at the $r \sim 10$ $\mu$m (1 $\mu$m) level for the uncertainties on luminosity corresponding to the systematic uncertainties on luminosity of $\Delta L/L \sim 10^{-3}$ ( $10^{-4}$) of the machine operation at the Higgs ($Z$-pole) energies;

2. monitoring of the detector alignment and calibration of detector position by tracking of Bhabha electrons with upstream detectors;

3. energy resolution and separation of $e/\gamma$ for measurements of single photons and radiative Bhabha events;

4. maximum coverage and segmentation of the LumiCal to accommodate the dual beampipe and the beam crossing of 33 mrad;

5. minimizing shower leakage into the central tracking volume.

The LumiCal detector option that can reach the 1 $\mu$m precision on electron impact position is very much limited silicon detectors segmented in strips or pixels. Silicon strip detectors of 50 $\mu$m readout pitch is commonly reaching a resolution of $\sigma \sim 5$ $\mu$m. The uncertainty on the mean ($\bar{\sigma} = \sigma/\sqrt{n}$) would be much smaller. The selection of Bhabha events is set on a fiducial edge of $\theta_{min}$, for example, by choosing the center region in the gap between two silicon strips. The systematic uncertainty is therefore the number of events being selected with an uncertainty on $\bar{\sigma}$ despite the detector resolution, and would be relatively small. The uncertainty on $\theta_{min}$ is indicated by the Gaussian curve in Figure 9.9(a). The alignment of the detector position would be the major systematic requirement for an absolute precision of 1 $\mu$m.

A conceptional Luminosity detector is illustrated in Figure 9.10 for the combination of a silicon detector and a calorimeter around the beam pipe for measurement of the electron impact position and energy. The segmentation of the calorimeter is considered for the back-to-back resolution detecting a pair of Bhabha electrons, and for separation of $e/\gamma$ in case of a radiative photon accompanied with the electron or from beam background. The thickness is determined for the energy resolution favorable of $> 20X_0$ for shower containment of a 50 GeV electron. The option on the calorimeter is limited by the space available. The traditional crystal or scintillator-based calorimeter will require more than 20 cm in length for $> 20X_0$. The most compact design would be a sandwiched stack of Silicon samplers with Tungsten layers of $1X_0$ (3.5 mm thick), to a total of about 10 cm that weights about 400 kg.

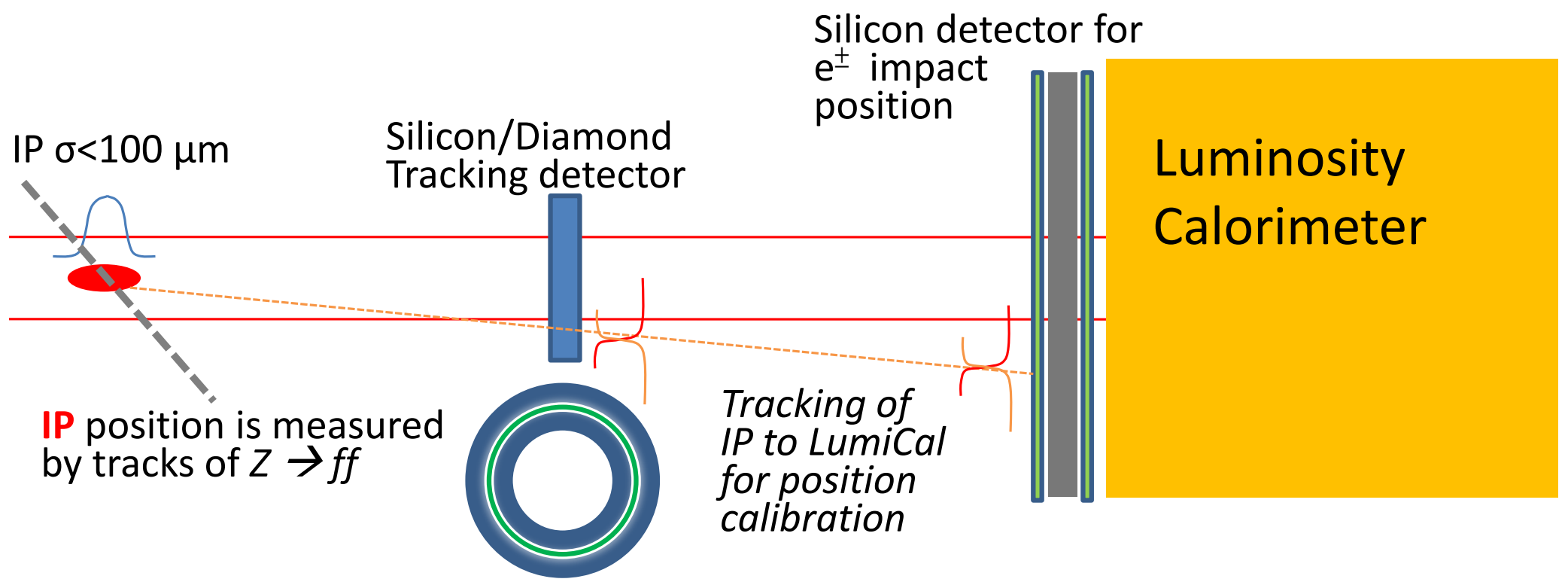

**Figure 9.10:** A conceptional luminosity detector configuration is illustrated utilizing a upstream silicon/diamond detector for calibration of detector position. The tracking with the interaction point (IP), upstream silicon/diamond detector and the front layer of luminosity detector may suffice with a dedicated outer radius region to provide "cross-hair" references for the offset in x-y plane and in z axis.

The alignment precision of the front-layer Silicon detector is the most critical issue to reach 1 μm in radius for the luminosity measurement of $10^{-4}$. At the 1 μm level, a monitoring system with laser alignment is required to calibrate the detector position. The $\theta$ angle of a detected electron impact position is calculated assuming an IP position measured by the beam steering and the central tracking system. The IP position relative to the luminosity detector could be limited to survey relative to the central tracking devices or beam pipe. If feasible, a tracking system on the Bhabha electrons will improve the measurement precision of the electron theta angle. This is illustrated in Figure 9.10 for the option with a ring of silicon or diamond detector mounted in front of the Luminosity detector. In this configuration, an electron track is measured with respect to the IP, the ring detector, and the LumiCal impact positions. The ring detector offers a second survey, and by extrapolation, to calibrate the LumiCal position.

The front silicon layer of the luminosity detector will measure electron impact positions to a few micron. If this will be a fine-pitch strip detector, the position is measured by strips collecting the ionization charges generated by a traversing electron. In Figure 9.11, the charge sharing is illustrated for $\eta = Q_r/(Q_r + Q_l)$ with the ionization charges collected by the strips on the right (left) of the impact position. The distribution is collected for a test device having strips implanted in 25 μm pitch, and the readout in 50 μm pitch by wire bonding to every other strips. The floating strip between two readout strips attracts charges drifting towards it and results to the bump at $\eta \sim 0.5$, in particular for a wide cluster of charges collected by three strips (dotted line). The impact position of a particle is approximated by its center-of-gravity weighted on the charges between two strips. With the $\eta$ distribution, the non-linear distribution can be corrected to achieve a position resolution of better than $\sim 5$ μm for the readout pitch of 50 μm. With the strip detectors placed in a magnetic field, the ionization charge in the silicon wafer is drifted toward one side, and therefore the $\eta$ distribution is tilted un-evenly. Without a proper correction for the $\eta$, the off-set to the true impact position can be as large as half the readout pitch.

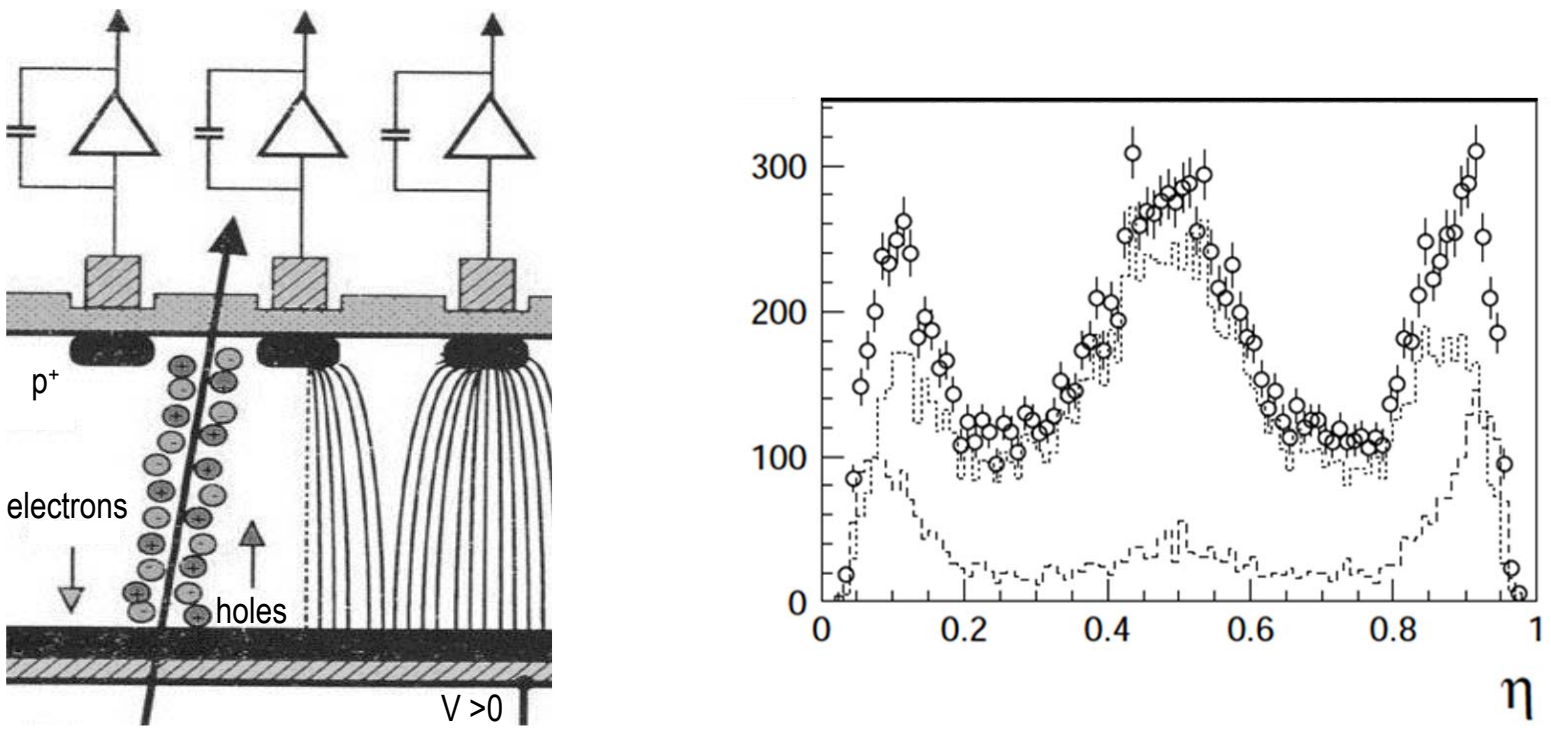

**Figure 9.11:** Charge collection by silicon strips is illustrated for ionization charges generated by a traversing particle. The $\eta = Q_r/(Q_r + Q_l)$ distributions are made for charge sharing to left and right strips to the impact position, for a test device with strip implementation in 25 μm pitch and the readout of every other strips in 50 μm pitch. The $\eta$ distributions are also plotted for contents with charges collected by two-strip (dotted) and three-strip (dashed) cases. The middle bump corresponds to the position of the floating strip between two readout strips.

If the luminosity detector will be a sandwiched silicon-tungsten calorimeter. The silicon wafer for shower sampling may be segmented like the case of the OPAL Lumi-Cal [17], applying 2.5 mm wide strips in circular span of 11.25°. The resolution on detection of an electron position, as well as for $e/\gamma$ separation is at the 1 mm level. Assuming that the Bhabha electrons has the fiducial edge, $\theta_{min}$, chosen at the middle between two strips, and the events are evenly divided to left and right strips without charge sharing. The systematic uncertainty to luminosity measurement is by the alignment uncertainty of the strip position of a few microns, and is not by the resolution.

Charge sharing between the gap of two-strips have been studied with prototype wafers[18] shown in Figure 9.12. The wafer dimension is $65 \times 65$ mm$^2$ implemented with 2 mm wide strips and the gaps from 50 μm to 160 μm. The beam test was conducted with a set of fine-pitched strip detectors as a telescope to provide reference positions of incident electrons scattered across strips and gaps. The charge sharing for electrons in the gaps are compared for $\eta$ distributions in Figure 9.12, which are found compatible for the different gap widths. Charge collection shows no loss, and are drifted toward the near strips with the $\eta$ peaking at the edges. The dispelling charges in the middle of a gap is difficult for detecting the position of an incident electron in the gap. But, it does divide the event fraction cleanly to the near side of the strips.

The double ring configuration of the CEPC machine design at the interaction point has a beam crossing angle of 33 mrad. The effect to the electrons of Bhabha interaction is a boost off the accelerator ring center, by a maximum 16.5 mrad in the horizontal direction. Electrons-positron pairs of the BHLUMI simulation in center-of-mass frame were boosted and the distributions in opening angles and impact positions at the LumiCal front-layer at $z = \pm 100$ cm are plotted in Figure 9.13. The boost is toward +x direction of the laboratory frame. The opening angle shifts from back-to-back of $\pi$ to $\pi - 33$ mrad. The impact positions of the electron-positron pairs overlapped on x-y plane are shown in slices of every 45 degrees to illustrate the offset in back-to-back positions due to the boost and the dependence on pT direction. The maximum offset is 33 mm on x-axis for electrons and positions scattered on horizontal direction.

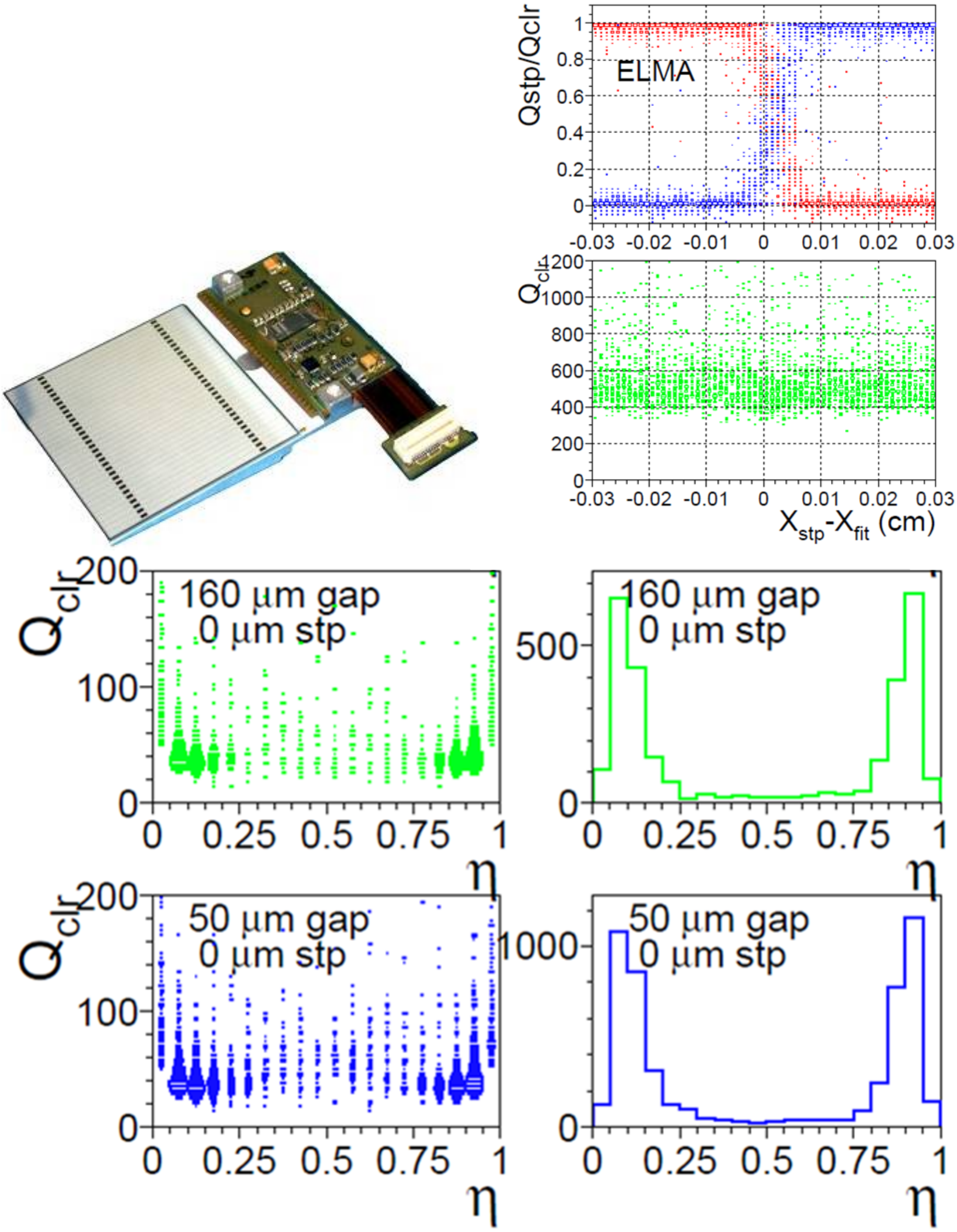

**Figure 9.12:** Beam tests using prototype silicon wafer of the CMS pre-shower detector (top-right) were conducted for collection of ionization charges generated by traversing particles across the gap between strips. The charge sharing by adjacent strips are plotted (top-left) to the reference impact position ($X_{\rm fit}$ extrapolation of a upstream telescope). The sum strip charges is compatible to the hits on a strip. The charge sharing in $\eta = Q_r/(Q_r + Q_l)$ peaks near 0 and 1 (bottom), indicating non-linear response to the randomly distributed beam particles across the gap.

The beam-pipes are centered at $x = \pm 16.5$ mm assuming a 20 mm radius at $z = \pm 100$ cm. In Figure 9.13 the green lines indicate the beam-pipe enclose and a tentative coverage of the LumiCal in circular and rectangular segments. The impact positions are illustrated for both electron and positron detected. Electrons of low scattering angles in -x direction, are lost into beam-pipe due to the boost. The loss of events in vertical direction is much less. The LumiCal acceptance is favorable for the vertical dimension to be as low as possible for a total integrated Bhabha cross section larger than 50 nb.

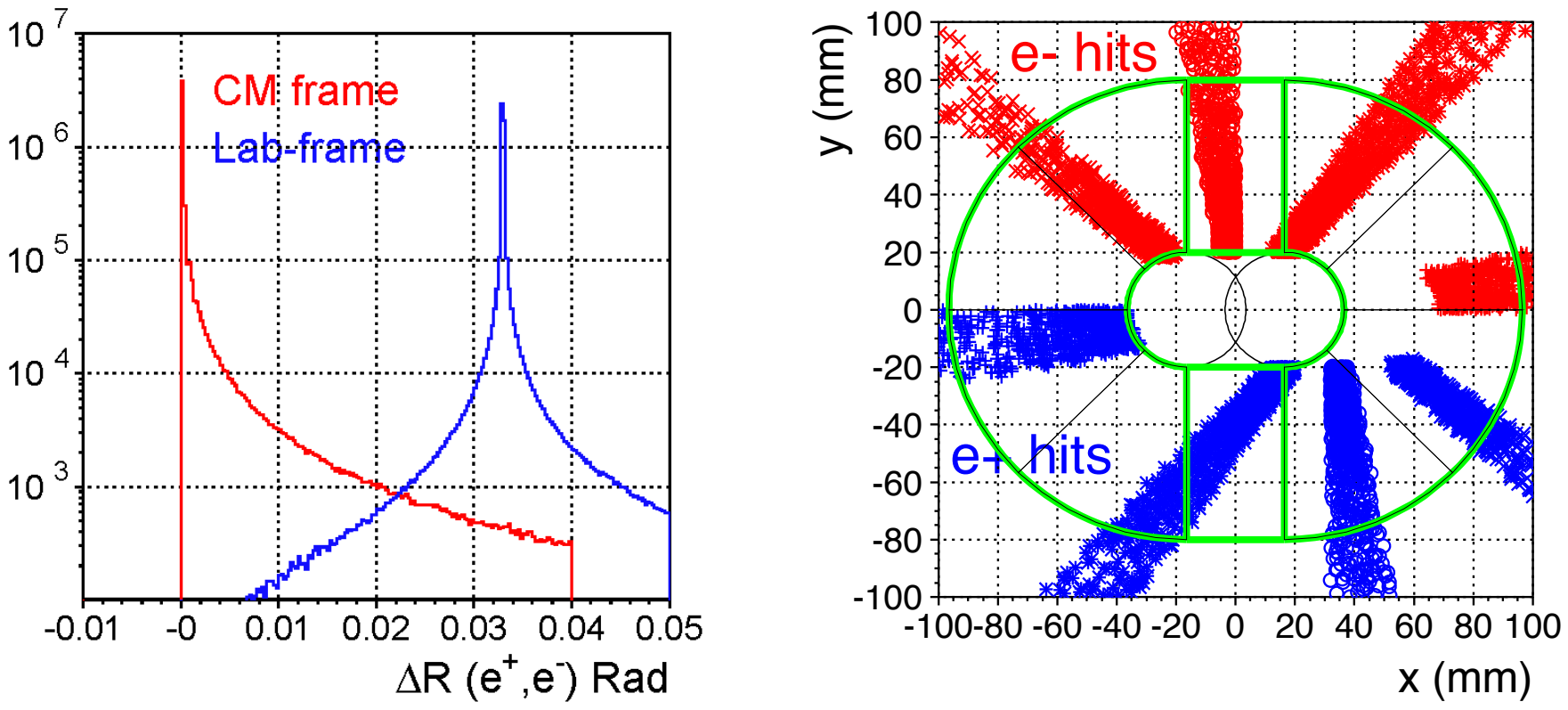

**Figure 9.13:** Bhabha events at the $Z$-pole were simulated with the BHLUMI in the center-of-mass frame. The back-to-back opening angle of scattered electron-position pairs is illustrated (red curve, left plot). The beam-crossing of 33 mrad corresponds to a boost in the $+x$ direction to electrons. The back-to-back distribution of Bhabha events is shifted to $\pi - 33$ mrad (blue curve). The offsets to the impact positions of the electron-positron pairs on the LumiCal surface are plotted in slides of $\phi$ angles every 45 degrees (right plot) requiring both electron and positron observed out of the beam-pipe enclose (green lines) of 20 mm radius for the in-coming (out-going) beams centered at $x$ = -(+)16.5 mm. Events on $x$-axis are lost into the beam-pipe due to the boost.

The LumiCal mounted in front of the quadruple magnet at $z = \pm 100$ cm is half way in the tracking volume of $z = \pm 200$ cm. Shower leakage of electrons at the edge of LumiCal is investigated with a GEANT simulation with parameters cross-checked with a lateral shower study [19]. The LumiCal is configured assuming a sandwiched Silicon-Tungsten calorimeter stacked in twenty decks of 2 mm air-gaps and $1X_0$ thick tungsten layers. Each air-gap has a layer of silicon wafer of 0.3 mm thick. The front layer of the LumiCal is positioned at $z = \pm 100$ cm. The geometry of the LumiCal is tested in two configurations: a TUBE with uniform inner and outer radii of 25 and 100 mm, respectively; and a CONE shape with the outer edge at a constant angle of $\arctan 0.1$ to the interaction point. The CONE shape is intended for well separated absorption of electron shower at a $\theta$ threshold. Illustrated in Figure 9.14 are the event display of the simulations. Out of the LumiCal, a 5 mm iron cone at $|\cos\theta| = 0.992$ is implemented for absorption of low energy shower secondaries traversing into the center tracking volume.

The TUBE configuration has a corner of about 5 mrad on the outer edge to the IP, where energetic shower secondaries can penetrate and leak to detector endcap region. The CONE shape allows the shower fully developed once the electron enters the calorimeter coverage. The shower leakage reaching the Fe-cone is recorded for the particle energies arriving and penetrating through. The statistics are listed in Table 9.5 for 50 GeV and 125 GeV electrons. When the shower is well contained, the leakage is just a few dozens of less than 30 MeV particles. A shower on the edge creates up to 3k secondaries into the tracking volume mostly of less than 100 MeV. The 5 mm iron layer can filter a large fraction of them, to less than 1k particles traversing through.

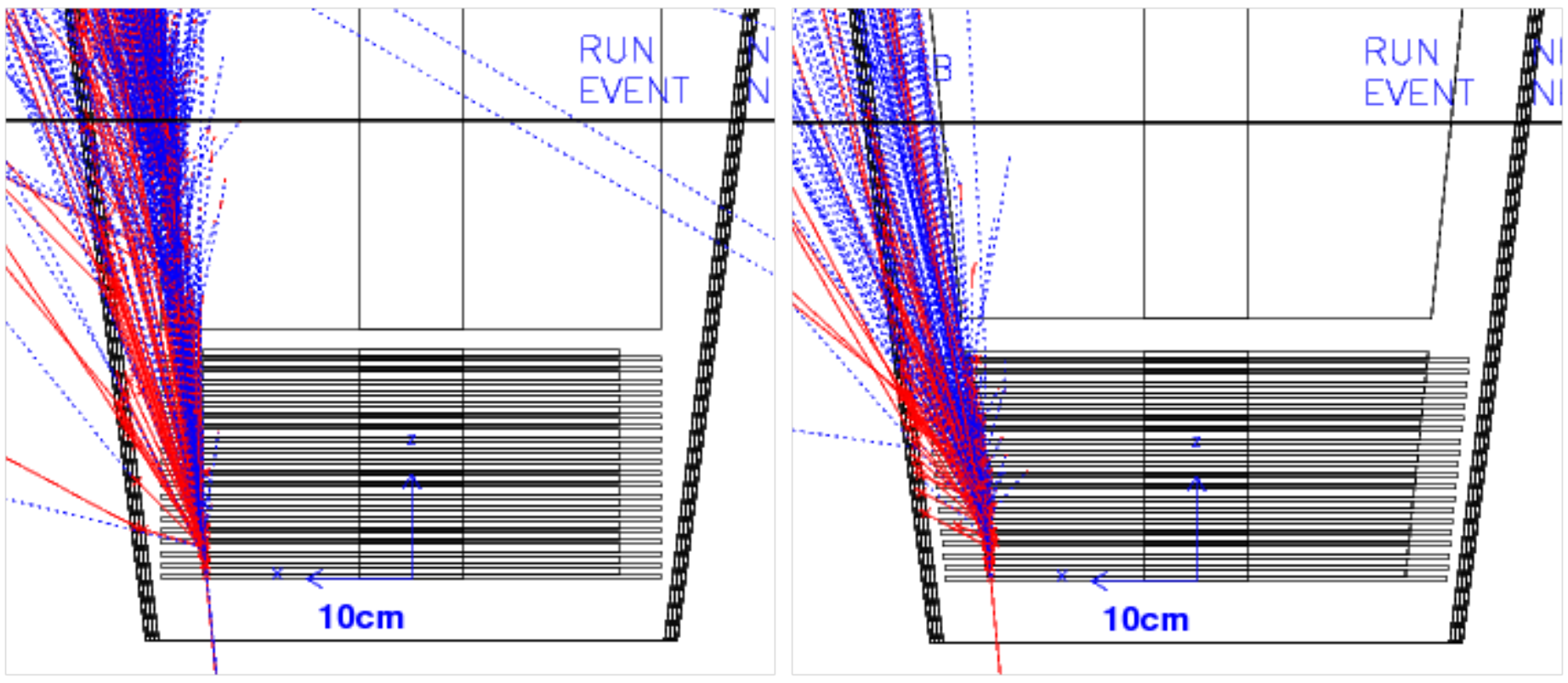

**Figure 9.14:** Event display of a GEANT simulation for electron shower on the LumiCal configuration stacked with 20 decks of silicon and Tungsten layers in TUBE (left) and CONE (right) shapes.

### 9.5.2 SYSTEMATIC EFFECTS

The measurement of luminosity is by counting of Bhabha events $N_{\mathrm{Bh}}$ detected in coincidence in the two halves of the luminosity calorimeter LumiCal. The luminosity figure is then obtained from the equation of $\mathcal{L} = N_{acc}/(\epsilon\sigma^{vis})$. The visible cross section for the Bhabha process, $\sigma_{vis}$, should be integrated over the same phase space as used for the counting of Bhabha events. The limited precision with which the experimental acceptance region is defined gives rise to a number of systematic effects. Further, other processes misidentified as Bhabha and the limited accuracy of the theoretical calculation [20–23] of $\sigma_{vis}$ contribute to the overall systematic uncertainty.

| | **50 GeV electrons** | | **125 GeV electrons** | |
|---|---|---|---|---|
| | **TUBE** | **CONE** | **TUBE** | **CONE** |
| $\theta$ (mrad) | $N_{\mathrm{enter}}/N_{\mathrm{pass}}$ | $N_{\mathrm{enter}}/N_{\mathrm{pass}}$ | $N_{\mathrm{enter}}/N_{\mathrm{pass}}$ | $N_{\mathrm{enter}}/N_{\mathrm{pass}}$ |
| 40 | 15.4/5.6 | 13.6/5.8 | 38.0/16.0 | 35.8/14.7 |
| 90 | 392/155 | 173/76 | 1028/399 | 434/19.7 |
| 95 | 501/290 | 367/152 | 2389/720 | 937/382 |
| 98 | 762/216 | 860/284 | 1718/473 | 2176/725 |
| 99 | 553/140 | 1331/367 | 1102/273 | 3306/915 |

**Table 9.5:** Number of particles leaking out of the LumiCal outer radius ($N_{\mathrm{enter}}$) and number of particles passing through the Fe-cone ($N_{\mathrm{pass}}$). Two different detector designs (TUBE and CONE) and two shower energies (50 GeV and 125 GeV) are simulated. A shower on the edge creates up to 3k secondaries toward the tracking volume, which are mostly of less than 100 MeV and are filtered by the 5 mm thick Fe-cone.

A generator-level study was performed to assess the effects related to the precision of the Bhabha acceptance region on Bhabha counting. An underlying assumption of the study is that the LumiCal is centered on the outgoing beam axis. This assumption is essential for data-driven control of the radial offset of LumiCal with respect to the IP, as well as for Bhabha event counting based on the mirrored asymmetric polar-angle acceptance regions on the left and right side of the detector [17] (in further text, *OPAL-style selection*). OPAL-style counting cancels out biases due to left-right asymmetries of the experimental angular acceptance. It is further assumed that for the final state particles hitting the radial region between 50 mm and 75 mm, corresponding to the detector Fiducial Volume (FV), shower leakage has a negligible effect on the reconstruction of the polar angle and the energy.

Bhabha event samples are generated using the BHLUMI generator [16]. Center-of-mass energy of 240 $\mathrm{GeV}$ is assumed, corresponding to approximately the energy of the maximum Higgs boson production cross section. The particles are generated in the range of polar angles including a $\sim$ 7 mrad margin outside the FV to allow non-collinear final state radiation (FSR) to contribute to the events. After event generation, smearing is applied to the final particle vertices and momenta according to the nominal CEPC parameters. Additional smearing or bias is then applied according to one systematic effect at a time. Four momenta of close-by particles are summed up to account for cluster merging in LumiCal. The selection criteria to count an event consist of the OPAL-style angular selection and the requirement that the energy of both detected showers is above 50$\%$ of the nominal beam energy. The relative acceptance bias is determined as the relative difference between the Bhabha count $N_{\mathrm{Bh},i}$ obtained with the inclusion of the considered effect $i$ and $N_{\mathrm{Bh}}$ obtained with the nominal set of parameters.

Table 9.6 lists the requirements on beam delivery, MDI and LumiCal installation, needed to limit individual systematic effects in the luminosity measurement to $1 \times 10^{-3}$, such as required for the Higgs boson physics program at the CEPC. Parameters influencing the integral luminosity precision are given as follows:

- $\Delta E_{\mathrm{CM}}$, uncertainty of the available center-of-mass energy affecting the Bhabha cross-section,
- $E_{e^+} - E_{e^-}$, asymmetry of the incident beam energies resulting in a net longitudinal boost of the event,
- $\frac{\delta\sigma_{E_{beam}}}{\sigma_{E_{beam}}}$, uncertainty of the beam energy spread,
- $\Delta x_{\mathrm{IP}}$ and $\Delta z_{\mathrm{IP}}$, radial and axial offsets of the IP w.r.t. the LumiCal,
- Beam synchronization, resulting in axial offset of the IP w.r.t. the LumiCal,
- $\sigma_{x_{\mathrm{IP}}}$ and $\sigma_{z_{\mathrm{IP}}}$, radial and axial fluctuations of the scattering position,
- $r_{in}$, inner radius of the LumiCal acceptance region,
- $\sigma_{r_{\mathrm{shower}}}$, reconstruction precision of the radial shower coordinate,
- $\Delta d_{\mathrm{IP}}$, uncertainty of the distance between the LumiCal halves.

Most requirements are technically feasible with the present state of the art of accelerator and detector technology. The most important challenge identified is the precision of the inner acceptance radius $r_{\text{in}}$ of LumiCal. In order to keep the luminosity precision of 1 per mille, $r_{\text{in}}$ must be known to within 10 μm. The precision requirement of $r_{\text{in}}$ scales linearly with the required luminosity precision, implying a correspondingly stricter requirement for the $Z$-pole run.

| Parameter | Unit | Limit |
|---|---|---|
| $\Delta E_{\text{CM}}$ | MeV | 120 |
| $E_{e^+} - E_{e^-}$ | MeV | 240 |
| $\frac{\delta\sigma_{E_{beam}}}{\sigma_{E_{beam}}}$ | | effect canceled |
| $\Delta x_{\text{IP}}$ | mm | <1 |
| $\Delta z_{\text{IP}}$ | mm | 10 |
| Beam synchronization | ps | 7 |
| $\sigma_{x_{\text{IP}}}$ | mm | 1 |
| $\sigma_{z_{\text{IP}}}$ | mm | 10 |
| $r_{in}$ | $\mu$m | 10 |
| $\sigma_{r_{\text{shower}}}$ | mm | 1 |
| $\Delta d_{\text{IP}}$ | μm | 500 |

**Table 9.6:** Requirements on beam delivery, MDI and LumiCal installation, needed to limit individual systematic effects to $< 1 \times 10^{-3}$.

### 9.5.3 SUMMARY

Instrumentation of the very forward region is very important for the realization of the CEPC physics program. Several technology options are under consideration. Some of them have been successfully applied at LEP or are under study for other future projects. A tracker placed in front of the LumiCal can improve polar angle measurement accuracy, facilitate LumiCal alignment and enable electron-photon separation. LumiCal centered on the outgoing beam axis is studied for the systematic effects at the required level. Precision requirements on beam delivery, MDI and LumiCal installation have been addressed by simulation, and proven to be feasible with the present state-of-the-art of accelerator and detector technology.

## 9.6 DETECTOR INTEGRATION

Both QD0 and QF1 are located inside the detector, which drastically complicates the support and alignment of the detector and machine components in the interaction region. As shown in Figure 9.15, the first attempt has been made as inspired by the remote vacuum connection (RVC) scheme developed at the SuperKEKB and Bell II [24]. With proper installation tools, it might be feasible to attach the heavy LumiCal to the final focus magnets, which are mounted on a dedicated supporting structure extended from a pillar outside the detector and suspended from the solenoid cryostat. They will be integrated together be-fore being pushed into the interaction region. Both magnets and LumiCal can be aligned with dedicated laser trackers [25] to achieve their required installation precisions. However, the additional material introduced in front of the LumiCal can impair the LumiCal performance to a large extent and must be minimized. Alternative installation schemes are also under investigation but might introduce more complexities to the detector integration. Furthermore, the shaped beam pipe and surrounded silicon detectors will possibly be supported from a structure of carbon fiber reinforced plastic, which can hang at the flanges of the field cage of the Time Projection Chamber (TPC). Significant effort is required to realize a solid mechanical design and to define a reasonable procedure for the detector and machine installation.

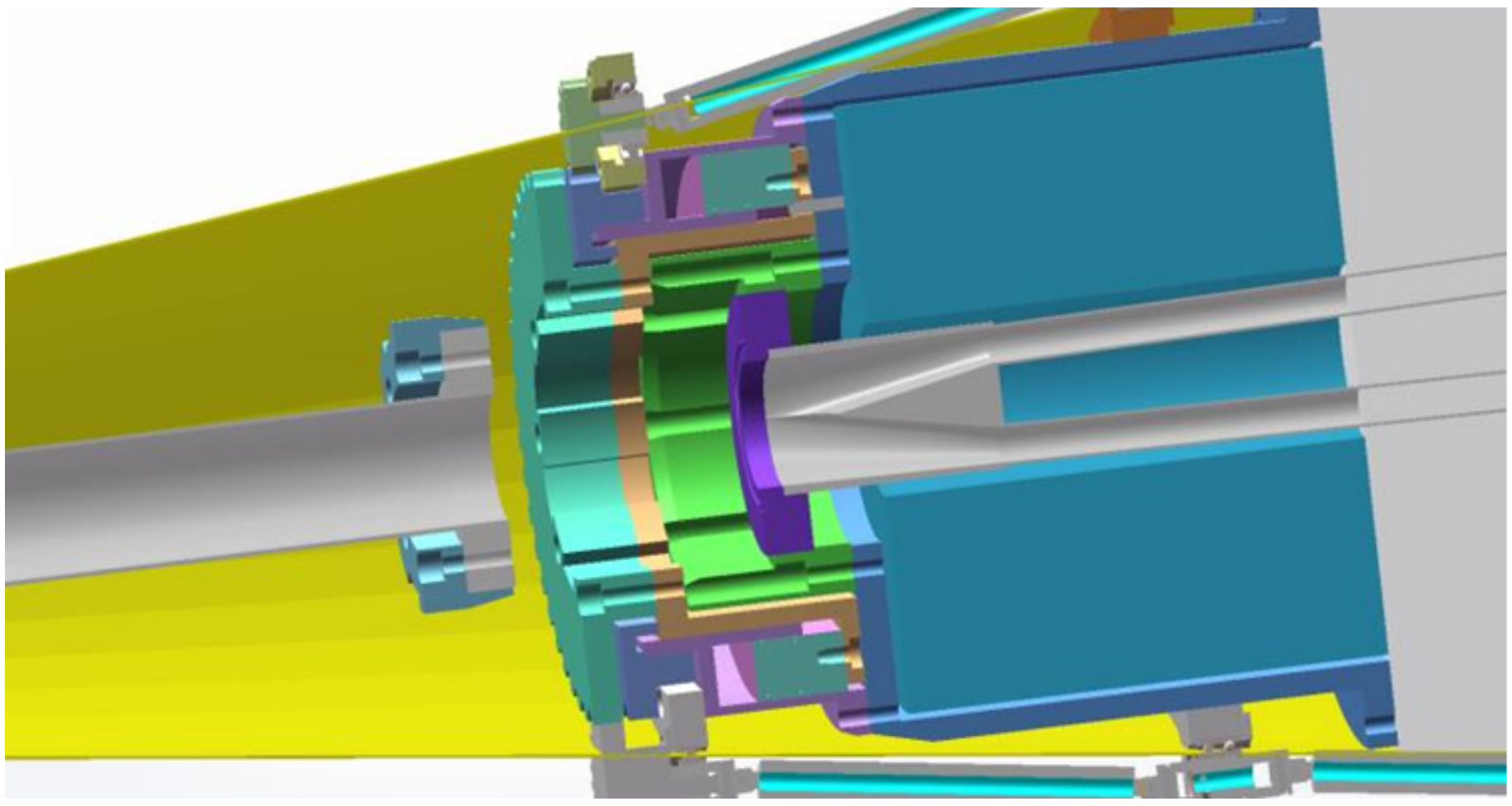

**Figure 9.15:** Illustration of one possible installation scheme as inspired the remove vacuum connection (RVC) scheme developed at the SuperKEKB/Belle II. The LumiCal will be first attached to the cryostat of the magnets and then pushed together into the interaction region.

CHAPTER 10

# SIMULATION, RECONSTRUCTION AND PHYSICS OBJECT PERFORMANCE

This chapter summarizes the expected performance of the CEPC baseline detector concept based on the Monte Carlo (MC) simulation studies. Section 10.1 describes the software and algorithm tools, the event generation, and the detector simulation as well as the event reconstruction. Section 10.2 presents the performance for identifying and measuring basic physics objects that are building blocks for physics analyses, such as leptons, photons, jets and their flavors. The results presented here represent the first attempt to understand the performance of the CEPC baseline detector. They will likely improve with further studies and optimization.

## 10.1 EVENT SIMULATION AND RECONSTRUCTION

The simulation of physics events and detector responses and the reconstruction of the raw detector information are vital for high energy physics experiments. Figure 10.1 is a flow chart of the event simulation and reconstruction. In this section, the functionalities of key components of the chart are described.

### 10.1.1 EVENT SIMULATION

For the studies of the CEPC physics performance, the Whizard package [1] is used as the main event generator to produce physics events. Collaborating with the Whizard team, a dedicated CEPC beam parametrization has been implemented in its official release. The Whizard generator is used to simulate SM processes, including both the Higgs boson signal and all its SM background samples. Additionally, Madgraph [2] and Pythia [3] generators are used to produce samples from Beyond the Standard Model (BSM) physics.

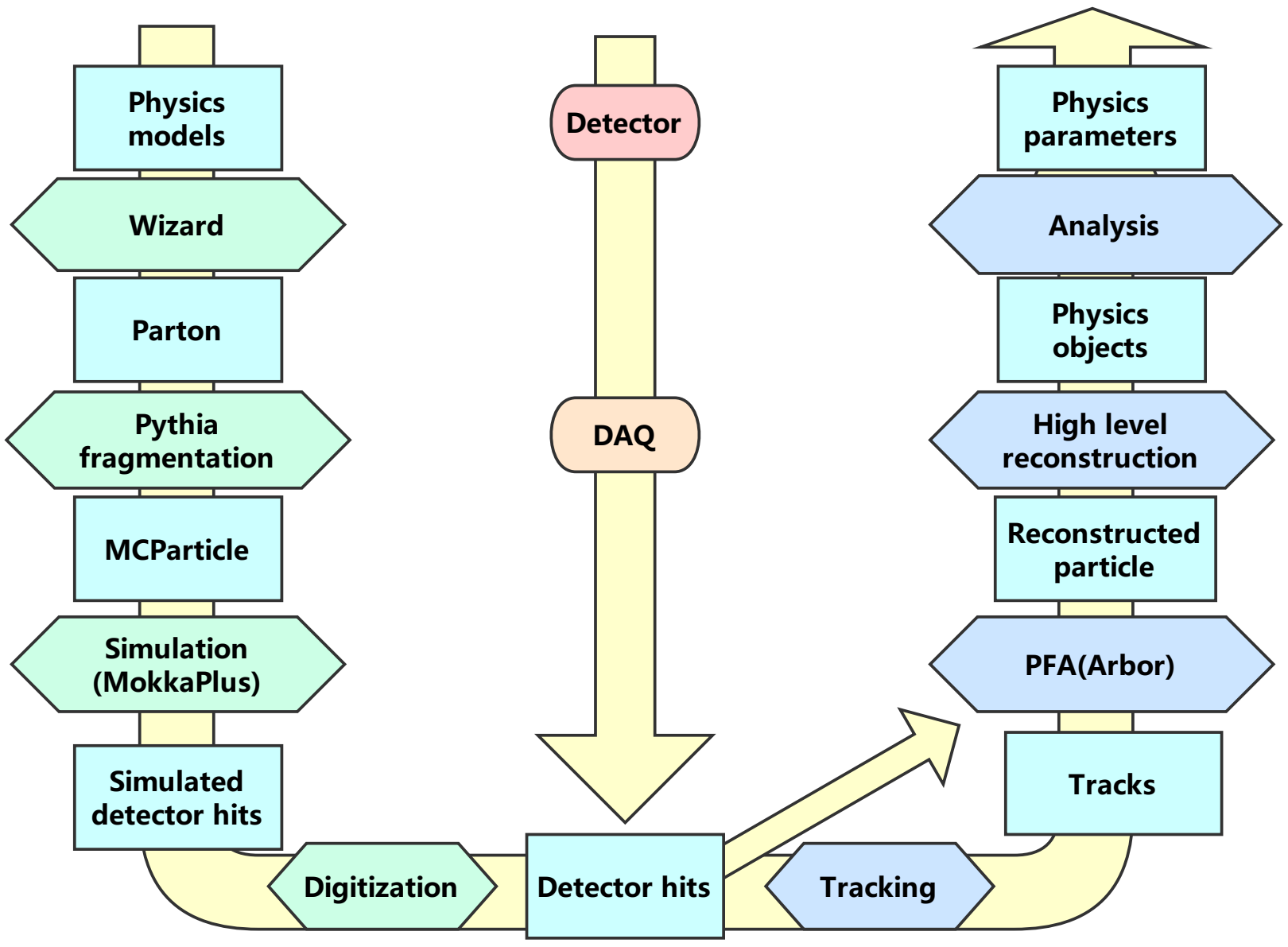

**Figure 10.1:** The flow chart of the CEPC simulation studies.

A GEANT4-based detector simulation framework, MokkaPlus, is used for the CEPC detector simulation. MokkaPlus is a virtual geometry constructor that compiles with the GEANT4 libraries [4] and a MySQL database [5]. It is an improved version of Mokka [6], a simulation framework used for early linear collider studies. The digitization of simulated energy deposits in the detector are performed using a general algorithm that reproduces the test beam results [7] of the calorimeter and an ilcsoft scheme [8] for the tracking detectors. The parameter values of the `ilcsoft` scheme are tuned to match the CEPC detector design. In addition, a fast simulation based on the efficiency and resolution parametrization derived from the full simulation has also been developed. The fast simulation is used to produce most of the background samples for studies presented in this report.

### 10.1.2 EVENT RECONSTRUCTION

The event reconstruction chain starts with the track reconstruction, followed by the particle flow interpretation of tracks and calorimeter hits and finally the reconstruction of compound physics objects such as converted photons, $K_S$'s, $\tau$-leptons and jets.

Tracks are reconstructed from hits in the tracking detectors by a tracking module. The module is currently based on the Clupatra module [9] of ilcsoft which has been shown to have excellent performance. A CEPC-specific tracking module with the flexibility of geometry modification is under development.

A dedicated particle flow reconstruction toolkit, ARBOR [10, 11], has been developed for the CEPC baseline detector concept. ARBOR is composed of a clustering module and a matching module. The clustering module reads the calorimeter hits and forms clusters of hits (also called branches) which are then arranged into a tree topology as illustrated in Figure 10.2 for the 3-prong decay of a $\tau$-lepton. The matching module iden-tifies calorimeter clusters with matching tracks and builds reconstructed charged particles. The remaining clusters are reconstructed into photons, neutral hadrons and unassociated fragments. From this unique list of particles, simple particles such as electrons, muons, photons, charged pions and kaons can then be identified.

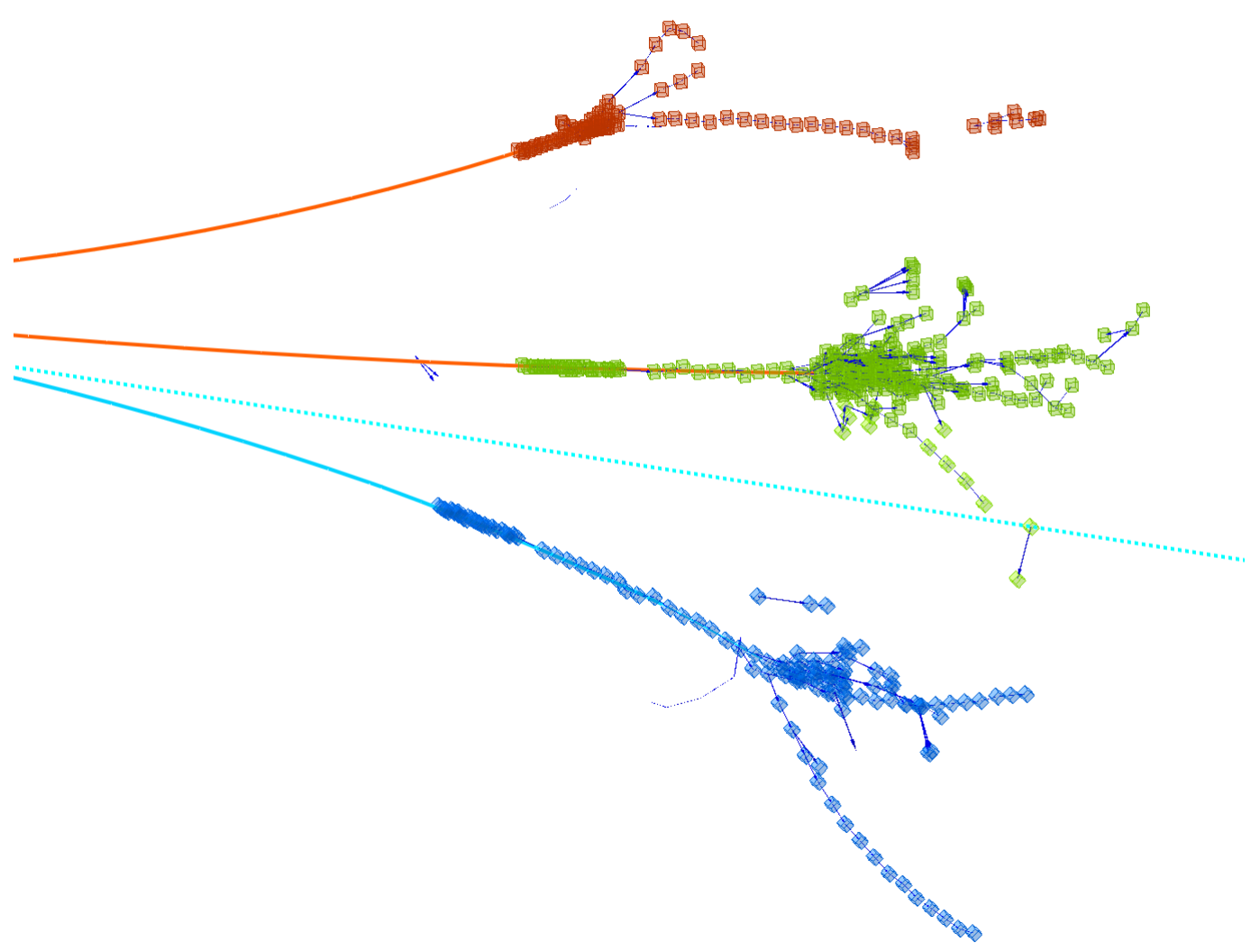

**Figure 10.2:** An illustration of particle flow reconstruction: the 3-prong decay of a $\tau$-lepton from $Z \to \tau^+\tau^-$ reconstructed by the ARBOR algorithm. Three branches of calorimeter clusters correspond to the three reconstructed charged particles of the $\tau$-lepton decay: a 5.7 GeV $\pi^+$ (red), a 27.4 GeV $\pi^+$ (green) and a 10.3 GeV $\pi^-$ (blue). Also shown are lines representing Monte Carlo truth information: two $\pi^+$'s (red), a $\pi^-$ (cyan) and a $\bar{\nu}_\tau$ (cyan dashed).

The particle flow reconstruction provides a coherent interpretation of an entire physics event and, therefore, is well suited for the reconstruction of compound physics objects such as converted photons, $K_S$'s, $\tau$-leptons and jets. The reconstruction of $\tau$-leptons and jets are described in Section 10.2. CORAL, an algorithm that targets the reconstruction of converted photons, $\pi^0$'s and $K_S$'s, is being developed.

#### 10.1.2.1 TRACK RECONSTRUCTION

The CEPC baseline tracker consists of a silicon tracking system and a barrel TPC. The two subsystems play complementary roles. The silicon system provides high precision spatial point measurements whereas the TPC has 220 radial layers which simplifies the track finding of the detector. In addition, the silicon system includes a forward tracking system that extends the solid angle coverage of the tracker.

The performance of the CEPC tracker is studied using two samples: a single muon particle sample and an $e^+e^- \to Z \to \tau^+\tau^-$ sample at $\sqrt{s} = 91.2$ GeV. The single muon sample is used to characterize the tracking efficiency and momentum resolution for isolated tracks while the $Z \to \tau^+\tau^-$ sample, with one of the $\tau$-leptons decays into a 3-prong, provides a test for reconstructing closely spaced tracks.

The single muon sample covers a momentum range of 0.1 GeV to 100 GeV and the full angular range. Figure 10.3 shows the extracted efficiency and momentum reso-lution as a function of the polar angle for different momentum bins. For muons in the tracking fiducial volume of $|\cos\theta| < 0.985$ and with momentum above 0.5 GeV, the reconstruction efficiency is nearly 100%. The momentum resolution reaches per mille level for the momentum range of 10–100 GeV in the barrel region. The resolution is limited by material-induced multiple scatterings at low momentum and by the magnetic field, the level-arm and the precision of spatial measurements at high momentum, consistent with the design goal outlined in Chapter 3. The $\tau$-leptons from $Z \to \tau^+\tau^-$ are highly boosted and can lead to three closely spaced charged particles in their 3-prong decays, see Figure 10.2. For this sample, the efficiency for reconstructing all three tracks of the $\tau$ decays is found to be close to 100% .

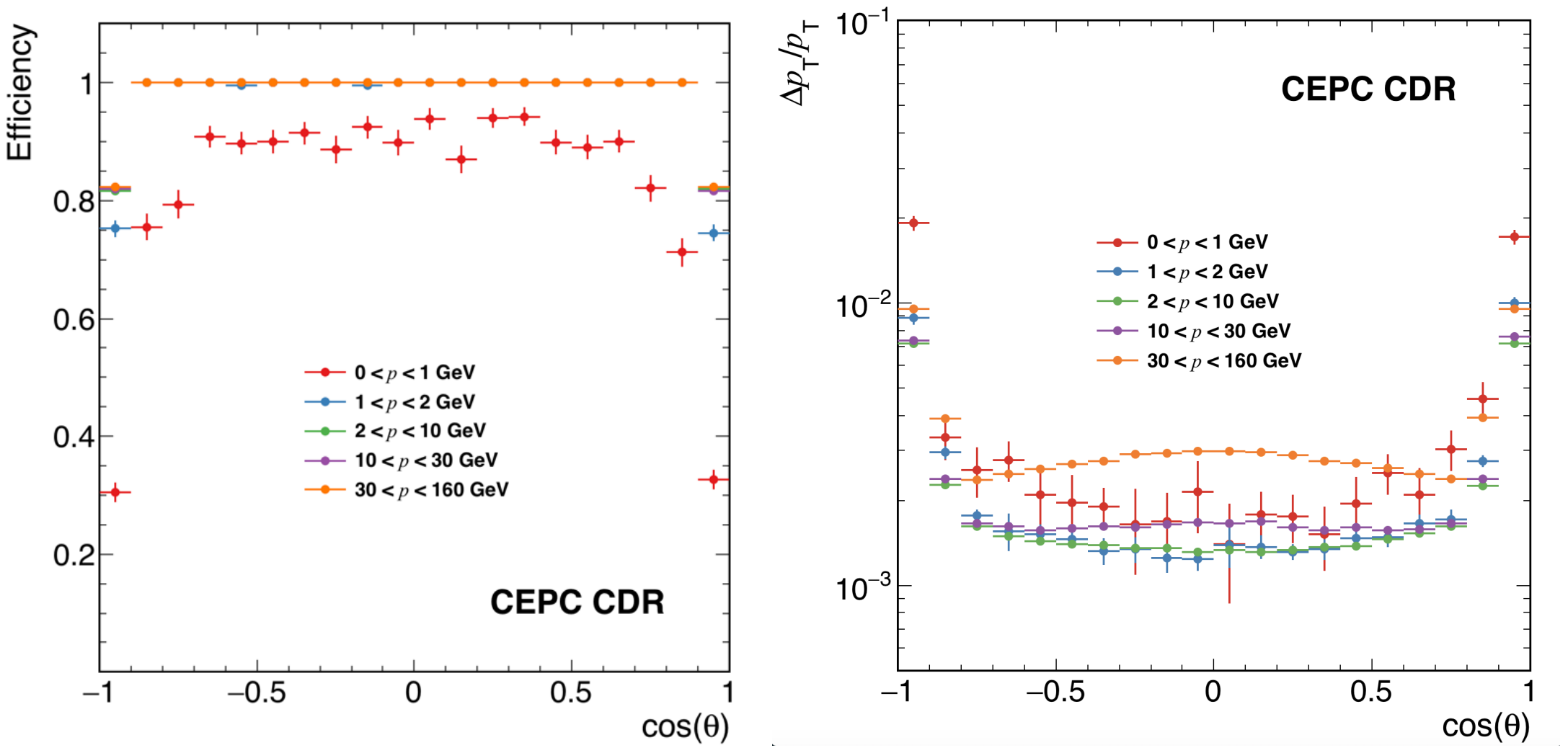

(a) (b)

**Figure 10.3:** Single track reconstruction: (a) efficiency and (b) momentum resolution as a function of the cosine of the polar angle in different momentum bins.

## 10.1.2.2 CLUSTER RECONSTRUCTION

The high granularity calorimetry of the CEPC baseline detector concept are well suited for reconstructing clusters of energy deposits by traversing particles. The fine segmentation allows the reconstruction of individual particles produced in shower cascades, see Figure 10.2.

Two relevant performance measures of the cluster reconstruction are the energy collection efficiency for single neutral particles and spatial separation capability for two closely spaced neutral particles. For photons with energy above 5 GeV, ARBOR is able to collect more than 99% of the energy deposited in the calorimeter while keeping the mis-clustering rate small. Good cluster spatial separation capability is essential for the reconstruction of compound particle objects such as $\pi^0$'s and $\tau$-leptons. Figure 10.4(a) is a demonstration of the reconstructed clusters from two closely spaced photons from a $\pi^0$ decay. The efficiencies for successfully reconstructing two photon clusters as functions of their separation at the calorimeter entry points are shown in Figure 10.4(b) for three ECAL cell sizes. The critical distance, defined as the separation at which the efficiency for reconstructing two photon clusters is 50%, is found to be 16 mm for the baseline de-sign of the ECAL cell size of $10 \times 10$ mm$^2$. This corresponds to an average efficiency of 50% for reconstructing two photons as two separate clusters from a 30 GeV $\pi^0$ decay with equal energy. For neutral hadrons with energy above 5 GeV, the energy collection efficiency is roughly 90%.

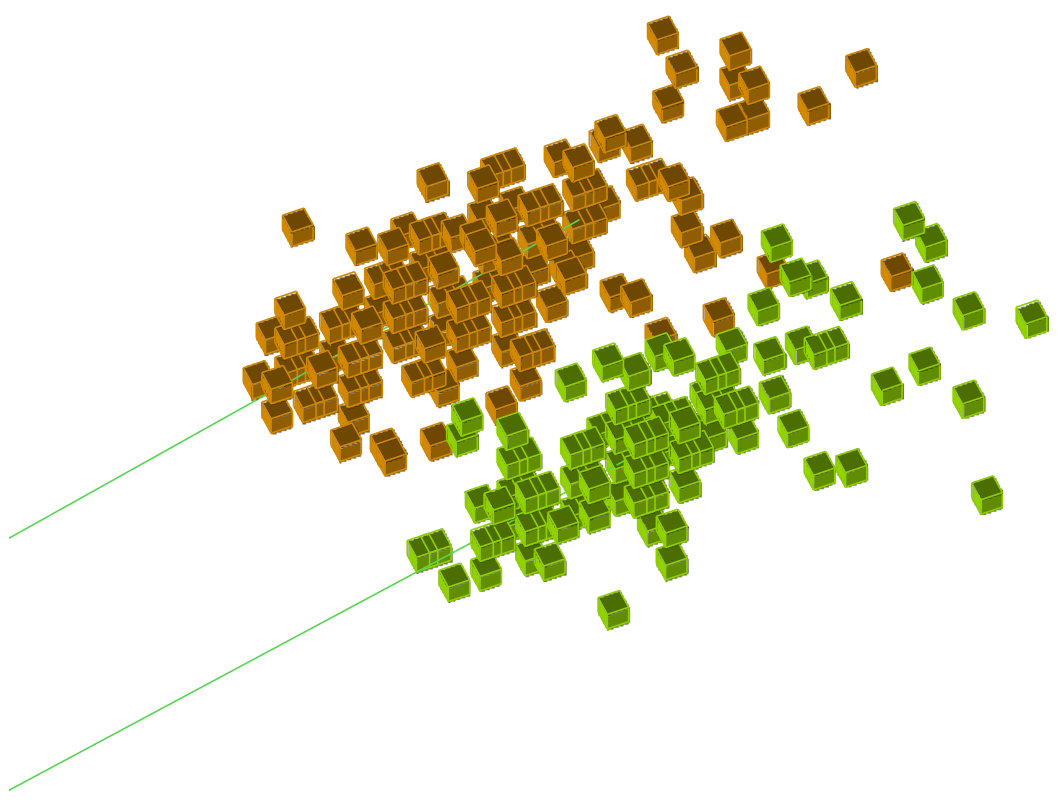

(a)

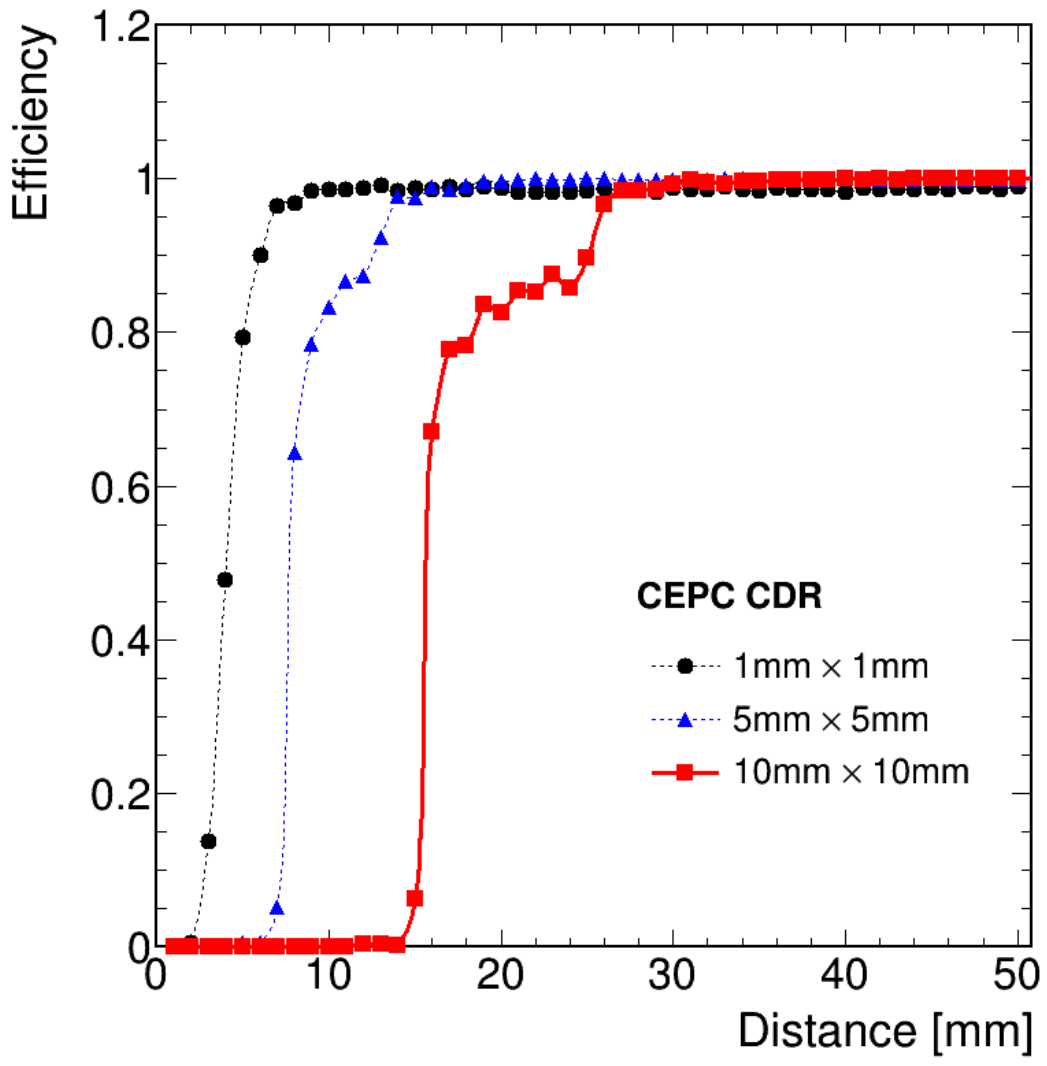

(b)

**Figure 10.4:** (a) Clusters of the two photons of energies of 6.2 GeV and 5.8 GeV from a 10 GeV $\pi^0$ decay reconstructed in the Silicon-Tungsten ECAL with a $10\times10$ mm$^2$ cell size. (b) Reconstruction efficiencies of two photons as two separate clusters as functions of the distance between their calorimeter impact points for three ECAL cell sizes. The ECAL cell size of the CEPC baseline detector is chosen to be $10 \times 10$ mm$^2$. The secondary step structures in the efficiency curves reflect the finite granularity of the calorimeter.

## 10.2 OBJECT IDENTIFICATION AND PERFORMANCE

Particle flow reconstruction leads to a unique list of particles from which electrons, muons, photons, $\tau$-leptons, and jets etc., the physics objects as they are customarily called, can be identified or built. These objects serve as building blocks for further physics analyses as presented in Chapter 11. In this section, their general identification and the expected performance are described. For analyses of specific processes, the identification and performance can often be improved by utilizing the unique topologies of the events under study.

### 10.2.1 LEPTONS

Leptons ($\ell$, $\ell = e, \mu$)[1] are bedrocks to the CEPC physics program. $Z \to e^+e^-$ and $Z \to \mu^+\mu^-$ decays are indispensable for electroweak measurements and for the model-independent identification of the Higgs boson through the recoil mass method. A large fraction of Higgs bosons decay, directly or via cascade, into final states with electrons and muons.

[1]Unless otherwise noted, leptons refer to electrons, muons or their antiparticles.

The particle-flow oriented baseline detector, particularly its finely-segmented calorimetry system, provides enormous information for the lepton identification. High energy electrons and hadrons will likely induce thousands of hits whereas muons deposit little energy in the calorimeter. Electrons can be identified from their pencil-like electromagnetic shower development in the ECAL matched with tracks in the tracker. Muons exhibit themselves as minimum ionizing particles in the calorimeter matched with tracks in the tracker as well as in the muon detector. Moreover, the $dE/dx$ measurements in the TPC could provide additional discrimination of electrons from muons and hadrons for energies up to 10 GeV.

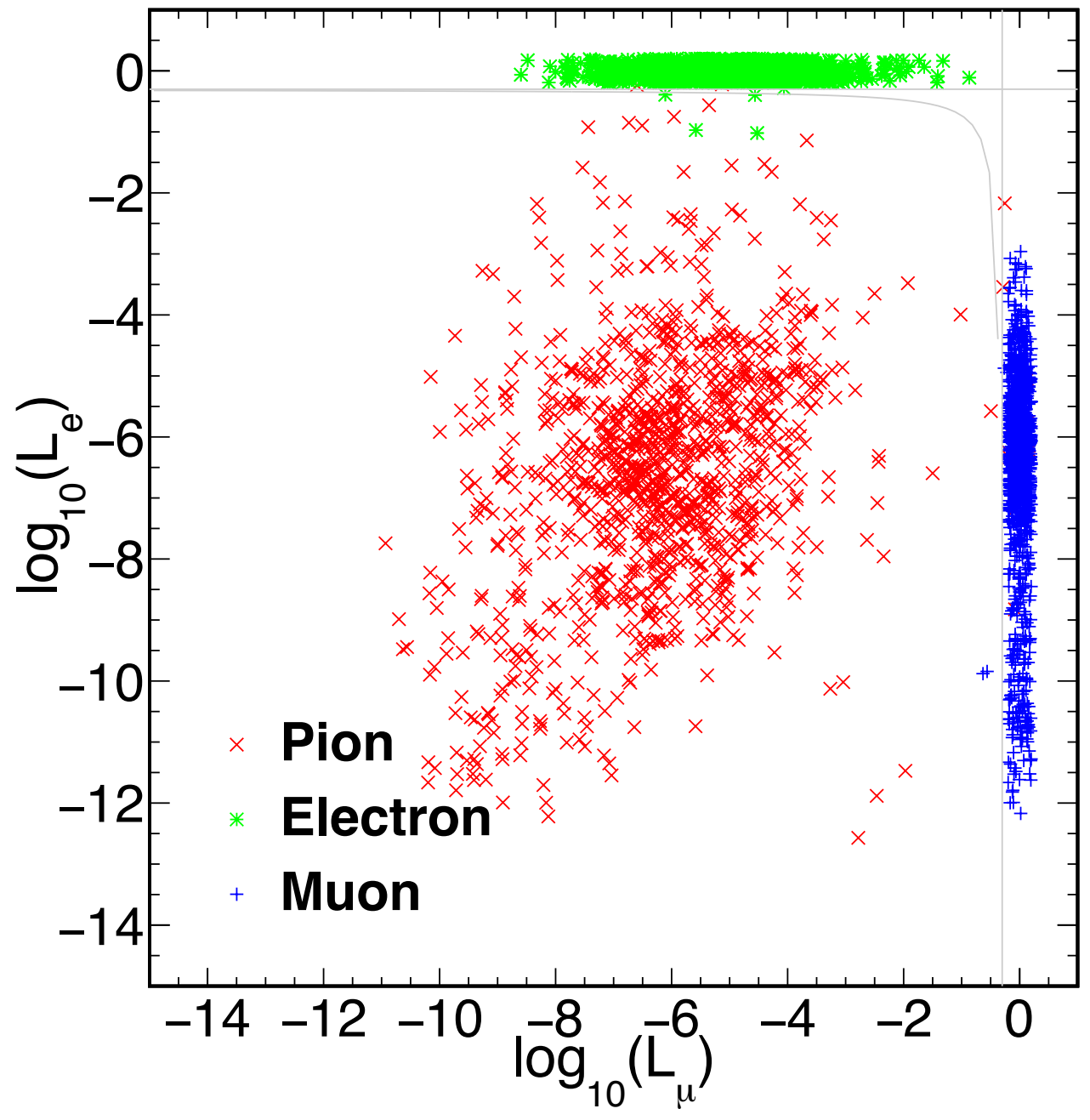

**Figure 10.5:** Distributions of the logarithm of $e$-likeliness $L_e$ and $\mu$-likeliness $L_\mu$ expected from 40 GeV electrons, muons and charged pions in the barrel calorimeter ($|\cos\theta| < 0.7$).

A lepton identification algorithm, LICH [12], has been developed and implemented in ARBOR. LICH combines more than 20 discriminating variables from the detector to build lepton-likelihoods, $e$-likeliness ($L_e$) and $\mu$-likeliness ($L_\mu$), using a multivariate technique. Figure 10.5 compares the two-dimensional distributions of $L_e$ and $L_\mu$ expected from single electrons, muons and charged pions, showing clear separations among these particles. For leptons above 2 GeV, an identification efficiency better than 99.5% and a mis-identification probability from hadrons smaller than 1% can be achieved. The main sources of mis-identification are irreducible backgrounds from the $\pi^\pm \to \mu^\pm$ decays for muons and highly electromagnetic like $\pi^\pm$ clusters ($\pi^0$ produced in pion-nucleon interactions) for electrons. The momentum resolution of the tracker (see Section 10.1.2.1) largely determines the resolutions for both electrons and muons. However, a degradation in the resolutions is expected from Bremsstrahlung radiation, most importantly for electrons and to a lesser extent for muons. Recovering the radiation energy losses using the ECAL measurements should improve the resolutions. However, this is not implemented for the current studies.

For complex physics events, lepton identification will be affected by the limited spatial separation capability of the detector. For example, the efficiency for successfully identifying two leptons with opposite charges is found to be 97–98% for the $e^+e^- \to ZH \to \ell^+\ell^- H$ events. The small loss of the efficiency can be attributed to overlapping clusters in the calorimeter.

Figure 10.6 shows the reconstructed recoil mass[2] distributions of the $Z \to \mu^+\mu^-$ and $Z \to e^+e^-$ decays from the $e^+e^- \to ZH$ process, and Figure 10.7 is the dimuon invariant mass distribution of $H \to \mu^+\mu^-$ again from $e^+e^- \to ZH$. The sharp peak at the Higgs boson mass demonstrates the excellent lepton energy/momentum and angular resolutions. The tails are due to radiation effects. The recoil mass distributions are critical for the model-independent identification of the Higgs boson and a good dimuon mass resolution is essential for identifying $H \to \mu^+\mu^-$ decays, see Section 11.1.

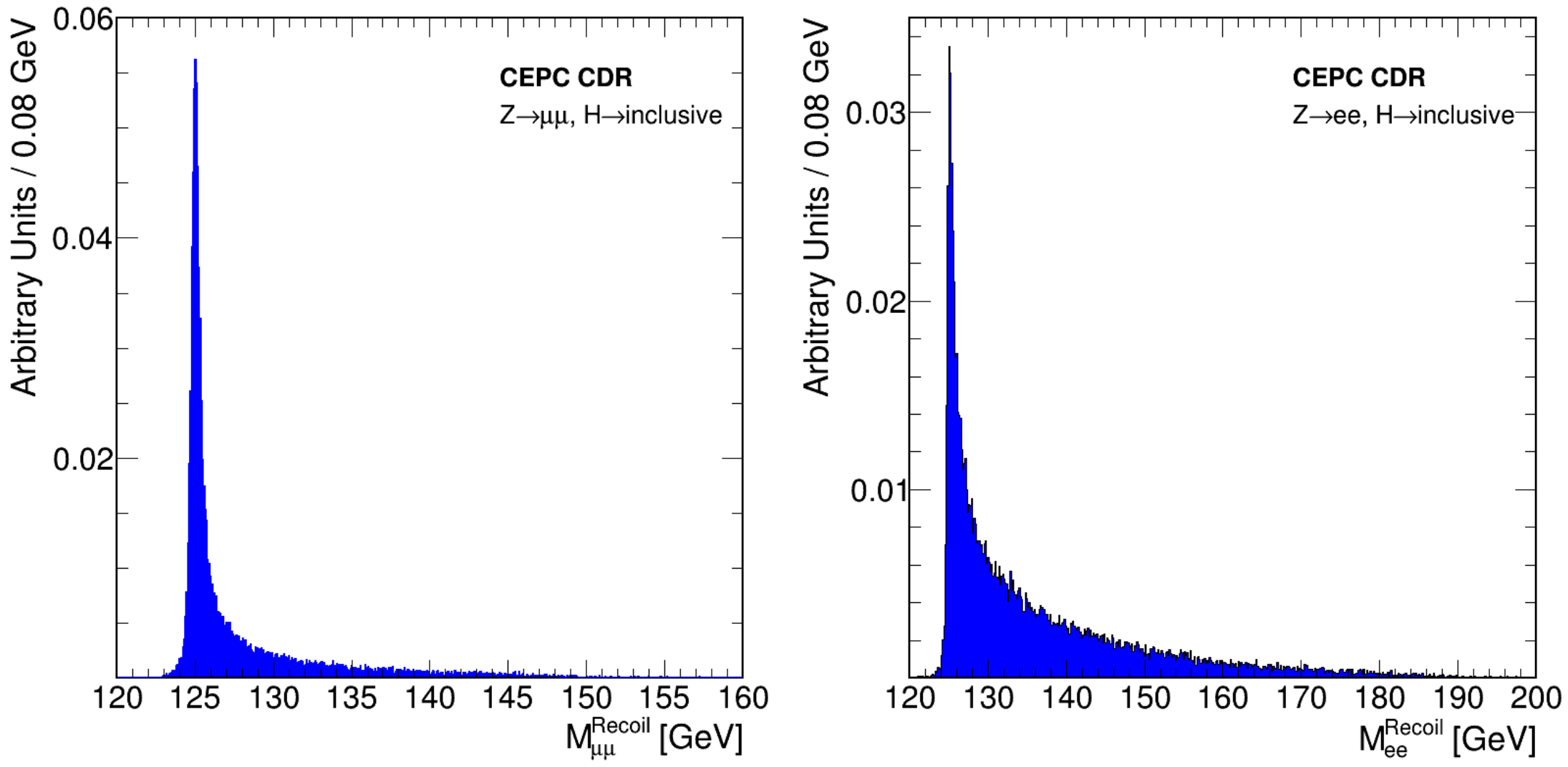

(a) (b)

**Figure 10.6:** The reconstructed dilepton recoil mass distributions of $\mu^+\mu^- H$ and $e^+e^- H$ events, both normalized to unit area. The distributions peak strongly at the Higgs boson mass. The high-mass tails are results of radiation effects (for example the initial- and final-state radiations, the beamstrahlung and the Bremsstrahlung effects). Fit with double-sided crystal ball functions, each distribution exhibits a core width of 200–300 MeV. The $e^+e^- H$ has a much more significant high mass tail, as the electrons have much stronger radiation effects compared to the muons.

## 10.2.2 PHOTONS

Photons can be produced from either initial- and final-state radiation or decays of unstable particles. Precise photon measurements are essential, for example, for studying the $H \to \gamma\gamma$ decay and counting neutrino species. Moreover, photons are a large part of secondary particles that form jets and have an important role in the $\tau$-lepton identification, they impact all aspects of the physics at the CEPC.

Photons have similar signatures as electrons in the calorimeter, but in general without matching tracks in the tracker. However, 5–10% of photons in the central region and $\sim$25% of photons in the forward region convert to $e^+e^-$ pairs through their interaction with the materials in front of the calorimeter. Some of these converted photons will have reconstructed matching tracks. Figure 10.8(a) shows the amount of material in the unit of radiation length, and Figure 10.8(b) shows the photon conversion rates at different polar angles. For unconverted photons of energies above 5 GeV, the identification efficiency is nearly 100% with more than 99% of their energy reconstructed. For the current studies, a simplistic algorithm has been used to identify converted photons. Approximately 80% of the converted photons are recovered using this algorithm. The rate of misidentifying a hadronic jet as a photon is found to be negligible.

[2]See Section 11.1.2 for the definition.

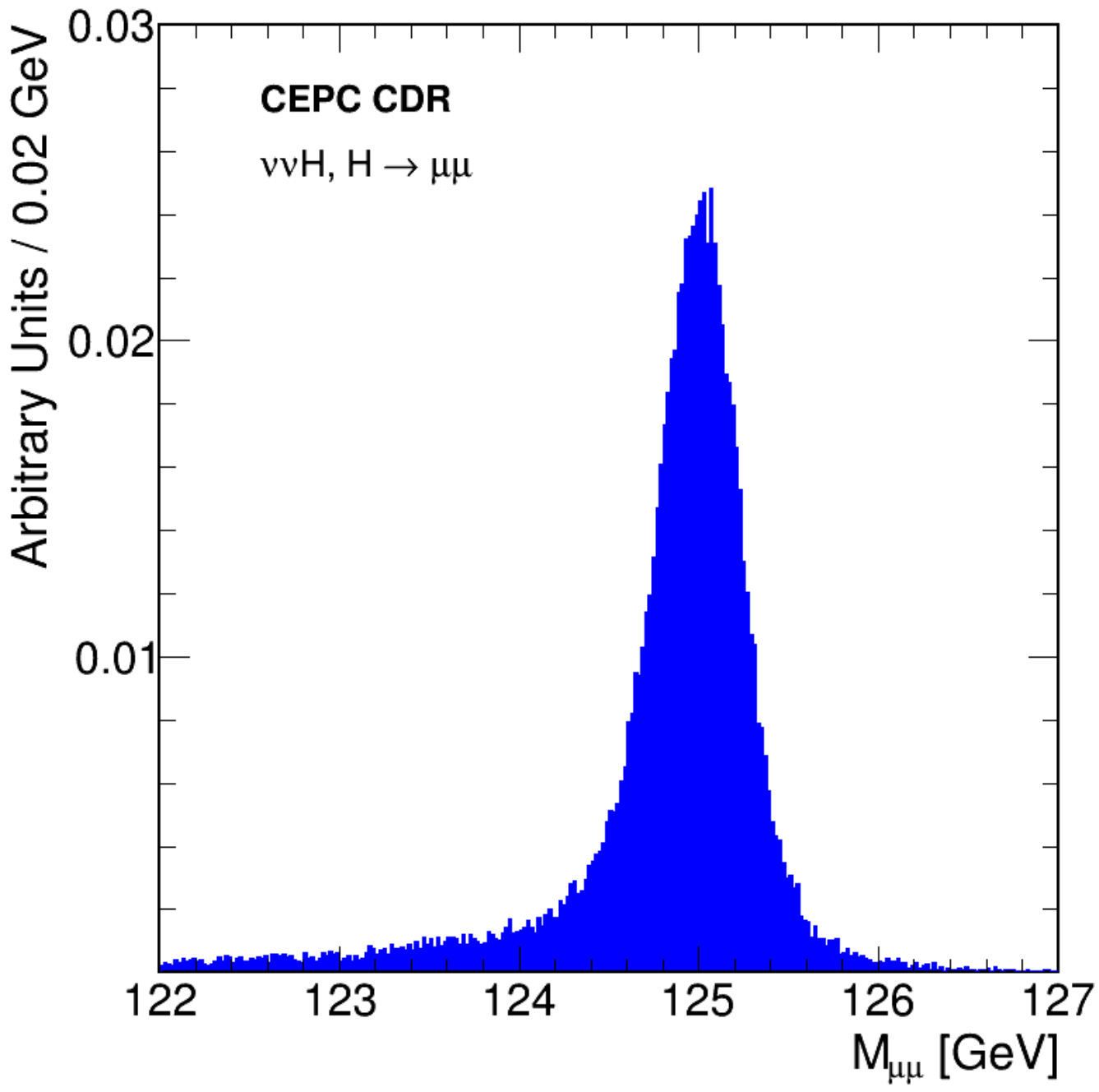

**Figure 10.7:** The reconstructed invariant mass distributions of dimuons from the $H \to \mu^+\mu^-$ decay produced in the $e^+e^- \to ZH$ process. The distribution shows a relative mass resolution of 0.19%.

Figure 10.9(a) compares the energy resolution of unconverted photons of the baseline detector concept with the intrinsic resolution of the calorimeter. The intrinsic resolution is obtained from MC simulation without material in front and gaps between modules and is consistent with the CALICE test beam result [13]. It represents the ultimate resolution of the detector. The material in the tracker and geometric inhomogeneities are the main causes for the degradation in the resolution of the baseline detector. These effects are currently not taken into account in the calibration. The resolution, both the sampling and constant terms, is expected to improve significantly once geometry dependent calibration is implemented. The photon energy resolution can be benchmarked using the diphoton mass distribution of the $H \to \gamma\gamma$ decay as shown in Figure 10.9(b). The width of the mass distribution is dominated by the energy resolution effect because of the narrow intrinsic Higgs boson width. The current diphoton mass resolution is approximately 2.5% compared with 1.7% of the intrinsic ECAL resolution.

### 10.2.3 TAU LEPTONS

As the heaviest lepton, $\tau$-leptons have a unique role in studying Higgs boson physics. Leptonic decays of $\tau$-leptons, $\tau \to e\nu\nu$ and $\tau \to \mu\nu\nu$, are indistinguishable from electrons or muons. Hadronic decays of the $\tau$-leptons appear in the detector as narrow pencil-shaped hadronic jets with low particle multiplicity. A basic $\tau$-lepton identification algorithm has been developed for hadronic decays. The algorithm starts with a seed track with its energy above 1.5 GeV and clusters charged and neutral particles in a small cone of radius of 0.12 radians around it to form the $\tau$-lepton candidate. The invariant mass formed by the particles in the cone is required to be in the range of 0.01–2 GeV, consistent with the $\tau$-lepton mass. Furthermore, a discriminant variable based on the longitudinal and transverse impact parameters of the leading track is constructed and the variable is required to be consistent with the non-zero lifetime of the $\tau$-lepton. Finally, the $\tau$-lepton candidate is required to be isolated. The total energy in an annular cone or radius between 0.12–0.31 radians is required to be less than 8% of the $\tau$-lepton candidate energy. The main backgrounds are hadronic jets.

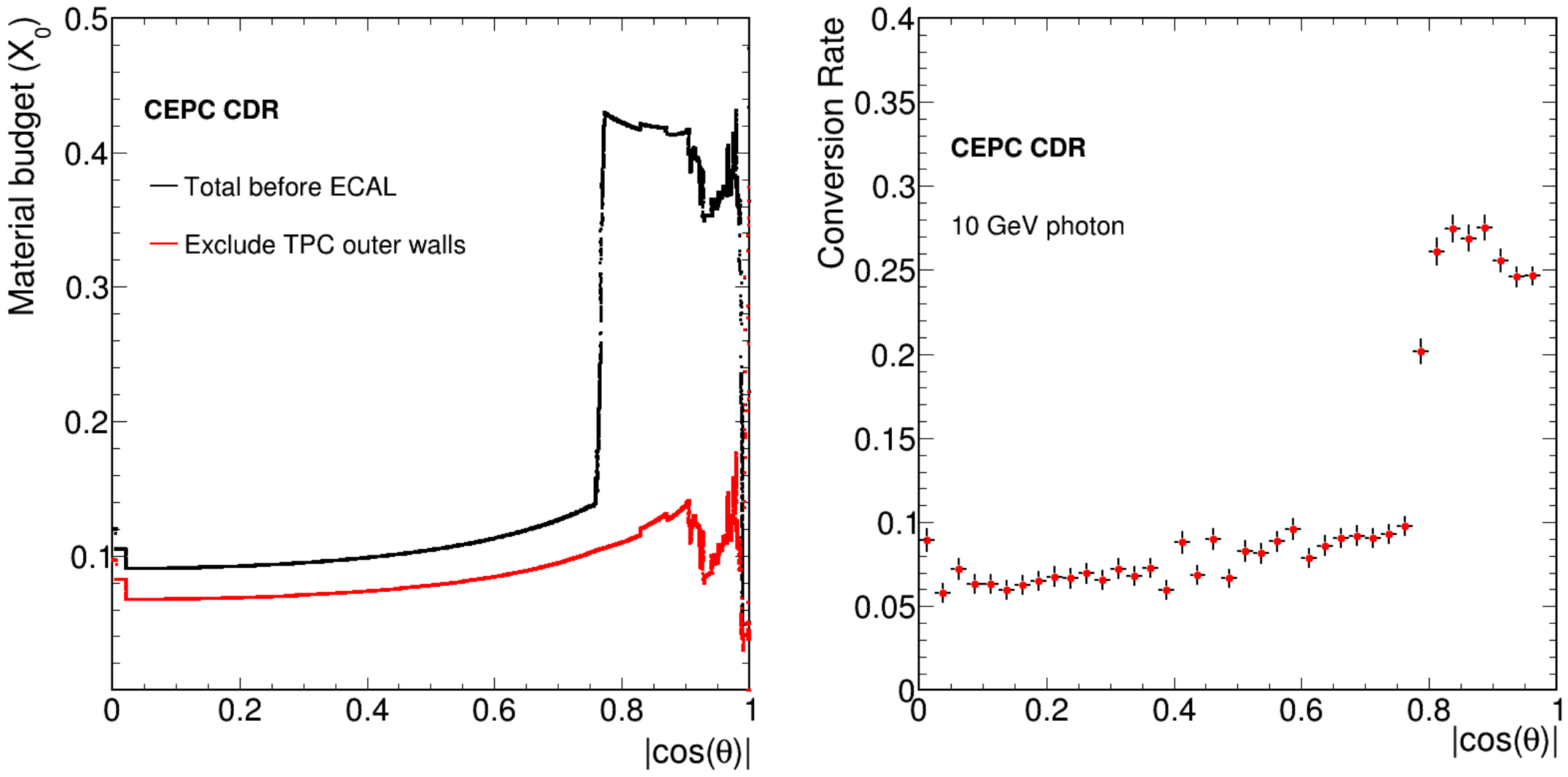

(a) (b)

**Figure 10.8:** (a) The amount of material in the unit of radiation length inside the tracker and (b) the conversion rate of 10 GeV photons for different polar angles.

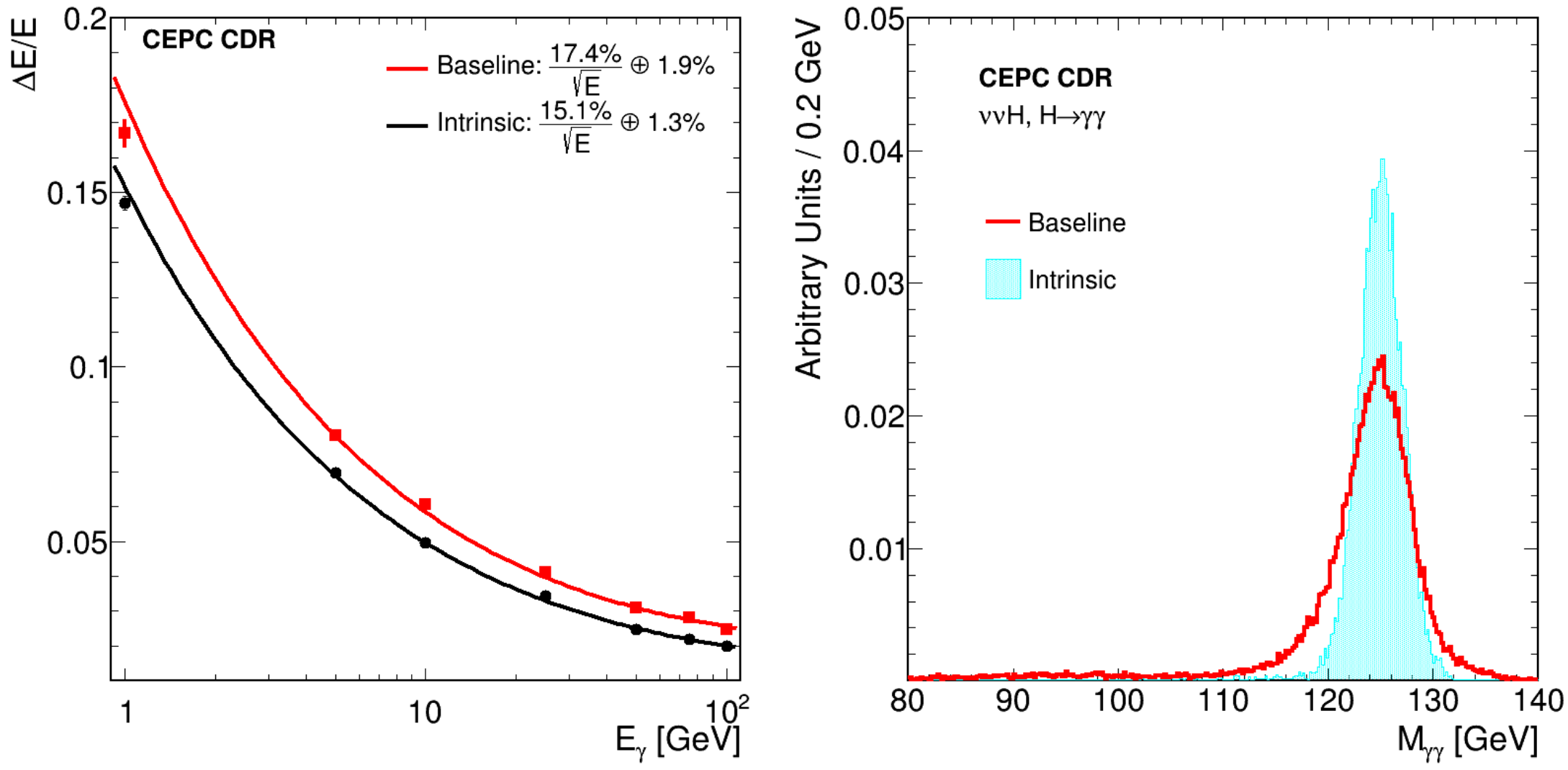

(a) (b)

**Figure 10.9:** (a) The energy resolution of unconverted photons as a function of energy of the baseline detector compared with the intrinsic resolution of the calorimeter and (b) the invariant mass distribution of diphotons from the $H \to \gamma\gamma$ decay of the $e^+e^- \to ZH$ process. The resolution of the baseline detector is expected to improve with further optimizations and geometry-dependent calibrations which should bring it closer to the intrinsic resolution of the calorimeter.

Figure 10.10(a) is a graphic representation of the $\tau$-lepton identification. The efficiency and the purity as functions of the visible energy of the $\tau$-lepton candidate are shown in Figure 10.10(b), measured from the $e^+e^- \to ZH$ events with the $Z \to q\bar{q}$ and $H \to \tau^+\tau^-$ decays. For visible energy between 20–80 GeV, the efficiency is above 80% and the purity is close to 90%. The loss of efficiency is largely due to the large cone size used for the isolation requirement. Improvement in performance can be expected from further optimizations.

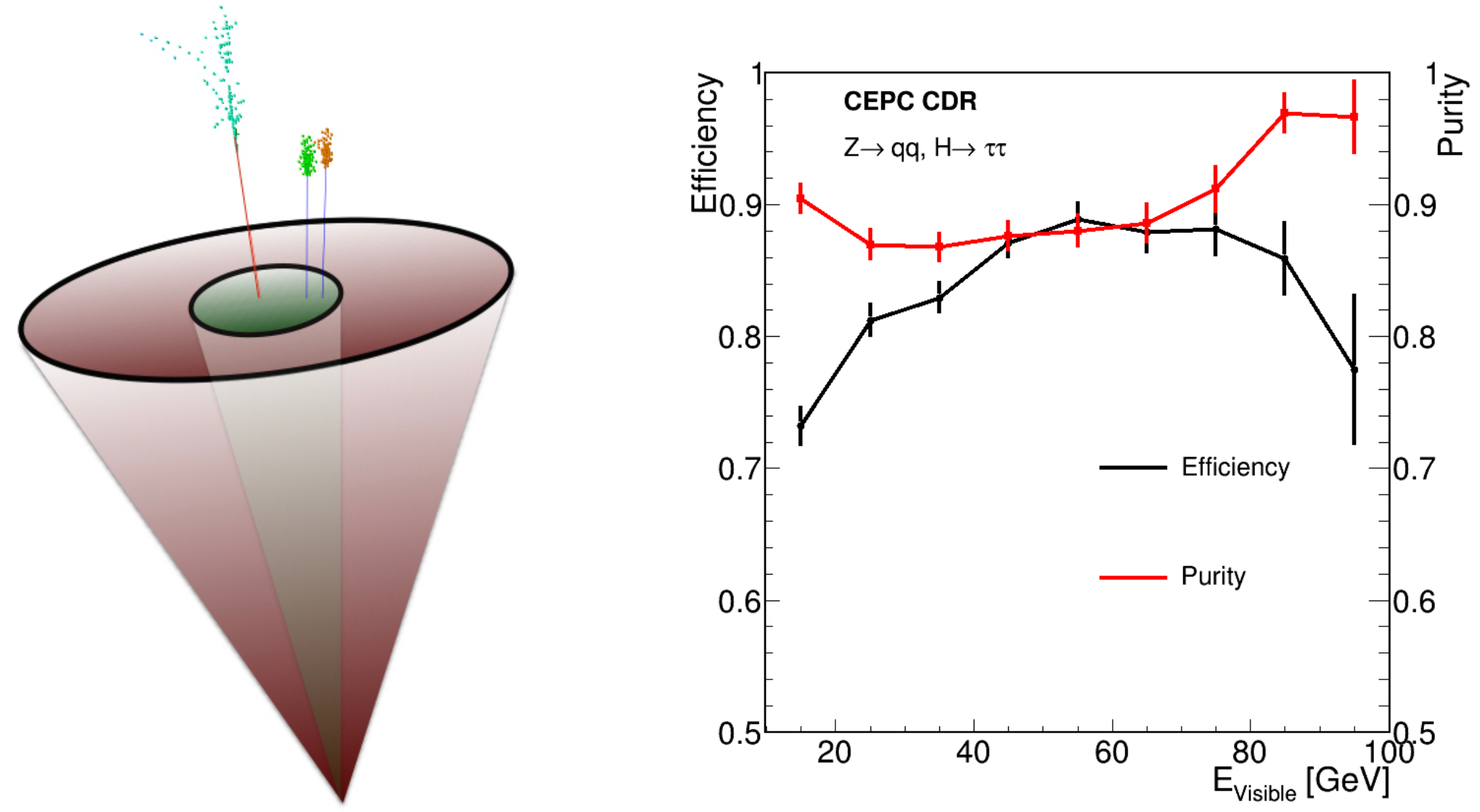

(a) (b)

**Figure 10.10:** (a) Illustration of the $\tau$-lepton identification and (b) the efficiency and purity as functions of the visible energies of the $\tau$-lepton candidates. Both the efficiency and purity are determined from $e^+e^- \to ZH \to q\bar{q}\tau^+\tau^-$ events at $\sqrt{s} = 240$ GeV.

### 10.2.4 JETS

The vast majority of the events produced at the CEPC have hadronic jets in their final states. For example, 70% of the Higgs bosons decay directly to a pair of jets and another 20% decay indirectly to jets through intermediate $W$ or $Z$ bosons. Coincidentally, about 70% of the $W$ or $Z$ bosons each decays to dijets. Thus the impact of jets to the CEPC physics program cannot be overstated.

Jets are formed from particles reconstructed by ARBOR using the Durham clustering algorithm [14]. The ambiguity in clustering is the leading source of uncertainty in jet reconstruction and measurements, particularly in events with closely spaced physics objects.

Jet energies are foreseen to be calibrated through a two-step process. First, calibrations are applied to particles identified by ARBOR. While the energies of the charged particles are determined by their track momenta, the energies of neutral particles are currently calibrated using MC simulation and can be calibrated using the test beam or collision data when they are available. Approximately 35% of the jet energy is carried by neutral particles. In the second step, the jet energy are calibrated using physics events. At the CEPC, $W$ and/or $Z$ bosons are copiously produced and can be identified with high efficiency and purity. Thus $W \to q\bar{q}$ and $Z \to q\bar{q}$ decays serve as standard candles for the jet energy calibration. Clean samples of $WW \to \ell\nu q\bar{q}$ of the Higgs factory and $WW$ threshold scan operations, $ZZ \to \nu\bar{\nu}q\bar{q}$ of the Higgs factory operation, and $Z \to q\bar{q}$ of the $Z$ factory operation can be selected. The enormous statistics allows the jet response to be characterized in detail.

Figure 10.11(a) shows energy ratios between the reconstructed jets and MC particle jets for different polar angles derived from the simulated $ZZ \to \nu\bar{\nu}q\bar{q}$ events. The ratios are close to unity and the corrections are $< 1\%$. The jet energy resolution is shown in Figure 10.11(b) as a function of jet energy for different jet flavors. For light jets, the resolution ranges from 6% at 20 GeV to 3.6% at 100 GeV. The resolutions for heavy-flavor jets are poorer as expected because of neutrinos in their decays. Major factors affecting the jet energy scale and/or resolution are jet flavor composition, shower fluctuations, clustering algorithm as well as the stability and uniformity of the detector responses. Their impacts can be minimized by detailed studies and calibrations. A sub-percent level jet energy scale precision and a jet energy resolution of 3–5% for the jet energy range of 20–100 GeV should be achievable.

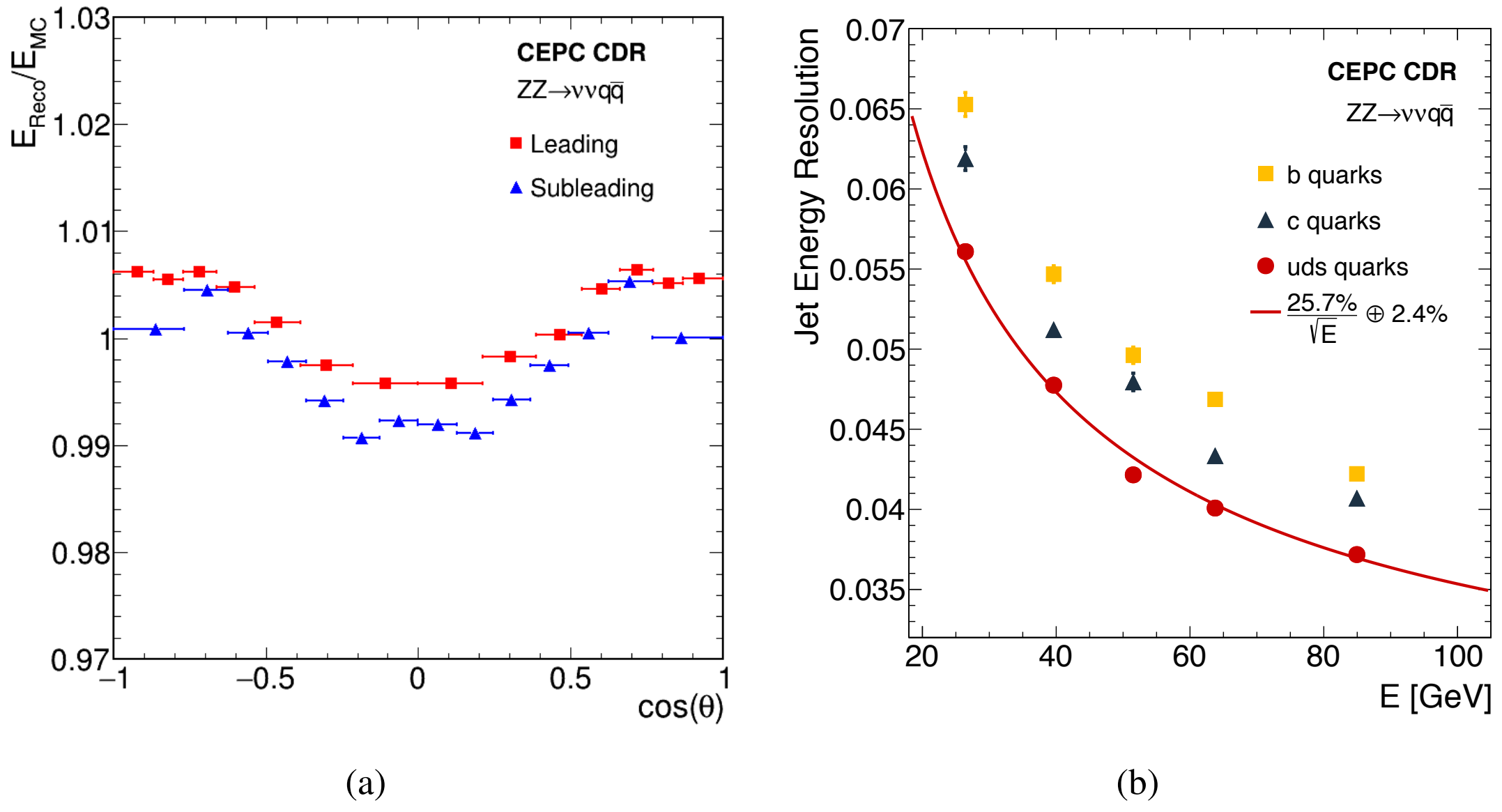

**Figure 10.11:** (a) The energy ratios between the reconstructed jets and MC particle jets as functions of cosine of their polar angles and (b) jet energy resolution as a function of jet energy for different jet flavor categories. These distributions are derived from the simulated $ZZ \to \nu\bar{\nu}q\bar{q}$ events. The energy ratios are shown for leading and subleading jets separately.

One key jet performance measure is the ability to separate hadronic decays of $W$, $Z$, and Higgs bosons. Figure 10.12(a) compares the reconstructed dijet invariant mass distributions from $W \to q\bar{q}$, $Z \to q\bar{q}$ and $H \to b\bar{b}/c\bar{c}/gg$ decays of $WW \to \ell\nu q\bar{q}$, $ZZ \to \nu\bar{\nu}q\bar{q}$ and $ZH \to \nu\bar{\nu}(b\bar{b}/c\bar{c}/gg)$ processes, respectively. Compared with $W \to q\bar{q}$, the $Z \to q\bar{q}$ and $H \to b\bar{b}/c\bar{c}/gg$ distributions have longer low-mass tails. These tails are from the heavy-flavor jets as demonstrated in Figure 10.12(b) where the distributions from $H \to b\bar{b}$, $H \to c\bar{c}$ and $H \to gg$ decays are separately shown and compared. The $H \to gg$ distribution is symmetric and has the best mass resolution (at approximately 3.8%) whereas the $H \to b\bar{b}$ decay has a long asymmetric low-mass tail and therefore degraded mass resolution. The degradations in resolution and the distortions in the mass distributions for the $H \to b\bar{b}$ and $H \to c\bar{c}$ decay are expected from neutrinos produced in semi-leptonic decays of $b$- and $c$-quarks. The mass resolutions for $W \to q\bar{q}$ and $Z \to q\bar{q}$ are 4.4%, leading to an average separation of $2\sigma$ or better for the hadronically decaying $W$ and $Z$ bosons.

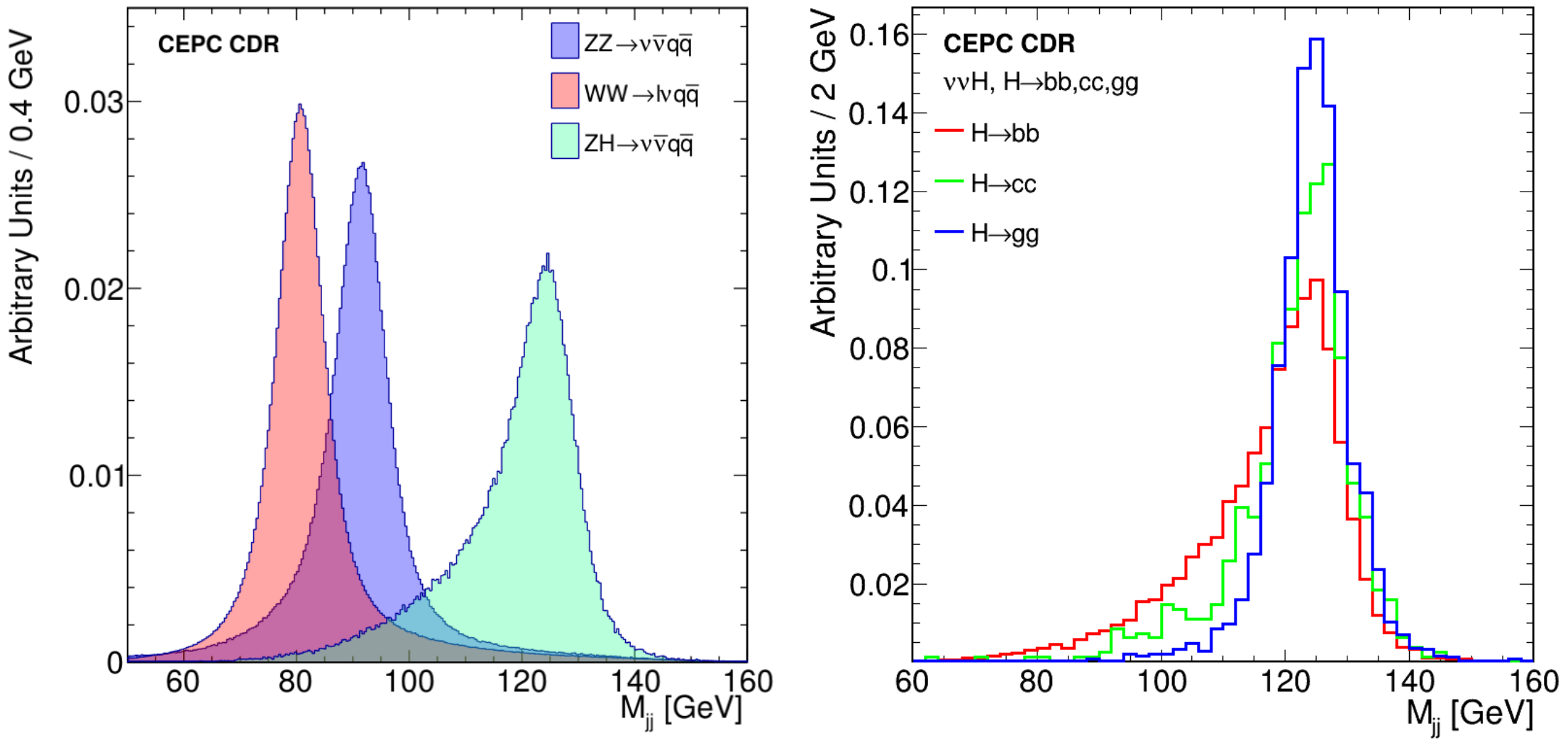

**Figure 10.12:** Reconstructed dijet mass distributions of (a) the $W \to q\bar{q}$, $Z \to q\bar{q}$ and $H \to b\bar{b}/c\bar{c}/gg$ decays from the $ZZ \to \nu\bar{\nu}q\bar{q}$, $WW \to \ell\nu q\bar{q}$ and $ZH \to \nu\bar{\nu}(b\bar{b}, c\bar{c}, gg)$ processes, respectively and (b) the separate $H \to b\bar{b}$, $H \to c\bar{c}$ and $H \to gg$ decays from the $ZH \to \nu\bar{\nu}(b\bar{b}/c\bar{c}/gg)$ process. All distributions are normalized to unit area.

## 10.2.5 JET FLAVOR TAGGING

Identification, *i.e.* tagging, of jet flavors is essential for the measurements of the Higgs boson couplings and the electroweak observables at the CEPC. Heavy-flavor quarks ($b$ and $c$) from $W$, $Z$ or Higgs boson decays hadronize quickly to form heavy bottom and charm hadrons ($B^0$, $B^{\pm}$, $B_s$, $D^0$, $D^{\pm}$, ...). Those hadrons are short-lived and have typical decay distances of a few millimeters. Therefore, the reconstruction of their decay vertices, often referred to as secondary vertices, is an important tool for tagging jet flavors. Other information such as jet and vertex mass, impact parameters and leptons inside the jets, are also frequently used to differentiate heavy-flavor jets from light-quark and gluon jets. For example, the excellent impact parameter resolution shown in Figure 4.3 can significantly improve the detector's capability for jet flavor tagging.

The jet flavor tagging is performed using LCFIPlus [15], the tagging algorithm used for linear collider studies. LCFIPlus reconstructs secondary vertices from the final-state particles identified by ARBOR. It combines more than 60 discriminant variables to calculate the $b$-likeliness ($L_B$) and $c$-likeliness ($L_C$) using a Boosted Decision Tree [16] method. Compared with the $b$-jet tagging, $c$-jet tagging is particularly challenging as charm hadrons have shorter lifetimes than bottom hadrons and therefore suffers more from backgrounds from light-quark and gluon jets. Benefiting from the high precision vertex system, the CEPC detector provides reasonable separation of $c$-jets from other flavor jets. Figure 10.13 shows the $b$-jet tagging efficiencies for different rejections of background jets, measured from a $Z \to q\bar{q}$ sample of the $Z$ factory operation. For this sample, $b$-jets can be tagged with an efficiency of 80% and a purity of $90\%$. Similarly, an efficiency of $60\%$ and a purity of $60\%$ can be achieved for the $c$-jet tagging. Purities can be improved by tightening the tagging requirements at the expense of reduced efficiencies. Figure 10.14 is a demonstration of the $b/c$-likeliness distributions of the $b$, $c$ and gluon jets from the $H \to b\bar{b}/c\bar{c}/gg$ decays, showing good separations between jets of different flavors.

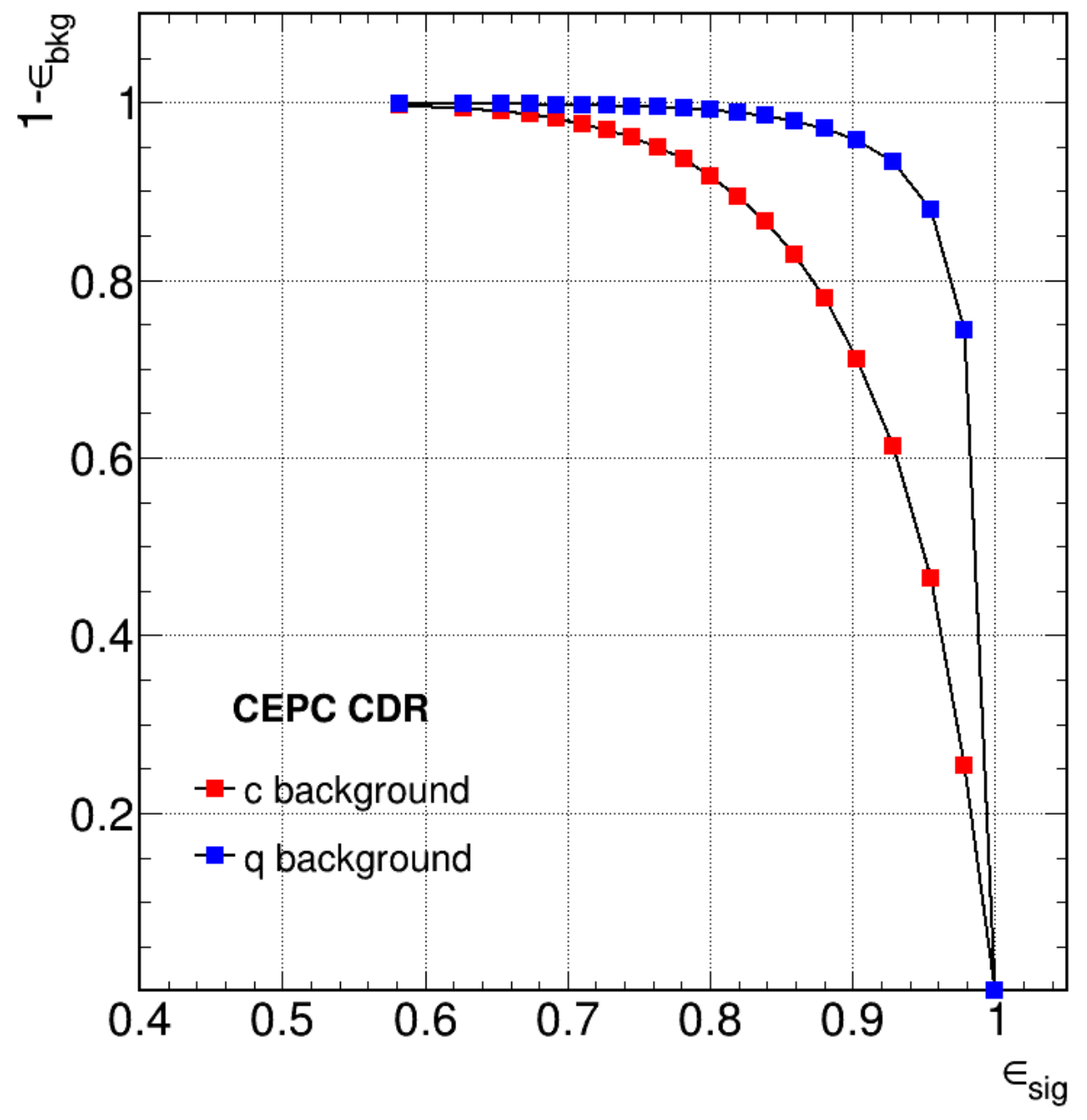

**Figure 10.13:** Efficiencies for tagging $b$-jets versus rejection of background jets, determined from an inclusive $Z \to q\bar{q}$ sample at the $Z$ factory operation.

### 10.2.6 MISSING ENERGY, MOMENTA AND MASSES

Neutrinos interact weakly with the detector and for all practical purposes escape detection without traces. The same is true for the hypothesized dark matter particles. However, their existences can be inferred from detectable ("visible") particles. The total energy and momentum of these "missing" particles, missing energy and momentum as they are usually called, can be calculated from the energies and momenta of visible particles through the energy-momentum conservation. In spite of their elusive nature, neutrinos are as important as visible particles for the CEPC physics program. About $20\%$ of the $Z$ bosons and $30\%$ of the $W$ bosons decay directly into final states with neutrinos. Searching for Higgs boson decays to dark matter particles is a key physics goal of the Higgs factory.

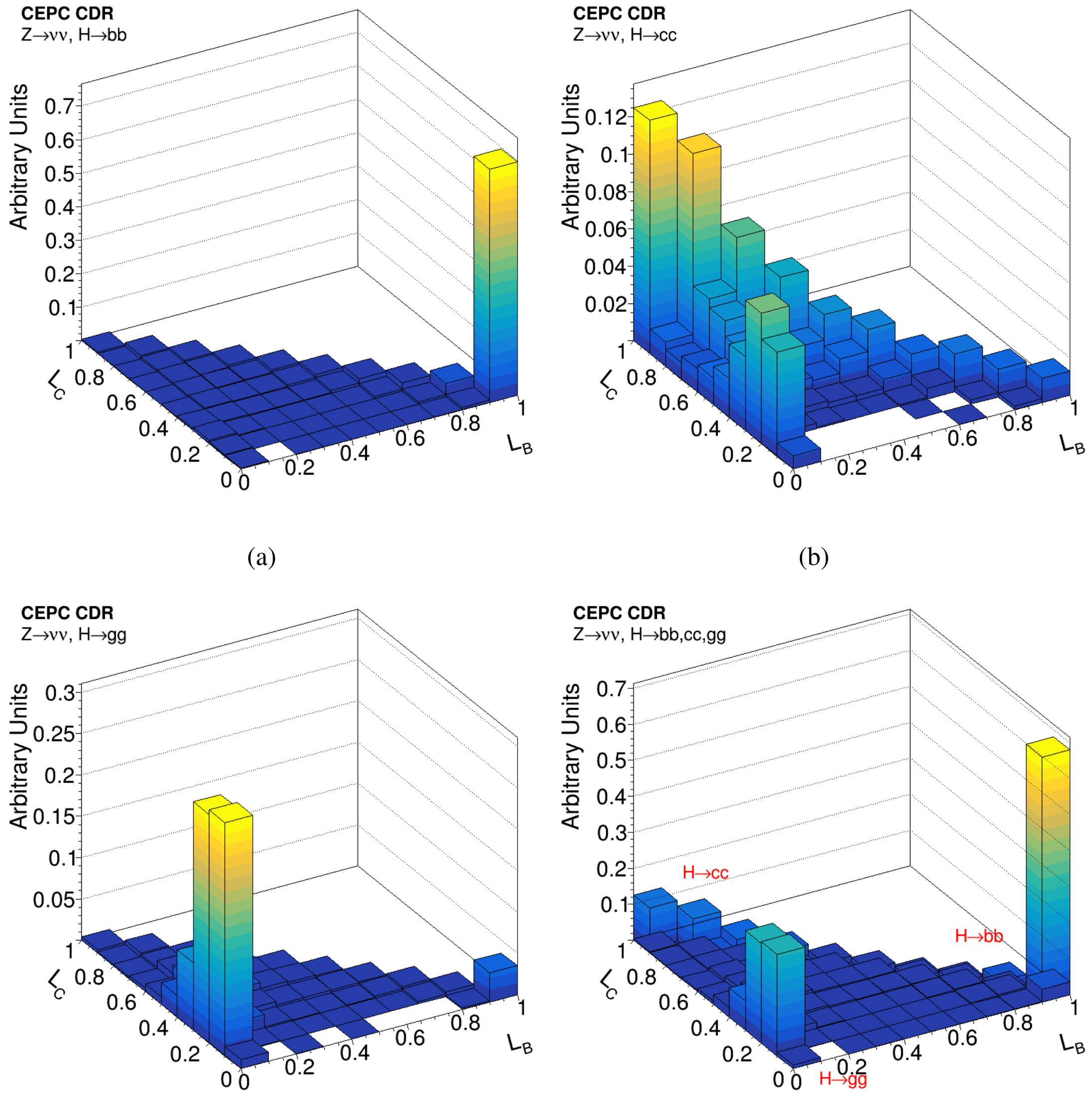

(a) (b)

(c) (d)

**Figure 10.14:** The two-dimensional distributions of $b$-likeliness $L_B$ and $c$-likeliness $L_C$ of jets from the $H \to b\bar{b}$, $H \to c\bar{c}$ and $H \to gg$ decays showing separately (a,b,c) and combined (d). Distributions of individual decays (a,b,c) are normalized to unit volume while the combined distribution is the sum of the three individual distributions, weighted by their branching ratios.

The excellent energy and momentum resolutions of the CEPC baseline conceptual detector for visible particles allow for the determinations of missing energy and momentum with good precision. This is demonstrated using $e^+e^- \to ZH$ events in Figure 10.15 which shows the missing mass distributions of events from, respectively, $(Z \to q\bar{q},\ H \to \mathrm{inv})$ and $(Z \to \nu\bar{\nu},\ H \to b\bar{b}/c\bar{c}/gg)$ decays. The missing mass, calculated from the missing energy and momentum, is the invariant mass of the system of undetected particles. The missing mass distribution peaks at the Higgs boson mass for the $H \to \mathrm{inv}$ decay and at the $Z$ boson mass for the $Z \to \nu\bar{\nu}$ decay, as expected. Contributions from different jet flavors are shown separately. The Higgs and $Z$ boson masses can be determined from missing masses with good precision, allowing their identification without direct detection.

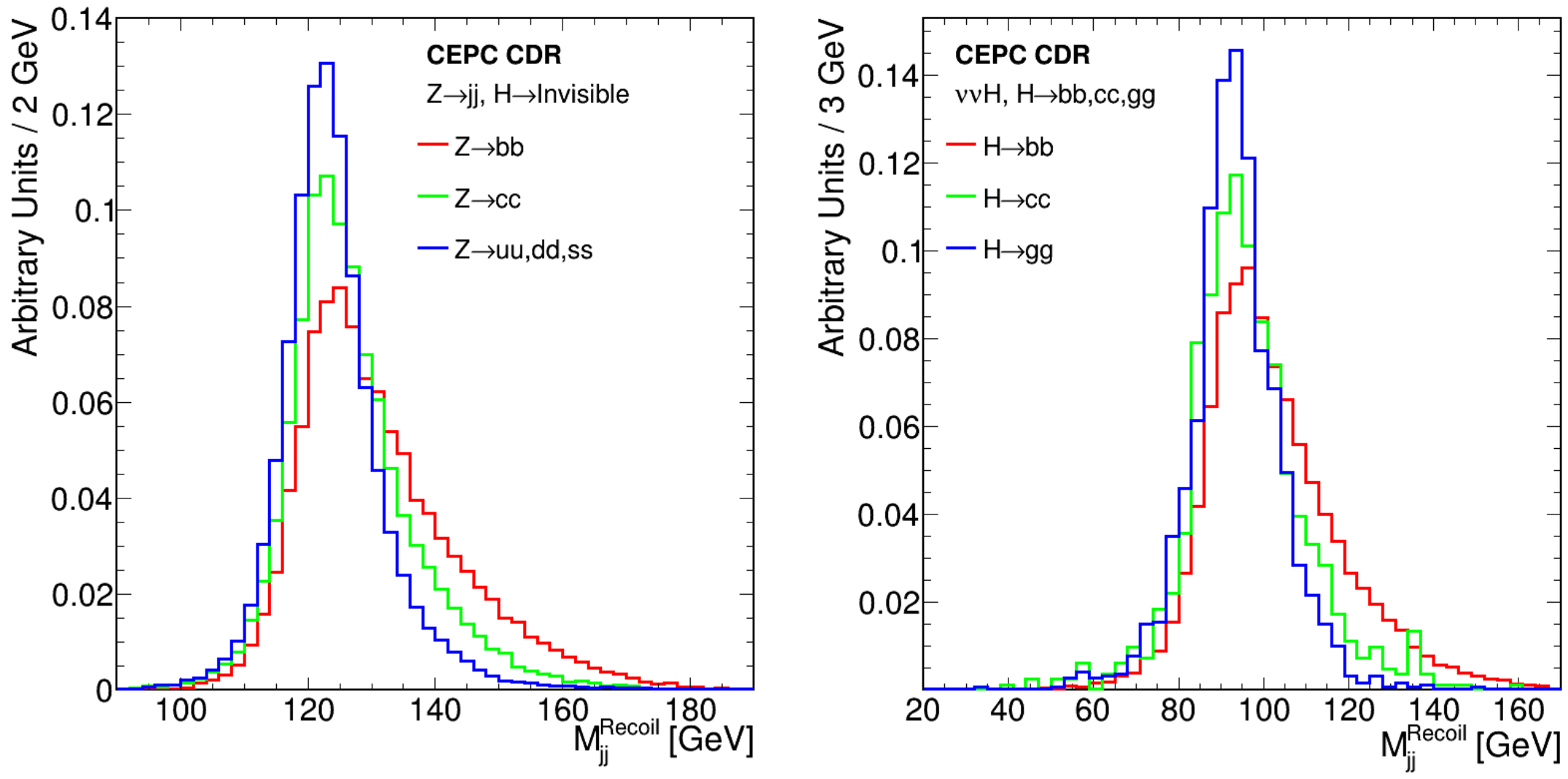

(a) (b)

**Figure 10.15:** The dijet recoil mass (*i.e.*, the missing mass) distributions of $e^+e^- \to ZH$ events in the (a) $Z \to q\bar{q}$, $H \to$ inv and (b) $Z \to \nu\bar{\nu}$, $H \to b\bar{b}/c\bar{c}/gg$ decay, separately for different jet flavors. All distributions are normalized to unit area. The light-jet distribution in (a) has a Gaussian core width of approximately 6 GeV and the corresponding width of the $H \to gg$ distribution in (b) is 8 GeV.

### 10.2.7 KAON IDENTIFICATION

Successful identification of charged kaons will greatly benefit the flavor physics program and aid in the determination of jet flavor as well as jet charge. The $dE/dx$ information from the TPC can be used to separate kaons from pions. Assuming a relative $dE/dx$ resolution of 5%, the measurement could lead to 2–4 $\sigma$ separation of $K/\pi$ for momentum between 2–20 GeV as shown in Figure 10.16.

The discriminating power of $dE/dx$ vanishes for pions and kaons with their momenta around 1 GeV. Meanwhile, a significant portion of the charged particles has an energy smaller than 2 GeV at the CEPC. To aid the separation of these low momentum charged particles, it has been proposed to add a ToF capability with a 50 ps resolution to the detector design. The ECAL could be instrumented with a few layers of time sensitive readout to provide the ToF information. Using both the ToF and $dE/dx$ information, a separation better than $2\sigma$ could be achieved for charged particles with momenta up to 20 GeV in the conservative scenario as shown in Figure 10.16(b). For the inclusive $Z \to q\bar{q}$ sample, charged kaons can be identified with an efficiency of 91% and a purity of 94%, integrated over the momentum range of 2–20 GeV.

## 10.3 SUMMARY

Precise measurements of the Higgs boson properties and the electroweak observables at the CEPC place stringent requirements on the performance of the CEPC detector to iden-

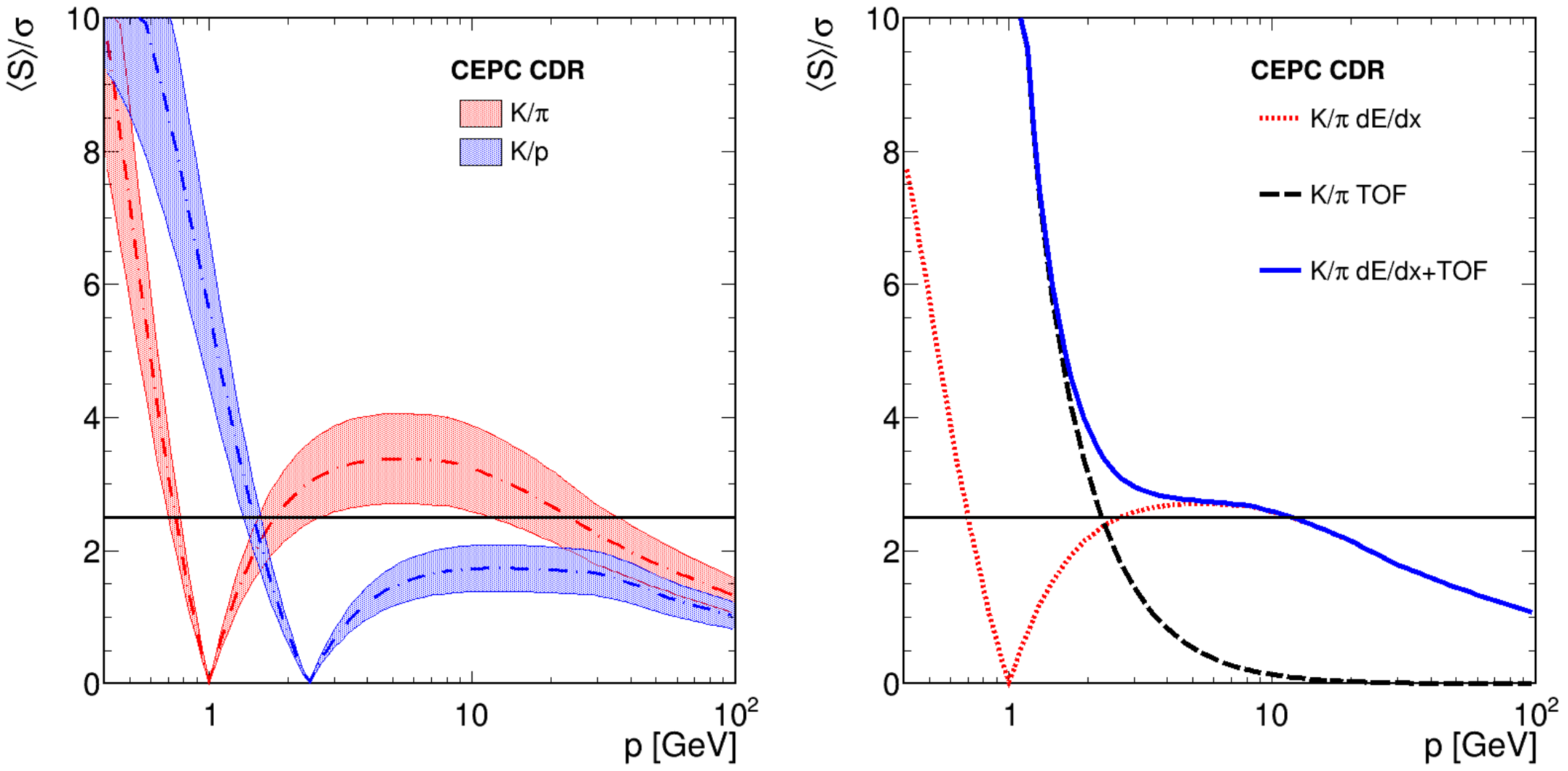

**Figure 10.16:** Charged particle identification: (a) $K/\pi$ and $K/p$ separations from the $dE/dx$ measurement in the TPC and (b) the $K/\pi$ separation from both the $dE/dx$ measurement and the proposed ToF information. The upper boundaries of the bands in (a) are the ideal separations predicted by the GEANT4 simulation while the lower boundaries correspond to conservative estimates with a 50% degradation in performance.

tify and measure physics objects such as leptons, photons, $\tau$-leptons, jets and their flavors with high efficiencies and purities as well as high precision. The performance of the CEPC baseline detector have been investigated with full simulation. Benchmark performance are described above and are briefly summarized below:

1 Leptons: an efficiency of $> 99.5\%$ with a mis-identification rate of $< 1\%$ for electrons and muons with momenta above 2 GeV, a relative mass resolution of 0.19% for the $H \to \mu^+\mu^-$ decay;

2 Photons: an efficiency of nearly 100% with negligible mis-identification rate from hadronic jets for unconverted photons above 5 GeV, relative mass resolution of 2.5% for the $H \to \gamma\gamma$ decay;

3 $\tau$-leptons: an efficiency of 80% or higher and a purity close to 90% measured from the $e^+e^- \to ZH \to q\bar{q}\tau^+\tau^-$ events at $\sqrt{s} = 240$ GeV;

4 Jet energy scale and resolution: the jet energy scale can be measured with a sub-percent accuracy and a jet energy resolution of 3–5% is achievable for the energy range relevant at the CEPC, enabling a $2\sigma$ or better separation of the $W \to q\bar{q}$ and $Z \to q\bar{q}$ decays;

5 Jet flavor tagging: efficiency/purity of 80%/90% for the $b$-jets tagging and 60%/60% for the $c$-jets tagging can be achieved for the $Z \to q\bar{q}$ sample of the $Z$ factory operation;

6 $K^\pm$ identification: kaons can be separated from pions at $2\sigma$ for momentum up to 20 GeV, corresponding to efficiency/purity of 95%/95% for identifying kaons in the $Z \to q\bar{q}$ sample integrated over the momentum range of 2–20 GeV.

Though significant progress has been made in understanding and characterizing the detector performance, the performance results can be further enhanced with improved algorithms and better calibrations. Nevertheless, the performance results as currently understood are sufficient to fulfill the requirements laid out in Chapter 3 and to meet the physics analysis needs as presented in Chapter 11.

**CHAPTER 11**

# PHYSICS PERFORMANCE WITH BENCHMARK PROCESSES

The historic discovery of a Higgs boson in 2012 by the ATLAS and CMS collaborations [1, 2] and the subsequent studies of the properties of the particle [3–9] indicate compatibility with the SM predictions within the current measurement uncertainties. The Higgs boson completes the list of fundamental particles in the SM. However, the discovery creates an inexplicable foundation for the SM theory. The origin and stability of the vast difference between the Planck and ElectroWeak (EW) scales, the nature of the electroweak phase transition, whether the Higgs boson couples to dark matter, and other fundamental questions remain to be understood in the future. The attempt to further address those questions will involve new physics beyond the SM which could lead to deviations from SM expectations when tested with precision measurements. A circular electron positron collider will provide a unique opportunity to perform precision measurements of the properties of the Higgs, $W$ and $Z$ bosons.

The CEPC will produce and record a huge number of Higgs, $W$ and $Z$ bosons. In this Chapter, its physics potential is demonstrated using two different classes of physics benchmarks, Higgs boson physics and precision EW physics. Using the software tools introduced in Section 10.1, the potential for Higgs boson physics is examined using simulation studies, see Section 11.1. The accuracy on the EW precision measurements is mainly limited by systematic uncertainties, which are examined in Section 11.2. The synergies of these different physics measurements, the complimentarity and comparison to the HL-LHC and other future high energy physics programs are discussed as well.

## 11.1 HIGGS BOSON PHYSICS

The Higgs boson is responsible for the electroweak symmetry breaking. It is the only fundamental scalar particle observed so far. The discovery of such a particle at the LHC was a major theoretical and experimental breakthrough. However, the SM is likely only an effective theory at the electroweak scale. To explore potential new physics at the electroweak scale and beyond, complementary approaches of direct searches at the energy frontier as well as precision measurements will be needed. The current LHC and the planned HL-LHC have the potential to significantly extend its new physics reach and to measure many of the Higgs boson couplings with precision of a few percent in a model-dependent way.

In contrast to the LHC, Higgs boson candidates can be identified through a technique known as the recoil mass method without looking at the Higgs boson decays themselves at the CEPC. Therefore, Higgs boson production can be disentangled from its decay in a model independent way. Moreover, the cleaner environment at a lepton collider allows a much better exclusive measurement of Higgs boson decay channels. All of these give the CEPC an impressive reach in probing Higgs boson properties. In this section, the results of the current CEPC simulation studies on the precision of the Higgs boson property measurements are summarized. In addition, the potential sensitivity to the $CP$ properties of the Higgs boson is also discussed. More details can be found in Ref. [10].

### 11.1.1 HIGGS BOSON PRODUCTION AND DECAY

Production processes for a 125 GeV SM Higgs boson at the CEPC are $e^+e^- \to ZH$ ($ZH$ or Higgsstrahlung), $e^+e^- \to \nu_e\bar{\nu}_e H$ ($\nu\bar{\nu}H$ or $W$ fusion) and $e^+e^- \to e^+e^-H$ ($eeH$ or $Z$ fusion) as illustrated in Figure 11.1. In the following, the $W$ and $Z$ fusion processes are collectively referred to as the vector-boson fusion (VBF) production.

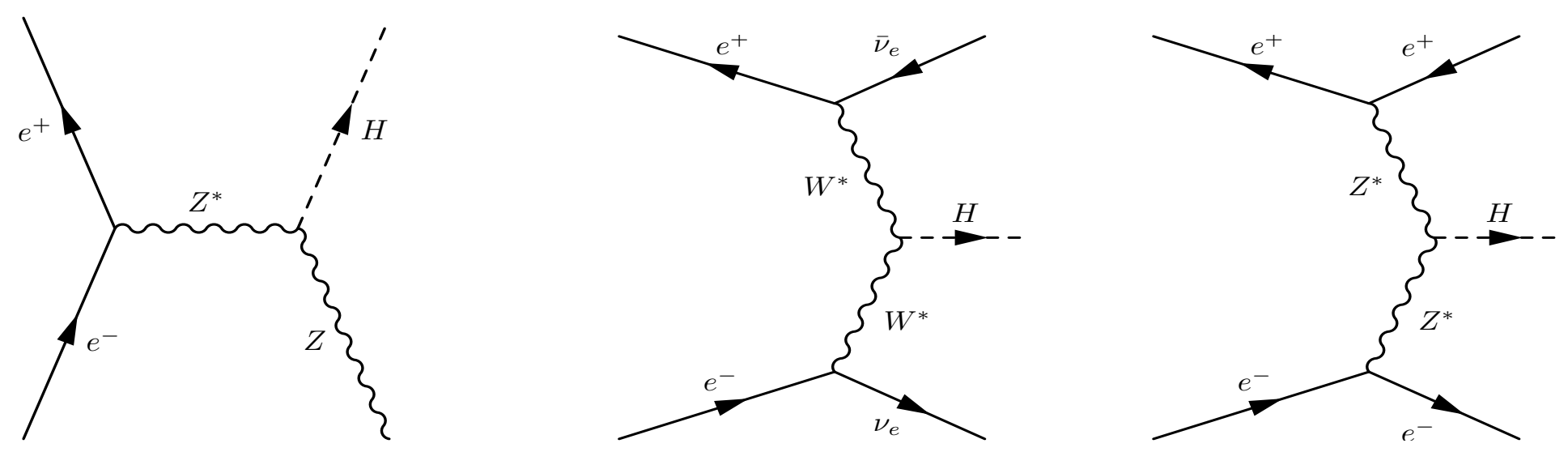

(a) (b) (c)

**Figure 11.1:** Feynman diagrams of the Higgs boson production processes at the CEPC: (a) $e^+e^- \to ZH$, (b) $e^+e^- \to \nu_e\bar{\nu}_e H$ and (c) $e^+e^- \to e^+e^-H$.

The SM Higgs boson production cross sections as functions of the center-of-mass energy are shown in Figure 11.2, assuming a Higgs boson mass of 125 GeV. Similarly, the Higgs boson decay branching ratios and natural width are shown in Table 11.1. As an $s$-channel process, the cross section of the $e^+e^- \to ZH$ process reaches its maximum at $\sqrt{s} \sim 250$ GeV, and then decreases asymptotically as $1/s$. The VBF production process proceeds through $t$-channel exchange of vector bosons and its cross section increases logarithmically as $\ln^2(s/M_V^2)$. Because of the small neutral-current $Zee$ coupling, the VBF cross section is dominated by $W$ fusion.

Numerical values of these cross sections at $\sqrt{s} = 240$ GeV are listed in Table 11.2. Because of the interference effects between $e^+e^- \to ZH$ and $e^+e^- \to \nu_e\bar{\nu}_e H$ for the $Z \to \nu_e\bar{\nu}_e$ decay and between $e^+e^- \to ZH$ and $e^+e^- \to e^+e^-H$ for the $Z \to e^+e^-$ decay, the cross sections of these processes cannot be separated. The breakdowns in Figure 11.2 and Table 11.2 are for illustration only. The $e^+e^- \to ZH$ cross section shown is from Figure 11.1(a) only whereas the $e^+e^- \to \nu_e\bar{\nu}_e H$ and $e^+e^- \to e^+e^-H$ cross sections include contributions from their interferences with the $e^+e^- \to ZH$ process.

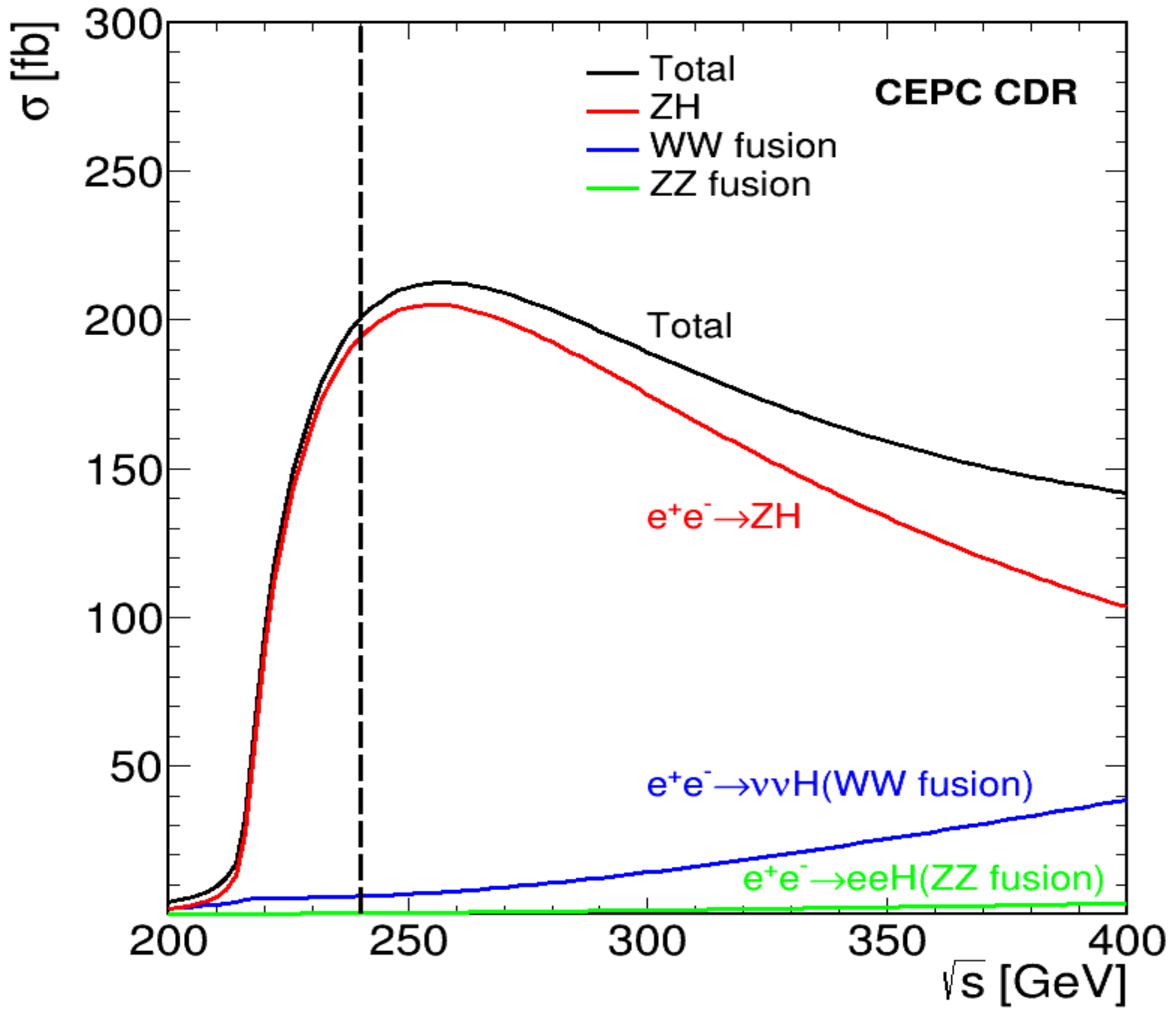

**Figure 11.2:** Production cross sections of $e^+e^- \to ZH$ and $e^+e^- \to (e^+e^-/\nu\bar{\nu})H$ as functions of $\sqrt{s}$ for a 125 GeV SM Higgs boson. The vertical dashed line indicates $\sqrt{s} = 240$ GeV, the nominal energy of the CEPC running as a Higgs factory.

The CEPC as a Higgs factory is designed to deliver a combined integrated luminosity of $5.6\,\text{ab}^{-1}$ to two detectors in 7 years. Over $10^6$ Higgs boson events will be produced during this period. The large statistics, well-defined event kinematics and clean collision environment will enable the CEPC to measure the Higgs boson production cross sections as well as its properties (mass, decay width and branching ratios, etc.) with precision far beyond those achievable at the LHC. In contrast to hadron collisions, $e^+e^-$ collisions are unaffected by underlying events and pile-up effects. Theoretical calculations are less dependent on higher order QCD radiative corrections. Therefore, more precise tests of theoretical predictions can be performed at the CEPC. The tagging of $e^+e^- \to ZH$ events using the recoil mass method (see Section 11.1.2), independent of the Higgs boson decay, is unique to lepton colliders. It provides a powerful tool to perform model-independent measurements of the inclusive $e^+e^- \to ZH$ production cross section, $\sigma(ZH)$, and of the Higgs boson decay branching ratios. Combinations of these measurements will allow for the determination of the total Higgs boson decay width and the extraction of the Higgs boson couplings to fermions and vector bosons.These measurements will provide sensitive probes to potential new physics beyond the SM.

| Decay mode | Branching ratio | Relative uncertainty |
|---|---|---|
| $H \to b\bar{b}$ | 57.7% | +3.2%, −3.3% |
| $H \to c\bar{c}$ | 2.91% | +12%, −12% |
| $H \to gg$ | 8.57% | +10%, −10% |
| $H \to \tau^+\tau^-$ | 6.32% | +5.7%, −5.7% |
| $H \to \mu^+\mu^-$ | $2.19 \times 10^{-4}$ | +6.0%, −5.9% |
| $H \to WW^*$ | 21.5% | +4.3%, −4.2% |
| $H \to ZZ^*$ | 2.64% | +4.3%, −4.2% |
| $H \to \gamma\gamma$ | $2.28 \times 10^{-3}$ | +5.0%, −4.9% |
| $H \to Z\gamma$ | $1.53 \times 10^{-3}$ | +9.0%, −8.8% |
| $\Gamma_H$ | 4.07 MeV | +4.0%, −4.0% |

**Table 11.1:** SM predictions of the decay branching ratios and natural width for a 125 GeV Higgs boson from Refs. [11–13]. The quoted uncertainties include contributions from both theoretical and parametric sources.

SM background processes include $e^+e^- \to e^+e^-$ (Bhabha), $e^+e^- \to Z\gamma$ ($Z$ radiative return), $e^+e^- \to WW/ZZ$ (diboson) as well as the single boson production of $e^+e^- \to e^+e^-Z$ and $e^+e^- \to e^+\nu W^-/e^-\bar{\nu}W^+$. Their cross sections and expected numbers of events for an integrated luminosity of $5.6\,\text{ab}^{-1}$ at $\sqrt{s} = 240$ GeV are shown in Table 11.2 as well. The energy dependence of the signal and background production cross sections is shown in Figure 3.1. Note that many of these processes can lead to identical final states after the decays of the $W$ or $Z$ bosons, and therefore can interfere. For example, $e^+e^- \to e^+\nu_e W^- \to e^+\nu_e e^-\bar{\nu}_e$ and $e^+e^- \to e^+e^-Z \to e^+e^-\nu_e\bar{\nu}_e$ have the same final state. Unless otherwise noted, these processes are simulated together to take into account interference effects for the studies presented in this report. Similar to the Higgs boson processes, the breakdowns shown in the table and figure assume the $W$ and $Z$ bosons are stable and are, therefore, for illustration only.

### 11.1.2 HIGGS BOSON TAGGING

Perhaps the most striking difference between hadron-hadron and $e^+e^-$ collisions is that electrons are fundamental particles whereas hadrons are composite. Consequently the energy of $e^+e^-$ collisions is known. Therefore through conservation laws, the energy and momentum of a Higgs boson can be inferred from other particles in an event without examining the Higgs boson itself. For a Higgsstrahlung event where the $Z$ boson decays to a pair of visible fermions ($ff$), the mass of the system recoiling against the $Z$ boson, commonly known as the recoil mass, can be calculated assuming the event has a total energy $\sqrt{s}$ and zero total momentum:

$$M^2_{\text{recoil}} = (\sqrt{s} - E_{ff})^2 - p^2_{ff} = s - 2E_{ff}\sqrt{s} + m^2_{ff}. \quad (11.1)$$

Here $E_{ff}$, $p_{ff}$ and $m_{ff}$ are, respectively, the total energy, momentum and invariant mass of the fermion pair. The $M_{\text{recoil}}$ distribution should show a peak at the Higgs boson mass $m_H$ for $e^+e^- \to ZH \to ffH$ and $e^+e^- \to e^+e^-H$ processes, and is expected to be smooth without a resonance structure for background processes in the mass region around 125 GeV. Two important measurements of the Higgs boson can be performed from the $M_{\text{recoil}}$ mass spectrum. The Higgs boson mass can be determined from the position of the resonance in the spectrum. The width of the resonance structure is dominated by the beam energy spread (including ISR effects) and energy/momentum resolution of the detector if the Higgs boson width is only 4.07 MeV as predicted in the SM. The best precision of the mass measurement can be achieved from the leptonic $Z \to \ell\ell\,(\ell = e, \mu)$ decays. The height of the resonance is proportional to the Higgs boson production cross section $\sigma(ZH)$.[1] Through a fit to the $M_{\text{recoil}}$ spectrum, the $e^+e^- \to ZH$ event yield, and therefore $\sigma(ZH)$, can be extracted, independent of the Higgs boson decays. The Higgs boson decay branching ratios can then be determined by measuring the $ZH$ cross sections of the individual Higgs boson decay modes. The recoil mass spectrum has been investigated for both leptonic and hadronic $Z$ boson decays as presented below.

| Process | Cross section | Events in $5.6\,\text{ab}^{-1}$ |
|---|---|---|
| Higgs boson production, cross section in fb | | |
| $e^+e^- \to ZH$ | 196.2 | $1.10 \times 10^6$ |
| $e^+e^- \to \nu_e\bar{\nu}_e H$ | 6.19 | $3.47 \times 10^4$ |
| $e^+e^- \to e^+e^- H$ | 0.28 | $1.57 \times 10^3$ |
| Total | 203.7 | $1.14 \times 10^6$ |
| Background processes, cross section in pb | | |
| $e^+e^- \to e^+e^- (\gamma)$ (Bhabha) | 930 | $5.2 \times 10^9$ |
| $e^+e^- \to q\bar{q}\,(\gamma)$ | 54.1 | $3.0 \times 10^8$ |
| $e^+e^- \to \mu^+\mu^- (\gamma)$ [or $\tau^+\tau^- (\gamma)$] | 5.3 | $3.0 \times 10^7$ |
| $e^+e^- \to WW$ | 16.7 | $9.4 \times 10^7$ |
| $e^+e^- \to ZZ$ | 1.1 | $6.2 \times 10^6$ |
| $e^+e^- \to e^+e^- Z$ | 4.54 | $2.5 \times 10^7$ |
| $e^+e^- \to e^+\nu W^-/e^-\bar{\nu}W^+$ | 5.09 | $2.6 \times 10^7$ |

**Table 11.2:** Cross sections of Higgs boson production and other SM processes at $\sqrt{s} = 240$ GeV and numbers of events expected in $5.6\,\text{ab}^{-1}$. Note that there are interferences between the same final states from different processes after the $W$ or $Z$ boson decays. Their treatments are explained in the text. With the exception of the Bhabha scattering process, the cross sections are calculated using the Whizard program [14]. The Bhabha scattering cross section is calculated using the BABAYAGA event generator [15] requiring final-state particles to have $|\cos\theta| < 0.99$. Photons, if any, must have $E_\gamma > 0.1$ GeV and $|\cos\theta_{e^\pm\gamma}| < 0.99$.

The leptonic $Z$ boson decay is ideal for studying the recoil mass spectrum of the $e^+e^- \to ZX$ events. The decay is easily identifiable and the lepton momenta can be precisely measured. Figure 11.3 shows the reconstructed recoil mass spectra of $e^+e^- \to$

[1]For the $Z \to e^+e^-$ decay, there will be a small contribution from the $e^+e^- \to e^+e^-H$ production.

$ZX$ candidates for the $Z \to \mu^+\mu^-$ and $Z \to e^+e^-$ decay modes. The analyses are based on the full detector simulation for the signal events and on the fast detector simulation for background events. The event selections are entirely based on the information of the two leptons, independent of the final states of Higgs boson decays. This approach is essential for the measurement of the inclusive $e^+e^- \to ZH$ production cross section and the model-independent determination of the Higgs boson branching ratios. The SM processes with at least 2 leptons in their final states are considered as backgrounds. As shown in Figure 11.3, the analysis has a good signal-to-background ratio. The long high-mass tail is largely due to the initial-state radiation. Leading background contributions after the selection are from $ZZ$, $WW$ and $Z\gamma$ events. Compared to the $Z \to \mu^+\mu^-$ decay, the analysis of the $Z \to e^+e^-$ decay suffers from additional and large background contributions from Bhabha scattering and single boson production.

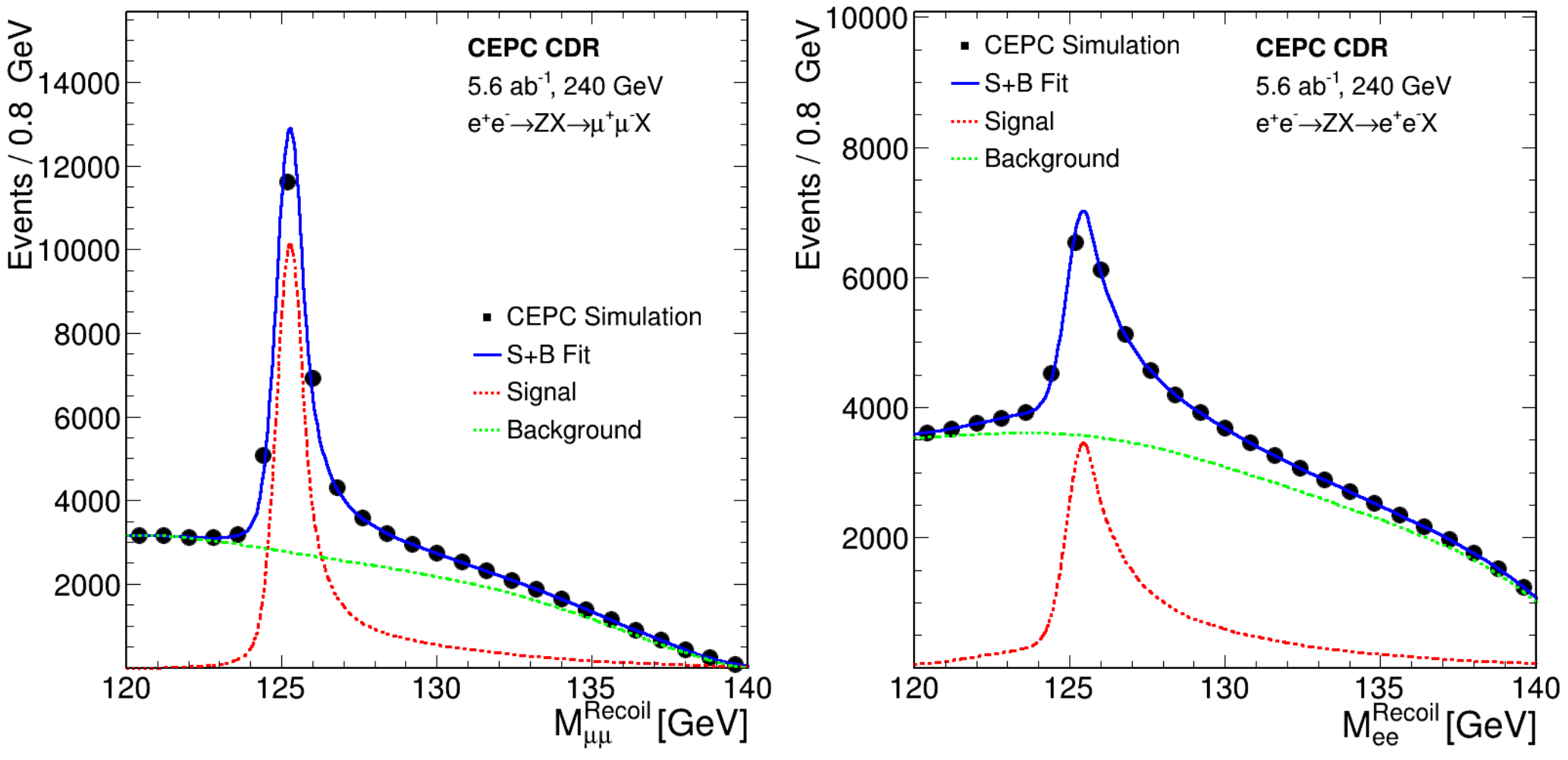

(a) (b)

**Figure 11.3:** The inclusive recoil mass spectra of $e^+e^- \to ZX$ candidates of (a) $Z \to \mu^+\mu^-$ and (b) $Z \to e^+e^-$. No attempt to identify $X$ is made. The markers and their uncertainties represent expectations from a CEPC dataset of $5.6\,\mathrm{ab}^{-1}$, whereas the solid blue curves are the signal-plus-background fit results. The dashed curves are the signal and background components.

The recoil mass technique can also be applied to the hadronic $Z$ boson decays ($Z \to q\bar{q}$) of the $e^+e^- \to ZX$ candidates. This analysis benefits from a larger $Z \to q\bar{q}$ decay branching ratio, but suffers from worse jet energy resolution compared with the track momentum. In addition, ambiguity in selecting jets from the $Z \to q\bar{q}$ decay, particularly in events with hadronic decays of the Higgs boson, can degrade the analysis performance and also introduce some model dependence. Therefore, the measurement is highly dependent on the detector performance and the jet clustering algorithm. Following the same approach as the ILC study [16], an analysis based on the fast simulation has been performed. After the event selection, main backgrounds arise from $Z\gamma$'s and $WW$ production.

### 11.1.3 MEASUREMENTS OF $\sigma(ZH)$ AND THE HIGGS BOSON MASS

Both the inclusive $e^+e^- \to ZH$ production cross section $\sigma(ZH)$ and the Higgs boson mass $m_H$ can be extracted from fits to the recoil mass distributions of the $e^+e^- \to ZX \to$ $(\ell^+\ell^-/q\bar{q})X$ candidates. For the leptonic $Z \to \ell\ell$ decays, the recoil mass distribution of the signal process $e^+e^- \to ZH$ (and also $e^+e^- \to e^+e^-H$ in case of the $Z \to e^+e^-$ decay) is modeled with a Crystal Ball function [17], whereas the total background is modeled with a polynomial function. As noted above, the recoil mass distribution is insensitive to the intrinsic Higgs boson width should it be as small as predicted by the SM. The Higgs boson mass can be determined with precision of 6.5 MeV and 14 MeV from the $Z \to \mu^+\mu^-$ and $Z \to e^+e^-$ decay modes, respectively. After combining all channels, an uncertainty of 5.9 MeV can be achieved. The $e^+e^- \to ZX \to q\bar{q}X$ process contributes little to the precision of the Higgs boson mass measurement due to the poor $Z \to q\bar{q}$ mass resolution, but dominates the sensitivity to the $e^+e^- \to ZH$ cross section measurement because of the larger event sample. A simple event counting analysis shows that the expected relative precision on $\sigma(ZH)$ is 0.6%. In comparison, the corresponding relative precision from the $Z \to e^+e^-$ and $Z \to \mu^+\mu^-$ decays is estimated to be 1.4% and 0.9%, respectively. The combined relative precision of the three measurements is 0.5%.

For the model-independent measurement of $\sigma(ZH)$, event selections independent of the Higgs boson decays are essential. However, additional selection criteria using the Higgs boson decay information can be applied to improve the Higgs boson mass measurement. This will be particularly effective in suppressing the large backgrounds in the $Z \to e^+e^-$ and $Z \to q\bar{q}$ decay modes. These improvements are not implemented in the current study.

### 11.1.4 ANALYSES OF THE INDIVIDUAL HIGGS BOSON DECAY MODES

Different decay modes of the Higgs boson can be identified through their unique signatures, enabling the measurements of production rates for these decays. Simulation studies of the CEPC baseline conceptual detector have been performed for the Higgs boson decay modes of $H \to b\bar{b}/c\bar{c}/gg$, $H \to WW^*$, $H \to ZZ^*$, $H \to \gamma\gamma$, $H \to Z\gamma$, $H \to \tau^+\tau^-$, $H \to \mu^+\mu^-$ and Higgs boson to invisible particles ($H \to \mathrm{inv}$). The large number of the decay modes of the $H$, $W$ and $Z$ boson as well as the $\tau$-lepton leads to a very rich variety of event topologies. This complexity makes it impractical to investigate the full list of final states stemming from the Higgs boson decays. Instead, a limited number of final states of the individual Higgs boson decay modes have been considered. For some Higgs boson decay modes, the chosen final states may not be the most sensitive ones, but are nevertheless representatives of the decay mode. In most cases, the dominant backgrounds come from the SM diboson production and the single $Z$ production with the initial and final state radiation.

The studies are optimized for the dominant $ZH$ process, however, the $e^+e^- \to \nu_e\bar{\nu}_eH$ and $e^+e^- \to e^+e^-H$ processes are included whenever applicable. The production cross sections of the individual decay modes, $\sigma(ZH) \times \mathrm{BR}$, are extracted. Combined with the inclusive $\sigma(ZH)$ measurement, these measurements will allow the determination of the Higgs boson decay branching ratios in a model-independent way. The main features of these studies are described below and their results are presented in Section 11.1.5.

For a SM Higgs boson with a mass of 125 GeV, nearly 70% of all Higgs bosons decay into a pair of jets: $b$-quarks (57.7%), $c$-quarks (2.9%) and gluons (8.6%). While the $H \to b\bar{b}$ decay has been observed at the LHC [18, 19], the $H \to c\bar{c}$ and $H \to gg$ decays are difficult, if not impossible, to be conclusively identified even at the HL-LHC due to the large backgrounds. In comparison, these three decays can be isolated and studied at the CEPC in detail. This is important as the $H \to c\bar{c}$ decay is the only promising channel to investigate the Higgs boson couplings to the second-generation quarks. The study considers all $Z$ boson decay modes except $Z \to \tau^+\tau^-$. The $H \to b\bar{b}/c\bar{c}/gg$ candidates are identified through the dijet invariant mass, or the recoil mass of the visible $Z$ boson decays, or both. Jet flavor tagging is employed to separate $H \to b\bar{b}, c\bar{c}, gg$ contributions. Figure 11.4(a) shows the reconstructed recoil mass distribution of the $Z \to \mu^+\mu^-$ decay. Compared with the distribution of inclusive Higgs boson decays shown in Figure 11.3(a), the background is significantly reduced through the identification of specific Higgs boson decay modes. Figure 11.4(b) is the dijet mass distribution of the $Z \to \nu\bar{\nu}$ decay, showing excellent signal-to-background ratio and good dijet mass resolution.

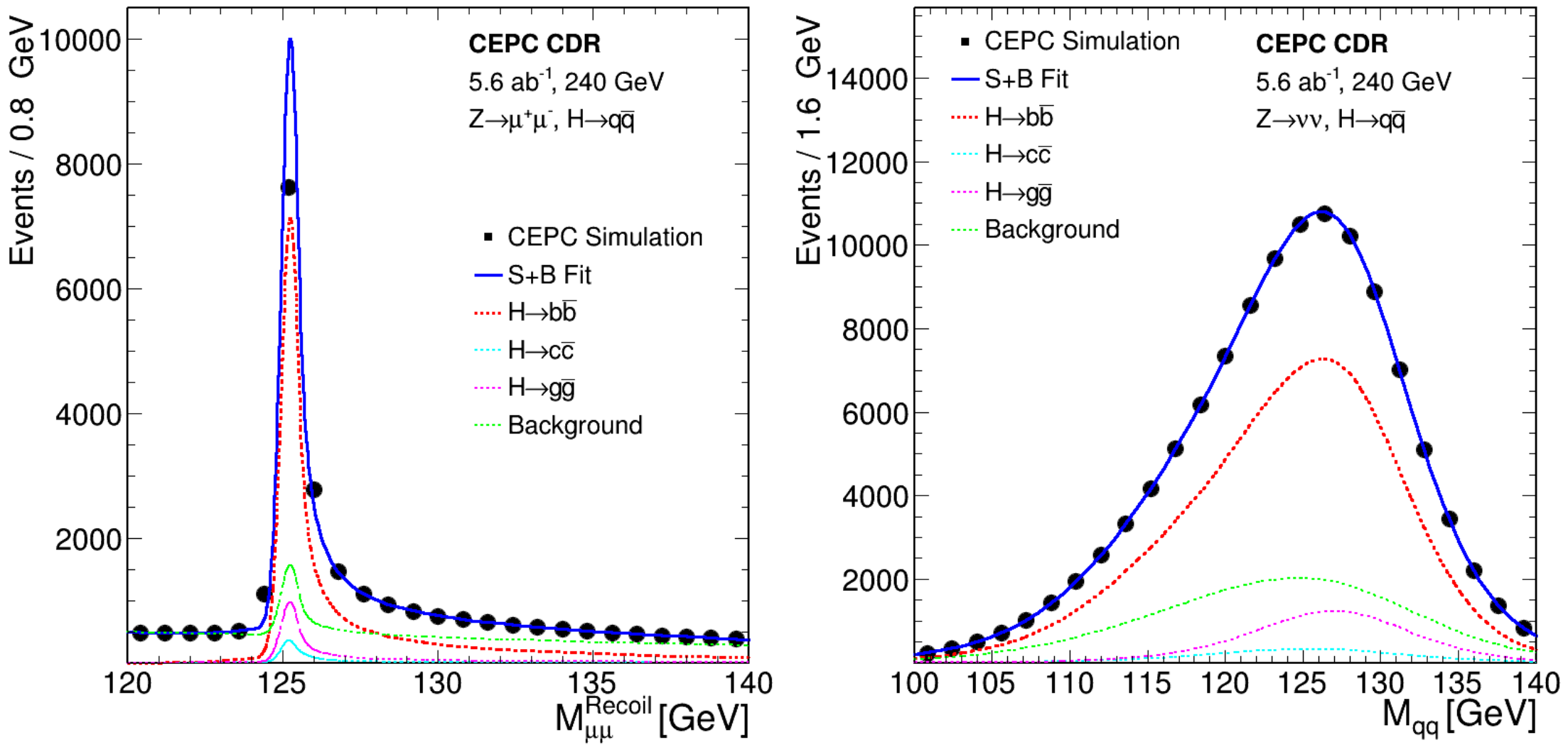

(a) (b)

**Figure 11.4:** (a) $e^+e^- \to ZH$ production with $H \to b\bar{b}/c\bar{c}/gg$: distributions of (a) the recoil mass of $Z \to \mu^+\mu^-$ and (b) the dijet invariant mass distribution for the $Z \to \nu\bar{\nu}$ decay. The markers and their uncertainties represent expectations from a CEPC dataset of 5.6 ab$^{-1}$, whereas the solid blue curves are the signal-plus-background fit results. The dashed curves are the signal and background components. Contributions from other Higgs boson decays are included in the background.

The $W$-fusion process $e^+e^- \to \nu_e\bar{\nu}_e H$ a cross section is only 3.2% of that of the $ZH$ process at $\sqrt{s} = 240$ GeV in the SM. This process has been explored with the $H \to b\bar{b}$ decay mode. The analysis suffers from large backgrounds from $ZH \to \nu\bar{\nu}b\bar{b}$ as it has the same final state. However, the $\nu\bar{\nu}H$ and $Z(\nu\bar{\nu})H$ contributions can be separated through the exploration of their kinematic differences. Higgs bosons are produced with different polar angle distributions. Moreover, the recoil mass distribution of the $b\bar{b}$ system should exhibit a resonance structure at the $Z$ boson mass for $Z(\nu\bar{\nu})H$ and show a continuum spectrum for $e^+e^- \to \nu_e\bar{\nu}_e H$. The $\nu\bar{\nu}H$ contribution is extracted through a fit to the two-dimensional distribution of the cosine of the polar angle and the recoil mass of the $b\bar{b}$ system.

The $H \to WW^*$ and $H \to ZZ^*$ decays were among the first decay modes studied at the LHC and were crucial for the discovery of the Higgs boson through their clean leptonic final states. However, due to the large backgrounds, hadronic final states of the $H \to WW^*$ and $H \to ZZ^*$ decays are out of reach at the LHC despite of their large branching ratios. This is not the case at the CEPC. In fact, most of the expected sensitivity at the CEPC to these two Higgs boson decay modes comes from final states with one or both vector bosons decay hadronically. A number of selected final states has been studied. For $H \to WW^*$, the final states included are $Z \to \ell\ell$, $H \to WW^* \to \ell\nu\ell\nu, \ell\nu q\bar{q}$; $Z \to \nu\bar{\nu}$, $H \to WW^* \to \ell\nu\ell\nu, q\bar{q}q\bar{q}$ and $Z \to q\bar{q}$, $H \to WW^* \to q\bar{q}q\bar{q}$. For $H \to ZZ^*$, they are $Z \to \mu^+\mu^-$, $H \to ZZ^* \to \nu\bar{\nu}q\bar{q}$ and $Z \to \nu\bar{\nu}$, $H \to ZZ^* \to \ell\ell q\bar{q}$. A combination of the recoil mass, the invariant mass of the $W \to q\bar{q}$ and $Z \to q\bar{q}$ decay as well as the leptonic decay signatures of $W$ and $Z$ bosons are used to identify $ZH$ events. Some of these analyses suffer from large backgrounds as shown, for example, in Figure 11.5(a), while others are almost background free as illustrated in Figure 11.5(b).

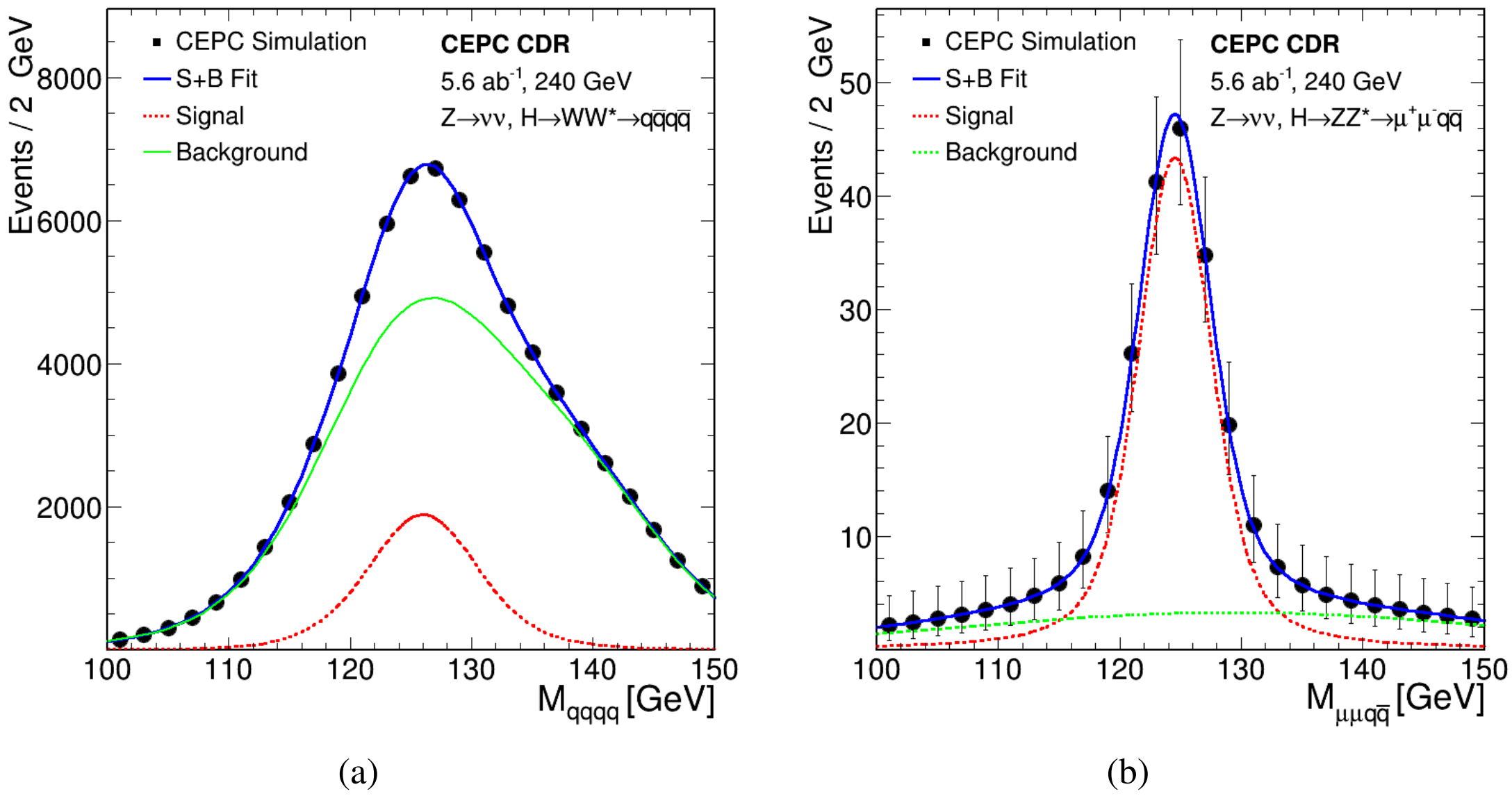

**Figure 11.5:** (a) $e^+e^- \to ZH$ production with $H \to WW^* \to q\bar{q}q\bar{q}$ and $Z \to \nu\bar{\nu}$: the invariant mass of the 4-jet system. (b) $e^+e^- \to ZH$ production with $H \to ZZ^* \to \mu^+\mu^-q\bar{q}$ and $Z \to \nu\bar{\nu}$: the invariant mass distribution of the dimuon and dijet system. The markers and their uncertainties represent expectations from a CEPC dataset of $5.6\,\text{ab}^{-1}$, whereas the solid blue curves are the signal-plus-background fit results. The dashed curves are the signal and background components. Contributions from other Higgs boson decays are included in the background.

The $H \to \gamma\gamma$ and $H \to Z\gamma$ decays have small branching ratios in the SM as they proceed through the $W$ boson and top quark loops. The CEPC sensitivity to these two decay modes has been examined. The analysis of the $ZH$ production in the $H \to \gamma\gamma$ channel suffers from large $e^+e^- \to (Z/\gamma^*)\gamma\gamma$ background where the photons arise from the initial and final state radiation. All $Z$ boson decay modes other than the $Z \to e^+e^-$ decay are considered for the $H \to \gamma\gamma$ studies. The $ZH$ production with $Z \to e^+e^-$ has an additional large background source of the Bhabha scattering process. As shown in Figure 11.6(a), the $H \to \gamma\gamma$ signal is expected to appear as a resonance over a smooth background in the diphoton mass distribution. $ZH$ production with $H \to Z\gamma$ decay will lead to events with two on-shell $Z$ bosons and one photon. The $H \to Z\gamma$ study targeted the signal process of $ZH \to ZZ\gamma \to \nu\bar{\nu}q\bar{q}\gamma$.[2] In this final state, the energy and mo-mentum of the $\nu\bar{\nu}$ system can be calculated from the visible energy and momentum of the event. The mass difference between the Higgs boson candidate and the candidate of the associated $Z$ boson can then be calculated. For signal events, this mass difference is expected to be $m_H - m_Z \sim 35$ GeV for the correct combinations as shown in Figure 11.7(b). For background events and the wrong combinations of signal events, the distribution should be smooth.

[2]Both $ZZ\gamma \to \nu\bar{\nu}q\bar{q}\gamma$ and $ZZ\gamma \to q\bar{q}\nu\bar{\nu}\gamma$ are considered in the study.

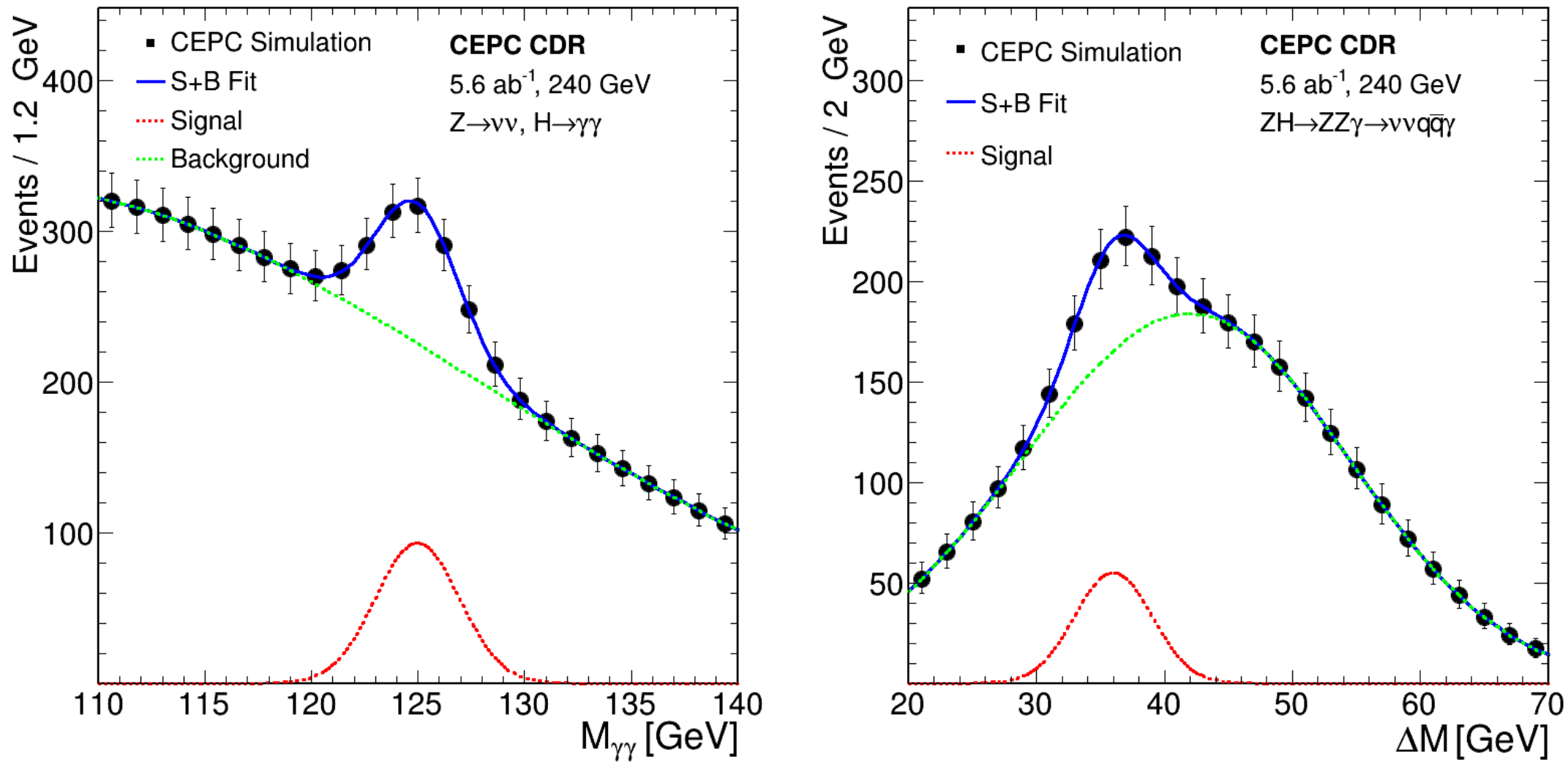

(a) (b)

**Figure 11.6:** (a) $e^+e^- \to ZH$ production with $H \to \gamma\gamma$: the invariant mass distribution of the selected photon pairs for $Z \to \nu\bar{\nu}$. (b) $e^+e^- \to ZH$ production with $H \to Z\gamma$: the distribution of the mass difference between the reconstructed $Z\gamma$ and $Z$ system including contributions from both $M(qq\gamma) - M(qq)$ and $M(\nu\bar{\nu}\gamma) - M(\nu\bar{\nu})$. The markers and their uncertainties represent expectations from a CEPC dataset of $5.6\,\text{ab}^{-1}$, whereas the solid blue curves are the signal-plus-background fit results. The dashed curves are the signal and background components.

The leptonic Higgs boson decays are accessible for $H \to \tau^+\tau^-$ and $H \to \mu^+\mu^-$ at the CEPC. Simulation studies of $ZH$ production with the $H \to \tau^+\tau^-$ decay have been performed for all $Z$ boson decay modes except $Z \to e^+e^-$. A boosted decision tree utilizing the reconstructed particle multiplicity and their separations is used to select ditau candidates from $H \to \tau^+\tau^-$. An impact-parameter based variable of the leading track of the ditau candidate is used as the final discriminant for the signal extraction. An example distribution of this variable for $Z \to \nu\bar{\nu}$ is shown in Figure 11.7(a). Similar to $H \to \gamma\gamma$, the $H \to \mu^+\mu^-$ decay also allows for the reconstruction of the Higgs boson decay with high precision. The signal is expected to appear as a resonance structure at $m_H$ over the smooth background in the dimuon mass spectrum. For this study, all $Z$ boson decay modes are considered. Figure 11.7(b) shows the dimuon mass distribution combining all $Z$ boson decay modes.

In the SM, the Higgs boson can decay invisibly via $H \to ZZ^* \to \nu\bar{\nu}\nu\bar{\nu}$ with a branching ratio of $1.06 \times 10^{-3}$. In many extensions to the SM, the Higgs boson can decay directly to invisible particles with a significantly higher branching ratio. At the CEPC, the invisible decay of the Higgs boson ($H \to \text{inv}$) can be directly identified using the information of missing energy/momentum and the recoil mass of the visible $Z$ boson decays. The sensitivity to $ZH$ production with $H \to \mathrm{inv}$ is estimated for $Z \to \ell\ell$ and $Z \to q\bar{q}$ decays. The SM $H \to ZZ^* \to \nu\bar{\nu}\nu\bar{\nu}$ decay is used to model the $H \to \mathrm{inv}$ decay both in the context of the SM and beyond. This is made possible by the fact that the Higgs boson is narrow scalar in the SM so that its production and decay can be treated separately. The upper limit on the BSM contribution to $\mathrm{BR}(H \to \mathrm{inv})$, $\mathrm{BR}^{\mathrm{BSM}}_{\mathrm{inv}}$, can then be estimated.

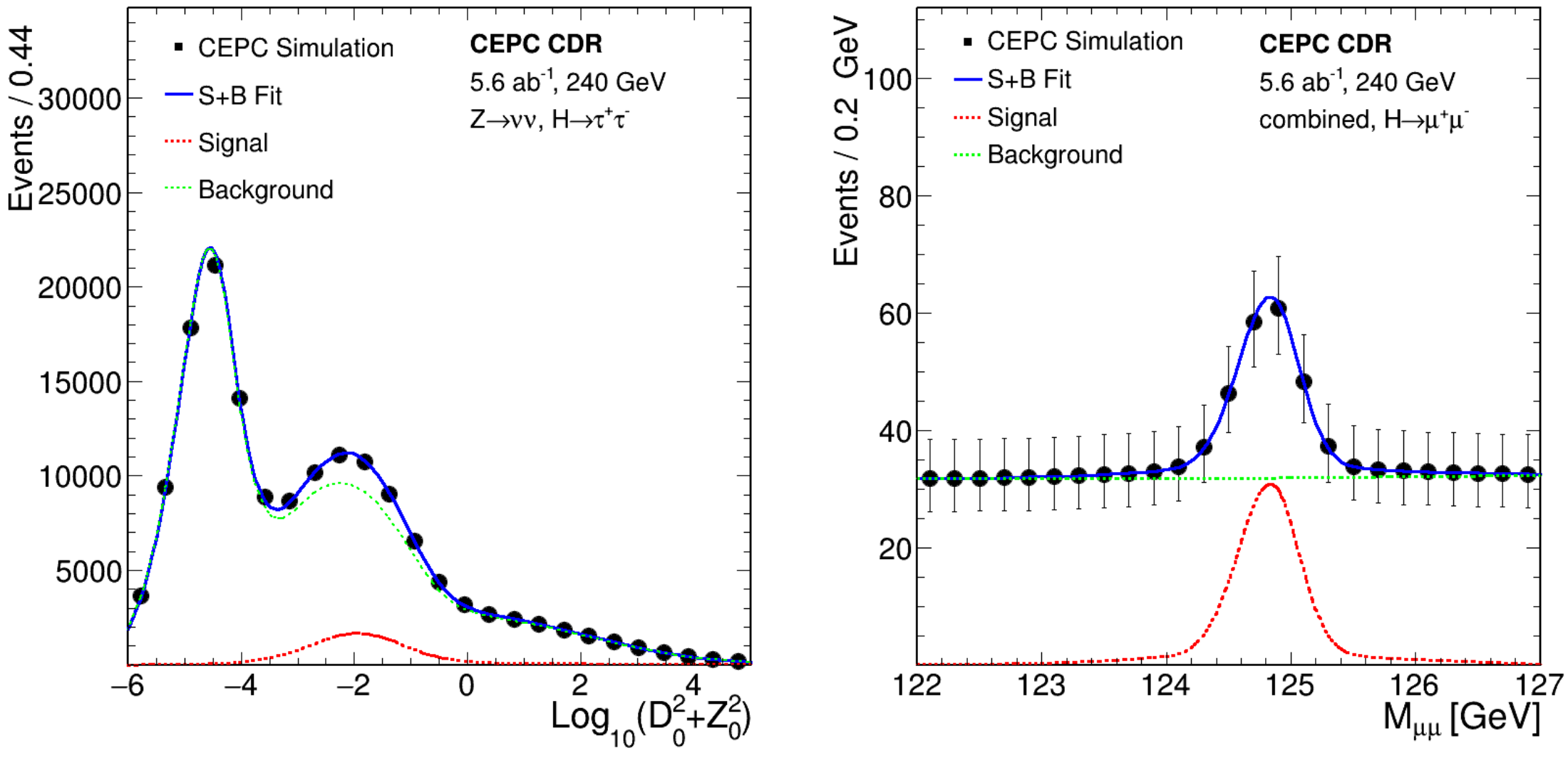

(a) (b)

**Figure 11.7:** (a) $e^+e^- \to ZH$ production with $H \to \tau^+\tau^-$: the distribution of impact parameter variable of the leading track of the ditau candidates for the $Z \to \nu\bar{\nu}$ decay mode. Contributions from other Higgs boson decays are included in the background. (b) $e^+e^- \to ZH$ production with $H \to \mu^+\mu^-$: the invariant mass distribution of the selected muon pairs combining all $Z$ boson decay modes. The markers and their uncertainties represent expectations from a CEPC dataset of $5.6\,\mathrm{ab}^{-1}$, whereas the solid blue curves are the signal-plus-background fit results. The dashed curves are the signal and background components.

### 11.1.5 COMBINATION OF THE INDIVIDUAL ANALYSES

With the measurements of inclusive cross section $\sigma(ZH)$ and the cross section times the branching ratio $\sigma(ZH) \times \mathrm{BR}$ for the individual Higgs boson decay modes, the Higgs boson decay branching ratio $\mathrm{BR}$ can be extracted. Most of the systematic uncertainties associated with the measurement of $\sigma(ZH)$ cancel in this procedure. A maximum likelihood fit is used to estimate the precision on the $\mathrm{BR}$s. For a given Higgs boson decay mode, the likelihood has the form:

$$L(\mathrm{BR}, \theta) = \mathrm{Poisson}\left[N^{\mathrm{obs}} \middle| N^{\mathrm{exp}}(\mathrm{BR}, \theta)\right] \cdot G(\theta), \tag{11.2}$$

where $\mathrm{BR}$ is the parameter of interest and $\theta$ represents nuisance parameters associated with systematic uncertainties. The number of observed events is denoted by $N^{\mathrm{obs}}$, $N^{\mathrm{exp}}(\mathrm{BR}, \theta)$ is the expected number of events, and $G(\theta)$ is a set of constraints on the nuisance parameters due to the systematic uncertainties. The number of expected events is the sum of signal and background events. The number of signal events is calculated from the integrated luminosity, the $e^+e^- \to ZH$ cross section $\sigma(ZH)$ measured from the recoil mass method, Higgs boson branching ratio $\mathrm{BR}$, the event selection efficiency, $\epsilon$. The number of the expected background events, $N^b$, is estimated using Monte Carlo samples. Thus:

$$N^{\exp}(\mathrm{BR}, \theta) = \mathrm{Lumi}(\theta^{\mathrm{lumi}}) \times \sigma_{ZH}(\theta^{\sigma}) \times \mathrm{BR} \times \epsilon(\theta^{\epsilon}) + N^b(\theta^b), \tag{11.3}$$

where $\theta^X$ ($X = \mathrm{lumi}, \sigma$, $\epsilon$ and $b$) are the nuisance parameters of their corresponding parameters or measurements. Even with $10^6$ Higgs boson events, statistical uncertainties are expected to be dominant and thus systematic uncertainties are not taken into account for the current studies. The nuisance parameters are therefore fixed to their nominal values.

| Property | Estimated Precision |
|---|---|
| $m_H$ | 5.9 MeV |
| $\Gamma_H$ | 3.1% |
| $\sigma(ZH)$ | 0.5% |
| $\sigma(\nu\bar{\nu}H)$ | 3.2% |

| Decay mode | $\sigma(ZH) \times \mathrm{BR}$ | BR |
|---|---|---|
| $H \to b\bar{b}$ | 0.27% | 0.56% |
| $H \to c\bar{c}$ | 3.3% | 3.3% |
| $H \to gg$ | 1.3% | 1.4% |
| $H \to WW^*$ | 1.0% | 1.1% |
| $H \to ZZ^*$ | 5.1% | 5.1% |
| $H \to \gamma\gamma$ | 6.8% | 6.9% |
| $H \to Z\gamma$ | 15% | 15% |
| $H \to \tau^+\tau^-$ | 0.8% | 1.0% |
| $H \to \mu^+\mu^-$ | 17% | 17% |
| $H \to \mathrm{inv}$ | – | $< 0.30\%$ |

**Table 11.3:** Estimated precision of Higgs boson property measurements expected from a CEPC dataset of 5.6 ab$^{-1}$ at $\sqrt{s} = 240$ GeV. All precision are relative except for $m_H$ and $\mathrm{BR}(H \to \mathrm{inv})$ for which $\Delta m_H$ and the 95% confidence level upper limit on $\mathrm{BR}_{\mathrm{inv}}^{\mathrm{BSM}}$ are quoted respectively. The $e^+e^- \to e^+e^-H$ cross section is too small to be measured with a reasonable precision.

Table 11.3 summarizes the estimated precision of Higgs boson property measurements, combining all studies described above and taking into account cross-feeds among the different Higgs boson production processes and decay modes. For the leading Higgs boson decay modes, namely $b\bar{b}$, $c\bar{c}$, $gg$, $WW^*$, $ZZ^*$ and $\tau^+\tau^-$, percent level precision is expected. The best achievable statistical uncertainties for a dataset of 5.6 ab$^{-1}$ are 0.27% for $\sigma(e^+e^- \to ZH) \times \mathrm{BR}(H \to b\bar{b})$ and 0.5% for $\sigma(e^+e^- \to ZH)$. Even for these measurements, statistics is likely to be the dominant uncertainty source. Systematic uncertainties due to the acceptance of the detector, the efficiency of the object reconstruction/identification, the luminosity and the beam energy determination are expected to be small. The integrated luminosity can be measured with a 0.1% precision, a benchmark already achieved at the LEP [20], and can be potentially improved in the future. The center-of-mass energy will be known better than 1 MeV, resulting negligible uncertainties on the theoretical cross section predictions and experimental recoil mass measurements.

The estimated precision is expected to improve as more final states are explored and analyses are improved. This is particularly true for $ZH \to ZWW^*$ and $ZH \to ZZZ^*$ with complex final states. Therefore, Table 11.3 represents conservative estimates for many Higgs boson observables.

### 11.1.6 HIGGS BOSON WIDTH

The Higgs boson width is of special interest as it is sensitive to BSM physics in Higgs boson decays that are not directly detectable or searched for. However, the 4.07 MeV natural width predicted by the SM is too small to be measured with a reasonable precision from the distributions of either the invariant mass of the Higgs boson decay products or the recoil mass of the system produced in association with the Higgs boson. In a procedure unique to lepton colliders, the width can be determined from the measurements of Higgs boson production cross sections and its decay branching ratios. This is because the inclusive $e^+e^- \to ZH$ cross section $\sigma(ZH)$ can be measured from the recoil mass distribution, independent of the Higgs boson decays.

By combining the measurements of the Higgs boson production cross section and decay branching ratios, the Higgs boson width can be calculated in a model-independent way:

$$\Gamma_H = \frac{\Gamma(H \to ZZ^*)}{\mathrm{BR}(H \to ZZ^*)} \propto \frac{\sigma(ZH)}{\mathrm{BR}(H \to ZZ^*)} \tag{11.4}$$

where $\Gamma(H \to ZZ^*)$ is the partial width of the $H \to ZZ^*$ decay. Because of the small expected $\mathrm{BR}(H \to ZZ^*)$ value for a 125 GeV Higgs boson (2.64% in the SM), the precision of $\Gamma_H$ is limited by the $H \to ZZ^*$ analysis statistics. It can be improved by including final states with larger branching ratios, e.g. the $H \to b\bar{b}$ decay:

$$\Gamma_H = \frac{\Gamma(H \to b\bar{b})}{\mathrm{BR}(H \to b\bar{b})} \tag{11.5}$$

where the partial width $\Gamma(H \to b\bar{b})$ can be independently extracted from the cross section of the $W$ fusion process:

$$\sigma(\nu\bar{\nu}H \to \nu\bar{\nu}\, b\bar{b}) \propto \Gamma(H \to WW^*) \cdot \mathrm{BR}(H \to b\bar{b}) = \Gamma(H \to b\bar{b}) \cdot \mathrm{BR}(H \to WW^*) \tag{11.6}$$

Thus, the Higgs boson total width is:

$$\Gamma_H = \frac{\Gamma(H \to b\bar{b})}{\mathrm{BR}(H \to b\bar{b})} = \frac{\Gamma(H \to WW^*)}{\mathrm{BR}(H \to WW^*)} \propto \frac{\sigma(\nu\bar{\nu}H)}{\mathrm{BR}(H \to WW^*)} \tag{11.7}$$

where $\mathrm{BR}(H \to b\bar{b})$ and $\mathrm{BR}(H \to WW^*)$ are measured from the $e^+e^- \to ZH$ process. The limitation of this method is the precision of the $\sigma(e^+e^- \to \nu_e\bar{\nu}_e H)$ measurement.

The expected precision on $\Gamma_H$ is 5.1% from the measurements of $\sigma(ZH)$ and $\mathrm{BR}(H \to ZZ^*)$ and is 3.5% from the measurements of $\sigma(\nu\bar{\nu}H)$ and $\mathrm{BR}(H \to WW^*)$. The quoted precision is limited by the $\mathrm{BR}(H \to ZZ^*)$ measurement for the former case and the $\sigma(\nu\bar{\nu}H)$ measurement for the latter case. The combined $\Gamma_H$ precision of the two measurements is 2.8%. The combination takes into account the correlations between the two measurements.

### 11.1.7 HIGGS BOSON COUPLING MEASUREMENTS

To understand the implications of the estimated CEPC precision shown in Table 11.3 on possible new physics models, the results need to be interpreted in terms of constraints on the parameters in the Lagrangian. This is often referred to as the "Higgs boson coupling measurements", even though the term can be misleading as discussed below.

There is no unique way to present the achievable precision on the couplings. Before going into the discussion of the CEPC results, we briefly comment on the choices made here. The goal of the theory interpretation here is to obtain a broad idea of the CEPC sensitivity to the Higgs boson couplings. The interpretation should be simple with intuitive connections between the models and the experimental observables. Ideally, it should have as little model assumptions as possible. Furthermore, it would be convenient if the results can be interfaced directly with the higher order theoretical calculations, renormalization group equation evolutions, etc. Unfortunately, it is impossible to achieve all of these goals simultaneously.

Two popular frameworks are, instead, chosen for the interpretation of the CEPC results: the so-called $\kappa$-framework [21–30] and the effective field theory (EFT) frameworks [31–51]. As discussed in more detail later, none of these is perfect. But neither is wrong either as long as one is careful not to over interpret the results. Another important aspect of making projections on the physics potential of a future experiment is that they need to be compared with other experiments. The choices made here follows the most commonly used approaches to facilitate such comparisons. In the later part of this section, the Higgs boson physics potential beyond coupling determination is also discussed.

#### 11.1.7.1 COUPLING FITS IN THE $\kappa$-FRAMEWORK

The Standard Model makes specific predictions for the Higgs boson couplings to the SM fermions, $g_{\rm SM}(Hff)$, and to the SM gauge bosons, $g_{\rm SM}(HVV)$. In the $\kappa$-framework, the potential deviations from the SM are parametrized using the $\kappa$ parameters defined as:

$$\kappa_f = \frac{g(Hff)}{g_{\rm SM}(Hff)}, \quad \kappa_V = \frac{g(HVV)}{g_{\rm SM}(HVV)}, \tag{11.8}$$

with $\kappa_i = 1$ being the SM prediction. The rates of the Higgs boson production and decays are modified accordingly. For example,

$$\begin{aligned} \sigma(ZH) &= \kappa_Z^2 \cdot \sigma_{\rm SM}(ZH) \\ \sigma(ZH) \times {\rm BR}(H \to ff) &= \frac{\kappa_Z^2 \kappa_f^2}{\kappa_\Gamma^2} \cdot \sigma_{\rm SM}(ZH) \times {\rm BR}_{\rm SM}(H \to ff) \end{aligned} \tag{11.9}$$

Here $\kappa_\Gamma^2 (\equiv \Gamma_H/\Gamma_H^{\rm SM})$ parametrizes the change in the Higgs boson width due to both coupling modifications and BSM decays.

Apart from the tree-level couplings, there are also loop-level couplings of $Hgg$, $H\gamma\gamma$ and $HZ\gamma$ in the SM. In the absence of new physics, the couplings through loops, often referred to as the effective couplings, can be expressed using the $\kappa$ parameters, described above.. BSM particles in the loops can also contribute to these effective couplings. For this reason, three additional $\kappa$ parameters: $\kappa_g$, $\kappa_\gamma$ and $\kappa_{Z\gamma}$ are introduced to parametrize the potential deviations from the SM for the three effective $Hgg$, $H\gamma\gamma$ and $HZ\gamma$ couplings, respectively.

It is possible that the Higgs boson can decay directly into new particles or have BSM decays to SM particles. In this case, two types of new decay channels should be distinguished:

1. Invisible decay. This is a specific channel in which Higgs boson decay into new physics particles that are “invisible” in the detector. Such decays can be specifically searched for. If detected, its rate can be measured. The CEPC sensitivity to this decay channel is quantified by the upper limit on $\mathrm{BR}^{\mathrm{BSM}}_{\mathrm{inv}}$.

2. Exotic decays. These include all the other new physics channels. Whether they can be observed, and, if so, to what precision, depends sensitively on the final states. In one extreme, the final states can be very distinct, and the rate can be well measured. In the another extreme, they can be completely swamped by the background. Without the knowledge of the final states and their expected precision, the exotic decays are accounted for by treating the Higgs boson width $\Gamma_H$ as an independent free parameter in the interpretation.

In general, possible deviations of all SM Higgs boson couplings should be considered. However, in the absence of obvious light new physics states with large couplings to the Higgs boson or to other SM particles, a very large deviation ($> \mathcal{O}(1)$) is unlikely. For smaller deviations, the Higgs phenomenology is not sensitive to the deviations of $\kappa_e$, $\kappa_u$, $\kappa_d$ and $\kappa_s$ as the Higgs boson couplings to these particles are negligible compared with the couplings to other particles [52]. Therefore, these $\kappa$ parameters are set to unities, their SM values.

The CEPC will not be able to directly measure the Higgs boson coupling to top quarks. A deviation of this coupling from its SM value does enter the $Hgg$, $H\gamma\gamma$ and $HZ\gamma$ amplitudes. However, this effect is parametrized by $\kappa_g$, $\kappa_\gamma$ and $\kappa_{Z\gamma}$ already. Therefore, $\kappa_t$ is not considered as an independent parameter. For simplicity, previous studies often do not include $\kappa_{Z\gamma}$ in the fit[3]. We will follow this approach here. This leaves the following set of 10 independent parameters:

$$\kappa_b,\ \kappa_c,\ \kappa_g,\ \kappa_W,\ \kappa_\tau,\ \kappa_Z,\ \kappa_\gamma,\ \kappa_\mu,\ \mathrm{BR}^{\mathrm{BSM}}_{\mathrm{inv}},\ \Gamma_H. \tag{11.10}$$

Additional assumptions can be made to reduce the number of parameters [13, 53]. For example, it can be reduced to a 7-parameter set, by assuming lepton universality, and the absence of exotic and invisible decays (excluding $H \to ZZ^* \to \nu\bar{\nu}\nu\bar{\nu}$) [21, 53]:

$$\kappa_b,\ \kappa_c,\ \kappa_g,\ \kappa_W,\ \kappa_Z,\ \kappa_\gamma,\ \kappa_\tau = \kappa_\mu. \tag{11.11}$$

This is useful for studies at hadron colliders as the Higgs boson total width cannot be measured with good precision. The interpretation of the CEPC results is also performed using this reduced set to allow for direct comparisons with the expected HL-LHC sensitivity.

The $\kappa_i$ parameters give a simple and intuitive parametrization of the potential deviations. It has a direct connection with the observables shown in Table 11.3 and does cover many possible modifications of the couplings. However, the $\kappa$-framework has its limitations as well. Strictly speaking, it should not be understood as the modification of the SM renormalizable couplings by a multiplicative factor. For instance, some of such $\kappa$ modifications violate gauge invariance. Higher order corrections in the $\kappa$-framework cannot be easily defined. Moreover, the $\kappa_i$ parameters do not parametrize all possible effects of new physics either. For example, apart from the overall size, potential new physics can also introduce form factors which can change the kinematics of particles that couple to a particular vertex. Manifestations of this effect can be seen in the EFT analysis. It is useful to compare with the EFT analysis discussed in the next subsection. The EFT relates $\kappa_Z$ and $\kappa_W$, and further expands them into three different Lorentz structures. Moreover, some of these higher dimensional $HVV$ couplings are also connected with $\kappa_\gamma$ and anomalous trilinear gauge couplings. The current EFT analysis does not include any new light degrees of freedom, in contrast to the $\kappa$-framework with independent parameters $\mathrm{BR}_{\mathrm{inv}}^{\mathrm{BSM}}$ and $\Gamma_H$. Overall, $\kappa$-framework does capture the big picture of the CEPC capability in precision Higgs boson measurements. It is useful as long as its limitations are understood.

[3] Adding $\kappa_{Z\gamma}$ back would only lead to completely negligible changes in the projection for other parameter and the precision $\kappa_{Z\gamma}$ itself is 8%.

The LHC and especially the HL-LHC will provide valuable and complementary information about the Higgs boson properties. For example, the LHC is capable of directly measuring the $t\bar{t}H$ process [54, 55]. It can also use differential cross sections to differentiate contributions between the top-quark and other heavy particle states in the loop of the $Hgg$ vertex [56–59]. Moreover, it can separate contributions from different operators in the couplings between the Higgs and vector bosons [60]. For the purpose of the coupling fit in the $\kappa$-framework, the LHC, with its large statistics, improves the precision of rare decays such as $H \to \gamma\gamma$. Note that a large portion of the systematic uncertainties intrinsic to a hadron collider can be canceled by taking ratios of measured cross sections. For example, combining the ratio of the rates of $pp \to H \to \gamma\gamma$ and $pp \to H \to ZZ^*$ at the LHC and the measurement of the $HZZ$ coupling at the CEPC can significantly improve the $\kappa_\gamma$ precision. These are the most useful inputs from the LHC to combine with the CEPC. Similar studies of combination with the LHC for the ILC can be found in Refs. [23, 24, 46, 61, 62].

The results of the 10-parameter and the 7-parameter fits for the CEPC with an integrated luminosity of $5.6\,\mathrm{ab}^{-1}$ are shown in Table 11.4. The combined precision with the HL-LHC estimates [63] are also shown. The combinations assume no associated theoretical uncertainties and, hence, represent the aggressive use of the HL-LHC projection.[4] It is assumed that the HL-LHC will operate at $\sqrt{s} = 14$ TeV and accumulate an integrated luminosity of 3000 $\mathrm{fb}^{-1}$. For the 7-parameter fit, the Higgs boson width is a derived quantity, not an independent parameter. Its precision, derived from the precision of the fitted parameters, is 2.4% for the CEPC alone and 1.8% when combined with the HL-LHC projection.

The CEPC Higgs boson property measurements mark a giant step beyond the HL-LHC. First of all, in contrast to the LHC, a lepton collider Higgs factory is capable of measuring the Higgs boson width and the absolute coupling strengths to other particles. A comparison with the HL-LHC is only possible with model dependent assumptions. One of such comparisons is within the framework of the 7-parameter fit, shown in Figure 11.8. Even with this set of restrictive assumptions, the advantage of the CEPC is still significant. The measurement of $\kappa_Z$ is more than a factor of 10 better. The CEPC can also improve significantly the precision on a set of $\kappa$ parameters that are affected by large backgrounds at the LHC, such as $\kappa_b$, $\kappa_c$, and $\kappa_g$. Note that this is in comparison with the HL-LHC projection with large systematic uncertainties. Such uncertainties are typically under much better control at lepton colliders. Within this 7-parameter set, the only coupling that the HL-LHC can give a competitive measurement is $\kappa_\gamma$, for which the CEPC sensitivity is statistically limited. This is also the most valuable input that the HL-LHC can give to the Higgs boson coupling measurements at the CEPC, which underlines the importance of combining the results from these two facilities.

[4]Note that the LHC and the CEPC have different sources of theoretical uncertainties, for detailed discussion, see Refs. [13, 21, 64–66].

| Relative coupling measurement precision and the 95% CL upper limit on $BR^{BSM}_{inv}$ | | | | |
|---|---|---|---|---|
| | 10-parameter fit | | 7-parameter fit | |
| Quantity | CEPC | CEPC+HL-LHC | CEPC | CEPC+HL-LHC |
| $\kappa_b$ | 1.3% | 1.0% | 1.2% | 0.9% |
| $\kappa_c$ | 2.2% | 1.9% | 2.1% | 1.9% |
| $\kappa_g$ | 1.5% | 1.2% | 1.5% | 1.1% |
| $\kappa_W$ | 1.4% | 1.1% | 1.3% | 1.0% |
| $\kappa_\tau$ | 1.5% | 1.2% | 1.3% | 1.1% |
| $\kappa_Z$ | 0.25% | 0.25% | 0.13% | 0.12% |
| $\kappa_\gamma$ | 3.7% | 1.6% | 3.7% | 1.6% |
| $\kappa_\mu$ | 8.7% | 5.0% | – | – |
| $BR^{BSM}_{inv}$ | $< 0.30\%$ | $< 0.30\%$ | – | – |
| $\Gamma_H$ | 2.8% | 2.3% | – | – |

**Table 11.4:** Coupling measurement precision from the 10-parameter fit and 7-parameter fit described in the text for the CEPC, and corresponding results after combination with the HL-LHC. All the numbers refer to are relative precision except for $BR^{BSM}_{inv}$ for which the 95% CL upper limit are quoted respectively. Some entries are left vacant for the 7-parameter fit as they are not dependent parameters under the fitting assumptions.

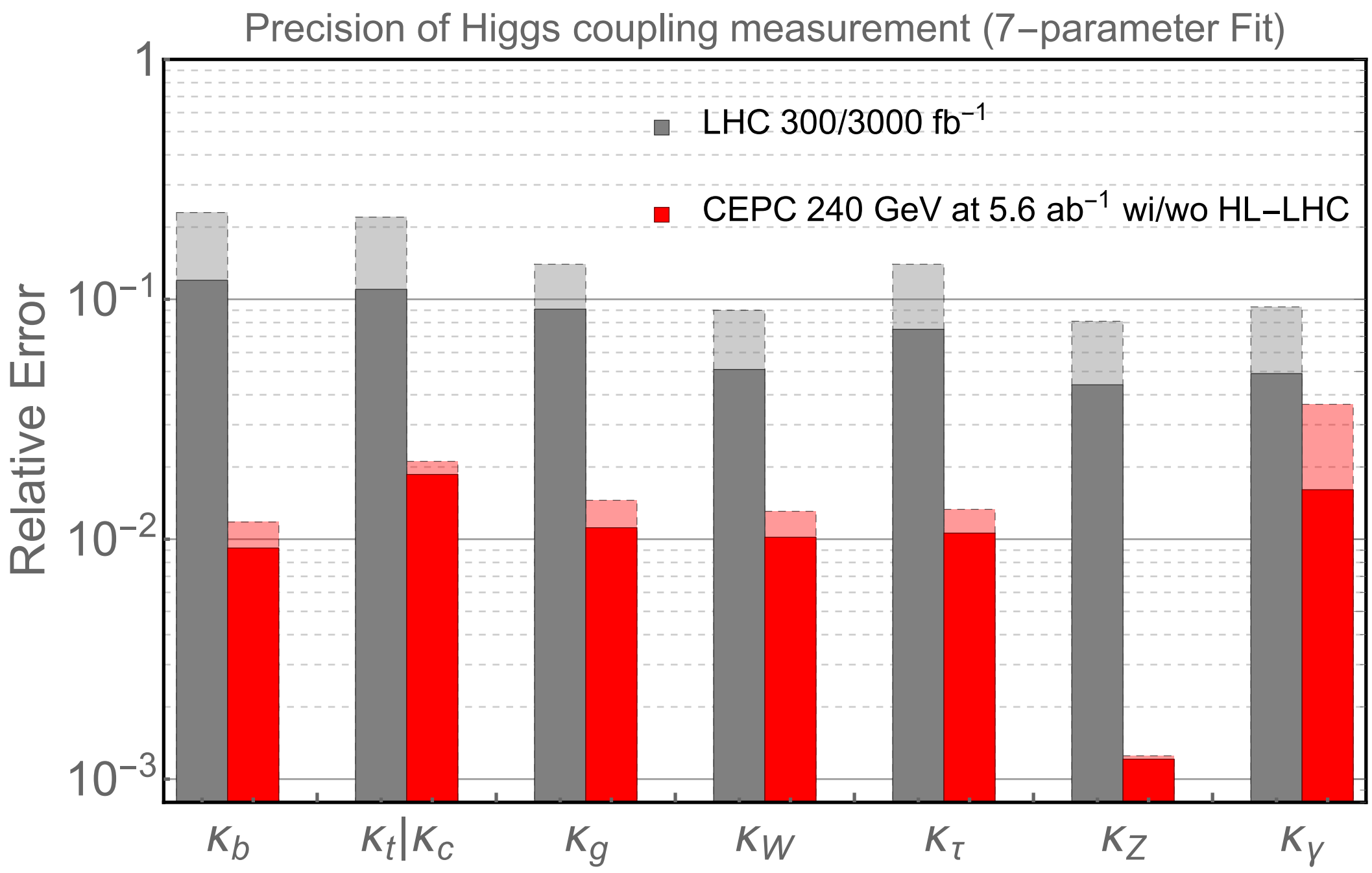

**Figure 11.8:** The results of the 7-parameter fit and comparison with the HL-LHC [63]. The projections for the CEPC at 240 GeV with an integrated luminosity of $5.6\,\text{ab}^{-1}$ are shown. The CEPC results without combination with the HL-LHC input are shown as light red bars. The LHC projections for an integrated luminosity of 300 $\text{fb}^{-1}$ are shown in light gray bars.

The direct search for Higgs boson decay into invisible BSM particles is well motivated and closely connected to the dark sectors. The CEPC with an integrated luminosity of $5.6\,\mathrm{ab}^{-1}$ has a sensitivity of 0.30% expressed in terms of the 95% CL upper limit on the decay branching ratio, as shown in Table 11.4. The HL-LHC, on the other hand, has a much lower sensitivity of 6–17% [21] while optimistically may reach 2–3.5% [67].

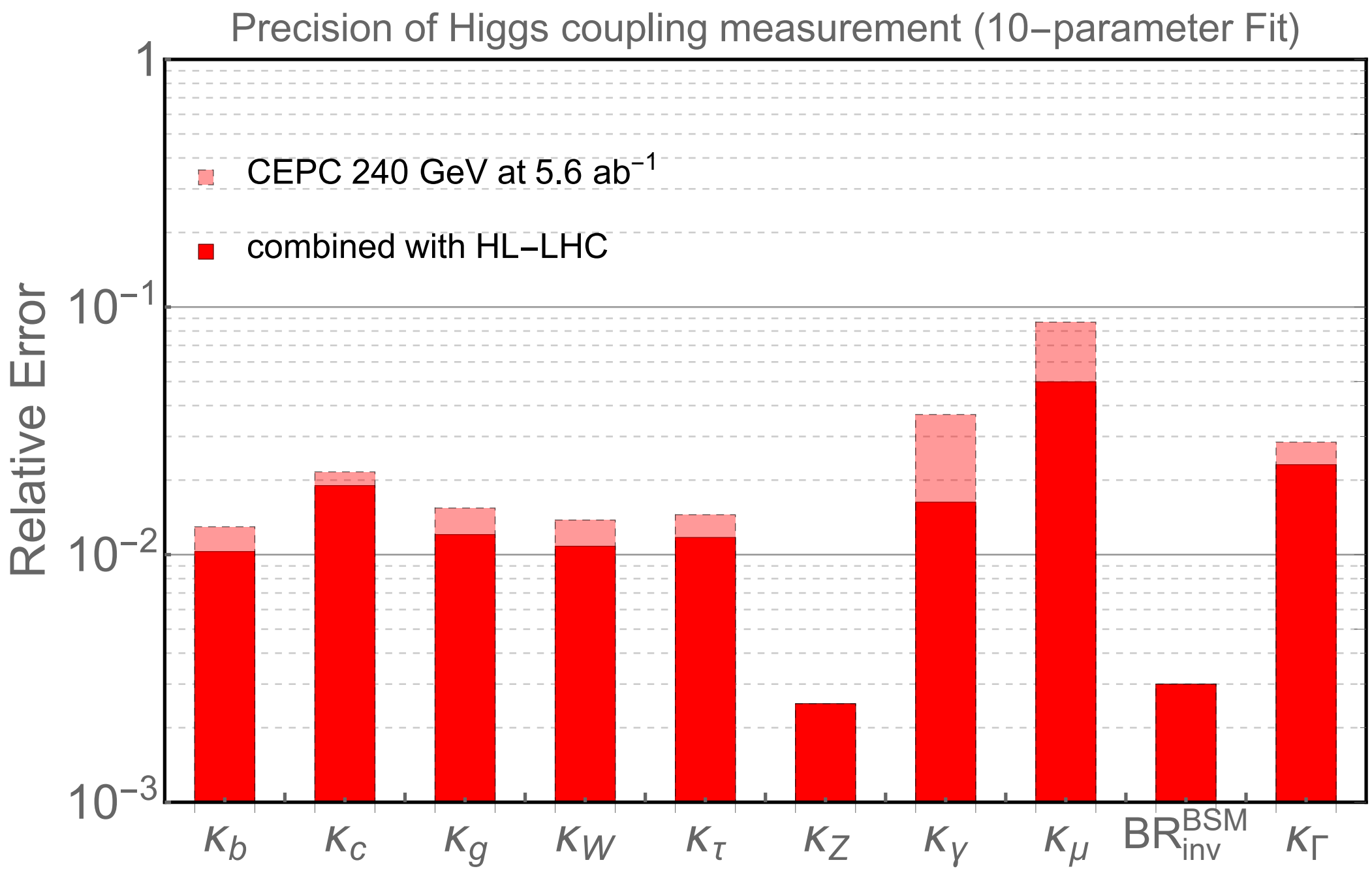

**Figure 11.9:** The 10 parameter fit results for the CEPC at 240 GeV with an integrated luminosity of $5.6\,\mathrm{ab}^{-1}$ (light red bars) and for the combination with the HL-LHC inputs (dark red bars). All the numbers are relative precision except for $\mathrm{BR}_{\mathrm{inv}}^{\mathrm{BSM}}$ for which the 95% CL upper limit are quoted.

As discussed above, one of the greatest advantages of a lepton collider Higgs factory is its capability to measure the Higgs boson width and couplings in a *model-independent* way. The projection of such a determination at the CEPC is shown in Figure 11.9. For most of the measurements, an order of magnitude improvements over the HL-LHC are expected. The CEPC has a clear advantage in the measurement of $\kappa_Z$. It can also set a much stronger constraint on $\mathrm{BR}_{\mathrm{inv}}^{\mathrm{BSM}}$.

#### 11.1.7.2 EFFECTIVE-FIELD-THEORY ANALYSIS

With the assumption that the scale of new physics is higher than the relevant energy scale directly accessible at the Higgs factory, the effect of the new physics can be characterized within the EFT framework. In this framework, operators with dimension larger than four supplement the SM Lagrangian. Imposing baryon and lepton numbers conservation, all higher dimensional operators are of even dimension:

$$\mathcal{L}_{\rm EFT} = \mathcal{L}_{\rm SM} + \sum_i \frac{c_i^{(6)}}{\Lambda^2}\mathcal{O}_i^{(6)} + \sum_j \frac{c_j^{(8)}}{\Lambda^4}\mathcal{O}_j^{(8)} + \cdots \tag{11.12}$$

The leading new physics effects at the electroweak scale would be from the dimension-six operators. To obtain robust constraints on the Wilson coefficients, $c_i$, a global analysis is required, which includes contributions from all possible dimension-six operators. While a large number of dimension-six operators can be written down, only a subset of them contribute to the Higgs boson processes at the leading order. Among these operators, some are much better constrained by other measurements. It is thus reasonable to focus on the operator that primarily contribute to the Higgs boson processes and therefore reduce the parameter space by making appropriate assumptions, as done in the recent studies for future lepton colliders [42, 44–49]. Following these studies, the $CP$-violating operators as well as the ones that induce fermion dipole interactions are discarded. At the leading order, $CP$-violating operators do not have linear contributions to the rates of the Higgs boson processes. While they do contribute to angular observables at the leading order [40, 41], these operators are usually much better constrained by the Electric Dipole Moment (EDM) experiments [68–70], though some parameter space is still available for the $CP$-violating couplings of the Higgs boson to heavy flavor quarks and leptons [71, 72]. The interference between the fermion dipole interactions with SM terms are suppressed by the fermion masses. The corresponding operators also generate dipole moments, which are stringently constrained, especially for light fermions. For the operators that modify the Yukawa coupling matrices, only the five diagonal ones that correspond to the top, charm, bottom, tau, and muon Yukawa couplings are considered, which are relevant for the Higgs boson measurements at the CEPC.

Before presenting the projections, some brief comments on the EFT framework are in order. In comparison with the $\kappa$-framework, a significant advantage of the EFT is that it gives physical parametrization of potential new physics effects. EFT operators can be used directly in computations. The EFT framework also allows for a natural inclusion of new observables, with possible correlations automatically taken into account. At the same time, the connections with experimental observables are less direct and intuitive. Sometimes, the EFT approach is referred to as model-independent. This is only accurate to a certain extent. It assumes that there are no new light degrees of freedom. In practice, assumptions are often made to simplify the set of EFT operators, as also done here.

The electroweak precision observables are already tightly constrained by the LEP $Z$-pole and $W$ mass measurements. The CEPC $Z$-pole run can further improve the constraints set by the LEP, thanks to the enormous amount ($\sim 10^{11}$–$10^{12}$) of $Z$ bosons. The $W$ mass can also be measured with a precision of a few MeVs at the CEPC even without a dedicated $WW$ threshold run. Given that the expected precision of the $Z$-pole observables and the $W$ mass are much higher than the ones of Higgs boson observables, it is assumed that the former ones are perfectly constrained, which significantly simplifies the analysis. In particular, in a convenient basis all the contact interaction terms of the form $HVf\bar{f}$ can be discarded since they also modify the fermion gauge couplings. Realistic $Z$-pole constraints have also been considered in recent studies [46, 47, 49], but certain assumptions (such as flavor-universality) and simplifications are made. Future studies with more general frameworks are desired to fully determine the impact of the $Z$-pole measurements on the Higgs boson analysis.

| CEPC 240 GeV ($5.6\,\mathrm{ab}^{-1}$) | | | | |
|---|---|---|---|---|
| | uncertainty | correlation matrix | | |
| | | $\delta g_{1,Z}$ | $\delta\kappa_\gamma$ | $\lambda_Z$ |
| $\delta g_{1,Z}$ | $1.2 \times 10^{-3}$ | 1 | 0.08 | -0.90 |
| $\delta\kappa_\gamma$ | $0.9 \times 10^{-3}$ | | 1 | -0.42 |
| $\lambda_Z$ | $1.3 \times 10^{-3}$ | | | 1 |

**Table 11.5:** The estimated constraints on aTGCs from the measurements of the diboson process $e^+e^- \to WW$ in the semi-leptonic channel at the CEPC 240 GeV with $5.6\,\mathrm{ab}^{-1}$ data and unpolarized beams. All angular distributions are used in the fit. Only the statistical uncertainties of the signal events are considered, assuming a selection efficiency of 80%.

The measurements of the Triple Gauge boson Couplings (TGCs) from the diboson process $e^+e^- \to WW$ play an important role in the Higgs boson coupling analysis under the EFT framework. Focusing on $CP$-even dimension-six operators, the modifications to the triple gauge vertices from new physics can be parametrized by three anomalous TGC parameters (aTGCs), conventionally denoted as $\delta g_{1,Z}$, $\delta\kappa_\gamma$ and $\lambda_Z$ [73, 74]. Among them, $\delta g_{1,Z}$ and $\delta\kappa_\gamma$ are generated by operators that also contribute to the Higgs boson processes. At 240 GeV, the $e^+e^- \to WW$ cross section is almost two orders of magnitude larger than the $ZH$ cross section. The measurements of the diboson process thus provide strong constraints on the operators that generate the aTGCs. A dedicated study on the TGC measurements at the CEPC is not currently available. A simplified analysis is thus performed to estimate the aTGCs sensitivity. The results are shown in Table 11.5. The analysis roughly follows the methods in Refs. [45, 75]. Only the $WW$ events in the semi-leptonic (electron or muon) channel are used, which are easier to reconstruct and have a sizable branching ratio ($\approx 29\%$). In particular, the production polar angle, as well as the two decay angles of the leptonically decaying $W$ boson, can be fully reconstructed, which contain important information on the aTGCs. The two decay angles of the hadronically decaying $W$ boson can only be reconstructed with a two-fold ambiguity. A $\chi^2$ fit of the three aTGC parameters to the binned distribution of all five angles is performed, from which the one-sigma interval for each of the three aTGCs as well as the correlations among them are extracted. A signal selection efficiency of 80% is assumed. The effects of systematic uncertainties and backgrounds are not considered, assuming they are under control after the selection cuts.

Under these assumptions, the dimension-six operator contribution to the Higgs boson and diboson processes has a total of twelve degrees of freedom. While all non-redundant bases are equivalent, it is particularly convenient to choose a basis in which the twelve degrees of freedom can be mapped to exactly twelve operators, whereas the rest are removed by the assumptions. Two such bases are considered in this analysis. The first is defined by the set of dimension-six operators in Table 11.6. Among them, $\mathcal{O}_{3W}$ corresponds to the aTGC parameter $\lambda_Z$, $\mathcal{O}_{HW}$ and $\mathcal{O}_{HB}$ generate the aTGC parameters $\delta g_{1,Z}$ and $\delta\kappa_\gamma$ as well as Higgs boson anomalous couplings, while the rest operators can only be probed by the Higgs boson measurements at the leading order. The second basis is the so-called "Higgs basis," proposed in Ref. [76]. In the Higgs basis, the parameters are defined in terms of the mass eigenstates after the electroweak symmetry breaking, and can be directly interpreted as the size of the Higgs boson couplings. Different from the original Higgs basis, this analysis follows Ref. [45], with the parameters associated with the $Hgg$, $H\gamma\gamma$ and $HZ\gamma$ vertices normalized to the SM one-loop contributions, and denoted as $\bar{c}_{gg}$, $\bar{c}_{\gamma\gamma}$ and $\bar{c}_{Z\gamma}$ (as opposed to $c_{gg}$, $c_{\gamma\gamma}$ and $c_{Z\gamma}$ in Ref. [76]). The parameter $\bar{c}^{\text{eff}}_{gg}$ is further defined to absorb all contributions to the $Hgg$ vertex. With these redefinitions, the set of twelve parameters is given by

$$\delta c_Z\,,\;\; c_{ZZ}\,,\;\; c_{Z\Box}\,,\;\; \bar{c}_{\gamma\gamma}\,,\;\; \bar{c}_{Z\gamma}\,,\;\; \bar{c}^{\text{eff}}_{gg}\,,\;\; \delta y_t\,,\;\; \delta y_c\,,\;\; \delta y_b\,,\;\; \delta y_\tau\,,\;\; \delta y_\mu\,,\;\; \lambda_Z\,. \tag{11.13}$$

| | |
|---|---|
| $\mathcal{O}_H = \frac{1}{2}(\partial_\mu \lvert H^2 \rvert)^2$ | $\mathcal{O}_{GG} = g_s^2 \lvert H \rvert^2 G^A_{\mu\nu} G^{A,\mu\nu}$ |
| $\mathcal{O}_{WW} = g^2 \lvert H \rvert^2 W^a_{\mu\nu} W^{a,\mu\nu}$ | $\mathcal{O}_{y_u} = y_u \lvert H \rvert^2 \bar{Q}_L \tilde{H} u_R + \text{h.c.}$ $(u \to t, c)$ |
| $\mathcal{O}_{BB} = g'^2 \lvert H \rvert^2 B_{\mu\nu} B^{\mu\nu}$ | $\mathcal{O}_{y_d} = y_d \lvert H \rvert^2 \bar{Q}_L H d_R + \text{h.c.}$ $(d \to b)$ |
| $\mathcal{O}_{HW} = ig(D^\mu H)^\dagger \sigma^a (D^\nu H) W^a_{\mu\nu}$ | $\mathcal{O}_{y_e} = y_e \lvert H \rvert^2 \bar{L}_L H e_R + \text{h.c.}$ $(e \to \tau, \mu)$ |
| $\mathcal{O}_{HB} = ig'(D^\mu H)^\dagger (D^\nu H) B_{\mu\nu}$ | $\mathcal{O}_{3W} = \frac{1}{3!} g \epsilon_{abc} W^{a\,\nu}_\mu W^b_{\nu\rho} W^{c\,\rho\mu}$ |

**Table 11.6:** A complete set of $CP$-even dimension-six operators that contribute to the Higgs boson and TGC measurements, assuming there is no correction to the $Z$-pole observables and the $W$ mass, and also no fermion dipole interaction. $G^A_{\mu\nu}$, $W^a_{\mu\nu}$ and $B_{\mu\nu}$ are the field strength tensors for the SM $SU(3)_c$, $SU(2)_L$ and $U(1)_Y$ gauge fields, respectively. For $\mathcal{O}_{y_u}$, $\mathcal{O}_{y_d}$ and $\mathcal{O}_{y_e}$, only the contributions to the diagonal elements of the Yukawa matrices that corresponds to the top, charm, bottom, tau, and muon couplings are considered.

These parameters can be conveniently interpreted as the precision of the Higgs boson couplings analogous to those in the $\kappa$-framework. In particular, $\delta c_Z$, $\bar{c}_{\gamma\gamma}$, $\bar{c}_{Z\gamma}$, $\bar{c}^{\text{eff}}_{gg}$ and $\delta y_{t,c,b,\tau,\mu}$ modifies the sizes of the SM Higgs boson couplings to $ZZ$, $\gamma\gamma$, $Z\gamma$, $gg$ and fermions, respectively. The parameters $c_{ZZ}$ and $c_{Z\Box}$ parametrize the anomalous $HZZ$ couplings:

$$\mathcal{L} = \frac{h}{v}\left[c_{ZZ}\,\frac{g^2+g'^2}{4} Z_{\mu\nu}Z^{\mu\nu} + c_{Z\Box}\, g^2 Z_\mu \partial_\nu Z^{\mu\nu}\right] + \dots\,, \tag{11.14}$$

which are not present in the SM at the leading order. The $HWW$ couplings are written in terms of the parameters shown in Eq. 11.13 via gauge invariance and are not shown explicitly. For the three aTGC parameters, $\lambda_Z$ is kept in Eq. 11.13, while $\delta g_{1,Z}$ and $\delta\kappa_\gamma$ are written in terms of linear combinations of $c_{ZZ}$, $c_{Z\Box}$, $\bar{c}_{\gamma\gamma}$ and $\bar{c}_{Z\gamma}$. The exact definitions of the Higgs basis and the translation to the basis in Table 11.6 can be found in Ref. [45].

The estimated precision of all the Higgs boson rate measurements in Section 11.1.5 (Table 11.3), along with their correlations, are included as inputs to the EFT global analysis. In addition, the angular observables of the $e^+e^- \to ZH$, $Z \to \ell^+\ell^-$, $H \to b\bar{b}$ channel are included, following the studies in Refs. [40, 41]. This channel is almost background-free after the selection, with a signal selection efficiency of about 40%. For the TGC measurements, the results in Table 11.5 are used as inputs. The global $\chi^2$ is obtained by summing over the $\chi^2$ of all the measurements. Due to the high precision of the measurements, it is shown that for all observables, keeping only the linear terms of all EFT parameters gives a very good approximation [45]. This greatly simplifies the fitting procedure, as the total $\chi^2$ can be written as

$$\chi^2 = \sum_{ij} (c - c_0)_i \, \sigma_{ij}^{-2} \, (c - c_0)_j \, , \text{where } \sigma_{ij}^{-2} \equiv (\delta c_i \, \rho_{ij} \, \delta c_j)^{-1} \, , \tag{11.15}$$

where $c_i$'s are the EFT parameters, $c_0$'s are the corresponding central values which are zero by construction, as the measurements are assumed to be SM-like. The one-sigma uncertainties $\delta c_i$ and the correlation matrix $\rho$ can be obtained from $\sigma_{ij}^{-2} = \partial^2 \chi^2 / \partial c_i \partial c_j$.

For comparison, the sensitivities of the LHC 14 TeV with total luminosities of $300\,\text{fb}^{-1}$ and $3000\,\text{fb}^{-1}$ are also considered. These are combined with the diboson ($e^+e^- \to WW$) measurements at the LEP as well as the LHC 8 TeV Higgs boson measurements. For the LHC 14 TeV Higgs boson measurements, the projections by the ATLAS collaboration [63] are used, while the composition of each channel is obtained from Refs. [77–81]. The constraints from the LHC 8 TeV Higgs boson measurements and the diboson measurements at the LEP are obtained directly from Ref. [82]. While the LHC diboson measurements can potentially improve the constraints on aTGCs set by the LEP [38], they are not included in this analysis due to the potential issues related to the validity of the EFT [83, 84] and the assumption that the TGCs are dominated by the non-anomalous terms [85].

The results of the 12-parameter fit at the CEPC are shown in Figure 11.10 for the Higgs basis and Figure 11.11 for the basis in Table 11.6. The results from the LHC Higgs boson measurements (both $300\,\text{fb}^{-1}$ and $3000\,\text{fb}^{-1}$) combined with the LEP diboson measurements are shown in comparison. The results of the combination of the CEPC with the HL-LHC ($3000\,\text{fb}^{-1}$) are also shown in addition to the ones from the CEPC alone. In Figure 11.10, the results are shown in terms of the one-sigma precision of each parameter. The LHC results are shown with gray columns with $300\,\text{fb}^{-1}$ ($3000\,\text{fb}^{-1}$) in light (dark) bars, while the CEPC ones are shown with the red columns, with the CEPC-alone (combination with the HL-LHC) results shown in light (dark) bars. In Figure 11.11, the results are presented in terms of the sensitivity to $\Lambda/\sqrt{|c_i|}$ at 95% CL for each operator as defined in Eq. 11.12, where $\Lambda$ is the scale of new physics and $c_i$ is the corresponding Wilson coefficient. Four columns are shown separately for the LHC $300\,\text{fb}^{-1}$, the HL-LHC $3000\,\text{fb}^{-1}$, the CEPC alone and the CEPC combined with the HL-LHC. The results of the global fits, i.e. simultaneous fits to the 12 parameters, are shown with dark colored bars. The results from individual fits are shown with light colored bars, which are obtained by switching on one operator at a time with the rest fixed to zero.

It is transparent from Figure 11.10 that the CEPC can measure the Higgs boson couplings with precision that is one order of magnitude better than the LHC [63, 86]. For the parameters $\bar{c}_{\gamma\gamma}$, $\bar{c}_{Z\gamma}$ and $\delta y_\mu$, the clean signal and small branching ratios of the corresponding channels ($H \to \gamma\gamma/Z\gamma/\mu\mu$) makes the HL-LHC precision comparable to the CEPC. The combination with the LHC measurements thus provides non-negligible improvements, especially for those parameters. It should be noted that, while $\delta y_t$ modifies the $Hgg$ vertex via the top-quark loop contribution, the CEPC alone cannot discriminate it from the $Hgg$ contact interaction obtained from integrating out a heavy new particle in the loop. The parameter $\bar{c}_{gg}^{\text{eff}}$ absorbs both contributions and reflects the overall precision of the $Hgg$ coupling. The combination with the LHC $t\bar{t}H$ measurements can resolve this flat direction. The CEPC measurements, in turn, can improve the constraint on $\delta y_t$ set by the LHC by providing much better constraints on the other parameters that contribute to the $t\bar{t}H$ process. It should also be noted that the measurement of the charm Yukawa coupling is not reported in Ref. [63], while the projection of its constraint has a large variation among different studies and can be much larger than one [87–92]. Therefore, $\delta y_c$ is fixed to be zero for the LHC-only fits, as treating $\delta y_c$ as an unconstrained free parameter generates a flat direction in the fit which makes the overall sensitivity much worse. The CEPC, on the other hand, provides excellent measurements of the charm Yukawa coupling and can constrain $\delta y_c$ to about $\sim 2\%$.

Regarding the sensitivity to $\Lambda/\sqrt{|c_i|}$ in Figure 11.11, it is also clear that the CEPC has a significantly better performance than the LHC. If the couplings are naïvely assumed to be of order one ($c_i \sim 1$), the Higgs boson measurements at the CEPC would be sensitive to new physics scales at several TeV. While the individual sensitivity to some of the operators at the LHC can be comparable to the CEPC (*e.g.*, $O_{WW}$ and $O_{BB}$ from the measurement of $H \to \gamma\gamma$), the CEPC sensitivity is much more robust under a global framework. This is due to its comprehensive measurements of both the inclusive $ZH$ cross section and the exclusive rates of many Higgs boson decay channels. Operators $O_{GG}$ and $O_{y_t}$ both contribute to the $Hgg$ vertex. While the CEPC can provide strong constraints on either of them if the other is set to zero, they can only be constrained in a global fit if the $t\bar{t}H$ measurements at the LHC are also included. It is also important to note that the validity of EFT can be a potential issue for the LHC measurements [83]. Depending on the size of the couplings, the inferred bounds on the new physics scale $\Lambda$ can be comparable with or even smaller than the energy scale probed by the LHC. The CEPC has a smaller center of mass energy and much better precision, which ensures the validity of EFT for most new physics scenarios.

In Table 11.7, the numerical results of the global fit are presented for the CEPC in terms of the one-sigma interval of the 12 parameters and the correlations among them. The results assume an integrated luminosity of $5.6\,\mathrm{ab}^{-1}$ at 240 GeV with unpolarized beams, both without and with the combination with the HL-LHC ($3000\,\mathrm{fb}^{-1}$) Higgs boson measurements. With both the one-sigma bounds and the correlation matrix, the corresponding $\chi^2$ can be reconstructed, which can be used to derive the constraints in any other EFT basis or any particular model that can be matched to the EFT. This offers a convenient way to study the sensitivity to new physics models, as detailed knowledge of the experimental measurements are not required.

In the EFT framework, it is explicitly assumed that the Higgs boson width is the sum of all partial widths of its SM decay channels. This is because the EFT expansion in Eq. 11.12 relies on the assumption that the new physics scale is sufficiently high, while any potential Higgs boson exotic decay necessarily introduces light BSM particles, thus in direct conflict with this assumption. One can nevertheless treat the Higgs boson total width as a free parameter in the EFT global fit and obtain an indirect constraint of it, as done in Ref. [46]. With this treatment, the CEPC can constrain the Higgs boson width to a precision of $1.7\%$ ($1.6\%$ if combined with the HL-LHC). This result is significantly better than the one from the 10-parameter coupling fit in Table 11.4 ($3.4\%/2.6\%$). The improvement is mainly because the $HWW$ and $HZZ$ couplings are treated as being independent in the 10-parameter coupling fit, while in the EFT framework they are related to each other under gauge invariance and custodial symmetry. It should also be noted that the Higgs boson width determined using Eqs. 11.4 and 11.7 explicitly assumes that the $HWW$ and $HZZ$ couplings are independent of the energy scale. Such an assumption is not valid in the EFT framework with the inclusion of the anomalous couplings.

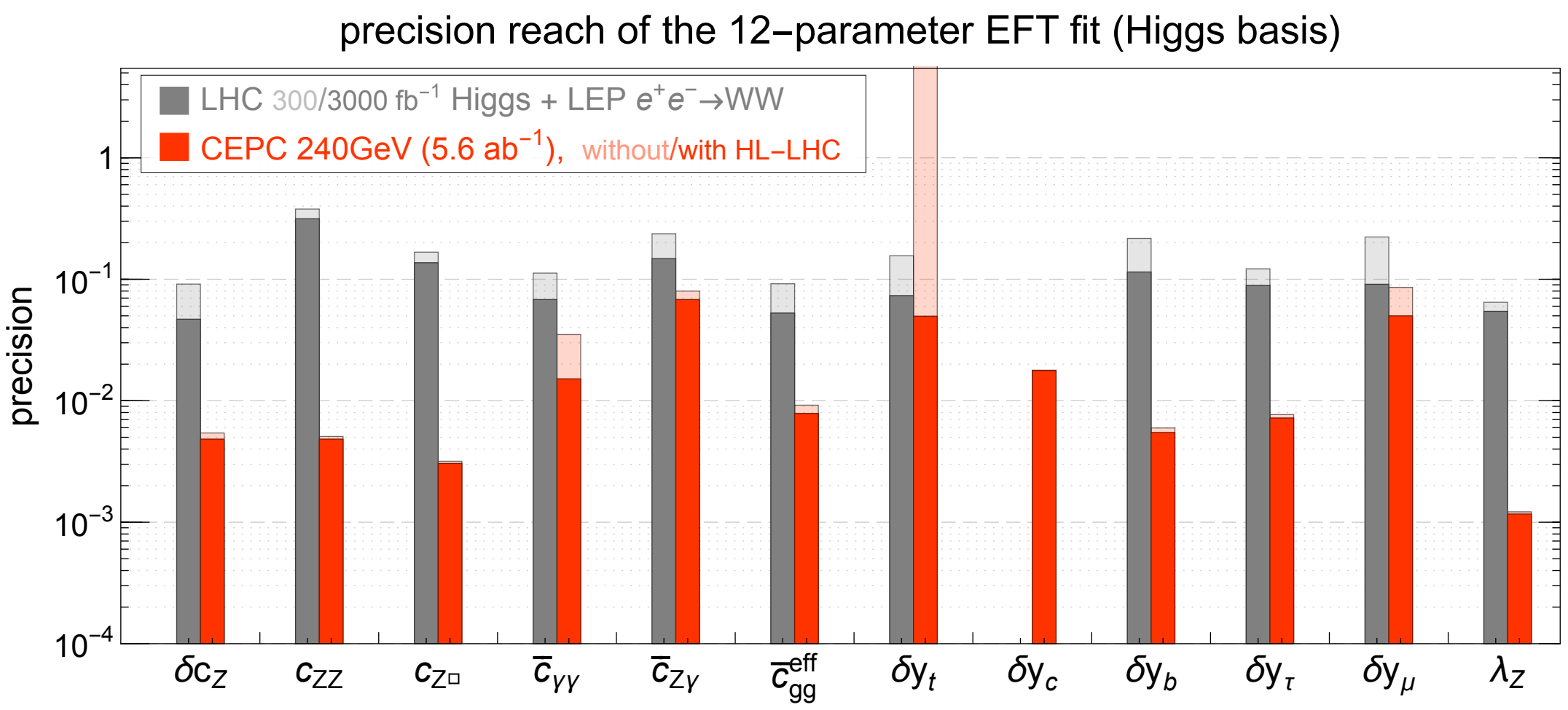

**Figure 11.10:** One-sigma precision of the twelve parameters in the Higgs basis. The first column shows the results from the LHC Higgs boson measurements with $300\,\text{fb}^{-1}$ (light gray bars) and $3000\,\text{fb}^{-1}$ (dark gray bars) combined with the LEP diboson ($e^+e^- \to WW$) measurement. The second column shows the results from the CEPC with $5.6\,\text{ab}^{-1}$ data collected at $240\,\text{GeV}$ with unpolarized beam. The results from the CEPC alone are shown in light red bars, and the ones from a combination of the CEPC and the HL-LHC are shown in dark red bars. For the LHC fits, $\delta y_c$ is fixed to zero.

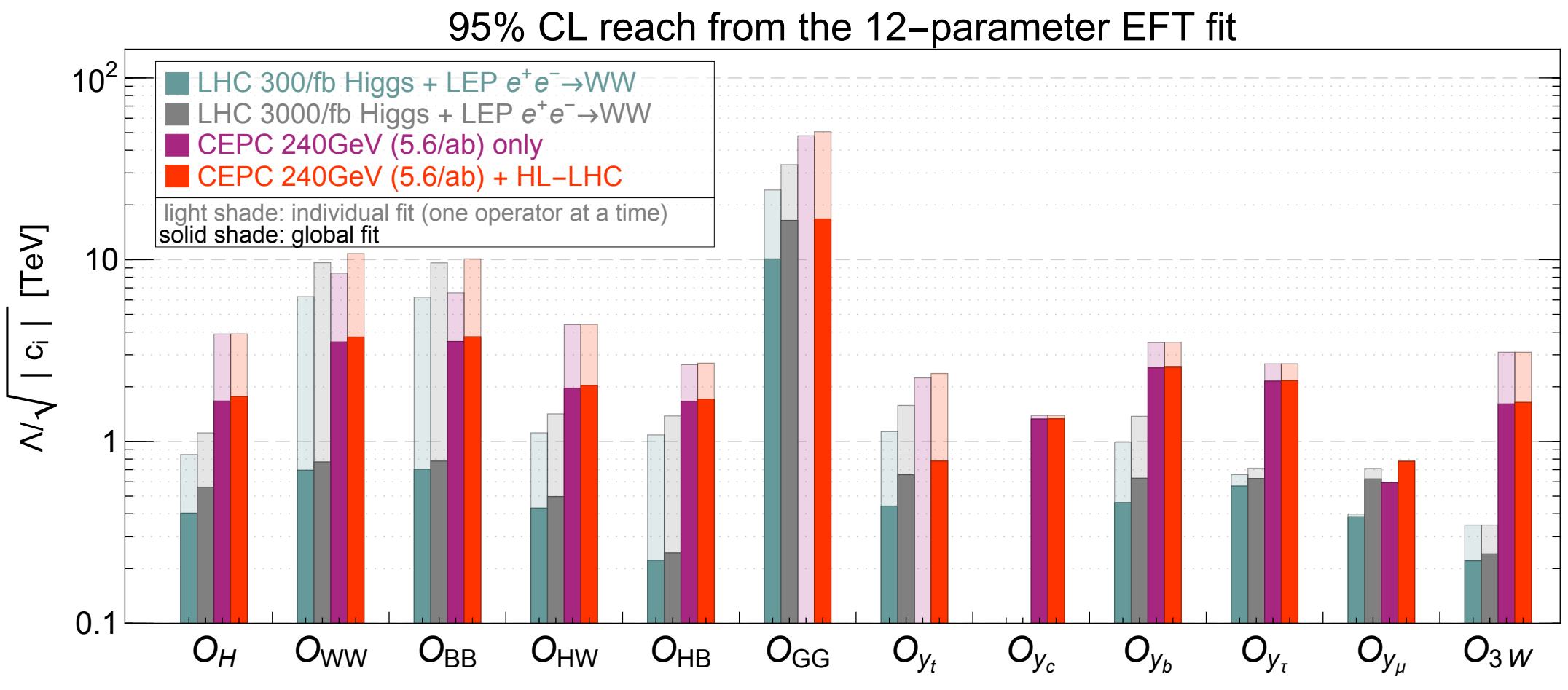

**Figure 11.11:** The 95% CL sensitivity to $\Lambda/\sqrt{|c_i|}$ for the operators in the basis defined in Table 11.6. The first two columns show the results from the LHC Higgs boson measurements with $300\,\text{fb}^{-1}$ and $3000\,\text{fb}^{-1}$ combined with the LEP diboson ($e^+e^- \to WW$) measurement. The last two columns show the results from the CEPC alone and the combination of the CEPC and the HL-LHC ($3000\,\text{fb}^{-1}$). The results of the global fits are shown as dark colored bars. The results from individual fits (by switching on one operator at a time) are shown with light colored bars. For the LHC fits, $\delta y_c$ is fixed to zero.

### 11.1.8 THE HIGGS BOSON SELF-COUPLING

The Higgs boson self-coupling is a critical parameter governing the dynamics of the electroweak symmetry breaking. In the SM, the Higgs boson trilinear and quadrilinear cou-plings are fixed once the values of the electroweak vacuum expectation value and the Higgs boson mass are known. Any deviation from the SM prediction is thus clear evidence of new physics beyond the SM. The Higgs trilinear coupling is probed at the LHC by the measurement of the di-Higgs production, $pp \to HH$. Current bounds on the Higgs trilinear coupling is at the $\mathcal{O}(10)$ level, while the HL-LHC is expected to improve the precision to the level of $\mathcal{O}(1)$ [93]. The prospects for extracting the Higgs boson quadrilinear coupling are much less promising, even for a 100 TeV hadron collider [94].

| Higgs basis | | | | | | | | | | | |
|---|---|---|---|---|---|---|---|---|---|---|---|
| $\delta c_Z$ | $c_{ZZ}$ | $c_{Z\Box}$ | $\bar{c}_{\gamma\gamma}$ | $\bar{c}_{Z\gamma}$ | $\bar{c}_{gg}^{\text{eff}}$ | $\delta y_t$ | $\delta y_c$ | $\delta y_b$ | $\delta y_\tau$ | $\delta y_\mu$ | $\lambda_Z$ |
| 0.0054 | 0.0051 | 0.0032 | 0.035 | 0.080 | 0.0092 | – | 0.018 | 0.0060 | 0.0077 | 0.086 | 0.0012 |
| 0.0048 | 0.0048 | 0.0030 | 0.015 | 0.068 | 0.0079 | 0.050 | 0.018 | 0.0055 | 0.0072 | 0.050 | 0.0012 |
| $c_i/\Lambda^2$ [TeV$^{-2}$] of dimension-six operators | | | | | | | | | | | |
| $c_H$ | $c_{WW}$ | $c_{BB}$ | $c_{HW}$ | $c_{HB}$ | $c_{GG}$ | $c_{y_t}$ | $c_{y_c}$ | $c_{y_b}$ | $c_{y_\tau}$ | $c_{y_\mu}$ | $c_{3W}$ |
| 0.18 | 0.040 | 0.040 | 0.13 | 0.18 | – | – | 0.28 | 0.077 | 0.11 | 1.4 | 0.19 |
| 0.16 | 0.035 | 0.035 | 0.12 | 0.17 | 0.0018 | 0.82 | 0.28 | 0.076 | 0.11 | 0.83 | 0.19 |

**Table 11.7:** The one-sigma uncertainties for the 12 parameters from the CEPC (240 GeV, $5.6\,\text{ab}^{-1}$) in the Higgs basis and the basis of dimension-six operators. For both cases, the upper (lower) row correspond to results without (with) the combination of the HL-LHC Higgs boson measurements.. Note that, without the $t\bar{t}H$ measurements, $\delta y_t$ can not be constrained in a global fit, thus $c_{GG}$ and $c_{y_t}$ can not be resolved.

To measure the di-Higgs production at a lepton collider, a sufficiently large center of mass energy ($\gtrsim 400\,\text{GeV}$) is required, which is likely to be achieved only at a linear collider. The CEPC, instead, can probe the Higgs boson trilinear coupling via its loop contributions to the single Higgs boson processes. This indirect approach, nevertheless, provides competitive sensitivity, since the loop suppression is compensated by the high precision of the Higgs boson measurements at the CEPC [95]. With a precision of $0.5\%$ on the inclusive $ZH$ cross section at 240 GeV, the Higgs boson trilinear coupling can be constrained to a precision of $35\%$, assuming all other Higgs boson couplings that contribute to $e^+e^- \to ZH$ are SM-like. [5] While this indirect bound is comparable to the direct ones at linear colliders, it relies on strong assumptions which are only applicable to some specific models. A more robust approach is to include all possible deviations on the Higgs boson couplings simultaneously and constrain the Higgs boson trilinear coupling in a global fit. The EFT framework presented in Section 11.1.7.2 is ideal for such an analysis. Under this framework, the one-loop contributions of the trilinear Higgs boson coupling to all the relevant Higgs boson production and decay processes are included, following Ref. [48]. The new physics effect is parametrized by the quantity $\delta\kappa_\lambda \equiv \kappa_\lambda - 1$, where $\kappa_\lambda$ is the ratio of the Higgs boson trilinear coupling to its SM value,

$$\kappa_\lambda \equiv \frac{\lambda_3}{\lambda_3^{\text{sm}}}\,, \qquad \lambda_3^{\text{sm}} = \frac{m_H^2}{2v^2}\,. \tag{11.16}$$

[5] A better precision can be obtained by using, in addition, exclusive channels, such as $\sigma(ZH) \times \text{BR}(H \to b\bar{b})$. However, this will require an even stronger assumption, i.e. that all Higgs boson couplings contributing to the branching ratios are also SM-like except for the Higgs boson trilinear coupling.

The global fit is performed simultaneously with $\delta\kappa_\lambda$ and all the 12 EFT parameters defined in Section 11.1.7.2. The results are presented in Table 11.8. The results for the HL-LHC are also shown, which were obtained in Ref. [96] under the same global framework. For the CEPC 240 GeV, the one-sigma bound on $\delta\kappa_\lambda$ is around $\pm 3$, significantly worse than the $35\%$ in the $\delta\kappa_\lambda$-only fit. This is a clear indication that it is difficult to resolve the effects of $\delta\kappa_\lambda$ from other Higgs boson couplings. For the HL-LHC, the sensitivity to $\delta\kappa_\lambda$ is still dominated by di-Higgs production. However, as a result of the destructive interferences among diagrams, di-Higgs production at the LHC cannot constrain $\delta\kappa_\lambda$ very well on its positive side, even with the use of differential observables [97]. The combination of the HL-LHC and the CEPC 240 GeV thus provides a non-trivial improvement to the HL-LHC result alone, in particular for the two-sigma bound on the positive side, which is improved from $+6.1$ to $+2.7$. This is illustrated in Figure 11.12, which plots the profiled $\chi^2$ as a function of $\delta\kappa_\lambda$ for the two colliders.

| Bounds on $\delta\kappa_\lambda$ | $\Delta\chi^2 = 1$ | $\Delta\chi^2 = 4$ |
|---|---|---|
| CEPC 240 GeV (5.6 ab$^{-1}$) | $[-3.0,\ +3.1]$ | $[-5.9,\ +6.2]$ |
| HL-LHC | $[-0.9,\ +1.3]$ | $[-1.7,\ +6.1]$ |
| HL-LHC + CEPC 240 GeV | $[-0.8,\ +1.0]$ | $[-1.5,\ +2.7]$ |

**Table 11.8:** The $\Delta\chi^2 = 1$ (one-sigma) and $\Delta\chi^2 = 4$ (two-sigma) bounds of $\delta\kappa_\lambda$ for various scenarios, obtained in a global fit by profiling over all other EFT parameters. The results for the HL-LHC are obtained from Ref. [96].

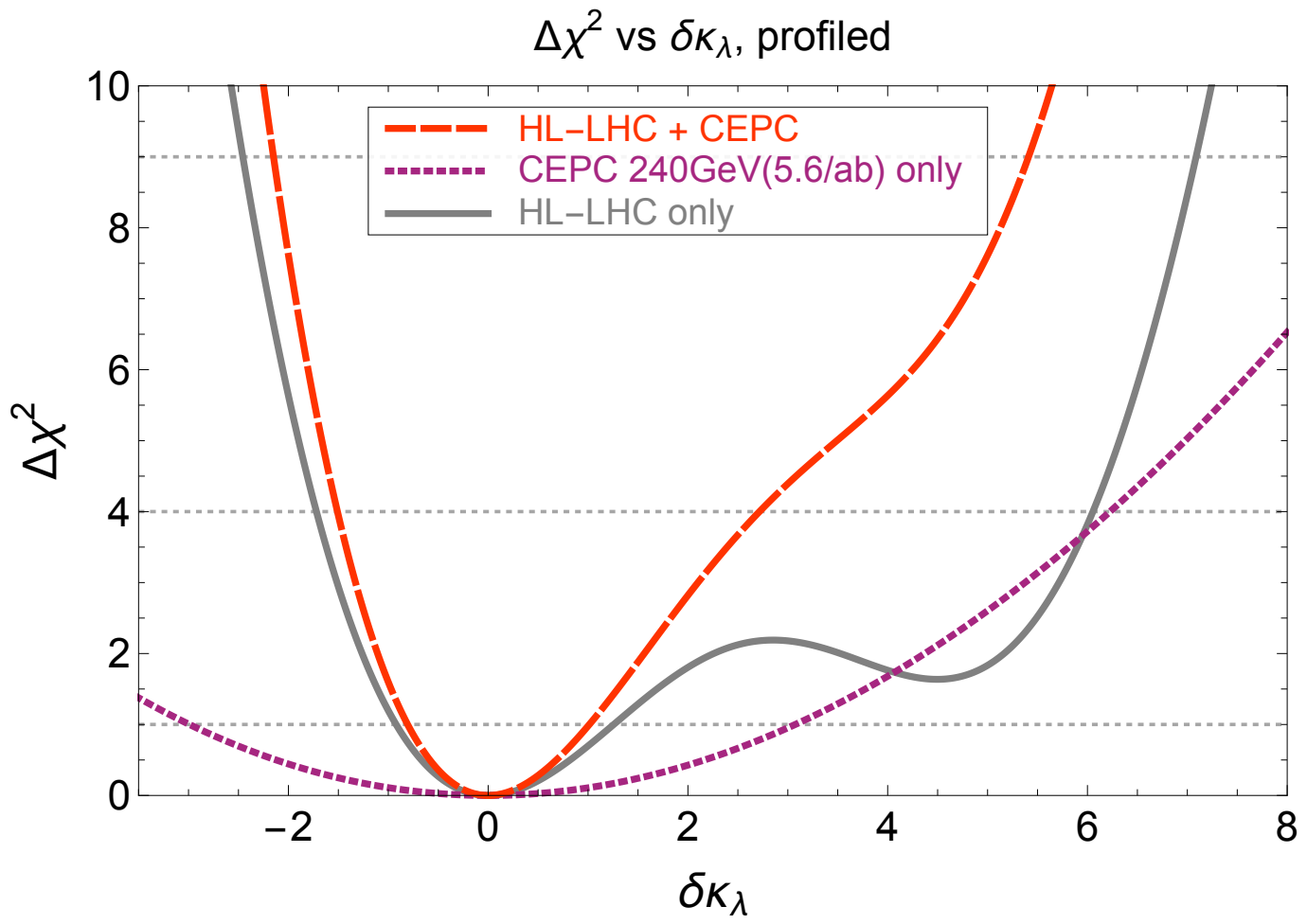

**Figure 11.12:** Chi-square as a function of $\delta\kappa_\lambda$ after profiling over all other EFT parameters for the HL-LHC, the CEPC and their combination. The results for the HL-LHC are obtained from Ref. [96].

### 11.1.9 HIGGS BOSON AND TOP-QUARK COUPLINGS

Interactions of the Higgs boson with the top quark are widely viewed as a window to new physics beyond the SM. The CEPC potential on the interactions between the Higgs boson and the top quark can be evaluated [98–101] by parametrizing these interactions in terms of dimension-six gauge-invariant operators [102, 103]. This EFT basis enlarges the Higgs basis EFT considered above. Moreover, the *CP* violation effects in the third generation Yukawa couplings are reflected in the imaginary parts of the Wilson coefficients of operators $\mathcal{O}_{y_t}$ and $\mathcal{O}_{y_b}$,

$$\Delta y_t = y_t^{\rm SM}\left(\Re[C_{y_t}]\frac{v^3}{2m_t\Lambda^2} + i\Im[C_{y_t}]\frac{v^3}{2m_t\Lambda^2}\right) \tag{11.17}$$

$$\Delta y_b = y_t^{\rm SM}\left(\Re[C_{y_b}]\frac{v^3}{2m_b\Lambda^2} + i\Im[C_{y_b}]\frac{v^3}{2m_b\Lambda^2}\right). \tag{11.18}$$

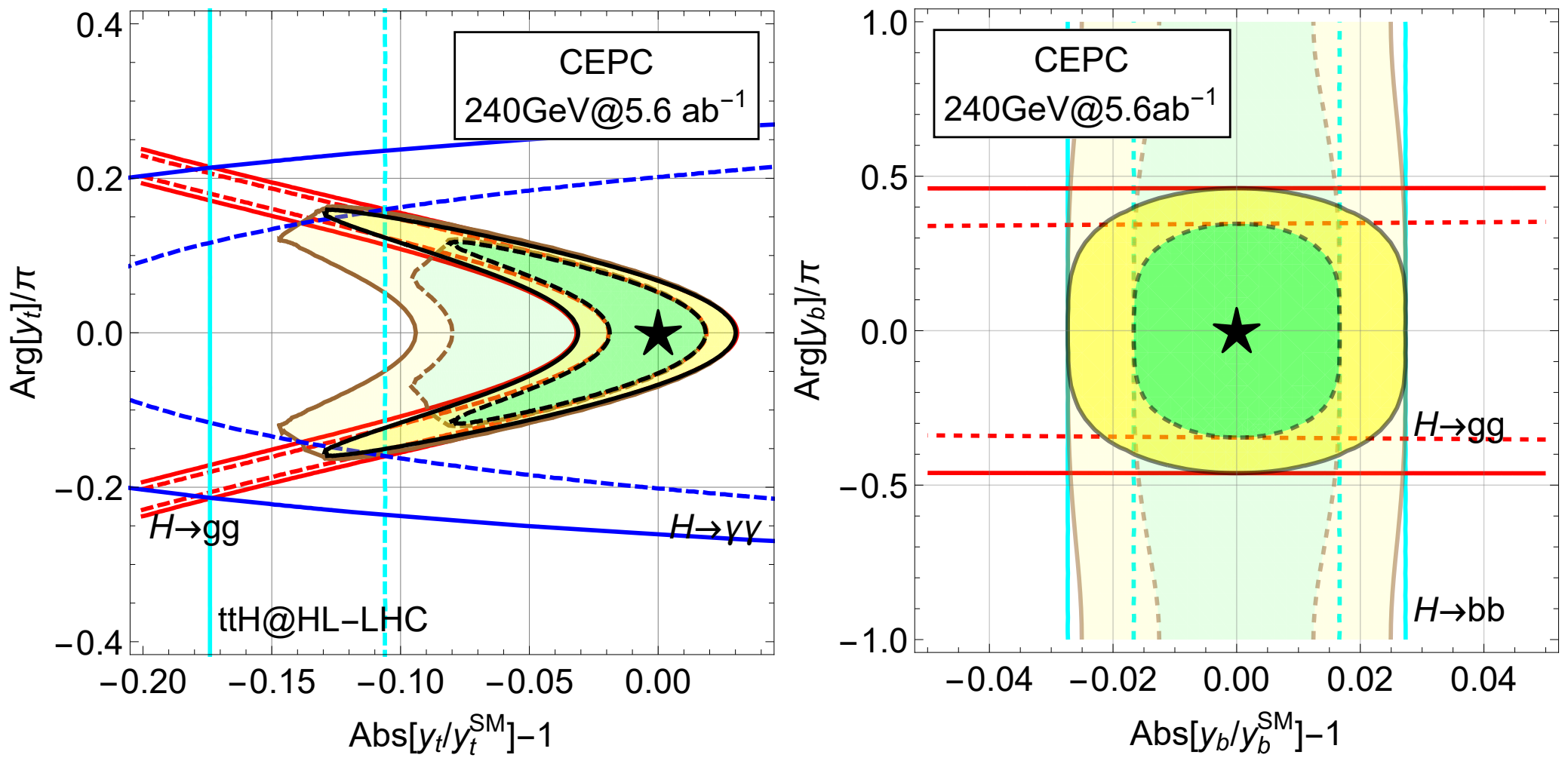

(a) (b)

**Figure 11.13:** Results for analysis on $C_{y_t}$ and $C_{y_b}$ in the projected allowed regions for modification to the top-quark and bottom-quark Yukawa coupling magnitude and *CP* phase at 68% and 95% CL. The combined results for the CEPC are shown in black curves. The source of individual constraints for the single operator analysis are labeled correspondingly. For a joint analysis of simultaneous appearance of both $\mathcal{O}_{y_t}$ and $\mathcal{O}_{y_b}$ operators, the results for the CEPC are shown in the enlarged yellow (95% CL) and green regions (68% CL) with thick brown boundary lines.

In this section, the effect of introducing *CP* phases in the Yukawa operators in Higgs boson physics is discussed. For more detailed discussion on a complete set of Higgs boson and Top quark operators, see Ref. [98]. The dominant constraints come from $H \to \gamma\gamma$ and $H \to gg$ for $\mathcal{O}_{y_t}$, and from $H \to gg$ and $H \to b\bar{b}$ for $\mathcal{O}_{y_b}$. Given that $H \to gg$ measurements are sensitive to both operators, a joint analysis of $\mathcal{O}_{y_t}$ and $\mathcal{O}_{y_b}$ will yield a significantly different result comparing to individual operator analysis. A joint analysis for these two operators in terms of Yukawa coupling strengths and the associated *CP* phases is performed at the CEPC. The important physics cases for such considerations are highlighted.

Constraints on the top-quark and bottom-quark Yukawa couplings, including their *CP* phases, are presented, respectively, in the left and right panels of Figure 11.13, respectively. The 68% and 95% CL exclusion bands are shown in solid and dashed lines. The limits for the CEPC are shown in *bright* black and magenta lines for individual operator analysis and the *bright* green and yellow shaded regions representing the allowed parameter space at 68% and 95% CL, respectively. The *dimmed* thick black curves rep-resent the results after turning on both operators $\mathcal{O}_{tH}$ and $\mathcal{O}_{bH}$ at the same time, using a profile-likelihood method profiling over other parameters. Furthermore, in the left panel the cyan band represents constraints from the HL-LHC $t\bar{t}H$ measurements, red bands are constraints from the CEPC $H \to gg$ measurements and blue bands are constraints from the CEPC $H \to \gamma\gamma$ measurements. Similarly, in the right panel, the cyan bands are constraints from $H \to b\bar{b}$ and the red bands are constraints from $H \to gg$ at the CEPC.

The left panel of Figure 11.13 shows that the expected sensitivity on the modification in the magnitude of top-quark Yukawa coupling is around $\pm 3\%$ for the single operator analysis. This is relaxed to $[-9.5\%, +3\%]$ assuming zero $CP$ phase for the top-quark Yukawa coupling and allowing the bottom-quark Yukawa coupling and its phase to vary freely. The phase of the top-quark Yukawa coupling can be constrained to $\pm 0.16\pi$. This constraint is driven by the $H \to \gamma\gamma$ measurement, where a sizable phase shift will enlarge the $H \to \gamma\gamma$ decay rate via reducing the interference with the SM $W$ boson loop. The constraint on the magnitude of the top-quark Yukawa coupling is driven by the $H \to gg$ measurement which is dominated by the top-quark loop contribution. Note that constraints from the $H \to gg$ measurement are not constant with respect to the Yukawa coupling magnitude. This is due to the different sizes of the top-quark loop contribution to $Hgg$ through scalar and pseudoscalar couplings. Similarly, as shown in the right panel of Figure 11.13 for the bottom-quark Yukawa coupling, the constraint for the magnitude is $\pm 2.5\%$. For the $CP$ phase, the constraint changes from $\pm 0.47\pi$ to zero when the top-quark Yukawa coupling is left free.

### 11.1.10 TESTS OF HIGGS BOSON SPIN/$CP$

The $CP$ properties of the Higgs boson and, more generally, its anomalous couplings to gauge bosons in the presence of BSM physics, can be measured at the CEPC using the $e^+e^- \to Z^* \to ZH \to \mu^+\mu^- b\bar{b}$ process. It is convenient to express the effects of the anomalous couplings in terms of the fractions of events from the anomalous contribution relative to the SM predictions. These fractions are invariant under the independent rescalings of all couplings, see Refs. [104–106].

Two of the anomalous $HZZ$ coupling measurements are of particular interest at the CEPC: the fraction of the high-order $CP$-even contribution due to either SM contribution or new physics, $f_{a2}$, and the fraction of a $CP$-odd contribution due to new physics, $f_{a3}$. The following two types of observables can be used to measure these anomalous couplings of the Higgs bosons.

1. The dependence of the $e^+e^- \to Z^* \to ZH$ cross section on $\sqrt{s}$ is different for different $CP$ property of the Higgs boson [106]. Therefore, measurements of the cross section at several different energies will yield useful information about anomalous $HZZ$ couplings. However this has non-trivial implications to the accelerator design and is not included in this study as a single value of $\sqrt{s}$ is assumed for the CEPC operating as a Higgs boson factory.

2. Angular distributions, $\cos\theta_1$ or $\cos\theta_2$ and $\Phi$ as defined in Figure 11.14. These angles are also sensitive to interference between $CP$-even and $CP$-odd couplings. In particular forward-backward asymmetry with respect to $\cos\theta_1$ or $\cos\theta_2$ and non-trivial phase in the $\Phi$ distributions can lead to an unambiguous interpretation of $CP$ violation.

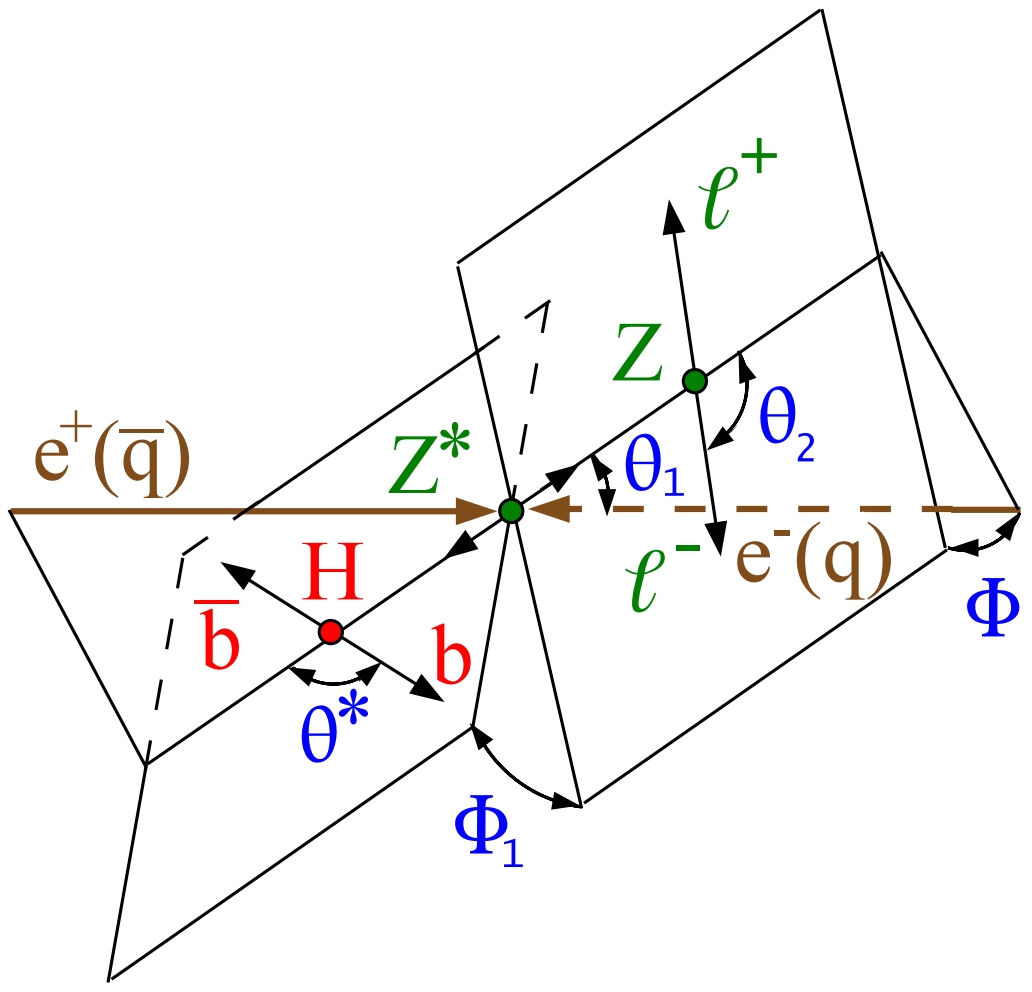

**Figure 11.14:** Higgs boson production and decay angles of the $e^+e^- \to Z^* \to ZH \to \mu^+\mu^-b\bar{b}$ process [106].

To estimate the sensitivity on the anomalous couplings, a maximum likelihood fit [106] is performed to quantify the compatibility of the observed angular distributions to the theory predictions, including both signal and background processes. In this likelihood fit, the signal probability density functions are taken from analytical predictions that are validated using a dedicated MC program, the JHU generator [104, 105], which incorporates all the anomalous couplings, spin correlations, and the interference of all contributing amplitudes. The background probability density function is modeled using simulation based on $e^+e^- \to ZZ \to e^+e^-b\bar{b}$ process in MadGraph [107].

Several thousand statistically independent pseudo-experiments are generated and fitted to estimate the sensitivity to $f_{a2}$ and $f_{a3}$, defined as the smallest values that can be measured with $3\sigma$ away from 0. All other parameters in the fit, including the number of expected signal and background events, are fixed. The expected sensitivity of $3\sigma$ discovery is estimated to be $0.018$ for $f_{a2}$ and $0.007$ for $f_{a3}$. Figure 11.15(a,b) show the distributions of the fitted values of $f_{a2}$ and $f_{a3}$ from the pseudo-experiments expected for $f_{a2} = 0.018$ and $f_{a3} = 0.008$, respectively. A simultaneous fit of $f_{a2}$ and $f_{a3}$ is also performed with the 68% and 95% CL contours shown in Figure 11.15(c).

The sensitivities of $f_{a2}$ and $f_{a3}$ are then converted to the corresponding parameters defined for the on-shell $H \to ZZ^*$ decays, $f_{a2}^{\rm dec}$ and $f_{a3}^{\rm dec}$, in order to compare with the sensitivities from the LHC experiments as described in Ref. [106]. The corresponding sensitivities of $f_{a2}^{\rm dec}$ and $f_{a3}^{\rm dec}$ are $2 \times 10^{-4}$ and $1.3 \times 10^{-4}$, respectively. The much smaller values in the $f_{a2,a3}^{\rm dec}$ are due to the much smaller $m_{Z^*}^2$ in the $H \to ZZ^*$ decay compared to the value in the $Z^* \to ZH$ production.

Compared to the ultimate sensitivity of HL-LHC as shown in Ref. [106], the sensitivities in the $f_{a2}$ and $f_{a3}$ at the CEPC are a factor of 300 and 3 better. Further improvements can be achieved by exploring kinematics in the $H \to b\bar{b}$ decays, including other $Z$ decay final states, and combining with the overall cross-section dependence of the signal as obtained by a threshold scan in $\sqrt{s}$.

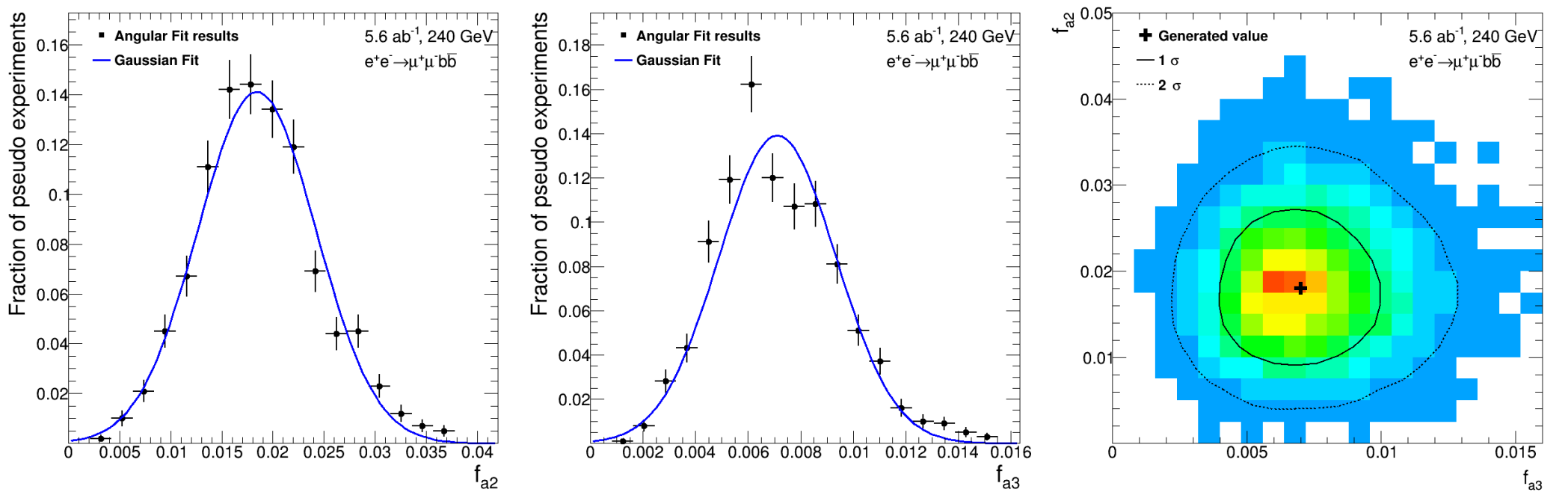

(a) (b) (c)

**Figure 11.15:** Distributions of fitted values of (a) $f_{a2}$ and (b) $f_{a3}$ in a large number of pseudo-experiments. Only the parameter shown is floated in these fits. Other parameters are fixed to their SM values. (c) Simultaneous fit of non-zero $f_{a2}$ and $f_{a3}$, with 68% and 95% confidence level contours shown.

### 11.1.11 SUMMARY

Many new physics models predict Higgs boson coupling deviations at the sub-percent level, beyond those achievable at the LHC. The CEPC complements the LHC and will be able to study the properties of the Higgs boson in great detail and with unprecedented precision. At the CEPC, most of Higgs boson couplings can be measured with precision at a percent level or better, in particular the coupling to the $Z$ boson can be determined with a precision of 0.25%. More importantly, the CEPC will be able to measure many of the key Higgs boson properties such as the total width and decay branching ratios in a model-independent way, greatly enhancing the coverage of new physics searches. Furthermore, the clean event environment of the CEPC will allow the identification of potential unknown decay modes, which are very challenging at the LHC.

This section provides a snapshot of the current studies. Many of these are ongoing and more analyses are needed to fully understand the physics potential of the CEPC. Nevertheless, the results presented here have already built a strong case for the CEPC as a Higgs factory. The CEPC has the potential to characterize the Higgs boson in the same way LEP did with the $Z$ boson, and potentially shed light on new physics.

## 11.2 $W$ AND $Z$ BOSON PHYSICS

With high production cross sections and large integrated luminosity, the CEPC will reach a new level of precision for the measurements of the properties of the $W$ and $Z$ bosons. Precise measurements of the $W$ and $Z$ boson masses, widths, and couplings are critical to test the consistency of the SM [108]. In addition, many BSM models predict new couplings of the $W$ and $Z$ bosons to other elementary particles. Precise electroweak (EW) measurements performed at the CEPC could discover deviations from the SM predictions and reveal the existence of new physics that are beyond the reach of direct searches at the current experiments.

Significant improvements are expected from the CEPC measurements. Table 11.9 lists the expected precision from the CEPC compared to the achieved precision from the LEP experiments for various measurements. Details about the estimation of these uncertainties are described in the following.

| Observable | LEP precision | CEPC precision | CEPC runs | CEPC $\int \mathcal{L}dt$ |
|---|---|---|---|---|
| $m_Z$ | 2.1 MeV | 0.5 MeV | $Z$ pole | 8 ab$^{-1}$ |
| $\Gamma_Z$ | 2.3 MeV | 0.5 MeV | $Z$ pole | 8 ab$^{-1}$ |
| $A_{FB}^{0,b}$ | 0.0016 | 0.0001 | $Z$ pole | 8 ab$^{-1}$ |
| $A_{FB}^{0,\mu}$ | 0.0013 | 0.00005 | $Z$ pole | 8 ab$^{-1}$ |
| $A_{FB}^{0,e}$ | 0.0025 | 0.00008 | $Z$ pole | 8 ab$^{-1}$ |
| $\sin^2\theta_W^{\rm eff}$ | 0.00016 | 0.00001 | $Z$ pole | 8 ab$^{-1}$ |
| $R_b^0$ | 0.00066 | 0.00004 | $Z$ pole | 8 ab$^{-1}$ |
| $R_\mu^0$ | 0.025 | 0.002 | $Z$ pole | 8 ab$^{-1}$ |
| $m_W$ | 33 MeV | 1 MeV | $WW$ threshold | 2.6 ab$^{-1}$ |
| $m_W$ | 33 MeV | 2–3 MeV | $ZH$ run | 5.6 ab$^{-1}$ |
| $N_\nu$ | 1.7% | 0.05% | $ZH$ run | 5.6 ab$^{-1}$ |

**Table 11.9:** The expected precision in a selected set of EW precision measurements at the CEPC and the comparison with the precision from the LEP experiments. The CEPC accelerator running mode and total integrated luminosity expected for each measurement are also listed. Relative uncertainties are quoted for $N_\nu$ measurements.

## 11.2.1 $Z$ POLE MEASUREMENTS

The CEPC offers the possibility of dedicated low-energy runs at the $Z$ pole for at least two years with a high instantaneous luminosity ($1.6 \times 10^{35}$ cm$^{-2}$s$^{-1}$). The expected integrated luminosity for the CEPC $Z$ pole runs is more than 8 ab$^{-1}$, corresponding to $10^{11}$–$10^{12}$ $Z$ bosons. These runs allow for high precision electroweak measurements of the $Z$ boson properties, such as mass, total width and partial widths, and the parameters like the ratios $R_b = \Gamma_{Z\to b\bar{b}}/\Gamma_{\rm had}$ and $R_\ell = \Gamma_{\rm had}/\Gamma_{Z\to\ell\bar{\ell}}$.[6] It would also perform high precision measurements of the forward-backward charge asymmetry ($A_{FB}$) and polarization asymmetries, which allow the determination of the effective weak mixing angle ($\sin^2\theta_W^{\rm eff}$). Another important quantity, which can be determined from the hadronic cross section measurement at the $Z$ peak, is the number of light neutrino species ($N_\nu$). It is also possible to perform some measurements with the $Z$ boson without these dedicated low-energy runs near or at the $Z$ pole. For example, the direct measurement of the number of light neutrino species can also be performed in $ZH$ run at 240 GeV.

### 11.2.1.1 $Z$ BOSON MASS AND WIDTH MEASUREMENTS

The mass $m_Z$, together with its total width $\Gamma_Z$, is a fundamental parameter in the SM and was determined with an overall uncertainty of 2 MeV by the four LEP experiments [109–112]. The lineshape scan around the $Z$ boson peak was performed from 87.9 GeV to 94.3

[6]Here $R_\ell$ is defined as the ratio to any *one* charged lepton flavor, assuming lepton universality, not the ratio to the sum of all lepton flavors.

GeV. The $Z$ boson mass and widths were measured by a combined fit to the hadronic and leptonic cross sections in the on-peak and off-peak datasets. Most of the information on $M_Z$ and $\Gamma_Z$ is extracted from the off-peak runs. The main uncertainty of $m_Z$ includes the statistical uncertainty (1.2 MeV), and the LEP beam energy uncertainty (about 1.7 MeV).

| $\sqrt{s}$ (GeV) | Luminosity (ab$^{-1}$) |
|---|---|
| 87.9 | 0.25 |
| 90.2 | 0.25 |
| 91.2 | 7 |
| 92.2 | 0.25 |
| 94.3 | 0.25 |

**Table 11.10:** The proposed five $e^+e^- \to Z$ resonance scan runs and their integrated luminosity, for a total integrated luminosity of 8 ab$^{-1}$.

A precision of 0.5 MeV in $m_Z$ and $\Gamma_Z$ can be achieved at the CEPC. The lineshape scan around the $Z$ peak is the key for improving the $m_Z$ measurement. The LEP measurement was limited by the statistics in their off-peak runs, therefore the luminosity in the $Z$ off-peak runs plays an important role in the $m_Z$ measurement. We propose four off-peak runs and one on-peak run for the CEPC $Z$ mass scan, as listed in Table 11.10. The expected $m_Z$ uncertainty at the CEPC due to statistics is below 0.1 MeV.

The major systematic uncertainty is the beam energy scale uncertainty. The beam energy is expected to be measured by the depolarizing resonance method, developed at the LEP [113]. A precision of 0.1 MeV was achieved by LEP [113] using this method. However these measurements were not performed all the time in LEP, and the major experimental systematic uncertainty due to extrapolation of these precise values of the beam energy at a particular time to the full set of data. In order to reduce the systematic due to the extrapolation, CEPC is foreseen to operate with colliding and non-colliding bunches in Z pole and WW threshold scan runs. In this operation mode, beam polarization can be built up in non-colliding bunches for beam momentum measurement with depolarizing resonance method. The systematics due to extrapolation of the measurements from non-colliding bunches to colliding bunches is expected to less than 500 MeV.

Another source of systematic uncertainty to be considered is the beam energy spread $\delta E$, whose effect on the cross section is proportional to $d^2\sigma/dE^2\delta E^2$. At LEP an energy spread of about $50$ MeV (with an uncertainty of $1$ MeV) reflected in correction factors of 0.16% for the $Z$ peak cross section. The beam energy spread does not affect $M_Z$. As described in Section 3, the energy spread of about $36$ MeV is expected on $Z$ pole runs in CEPC (with an uncertainty of about $0.4$ MeV).

The uncertainty in the luminosity measurement is expected to be the subleading systematic uncertainty. As described in Section 9.5.2, the uncertainty of absolute luminosity measurement is about 0.05%. $m_Z$ and $\Gamma_Z$ measurements does not depend on the absolute luminosity measurement, but on relative luminosity measurements between scan points. The ratio of the integrated luminosity between each off-peak run and Z-pole run are expected to be measured with a 0.01% level precision, and the corresponding uncertainty in $m_Z$ measurement is less than 0.1 MeV.

The $Z$ peak scan is also needed for the $Z$ lineshape studies. The lineshape is strongly affected by the emission of initial state photon radiation (ISR), which shifts the peak position by about 100 MeV and decreases the cross section by about 25%. Moreover the emission of ISR creates a distortion of the shape with respect to a Breit-Wigner form, with the appearance of a typical radiative tail for $\sqrt{s} \sim 93$ GeV. An efficient way to account for ISR is to convolute a QED radiator function with the kernel cross section $\hat{\sigma}$ evaluated at the proper reduced center-of-mass energy, according to

$$\sigma(s) = \int_{z_0}^{1} dz H(z; s)\hat{\sigma}(zs)\,, \tag{11.19}$$

where $z_0$ depends on the event selection. The radiator function $H(z; s)$ is known at $\mathcal{O}(\alpha^2)$ and including leading terms of $\mathcal{O}(\alpha^3 L^3)$ [114, 115], where $L = \log(s/m_e^2)$. Such calculations have been implemented in two independent codes, TOPAZO [116–119] and ZFITTER [120–122], and are used to estimate the impact of the residual QED uncertainty on the $Z$ mass and width measurements at the level of 0.1 MeV [123–125]. The uncertainty on the cross sections due to QED was estimated below the 0.01% level. The kernel cross section in the SM is composed of three contributions: $Z$ exchange, $\gamma$ exchange and their interference, which are calculated perturbatively with NLO precision supplemented with higher order terms from running couplings and QCD corrections, which guarantees a theoretical uncertainty on the observables at the 0.01% level. Aiming at a model-independent parametrization of the $Z$ lineshape around the $Z$ resonance [126], the $Z$ exchange contribution can be parametrized in the following way:

$$\sigma^Z_{f\bar{f}} = \sigma^{\text{peak}}_{f\bar{f}} \frac{s\Gamma_Z^2}{(s - m_Z)^2 + s^2\Gamma_Z^2/m_Z^2}$$

$$\sigma^{\text{peak}}_{f\bar{f}} = \frac{\sigma^0_{f\bar{f}}}{R_{\text{QED}}} \quad \text{with} \quad \sigma^0_{f\bar{f}} = \frac{12\pi}{m_Z^2}\frac{\Gamma_{ee}\Gamma_{f\bar{f}}}{\Gamma_Z^2}\,,$$

where $R_{\text{QED}}$ corrects for the final state QED radiation. The two contributions which are not factorizable on the $Z$ exchange are calculated in the SM, thus introducing a dependence on the SM parameters, such as the mass of the top quark, the Higgs boson mass and the strong coupling constant. The SM model parameter dependence has been estimated at the end of LEP1 operations to be below 0.1% level, which was the target precision. After subtraction of the non-factorizable terms, the resonance parameters are extracted from the lineshape data by means of a global fit. In particular, assuming lepton universality, four quantities are extracted from the fit: $m_Z$, $\Gamma_Z$, $R_\ell$ and the hadronic peak cross section $\sigma_h$. The $Z$ resonance parameter measurements can be directly compared with the SM predictions. The latter have been completed very recently at complete two-loop electroweak accuracy, including corrections due to the top-quark Yukawa coupling up to $\mathcal{O}(\alpha_t^3)$ and mixed $\mathcal{O}(\alpha\alpha_s)$ and $\mathcal{O}(\alpha_t\alpha_s^n)$ contributions with $n = 2, 3$ and $\mathcal{O}(\alpha_t^2\alpha_s)$ terms (see Ref. [127]). This level of the precision of the perturbative calculations guarantees that the remaining theoretical uncertainties in the range 0.005%–0.05% for the various $Z$ resonance parameters (including ratios of partial widths), which is, depending on the quantity, comparable or slightly larger than the projected experimental precision discussed in the following sections. Further theoretical efforts would be needed to obtain negligible theoretical systematic uncertainties.

An alternative, more model independent, approach to fit the $Z$ lineshape data is the one proposed in Refs. [128–130], where the $Z$-$\gamma$ interference is obtained by fitting addi-tional parameters, $j_Z^i$ (one for each measured $f\bar{f}$ channel), to off-peak data. At the CEPC this method could be adopted at the $ZH$ and $WW$ threshold runs. Similar measurements were performed at LEP2 for the hadronic parameter $j_Z^{\mathrm{had}}$, although they were limited by the available statistics [131–134].

#### 11.2.1.2 $R_B$

The partial width of the $Z$ boson to each individual decay channel is proportional to the square of the fundamental $Z$-boson fermion couplings. The ratio of the partial widths $R_b$ is sensitive to electroweak radiative corrections from new particles, where the $R_b = \Gamma(Z \to b\bar{b})/\Gamma(Z \to hadrons)$. For example, the existence of scalar top quarks or charginos in supersymmetry could lead to an observable change of $R_b$ from the SM prediction. Precise measurements of $R_b$ have been made by the LEP collaborations [135–139] and by the SLD collaboration [140] using hadronic $Z$ events.

Decays of $B$-hadrons were tagged using tracks with large impact parameters and/or reconstructed secondary vertices, complemented by event shape variables. The combination of LEP and SLD measurements yields a value of $0.21629 \pm 0.00066$ for $R_b$ [141]. The statistical uncertainty of $R_b$ is 0.00044.The dominant systematic error is the uncertainty due to hemisphere tag correlations for $b$ events. The secondary uncertainty is the uncertainty in gluon splitting (0.00023).

A precision of $4 \cdot 10^{-5}$ can be achieved for the measurement of $R_b$ at the CEPC, which will improve the current precision in experimental measurement by one order of magnitude. The main systematic uncertainty will be due to the hemisphere tagging correlations in $Z \to b\bar{b}$ events, which can be reduced to a level of $4 \cdot 10^{-5}$, a factor of ten lower than previous measurements, from the expected improvement in the $b$-tagging performance of the CEPC detector. The improvement in $b$-tagging efficiency ($\epsilon_b$) is the key to reduce this uncertainty. The dependence on this uncertainty to $\epsilon_b$ is non-linear. In this study, this uncertainty is assumed to be proportional to $(1-\epsilon_b)^2$, and it becomes irrelevant in the limit of a 100% $b$-tagging efficiency. By taking advantage of the next-generation vertex detector, the $b$-tagging efficiency at the CEPC is expected to be around 70% with a purity of 95% as shown in Figure 10.13. This efficiency in CEPC experiment is about 40%–50% higher than the efficiency achieved in LEP experiments at similar $b$-jet purity, therefore the uncertainty due to hemisphere tagging correlations is expected to improve by one order of magnitude.

The second largest systematic error due to the uncertainty in probability of gluon splitting into a $b\bar{b}$ pair ($g_{b\bar{b}}$). This quantity was meaured by ALEPH, DELPHI, OPAL and SLD collaborations in secondary production process of bottom quarks ($e^+e^- \to q\bar{q}g, g \to b\bar{b}$) [142–144], and in both primary and secondary production process of b quarks ( $e^+e^- \to b\bar{b}b\bar{b}$) [145]. These measurements were limited by MC statistics and b-jet tagging performance. However both issue are not the limiting factors in CEPC. Dedicated $g_{b\bar{b}}$ measurements in $e^+e^- \to b\bar{b}b\bar{b}$ and $e^+e^- \to q\bar{q}b\bar{b}$ final state are expected to perform in CEPC, the uncertainty in $R_b$ measurement due to $g_{b\bar{b}}$ is expected to be less than $1 \cdot 10^{-5}$.

#### 11.2.1.3 THE PARTIAL DECAY WIDTH OF $Z \to \mu^+\mu^-$

The $\mu^+\mu^-$ channel provides the cleanest leptonic final state. The combination of LEP measurements [109–112] yields a value of $R_l^0 = 20.767 \pm 0.025$, where $R_l^0 = \Gamma(Z \to hadrons)/\Gamma(Z \to l^+l^-)$. Major systematic uncertainties includes the uncertainty in the modeling of $Z \to \mu^+\mu^-\gamma$ events (0.01) and the uncertainty due to photon trigger and identification efficiency (0.01).

A precision of 0.002 in $R_l^0$ can be achieved at the CEPC. The main systematic is expected to be the uncertainty in modeling the $Z \to \mu^+\mu^-\gamma$ events. About 2% of the $Z \to \mu^+\mu^-$ sample are classified as $Z \to \mu^+\mu^-\gamma$ events with a photon detected in ECAL. For this class of events, the most critical issue is to reconstruct and identify the low energy photon object with high efficiency. Benefiting from high granularity of the CEPC ECAL ($10 \times 10\ \mathrm{mm}^2$ as shown in Chapter 5), we expect to have close to 100% efficiency for a photon with $E > 5$ GeV, as shown in Section 10.2.2. Another challenge in this measurement is to reduce the systematic uncertainty due to QED ISR events. Detailed studies of radiative events in $Z$ off-peak runs are expected, especially the $Z$ off-peak runs at $\sqrt{s} = 92.2$ GeV. Benefiting from high statistics in $Z$ off-peak runs ($> 1\,\mathrm{ab}^{-1}$), the uncertainty in the modeling $Z \to \mu^+\mu^-\gamma$ events is expected to reduced to 0.002.

#### 11.2.1.4 FORWARD-BACKWARD ASYMMETRY AND WEAK MIXING ANGLE MEASUREMENTS AT THE $Z$ POLE

Another important class of measurements for the study of the chiral couplings of the $Z$ boson to fermions is given by the forward-backward asymmetry in the processes $e^+e^- \to f\bar{f}$:

$$A_{FB} = \frac{\sigma_F - \sigma_B}{\sigma_F + \sigma_B}\,, \tag{11.20}$$

which are optimal observables to quantify the parity violation of the neutral currents. In fact, the differential tree-level cross section at the $Z$ peak, for unpolarized incoming beams and including only the $Z$ exchange contribution, can be written as

$$\frac{d\sigma_{f\bar{f}}}{d\cos\vartheta} = \frac{3}{8}\sigma^{\mathrm{tot}}\left[1 + \cos^2\vartheta + 2\mathcal{A}_e\mathcal{A}_f\cos\vartheta\right]\,, \tag{11.21}$$

where the coefficients $\mathcal{A}_i$ can be expressed in terms of the left/right couplings $g_{\mathrm{L/R},i}$ or the vectorial and axial-vector couplings $g_V^i$ and $g_A^i$:

$$\mathcal{A}_f = \frac{g_{\mathrm{L},f}^2 - g_{\mathrm{R},f}^2}{g_{\mathrm{L},f}^2 + g_{\mathrm{R},f}^2} = \frac{2g_{\mathrm{V}}^f g_{\mathrm{A}}^f}{(g_{\mathrm{V}}^f)^2 + (g_{\mathrm{A}}^f)^2} = \frac{\frac{g_V^f}{g_A^f}}{1 + \left(\frac{g_V^f}{g_A^f}\right)^2}. \tag{11.22}$$

In particular, the last expression of Equation 11.22 shows that the observable $A_{FB}$ provides information that is complementary to the partial widths. This is because it is sensitive to the ratio $g_V^f/g_A^f$, whereas the partial widths are proportional to $(g_{\mathrm{V}}^f)^2 + (g_{\mathrm{A}}^f)^2$. The same expression is used to define the parameter $\sin^2\vartheta_{\mathrm{eff}}^f$

$$4|Q_f|\sin^2\vartheta_{\mathrm{eff}}^f = 1 - \frac{g_V^f}{g_A^f}\,, \tag{11.23}$$

which, at tree-level, coincides with $1 - (M_W/M_Z)^2$. The relation between $A_{FB}$ and $\mathcal{A}_i$ at the $Z$ pole, for unpolarized beams, is the following:

$$A_{FB}^{0,f} = \frac{3}{4}\mathcal{A}_e\mathcal{A}_f\,. \tag{11.24}$$

Since $\sin^2\vartheta_{\mathrm{eff}} \sim 1/4$, $\mathcal{A}_\ell$ is close to $0.1$ for leptons while for quarks $\mathcal{A}_u \sim 0.7$ and $\mathcal{A}_d \sim 0.9$. This, associated to the fact that the asymmetry measurements at LEP were limited by statistics, implied that $A_{FB}^{0,b}$ offered the most precise determination of $\sin^2\vartheta_{\mathrm{eff}}$.

$\boldsymbol{A_{FB}^{0,b}}$ Measurements have been made at LEP experiments [146–153]. The combination of LEP measurements yields a value of $0.0992 \pm 0.0016$ for $A_{FB}^{0,b}$ [141].

$Z \rightarrow b\bar{b}$ events were identified by tagging two $b$ jets. Each event was divided into forward and backward categories by the plane perpendicular to the thrust axis, which contains the interaction point. The knowledge of the direction of the b-quark from from $Z \rightarrow b\bar{b}$ day is required for this measurement. Two methods have been used in LEP experiments. The first method is to reconstruct jet and vertex charge for identifying b-quark against anti-b-quark in inclusive b-hadron decays final state [146–149]. This method suffers from limited jet charge tagging performance due to small jet charge separation between a b-quark and anti-quark jet. The second method is to use the charge of the electrons and muons from semileptonic decays to identify the quark charge (lepton charge tag) [150–153]. This method suffers from limited statistics in semileptonic decays, uncertainties in the semileptonic branching ratios BR( $b \rightarrow l$ ), BR($b \rightarrow c \rightarrow l$) and BR( $c \rightarrow l$) and the MC modelling uncertainties in semileptonic decay.

A precision at the level of $1 \cdot 10^{-4}$ can be achieved for the measurement of $A_{FB}^{0,b}$ at the CEPC. As statistics will not be a limiting factor at CEPC, the lepton charge method in semileptonic decays final state will be used for CEPC measurement. In order to select pure signal sample from $b \rightarrow l$ final state, high transverse momentum lepton is selected in CEPC $A_{FB}^{0,b}$ measurement to reduce the background from $b \rightarrow c \rightarrow l$ process with the wrong sign correlation (opposite sign) with respect to the primary b quark, as well as $Z \rightarrow c\bar{c}$ background. The MC modelling and branching ratio uncertainties in semileptonic decays of b-hadron are expected to reduce by at least one order of magnitude based on BELLE II measurements in the coming years [154].

$\boldsymbol{A_{FB}^{0,l}}$ To determine the weak mixing angle, the most precise measurement comes from the inclusive left-right beam-polarization asymmetry ($A_{LR}^{l}$) [155], however this measurement requires large fraction of longitudinal beam polarization. As beam polarization design of the CEPC accelerator has not concluded yet, the measurements are assumed to perform without longitudinal polarized beams, which is similar to the $A_{FB}^{0,l}$ measurements at LEP [109–112].

A precision at the level of $5 \cdot 10^{-5}$ ($8 \cdot 10^{-5}$) can be achieved for the measurement of $A_{FB}^{0,\mu}$ ($A_{FB}^{0,e}$) at the CEPC. The dominant systematic uncertainty are expected to be the uncertainties in the geometrical acceptance, including alignment precision of tracker and alignment precision between tracker and calorimeters. The $A_{FB}^{0,e}$ measurement is also subjected to the accuracy of the correction for the $t$-channel contribution.

**Tau polarization** An independent determination of $\sin^2 \vartheta_{\rm eff}$ can be achieved through the measurement of the $\tau$ average polarization, defined as [141]

$$\langle P_\tau \rangle = \frac{\sigma_R - \sigma_L}{\sigma_R + \sigma_L} \,. \tag{11.25}$$

The $\tau$ polarization can be accessed because the $\tau$ lepton decays in a parity violating manner within the detector, so that the polarization can be obtained by measuring the angular distributions of the $\tau$ decay products. In particular, the net $\tau$ polarization as a function of the scattering angle can be written as

$$P_\tau(\cos\vartheta) = -\frac{\mathcal{A}_\tau(1+\cos^2\vartheta) + 2\mathcal{A}_e\cos\vartheta}{(1+\cos^2\vartheta) + 2\mathcal{A}_\tau\mathcal{A}_e\cos\vartheta} \tag{11.26}$$

Thus a measurement of the differential $\tau$ polarization $P_\tau$ allows a determination of $\mathcal{A}_\tau$ and $\mathcal{A}_e$ and, consequently, a determination of $\sin^2\vartheta_{\rm eff}$. Contrary to LEP, statistics will not be a limiting factor at CEPC. The knowledge of the $\tau$ branching fractions and decay modelling, which were important sources of systematics at LEP, will benefit from the measurement performed in the past at $B$-factories and in the near future at Belle2. Another positive aspect of the $\tau$ polarization measurement is its reduced dependence on the beam energy with respect to the forward-backward asymmetries discussed above. A more quantitative discussion on the conceivable precision on $\sin^2\vartheta_{\rm eff}$ from $\tau$ polarization measurement will be the subject of future work.

**$\sin^2\theta_W^{\rm eff}$** The absolute precision of the effective weak mixing angle measurement at the CEPC using $Z \to b\bar{b}$, $Z \to l^+l^-$ events is expected to be at the level of $1 \cdot 10^{-5}$.

The theory uncertainty of the SM prediction is below $1 \cdot 10^{-5}$ threshold, thanks to the recent complete two-loop electroweak calculation of Ref. [156].

The energy dependence of the asymmetries around the $Z$ resonance is sensitive to the interference between the $Z$ and $\gamma$ exchange. It allows to measure the values of the axial-vector couplings.

#### 11.2.1.5 LIGHT NEUTRINO SPECIES COUNTING

Two different methods have been used to determine the number of light neutrino species ($N_\nu$) at the LEP. The first one is the indirect method using the analysis of the $Z$ lineshape, and it uses the data collected by the $Z$ resonance scan runs. The $Z$ peak scan at the CEPC can improve the LEP determination of $N_\nu$ by a factor of three. The second method is a direct measurement, which is based on the measurement of the cross section for the radiative process $e^+e^- \to \nu\nu\gamma$. The second method at the CEPC is supposed to use the $ZH$ runs and improve the LEP direct determination by a factor of ten.

The systematic uncertainties of theoretical origin associated with the two methods are completely different: the indirect one relies on the precision calculation of the $Z$ partial decay widths, while the direct one needs the calculation of higher order radiative corrections for the process $e^+e^- \to \nu\bar{\nu}\gamma$. Moreover the two methods use completely different datasets, therefore they are independent and complementary. The sensitivity to new physics will be different for these two methods. A deviation of $N_\nu$ from an integer value would signal the presence of new physics. Possible contributions include WIMP dark matter particles, and other weakly coupled particles such as exotic neutrinos, gravitinos, or KK gravitons in theories with large extra dimensions. Thus, when we refer to the number of neutrino species, we actually include any number of possible invisible particles other than neutrinos. The subprocess $e^+e^- \to \nu_e\bar{\nu}_e\gamma$ is particularly important because it will allow to investigate possible deviations with respect to the SM in the vertex $\gamma W^+W^-$, in a complementary way with respect to the $W^+W^-$ production cross section, where both $\gamma W^+W^-$ and $ZW^+W^-$ vertices appear in the matrix element.

**Indirect method from Z line shape** The indirect method assumes all contributions from invisible channels come from the $Z \to \nu\bar{\nu}$ decay, assuming that the total $Z$ boson width does not receive additional beyond the SM contributions. The method uses the analysis of the $Z$ boson lineshape, subtracting the visible partial widths of the hadrons ($\Gamma_{\rm had}$), and the partial widths of the leptons ($\Gamma_\ell$) from the total width $\Gamma_Z$. The invisible width $\Gamma_{\rm inv}$ can

be written as:

$$\Gamma_{\rm inv} = N_\nu \Gamma_\nu = \Gamma_Z - \Gamma_{\rm had} - 3\Gamma_\ell. \tag{11.27}$$

We take as our definition of the number of neutrinos $N_\nu = \Gamma_{\rm inv}/\Gamma_\nu$, i.e. the ratio of the invisible width to the Standard Model expectation for the partial width to a single neutrino species.

In the context of the SM, equation 11.27 can be written as follows:

$$N_\nu = \frac{\Gamma_\ell}{\Gamma_\nu}\left(\sqrt{\frac{12\pi R_\ell}{m_Z^2 \sigma^0_{\rm had}}} - R_\ell - 3\right). \tag{11.28}$$

The final LEP1 result was $N_\nu = 2.9840 \pm 0.0082$ [141]. As shown in equation 11.28, the precision of $N_\nu$ depends on the the lepton partial width $R_\ell$ measurement, the $Z$ mass measurement, and the hadronic cross section of the $Z$ boson on its mass peak ($\sigma^0_{\rm had}$). The decomposition of the error on $N_\nu$ is given by [141]

$$\delta N_\nu \simeq 10.5\frac{\delta n_{\rm had}}{n_{\rm had}} \oplus 3.0\frac{\delta n_{\rm lep}}{n_{\rm lep}} \oplus 7.5\frac{\delta\mathcal{L}}{\mathcal{L}}, \tag{11.29}$$

where $\delta n_{\rm had}/n_{\rm had}$, $\delta n_{\rm lep}/n_{\rm lep}$ and $\delta\mathcal{L}/\mathcal{L}$ represent the total errors on the number of selected hadronic and leptonic events and on the luminosity determination, respectively. The symbol $\oplus$ denotes the summation in quadrature. The final theoretical uncertainty of 0.061% [157] (0.054% [158, 159]), available at the end of LEP operation [141] for the small angle Bhabha process is reflected in a systematic uncertainty of 0.15%, i.e. $\sim$ 50% of the total uncertainty of 0.27%, on $N_\nu$.

A relative precision of 0.1% in the $N_\nu$ measurement with the indirect method can be achieved at the CEPC, improving the current precision by a factor of three. Benefiting from the recent development in the luminosity detector technology, the relative uncertainty due to luminosity can be reduced to 0.05%. The theoretical uncertainty of predictions for the small angle Bhabha scattering process can be reduced, conservatively, to 0.05% or below, mainly due to the recent progress in the evaluation of the hadronic contribution to the photon vacuum polarization [160–162]. This uncertainty can be pushed down even further, close to the 0.01% scale, once the NNLO QED predictions are matched to higher order soft/collinear contributions [163].

**Direct method using $e^+e^- \to \nu\bar{\nu}\gamma$ events** The direct method is based on the process $e^+e^- \to \nu\bar{\nu}\gamma$, whose cross section is proportional to $N_\nu$. The typical signature is a single photon in the detector with energy $E_\gamma = (s - m_Z^2)/(2\sqrt{s})$. The most precise direct $N_\nu$ measurements at LEP were carried out by L3 [164] and DELPHI [165] Collaborations. By combining the direct measurements at LEP, the current experimental result is $N_\nu = 2.92 \pm 0.04$.

The statistical uncertainty of $N_\nu$ in the previous measurement is 1.7%. The main systematic uncertainty from the L3 measurement includes the uncertainty in single photon trigger efficiency (0.6%), and photon identification efficiency (0.3%), and the uncertainty in identifying the converted photons (0.5%). The systematic uncertainty of theoretical origin is due to the knowledge of higher order radiative corrections to the process $e^+e^- \to \nu\bar{\nu}\gamma$, within the SM. At LEP an uncertainty at the percent level was achieved through complete tree-level matrix elements for $e^+e^- \to \nu\bar{\nu}\gamma$ and $e^+e^- \to \nu\bar{\nu}\gamma\gamma$, properly combined with higher order initial state multiphoton radiation [166]. The bulk of the electroweak corrections were accounted for through running couplings on top of the tree-level matrix elements or through $\mathcal{O}(\alpha)$ corrections to $Z\gamma$ production [166–171]. A first calculation including one-loop electroweak corrections appeared in Ref. [172], with an estimated uncertainty of the order of 1%.

An overall relative precision of 0.05% can be achieved for the direct measurement of $N_\nu$ at the CEPC.Benefiting from the excellent performance of the CEPC inner tracker, the uncertainty due to the converted photon selection efficiency is expected to be negligible. The granularity of the CEPC ECAL is expected to be 10 to 100 times better than the detectors at LEP. Therefore photons can be identified with high purity with loose electromagnetic shower shape requirements. The uncertainty of photon efficiency can be reduced to less than 0.05%. On the theory side, the understanding of the electroweak corrections is expected to reach the two-loop level, in addition to the matching to higher order QED corrections. Given the recent progress in the calculation of NNLO corrections for $2 \to 3$ processes at the LHC, the program looks feasible. It would be also worth to investigate the ratio $\sigma(e^+e^- \to \nu\bar{\nu}\gamma)/\sigma(e^+e^- \to \mu^+\mu^-\gamma)$, where (large) part of the ISR radiative corrections are expected to cancel, provided the luminosity allows enough statistics for the process $e^+e^- \to \mu^+\mu^-\gamma$.

### 11.2.2 MEASUREMENT OF THE $W$ BOSON MASS

In $e^+e^-$ collisions, $W$ bosons are mainly produced in pairs, through the reaction $e^+e^- \to W^+W^-$. At the reaction threshold, $\sqrt{s} \sim 2m_W$, the cross section of this process is very sensitive to $m_W$, providing a natural method for the measurement of this parameter. At center-of-mass energies above the $W^+W^-$ production threshold, $m_W$ can be determined from the peak of the invariant mass distribution of its decay products. Both methods are very complementary: while the former requires an accurate theoretical prediction of the $W^+W^-$ production cross section as a function of $m_W$ and a precise determination of the collider luminosity, the latter mostly relies on a good resolution in the reconstruction of the hadronic invariant mass, and a precise control of the detector calibration.

Both methods have been used at LEP [173–176]. With only about 40 pb$^{-1}$ collected by the four LEP experiments at $\sqrt{s} \sim 161.3$ GeV and given the low cross section at threshold, the threshold method is limited by a significant statistical uncertainty of about 200 MeV. The final state reconstruction method exploited the full LEP2 dataset, about 2.6 fb$^{-1}$ collected between $\sqrt{s} \sim 161.3$ GeV and 206 GeV, and achieved a total uncertainty of 33 MeV. While this measurement used both the $W^+W^- \to \ell\nu qq$ and $W^+W^- \to qqqq$ channels, the fully hadronic channel is limited by uncertainties in the modeling of hadronization and interactions between the decaying $W$ bosons, and the semi-leptonic final state dominates the precision of the final result.

Accounting for results from LEP [173–176], results from the CDF and D0 experiments at the Tevatron [177, 178], and from ATLAS at the LHC [179], the present world-average value of $m_W$ has an uncertainty estimated between 12 and 13 MeV. The uncertainty is expected to fall below 10 MeV when including the final LHC results. A natural goal for CEPC is thus to reach a precision well below 5 MeV, making optimal use of $W^+W^-$ cross section data around $\sqrt{s} \sim 161$ GeV, and of the final state invariant mass distributions at $\sqrt{s} \sim 240$ GeV. The achievable precision of both methods is described below.

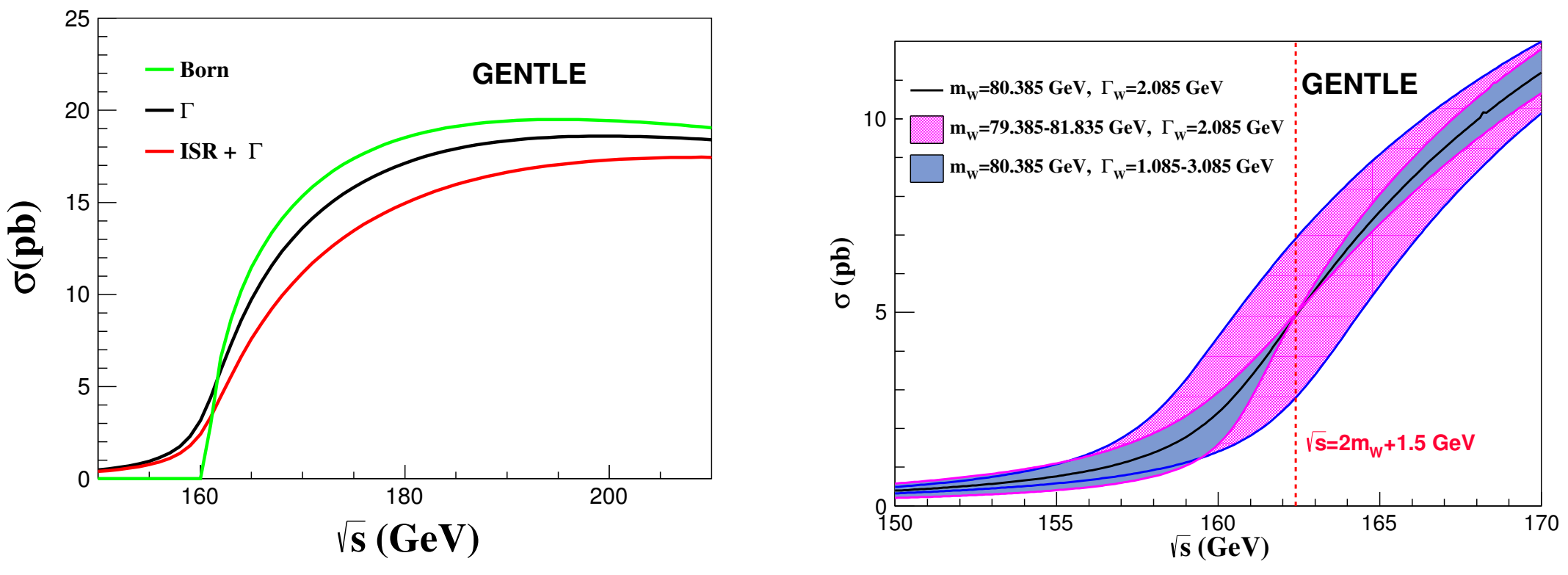

(a) (b)

**Figure 11.16:** $W^+W^-$ production as a function of $\sqrt{s}$, (a) at Born level, including finite width effects, and including initial state radiation corrections; and (b) for a range of values of $m_W$ and $\Gamma_W$.

## DETERMINATION OF $M_W$ AND $\Gamma_W$ FROM THE $W^+W^-$ PRODUCTION CROSS SECTION

In this section, the possibility of extracting the W boson mass and width from the production cross section is explored. The study assumes a total integrated luminosity of $L = 3.2$ ab$^{-1}$, which can be collected in one year, assuming an instantaneous luminosity of 2.5 ab$^{-1}$. For this study, the GENTLE program version 2.0 [180] is used to calculate $\sigma_{WW}$ as a function of the center-of-mass energy, $m_W$ and $\Gamma_W$. The behavior of the cross section as a function of the center-of-mass energy, $\sqrt{s}$, is illustrated in Figure 11.16.

The measurement sensitivity is optimized by taking into account the following:

- the integrated luminosity target for up to three values for $\sqrt{s}$;
- in the case of two $\sqrt{s}$ values, a three-dimensional optimization is performed, scanning both values $\sqrt{s}$ in steps of 100 MeV, and the fraction of integrated luminosity spent at each point in steps of 5%;
- in the case of three $\sqrt{s}$ values, a five-dimensional optimization is performed in a similar way.

The systematic uncertainties are listed below, most of the systematic uncertainties are correlated among the different $\sqrt{s}$ WW threshold scan runs.

- Beam energy measurement: The analysis assumes an uncertainty of 0.5 MeV in the beam energy measurement as described in Section 11.2.1.
- Beam energy spread: The energy spread of about 80 MeV is expected on $WW$ threshold scan runs in CEPC with an uncertainty of less than 0.8 MeV.
- Overall normalization uncertainties: Integrated luminosity, object reconstruction and identification and the theoretical calculation of the $e^+e^- \to W^+W^-$ cross section. It is assumed that these sources sum up to a total relative uncertainty of $2 \times 10^{-4}$ on the ratio between measured and predicted cross sections.

The result of the statistical optimization leads to a three-point scenario, with most of the data collected at energies of 157.5 and 162.5 GeV.

Another run at energy of 172 GeV is also needed for $\alpha_S(m_W{}^2)$ measurement. Ref. [181] shows that high-luminosity runs at $sqrts = 2m_W$ will allow for an $\alpha_S$ determination with uncertainties as low as 0.2%. A summary of given in Table 11.11. The final measurement uncertainties, assuming this optimal scenario and systematic uncertainties are described above, are collected in Table 11.12. We conclude that an uncertainty of about 1 MeV can be achieved for $m_W$, and 3 MeV for $\Gamma_W$. Both $m_W$ and $\Gamma_W$ are expected to be dominated by statistical uncertainties. The major systematic uncertainty for $m_W$ measurement is expected to be the uncertainty in beam energy calibration, while the $\Gamma_W$ measurement is significantly affected by the beam energy spread.

| $\sqrt{s}$ (GeV) | Luminosity (ab$^{-1}$) |
|---|---|
| 157.5 | 0.5 |
| 161.5 | 0.2 |
| 162.5 | 1.3 |
| 172.0 | 0.6 |

**Table 11.11:** The proposed $e^+e^- \to W^+W^-$ threshold scan runs and their integrated luminosity, for a total integrated luminosity of 2.6 ab$^{-1}$.

| Observable | $m_W$ | $\Gamma_W$ |
|---|---|---|
| Source | Uncertainty (MeV) | |
| Statistics | 0.8 | 2.7 |
| Beam energy | 0.4 | 0.6 |
| Beam spread | – | 0.9 |
| Corr. syst. | 0.4 | 0.2 |
| Total | 1.0 | 2.8 |

**Table 11.12:** Dominant systematic uncertainties in the measurement of $m_W$ and $\Gamma_W$, using the production cross section at threshold at CEPC.

## DETERMINATION OF $M_W$ BY KINEMATIC RECONSTRUCTION

According to LEP experience, the fully hadronic final state is limited by systematic uncertainties that are difficult to control using data. The present section therefore concentrates on the semi-leptonic final states, where one $W$ boson decays to an electron or a muon, while the other decays hadronically. An estimate of the $m_W$ measurement potential is presented based on $WW \to \ell\nu qq$ events ($\ell = e, \mu$), and the potential of hadronic $Z$ boson decays to calibrate the measurement of the hadronic invariant mass is evaluated.

The $W^+W^-$ cross section at $\sqrt{s} = 240$ GeV is about 17 pb. For an integrated luminosity of 5.6 ab$^{-1}$, this corresponds to a sample of about $95 \times 10^6$ $W$ boson pairs, and $2.8 \times 10^7$ $WW \to \ell\nu qq$ events. For the $ZZ$ production, the cross section is about 1 pb, yielding about $5.6 \times 10^6$ $Z$ boson pairs, and $1.6 \times 10^6$ $ZZ \to \nu\bar{\nu}q\bar{q}$ events. Benefiting from the much better known $Z$ boson mass compared to the $W$ boson mass, the $Z \to q\bar{q}$ resonance is expected to provide a benchmark to check the detector calibration. However, it should be noted that this sample is small compared to the $W \to q\bar{q}$ one and the presence of heavy quarks in the $Z$ boson decays has to be accounted for when deriving constraints on the hadronic response in $W$ events.

The $W^+W^-$ event selection criteria require the presence of one reconstructed electron or muon with energy greater than 10 GeV, and missing transverse momentum greater than 10 GeV. The invariant mass of all reconstructed final state particles must be greater than 50% of the center-of-mass energy; the hadronic system, $i.e.$ the set of all particles excluding the selected lepton, is clustered into two jets and its invariant mass distribution is used to probe the $W$ boson mass. A $b$-tag veto can be applied to enrich the selected samples in light-quark decays, and reduce the systematic differences between the $W$ and $Z$ boson samples. In the $\mu\nu q\bar{q}$ channel, the efficiency of these criteria is 71.3%, as shown in Table 11.13. Corresponding selection efficiencies for $ZZ \to \nu\bar{\nu}q\bar{q}$ events are shown in Table 11.14. The corresponding hadronic invariant mass distributions are shown in Figure 11.17. After these selections, the backgrounds are expected to be too small to play any non-negligible role in the measurement.

| Selection | Efficiency (%) | Nb. of events |
|---|---|---|
| $E_\mu > 10$ GeV, $\lvert\cos(\theta_\mu)\rvert < 0.995$ | 85.4 | $11.9 \times 10^6$ |
| $p_T^{miss} > 10$ GeV | 82.0 | $11.5 \times 10^6$ |
| $m_{vis} > 0.5 \times \sqrt{s}$ | 75.6 | $10.6 \times 10^6$ |
| $b$-tag score $< 0.5$ | 71.3 | $10.0 \times 10^6$ |

**Table 11.13:** Efficiency of the event selection criteria in the $WW \to \mu\nu qq$ channel.

| Selection | Efficiency (%) | Number of events |
|---|---|---|
| Missing Energy $> 35$ GeV | 69.9 | $1.12 \times 10^6$ |
| $m_{vis} > 0.2 \times \sqrt{s}$ | 66.4 | $1.06 \times 10^6$ |
| $b$-tag score $< 0.5$ | 50.1 | $0.80 \times 10^6$ |

**Table 11.14:** Efficiency of the event selection criteria in the $ZZ \to \nu\nu qq$ channel.

Given the large expected statistics, the inclusion of the $e\nu qq$ channel, and the good resolution in the invariant mass distribution, the statistical sensitivity of the $m_W$ measurement becomes better than 1 MeV. Using the $ZZ \to \nu\nu qq$ sample alone, the detector calibration can be checked to about 6 MeV. Further calibration samples can be extracted from radiative return events ($e^+e^- \to Z\gamma$). In addition, runs at $\sqrt{s} = 91.2$ GeV will be required for general detector alignment, monitoring and calibrations; these runs will provide copious samples of hadronic $Z$ boson decays that will further constrain the hadronic calibration. Combining all information, the statistical precision of the calibration samples will match that of the $W$ boson decays.

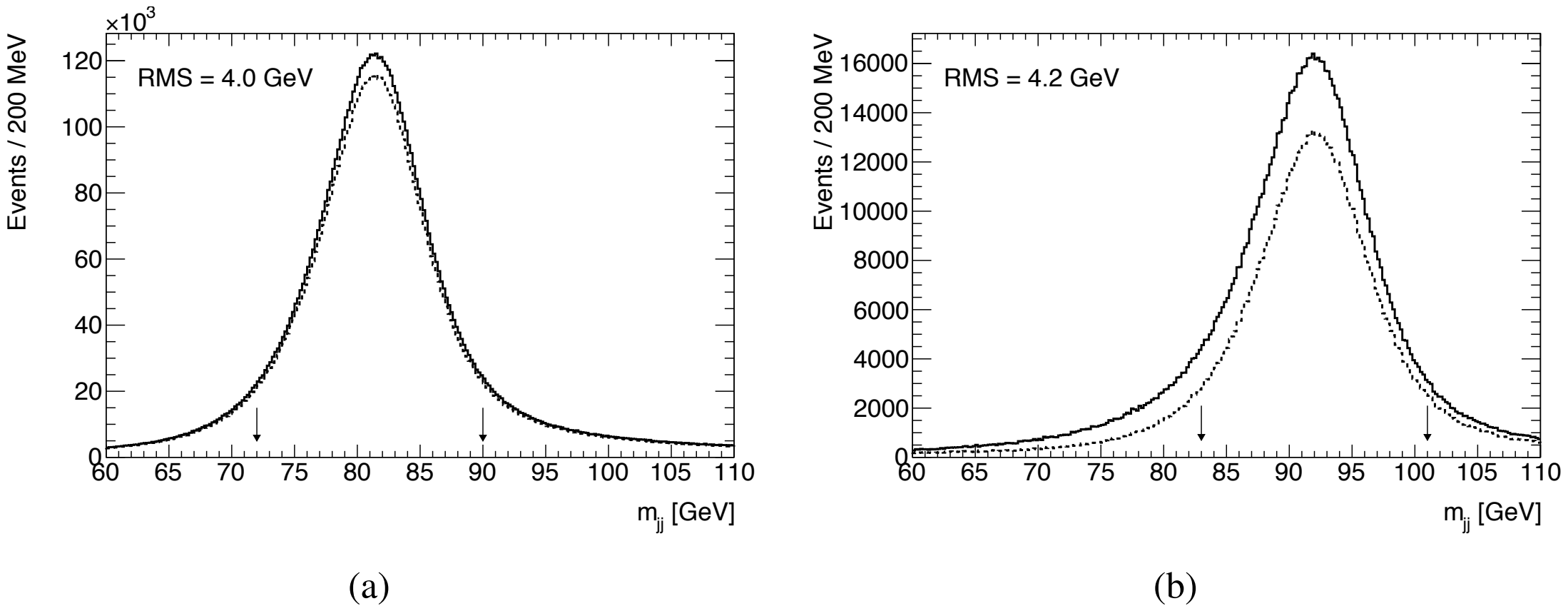

**Figure 11.17:** Dijet invariant mass distributions for (a) $WW \to \mu\nu qq$ events, without and with a $b$-jet veto cut, and correspondingly for (b) $ZZ \to \nu\nu qq$ events. The RMS of the distributions are quoted for the interval indicated by the arrows.

The sensitivity can be further enhanced using kinematic fits to constrain the reconstructed lepton and jet momenta to match the known center-of-mass energy ($\Sigma_i E_i = \sqrt{s}$) and total event momentum ($\Sigma_i \vec{p_i} = \vec{0}$). This method was routinely used at LEP, gaining a factor of about 3 in the statistical precision, at the expense of an explicit dependence of the measurement on the beam energy. Given the expected statistical precision at CEPC, this refinement seems unnecessary here. At the CEPC, the beam energy calibration and initial state radiation are expected to contribute less than 1 MeV to the measurement uncertainty. Further significant sources of systematic uncertainty include the lepton momentum scale, which can be reduced using $Z$ boson decays as discussed above, and the modeling of hadronization. The latter can be strongly reduced using measurements of rates and distributions of identified particles, in both $Z$ and $W$ boson decays.

The primary sources of uncertainty are summarized in Table 11.15, comparing LEP and CEPC. A total uncertainty at the level of 3 MeV seems reachable.

### 11.2.3 OBLIQUE PARAMETERS

Using the estimated experimental capabilities of CEPC, we carry out a fit to determine the sensitivity of CEPC to the oblique electroweak parameters $S$ and $T$ [182, 183]. We omit the parameter $U$ that is often included in fits as it arises from a dimension-8 operator in theories with a weakly coupled Higgs boson [184], and so is expected to be much smaller than $S$ and $T$ which arise at dimension 6. In the electroweak fit we treat the following five well-measured observables as parameters, from which the Standard Model prediction for all of the other observables can be computed:

$$\alpha_s(m_Z^2), \Delta\alpha_{\text{had}}^{(5)}(m_Z^2), m_Z, m_t, m_H. \tag{11.30}$$

Of these parameters, CEPC is expected to significantly improve our knowledge of $m_Z$. The primary power of CEPC is in improving the precision of measurements of other observables, including $m_W$ and $\sin^2\theta_{\text{eff}}^{\ell}$, which may be derived from these parameters. Readers interested in more background information may find a thorough and up-to-date review of the status of electroweak precision in Ref. [185].

| Collider | LEP | CEPC |
|---|---|---|
| $\sqrt{s}$ (GeV) | 180–203 | 240 |
| $\int \mathcal{L}dt$ | 2.6 fb$^{-1}$ | 5.6 ab$^{-1}$ |
| Channels | $\ell\nu qq, qqqq$ | $\ell\nu qq$ |
| Source | Uncertainty (MeV) | |
| Statistics | 25 | 1.0 |
| Beam energy | 9 | 1.0 |
| Hadronization | 13 | 1.5 |
| Radiative corrections | 8 | 1.0 |
| Detector effects | 10 | 1.5 |
| Total | 33 | 3.0 |

**Table 11.15:** Dominant systematic uncertainties in the measurement of $m_W$ using direct reconstruction, as achieved at LEP, and expected at CEPC.

The inputs to the fit are listed in Table 11.16. Notice that we have performed the fit directly using forward-backward asymmetry parameters $A_{\rm FB}^{0,f}$ as inputs, rather than the derived quantities $\sin^2\theta_{\rm eff}^f$ that were used in earlier work [193, 194]. The forward-backward asymmetries more directly reflect the experimental measurements; on the other hand, theoretical predictions are often expressed in terms of the effective weak mixing angles [156, 190]. They are related through the asymmetry parameters $A_f$:

$$A_f = \frac{1-4|Q_f|\sin^2\theta_{\rm eff}^f}{1-4|Q_f|\sin^2\theta_{\rm eff}^f + 8|Q_f|^2\sin^4\theta_{\rm eff}^f}, \tag{11.31}$$

$$A_{\rm FB}^{0,f} = \frac{3}{4}A_e A_f. \tag{11.32}$$

There is an extensive literature on the computation of the $S$ and $T$ dependence of observables (e.g. [182, 183, 195]); a convenient tabulation of the results may be found in Appendix A of [196]. Assembling these results, we obtain a prediction of the observables in terms of the five input parameters, $S$, and $T$. In the fit we compute a profile likelihood, floating the five parameters to obtain the maximum likelihood for given $S$ and $T$.

The fit is performed following [193] (which in turn relied on [197–199]): in constructing a likelihood we treat experimental uncertainties as Gaussian but theory uncertainties as a flat prior, leading to an effective $\chi^2$ function

$$\chi^2_{\rm mod} = \sum_j \left[-2\log\left(\mathrm{erf}\left(\frac{M_j-O_j+\delta_j}{\sqrt{2}\sigma_j}\right) - \mathrm{erf}\left(\frac{M_j-O_j-\delta_j}{\sqrt{2}\sigma_j}\right)\right) - 2\log\left(\sqrt{2\pi}\sigma_j\right)\right], \tag{11.33}$$

with $M_j$ the measured value, $O_j$ the prediction for the observable, $\sigma_j$ the experimental uncertainty, and $\delta_j$ the theory uncertainty.

Our estimates of theory uncertainties assume that full three-loop computations of the parametric dependence of observables in the Standard Model will be completed. The remaining uncertainties are estimated based on [189–192]. In the case of the $W$ mass measurement, an uncertainty of 1 MeV from the computation of the near-threshold $WW$ cross section is added in quadrature with the estimated four-loop theory uncertainty in the observable itself.

| Observable | Value | Exp. Uncertainty | Th. Uncertainty |
|---|---|---|---|
| $\alpha_s(m_Z^2)$ | $0.1185$ | $1.0 \times 10^{-4}$ [66] | $1.5 \times 10^{-4}$ |
| $\Delta\alpha_{\rm had}^{(5)}(m_Z^2)$ | $276.5 \times 10^{-4}$ | $4.7 \times 10^{-5}$ [186] | – |
| $m_Z$ [GeV] | $91.1875$ | **0.0005** | – |
| $m_t$ [GeV] (pole) | $173.34$ | $0.6$ [187] | $0.25$ [188] |
| $m_H$ [GeV] | $125.14$ | $0.1$ [186] | – |
| $m_W$ [GeV] | $80.358617$ [189] | **0.001** | $1.4 \times 10^{-3}$ |
| $A_{\rm FB}^{0,b}$ | $0.102971$ [156, 190] | $\mathbf{1.0 \times 10^{-4}}$ | $8.3 \times 10^{-5}$ |
| $A_{\rm FB}^{0,\mu}$ | $0.016181$ [190] | $\mathbf{4.9 \times 10^{-5}}$ | $2.6 \times 10^{-5}$ |
| $A_{\rm FB}^{0,e}$ | $0.016181$ [190] | $\mathbf{8.1 \times 10^{-5}}$ | $2.6 \times 10^{-5}$ |
| $\Gamma_Z$ [GeV] | $2.494682$ [127] | **0.0005** | $2 \times 10^{-4}$ |
| $R_b \equiv \Gamma_b/\Gamma_{\rm had}$ | $0.2158459$ [127] | $\mathbf{4.3 \times 10^{-5}}$ | $7 \times 10^{-5}$ |
| $R_\ell \equiv \Gamma_{\rm had}/\Gamma_\ell$ | $20.751285$ [127] | $\mathbf{2.1 \times 10^{-3}}$ | $1.5 \times 10^{-3}$ |
| $\Gamma_{Z\to{\rm inv}}$ [GeV] | $0.167177$ [127] | $\mathbf{8.4 \times 10^{-5}}$ | – |

**Table 11.16:** Inputs to the CEPC fit. Numbers in bold are expected experimental uncertainties from CEPC measurements. Other entries reflect anticipated uncertainties at the time of CEPC operation. The numbers in the "Value" column for the first five parameters are current measurements; those below the horizontal line give the Standard Model calculated value as a function of the five parameters. Theory uncertainties are future projections assuming complete 3-loop calculations, based on estimates in Refs. [189–192].

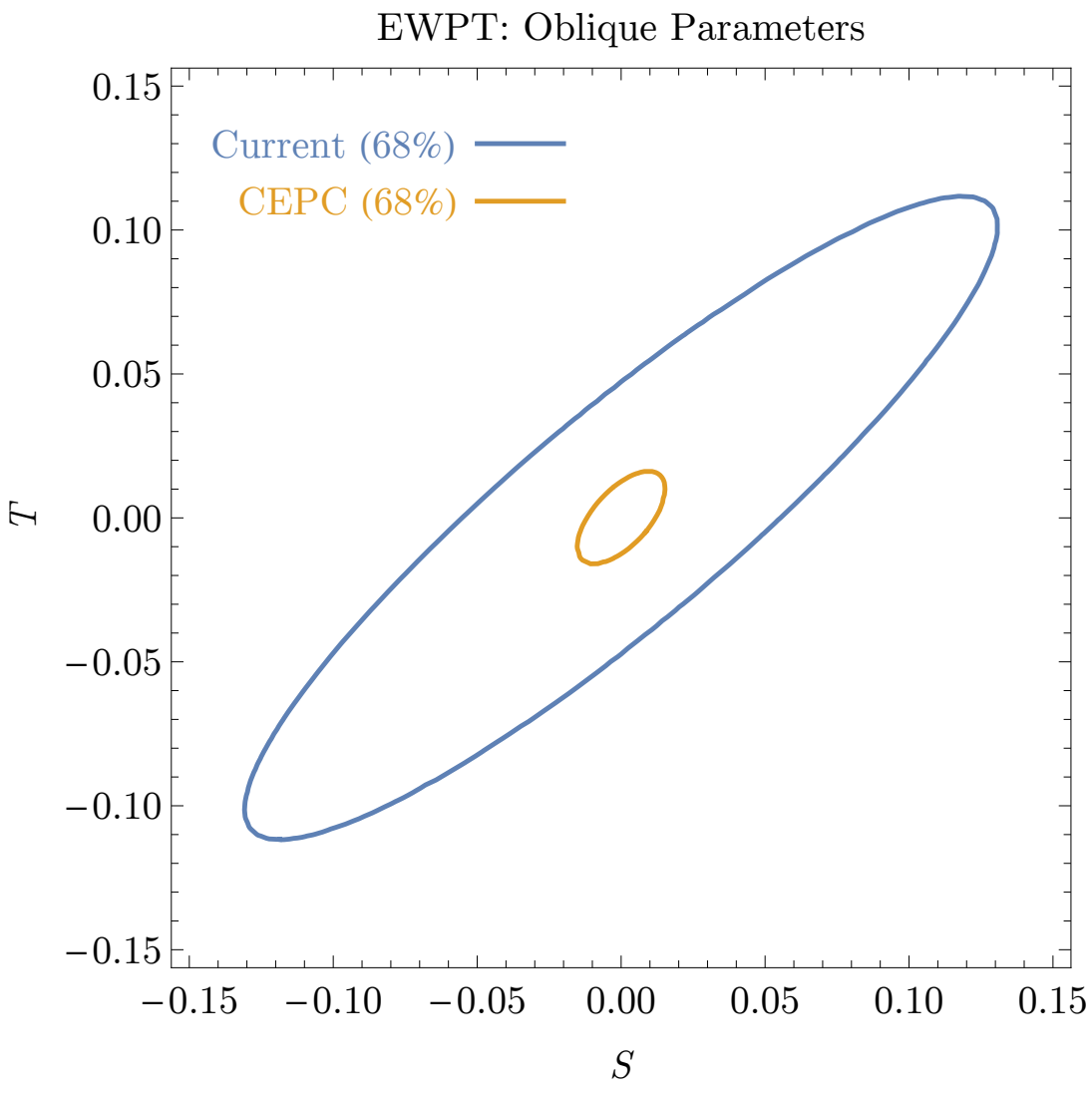

(a)

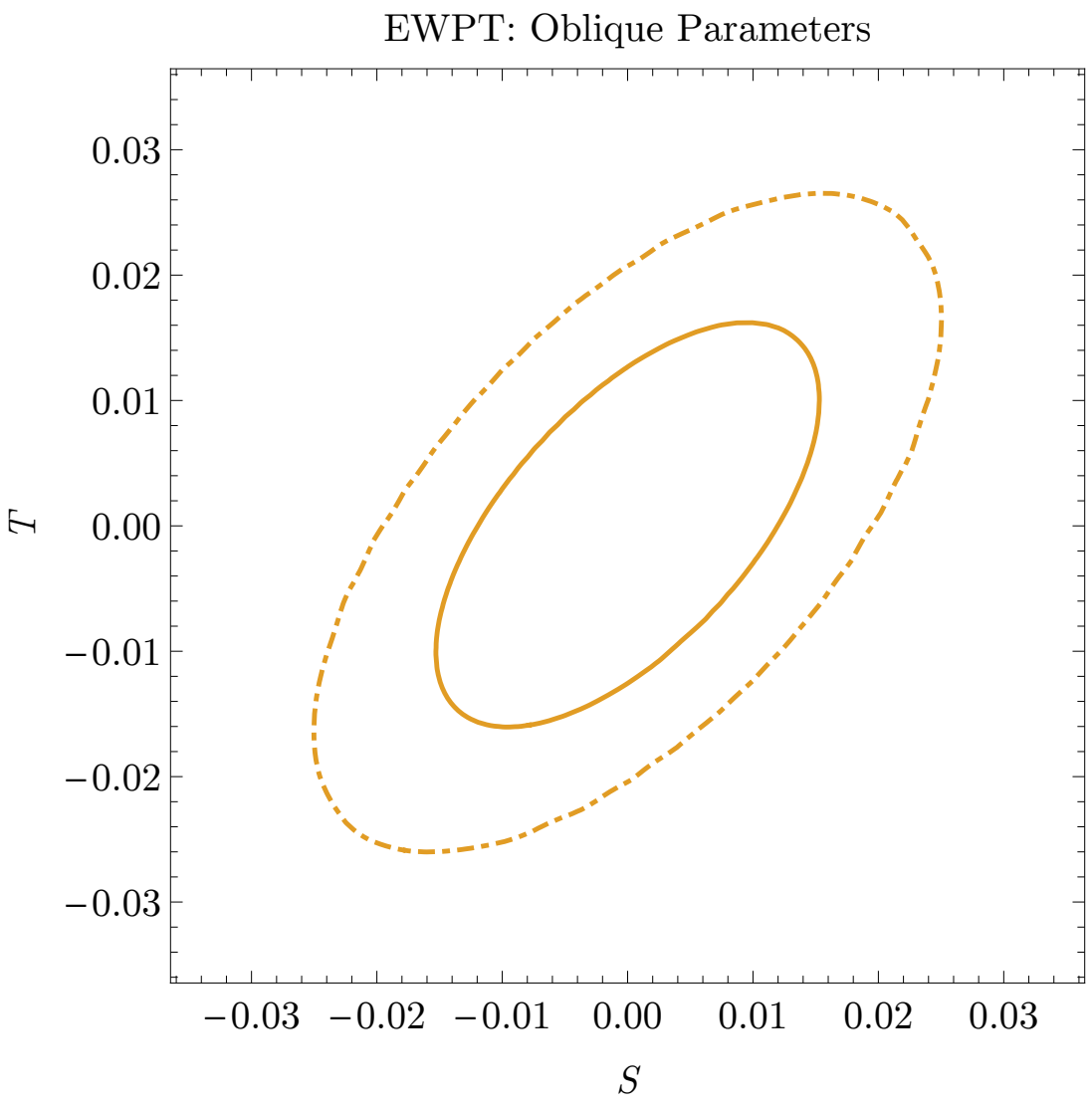

(b)

**Figure 11.18:** CEPC constraints on the oblique parameters $S$ and $T$. (a):comparison of CEPC projection (orange) to current constraints (blue). Contours are 68% confidence level. (b): a closer look at the CEPC fit, showing 68% confidence level (solid) and 95% confidence level (dashed).

The results of the fit are depicted in Figure 11.18. Solid contours are 68% confidence level curves, meaning $\Delta\chi^2_{\rm mod} = 2.30$; the dashed contour is 98% C.L. ($\Delta\chi^2_{\rm mod} = 6.18$). For clarity we have assumed that the measured central values will precisely agree with Standard Model predictions. In particular, the contour depicting current constraints is artificially displaced to be centered at the origin, though it accurately reflects the size of the uncertainties in current data. From the figure, we see that the results of CEPC will significantly shrink the error bars on the $S$ and $T$ parameters relative to currently available data.

By fixing $T = 0$ or $S = 0$, we can also obtain the projected one-parameter 68% C.L. bounds on $S$ and $T$. As one-parameter fits these correspond to $\Delta\chi^2_{\rm mod} = 1.0$. We obtain:

$$|S| < 3.6 \times 10^{-2} \text{ (current)}, \quad 7.9 \times 10^{-3} \text{ (CEPC projection)}, \tag{11.34}$$
$$|T| < 3.1 \times 10^{-2} \text{ (current)}, \quad 8.4 \times 10^{-3} \text{ (CEPC projection)}. \tag{11.35}$$

Thus CEPC will achieve about a factor of four improvements in the precision of both oblique parameters that are considered here.

CHAPTER 12

# FUTURE PLANS AND R&D PROSPECTS

One of the main goals of this CDR is to demonstrate that the proposed conceptual detectors can meet the physics requirements and are feasible to be built in the timescale of the CEPC project. Compared with the pre-CDR, the baseline detector concept has been reoptimized and its performance has been investigated with more sophisticated simulation and reconstruction tools. A variant of the baseline concept with a full silicon tracker has also been introduced. In addition, an alternative concept based on a drift chamber tracker inside a 2 T solenoid and dual-readout calorimetry has been proposed. Initial studies, presented in Chapters 10 and 11, show that the baseline concept detector is capable of achieving the physics goal of the CEPC. Moving forward, more in-depth R&D and studies are required for the preparation of the Technical Design Report. These include the sharpening the physics case particularly for the electroweak and flavor physics, finalizing detector technological choices, further design optimization to improve performance and reduce cost, designing mechanical, electrical and thermal systems, and developing installation and integration schemes. This Chapter gives a brief summary of the planned future work.

## 12.1 PROJECT TIMELINE

The current timeline for the CEPC project is shown in Figure 12.1. It assumes steady progress and favorable funding decisions at the earliest possible dates. The first stage of the project was the completion of the Preliminary CDR (Pre-CDR) in 2015, and this CDR in 2018. The publication of this document marks the conclusion of this stage. The next stage is the 5-year period from 2018 to 2022 for detector R&D towards the completion of the Technical Design Report (TDR). The formation of international collaborations envi-sioned for the two experiments will also take place during this period. The longer term stages are:

**2022-2030:** The construction of both the accelerator and the detectors will commence in 2022 during the Chinese government's $14^{th}$ Five-Year Plan period, continue in the $15^{th}$ Five-Year Plan period, and be completed by 2030.

**2030-2040:** Data-taking will begin as early as in 2030, the start of the $16^{th}$ Five-Year Plan period, and continue for about 10 years till 2040.

**2040:** Superconducting magnets for the Super Proton Proton Collider (SPPC) project are expected to be ready for installation, and the SPPC era will begin.

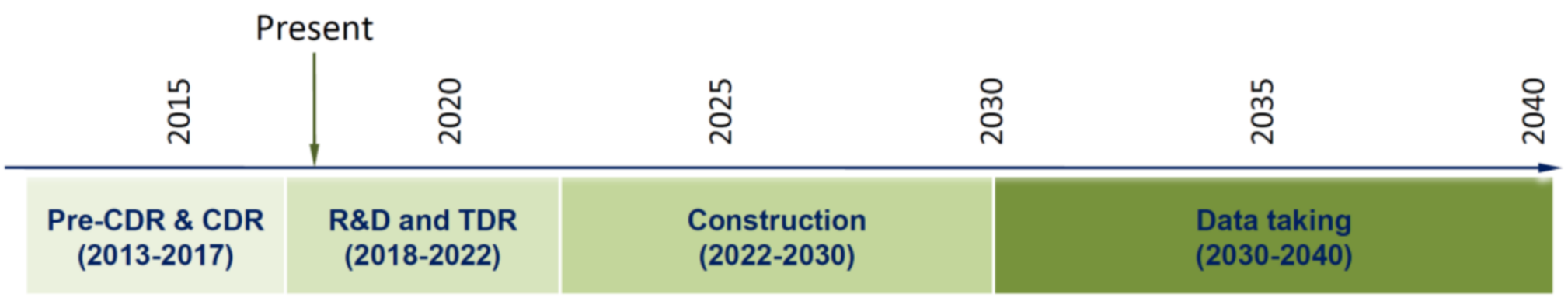

**Figure 12.1:** An optimistic timeline for the CEPC project: the Technical Design Report are to be delivered by 2022, the construction takes place between 2022–2030, and the data-taking starts soon after.

The realization of such an ideal timeline depends on many factors, but the project approval is the single most critical one. After the completion of this CDR, the focus will turn into the formation of two international collaborations that will finalize the detector designs and prepare the TDRs. The targeted R&D of key technologies will need to be completed before the final design choices are made. These R&D and future work are briefly described below.

## 12.2 SIMULATION, RECONSTRUCTION AND OBJECT PERFORMANCE

Offline algorithms such as simulation, reconstruction and calibration play important roles in maximizing the potential of the CEPC detector. The performance of physics objects depend on these algorithms. Though basic understanding of their performance has been achieved and documented in this report, more need to be done to optimize existing algorithms and to develop new ones with the goal to improve their performance for physics analyses.

The first priority of the algorithm development is to assist the detector design optimization and to adapt to the changes in detector technologies and geometries. This work will continue until the design is finalized. Other works include:

- Develop sub-detector digitization algorithms;
- Develop calibration methods including in-situ calibrations;
- Improve the reconstruction of events with complex topologies;
- Improve the identification of particles such as $\tau$-leptons and develop algorithms to identity converted photons;

- Develop Bremsstrahlung photon recovery algorithms to improve the energy measurements of electrons and muons;
- Improve and characterize in detail the detector performance;
- Investigate advanced reconstruction algorithms and pattern recognitions.

## 12.3 STUDIES OF PHYSICS POTENTIALS

The physics case for the CEPC summarized in this CDR covers the main physics motivations. The goal of the future study is to strengthen it. Several case studies, based on some of the most prominent new physics scenarios, are presented. In the future, such studies need to be expanded to include more new physics scenarios and take into account the latest developments.

Another major direction for the future is the development of state of art Monte Carlo tools, which implement the most precise calculation available and are capable of making both inclusive and differential predictions.

Higgs boson physics is the primary physics target for the CEPC. The studies presented in the CDR gives a leading order assessment of the capability of the CEPC. More studies are needed to fully understand the potential and the limitation of the CEPC. A few lines of studies are foreseen:

- The inclusive $e^+e^- \to ZH$ cross section is an anchor to all Higgs boson coupling measurements. Its precision is largely determined by the $Z \to q\bar{q}$ final state which suffers from large diboson backgrounds. It remains to be confirmed that the systematic uncertainty is indeed negligible.
- The Higgs boson width affects all Higgs boson coupling measurements as well. Its precision can be improved by study better separation between $\nu\bar{\nu}H(b\bar{b})$ and $Z(\nu\bar{\nu})H(b\bar{b})$ events.
- The precision of individual decay modes can be improved by either improving existing analyses using advanced analysis techniques or analyzing many unexplored final states.
- One of the greatest advantages of the CEPC is its ability to search for exotic or unknown Higgs boson decays. More work is clearly needed to take advantage of this unique opportunity.

The electroweak precision measurement is an important component of the CEPC physics program. The most important factor is the estimation of systematical uncertainties. More realistic detector effects need to be taken into account. Some of the projections, such as the precision in the $Z$ mass measurement, should be assessed once more detailed design of the accelerator has been made. The results should also be updated when more accurate theoretical computation and estimate of the theoretical uncertainty become available. The CDR also presents some simple estimates for the reach of exotic decays of the $Z$ boson, taking advantage of the high statistics at the Tera-Z factory. Such estimates need to be sharpened with more detailed studies.

Flavor physics and QCD tests/measurements haven't received the attention they deserve. With billions of $b$-quark, $c$-quark and $\tau$-leptons, the CEPC is a gold mine for flavor physics. Precise tests and measurements of QCD over a wide range of $e^+e^-$ collision energies is another front to explore. The CDR only contains very preliminary projections based on simple simulation and extrapolation. They shall benefit greatly from more detailed and realistic simulation.

## 12.4 TRACKING SYSTEM

The tracking system is one of the detector sub-systems that requires the most research and development. The CEPC physics goals are highly demanding on the tracking system, requiring momentum resolutions and material budget better than existing detectors. Significant amount of research work is still necessary to define the best detector solutions among the ones presented in this CDR.

### 12.4.1 VERTEX DETECTOR

The vertex detector has to fulfill highly demanding requirements imposed by the CEPC physics program. Meeting simultaneously the single-point resolution, low material budget, fast readout, low power consumption and radiation tolerance requirements, demands new developments only achievable with a coherent set of R&D activities:

- Enhancement of density, radiation hardness and ultra-light module assembling;
- Explore smaller production line for TowerJazz and LAPIS in conjunction with the Nano-particle Deposition (NpD) technique;
- Improve the charge collection efficiency of the TowerJazz process by N-type implant;
- Improve the radiation hardness and low power design for SOI process;
- Sensor thinning for CMOS and SOI;
- Detailed designs for mechanical supports to enable cooling, cabling and power conservation.

These R&D studies towards the final vertex detector will be supported and verified through prototyping of detector sensors and mechanical structures.

### 12.4.2 SILICON TRACKER

The silicon tracker is designed to have low material and provide excellent spatial measurements with high efficiencies for charged particles. Through preliminary investigations, several critical R&D items have been identified for the TDR phase:

- Alternative pixelated strip sensors with CMOS technologies;
- $p^+$-on-n silicon microstrip sensors with slim-edge structure;
- Front-end electronics with low power consumption and low noise; fabricated with CMOS technologies of small feature size;
- Efficient powering with low material budget and $CO_2$ cooling techniques;
- Lightweight but robust support structure and related mechanics;
- Detector layout optimization, in particular in the forward region.

### 12.4.3 TIME PROJECTION CHAMBER

The TPC is the CDR baseline-detector central tracker. A modularized design with gas amplification and readout pad optimization has been investigated. Low power consumption electronics and ion backflow have also been considered. Significant R&D is however still required to finalize its design and optimize its operation. The TPC position measurement is sensitive to ions in the gas volume, and more studies regarding operation at the $Z$-pole run are necessary. The future R&D consists of:

- Further development of the hybrid structure module;
- Laser calibration and alignment system development;
- Study high-rate operation at the $Z$-pole in detailed MC simulations including beam backgrounds;
- Develop a more realistic estimate of the effects of ion backflow on the position measurements.

### 12.4.4 FULL-SILICON TRACKER

The full-silicon tracker is a viable option for the CEPC tracking system under the same boundary conditions of the baseline detector concept. In this option, the TPC is replaced with additional silicon detector layers and the layout of the tracker is re-optimized. To explore the full potential of the all silicon tracker, possible improvements are:

- Further optimization of the layout for better performance and lower cost;
- Study track reconstruction performance in busy events;
- Address questions regarding low-power consumption, cooling and low-material budget for large-area silicon detectors

The full-silicon tracker will benefit from the research and development being carried out for the vertex and silicon tracker detectors. A coordinated effort between those teams, the layout optimization, and engineering will be necessary.

### 12.4.5 DRIFT CHAMBER TRACKER

The drift chamber tracker is an option to provide good tracking, high precision momentum measurement and excellent particle identification. The R&D work to be pursued in the near future includes:

- Final layout optimization of the full tracking system, which, besides the drift chamber, includes the vertex detector and the silicon micro-strip detector layers (the silicon wrapper, as indicated in Figure 3.10).
- Studies of non-flammable gas mixtures alternative to He/Isobutane 90/10.
- Particle identification performance with cluster counting, to be assessed in beam tests of realistic drift chamber prototypes.
- Full length ($> 4$ m) prototypes to establish limits of the wires electrostatic stability.

## 12.5 CALORIMETRY

Two technology options based on, respectively, the PFA concept and the dual-readout concept are being explored for the CEPC calorimetry system. The PFA approach aims to develop high-granularity electromagnetic and hadronic calorimeters capable of measuring individual particles in a jet, whereas the dual-readout approach aims for a combined and homogeneous calorimeter with excellent performance for both electromagnetic and hadronic particle showers.

### 12.5.1 PARTICLE FLOW ORIENTED ELECTROMAGNETIC CALORIMETER

The CEPC ECAL is required to have a good intrinsic energy resolution for precise energy measurement of electrons and photons, as well as excellent shower imaging capability that would allow to identify photons from close-by showers, reconstruct detailed properties of a shower and distinguish electromagnetic showers from hadronic ones effectively. Extensive and focused R&D will be conducted on developing a tungsten-based high-granularity sampling calorimeter with either silicon or scintillator as active medium to address the above requirements. Future study work towards the TDR will involve detector design optimization and critical common R&D in view of the CEPC experimental conditions, particularly the continuous operation mode imposed by the CEPC accelerator, and a great deal of prototyping for the two ECAL technology options:

- Design optimization and common R&D
  - Optimization of primary detector parameters with full detector simulation;
  - Thermal studies of detector and electronics components through both simulation and prototyping;
  - Cooling design based on the above studies and prototyping;
  - Development of technological prototypes to address power, cooling and front-end electronics issues;
  - Design of detector modules. Development of technology for fabricating large-size detector modules.
- Silicon-Tungsten ECAL
  - Full characterization of a physics Silicon-Tungsten ECAL prototype using its existing test beam data.
- Scintillator-Tungsten ECAL
  - Development of a SiPM-scintillator coupling scheme that allows very large dynamic range;
  - Development of technology of fabricating high-quality scintillator strips with required fine structures;
  - Design, construction and characterization of a small-size physics prototype.

### 12.5.2 PARTICLE FLOW ORIENTED HADRONIC CALORIMETER

High granularity hadronic calorimeter is an essential part of the PFA-based CEPC calorimetry system. Currently R&D activities for HCAL system includes sampling calorimeters with stainless steel as the absorber and gaseous detector (RPC or GEM) or scintillator tiles with embedded electronics as active sensors. Hence, two technology options are considered, one is SDHCAL based on gaseous detector and the other is AHCAL based on scintillator with SiPM. The future R&D plans for HCAL system include design, construction and performance studies of various prototypes with the CALICE collaboration.

- Make performance study of SDHCAL technological prototype based on RPC using MC and test beam data samples;
- Design and construction of a small-size AHCAL prototype using scintillator and SiPM, make performance study with test beam and MC samples;
- Optimize the geometry and cell size of SDHCAL and AHCAL designs;
- Design active cooling system for both SDHCAL and AHCAL prototypes to address ASIC and frond-end electronics heating issues;
- Develop large size RPC (e.g. 2-3 $m^2$), optimize gas distribution and circulation system to improve gas uniformity;
- Develop Multi-layer RPC with excellent time resolution (eg. 50 ps) which may help to identify showers from neutrons;
- Explore ASIC chips with time information (e.g. PETIROC), design and develop corresponding PCB and front-end electronics;
- Develop THGEM with very compact structure and stable operation.
- Explore new types of SiPM (eg. NDL SiPM) with better performance/price ratio.

### 12.5.3 DUAL-READOUT CALORIMETER

Concerning the dual-readout calorimeter, the following R&D program is being pursued in order to address and clarify several issues before the TDR:

- Absorber material choice, current candidates are lead, brass and iron;
- Machining and assembly procedure for modules of $\sim 10 \times 10\ \mathrm{cm}^2$ cross section;
- Development of a modular, projective solution for a $4\pi$ calorimeter concerning both the construction of single modules and the design and construction of a full detector;
- Identification of adequate solid-state photo-sensors in order to independently optimize both Čerenkov and scintillation light detection (with respect to PDE, linearity, dynamic range and cross-talk performance);
- Readout granularity (i.e. identify the optimal fiber grouping into a single readout channel);
- Identification of a tailored front-end electronics, likely composed by an ASIC and an FPGA chip, in order to extract in real time both charge and time information (in principle, a time resolution of 100 ps should allow to identify the shower starting point inside the calorimeter with a precision of about 6 cm);

- Particle ID performance with Particle Flow Algorithms, with and without a longitudinal segmentation;
- Development and validation of full and fast simulations of both test beam modules and an integrated $4\pi$ detector;
- Assessment of the performance for the most relevant physics channels (such as $W$, $Z$, $H$ decays).

## 12.6 MAGNET

The two experimental concepts proposed in this CDR include a large superconducting solenoid with 2-3 Tesla central field. They require a complicated cancellation-magnet scheme to minimize the disturbance of the beam. Towards TDR, the following R&D work will be conducted:

- Longer conductor with higher $I_c$: A 10 m long NbTi Rutherford cable embedded inside stabilizer has been developed, which provides 5 kA $I_c$ at 4 T background magnetic field. In the future, a longer conductor with a higher 15 kA $I_c$ at 4 T background is in our R&D plan. Meanwhile the measurement of the material properties and the tensile stress of the cable will be a necessary part for the structural simulation of the solenoid coil and the conductor development.
- YBCO stacked-tape cable and cryogenics: Continue R&D on the large HTS solenoid concept proposed for the IDEA detector. The HTS solenoid is supposed to use YBCO stacked-tape cable as conductor. The operation temperature of the cold mass would be raised to 20 K. Research will continue both on the YBCO stacked-tape cable and the cryogenics.
- A 1:10 scale thermosyphon circuit model: In order to study the phase transition process of helium in the circuit, the changes of the temperature distribution and the density distribution over the time, a 1:10 scale thermosyphon circuit will be established for simulation and experiment.

## 12.7 MUON DETECTOR SYSTEM

Located within the solenoid flux return yoke, the muon detector system is required to identify muons with very high efficiency. Both resistive plate chamber and micro pattern gas detector are considered in this CDR. Future R&D requires detailed studies of different technologies and further optimization of baseline design parameters. Several critical R&D items have been identified, including:

- Integrate muon system in full reconstruction and study the muon system physics impact in Monte Carlo simulations.
- Long-lived particle optimization: Explore new physics scenario of long-lived particles and exotic decays. Optimize detector parameters and technologies.

- Layout and geometry optimization: Detailed studies on the structure of the segments and modules need to be carried out to minimize the dead area and to optimize the interface for routing, support and assembly.
- Detector optimization: Study aging effects, improve long-term reliability and stability, readout technologies.
- Detector industrialization: Improve massive and large area production procedures for all technologies.

## 12.8 READOUT ELECTRONICS, TRIGGER AND DATA ACQUISITION

New technologies will emerge before the CEPC DAQ system has to be built. Attention will be made to follow and explore the ongoing improvements of data communication and processing technologies. The DAQ requirements will become more clear as the subdetector options mature. In particular the following areas will be addressed:

- High-speed and low-latency communication technology for data readout.
- A high-efficient data flow distribution schema for data dispatching to large computing farm, where data will be concentrated, re-formatted, possibly zero-suppressed, assembled into full events and filtered.
- Online software trigger and data compression algorithms will be studied and provided based on physics and detector requirements. The implementation of the data processing inside the online farm will then be pursued.

## 12.9 MACHINE DETECTOR INTERFACE

Machine-Detector Interface (MDI) represents one of the most challenging tasks that affects both the accelerator and the detector. The interaction region (IR) has to focus both electron and positron beams to small spot sizes to maximize the machine luminosity. The following R&D will be carried out during the next TDR phase:

- Interaction region layout re-design/optimization;
- Background models validation with experimental data, e.g. SuperKEKB/Belle II;
- Beam pipe design together with SR photon protection, HOM absorber and cooling if needed;
- Installation scheme that involves both LumiCal and final focusing magnets;
- Prototyping R&D on LumiCal, and demonstration of alignment of desired precision with laser and optical devices.

## 12.10 ENGINEERING AND DETECTOR INTEGRATION

In the next few years, building up to the TDR and detector construction, the overall engineering effort needs to ramp up considerably to ensure that the detectors concepts developed can indeed be built. This effort needs to take into account the detector integration and consider all aspects of every detector subsystem and their services. One major issue is the interaction region. Having the last focusing quadrupole inside the detector volume brings severe constraints to the detector design, and requires detailed engineering design. The installation scheme that involves both LumiCal and final focusing magnets needs to be devised. Engineering studies of the overall mechanical stability of the central beam pipe need to be pursued.

Careful studies are also required to design the solenoid magnet and study the effect of the solenoid magnetic field on the Booster accelerator. Development of detector alignment techniques and deformation monitoring are essential to obtain the optimal tracking resolution, specially with the light mass requirements for the track systems, and need to be developed before the TDR. Small scale prototypes of several detector subsystems will be built. These include prototypes for the HCAL, ECAL and dual-readout calorimeters, and for the tracking systems: Time Projection Chamber, Pixel detector, and drift chamber. Engineering studies for the calorimeters mechanical support and installation are required, as well as for the mounting of the tracker inside the calorimeters. Finally the cooling and power systems need to be designed and guaranteed to be sufficient.

**CHAPTER 13**

# SUMMARY

The discovery of the Higgs boson in 2012 at the LHC opened up the physics case for a circular electron-positron collider operating near the Higgs-strahlung threshold at $\sqrt{s} \sim 240$ GeV. The CEPC accelerator and detectors, described in these two volumes of the Conceptual Design Report, will deliver milestone physics results at a relevant timescale. The design is based on achievable technology, while keeping the construction cost and power consumption affordable. The CEPC Higgs measurements will by far surpass the precision of those achievable at the LHC. The ability to run the CEPC at the $Z$-pole and $WW$ threshold energies will allow extremely precise measurements of the $W$ and $Z$ bosons. These measurements will provide stringent tests of the underlying fundamental physics principles of the SM. Together with the Higgs measurements, they will be instrumental in the exploration of new physics beyond the SM.

The CEPC accelerator operations in rather different conditions required for the CEPC physics program place demanding requirements on its detectors. The detector concepts proposed here to meet those requirements are based on complementary strategies and technologies. Together they demonstrate the capability of delivering the detector performance essential for the physics program. Further advances in the state-of-art detector technologies from now, through the Technical Design Report, and until the detector construction, are expected to make these capabilities even more feasible.

All detector concepts proposed realize the need for high-granularity calorimeters and high-precision tracking, albeit using different technologies. Two different calorimetry approaches are presented: a particle flow approach utilizing the detailed particle shower information for the jet reconstruction; or a dual-readout approach with a combined, homogeneous, detector capable of measuring simultaneously both electromagnetic and hadronic particle showers. The magnetic field strength of the detector solenoid is another differen-tiating factor. It will need to be carefully chosen to maximize the physics output of the CEPC, balancing between the maximum luminosity achievable for the $Z$-pole operation and the need for high precision track momentum measurements.

While only the CEPC baseline detector concept has been studied in detail through realistic simulation, different options presented are expected to have similar capabilities. The performance results summarized in Chapters 10 and 11 demonstrate that it is possible to achieve the physics goals of the CEPC. In the coming years, the detector simulation studies will be enhanced to address specific issues requiring further optimization, while physics activities will be broadened to include flavor and QCD physics as well as new physics processes.

The detector R&D has started and will expand both in China and abroad. The international particle physics community is expected to coalesce into two collaborations to further develop the detector designs that will eventually evolve into the two experiments at the CEPC. The ultimate designs will not necessarily be the concepts proposed in this CDR, but will likely incorporate many of the detector technologies described here.

The CEPC has been identified as the top priority for the next high-energy physics (HEP) project in China by the HEP Division of the Chinese Physical Society. It has received strong supports from particle physicists at home and abroad. A significant amount of R&D funds has been received from the Chinese government for the 5-year R&D plan as described in both volumes of this CDR. During the past several years, a core team of accelerator physicists and engineers has been formed to steer the CEPC accelerator project from the R&D, through the final design, to the construction and operation. A number of candidate sites have been identified and the final site will be selected after the project receiving government approval. The particle physics community in China is steadily growing, in part through participations in both domestic and international HEP projects. The technological knowledge acquired through these participations will be invaluable for the design, construction, and operation of the CEPC detectors.

For a project of this scale, international collaboration is a key for its success. IHEP has a long history collaborating with other laboratories and universities both domestically and internationally. This CDR itself is the result of collaborative work by particle physicists around the world. The work presented has also benefited from the tools and methodologies developed in the frameworks for other international HEP programs, such as the ILC. Many of the concepts and technologies proposed for the CEPC detectors are evolutions from the ILC detector designs, adapted for a circular collider. Moreover, the detector technologies proposed benefit greatly from the R&D performed for the ILC, the HL-LHC and others.

This project aims to grow into a truly worldwide endeavor, with international collaboration for both the accelerator and the detectors. A CEPC International Advisory Committee consisting of two dozen world renowned HEP physicists meets once a year in Beijing to offer valuable advices about the project. The collaboration with several other proposed future collider projects – the ILC, CLIC and FCC – includes visitor exchanges, jointly organized workshops (e.g., the ICFA eeFACT workshop series on high luminosity circular $e^+e^-$ colliders), and accelerator schools.

The Chinese government has announced an ambitious plan to establish several large international scientific research centers in China in the coming decade. The CEPC is a strong and viable candidate. The Chinese particle physics community together with international collaborators will propose the CEPC to the Chinese government to compete for a place in its plan.

# Glossary

$CP$ Charge-Parity 57, 58

**ADC** Analog-to-Digital Convert 155

**AFE** Analog Front-End 155

**AHCAL** Analog HCAL 194, 208

**ALP** Axion-Like Particle 29, 33

**ASIC** Application Specific Integrated Circuits 155

**ASU** Active Sensor Unit 200

**ATCA** Advanced Telecommunications Computing Architecture 269

**BAU** Baryon Asymmetry of the Universe 51, 57

**BE** Back-end Electronics 206

**BMR** Boson Mass Resolution 124

**BSM** Beyond the Standard Model 297

**CDR** Conceptual Design Report 1, 2, 116, 388

**CEPC** Circular Electron Positron Collider 1

**CMOS** Complementary Metal-Oxide-Semiconductor 146

**CPS** CMOS Pixel Sensors 167

**DAQ** Data AcQuisition 267

**DCH** Drift CHamber 175

**DEPFET** Depleted P-channel Field Effect Transistor 147

**DHCAL** Digital HCAL 194

**DLC** Diamond Like Carbon 262

**DM** Dark Matter 35

**DR** Dual Readout 224

**DSP** Digital Signal Processors 155

**EB** Event Builder 272

**EBM** Event Building Manager 272

**ECAL** Electromagnetic Calorimeter 130, 300, 304

**EDM** Electric Dipole Moment 333

**EEC** Energy-Energy Correlation 65

**EFT** Effective Field Theory 10

**EIT** Endcap Inner Tracker 170

**EOT** Endcap Outer Tracker 170

**EQR SiPM** SiPM with Epitaxial Quenching Resistors 222

**ETD** Endcap Tracking Detector 165

**EW** ElectroWeak 315

**EWPT** ElectroWeak Phase Transition 20, 21

**EWSB** ElectroWeak Symmetry Breaking 20

**FE** Front-End 206

**FEA** Finite Element Analysis 250, 251, 256

**FPGA** Field-Programmable Gate Array 222, 253

**FST** Full-Silicon Tracker 115

**FTD** Forward Tracking Detector 165

**FV** Fiducial Volume 290

**GEM** Gas Electron Multiplier 153, 160, 161

**GEM-MM** hybrid Gas Electron Multiplier and Micro-Mesh Gaseous Structure 153, 160, 161, 162, 163

**GRPC** Glass Resistive Plate Chambers 130, 191, 194, 210

**HCAL** Hadronic CALorimeter 130, 207

**HL-LHC** High-Luminosity LHC ix

**HOM** High Order Mode 274

**HTS** High Temperature Superconductor 255

**IBF** Ion Backflow 154, 159

**IDEA** Innovative Detector for Electron-positron Accelerator 132

**IGBT** Insulated Gate Bipolar Transistor 253

**ILD** International Large Detector 115

**IP** Interaction Point 139, 144, 170, 175

**IR** Interaction Region 273

**LHe** Liquid Helium 249, 251, 257

**LNV** Lepton-Number Violating 52

**LR** Left-Right 52

**LTS** Low Temperature Superconductor 255

**LumiCal** Luminosity Calorimeter 132

**MC** Monte Carlo 120, 145

**MDI** Machine-Detector Interface 132, 273

**MDT** Monitored Drift Tubes 259

**MicroMegas** Micro-Mesh Gaseous Structure 153, 160, 161

**MIP** Minimum Ionizing Particle 177, 197

**MPGD** Micro-Pattern Gas Detector 150, 152, 153, 160

**MPV** Most Probable Value 201

**MSSM** Minimal Supersymmetric Standard Model 13

**MWPC** Multi-Wire Proportional Chamber 159

**NGL** Non-Global Logarithms 66

**NIEL** Non-Ionizing Energy Loss 276

**NpD** Nano-particle Deposition 380

**PCB** Printed Circuit Board 221

**PDF** Parton Distribution Functions 63

**PFA** Particle Flow Algorithm 129, 194, 206, 207

**PID** Particle IDentification 224

**PM** PhotoMultiplier 224

**QCD** Quantum ChromoDynamics 8

**QED** Quantum ElectroDynamics 62

**ROS** Read Out System 272

**RPC** Resistive Plate Chambers 132

**S/N** Signal-to-Noise 199

**SAR** Successive Approximation Register 155

**SCA** Switched-Capacitor-Array 207

**SDHCAL** Semi-Digital Hadronic CALorimeter 209, 210

**SET** Silicon External Tracker 165

**SiC** Silicon Carbide 230

**SiLC** Silicon tracking collaboration for the Linear Collider 168

**SiPMs** Silicon PhotoMultipliers 194

**SIT** Silicon Inner Tracker 165

**SM** Standard Model 115, 297

**SNR** Signal to Noise Ratio 202

**SOI** Silicon On Insulator 146, 380

**SOT** Silicon Outer Tracker 170

**SR** Synchrotron Radiation 116

**TDR** Technical Design Report 377

**TGC** Triple Gauge boson Coupling 334

**THGEM** Thick Gas Electron Multiplier detectors 210

**TID** Total Ionizing Dose 145

**ToF** Time-of-Flight 130, 311

**TPC** Time Projection Chamber 115, 150, 380

**TQFP** Thin Quad-Flat Packaging 218

**VBF** Vector Boson Fusion 53

**VDSM** Very Deep Sub-Micro 147

**VEV** Vacuum Expectation Value 21

**YBCO** Yttrium Barium Copper Oxide, $YBa_2Cu_3O_7$ 255

# AUTHOR LIST

The following list includes the signatories for both Volume I and Volume II of the CEPC Conceptual Design Report.

Marcello Abbrescia[111], Muhammad Ahmad[26], Xiaocong Ai[26], Sebastiano Albergo[114], Muhammad Abid Aleem[142], Maxim Alexeev[128], Malik Aliev[146], Wolfgang Altmannshofer[176], Fábio Alves[37], Fenfen An[26,193], Guangpeng An[26,221], Haipeng An[59], Qi An[63,221], Rui An[187], Attilio Andreazza[119], Marc Anduze[93], Massimiliano Antonello[118], Juan Antonio Garcia Pascual[26], Stefan Antusch[154], Abdesslam Arhrib[139], Nima Arkani-Hamed[191], Kirk Arndt[167], Patrizia Azzi[122], Paolo Azzurri[125], Robert B. Palmer[220], Bowen Bai[26], Sha Bai[26], Yang Bai[219], Yu Bai[55], Vladislav Balagura[93], Ilaria Balossino[115], Yong Ban[45], Vernon Barger[219], Timothy Barklow[208], João Barreiro Guimarães da Costa[26], Paolo Bartalini[6], Francesco Becattini[116], Franco Bedeschi[125], Lorenzo Bellagamba[112], Alberto Belloni[196], Giovanni Bencivenni[110], J. Scott Berg[220], Catrin Bernius[208], Monica Bertani[110], Claudia Bertella[26], Michele Bertucci[121], XiaoJun Bi[26], Yuanjie Bi[10], Ligong Bian[14], Tianjian Bian[26], Fabrizio Bianchi[128], Ikaros I. Bigi[202], Michela Biglietti[126], Andrea Bignami[121], Gianmario Bilei[124], Maarten Boonekamp[90], Lisa Borgonovi[112], Daniela Bortoletto[167], Alberto Bortone[128], Davide Boscherini[112], Angelo Bosotti[121], Vincent Boudry[93], Sylvie Braibant[112], Joseph Bramante[85], Paolo Branchini[126], Jean-Claude Brient[93], Massimo Caccia[118], Chengfeng Cai[57], Hao Cai[65], Wenyong Cai[81], Xiao Cai[26,221], Yiming Cai[59], Yuchen Cai[37], Yunhai Cai[208], Zhiqiang Cai[75], Marc Cano Bret[47], Bo Cao[41], Dewen Cao[26], Jianshe Cao[26], Junjie Cao[22], Qing-Hong Cao[45], Carlo Michel Carloni Calame[123], Raimon Casanova[150], Vincenzo Cavasinni[125], Junying Chai[128], Weiping Chai[27], NingBo Chang[67], Qin Chang[22], Spencer Chang[203], We-Fu Chang[161], Xuejun Chang[81], Yuan-Hann Chang[158], Sergei Chekanov[168], Bin Chen[26], Chunhui Chen[193], Fusan Chen[26], Gang Chen[11], Gang Chen[26], Gang Chen[153], Guoming Chen[26], Huan Chen[11], Huirun Chen[37], Lei Chen[10], Liejian Chen[26], Mingjun Chen[26], Mingshui Chen[26], Nian Chen[63], Ning Chen[39], Shanhong Chen[26], Shanzhen Chen[113], Shao-Long Chen[6], Shaomin Chen[59], Shenjian Chen[37], Shi Chen[60], Wei Chen[57],

Wen Chen[86], Xin Chen[59], Xun Chen[47], Xurong Chen[27], Ye Chen[26], Yu Chen[57], Yuan Chen[26], Yuanbai Chen[26,221], Yukai Chen[26], Zhenxing Chen[45], Hao Cheng[81], Hsin-Chia Cheng[172], Huajie Cheng[26,60], Jian Cheng[26], Shan Cheng[24], Tongguang Cheng[206], Weishuai Cheng[128], Sanha Cheong[208,208], Yunlong Chi[26], Gianluigi Chiarello[127], Mauro Chiesa[98], Wen Han Chiu[178], Weiren Chou[184], Chungming Paul Chu[26], Ming-chung Chu[104], Xiaotong Chu[26], Zhaolin Chu[47], Chun-Khiang Chua[157], Gianluigi Cibinetto[115], Marco Ciuchini[126], Marina Cobal[129], James Cochran[193], Rafael Coelho Lopes de Sa[197], Fabio Cossio[128], Nathaniel Craig[175], Han Cui[26], Hanhua Cui[26], Xiaohao Cui[26], Zhaoyuan Cui[169], David Curtin[87], Raffaele Tito D'Agnolo[208], Jian-Ping Dai[47], Jianping Dai[26], Jin Dai[26], Lei Dai[4], Wei Dai[11], Xuwen Dai[26], Stefania De Curtis[116], Nicola De Filippis[111], Erika de Lucia[110], Francesca De Mori[128], Antonio Delgado[202], Marcel Demarteau[168], Changdong Deng[26], Wei-Tian Deng[23], Zhi Deng[59], Marco Destefanis[128], P. S. Bhupal Dev[217], Biagio Di Micco[126], Ran Ding[45], Xuefeng Ding[109], Yadong Ding[26], Manuel Dionisio da Rocha Rolo[128], Danilo Domenici[110], Bingbing Dong[47], Chongmin Dong[81], Dong Dong[26], Haiyi Dong[26], Jianing Dong[46], Jing Dong[26,221], Lan Dong[26], Mingyi Dong[26,221], Jens Dopke[163], Milos Dordevic[147], Tiago dos Santos Ramos[197], Patrick Draper[189], Marco Drewes[84], Mingxuan Du[37], Guang Hua Duan[31], Lars Eklund[164,60], Sarah Eno[196], Jens Erler[138], Rouven Essig[211], Jiji Fan[171], Shenghong Fan[74], Wenjie Fan[74], Xiangning Fan[54], Shuangshi Fang[26], Yaquan Fang[26,60,221], Livio Fano'[124], Addolarata Farilla[126], Riccardo Farinelli[115], Angeles Faus-Golfe[91], Michael A. Fedderke[210,177], Giulietto Felici[110], Changqing Feng[63,221], Cunfeng Feng[46], Feng Feng[12], Jianxin Feng[63], Jun Feng[54], Tai-Fu Feng[21], Roberto Ferrari[123], Oliver Fischer[99], Luis Roberto Flores Castillo[104], Elisa Fontanesi[112], Ayres Freitas[204], M. Frotin[93,89], Claudia Frugiuele[108], Chengdong Fu[26], Jinyu Fu[26,221], Qibin Fu[58], Shinian Fu[26], Elina Fuchs[108], Shigeki Fukuda[130], Emidio Gabrielli[129], Luciano Gaido[128], Pingping Gan[45], Jie Gao[26], Jun Gao[47], Wu Gao[44], Yanyan Gao[165], Yu Gao[26], Yuanning Gao[45], Isabella Garzia[115], Gabriella Gaudio[123], Shao-Feng Ge[131,177], Chao-Qiang Geng[161], Huiping Geng[26], Lisheng Geng[2], Simonetta Gentile[127], Paolo Giacomelli[112], J. Giraud[95], Gian Giudice[156], Rohini M. Godbole[106], Dianjun Gong[26], Guanghua Gong[59], Hui Gong[59], Li Gong[39], Lingling Gong[26], Stefania Gori[176], Quanbu Gou[26], Francesco Grancagnolo[117], Mario Greco[126], Michela Greco[128], Sergei S. Gribanov[144], Sebastian Grinstein[150], D. Grondin[95], Jiayin Gu[97], Limin Gu[37], Pei-Hong Gu[47], Vincenzo Guidi[115], Fangyi Guo[63,26], Jingyuan Guo[57], Jun Guo[47], Lei Guo[14], Xin Guo[53], Yuanyuan Guo[27], Zhigang Guo[81], Ramesh Gupta[220], Chengcheng Han[131], Dejun Han[4], Jinzhong Han[72], Liangliang Han[37], Ran Han[3], Ruixiong Han[26], Shuang Han[65], Yanliang Han[91], Yubo Han[26], Yanfeng Hang[48,47], Jiankui Hao[45], Xiqing Hao[22], Dayong He[26], Hong-Jian He[47,48], Jibo He[60], Jun He[26], Min He[47], Xiang He[26], Xianke He[47], Xiaogang He[48,47], Yangle He[22], Zhenqiang He[26], Zhenqiang He[26], Sven Heinemeyer[151], Yuekun Heng[26,221], Daojin Hong[63], Jiangliu Hong[47], Yang Hong[63,26], YuenKeung Hor[57], J.-Y. Hostachy[95], Qingbo Hou[81], Suen Hou[160], Zhilong Hou[26], Shih-Chieh Hsu[216], Bitao Hu[34], Jifeng Hu[47], Jun Hu[26,221], Shouyang Hu[10], Shuyang Hu[47], Tao Hu[26,221], Yongcai Hu[44], Yu Hu[26], Zhen Hu[184], Zhongjun Hu[30], Chao-Shang Huang[31], Fapeng Huang[133], Guangshun Huang[63,221], Guo-yuan Huang[26], Jinghui Huang[11], Jinshu Huang[40], Junjie Huang[75], Liangsheng Huang[26], Rijun Huang[36], Shuhui Huang[57], Tongming Huang[26], Tuchen Huang[58], Xiaozhong Huang[26], Xingtao Huang[46], Xuguang Huang[17], Yanping Huang[26], Yongsheng Huang[26,221], Yuyan Huang[59], Ran Huo[174], Fedor V. Ignatov[144], Munawar Iqbal[141,143], Paul Jackson[83], Tahir Javaid[26], Ivanka Bozovic Jelisavcic[147], Daheng Ji[26], Qingping Ji[22], Xiaoli Ji[37], Jia Jia[44], Junji Jia[65], Yu Jia[26], Zihang Jia[37], Jiechen Jiang[26], Yun Jiang[88], Jianbin Jiao[46], Dapeng Jin[26,221], Mingjie Jin[26], Shan Jin[37], Song Jin[26], Yanli Jin[26], Yi Jin[61], Maoqiang Jing[26,62], Jaya John[167], Tim Jones[165], Xudong Ju[26], Adil Jueid[47], Sunghoon Jung[134], Susmita Jyotishmati[214], Goran Kacarevic[147], Gordon Kane[199], Wen Kang[26], Muge Karagoz[196], Chikuma Kato[48,47], Zhiyong Ke[26], Dmitri Kharzeev[211],

Valentin Khoze[162], Can Kilic[213], Ryuta Kiuchi[26], Pyungwon Ko[135], Tetsuya Kobayashi[130], Panyu Kong[55], Shibei Kong[26], Joachim Kopp[97], Ashutosh Kotwal[182], Jonathan Kozaczuk[189], Evgeny A. Kozyrev[144], Alexander Krasnov[153], Eric Kuflik[107], Chia-Ming Kuo[158], King Wai Kwok[103], Francois Lagarde[47], Pei-Zhu Lai[158], Imad Laktineh[94], Boyang Lan[55], Xiaofei Lan[13], Lia Lavezzi[128,26], Seung J. Lee[136], Ge Lei[26], Yongbin Leng[28], Sze Ching Leung[204], Bingzhi Li[47], Bo Li[68], Bo Li[26], Boyang Li[59], Changhong Li[69], Cheng Li[63], Congqiao Li[45], Dazhang Li[26], Dikai Li[11], Fei Li[11], Fei Li[26,221], Fei Li[54], Fengyun Li[45], Gang Li[1], Gang Li[26,acc], Gang Li[26,det], Gang Li[159], Gexing Li[26], Guangrui Li[26], Guangrui Li[26], Hai-Bo Li[26], Haifeng Li[46], Haoqing Li[47], Hengne Li[53], Honglei Li[61], Huijing Li[17], Jin Li[26], Jing Li[47], Jinmian Li[52], Jinyan Li[26], Jungang Li[26], Kang Li[19], Ke Li[208,26], Ke Li[26], Li Li[26], Liang Li[47], Lianming Li[54], Long Li[46], Mengran Li[26], Minxian Li[26], Peirong Li[34], Peiyu Li[10], Peng Li[27], Qiang Li[45], Qiaodan Li[37], Quansheng Li[81], Rui Li[57], Shaopeng Li[26], Shiyuan Li[46], Shu Li[48,47], TianJun Li[31], Tong Li[39], Weiguo Li[26], Wenjun Li[22], Xiaoling Li[46], Xiaomei Li[10], Xiaoping Li[26], Xin Li[63,221], Xingguo Li[47], Xin-Qiang Li[6], Yanwei Li[81], Yiming Li[26], Ying Li[68], Ying-Ying Li[103], Yufeng Li[26], Yulan Li[59], Zhao Li[26], Zhenghua Li[81], Zhihui Li[52], Zhongquan Li[26], Ziyuan Li[57], Chaohui Liang[80], Hao Liang[26,221], Jing Liang[26], Jinhan Liang[37], Zhijun Liang[26,221], Zuotang Liang[46], Hean Liao[79], Hongbo Liao[26], Jinfeng Liao[190], Libo Liao[26], Wei Liao[16], Yi Liao[39], Yunfeng Liao[25], Chuangxin Lin[57], Hai Lin[59], Haiying Lin[26], Jiajie Ling[57], Pan Ling[53], Mariangela Lisanti[205], Ao Liu[183], Baiqi Liu[26], Beijiang Liu[26], Bing Liu[47], Bo Liu[26,193], Dianyu Liu[47], Dong Liu[46], Fu-Hu Liu[51], Hongbang Liu[18], Hu Liu[26], Jia Liu[179], Jiaming Liu[26], Jian Liu[4], Jian Liu[46], Jianbei Liu[63,221], Jianfei Liu[28], Jiangtao Liu[26], Jie Liu[60], Jindong Liu[26], Kexin Liu[45], Ling Liu[34], Liqiang Liu[30,60], Ming Liu[73], Ning Liu[36], Peilian Liu[195], Qian Liu[60], Qingyuan Liu[46], Shubin Liu[63,221], shulin Liu[26,221], Tao Liu[103], Weitao Liu[80], Wendi Liu[10], Xiang Liu[34], Xiaohui Liu[4], Xin Liu[32], Xuesong Liu[59], Xuewen Liu[37], Xuyang Liu[26], Yandong Liu[4], Yang Liu[26,60], Yanlin Liu[63], Yi Liu[26], Yong Liu[26], Yudong Liu[26], Zengqiang Liu[81], Zeyuan Liu[60], ZhanFeng Liu[55], Zhaofeng Liu[26], Zhaofeng Liu[168], Zhen Liu[184,196], Zhen Liu[155], Zhenan Liu[26,221], Zhenchao Liu[26], Zhongxiu Liu[26], Zuowei Liu[37], Javier Llorente Merino[26], Cheuk Yee Lo[105], Ivan B. Logashenko[144], Andrew Long[207], Xinchou Lou[26,60,214,221], Ian Low[201,168], Matthew Low[191], Cai-Dian Lu[26], Wei Lu[59], Weiguo Lu[26,221], Yuanrong Lu[45], Yunpeng Lu[26,221], Zhijun Lu[26], Strahinja Lukic[147], Mingcheng Luo[26], Pengwei Luo[57], Qing Luo[63], Tao Luo[26], Xiaofeng Luo[6], Yanting Luo[81], Zhenhuan Luo[11], Kunfeng Lyu[103], Xiaorui Lyu[60], Huizhou Ma[26], Kai Ma[50], Lianliang Ma[46], Na Ma[26], Qiang Ma[26], Rui Ma[53], Wen-Gan Ma[63], Xiao Ma[81], Xiaotian Ma[54], Xinpeng Ma[26], Yanling Ma[81], YanQing Ma[45], Yongsheng Ma[26], Yue Ma[59], Zhongjian Ma[26], Marco Maggiora[128], Frédéric Magniette[93], Ernie Malamud[184], Michelangelo Mangano[156], Lijun Mao[27], Mingling Mao[81], Yajun Mao[45], Yanmin Mao[81], YanYan Mao[81], Adam Martin[202], Verena Ingrid Martinez Outschoorn[197], Shigeki Matsumoto[131], Matthew Mccullough[156], Steve McMahon[163], Patrick Meade[211], Barbara Mele[127], Bruce Mellado[149,148], Lingling Men[26], Cai Meng[26], Fanbo Meng[26], Giulio Mezzadri[115], Zhenghui Mi[26], Paolo Michelato[121], Tianjue Min[37], Lei Ming[37], Monika Mittal[47], James Molson[91], Laura Monaco[121], Guido Montagna[123], Gianfranco Morello[110], Mauro Moretti[115], Zhihui Mu[26], Jérome Nanni[93], Matthias Neubert[97], Oreste Nicrosini[123], Changshan Nie[81], Sergei Nikitin[144], Feipeng Ning[26], Guo-Zhu Ning[21], Nishu Nishu[47], Yazhou Niu[63], Isobel Ojalvo[205], Ivan Okunev[144], Rustem Ospanov[63], Qun Ouyang[26,221], Carlo Pagani[121], Stathes Paganis[159], Marco Panareo[117], Mila Pandurovic[147], Tong Pang[204], Giancarlo Panizzo[129], Rocco Paparella[121], Bibhuti Parida[146], Emilie Passemar[190], Guoxi Pei[26], Shilun Pei[26], Quanling Peng[26], Xiaohua Peng[26], Yuemei Peng[26], Alexey Petrov[218], Lorenzo Pezzotti[123], Fulvio Piccinini[123], Thomas Pierre-Émile[93], Rong-Gang Ping[26], Richard Plackett[167], Giacomo Polesello[123], Marco Poli Lener[110], Alessandro Polini[112], Alexandr S. Popov[144], Soeren Prell[193], Huirong Qi[26,221], Ming Qi[37], Jianming Qian[199], Sen Qian[26,221], Zhuoni Qian[133], Cong-Feng Qiao[60], Yusi Qiao[26], Guang-You Qin[6], Qin Qin[101], Qing Qin[26], Zhonghua Qin[26,221], Huamin Qu[26],

Harikrishnan Ramani[177,195], Michael Ramsey-Musolf[197], Stefano Redaelli[156], Matthew Reece[185], Jing Ren[26], Dirk Rischke[96], Angelo Rivetti[128], Chiara Roda[125], Luigi Rolandi[125], Nikolaos Rompotis[165], Manqi Ruan[26], Xifeng Ruan[149], Richard Ruiz[84,162], GianLuca Sabbi[195], Wen-Long Sang[56], Romualdo Santoro[118], Juan J. Sanz-Cillero[152], Matthias Schlaffer[108], Frank Schmid[156], Philip Schuster[208], Alex Schuy[216], Pedro Schwaller[97], Antonella Sciuto[114], Tanaji Sen[184], Daniele Sertore[121], Peng Sha[26], Lianyou Shan[26], Feng Shang[81], Dingyu Shao[156], Jianxiong Shao[34], Yaroslav Shashkov[145], Jessie Shelton[189], Cheng-Ping Shen[2], Peixun Shen[39], Qiuping Shen[47], Yang Shen[75], Yuqiao Shen[8], Zhongtao Shen[63], Can Shi[55], Haoyu Shi[26], Jingyuan Shi[55], Liaoshan Shi[57], Renjie Shi[81], Xin Shi[26,221], Yu-Ji Shi[47], Yukun Shi[63], Yulong Shi[81], Seodong Shin[137], Ian Shipsey[167], Gary Shiu[219], Guan Shu[26], Jing Shu[31], Zonguo Si[46], Luca Silvestrini[127], Sergey Sinyatkin[144], Hong Song[26], Mao Song[1], Weimin Song[163], Emannuel Stamou[178], Diktys Stratakis[220], Dong Su[208], Feng Su[26], Shufang Su[169], Wanyun Su[48,47], Wei Su[83], Yangjie Su[57], Yanfeng Sui[26], Michael Sullivan[208], Baogeng Sun[178], Daming Sun[70], Guoqiang Sun[81], Hao Sun[15], Junfeng Sun[22], Liang Sun[65], Peng Sun[4], Peng Sun[36], Qingfeng Sun[26], Shengsen Sun[26], Sichun Sun[127], Tingting Sun[11], Xiangming Sun[6], Xianjing Sun[26], Xilei Sun[26,221], Yanjun Sun[43], Yongzhao Sun[26], Raman Sundrum[196], Chuanxiang Tang[59], Guangyi Tang[26], Jian Tang[57], Jiannan Tang[47], Jingyu Tang[26], Songzhi Tang[81], Yi-Lei Tang[135], Jia Tao[26], Junquan Tao[26], Giovanni Tassielli[117], Roberto Tenchini[125], Dmitry Teytelman[181], Jesse Thaler[198], Saike Tian[26], Xingcheng Tian[26,221], Guido Tonelli[125], Xinyu Tong[5], Luca Trentadue[120], Yuhsin Tsai[196], Dmitri Tsybychev[211,146], Yanjun Tu[105], Christopher G. Tully[205], Brock Tweedie[204], James Unwin[188], Monica Verducci[126], Alessandro Vicini[119], Henri Videau[93], Georg Viehhauser[167], Iacopo Vivarelli[166], Joost Vossebeld[165], Natasa Vukasinovic[147], Xia Wan[49], Biao Wang[209], Bin Wang[26], Binlong Wang[60], Bo Wang[55], Chengtao Wang[26], Chengwei Wang[37], Chenliang Wang[47], Dayong Wang[45], Dou Wang[26], Fei Wang[71], Feng Wang[65], Gang Wang[173], Haijing Wang[26], Haiyun Wang[26], Hui Wang[188], Jian Wang[100], Jianchun Wang[26,212], Jianli Wang[26], Jianxiong Wang[26], Jiawei Wang[104], Jie Wang[63], Jin Wang[26], Jing Wang[26], Jinwei Wang[26], Ke Wang[26,221], Kechen Wang[66], Kunfu Wang[76], Liangliang Wang[26], Lian-Tao Wang[178], Lijiao Wang[26], Linlin Wang[26], Longge Wang[81], Lu Wang[81], Meifen Wang[26], Meng Wang[46], Na Wang[26,221], Pengcheng Wang[26], Qun Wang[63], Qunyao Wang[26], Ran Wang[54], Ren-Jie Wang[92], Rongkun Wang[63], Shaobo Wang[47], Shaozhe Wang[75], Shengchang Wang[26], Shuzheng Wang[26], Siguang Wang[45], Tianhong Wang[20], Tong Wang[26], Wei Wang[37], Wei Wang[47], Wei Wang[57], Weiping Wang[63], Wenyu Wang[5], Xi Wang[47], Xiangjian Wang[26], Xiangqi Wang[64], Xiaolong Wang[26], Xiaoning Wang[26], Xiaoping Wang[168], Xiaoping Wang[81], Xin Wang[26], Xin Wang[37], Xingze Wang[9], Xiongfei Wang[26], Yan Wang[26], Yaqian Wang[170], Yi Wang[59], Yifang Wang[26,60,221], Ying Wang[57], Yiwei Wang[26], Yong Wang[63], Youkai Wang[49], Yu Wang[11], Yu Wang[63], Yufeng Wang[92,62], Yuhao Wang[37], Zhaofeng Wang[26], Zhigang Wang[26,221], Zhipeng Wang[54], Zhiyong Wang[26], Zirui Wang[47], Zixun Wang[140], Zongyuan Wang[26], Runing Wei[37], Shujun Wei[60], Wei Wei[26,221], Xiaomin Wei[44], Yuanyuan Wei[26], Yuqian Wei[26], Shuopin Wen[26], Zhiwen Wen[27], Stephane Willocq[197], Holger Witte[220], Jinfei Wu[26], Juan Wu[11], Kewei Wu[26], Lei Wu[36], Linghui Wu[26], Mengqing Wu[26], Peiwen Wu[135], Tianya Wu[150,6], Wenhuan Wu[26,221], Xi Wu[26], Xiao-Hong Wu[16], Xing-Gang Wu[14], Xu Wu[54], Xueting Wu[26], Ye Wu[26], Yuwen Wu[26], Zhi Wu[26,221], Zhigang Wu[26], Wenhao Xia[26], Dao Xiang[47], Qian-Fei Xiang[45], Zhong-Zhi Xianyu[185], Bo-Wen Xiao[6], Dengjie Xiao[26], Ning Xiao[54], Ouzheng Xiao[26], Rui-Qing Xiao[47], Yu Xiao[81], Zhen-Jun Xiao[36], Ke-Pan Xie[134], Xinhai Xie[57], Yuehong Xie[6], Yuguang Xie[26,221], Zongtai Xie[26], Qingzhi Xing[59], Zhizhong Xing[26], Qinglei Xiu[26], Chang Xu[9], Da Xu[26,60], Fanrong Xu[33], Guanglei Xu[26], Guangzhi Xu[35], Haocheng Xu[37], Hongge Xu[11], Hongliang Xu[64], Hui Xu[37], Ji Xu[47], Nu Xu[6], Qing Xu[81], Qingjin Xu[26], Qingjun Xu[19], Wei Xu[26], Wei Xu[64], Yin Xu[39], Yongheng Xu[37], Zijun Xu[208], Wei Xue[156], Bin Yan[200], Jun Yan[47], Liang Yan[128], Mingyang Yan[26], Qi-Shu Yan[60], Tian Yan[26], Wenbiao Yan[63,221], Yingbing Yan[28], Bingfang Yang[22],

Haijun Yang[47,48], Huan Yang[26], Jiancheng Yang[27], Jianquan Yang[26], Jin Min Yang[31], Jing Yang[81], Junfeng Yang[63,221], Li Yang[81], Liu Yang[65], Mei Yang[26], Ping Yang[6], Qianwen Yang[42], Xingwang Yang[26], Xuan Yang[46], Ye Yang[59], Ying Yang[26], Yong Yang[47], Yongliang Yang[63], Yueling Yang[22], Zhenwei Yang[59], Weichao Yao[26], Wei-Ming Yao[195], Elena Yatsenko[47], Hanfei Ye[37], Jingbo Ye[209], Mei Ye[26,221], Qiang Ye[26], Ziping Ye[186], Fang Yi[26,60], Kai Yi[192], Xue Yilun[37], Hang Yin[6], Pengfei Yin[26], Xiangwei Yin[26], Ze Yin[81], Zhongbao Yin[6], Zhengyun You[57], Charles Young[208], Boxiang Yu[26,221], Chenghui Yu[26], Chunxu Yu[39], Dan Yu[26], Felix Yu[97], Fusheng Yu[34], Lingda Yu[26], Lu Yu[26], Yue Yu[59], Zhao-Huan Yu[57], Li Yuan[2], Siyu Yuan[73], Ye Yuan[26], Youjin Yuan[27], Zhiyang Yuan[26], Chongxing Yue[35], Junhui Yue[26], Qian Yue[59], Un-Nisa Zaib[26], Jian Zhai[81], Jiyuan Zhai[26], Baotang Zhang[26], Ben-Wei Zhang[6], Bo Zhang[63], Bowen Zhang[37], Cen Zhang[26], Chad Zhang[26], Chunlei Zhang[4], Di Zhang[26], Gang Zhang[59], Guangyi Zhang[22], Guoqing Zhang[26], Hai-Bin Zhang[21], Hao Zhang[26], Honghao Zhang[57], Hongyu Zhang[26,221], Huaqiao Zhang[26], Hui Zhang[81], Jian Zhang[26,221], Jian Zhang[59], Jianhui Zhang[102], Jianqin Zhang[26], Jielei Zhang[67], Jingru Zhang[26], Jinlong Zhang[168], Junrui Zhang[81], Kaili Zhang[26], Lei Zhang[37], Liang Zhang[46], Liming Zhang[59], Linhao Zhang[26], Ren-You Zhang[63,221], Rui Zhang[25], Sifan Zhang[37], Tianjiao Zhang[47], Wenchao Zhang[49], Xiangke Zhang[47], Xiaohui Zhang[81], Xinmin Zhang[26], Xinying Zhang[26], Xueyao Zhang[46], Yang Zhang[82], Yao Zhang[26], Yi Zhang[34], Ying Zhang[26,221], Yongchao Zhang[217], Yu Zhang[1], Yuan Zhang[26], Yuhong Zhang[194], Yu-Jie Zhang[2], Yulei Zhang[47], Yulian Zhang[38], Yumei Zhang[58], Yunlong Zhang[63,221], Yuxuan Zhang[213], Zhaoru Zhang[26], Zhen Zhang[11], Zhenyu Zhang[65], Zhiqing Zhang[91], Zhiyong Zhang[63], Chen Zhao[44], Hang Zhao[26], Jingxia Zhao[26], Jingyi Zhao[26], Ling Zhao[26], Mei Zhao[26,221], Minggang Zhao[39], Mingrui Zhao[10], Peng Zhao[7], Qiang Zhao[41], Shensen Zhao[63], Shuai Zhao[47], Shu-Min Zhao[21], Tongxian Zhao[26], Wei Zhao[26], Xianghu Zhao[26], Xiaoran Zhao[84], Xiaoyan Zhao[26], Ying Zhao[26], Yu Zhao[79], Yue Zhao[215], Zhengguo Zhao[63,221], Zhen-Xing Zhao[47], Zhuo Zhao[26], Bo Zheng[62], Hongjuan Zheng[26], Liang Zheng[11], Ran Zheng[44], Shuxin Zheng[59], Taifan Zheng[37], Xuxing Zheng[26], Ya-Juan Zheng[132], Yangheng Zheng[60], Yu Zhi[10], Bin Zhong[36], Yiming Zhong[170], Bing Zhou[199], Hai-Qing Zhou[55], Hang Zhou[36], Jia Zhou[123], Jianxin Zhou[26], Jing Zhou[10], Maosen Zhou[26], Nan Zhou[77], Ning Zhou[47], Ningchuang Zhou[26], Shiyu Zhou[59], Shun Zhou[26], Sihong Zhou[29], Siyi Zhou[103], Xiang Zhou[65], Yang Zhou[26,221], Yi Zhou[63], Yu-Feng Zhou[31], Zusheng Zhou[26], Changhe Zhu[76], Chengguang Zhu[46], Chenzheng Zhu[26,60], Dechong Zhu[26], Hongbo Zhu[26,221], Hongyan Zhu[26], Hua-Xing Zhu[70], Jiamin Zhu[78], Jiang Zhu[57], Jingya Zhu[65], Jingyu Zhu[26], Junjie Zhu[199], Kai Zhu[26], Kejun Zhu[26,221], Kun Zhu[45], Li Zhu[81], Ruilin Zhu[36], Xianglei Zhu[59], Xuezheng Zhu[26], Yifan Zhu[47], Yingshun Zhu[26], Yongfeng Zhu[26], Zian Zhu[26], Xuai Zhuang[26], Hongshi Zong[37], Cong Zou[81], Jiaheng Zou[26], Ye Zou[26], Jure Zupan[180].

[1] AnHui University, Hefei, Anhui.
[2] Beihang University, Beijing.
[3] Beijing Institute of Spacecraft Environment Engineering (BISEE), Beijing.
[4] Beijing Normal University, Beijing.
[5] Beijing University of Technology, Beijing.
[6] Central China Normal University, Wuhan, Hubei.
[7] Chang'an University, Xi'an, Shaanxi.
[8] Changzhou Institute of Technology, Changzhou, Jiangsu.
[9] China Academy of Space Technology, Beijing.
[10] China Institute of Atomic Energy, Beijing.
[11] School of Mathematics and Physics, China University of Geosciences, Wuhan, Hubei.
[12] China University of Mining and Technology, Beijing.
[13] China West Normal University, Nanchong, Sichuan.
[14] Department of Physics, Chongqing University, Chongqing.

[15] Department of Physics, Dalian University of Technology, Dalian, Liaoning.
[16] East China University of Science and Technology, Shanghai.
[17] Department of Physics, Fudan University, Shanghai.
[18] Guangxi University, Nanning, Guangxi.
[19] Department of Physics, Hangzhou Normal University, Hangzhou, Zhejiang.
[20] Harbin Institute of Technology, Haerbin, Heilongjiang.
[21] Department of Physics, Hebei University, Baoding, Hebei.
[22] Henan Normal University, Xinxiang, Henan.
[23] Huazhong University of Science and Technology, Wuhan, Hubei.
[24] School of Physics and Electronics, Hunan University, Changsha, Hunan.
[25] Institute of Electronics, Chinese Academy Of Sciences, Beijing.
[26] Institute of High Energy Physics, Chinese Academy of Sciences, Beijing.
[27] Institute of Modern Physics, Chinese Academy Of Sciences, Lanzhou, Gansu.
[28] Shanghai Institute of Applied Physics, Chinese Academy Of Sciences, Shanghai.
[29] Inner Mongolia University, Hohhot, Neimenggu.
[30] Technical Institute of Physics and Chemistry, Chinese Academy of Sciences, Beijing.
[31] CAS Key Laboratory of Theoretical Physics, Institute of Theoretical Physics, Chinese Academy of Sciences, Beijing.
[32] Department of Physics, Jiangsu Normal University, Xuzhou, Jiangsu.
[33] Department of Physics, Jinan University, Guangzhou, Guangdong.
[34] Lanzhou University, Lanzhou, Gansu.
[35] Department of Physics, Liaoning Normal University, Dalian, Liaoning.
[36] School of Physics and Technology, Nanjing Normal University, Nanjing, Jiangsu.
[37] Department of Physics, Nanjing University, Nanjing, Jiangsu.
[38] Nanjing University of Aeronautics and Astronautic, Nanjing, Jiangsu.
[39] Nankai University, Tianjin.
[40] Nanyang Normal University, Nanyang, Henan.
[41] North China Electric Power University, Beijing.
[42] Northeast Normal University, Changchun Jilin.
[43] Northwest Normal University, Lanzhou, Gansu.
[44] Northwesten Polytechnical University, Xi'an, Shaanxi.
[45] School of Physics, Peking University, Beijing.
[46] Institute of Frontier and Interdisciplinary Science and Key Laboratory of Particle Physics and Particle Irradiation, Shandong University, Qingdao, Shandong.
[47] School of Physics and Astronomy, Shanghai Jiao Tong University, Shanghai.
[48] Tsung-Dao Lee Institute, Shanghai Jiao Tong University, Shanghai.
[49] Shannxi Normal University, Xi'an, Shaanxi.
[50] Shannxi University of Technology, Hanzhong, Shaanxi.
[51] Shanxi University, Taiyuan, Shanxi.
[52] Sichuan University, Chengdu, Sichuan.
[53] South China Normal University, Guangzhou, Guangdong.
[54] School of Information Science and Engineering, Southeast University, Nanjing, Jiangsu.
[55] School of Physics, Southeast University, Nanjing, Jiangsu.
[56] School of Physical Science and Technology, Southwest University, Chongqing.
[57] School of Physics, Sun Yat-sen University, Guangzhou, Guangdong.
[58] Sino-French Institute of Nuclear Engineering and Technology, Sun Yat-sen University, Guangzhou, Guangdong.
[59] Center for High Energy Physics, Tsinghua University, Beijing.
[60] University of Chinese Academy of Sciences, Beijing.
[61] School of Physics and technology, University of Jinan, Jinan, Shandong.
[62] University of South China, Hengyang, Hunan.
[63] Department of Modern Physics, University of Science and Technology of China, Hefei, Anhui.
[64] Synchrotron Radiation Lab, University of Science and Technology of China, Hefei, Anhui.
[65] School of Physics and Technology, Wuhan University, Wuhan, Hubei.
[66] Department of Physics, School of Science, Wuhan University of Technology, Wuhan, Hubei.
[67] Xinyang Normal University, Xinyang, Henan.

[68] Department of Physics, Yantai University, Yantai, Shandong.
[69] Yunnan University, Kunming, Yunnan.
[70] Zhejiang University, Hangzhou, Zhejiang.
[71] Zhengzhou University, Zhengzhou, Henan.
[72] Zhoukou Normal Institute, Zhoukou, Henan.
[73] Angang Steel Company Limited, Anshan, Liaoning.
[74] Beijing Prodetec Technoloy Co.,Ltd, Beijing.
[75] GLVAC Industrial Technology Research Institute of High Power Devices, Kunshan, Jiangsu.
[76] Hanzhong Yuanhang Precision Machinery Co.,Ltd, Hanzhong, Shaanxi.
[77] Luvata Tube(Zhongshan)Ltd., Zhangshan, Guangdong.
[78] Shanghai Superconductor Technology Co., Ltd., Shanghai.
[79] Wuxi Toly Electric Works Co., Ltd., Wuxi, Jiangsu.
[80] Xi'an Superconducting Magnet Technology Co.,Ltd., Xi'an, Shaanxi.
[81] Yellow River Engineering Consulting Co., Ltd., Zhengzhou, Henan.
[82] Monash University, Melbourne.
[83] University of Adelaide, Adelaide, South Australia.
[84] Université catholique de Louvain, Louvain-la-Neuve.
[85] Perimeter Institute for Theoretical Physics, Waterloo, Ontario.
[86] University of Alberta,Edmonton, Alberta.
[87] University of Toronto, Toronto, Ontario.
[88] Niels Bohr Institute, University of Copenhagen, Copenhagen.
[89] GEPI, Meudon.
[90] IRFU, CEA, Universite Paris-Saclay, Paris.
[91] Laboratoire de l'Accélérateur Linéaire, Université Paris-Sud, CNRS/IN2P3, Université Paris-Saclay, Orsay.
[92] Laboratoire de Physique Nucléaire et de Hautes Energies (LPNHE), Sorbonne Université, Paris-Diderot Sorbonne Paris Cité, CNRS/IN2P3.
[93] Laboratoire Leprince-Ringuet (LLR), École polytechnique, CNRS-IN2P3.
[94] Lyon 1 University, IPNL.
[95] Univ. Grenoble Alpes, CNRS, Grenoble INP, LPSC-IN2P3, Grenoble.
[96] Frankfurt University, Frankfurt.
[97] Johannes Gutenberg University of Mainz, Mainz, Rhineland Palatinate.
[98] Julius-Maximilian-Universität Würzburg, Institut für Theoretische Physik und Astrophysik, Würzburg.
[99] Karlsruhe Institute of Technology, Karlsruhe, Baden-Württemberg.
[100] Technical University of Munich, Munich, Bavaria.
[101] Theoretische Physik 1, Naturwissenschaftlich-Technische Fakultät, Universität Siegen, Siegen.
[102] University of Regensburg, Regensburg, Bavaria.
[103] Department of Physics, Hong Kong University of Science and Technology.
[104] The Chinese University of Hong Kong.
[105] The University of Hong Kong.
[106] Indian Institute of Science, Bangalore, Karnataka.
[107] Hebrew University, Jerusalem.
[108] Weizmann Institute of Science, Rehovot.
[109] INFN - Gran Sasso Science Institute, GSSI.
[110] INFN - Laboratori Nazionali di Frascati.
[111] INFN - Sezione di Bari, University and Politecnico of Bari.
[112] INFN - Sezione di Bologna and University of Bologna.
[113] INFN - Sezione di Cagliari.
[114] INFN - Sezione di Catania and University of Catania.
[115] INFN - Sezione di Ferrara and University of Ferrara.
[116] INFN - Sezione di Firenze and University of Firenze.
[117] INFN - Sezione di Lecce and University of Lecce.
[118] INFN - Sezione di Milano and University of Insubria.
[119] INFN - Sezione di Milano and University of Milano.
[120] INFN - Sezione di Milano Bicocca and University of Parma.
[121] INFN - Sezione di Milano, Laboratoro LASA and University of Milano.

[122] INFN - Sezione di Padova and University of Padova.
[123] INFN - Sezione di Pavia and University of Pavia.
[124] INFN - Sezione di Perugia and University of Perugia.
[125] INFN - Sezione di Pisa, Universita' di Pisa and Scuola Normale Superiore.
[126] INFN - Sezione di Roma Tre and University of Roma Tre.
[127] INFN - Sezione di Roma1 and University of Roma "La Sapienza".
[128] INFN - Sezione di Torino and University of Torino.
[129] INFN - Sezione di Trieste and University of Udine.
[130] High Energy Accelerator Research Organization (KEK), Tsukuba, Ibaraki.
[131] Kavli IPMU (WPI), UTIAS, The University of Tokyo, Kashiwa, Chiba.
[132] Osaka University, Osaka.
[133] Center for Theoretical Physics of the Universe, Institute for Basic Science (IBS), Daejeon.
[134] Center for Theoretical Physics, Department of Physics and Astronomy, Seoul National University, Seoul.
[135] Korea Institute for Advanced Study, Seoul.
[136] Korea University, Seoul.
[137] Yonsei University, Seoul.
[138] National Autonomous University of Mexico, Mexico City.
[139] University Abdelmalek Essaadi, Faculty of sciences and techniques, Tangier.
[140] University of Groningen, Groningen.
[141] Pakistan Atomic Energy Commission, Islamabad.
[142] Pakistan Institute of Nuclear Science and Technology, Islamabad.
[143] CHEP, University of the Punjab, Quaid-e-Azam Campus, Lahore.
[144] Budker Institute of Nuclear Physics, Novosibirsk State University, Novosibirsk.
[145] National Research Nuclear University MEPhI, Moscow.
[146] Tomsk State University, Siberia.
[147] Vinca Institute of Nuclear Sciences, University of Belgrade, Belgrade.
[148] iThemba LABS, National Research Foundation, Somerset West.
[149] School of Physics and Institute for Collider Particle Physics, University of the Witwatersrand, Johannesburg, Wits.
[150] Institut de Fisica d'Altes Energies, Barcelona.
[151] Instituto de Fisica Teorica UAM-CSIC, Madrid.
[152] Universidad Complutense de Madrid, Madrid.
[153] Uppsala University, Uppsala.
[154] Basel University, Basel.
[155] DPNC, University of Geneva, Geneva.
[156] CERN, Geneva.
[157] Chung Yuan Christian University, Taoyuan.
[158] Department of Physics and Center for High Energy and High Field Physics, National Central University, Taoyuan.
[159] Department of Physics, National Taiwan University, Taipei.
[160] Institute of Physics, Academia Sinica, Taipei.
[161] National Tsing-Hua University, Hsinchu.
[162] IPPP, Durham University, Durham.
[163] Particle Physics Department, Rutherford Appleton Laboratory, Didcot.
[164] University of Glasgow, Glasgow, Scotland.
[165] Department of Physics, University of Liverpool, Liverpool.
[166] University of Sussex, Falmer, East Sussex.
[167] Department of Physics, University of Oxford, Oxford.
[168] High Energy Physics Division, Argonne National Laboratory, Argonne, IL.
[169] Department of Physics, University of Arizona, Tucson, AZ.
[170] Physics Department, Boston University, Boston, MA.
[171] Brown University, Providence, RI.
[172] University of California Davis, Davis, CA.
[173] University of California Los Angeles, Los Angeles, CA.
[174] University of California Riverside, Riverside, CA.

[175] University of California Santa Barbara, Santa Barbara, CA.
[176] University of California Santa Cruz, Santa Cruz, CA.
[177] Berkeley Center for Theoretical Physics, Department of Physics,University of California Berkeley, Berkeley, CA.
[178] Department of Physics, University of Chicago, Chicago, IL.
[179] Enrico Fermi Institute, University of Chicago, Chicago, IL.
[180] University of Cincinnati, Cincinnati, OH.
[181] Dimtel, Inc., San Jose, CA.
[182] Duke University, Durham, NC.
[183] Euclid Techlabs, Solon, OH.
[184] Fermi National Accelerator Laboratory, Batavia, IL.
[185] Harvard University, Cambridge, MA.
[186] Department of Physics, University of Houston, Houston, TX.
[187] Illinois Institute of Technology, Chicago, IL.
[188] University of Illinois at Chicago, Chicago, IL.
[189] University of Illinois at Urbana-Champaign, Urbana and Champaign, IL.
[190] Department of Physics, Indiana University, Bloomington, IN.
[191] Institute for Advanced Study, Princeton, NJ.
[192] University of Iowa, Iowa City, IA.
[193] Department of Physics and Astronomy, Iowa State University, Ames, IA.
[194] Thomas Jefferson National Accelerator Facility, Newport News, VA.
[195] Lawrence Berkeley National Labotatory, Berkeley, CA.
[196] University of Maryland, College Park, MD.
[197] University of Massachusetts Amherst, Amherst, MA.
[198] Massachusetts Institute of Technology, Cambridge, MA.
[199] Department of Physics, University of Michigan, Ann Arbor, MI.
[200] Michigan State University, East Lansing, MI.
[201] Northwestern University, Evanston, IL.
[202] Department of Physics, University of Notre Dame du Lac, Notre Dame, IN.
[203] University of Oregon, Eugene, OR.
[204] Department of Physics and Astronomy, University of Pittsburgh, Pittsburgh, PA.
[205] Princeton University, Princeton, NJ.
[206] Purdue University Northwest, Westville, IN.
[207] Rice University, Houston, TX.
[208] SLAC National Accelerator Laboratory, Menlo Park, CA.
[209] Department of Physics, Southern Methodist University, Dallas, TX.
[210] Stanford Institute for Theoretical Physics, Department of Physics, Stanford University, Stanford, CA.
[211] Stony Brook University, Stony Brook, NY.
[212] Department of Physics, Syracuse University, Syracuse, NY.
[213] University of Texas at Austin, Austin, TX.
[214] University of Texas at Dallas, Richardson, TX.
[215] University of Utah, Salt Lake City, UT.
[216] Department of Physics, University of Washington, Seattle, WA.
[217] Department of Physics and McDonnell Center for the Space Sciences, Washington University in St. Louis, St. Louis, MO.
[218] Wayne State University, Detroit, MI.
[219] University of Wisconsin-Madison, Madison, WI.
[220] Brookhaven National Laboratory, Upton, Suffolk County, NY.
[221] State Key Laboratory of Particle Detection and Electronics, Beijing and Hefei.